%% file: thesis.tex
\documentclass[12pt,twoside,a4paper,openright]{report}

\usepackage[utf8]{inputenc}
\usepackage{graphicx}
\usepackage{amsmath,amssymb,amsfonts,braket}
\usepackage{mathtools}
\usepackage[top=3cm, bottom=2.8cm, left=2.8cm, right=2.8cm]{geometry}
\usepackage{makeidx}
\usepackage[justification=centering]{caption}
\usepackage{subcaption}
\usepackage{float}
\usepackage{cancel}
\usepackage[hidelinks]{hyperref}
\usepackage{fancyhdr}
\usepackage{minitoc}
\usepackage{cite}
\usepackage[labelsep=endash,format=plain]{caption}
\usepackage{verbatim}
\usepackage{pdfpages}

\hypersetup{
    colorlinks,
    citecolor=blue,
    filecolor=black,
    linkcolor=blue,
    urlcolor=blue
}

\fancypagestyle{plain}{
 \fancyhf{} 
 \fancyhead[OR,EL]{\sffamily
 \bfseries \nouppercase \thepage} 
  }
\usepackage[utf8]{inputenc}
\usepackage[english]{babel}
\usepackage{graphicx}
\usepackage{color}

\usepackage{rotating}
\usepackage{float}
\usepackage{enumitem}
\usepackage{appendix}
\usepackage{geometry}
\usepackage{cite}
\usepackage{slashed}
\usepackage[labelfont=bf]{caption}
\usepackage{breqn}
\usepackage{booktabs}
\usepackage{makeidx}
\usepackage{lmodern}
\usepackage{slashed}

\usepackage{pgf}
\usepackage{multirow}
\usepackage[absolute]{textpos} 
\usepackage{multicol} 
\usepackage{calc} 
\usepackage{tikz}

\label{The thesis}
\newcommand{\PhDTitleFR}{Dark Matter phenomenology : from simplified WIMP models to refined alternative solutions}

\renewcommand{\d}[1]{\ensuremath{\operatorname{d}\!{#1}}}
\newcommand*\diff{\mathop{}\!\mathrm{d}}
\newcommand{\bea}{\begin{eqnarray}}
\newcommand{\eea}{\end{eqnarray}}
\newcommand{\beq}{\begin{eqnarray}}
\newcommand{\eeq}{\end{eqnarray}}

\newcommand{\be}{\begin{equation}}
\newcommand{\ee}{\end{equation}}

\newcommand{\wt}[1]{\widetilde{#1}}
\newcommand{\pmatr}[1]{\begin{pmatrix} #1 \end{pmatrix}}
\newcommand{\overbar}[1]{\mkern 1.5mu\overline{\mkern-1.5mu#1\mkern-1.5mu}\mkern 1.5mu}

\newcommand{\appropto}{\mathrel{\vcenter{
  \offinterlineskip\halign{\hfil$##$\cr
    \propto\cr\noalign{\kern2pt}\sim\cr\noalign{\kern-2pt}}}}}

\def\gsm{g_{\text{\tiny{SM}}}}
\def\gdm{g_{\text{\tiny{DM}}}}

\def\mh{m_{\tilde{h}}}
\def\Gh{\Gamma_{\tilde h}}

\def\to{\rightarrow}

\def\ra{\rangle}
\def\la{\langle}

 \label{Personal}
\newcommand{\PhDname}{M. Mathias Pierre} 
\newcommand{\NNT}{2018SACLS238} 
\newcommand{\ecodocnum}{576} 
\newcommand{\ecodoctitle}{Particules Hadrons Énergie et Noyau : Instrumentation, Image, Cosmos et Simulation} 
\newcommand{\PhDspeciality}{Physique des particules} 
\newcommand{\PhDworkingplace}{\`{a} l'Universit\'{e} Paris-Sud} 
\newcommand{\defenseplace}{Orsay} 
\newcommand{\defensedate}{25 septembre 2018}

\label{Jury}
\newcommand{\jurynameA}{Geneviève Bélanger}
\newcommand{\jurygenderA}{Mme.} 
\newcommand{\juryadressA}{\it LAPTh Annecy}
\newcommand{\jurygradeA}{Professeur}
\newcommand{\juryroleA}{Rapporteur}
\newcommand{\jurynameB}{Michel Tytgat}
\newcommand{\jurygenderB}{M.} 
\newcommand{\juryadressB}{\it ULB Bruxelles}
\newcommand{\jurygradeB}{Professeur}
\newcommand{\juryroleB}{Rapporteur}
\newcommand{\jurynameC}{Emilian Dudas}
\newcommand{\jurygenderC}{M.} 
\newcommand{\juryadressC}{\it CPHT Polytechnique}
\newcommand{\jurygradeC}{Professeur}
\newcommand{\juryroleC}{Examinateur}
\newcommand{\jurynameD}{Carlos Mu\~{n}oz}
\newcommand{\jurygenderD}{M.} 
\newcommand{\juryadressD}{\it IFT-UAM Madrid}
\newcommand{\jurygradeD}{Professeur}
\newcommand{\juryroleD}{Examinateur}
\newcommand{\jurynameE}{Keith Olive}
\newcommand{\jurygenderE}{M.} 
\newcommand{\juryadressE}{\it University of Minnesota}
\newcommand{\jurygradeE}{Professeur}
\newcommand{\juryroleE}{Examinateur}
\newcommand{\jurynameF}{Ulrich Ellwanger}
\newcommand{\jurygenderF}{M.}
\newcommand{\juryadressF}{\it LPT Orsay}
\newcommand{\jurygradeF}{Professeur}
\newcommand{\juryroleF}{Président du jury}
\newcommand{\jurynameG}{Yann Mambrini}
\newcommand{\jurygenderG}{M.}

\newcommand{\jurygradeG}{Directeur de recherche}
\newcommand{\juryroleG}{Directeur de th\`ese}

\newcommand*\xbar[1]{%
   \hbox{%
     \vbox{%
       \hrule height 0.5pt %
       \kern0.5ex
       \hbox{%
         \kern-0.05em
         \ensuremath{#1}%
         \kern-0.05em
       }%
     }%
   }%
}

\label{Document}

\begin{document}
\renewcommand{\arraystretch}{1.3}
\setcounter{tocdepth}{1}

\begingroup
\pagenumbering{gobble}
\thispagestyle{empty}
\fontsize{12pt}{14pt}\selectfont
\newgeometry{textheight=150ex,textwidth=40em,top=30pt,headheight=30pt,headsep=30pt,inner=80pt,left=2.8cm,right=2.8cm}
\centering
\input{style-pagedegarde} 
\restoregeometry
\endgroup

\newpage\null\thispagestyle{empty}\newpage

\dominitoc
\pagenumbering{roman}

\chapter*{Acknowledgements}
\addcontentsline{toc}{part}{Acknowledgements}
\input{parts/acknowledgments}

\chapter*{Résumé en français}
\addcontentsline{toc}{part}{Résumé en français}
\input{parts/resume}

{\hypersetup{linkcolor=black}
\pdfbookmark[0]{\contentsname}{toc}
\tableofcontents}

\newpage 
\chapter*{Introduction}
\addcontentsline{toc}{part}{Introduction}
\input{parts/introduction}

\cleardoublepage
\pagenumbering{arabic}

\part{The Dark Matter problem}

\chapter{Modern Cosmology}
{\hypersetup{linkcolor=black}
\pdfbookmark[0]{\contentsname}{toc}
\minitoc}

\input{parts/cosmo.tex}

\chapter{From early evidences to recent observations}
{\hypersetup{linkcolor=black}
\pdfbookmark[0]{\contentsname}{toc}
\minitoc}
\input{parts/evidences.tex}

\chapter{The particle hypothesis}
{\hypersetup{linkcolor=black}
\pdfbookmark[0]{\contentsname}{toc}
\minitoc}
\input{parts/particleDM.tex}

\input{parts/alternative.tex}

\chapter{Current status of Dark Matter searches}
{\hypersetup{linkcolor=black}
\pdfbookmark[0]{\contentsname}{toc}
\minitoc}
\input{parts/DMsearches.tex}

\part{Towards realistic WIMP models}

\chapter{A first approach: simplified models}
{\hypersetup{linkcolor=black}
\pdfbookmark[0]{\contentsname}{toc}
\minitoc}
\input{parts/simplifiedmodels.tex}

\chapter{$Z^\prime$ portal through Chern-Simons interaction}
{\hypersetup{linkcolor=black}
\pdfbookmark[0]{\contentsname}{toc}
\minitoc}
\input{parts/chernsimons.tex}

\chapter{Flavourful $Z'$ portal for Dark Matter and $R_{K^{(*)}}$}
{\hypersetup{linkcolor=black}
\pdfbookmark[0]{\contentsname}{toc}
\minitoc}
\input{parts/flavor.tex}

\part{Alternative Dark Matter production beyond the WIMP paradigm}
\chapter{Vector Strongly Interacting Massive Particles}
{\hypersetup{linkcolor=black}
\pdfbookmark[0]{\contentsname}{toc}
\minitoc}
\input{parts/VSIMP.tex}

\chapter{Spin-2 portal Dark Matter}
{\hypersetup{linkcolor=black}
\pdfbookmark[0]{\contentsname}{toc}
\minitoc}
\input{parts/spin2.tex}

\chapter*{Conclusion}
\addcontentsline{toc}{part}{Conclusion}
\input{parts/conclusion}
\addcontentsline{toc}{part}{Appendices}
\appendix
\chapter{The Boltzmann equation}
{\hypersetup{linkcolor=black}
\pdfbookmark[0]{\contentsname}{toc}
\minitoc}
\input{parts/appendices/boltzmann.tex}
\chapter{Chern-Simons couplings}
{\hypersetup{linkcolor=black}
\pdfbookmark[0]{\contentsname}{toc}
\minitoc}
\input{parts/appendices/computation_chernsimons.tex}

\chapter{Spin-2 portal: amplitudes and rates}
{\hypersetup{linkcolor=black}
\pdfbookmark[0]{\contentsname}{toc}
\minitoc}
\input{parts/appendices/spin2appendix.tex}

\addcontentsline{toc}{part}{References}
\bibliographystyle{jhep}
\bibliography{biblio}

\end{document}

%% file: style-pagedegarde.tex

\begin{tikzpicture}[remember picture,overlay,color=purple!85!black]
	\draw[ultra thick]
		([yshift=-1cm,xshift=1cm]current page.north west)--     
		([yshift=-1cm,xshift=-1cm]current page.north east)--    
		([yshift=1cm,xshift=-1cm]current page.south east)--      
		([yshift=1cm,xshift=1cm]current page.south west)--cycle; 
\end{tikzpicture}

\begin{textblock}{15}(1.6,2.2)
 \flushleft NNT : \NNT
\end{textblock}

\begin{textblock}{1}(1.65,1)
\includegraphics[height=1.6cm]{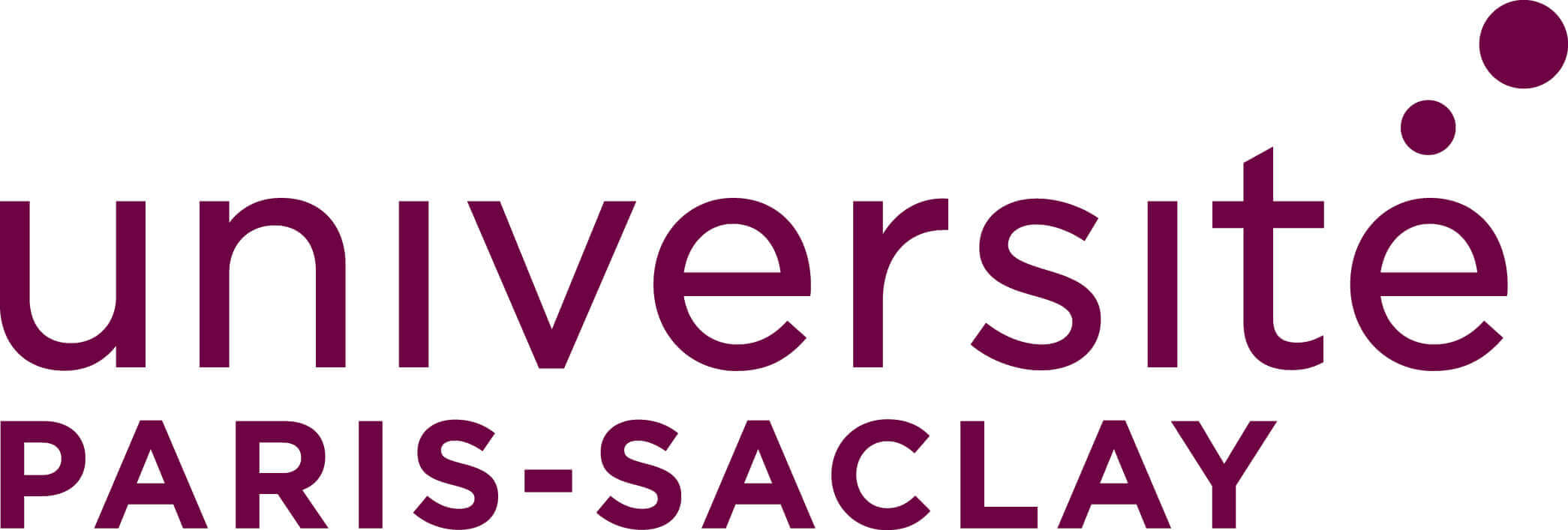} 
\end{textblock}

\begin{textblock}{1}(7,1)
\includegraphics[height=2.4cm]{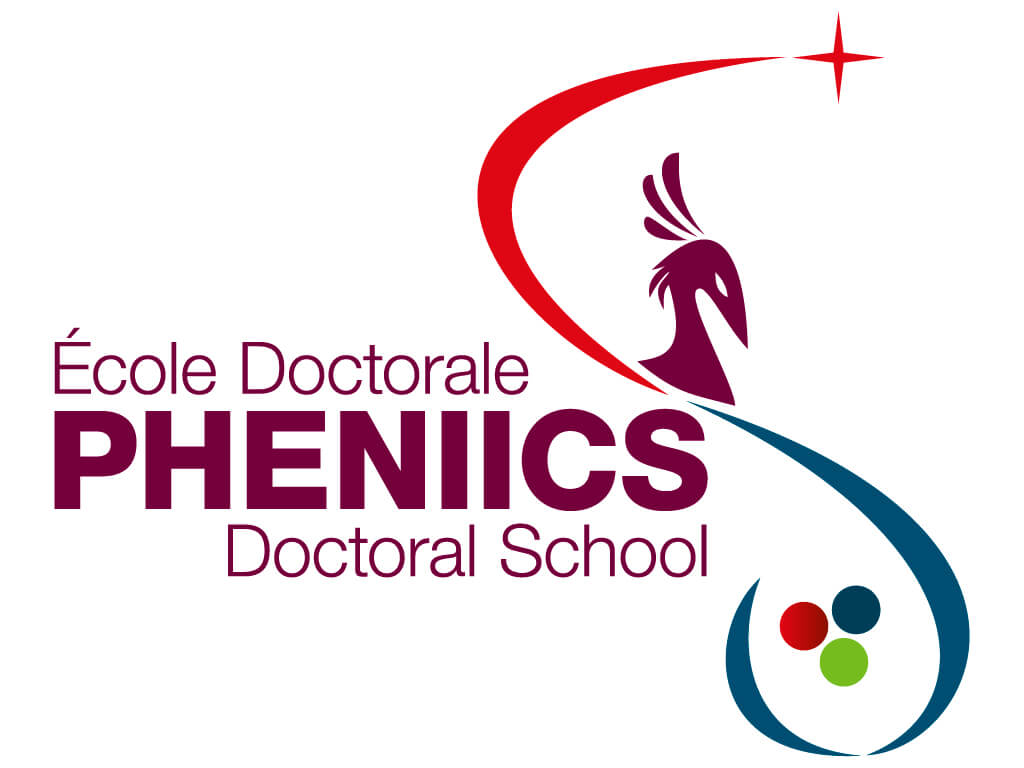} 
\end{textblock}

\begin{textblock}{1}(11,0.9)
\includegraphics[height=3.2cm]{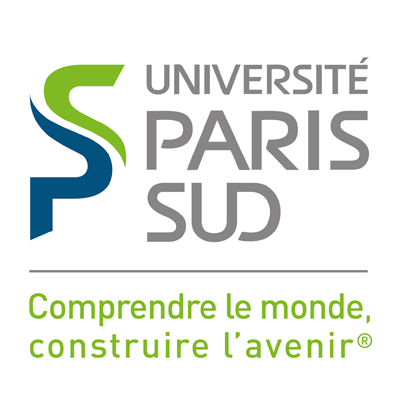} 
\end{textblock}

\vspace{3.4cm}
\color{purple!70!black} 
  \begin{center}    
    \Large\textsc{Th\`ese de doctorat\\ de l'Universit\'e Paris-Saclay} \\
    \Large{\textsc{pr\'epar\'ee \PhDworkingplace}} \\     \vspace{0.5cm}  
  \color{black} 
	\vfill
    \large{\'Ecole doctorale n$^{\circ}\ecodocnum$}\\ 
     \large{\ecodoctitle}  \\

     \large{Sp\'ecialit\'e de doctorat: \PhDspeciality\\} 
    \vspace{0.5cm}  
   \Large{par\\}
	\vfill
   \Large{\textbf{\PhDname}\\} 
    \vspace{1cm}  
    \Large{\textbf{\color{purple!70!black} 
\PhDTitleFR}\color{black}} 
    \vspace{1cm}  
\end{center}
\color{black}
\begin{flushleft}
\textit{Th\`ese pr\'esent\'ee et soutenue \`a \defenseplace, le \defensedate. \\}
\bigskip
Composition du Jury :
\end{flushleft}

\begin{center}
\begin{tabular}{llll}

    \jurygenderA & \textsc{\jurynameA}  & \jurygradeA & (\juryroleA) \\
    \null & \null & \juryadressA &\\   
   
    \jurygenderB & \textsc{\jurynameB}  & \jurygradeB & (\juryroleB) \\
    \null & \null & \juryadressB &\\ 
    
    \jurygenderC & \textsc{\jurynameC}  & \jurygradeC & (\juryroleC) \\
    \null & \null & \juryadressC &\\ 
    
    \jurygenderD & \textsc{\jurynameD}  & \jurygradeD & (\juryroleD) \\
    \null & \null & \juryadressD &\\ 
    
    \jurygenderE & \textsc{\jurynameE}  & \jurygradeE & (\juryroleE) \\
    \null & \null & \juryadressE &\\ 
    
    \jurygenderF & \textsc{\jurynameF}  & \jurygradeF & (\juryroleF) \\
    \null & \null & \juryadressF &\\ 
    
    \jurygenderG & \textsc{\jurynameG}  & \jurygradeG & (\juryroleG) \\
    \null & \null & \juryadressF &\\ 
   
  \end{tabular}    
\end{center}

\vspace{1.3cm}

%% file: parts/acknowledgments.tex
Je voudrais tout d'abord remercier toutes les personnes qui ont fait que ces trois dernières années au sein du Laboratoire de Physique Théorique d'Orsay se sont passées pour le mieux, à commencer par Sébastien Descotes-Genon pour son accueil au laboratoire et pour avoir été disponible et réactif pour toutes mes toutes demandes. Je remercie en particulier Philippe Molle et Jocelyne Raux pour s'être occupé de toutes les procédures administratives qui me dépassaient complètement et pour avoir accepté de valider mes ordres de mission à la dernière minute. Un grand merci également à Marie-Agnès Poulet pour sa bonne humeur et pour m'avoir beaucoup aidé, en particulier en fin de thèse.
\\

Je remercie Yann Mambrini, mon directeur de thèse pour avoir accepté de m'encadrer tout d'abord pendant mon stage de M2 et pendant ces trois dernières années, pour m'avoir permis d'apprendre une quantité astronomique de choses de par sa grand culture générale de notre domaine, qu'il s'agisse d'aspects très théoriques, phénoménologiques ou même historiques. Merci pour tous les conseils, de m'avoir transmis cette grande curiosité scientifique et de m'avoir fait découvrir cette incroyable communauté de physiciens, et également pour m'avoir poussé à présenter mes travaux et à échanger avec les membres de cette communauté.
\\

I would like to thank warmfully all the members of the jury : Geneviève Bélanger, Michel Tytgat, Carlos Munoz, Emilian Dudas, Keith Olive, Ulrich Ellwanger for being here on the 25/09/2018 and for taking the time to read this (very) long manuscript. All of you have helped me in some way during my PhD and I am really grateful for that.
\\

My curiosity for this field was initiated by my first supervisor, Jennifer Siegal-Gaskins, who welcomed me at Caltech for my first internship even though I did not know anything about dark matter and particle physics at that time. Thank you for giving me the opportunity to work on such an exiting topic in such an environment !
\\

The achievement of this PhD thesis would not have been possible without the substantial help from my "unofficial" supervisors, Giorgio Arcadi and Pradipta Ghosh, during the first and second year, who were present to help me with most of the technical issues that I faced during the early stages of my PhD and showed me the relevant tools and methods comonly employed in our field. Thank you for being so patient and helpful, I owe a substantial part of my current knowledge and understanding to both of you.
\\

One of the most interesting aspects of the work I did these past few years was the collaborative part for which I would like to warmfully all my collaborators from whom I really appreciated to work with :
Giorgio Arcadi, Nicolas Bernal, Gautam Bhattacharyya, Soo-Min Choi, Emilian Dudas, Maira Dutra, Miguel Escudero, Adam Falkowski, Pradipta Ghosh, Yonit Hochberg, Dan Hooper, Stephen F. King, Gordan Krnjaic, Eric Kuflik, Hyun Min Lee, Manfred Lindner, Yann Mambrini, Hitoshi Murayama, Keith Olive, Marco Peloso, Elena Perdomo, Stefano Profumo, Farinaldo S. Queiroz, Pat Scott, and Jennifer Siegal-Gaskins.
\\

Je souhaiterais également remercier Asmaa Abada pour les précieux conseils que j'ai reçu toutes ces années et pour m'avoir présenté le fantastique réseau Invisibles-Elusives qui m'aura permis de rencontrer plein d'étudiants et de professeurs incroyables.  
\\

L'aboutissement de cette thèse a grandement été facilitée par la présence d'étudiants et postdocs du LPT avec lesquels j'ai vraiment apprécié passer du temps, à commencer par mes co-bureaux : Matias Rodriguez-Vazquez et Luiz Vale-Silva, pour avoir fait de ce bureau sombre une pièce beaucoup plus agréable, ainsi que Andreï Angelescu, Debtosh Chowdhury, Hermès Bélusca,  Maira Dutra, Gabriel Jung, Luca Lionni, Florian Nortier, Timothé Poulain, Olcyr Sumensari, Hadrien Vroylandt et Xabier Marcano, pour leur bonne humeur et pour les bons moments partagés autour d'un café! Un grand merci également à Lucien Heurtier et Antoine Lehébel pour leurs commentaires et relecture de quelques chapitres de ce manuscrit en dernière minute!
\\

En dehors du temps passé au laboratoire, beaucoup de personnes ont contribué indirectement à la réussite de cette thèse, à commencer par mes colocataires les plus récents : Antoine, Eloïse, Jeremy, Maxime et Timothée, ainsi que des plus anciens : Amaudric, Camille et Maxence (et Séverin ?), avec lesquels j'aurai partagé de très bons moments à jouer au Bang!, 421 (surtout avec la chance de Jerem), à partager de nombreuses pintes, des raclettes, barbecues (surtout sans Antoine), pâtes carbo, à regarder l'intégrale des films de Dwayne Johnson, à jouer à smash bros, à tuer des zombies, pour les semaines au ski et pour avoit affronté les gyros infinissables de Milos, et j'en passe. Un grand merci à tous mes amis cachanais/physiciens pour les bons moments passés ces dernières années qui me paraissent compliqué de résumer sans ajouter 25 pages à ce manuscript : Adrien, Antoine R., Alizée, Audrey, Armand, Brigitte, Charlotte, Felix L. et Felix D-M., Géraldine, Jean, Mathis, Romain G., Simon et Starboule. Merci pour ses bonnes années passées sur le campus cachanais à jouer aux cartes (CPGB), au bière-pong, au babyfoot pendant les révisions des oraux d'agreg, à dégommer des kebabs au Beyti, et pour avoir été assez motivés pour partir en voyage à 13 en Ecosse en voiture. Merci aussi aux incontournables saccageurs du classico-hurlingo, Bebs, John Verns et Paps pour les multiples pintes et parties de fléchettes endiablées et m’avoir permis d’échouer chez vous à l’occasion ! Supplément John Verns et Guiboule pour les soirées jacuzzi à San Diego et virées à Vegas en Greyhound. Pour les bons moments passés avec toi et pour beaucoup de choses que je ne saurais résumer ici, merci Lucile.
\\

Un grand merci à mes amis non physiciens pour m'avoir permis de me déconnecter du monde de la matière noire à travers toutes ces soirées au théâtre des étoiles avec supplément pré-soirée chez Vlemma, sessions de renfo, fraco, sorties longues encadrées par notre coach préférée, vernissages, ciné, expos, concerts, restaurants, bars, ski, pour m'avoir motivé et accompagné à courir 42km en phase de rédaction du manuscrit, pour les weekends avec les lardons à manger gras, boire du génep et à entamer sévèrement notre capital de vie (et portefeuille..), pour les vacances incroyables à l'autre bout du monde à base de shortcut dans la jungle brésilienne avec supplément tostados et caïpis, pour s'être motivés à aller dormir par 0 degrés dans une cabane en pierre sur une ile écossaise (en phase de rédaction bien évidement), ou pour les courses en scooter au milieu des temples birmans : Benjamin, Candice, Camille, Céline, Emma, Chloé, Elodie, Henry, Louise, Matthieu A., Matthieu C. et Matthieu J., Nathalie, Nicolas, Paul, Rodolphe, Romain, Sylvain, Tony et Vlad! Merci et à bientôt pour allonger la liste ! 
\\

Pour finir, merci de tout cœur à ma famille : mes parents, pour m’avoir donné depuis tout petit la curiosité scientifique et l’envie d’apprendre toujours davantage, pour m'avoir toujours soutenu pendant ces longues études, à mes sœurs nées à Serre-les-Sapins Apolline et Lauriane, ainsi que Hélène et Benjamin B. pour les bons moments passés ensemble même si on ne se voit pas souvent !

%% file: parts/resume.tex
Plus de cent ans après le début de la révolution quantique, le 4 juillet 2012, les collaborations ATLAS et CMS ont annoncé l'observation d'une particule massive au LHC, le grand collisionneur de hadrons installé au CERN, avec une masse de l'ordre de 125 GeV. Cette particule, connue sous le nom de boson de Higgs, est la pierre angulaire et la dernière pièce manquante du modèle standard de la physique des particules, associant la relativité restreinte et la mécanique quantique au sein d'une théorie moderne des interactions fortes, faibles et électromagnétiques. Tout en faisant face à une pléthore de tests expérimentaux au cours des dernières décennies, le modèle standard fournit toujours une description précise de la physique en dessous de l'échelle atomique en permettant de prédire la constante de structure fine avec une précision relative incroyable de $ \sim 10 ^ {- 10 } $, probablement la quantité physique la plus précise jamais déterminée dans l'histoire de la science. \\

À l'autre extrémité des distances physiques, la dynamique des structures à grande échelle de l'univers composées de galaxies et d'amas de galaxies est régie par les lois de la relativité générale, la théorie géométrique de la gravité élaborée par Albert Einstein au début du XXe siècle. Des mesures cosmologiques récentes ont permis de déterminer la densité en matière-énergie de l’univers, intimement connectée à la structure géométrique de l’espace-temps selon la relativité générale. Le résultat est frappant. 70 $ \% $ de la densité d'énergie de l'univers est représenté par un terme cosmologique dans les équations d'Einstein appelé énergie sombre, les $ 30 \% $ restants sont composés de matière non-relativiste avec seulement 5 $ \% $ de matière baryonique ordinaire, que l'on peut directement observer. Environ 25$\%$ du budget énergétique de l'univers se présente sous la forme d'une composante invisible appelée matière noire.
Cent ans après la formulation de la relativité générale, l'une de ses plus importantes prédictions, l'émission d'ondes gravitationnelles par des systèmes binaires d'objets astrophysiques compacts, a été détecté pour la première fois par les collaborations LIGO et VIRGO confirmant la théorie relativité générale en tant que description précise des interactions gravitationnelles à ces échelles. \\

Cependant, le tableau dans sa globalité n’est pas complet, car une formulation quantique de tels effets gravitationnels n’est pas intégrée dans le modèle standard. Par conséquent, la nature mystérieuse de l’énergie sombre et de la matière noire représente l’un des plus grands défis de la physique moderne pour le siècle à venir. La question de la matière noire est en fait un problème de longue date qui a intrigué les astronomes au cours du siècle dernier et la solution la plus commune à cette problématique de substance invisible consiste à invoquer des particules exotiques présentes dans les grandes structures astrophysiques et peuplant l'univers. Une solution possible pour expliquer l’abondance de telles particules est de supposer que la matière sombre est constituée de particules massives à interaction faible (WIMP) jadis en contact thermique avec le plasma primordial constitué de particules du modèle standard qui se sont découplées à un stade précoce de l’univers. \\


La première partie de cette thèse est dévouée à l'introduction de certains éléments théoriques nécessaires pour comprendre comment la combinaison des observations cosmologiques, y compris (entre autres) les études du fond diffus cosmologique, les supernovae distantes, de grands échantillons d'amas de galaxies, les mesures d’oscillation acoustique baryonique et les lentilles gravitationnelles ont fermement établi le modèle standard de la cosmologie qui comprend une nouvelle forme de matière à découvrir, la matière noire, qui représente environ 85$\%$ du contenu en matière de l’univers et environ 27$\%$ du budget énergétique global.\\

Dans cette thèse, nous avons passé en revue les fondements théoriques du paradigme WIMP en tant que solution attrayante à cette problématique de masse invisible, qui suggère de manière élégante une correspondance entre l’abondance de la matière noire calculée à partie d'un modèle et sa valeur effectivement observée. L'avantage majeur de ce type de constructions et qu'elles ne nécessitent que peu d'hypothèses majeures, le contact thermique primordial entre le secteur sombre et le bain thermique du modèle standard ainsi que supposer que les paramètres de masse et sections efficaces d'annihilation des particules de matière noire sont typiquement de l'ordre de grandeur électrofaible du modèle standard.
En conséquence, des réalisations concrètes de modèles WIMP ont été développées dans différents cadres BSM ("Beyond the Standard-Model"), accessibles à plusieurs stratégies de recherche telles que les recherches directes, indirectes et auprès de collisionneurs de particules dont le statut actuel et les perspectives ont été discutés.\\

En particulier, la première partie de cette thèse a porté sur l'étude de constructions de modèles simplifiés de WIMPs, à savoir le portail Higgs et le portail $ Z $, et nous en avons conclu que ces modèles seront largement exclus, à l’exception du cas où la matière noire est une particule fermionique avec seulement des couplages axiaux avec le boson $ Z $ (par exemple, dans le cas d'un fermion de Majorana), en l'absence de signaux dans la prochaine génération d'expériences de détection directe. La tension avec les contraintes de détection directes peut être relaxée par exemple en introduisant un médiateur BSM scalaire s-channel. Cependant, l'introduction d'un tel degré supplémentaire de liberté semble assez artificielle et pourrait nécessiter une motivation théorique pour le justifier dans le cas d'une théorie minimaliste.
\begin{center}
\begin{figure}[h!]
  \begin{minipage}[l]{0.49\textwidth}
\includegraphics[width=\linewidth]{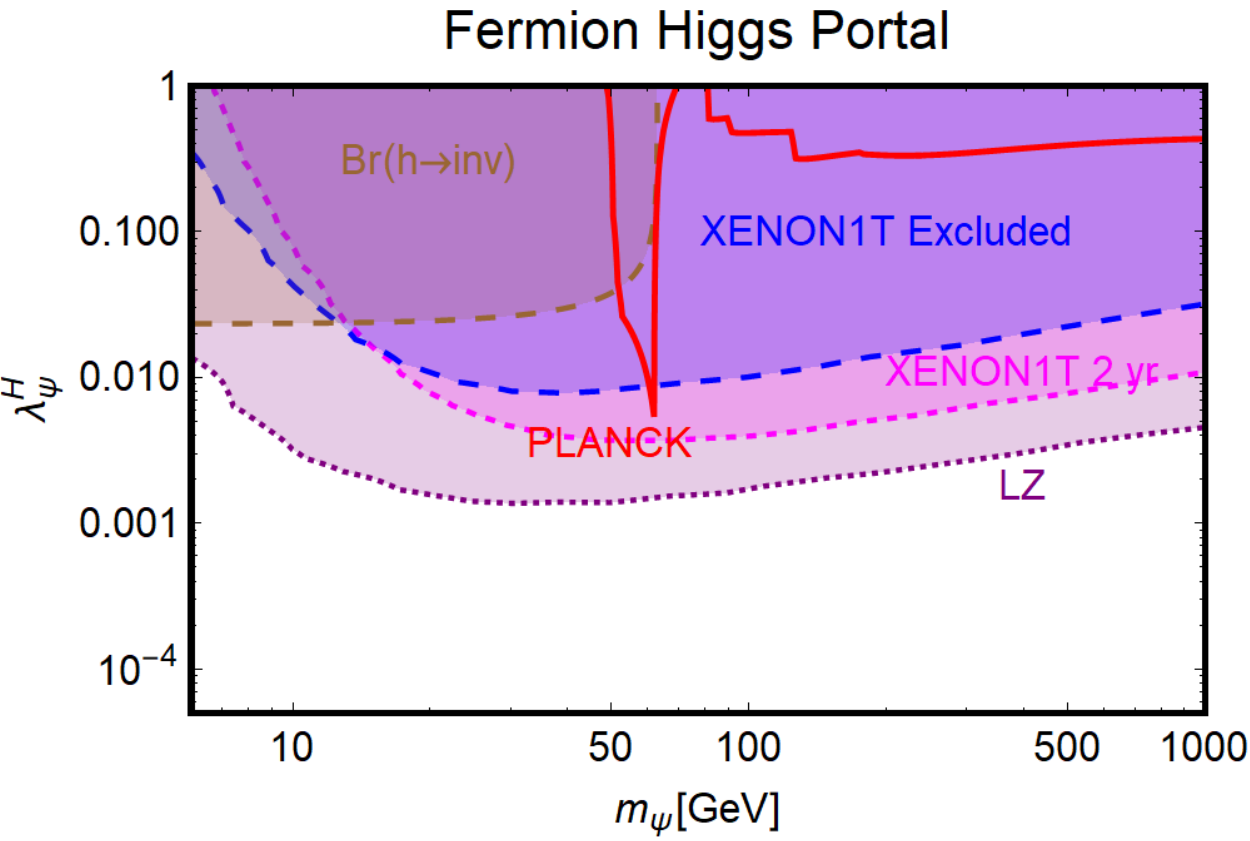}
   \end{minipage}\hfill
      \begin{minipage}[r]{0.49\textwidth}   
\includegraphics[width=\linewidth]{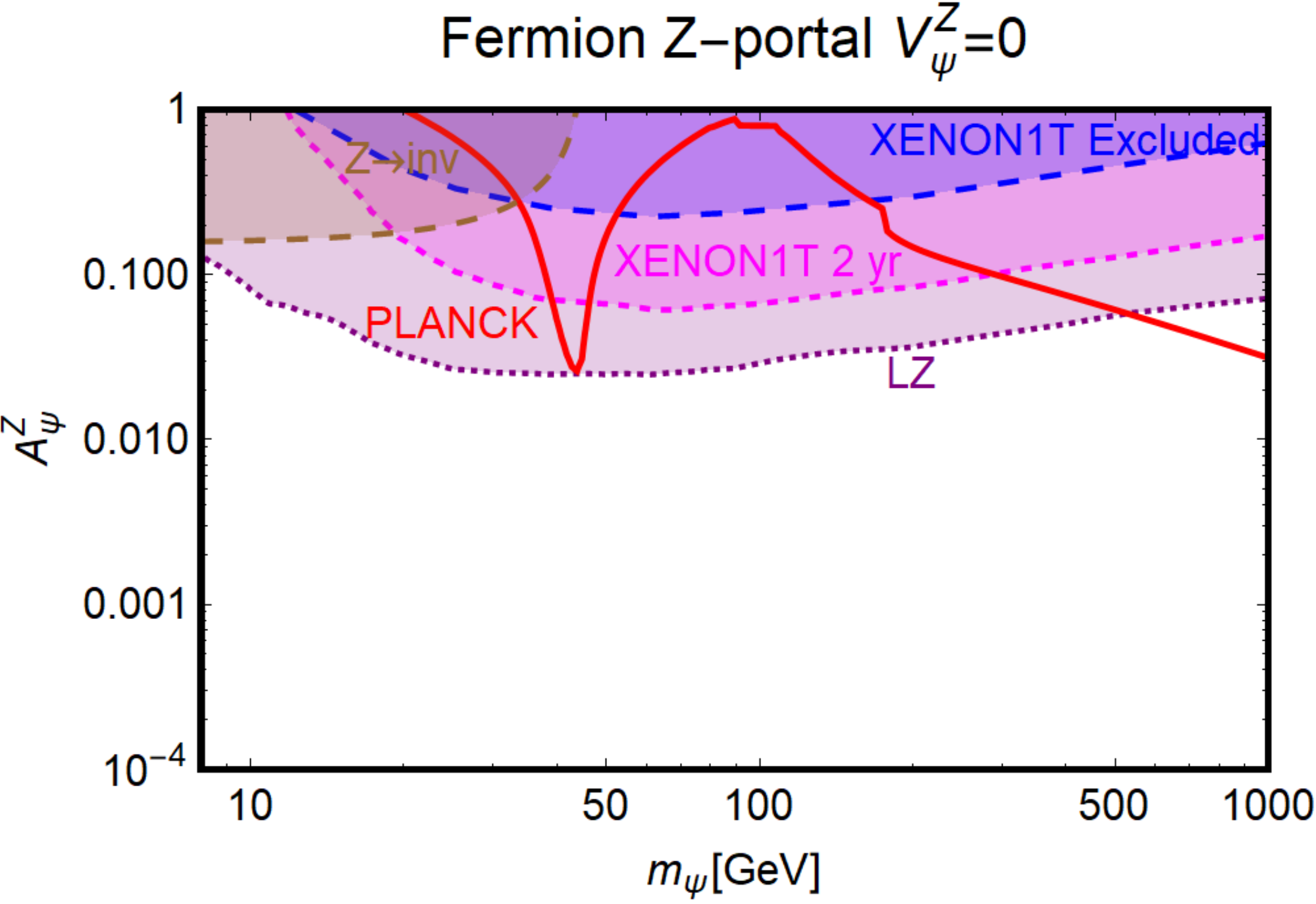}
   \end{minipage}
   \caption{ Illustration du portail Higgs et $Z$ dans le plan ou l'axe des ordonnées représente le couplage matière noire au mediateur et les ordonnées la masse des particules de matière noire dans le cas d'un candidat fermionique. Les courbes en rouge représentent l'espace des paramètrs compatibles avec les mesures de Planck de la densité produite de matière sombre. La région en bleu est exclue par les limites actuelles de détection directe. Les régions en magenta et violet seront exclues en l'absence d'observation de signal de la part des expériences XENON1T et LZ. La région en marron est exclue par la mesure précise du temps de demi-vie du médiateur dans chaque cas.}
\end{figure}
\end{center}
Le cadre BSM des symétries de jauge étendues est bien adapté au paradigme WIMP, car la stabilité de la matière noire et l'origine du médiateur peuvent naturellement apparaître dans de telles constructions.
Nous avons étudié des modèles spécifiques dans lesquels la relation entre un candidat de matière noire vectoriel et le champ de jauge massif d’un nouveau groupe de symétrie BSM brisé $ Z^\prime$ est médiée par un couplage de Chern-Simons (CS), dont l’origine est motivée par des mécanismes d’annulation des anomalies chirales. En particulier, nous avons exploré la possibilité de connecter le $ Z ^\prime $ au modèle standard via un terme de mélange cinétique avec le champ d'hypercharge $ B $ ainsi que la possibilité d'avoir un second terme de type CS impliquant le $ Z^\prime $ et le champ d'hypercharge. Nous avons effectué une analyse phénoménologique complète de tels modèles et proposé un cadre pour la génération du couplage de Chern-Simons à partir d'un modèle UV complet ainsi qu'une origine radiative du mélange cinétique.
Les résultats expérimentaux actuels et futurs favoriseraient une valeur du couplage CS difficile à expliquer
par une origine radiative sauf si plus de familles de fermions BSM sont inclus ou dans le cas où les couplages forts sont considérés. L'origine radiative du mélange cinétique entre le candidat matière noire vectoriel et le médiateur pourrait être évitée, afin de garantir la stabilité de la matière noire, en choisissant de manière appropriée l'assignation de charge des fermions lourds, mais serait assez artificielle.
L'analyse phénoménologique de ces modèles a montré la possibilité d'expliquer l'origine de la densité de matière noire dans notre univers basée sur le mécanisme WIMP. Cependant, une analyse théorique plus détaillée basée sur l'achèvement ultraviolet de ces considérations a montré que l'espace paramétrique viable compatible avec l'énoncé précédent nécessiterait un ajustement fin des paramètres de la théorie.\\


Dans le chapitre suivant, nous avons examiné un modèle savoureux, où une quatrième famille de VLF ("vector-like fermions") est introduite et chargée sous une symétrie supplémentaire $ U (1) ^ \prime $, en tant que solution possible aux anomalies de la saveur associées à l'observable $ R_ {K ^ {(* )}} $ constatée par l'expérience LHCb au cours des dernières années. Le boson massif $ Z ^ \prime $ associé à cette symétrie de jauge brisée joue le rôle de médiateur entre le modèle standard et les particules de matière noire, c’est-à-dire le neutrino droit singulet de la quatrième famille de LVF. En l'absence de mélange entre les quatre différentes familles de fermions, le $ Z ^ \prime $ est fermiophobe, sans couplage aux trois familles chirales, mais se couple à une quatrième famille de type vecteur. La présence de couplages de Yukawa entre certains nouveaux scalaires singlets chargés sous le groupe de symétrie $ U (1) ^ \prime $ et les quatre familles de fermions induit un mélange entre générations. De tels effets de mélange induisent des couplages du $ Z ^ \prime $ avec les doublets de lepton gauches de la deuxième famille et les doublets de quark gauches de la troisième famille. Ce modèle peut simultanément expliquer les ratios anomales associés à la désintégration des mésons $ B $ par l'observable $ R_ {K ^ {(*)}} $ et expliquer l'abondance de matière noire dans notre univers. Face à une pléthore de contraintes liées à la physique des saveurs, auprès des collisionneurs et de recherche sur la matière noire, cette analyse suggère un espace particulier des paramètres viable où le médiateur possède une masse de l'ordre de 500 GeV. La fenêtre autorisée peut être encore réduite par de meilleures mesures de précision du processus de production $ \nu_\mu + N \rightarrow \mu^+ + \mu^- + \nu_\mu + N $, et en considérant les améliorations futures de la détermination de la différence de masse des mésons $ B_s $. La gamme expérimentalement favorisée des masses et des couplages de matière noire pourrait être testée par la future génération d'expériences de détection directe.\\

\begin{figure}
\begin{minipage}[h!]{0.47\textwidth}
\includegraphics[height=4.5cm]{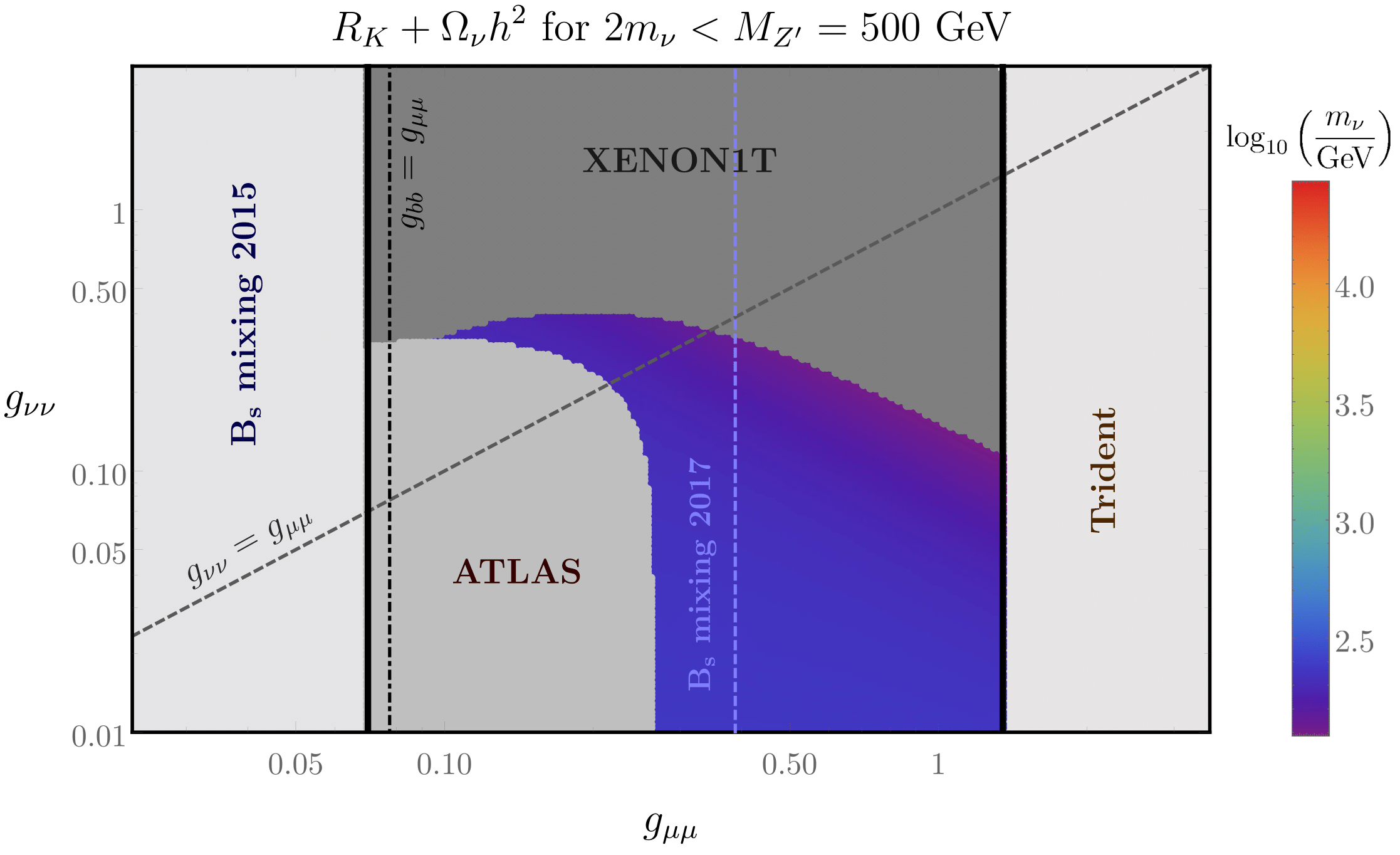}
\end{minipage}
\hfill
\begin{minipage}[h!]{0.52\textwidth}
\includegraphics[height=4.5cm]{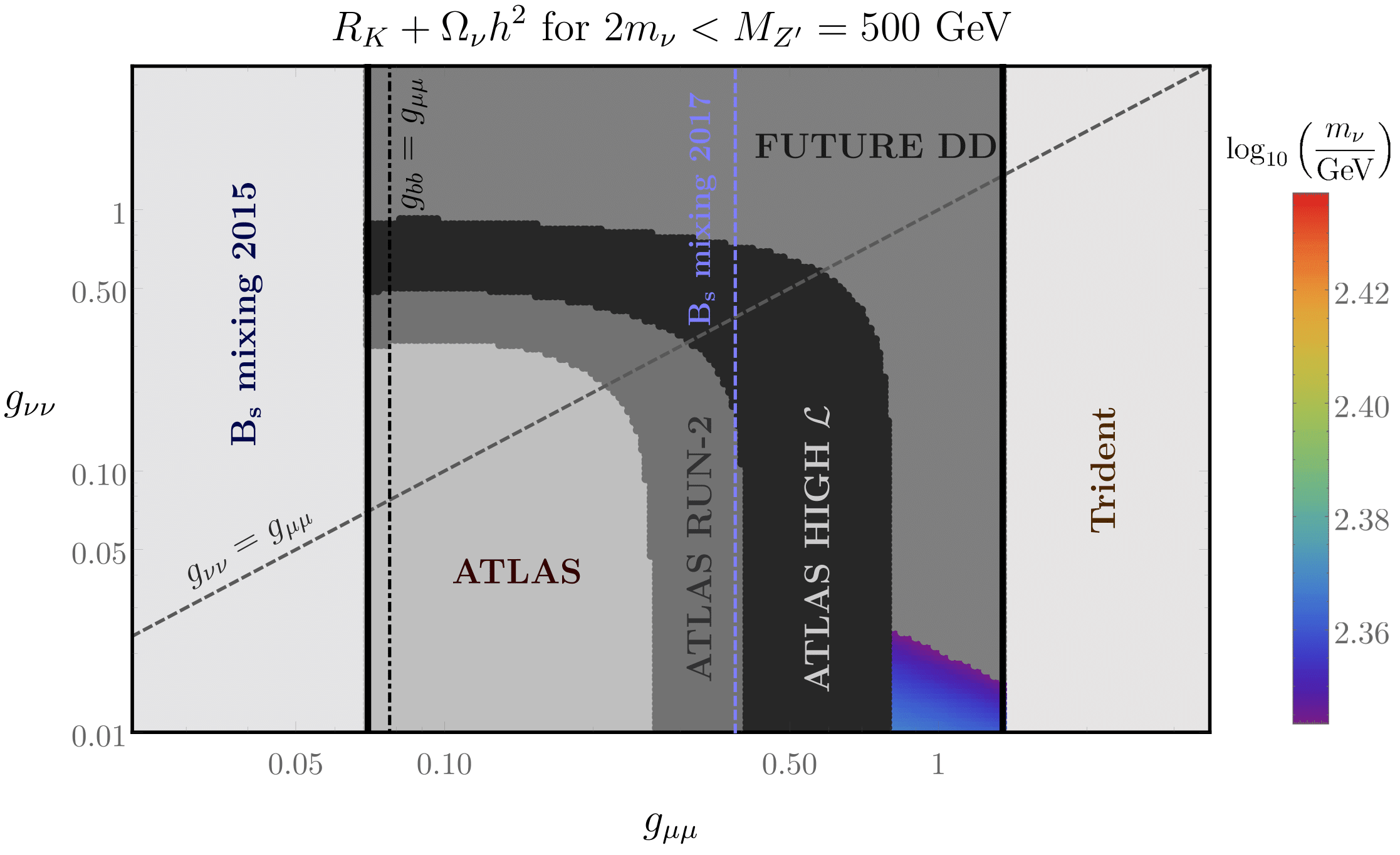}
\end{minipage}
\caption{
\label{fig:mzp500prospects}
Résumé de toutes les contraintes dans l'espace des paramètres pour une masses $M_{Z^\prime}=500~\text{GeV}$ on the left et prédictions compte tendu de la future sensibilité de collisioneurs à haute luminosité et futures expériences de détection directe.}
\end{figure}

Des mécanismes de production alternatifs de la matière sombre ont été discutés dans la troisième partie de cette thèse, tels que les mécanismes SIMP ("Strongly Interacting Massive Particles") et ELDER (ELastic Decoupling Relic). Dans ce type de mécanisme, le rôle des interactions au sein du secteur sombre ainsi que l'effet d'un découplage précoce entre la matière noire et le bain thermique du modèle standard impliquent une phénoménologie un peu différente du cas WIMP. En particulier, de tels cadres offrent la possibilité d’avoir une section efficace d’auto-interaction de la matière noire capable d’atténuer les tensions entre des observations astrophysiques et les simulations à N-corps basées sur une cosmologie $ \Lambda $CDM. Dans cette thèse, nous avons considéré une réalisation explicite du mécanisme SIMP sous la forme de SIMP vectoriels résultant d'une théorie de jauge non-abélienne $ SU (2) _X $, où la "custodial symmetry" accidentelle assure la stabilité des particules de matière sombre. Nous avons proposé plusieurs manières d’équilibrer cinétiquement les secteurs sombres et visibles dans ce contexte. En particulier, nous avons montré qu’un portail avec le Higgs sombre, responsable de la brisure de la symétrie non abélienne cachée, pouvait maintenir l’équilibre thermique entre les deux secteurs, de même qu’un portail vectoriel sombre massif se mélangeant cinétiquement avec l’hyperchage, avec des couplages Chern-Simons généralisés aux SIMP vectoriels, tout en restant cohérent avec les contraintes expérimentales actuelles.

\begin{center}
\begin{figure}[h!]
  \begin{minipage}[l]{0.49\textwidth}
\includegraphics[width=\linewidth]{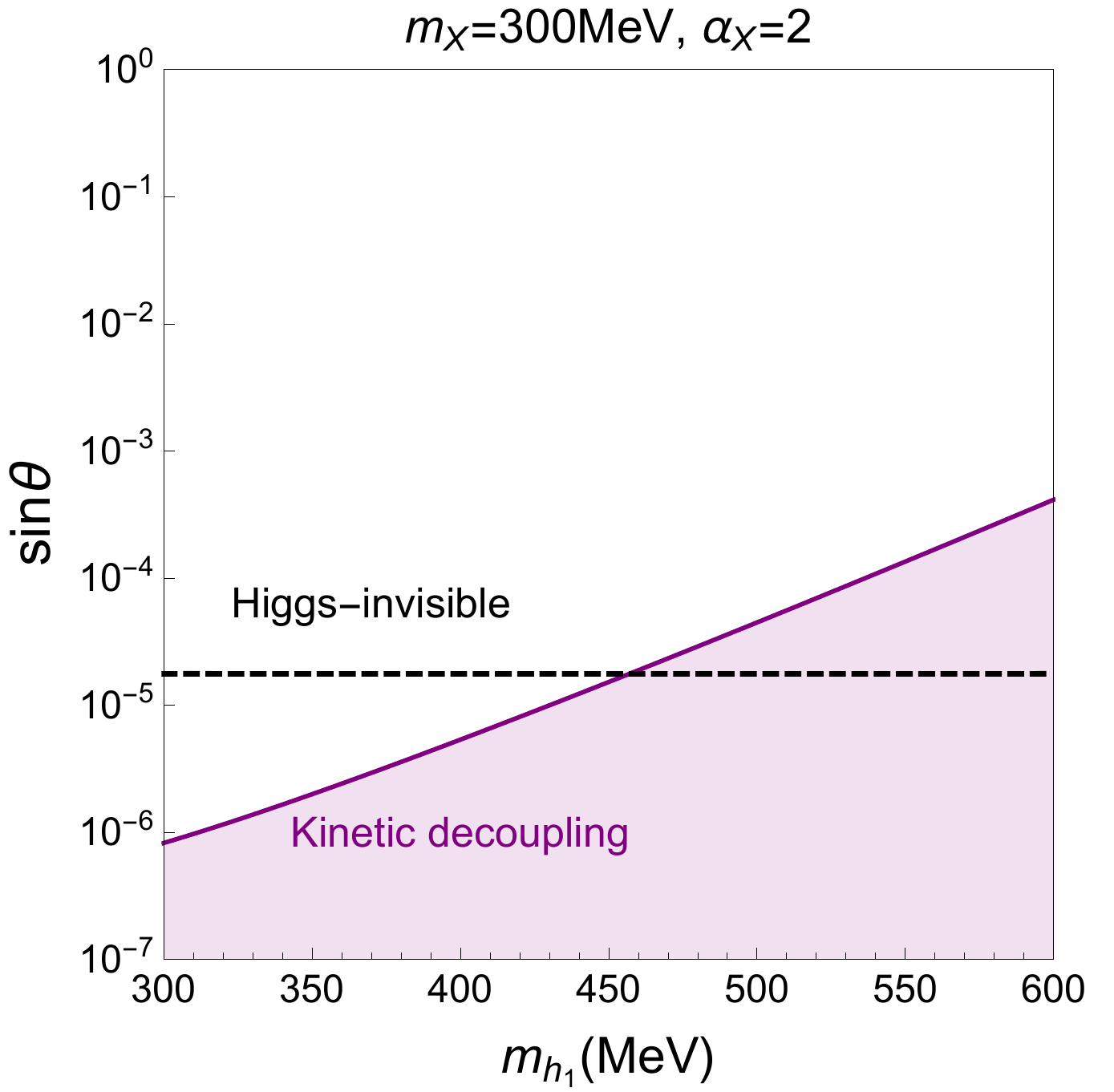}
   \end{minipage}\hfill
      \begin{minipage}[r]{0.49\textwidth}   
\includegraphics[width=\linewidth]{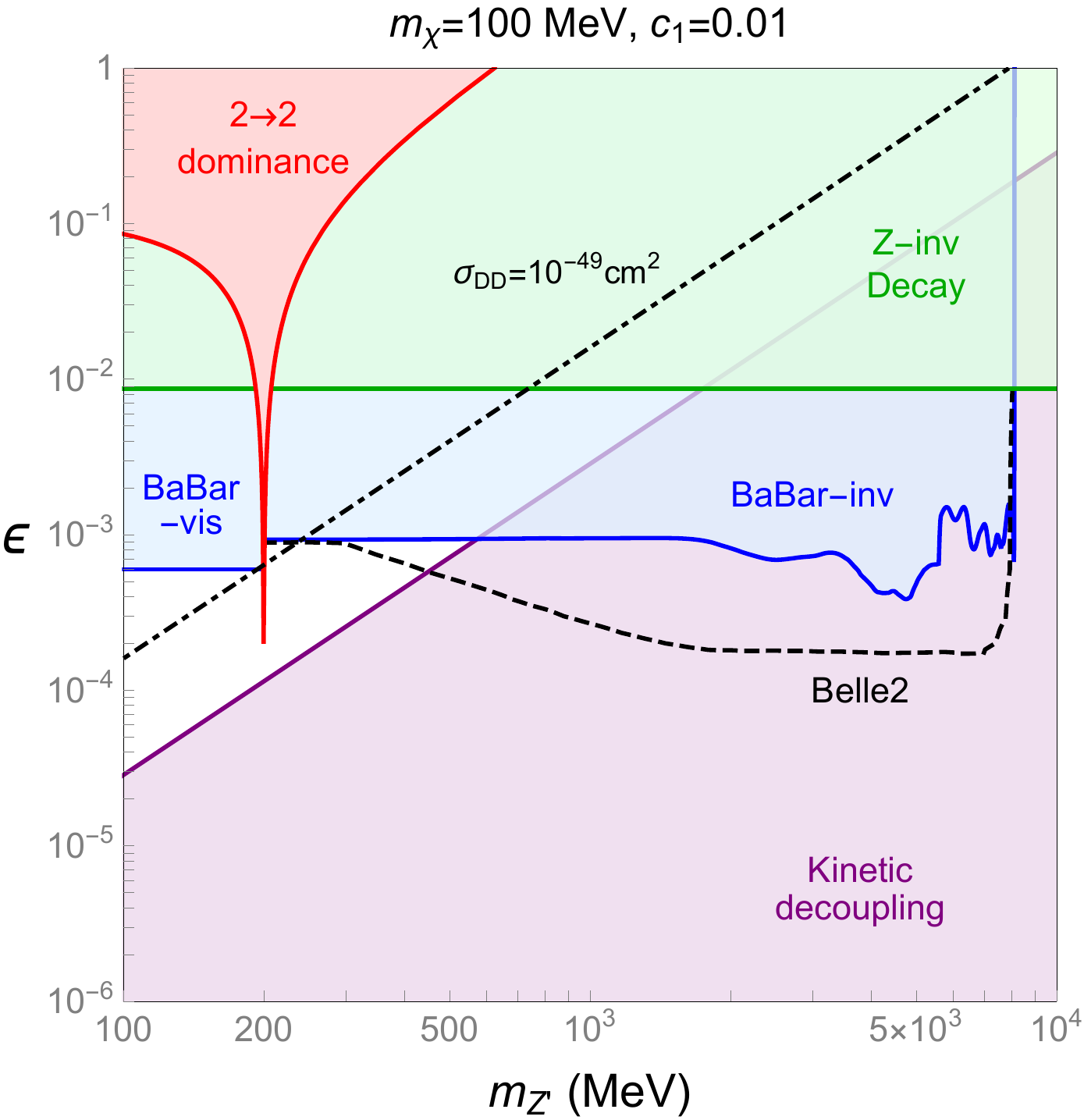}
   \end{minipage}
   \caption{Espace des paramètres viable dans le cas d'un portail avec le Higgs sombre, dans le plan angle de mélange avec le Higgs VS masse du higgs sombre, à gauche et dans le cas d'un portal avec le $Z^\prime$ à droite, dans le plan couplage du mélange cinétique en fonction de la masse du $Z^\prime$}
\end{figure}
\end{center}

Dans le dernier chapitre, nous avons discuté la possibilité de produire la matière noire de manière non-thermique à partir de l'annihilation ou de la désintégration de particules en contact avec le bain thermique du modèle standard. En particulier, nous avons étudié le cas où un messager de spin 2 massif peut jouer efficacement le rôle de portail entre les secteurs sombre et standard. Dans une grande partie de l'espace des paramètres compatible avec l'exigence de produire la bonne densité relique, la production par échange d'un médiateur de spin 2 massif domine les processus impliquant un graviton. Nous avons étudié l’impact de l’étape de réchauffement primordial de l’univers ("reheating") sur la production de la matière noire, et nous avons montré que nos résultats sont grandement influencés par la prise en compte des effets de réchauffage non-instantané. Non seulement nous constatons une augmentation des facteurs de production en raison de la grande dépendence en la température du taux de production, mais nous avons ont également montré que la présence d'un médiateur avec une masse de l'ordre de la température de réchauffage pourrait améliorer considérablement la production en raison des effets de résonance de la particule massive de spin 2.\\

Le travail présenté dans cette thèse visait à aborder l'une des questions ouvertes de la physique moderne en ce qui concerne notre compréhension des lois fondamentales de la nature. La nature précise de la matière noire reste encore inconnue à ce jour, mais il est clair que la plupart des modèles WIMP seront testés au cours des prochaines décennies. Cette thèse a mis en évidence le rôle déterminant de la prochaine génération d'expériences de détection directe et indirecte de matière sombre, et sa complémentarité avec des considérations théoriques concernant cette grande énigme qu'est la présence de matière noire dans notre univers.

%% file: parts/introduction.tex
More than one hundred years after the beginning of the quantum revolution, on the $4^{\rm th}$ of July 2012, the ATLAS and CMS collaborations announced the observation of a massive particle at the Large Hadron Collider with a mass of the order of 125 GeV. This particle, known as the Higgs boson, is the cornerstone and last missing piece of the Standard Model of particle physics, bringing together special relativity and quantum mechanics in the modern theory of the strong, weak and electromagnetic interactions. While facing a plethora of experimental tests over the past decades, the Standard Model still provides an accurate description of physics below the atomic scale by allowing to predict the value of the fine structure constant with an incredible relative precision of $\sim10^{-10}$, probably one the most precise quantity ever determined in the history of science.\\
On the other extremity of physical distances, the dynamics of large scale structures of the universe which are composed of galaxies and galaxy clusters, is governed by the laws of General Relativity, the geometrical theory of gravity elaborated by Albert Einstein at the beginning of the twentieth century. Recent cosmological measurements allowed to determine the matter-energy density of the universe, related to the geometrical structure of space-time according to General Relativity. The outcome is striking. 70$\%$ of the energy density of the universe is represented by a cosmological constant term in Einstein's equations known as Dark Energy, the remaining $30\%$ is composed of non-relativistic matter with only $5\%$ of ordinary baryonic matter. Around $25\%$ of the energy budget of the universe lies in the form of an invisible component called Dark Matter.
One hundred years after the formulation of General Relativity, one of its greatest predictions known as gravitational waves were detected for the first time by the LIGO and VIRGO collaborations confirming General Relativity as an accurate description of gravitational interactions. \\
However, the overall picture is not complete as a quantum formulation of such gravitational effects is not embedded in the Standard Model. Therefore, the obscure nature of Dark Energy and Dark Matter represents one of the strongest challenges of physics for the upcoming century. The Dark Matter issue is in fact a long-standing problem that has puzzled astronomers for the past century and the most common solution to this invisible substance issue is to invoke exotic particles present in large astrophysical structures. A possible solution to account for the abundance of such particles is to assume that the Dark Matter is made of Weakly Interacting Massive Particle (WIMP) formerly in thermal contact with the primordial plasma made of Standard Model particles that decoupled as some early stage of the universe.\\
In the first part of this thesis, we present the theoretical elements and experimental arguments required to understand the Dark Matter conundrum in its global picture. In a second part we attempt to tackle this issue by investigating beyond-the-Standard-Model realizations of the WIMP paradigm in the context of simplified models and in extended gauge structure models. In a last part we discuss alternative density production mechanisms and explore specific realizations. The appendices provide technical details regarding some computations performed in this work. The work presented in this thesis is based on the following publication list.

\section*{Articles published in peer-reviewed journals}

\begin{enumerate}

\item  \textbf{"Freezing-in Dark Matter through a heavy invisible $Z^\prime$"}~\cite{Bhattacharyya:2018evo}\\ G. Bhattacharyya, M. Dutra, Y. Mambrini and M. Pierre\\
$[$\href{https://arxiv.org/abs/1806.00016}{\textcolor{blue}{arXiv:1806.00016}}$]$\\
Phys. Rev. D \textbf{98}, 035038 (2018)

\item  \textbf{"Flavourful $Z'$ portal for vector-like neutrino Dark Matter and $R_{K^{(*)}}$"}~\cite{Falkowski:2018dsl}\\ A. Falkowski, S.F. King, E. Perdomo and M. Pierre\\
$[$\href{https://arxiv.org/abs/1803.04430}{\textcolor{blue}{arXiv:1803.04430}}$]$\\
JHEP \textbf{1808}, 061 (2018)

\item  \textbf{"Spin-2 Portal Dark Matter"}~\cite{Bernal:2018qlk}\\ N. Bernal, M. Dutra, Y. Mambrini, K.A. Olive, M. Peloso and M. Pierre\\
$[$\href{https://arxiv.org/abs/1803.01866}{\textcolor{blue}{arXiv:1803.01866}}$]$\\
Phys. Rev. D \textbf{97}, 115020 (2018)

\item  \textbf{"Vector SIMP Dark Matter"}~\cite{Choi:2017zww}\\ S-M. Choi, Y. Hochberg, E. Kuflik, H.M. Lee, Y. Mambrini, H. Murayama and M. Pierre\\
$[$\href{https://arxiv.org/abs/1707.01434}{\textcolor{blue}{arXiv:1707.01434}}$]$\\
JHEP \textbf{1710}, 162 (2017)

\item  \textbf{"$Z^\prime$ portal to Chern-Simons Dark Matter"}~\cite{Arcadi:2017jqd}\\ G. Arcadi, P. Ghosh, Y. Mambrini, M. Pierre and F.S. Queiroz\\
$[$\href{https://arxiv.org/abs/1706.04198}{\textcolor{blue}{arXiv:1706.04198}}$]$\\
JCAP \textbf{1711} no.11, 020 (2017) 

\item  \textbf{"GUT Models at Current and Future Hadron Colliders and Implications to Dark Matter Searches"}~\cite{Arcadi:2017atc}\\ G. Arcadi, M. Lindner, Y. Mambrini, M. Pierre and F.S. Queiroz\\
$[$\href{https://arxiv.org/abs/1704.02328}{\textcolor{blue}{arXiv:1704.02328}}$]$\\
Phys. Lett. B \textbf{771} 508-514 (2017) 

\item  \textbf{"The Waning of the WIMP? A Review of Models, Searches, and Constraints"}~\cite{Arcadi:2017kky}\\ G. Arcadi, M. Dutra, P. Ghosh, M. Lindner, Y. Mambrini, M. Pierre, S. Profumo and F.S. Queiroz\\
$[$\href{https://arxiv.org/abs/1703.07364}{\textcolor{blue}{arXiv:1703.07364}}$]$\\
Eur. Phys. J. C \textbf{78} no.3, 203 (2018)

\item  \textbf{"Scrutinizing a diphoton resonance at the LHC through Moscow zero"}~\cite{Arcadi:2016acg}\\ G. Arcadi, P. Ghosh, Y. Mambrini and M. Pierre\\
$[$\href{https://arxiv.org/abs/1608.04755}{\textcolor{blue}{arXiv:1608.04755}}$]$\\
JCAP \textbf{11}, 054 (2016)

\item  \textbf{"Re-opening Dark Matter windows compatible with a diphoton excess"}~\cite{Arcadi:2016dbl}\\ G. Arcadi, P. Ghosh, Y. Mambrini and M. Pierre\\
$[$\href{https://arxiv.org/abs/1603.05601}{\textcolor{blue}{arXiv:1603.05601}}$]$\\
JCAP \textbf{07}, 005 (2016)

\item  \textbf{"Sensitivity of CTA to Dark Matter signals from the Galactic Center"}~\cite{Pierre:2014tra}\\ M. Pierre, J.M. Siegal-Gaskins and P. Scott\\
$[$\href{https://arxiv.org/abs/1401.7330}{\textcolor{blue}{arXiv:1401.7330}}$]$\\
JCAP \textbf{10}, E01 (2014)

\end{enumerate}

\section*{Conference proceedings}

\begin{enumerate}[resume]

\item  \textbf{"	
Impact of Dark Matter Direct and Indirect Detection on Simplified Dark Matter Models"}~\cite{Arcadi:2015nea}\\ G. Arcadi, Y. Mambrini and M. Pierre\\
$[$\href{https://arxiv.org/abs/1510.02297}{\textcolor{blue}{arXiv:1510.02297}}$]$\\
PoS EPS-HEP2015 \textbf{396} (2015)

\item  \textbf{"	
Dark Matter in GUT inspired $Z^\prime$ portal scenarios"}~\cite{Pierre:2016wgb}\\ M. Pierre\\
$[$\href{https://arxiv.org/abs/1609.02424}{\textcolor{blue}{arXiv:1609.02424}}$]$\\
PoS CORFU2015 \textbf{061} (2016)

\end{enumerate}

%% file: parts/cosmo.tex
\vspace{0.3cm}

\noindent
The statement that a dark component is present in our universe is a conclusion of modern cosmology, a field that has been developed for the last century, which is based on many complex observations. In this chapter we present the basic tools required to describe our universe as a whole. We start by briefly describing the theory of General Relativity, emphasizing the connection between the matter-energy content of the universe to its geometrical structure, before deriving the conclusion that our universe is expanding and based on recent Supernovae results, that this expansion is in fact accelerating. In a second part, we give a qualitative description of the important stages in the history of the universe and we introduce thermodynamics quantities and tools, relevant in the context of Dark Matter phenomenology.

\section{The expanding universe}

\subsection{Elements of General Relativity}

The theory of General Relativity (GR) was elaborated at the beginning of the twentieth century and achieved in 1915 by the famous physicist Albert Einstein. This theory relates in a elegant formulation the geometry of a curved space-time to its matter and energy content through the Einstein's equation. In order to deal with distances in a curved space-time one has to introduce the metric tensor $g_{\mu \nu}$ related to a frame-invariant line element:
\begin{equation}
\d s^2=g_{\mu \nu} \d x^\mu \d x^\nu ~.
\end{equation}
In general relativity, the equations of motion (called \textit{the geodesic equations}) are expressed as a function of the metric tensor and its derivatives through the Christoffel symbol
\begin{equation}
\Gamma^\lambda_{\mu \nu}=\dfrac{1}{2}g^{\lambda \sigma}( \partial_\mu g_{\sigma \nu}+\partial_\nu g_{\mu \sigma} - \partial_\sigma g_{\mu \nu} )~.
\end{equation}
and take the following form
\begin{equation}
\dfrac{\diff^2 x^\mu}{\diff \lambda^2}+\Gamma^\mu_{\alpha \beta} \dfrac{\diff x^\alpha}{\diff \lambda}\dfrac{\diff x^\beta}{\diff \lambda}=0~.
\end{equation}
where $\lambda$ is an affine parameter of the trajectory, for instance the proper time. For a non-relativistic particle propagating in a weak gravitational potential $\phi(x)$, $\diff x^\alpha/\diff t \simeq (1,0,0,0)$ and $\Gamma^\mu_{00}\simeq \partial \phi/\partial x_{\mu}$, the geodesic equation is equivalent to Newton's relation
\begin{equation}
\dfrac{\diff^2 \vec{x}}{\diff t^2}=-\vec{\nabla}\phi~.
\end{equation}
In a more general case, Einstein's equations can be derived from the Hilbert-Einstein action using variational principle arguments
\begin{equation}
S=\dfrac{1}{16 \pi G_N}\int \diff^4 x \sqrt{-g} (R-2 \Lambda)+S_{ \text{mat}}~,
\label{eq:EHaction}
\end{equation}
where $g \equiv \text{det}(g_{\mu \nu})$, $\Lambda$ is the so-called cosmological constant introduced to describe the acceleration of the expansion of the universe as explained further on, $G_N$ is Newton's constant and $S_{\text{mat}}$ is the matter action.
The quantity $R$ is called the Ricci scalar and can be expressed as a function of the Riemann tensor constructed from contractions of Christoffel symbols
\begin{equation}
R^{\rho}_{\lambda \mu \nu}\equiv\partial_\mu \Gamma^\rho_{\lambda \nu}-\partial_\nu \Gamma^\rho_{\lambda \mu}+\Gamma^\sigma_{\lambda \nu} \Gamma^{\rho}_{\sigma \mu}-\Gamma^\sigma_{\lambda \mu} \Gamma^{\rho}_{\sigma \nu}~.
\end{equation}
From the Riemann tensor we can construct the Ricci tensor $R_{\mu \nu}$ contracted over two Lorentz indices and the Ricci scalar $R$ by taking the trace:
\begin{equation}
R_{\mu \nu}\equiv R^{\sigma}_{\mu \sigma \nu}~, \qquad R\equiv R_{\mu}^\mu~.
\end{equation}
Minimizing the action $\delta S=0$ leads to Einstein's equations:
\begin{equation}
R_{\mu\nu}-\dfrac{1}{2}g_{\mu\nu}R+\Lambda g_{\mu\nu}=8\pi G_N T_{\mu\nu}~.
\label{eq:Einstein}
\end{equation}
$T_{\mu \nu}$ is the energy-momentum tensor can be defined from the matter action:
\begin{equation}
T_{\mu \nu}\equiv -\dfrac{2}{\sqrt{-g}}\dfrac{\delta S_{\text{mat}} }{\delta g^{\mu \nu}}~.
\end{equation}
The compact form of Eq.~(\ref{eq:Einstein}) shows through the quantites defined above how any matter content, through the right-hand side, behaves in presence of curvature, through the left-hand side, and reciprocally how some matter-energy content would affect back the curvature of space-time. More than one hundred years after its formulation, the theory of General Relativity is still standing while facing the tremendous amount of data accumulated so far. Recently the collaborations LIGO and VIRGO detected for the first time gravitational waves~\cite{TheLIGOScientific:2017qsa}, one of the strongest prediction of general relativity, ruling out some possible extensions of this elegant theory.

\subsection{Distances in the universe}
The standard model of cosmology is based on the \textit{cosmological principle} which states that no observer occupies a preferred position in the universe and as a result, the universe must be homogeneous and isotropic on very large scales. This is obviously a strong assumption but the recent observations of the very large scale structures ($\sim 100~\text{Mpc}$) of the universe and the almost homogeneous Cosmic Microwave Background (CMB) map suggest that this hypothesis is effectively verified in our universe. Successively, several physicists derived the only possible form of the metric compatible with the assumptions implied by the cosmological principle. Friedmann, Lemaitre, Roberston and Walker in the 1920's-1930's showed that the metric, known as \textit{FLRW}, can be parametrized as
\begin{equation}
\d s^2=\d t^2-a^2(t)\Big[\frac{\d r^2}{1-kr^2}-r^2\d \theta^2-r^2\sin^2{\theta}\d \phi^2\Big]~,
\end{equation}
where $a(t)$ is the \textit{scale factor} describing how distances between two points at rest with respect to each other evolves in time due to the expansion of the universe.
$k$ is a parameter related to the intrinsic curvature of space and can take the following values:  
\begin{equation}
  k=\left\{
    \begin{split}
    & +1 \\ 
    & 0 \\
    & -1 \\
    \end{split}
\right.
\quad
\left.
    \begin{split}
    & \text{universe positively curved  : \underline{finite and closed}}\\ 
    & \text{universe flat : \underline{infinite and open}}\\
    & \text{universe negatively curved : \underline{infinite and open}}\\
    \end{split}
\right.
\end{equation}
A light-ray propagating in such a universe would feel a contraction or dilatation of space by a factor $a(t)$ and its wavelength would be affected by the same factor. Therefore we define the \textit{redshift} $z$ as the ratio of the scale factor at the present time $t_0$ over the scale factor at some time of emission $t$
\begin{equation}
1+z \equiv \dfrac{a(t_0)}{a(t)}=\dfrac{1}{a}~,
\end{equation}
where we used the standard convention $a(t_0)=1$.
It is convenient to rewrite the metric using the \textit{comoving coordinate} $\chi(r)$ defined as
\begin{equation}
\d \chi \equiv \dfrac{\d r}{\sqrt{1-kr^2}} ~,
\end{equation}
which takes the form:
\begin{equation}
\d s^2=\d t^2-a^2(t)\Big[\d \chi^2-r^2(\chi)(\d \theta^2+\sin^2{\theta}\d \phi^2 ) \Big]~,
\end{equation}
where we have the following relation between $r$ and $\chi$ depending on the curvature:
\begin{equation}
  r(\chi)=S_k(\chi)\equiv \left\{
    \begin{split}
    & \sin (\chi) \\ 
    & \chi \\
    & \sinh (\chi) \\
    \end{split}
\right.
\quad
\left.
    \begin{split}
    & ~\text{if}~k=+1 \\ 
    & ~\text{if}~k=0 \\ 
    & ~\text{if}~k=-1 \\ 
    \end{split}
\right.
\end{equation}
One can define the \textit{comoving distance} which corresponds to the distance traveled by light from $t_e$ to $t_0$ using the condition satisfied for light-like trajectories $\d s=0 $:
\begin{equation}
\chi(t_e) = \int_{t_e}^{t_0} \dfrac{\d t'}{a(t')}~.
\end{equation} 
Similarly one can express the so-called \textit{conformal time} $\eta$ which is the total distance traveled by light since $t=0$ defined as  
\begin{equation}
\eta (a) \equiv  \int_{0}^{t_0} \dfrac{\d t'}{a(t')} = \int_{a}^1 \dfrac{\d a'}{a^{\prime 2} H(a')}~,
\label{eq:conformaltime}
\end{equation}
where we introduced the expansion rate $H(t)$ called the \textit{Hubble parameter} defined as
\begin{equation}
H(t)\equiv \dfrac{\dot{a}}{a}~.
\end{equation}
Two important distances, relevant in the context of cosmological observations, are defined as follow\footnote{These definitions hold for a flat universe $k=0$. In a curved space, one has to redefine these quantities with the substitution $\chi \rightarrow S_k(\chi)$}:
\begin{itemize}
\item The luminosity distance $d_L$ can be expressed as a function of the flux $\mathcal{F}$ observed on a spherical shell at some distance $d_L$ of some object emitting light with an intrinsic luminosity $L$ at a given scale factor $a$ as:
\begin{equation}
\mathcal{F}\equiv\dfrac{L}{4\pi d_L^2(a)}~.
\end{equation}
Using conservation of the total number of photons from the emission time to the time of observation, the luminosity on the surface of a spherical shell located at a scale factor $a$ will be affected by a factor $a^2$ and the physical distance to this shell corresponds to the comoving distance $\chi(a)$ related to the observed flux:
\begin{equation}
\mathcal{F}=\dfrac{La^2}{4\pi \chi(a)^2}~,
\end{equation}
corresponding to a luminosity distance that can be expressed as:
\begin{equation}
d_L(a)=\dfrac{\chi(a)}{a}~.
\label{eq:dA}
\end{equation}
\item The angular distance $d_A$ of an object relates its physical length $D$ to its angular size in the sky $\theta$ with the following definition:
\begin{equation}
\theta \equiv \dfrac{D}{d_A(a)}~.
\end{equation}
We can write the angular size as the ratio of the comoving length of the object $D/a$ to the comoving distance from the observer
\begin{equation}
\theta=\dfrac{D/a}{\chi (a)}~,
\end{equation}
implying
\begin{equation}
d_A(a)=a\chi (a)~.
\end{equation}
\end{itemize}

\subsection{The Hubble parameter and energy content}

Following the cosmological principle, the energy-momentum tensor describing the cosmological fluid at large scales is the one of a perfect fluid with an equation of state $P=w\rho$ where $P$ is the pressure, $\rho$ is the energy density and $w$ a constant. The energy-momentum tensor of a perfect fluid has the following form:
\begin{equation}
T^{\mu\nu}=(\rho+P)u^{\mu}u^{\nu}-Pg^{\mu\nu}~,
\end{equation}
where $u^{\mu}$ is the four-velocity of the fluid. Using Einstein's equations one can derive the \textit{Friedmann equations} 

\begin{equation}
 H^2(t)=\frac{8\pi G_N}{3}\rho-\frac{k}{a^2}+\frac{\Lambda}{3}~,
\end{equation}

\begin{equation}
 \frac{\ddot{a}}{a}=-\frac{4\pi G_N}{3}(\rho+3P)+\frac{\Lambda}{3}~,
\end{equation}
A measurement of the value of Hubble parameter today from the CMB anisotropy map by the Planck collaboration gives $H_0=67.27 \pm 0.66~\text{km s}^{-1}~\text{Mpc}^{-1}$~\cite{Ade:2015xua}. For historical reasons it is convient to define the dimensionless parameter $h$ as:
\begin{equation}
H_0 \equiv 100~h ~\text{km s}^{-1}~\text{Mpc}^{-1}~,
\end{equation}
where the value of $h$ is of the order of unity $h\simeq 0.7$. Using energy-momentum conservation $\nabla_{\mu}T^{\mu\nu}=0$, we can derive the continuity relation expressing the time evolution of the energy density and pressure of the fluid to the Hubble parameter
\begin{equation}
\frac{\partial\rho}{\partial t}+3H(\rho+P)=0~.
\label{eq:continuityrelation}
\end{equation}
Considering the universe in an adiabatic expansion, the relation $\d(\rho a^3)=-P\d (a^3)$ holds and we can relate the evolution of the energy density of the fluid to the parameter $w$ of the equation of state $ \rho \propto a^{-3(1+w)}$. One can show that for non-relativistic matter ($m$)  $w=0$, leading to $\rho_{m} \propto a^{-3}$ where the numerical value of the exponent "3" corresponds to the dilution of space dimensions. For relativistic matter ($r$), the equation of state yields $w=1/3$ implying $\rho_{m} \propto a^{-4}$, where one factor of 3 is due to the space dilution and one factor of 1 for the redshift. In the case of an empty universe (without any matter content), only the cosmological constant term will impact the expansion of the universe with a parameter $w=-1$. We conveniently express the energy density in units of the \textit{critical density} defined as:
\begin{equation}
 \rho_{\text{crit}} \equiv \frac{3H^2}{8\pi G_N}~.
\end{equation}
The critical density corresponds to the energy density that the universe would possess in case of a vanishing curvature $k=0$. Then by further defining the \textit{cosmological parameters} $\Omega_i$ 
\begin{equation}
\Omega_i \equiv \dfrac{\rho_i}{\rho_{\text{crit}}} ,\qquad  \rho_{k}\equiv-\frac{3k}{8\pi G_N a^2}, \qquad \rho_\Lambda \equiv \frac{\Lambda}{8\pi G_N}~,
\end{equation}
and the total energy density of the universe 
\begin{equation}
\Omega_{\text{tot}}=\sum_{i \neq k} \Omega_i=1-\Omega_k~,
\end{equation}
one can rewrite the relation between the evolution of the Hubble parameter and the several components of the total energy density in the form:

\begin{equation}
H(a)=H_0 \sqrt{\Omega_{m,0}a^{-3}+\Omega_{r,0}a^{-4}+\Omega_{k,0}a^{-2}+\Omega_{\Lambda,0} }~.
\label{eq:Hofa}
\end{equation}
This equation explicits the evolution of the scale factor with time depending on the species dominating the energy density of the universe. For instance assuming that a flat universe ($k=0$) is dominated by a species with an equation-of-state parameter $w \geqslant -1$, the scale factor will always be growing with time and the universe will expand indefinitely:
\begin{equation}
a(t)\propto \left\{
    \begin{split}
    & e^{Ht} \\ 
    & t^{2/3(1+w)} \\    \end{split}
\right.
\quad
\left.
    \begin{split}
    & w=-1\\ 
    & w \neq -1
    \end{split}
\right.
\label{eq:aoft}
\end{equation}
Because the energy density associated to a cosmological constant term is not affected by the redshift caused by the expansion, if such a quantity is present in our universe, from Eq.~(\ref{eq:Hofa}) and Eq.~(\ref{eq:aoft}), independently of the initial matter content, the universe would inexorably lead to a cosmological constant domination at some stage, impliying an infinite exponential expansion of the universe, which seem to correspond to the present stage of our universe as discussed in the following.

\subsection{Measuring the expansion using candles}
\begin{center}
\begin{figure}[h!]
  \begin{minipage}[c]{0.5\textwidth}
\includegraphics[width=\linewidth]{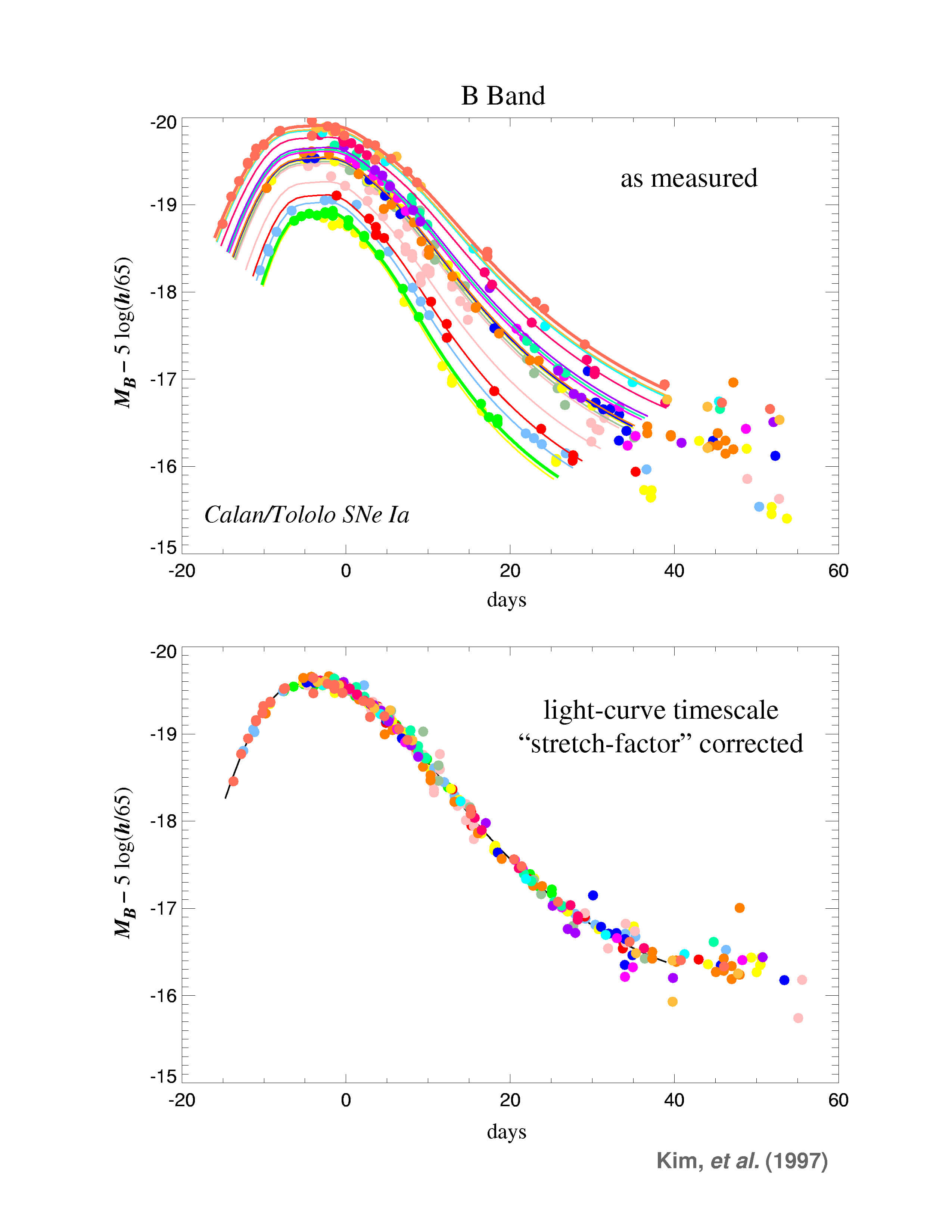}
   \end{minipage}\hfill
   \begin{minipage}[c]{0.5\textwidth}   
\includegraphics[width=0.8\linewidth]{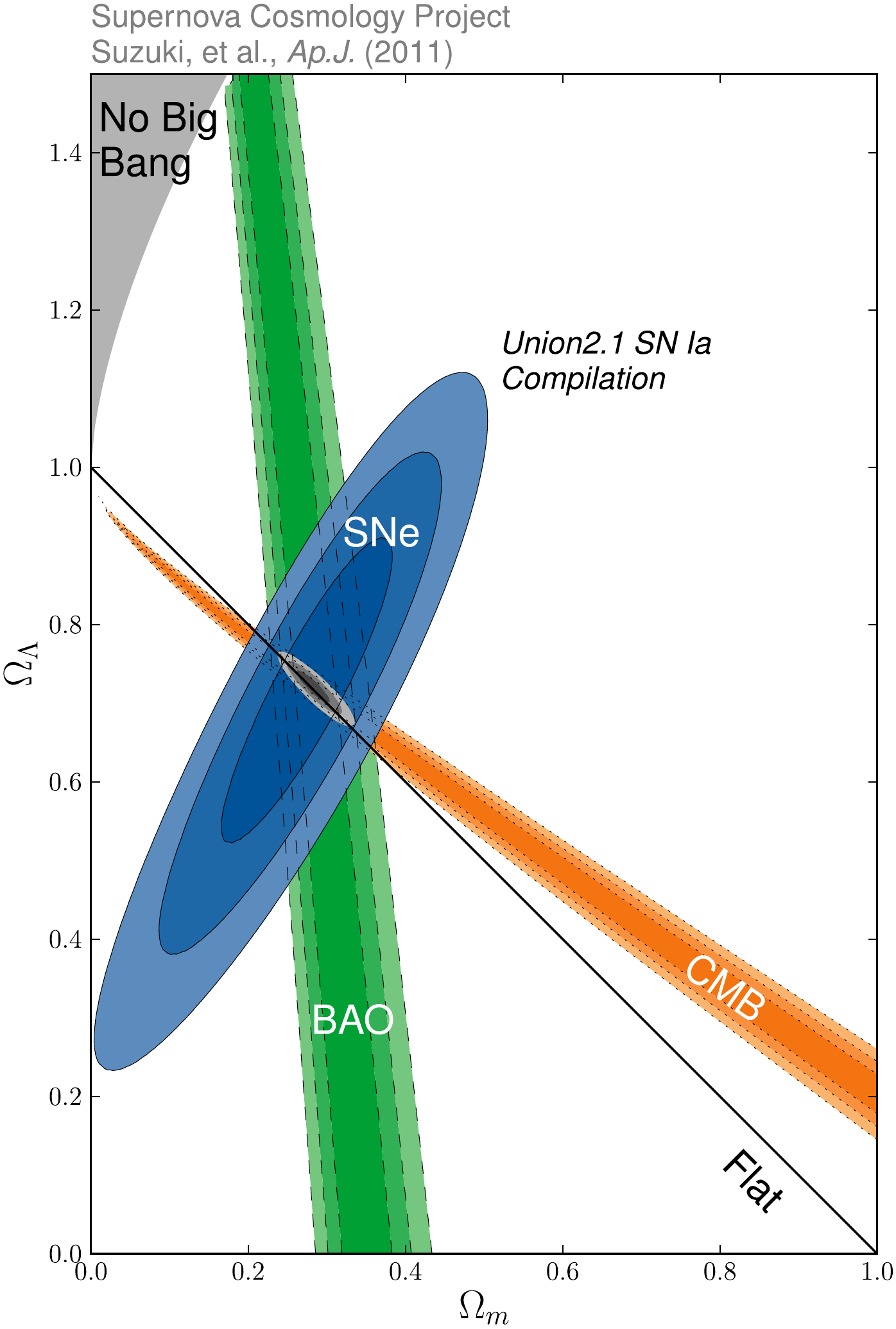}
   \end{minipage}
   \caption{\textbf{Left:} Illustration of the effect of the stretch normalization procedure: light curves of a sample of SNIa on the top pannel and same light curves corrected by the stretch factor on the bottom pannel. \textbf{Right:} Constraints on the cosmological parameters $\Omega_\Lambda$ and $\Omega_m$ in a $\Lambda$CDM universe from the Supernova Cosmology Project}
   \label{fig:stretchSNIa_cosmoparamSNIa}
\end{figure}
\end{center}

In the previous sections we exposed the relation between the energy density and the time evolution of the physical distances between objects in the universe. Notably, any physical event occuring with a strong luminosity at some time in the past should carry enough information for observers to understand how the energy density of the universe evolves along the line of sight from the time of emission until the present time. Type IA Supernovae (SNIa) are particularly interesting objects for this purpose as they are known to be extremely bright. They are typically formed from a white dwarf accretating matter off a nearby star and this brutal event would ignitiate nuclear reactions converting oxygen and carbon to iron, resulting in an explosion of the white dwarf and in one of the brightest astrophysical event that is observable nowadays. SNIa are considered as \textit{standard candles} which means that their maximum absolute magnitudes are nearly identical. Recently, a large sample of 740 supernovae was used in a joined analysis including Low-z SNIa, data from the SuperNovae Legacy Survey (SNLS), the Sloan Digital Sky Survey (SDSS) and the Hubble Space Telescope (HST)~\cite{Betoule:2014frx}. Light curves for each of this large sample of SNIa were measured covering the range between $z=0.01$ and $z=1.2$. For each supernovae the redshift can be deduced by using several possible techniques such as spectral line studies of the host galaxy. The distance modulus $\mu$ of a SNIA can be related to the luminosity distance which depends on the redshift $z$:
\begin{equation}
\mu \equiv m-M=5 \log_{10} \left( \dfrac{d_L(z)}{\text{pc}}\right)-5.00~,
\end{equation}
where $m$ is the apparent magnitude and $M$ the absolute magnitude. The luminosity distance can be related to the various form of energy density present in the universe as follow:
\begin{equation}
d_L(z)=\dfrac{c(1+z)}{H_0 \sqrt{|\Omega_k|}}S_k \left( H_0 \sqrt{|\Omega_k|} \int_{0}^z \dfrac{\d z^\prime}{\sqrt{\Omega_{m,0}(1+z^\prime)^{3}+\Omega_{k,0}(1+z^\prime)^{2}+\Omega_{\Lambda,0} }} \right)~.
\end{equation}
To be more precise, the dispersion of the peak luminosity of supernovae light curves is not as small as we implied, it is still within $\sim 50\%$. However it was realized that luminosity of brighter supernovae decreases slower with time ("\textit{brighter-slower}") and the light curves could be rescaled by taking into account this effect introducing a \textit{stretch parameter} $s$ which can be estimated from supernovae luminosity 15 days after their maximum value. Similarly, one can introduce a \textit{color} parameter $c$ to correct light curves of high or low redshift supernovae which are not measured using the same spectral filter. One can define an effective distance modulus $\mu^\star$ as 
\begin{equation}
\mu^\star=\mu+\alpha(s-1)-\beta c~,
\end{equation}
where $\alpha$ and $\beta$ are parameters. This procedure leads to a reduction of peak luminosity dispersion to $\sim 15\%$ as shown in Fig.~\ref{fig:stretchSNIa_cosmoparamSNIa}.

\begin{figure}[h!]
\begin{center}
\includegraphics[width=0.8\linewidth]{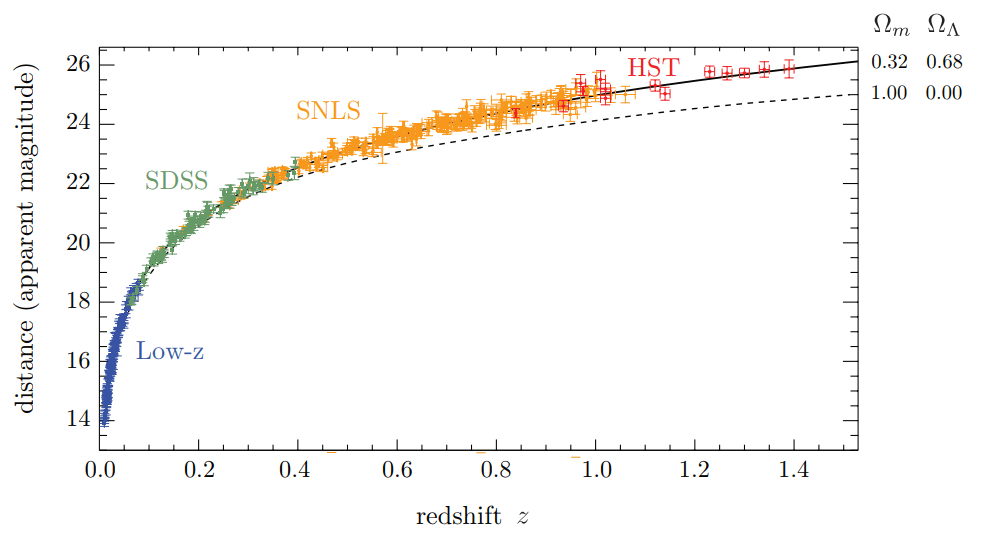}
\caption{Apparent magnitude of a large sample of SNIa, from low to high redshift data from the HST showing the expected behavior in a $100 \%$ matter dominated universe without cosmological contant and the best-fitting curve corresponding a large cosmological constant component.  Illustration taken from~\cite{Baumann}} 
\label{fig:magVSzSNIa}
\end{center}
\end{figure}

In order to derive an estimation of the cosmological parameters using SNIa, one can perform a minimization of the $\chi^2$ function defined as :
\begin{equation}
\chi^2 = \sum_{\text{SNIa}} \left( \dfrac{\mu^\star-5\log_{10}\Big( d_L(\Theta,z)/10~\text{pc}\Big)}{\sigma_{\mu^\star}} \right)^2,
\end{equation}
where $\Theta=\{\Omega_\Lambda,\Omega_m,\alpha,\beta\}$ and $\sigma_{\mu^\star}$ denotes the estimated error on the distance modulus. One of the most important experimental results of modern cosmology, which was derived using the procedure described previously, is the evidence for the domination of a cosmological constant term in the energy budget of the universe of the order of $\Omega_\Lambda\simeq 70\%$ and $\Omega_m \simeq 30\%$ at the present time as illustrated in Fig.~\ref{fig:magVSzSNIa} where we can explicitely see the non-negligible impact of $\Omega_\Lambda$ on the magnitude of large reshift SNIa. Fig.~\ref{fig:stretchSNIa_cosmoparamSNIa} shows the $1,2$ and $3 \sigma$ best-fitting contours in the $\{ \Omega_\Lambda,\Omega_m \}$ plane from the Supernova Cosmology Project (SCP) as well as results from the matter density spectrum and CMB measurements as discussed further on.

As a summary, SNIa light-curves measurements allow for an experimental probe of the acceleration of the expansion of the universe and for these reasons the Nobel Prize was awarded in 2011 to Saul Perlmutter for leading the SCP, and to Adam Riess and Bryan Schmidt for their leading work in the High-z Supernova Search Team (HZT). Results from SNIa are in agreement with estimations based on several other probes of the cosmological parameters and are perfectly compatible with a flat universe ($\Omega_k=0$). However results from SNIa cannot allow for the precise determination of the Dark Matter density but provide a precise estimation of the total matter density including baryons, which indirectly leads to constraints on the DM density when combined with other cosmological parameter estimations. \\
Therefore SNIa are used as a probe of the \textit{standard cosmological model} (alternatively called $\Lambda$CDM~\footnote{$\Lambda$ for a cosmological constant term and CDM for Cold Dark Matter.}) describing a flat universe composed of $\sim 70\%$ of dark energy in the form of a cosmological constant, $\sim 25\%$ of Dark Matter which behaves as non relativistic matter component but effectively does not interact with photons. The photon density only represents a small contribution $\Omega_r \simeq 10^{-4}$ while the remaining $\sim 5\%$ are composed of standard baryonic matter, i.e. the astrophysical structures and objects that are observable in the universe.

\section{Thermal history of the universe}
\subsection{The early universe in a nutshell}
In this section, some important stages of the evolution of the universe are briefly described. Stages denoted between parentheses correspond to theoretical ideas or main paradigm that have not been confirmed experimentally. Sections indicated in brackets will be discussed further on in the corresponding sections.

\paragraph{(Inflation [$T=?, z=?$] [Sec.~\ref{sec:endofinflation}])}was suggested as a solution to the so-called \textit{horizon problem}. This problem is related to the fact that the Cosmic Microwave Background temperature map is almost homogeneous, indicating that all the regions observed in the CMB map does seem to be causally connected at the time of emission\footnote{The CMB map cosmological implications are discussed in Sec.~\ref{sec:CMB}}. However such causal connections cannot be explained in the context of a $\Lambda$CDM universe and one has to invoke supplementary degrees of freedom in order to account for their origins. Causality between two events occuring at a precise space-time position can be quantified using the comoving particle horizon $\chi$ which depends of the conformal time $\eta$ defined as:
\begin{equation}
\chi(\eta)\equiv \int_{a_I}^{a(\eta)}\dfrac{1}{aH}\diff \log a~,
\end{equation}
where $a_I=0$ corresponds to the Big Bang singularity and $(aH)^{-1}$ is the \textit{comoving Hubble radius}. The solution arising from inflation scenarios is based on the fact that two events that does not appear as causally connected at the present time, could have been in the past only if the Hubble radius decreased at some point
\begin{equation}
\dfrac{\diff }{\diff t}(aH)^{-1}<0~.
\end{equation}
Therefore one could explain the apparent not-causally-connected regions of the CMB by ensuring that this condition is satified for some amount of time, causing an extremely fast expansion of the universe. In one of the simplest realization of inflation, one introduces a scalar field dominating the energy budget in the early stages of the universe while slowly rolling down to the minimum of its potential before decaying to Standard Model particles, that would reach a thermal equilibrium state almost immediately.

\paragraph{(Dark Matter freeze-out [$T\sim\text{GeV-TeV}$] [Sec.~\ref{sec:WIMP}])} In the WIMP paradigm, Dark Matter particles in a primordial thermal-equilibrium state with the Standard Model particle content decoupled from the thermal bath while becoming non-relativistic. In such a scenario, the Dark Matter comoving density remains "frozen" until the present epoch.

\paragraph{Electroweak phase transition [$T\sim 200~\text{GeV}$]} All the particles of the Standard Model, except photons, become massive. Subsequently, weak interactions are mediated by the massive $W^\pm$ and $Z$ bosons and therefore, are no longer long-range forces.

\paragraph{QCD phase transition [$T\sim 150~\text{MeV}$]} Strong interactions become effectively strong and reach their non-perturbative regime. Quarks and gluons bind together to form baryons and mesons becoming the relevant degrees of freedom afterwards.

\paragraph{Neutrino decoupling [$T\sim 1~\text{MeV}$] [Sec.~\ref{sec:kindec}]} Neutrino does not interact sufficiently with electrons compared to the Hubble expansion rate to ensure a thermal equilibrium state. This leads to neutrinos decoupling from the Standard Model thermal bath, while still being relativistic.

\paragraph{Electron-positron annihilation [$T\sim 500~\text{keV}$] [Sec.~\ref{sec:kindec}]} Electrons and positrons become non-relativistic while annihilating to photon pairs. Photons and neutrinos become the only relativistic species to propagate in the universe.

\begin{figure}[h]
\begin{center}
\includegraphics[width=0.4\linewidth]{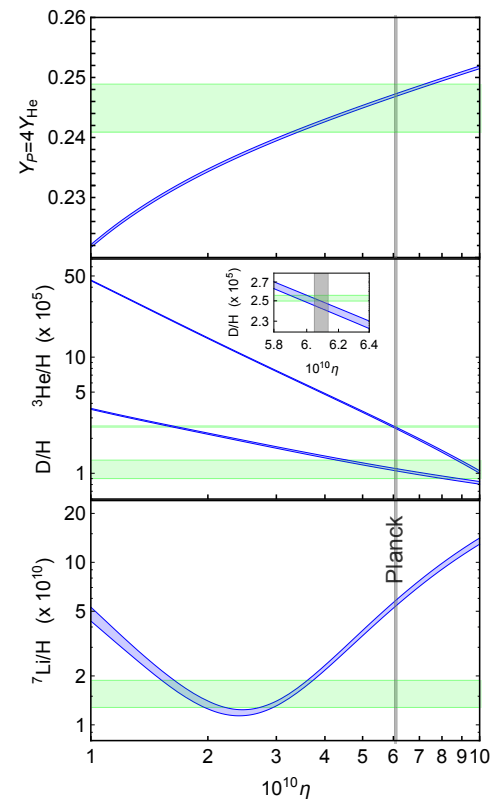}
\caption{Abundance of light nuclei in blue and red as function of the baryon-to-photon density ratio $\eta$ as predicted from BBN computations compared to observed values in green. Figure taken from~\cite{Pitrou:2018cgg}} 
\label{fig:BBN}
\end{center}
\end{figure}

\paragraph{Big Bang Nucleosynthesis [$T\sim 100~\text{keV}, z\sim4\times10^8$]} 
BBN is the stage of the universe corresponding to the formation of light nuclei such as ${}^2 \text{H}, {}^3 \text{H}, {}^3 \text{He}, {}^4 \text{He}, {}^7 \text{Li}$. Since only protons, neutrons, electrons and photons are present after neutrino decoupling, formation of these elements started at first with deuterium production through the only possible two-body process 
 \begin{equation}
 n+p \leftrightarrow {}^2\text{H} + \gamma
 \label{eq:deuteprod}
 \end{equation}
Even though the binding energy of the deuterium $B_{{}^2 \text{H}} \simeq 2.2~\text{MeV}$ is large compared to the temperature $T \lesssim 1~\text{MeV}$, this process is in equilibrium, compensated by the small baryon density relative to photon $\eta \equiv \rho_{\text{b}}/\rho_\gamma \ll 1$ because of the large Boltzmann suppression of the nucleon densities whose relative abundance is
\begin{equation}
\left.\dfrac{n_n}{n_p} \right|_{\text{eq}}\sim e^{-\Delta m/T}e^{-t/\tau_n}~,
\label{eq:neutronoverproton}
\end{equation}
where $\Delta m \equiv m_n-m_p\simeq1.29~\text{MeV}$ is the neutron-proton mass difference and $\tau_n\simeq 886~\text{s}$ is the neutron lifetime. The deuterium relative abundance can be expressed as
\begin{equation}
\left. \dfrac{n_{{}^2\text{H}}}{n_p} \right|_{\text{eq}}\simeq n_n^{\text{eq}}(m_p T)^{-3/2}e^{B_{{}^2\text{H}}/T}~.
\label{eq:deuteoverproton}
\end{equation}
For lower temperatures $T \sim 100~\text{keV}$, the process in Eq.~(\ref{eq:deuteprod}) is enhanced by the exponential factor of Eq.~(\ref{eq:deuteoverproton}) but suppressed by the density of neutrons that starts to efficiently decay. At some point, deuterium is sufficiently produced to initiate production of heavier hydrogen and lithium isotopes. Therefore, light elements production during BBN is very sensitive to the initial relative baryon-to-photon abundance $\eta$ as illustrated in Fig.~\ref{fig:BBN} showing recent BBN computations~\cite{Pitrou:2018cgg} compared to relative abundances measurements and Planck results. Measurement of relative abundances can be realized using several techniques such as observations of the interstellar medium around the Milky Way, studying the Lyman-$\alpha$ forest of distant quasars\footnote{see Sec.~\ref{sec:BAO} for more details.} or proto-stellar material in the Solar System. 
Beside the overall agreement between computations and observations, the only disagreement is present in the ${}^7\text{Li}$ abundance, known as the \textit{Lithium problem} which remains still unexplained nowdays.

\paragraph{Matter-radiation equality [$T\sim~0.7~\text{eV}, z\sim3400$]} Non-relativistic matter dominates the energy budget of the universe. The later period is called the \textit{dark ages}.

\paragraph{Recombination and Photon decoupling [$T\sim 0.3~\text{eV}$, $z\sim 1200$] [Sec.~\ref{sec:CMB}]} Electrons and $p^+$ recombine to form neutral hydrogen, the universe becomes transparent and the first image of the universe is emitted and propagates in every direction also known as the Comic Microwave Background.

\paragraph{Reionization [$T\sim5~\text{meV}, z\sim10$]} The first stars and galaxies form, heating and reonizing the hydrogen gas.

\paragraph{$\Lambda$-matter equality [$T\sim0.3~\text{meV},z\sim9$]} The energy density of the universe becomes dominated by dark energy, causing an acceleration of the expansion.

\paragraph{Present time [$T\sim0.24~\text{meV},z=0$]}

\subsection{Thermodynamics of the primordial thermal bath}
After the period of inflation that occured at early times, our expanding universe was filled with Standard Model particles interacting between them and forming a thermal bath. In the following section, after describing the last moments of the inflationary period, we introduce some thermodynamical definitions and quantities useful to describe the several consitutents of the thermal bath formed by these particles and discuss the notion of thermal decoupling, relevant for the following chapters.

\subsubsection{The end of inflation}
\label{sec:endofinflation}
\begin{center}
\begin{figure}[h!]
  \begin{minipage}[c]{0.5\textwidth}
\includegraphics[width=0.95\linewidth]{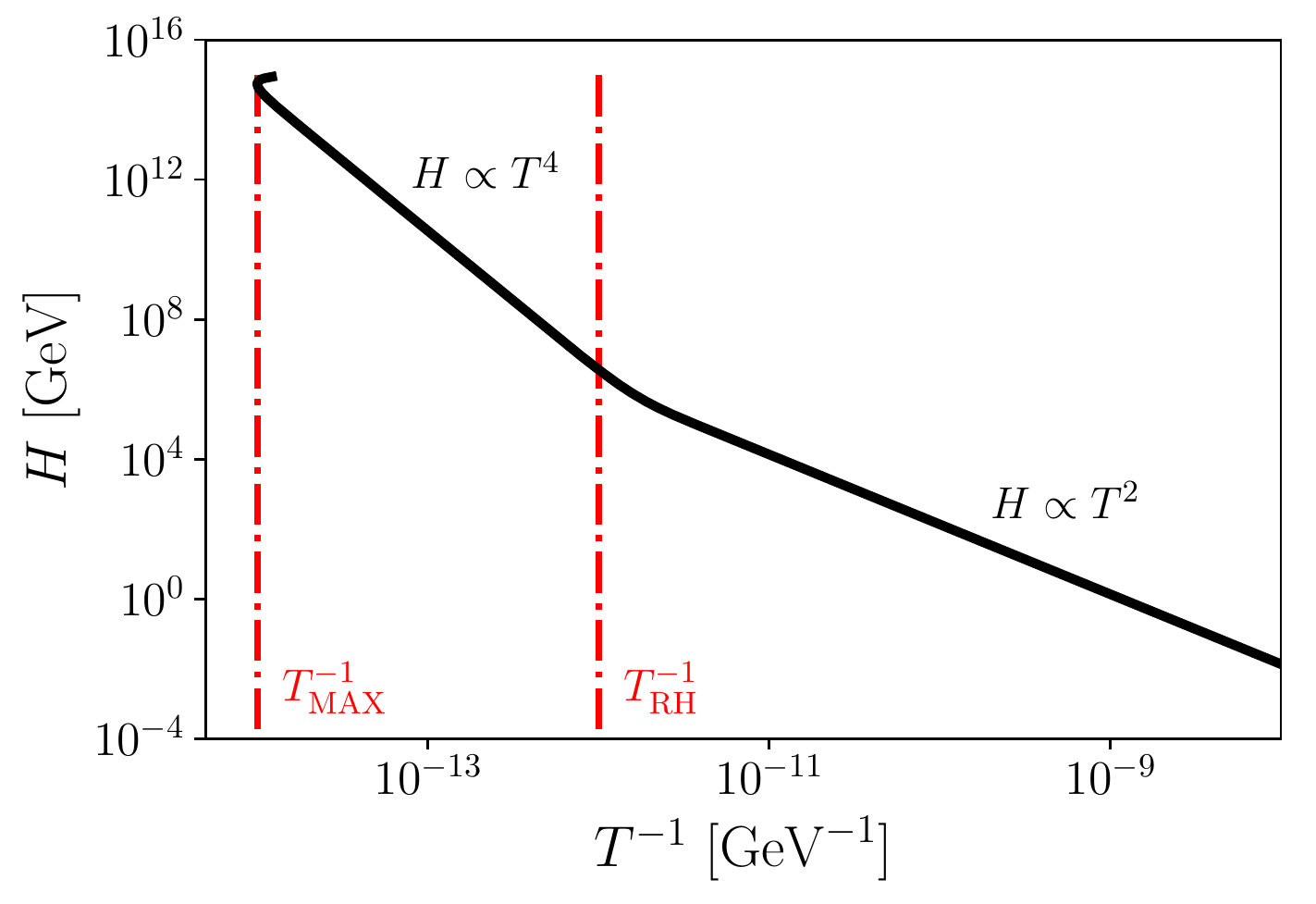}
   \end{minipage}\hfill
   \begin{minipage}[c]{0.5\textwidth}   
\includegraphics[width=0.95\linewidth]{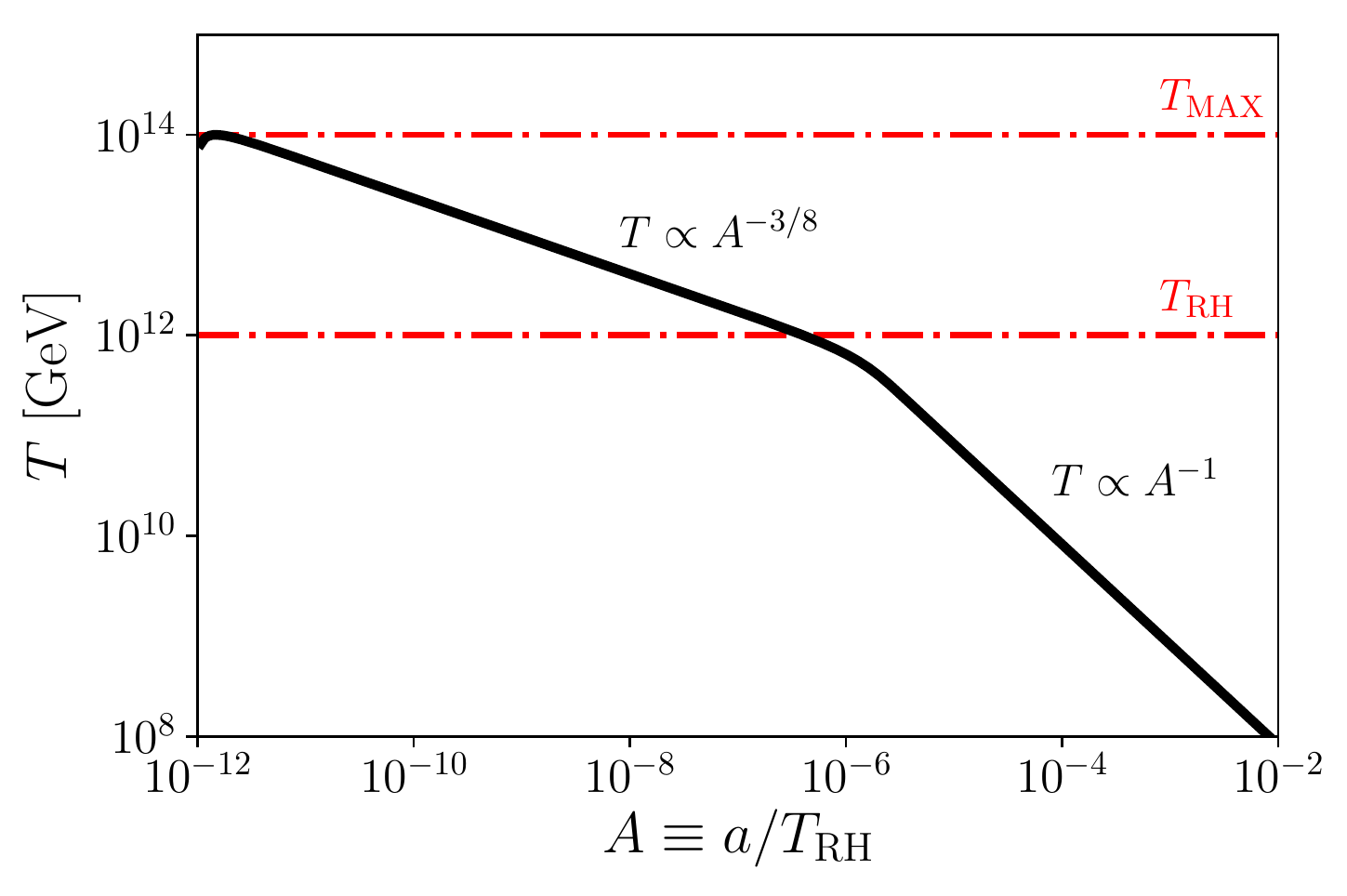}
   \end{minipage}
   \caption{Numerical solutions of the system of Boltzmann equations~(\ref{Eq:setboltzmannpreheating}) showing the relative behavior of the quantites $H, T$ and $a$ in the transition period between the inflationary epoch and radiation domination era.}
   \label{fig:reheatingBoltzmann}
\end{figure}
\end{center}

As discussed in the previous section, a scalar field dominating the energy budget of the universe is still nowadays the main paradigm invoked to explain inflation. It is commonly assumed that when the scalar field responsible for the inflation, the so-called \textit{inflaton} denoted by $\phi$, oscillates around the minimum of its potential, it would eventually decay to Standard Model particles in the simplest realization of inflation. In this case the evolution of the inflation energy density $\rho_\phi$ and energy density of the SM particle content $\rho_{\rm R}$\footnote{Here we assume that all the particles present in the Standard Model are relativistic at the scale considered.} are given by a set of coupled Boltzmann equations:
\begin{align}
\frac{\text{d}\rho_{\rm R}}{\d t}&=-4H\,\rho_{\rm R}+\Gamma_\phi\,\rho_\phi\,, \nonumber \\
\frac{\text{d}\rho_\phi}{\d t}&=-3H\,\rho_\phi-\Gamma_\phi\,\rho_\phi\,,
\label{Eq:setboltzmannpreheating}
\end{align}
where $\Gamma_\phi$ is the decay width of the inflaton. The Hubble expansion rate is given by:
\begin{equation}
H^2=\dfrac{8 \pi}{3 M_{\rm Pl}^2}(\rho_\phi+\rho_R)
\end{equation}
As discussed in the following sections, at such scales the Standard Model particle content reaches a thermal equilibrium state and $\rho_{\rm R}$ can be related to the temperature of the thermal bath. Based on a naive dimensional analysis, since the temperature is the only relevant scale we must have $\rho_{\rm R}\propto T^4$. In Fig.~\ref{fig:reheatingBoltzmann} a numerical solution of Eqs.~\ref{Eq:setboltzmannpreheating} is represented, showing that a maximal temperature $T_{\rm MAX}$ is reached at the early stages of the universe after inflation. We can distinguish two important regimes:
\begin{itemize}
\item \underline{The inflaton domination era:} At the earliest times, the inflaton is dominating the energy density of the universe and in this regime, one can show the following relations between the Hubble rate, the temperature and the scale factor $a$~\cite{Giudice:2000ex}
\begin{equation}
H(T)\propto T^{4}~, \qquad \text{and} \qquad T \propto a^{-3/8}~.
\end{equation}
\item \underline{The radiation domination era:} After the so-called \textit{reheating temperature} $T_{\rm RH}$, the Hubble expansion becomes dominated by the radiation energy density and the following relations hold:
\begin{equation}
H(T)\propto \rho_{\rm R}^{1/2} \propto T^{2}~, \qquad \text{and} \qquad T\propto a^{-1}~.
\end{equation}
\end{itemize} 
\subsubsection{Thermal equilibrium}
We want to describe the physics of the Standard Model thermal bath in the radiation domination era in order to understand the several steps of the formation of the present universe and emphasize the role of Dark Matter in this process. In order to do so, we need to discuss the notion of \textit{thermal equilibrium} which can be defined by the following two conditions : \textit{kinetic equilibrium} and \textit{chemical equilibrium}.\\ \newline
\underline{Kinetic equilibrium} between particles can be reached if the interaction rate of processes involving these particles (such as $e^+ \gamma \rightarrow e^+ \gamma$ for instance) is large enough such that their typical momenta will be redistributed homogeneously, allowing to describe them as a single system defined by a temperature $T$. If the kinetic equilibrium condition is satisfied, the phase space distribution $f_i$ of the species $i$ is described by a Fermi-Dirac (FD) or Bose-Einstein (BE) distribution depending on its spin:
\begin{equation}
f_i(p)=g_i ~ \frac{1}{e^{(E-\mu_i)/T}\pm 1}~,
\end{equation}
where $E$ is the energy of the particle with momentum $p$ and mass $m$ such that $E=\sqrt{m^2+p^2}$, $\mu_i$ is the chemical potential and $g_i$ the number of degree of freedom of the species $i$. ($+$) and  ($-$) correspond respectively to the FD and BE distributions. The phase space distribution depends only on the modulus of the momentum, according to the isotropy hypothesis, but not on position space because of the space homogeneity requirement.
Although not directly deductible from the previous expression, the phase space distribution depends actually on time through the time-evolution of the temperature as will be discussed further on. In the non-relativistic or classical limit, both can be approximated by the Maxwell-Boltzmann distribution : $f_i(p)\simeq g_i e	^{-\frac{E-\mu}{T}}$.\\ \newline
\underline{Chemical equilibrium:} Assuming number changing processes such as $A + B \leftrightarrow C + D $, the chemical equilibrium condition can be expressed as a relation on the chemical potential of the species involved:
\begin{equation}
\mu_A+\mu_B=\mu_C+\mu_D~.
\end{equation}
This condition must be satisfied for every number changing process involving $A, B, C$ and $D$ for the chemical equilibrium condition to hold. By definition, the chemical potential characterizes the modification of the energy density of a system when the number of particles is not conserved. Therefore one would expect that chemical equilibrium is reached when number-changing processes and their reverse processes occur at the same rate. For instance equilibration of the forward and backward rates of the processes $e^+ e^- \leftrightarrow 2 \gamma$ and $4\gamma \leftrightarrow 2 \gamma$ imply:
\begin{equation}
\mu_{e^+}+\mu_{e^-}=2\mu_\gamma \quad \text{and} \quad 4\mu_\gamma=2\mu_\gamma  \quad \rightarrow \quad \mu_\gamma=0 \quad \text{and} \quad \mu_{e^+}=-\mu_{e^-}~,
\end{equation}
which shows that the chemical potential of photons vanishes in a thermal bath and also that chemical potentials of particle and corresponding anti-particles are opposed.

\subsubsection{Thermal equilibrium condition}
The phase space density of Standard Model particles can be affected by two effects, the expansion of the universe and interactions between particles. Expansion tends to reduce the total number density by diluting space and interactions drive the distribution functions towards Bose-Einstein and Fermi-Dirac thermal distributions. If the expansion rate were too large, Standard Model particles would not have enough time to interact efficiently with each other and would not acquire thermal distributions. The Hubble expansion rate can be understood, at first order, as the inverse of the time $\Delta t$ taken for the universe to expand by a factor $\Delta a$ such that $H \simeq (\Delta a/a)(1/\Delta t)$. Taking $\Delta a \sim a$ yields $\Delta t \simeq H^{-1}$ which indicates that the Hubble rate is an estimator of the time needed for the size of the universe to be multipled by a factor 2 under the effect of expansion. Therefore, an interaction rate $\Gamma(T)$ larger than the Hubble rate implies typically that interactions cannot occur before the expansion drives the particles apart, forbidding them to redistribute their momenta and ensure a kinetic equilibrium state. This yields the following approximated kinetic equilibrium condition:
\begin{equation}
H(T) \lesssim \Gamma(T)~.
\label{eq:kineqcondition}
\end{equation}
Using naïve dimensional analysis, at high temperature the interaction rate will be the product of the number density $n$ by the typical scattering cross section $\sigma$:
\begin{equation}
\Gamma \sim \sigma n~,
\end{equation}
where $n$ scales as $n\sim T^3$, $\sigma\sim \alpha^2/T^2$ and $\alpha\sim 10^{-1}$ is roughly of the order of the fine structure constant. The Hubble expansion rate is given by
\begin{equation}
 H \sim \dfrac{T^2}{M_{\text{Pl}}}~,
\end{equation}
with $M_{\text{Pl}}$ the reduced Planck mass. Injecting the previous relations in the kinetic equilibrium condition of Eq.~(\ref{eq:kineqcondition}) gives :
\begin{equation}
T \lesssim \alpha^2 M_{\text{Pl}} \simeq 10^{16}~\text{GeV}
\end{equation}
As a result, kinetic equilibrium is guaranteed for temperatures much below the Planck scale $T \ll M_{\text{Pl}}$, regime which will be considered in the following. The chemical equilibrium condition will be satisfied as well as the typical cross section for number-changing processes will behave in the same way that scattering cross sections at high temperatures, therefore validating the thermal equilibrium condition for the Standard Model particles during the radiation domination era.

\subsubsection{Entropy and energy density}
\label{sec:entropyandenergydensity}
Provided that thermal equilibrium is satisfied and the notion of temperature can be defined, we can apply standard results of thermodynamics to describe the primordial Standard Model thermal bath. We can define the notion of entropy $S$ from the second principle of thermodynamics $T\diff S \equiv \diff U+P \diff V- \mu \diff N $ where $U$ and $N$ are respectively the total energy and number of particles in the system, $V$ the total volume and $P$ the pressure. Using the fact that $S$, $V$, $N$ and $U$ are extensive variables such that $\frac{\partial U}{\partial V}=\frac{U}{V}$, one can derive the following definition of the entropy density $s=S/V$:
\begin{equation}
s=\frac{\rho+P-\mu n}{T}~,
\end{equation}
where $\rho=U/V$ is the energy density, $P$ the pressure and $n \equiv N/V$ the number density. In absence of interactions, one can show that entropy density is conserved from the condition of energy-momentum conservation $\nabla_\mu T^{\mu \nu}=0$. Using the second principle of thermodynamics we can write the relation
\begin{equation}
\diff \rho=T \diff s +\dfrac{\diff V}{V}\Big( Ts - (P+\rho) \Big)=T \diff s~.
\end{equation}
Injecting the previous relation in the continuity relation~(\ref{eq:continuityrelation}) derived from energy-momentum conservation yields 
\begin{equation}
T \left( \dfrac{\diff s}{\diff t}+3Hs \right)=0 \quad \rightarrow \quad \dfrac{\diff (sa^3)}{\diff t}=0 \quad \rightarrow \quad \dfrac{\diff S}{\diff t}=0~.
\end{equation}
which demonstrates entropy conservation in absence of interactions between the thermal bath and external sources. The contribution of a species $i$ to the quantities $n$, $\rho$ and $P$ can be expressed as integrals of the phase space distribution over all possible momenta 
\begin{equation}
n_i=g_i \int f(p) \dfrac{\diff ^3 p}{(2\pi)^3}~, \qquad
\rho_i=g_i \int E f(p) \dfrac{\diff ^3 p}{(2\pi)^3}~, \qquad
P_i=g_i\int \dfrac{p^2}{3E}f(p) \dfrac{\diff ^3 p}{(2\pi)^3}~.
\end{equation}
One can derive analytical expressions of these quantities in some specific regimes :\\ \newline
\underline{The relativistic limit:} in the regime where masses and chemical potentials are negligible ($m \ll T$ and $\mu \ll T$) :
\begin{equation}
n_i=\left(\frac{3}{4}\right)\frac{\zeta(3)}{\pi^2}g_iT^3~,
\end{equation}
\begin{equation}
\rho_i=\left(\frac{7}{8}\right)\frac{\pi^2}{30}g_iT^4~,
\end{equation}
\begin{equation}
P_i=\dfrac{1}{3}\rho_i~,
\end{equation}
where the factors $(3/4)$ and $(7/8)$ are only present for fermions and should be substitued by $1$ for a bosonic species. \\ \newline
\underline{The non-relativistic limit:}($T \ll m$) densities of massive particles are exponentially suppressed at low temperatures
\begin{equation}
n_i=g_i\left(\frac{mT}{2\pi}\right)^{3/2}e^{-(m-\mu_i)/T}~,
\end{equation}
\begin{equation}
\rho_i = mn_i~,
\end{equation}
\begin{equation}
P_i=n_i T~,
\end{equation}
In order to derive the total radiation energy density of a system composed of several species, one has to sum over each contribution:
\begin{equation}
\rho=\sum_i \rho_i=\dfrac{\pi^2}{30}g_\star(T)T^4~,
\end{equation}
where $g_\star(T)$ is the effective number of relativistic degrees of freedom. Similarly one can define the total entropy as a sum of every possible contribution
\begin{equation}
s=\sum_i s_i=\dfrac{2\pi^2}{45}g_{\star,s}(T)T^3~,
\end{equation} 
The entropy conservation condition leads to the following result:
\begin{equation}
g_{\star,s}(T)T^3 a^3=~\text{constant}\quad  \rightarrow \quad a \propto \dfrac{1}{g_{\star,s}^{1/3}T}\simeq \dfrac{1}{T}~,
\end{equation}
where we assumed a mild temperature dependance of $g_{\star,s}$. This is an important result which allows to treat $1/T$ as a time parameter in the radiation era. \\
In the Standard Model, at high temperature, well above the electroweak symmetry breaking scale $v_{\text{EW}}$, all the particles are massless and the effective number of relativistic degrees of freedom is given by the sum of all the degrees of freedom present in the Standard Model which is:
\begin{equation}
g_\star(T>v_{\text{EW}})=\dfrac{7}{8}\left( N_F \times 2 \right)+N_V \times 2 + N_{\text{scalar}}=106.75~,
\end{equation}
where $N_F=45$ is the number of left and right-handed fermions, $N_V=12$ is the number of vector fields and $N_{\text{scalar}}=4$ the number of scalar fields present in the Standard Model. The factors of "2" denote the number of polarization states for fermions and massless vectors states.
\begin{figure}[h!]
\begin{center}
\includegraphics[width=0.7\linewidth]{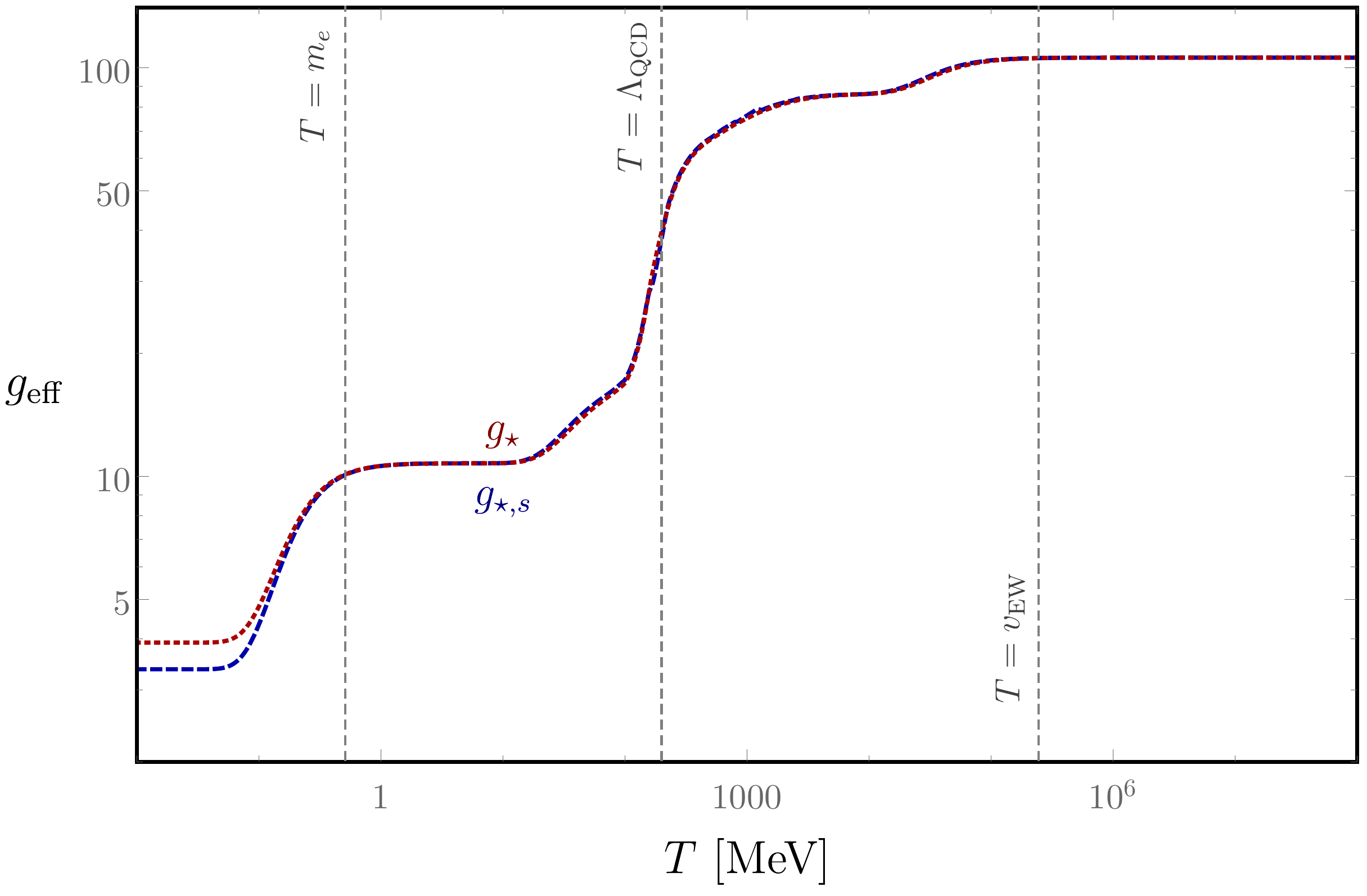}
\caption{Evolution of the effective relativistic degrees of freedom in the Standard Model with the temperature of the photon thermal bath.} 
\label{fig:gstarVST}
\end{center}
\end{figure}
In Fig.~\ref{fig:gstarVST}, the evolution of the effective number of degrees of freedom of the SM $g_{\text{eff}}=\{g_\star,g_{\star,s}\}$ with temperature is depicted. Starting from a high temperature $T \gg v_{\text{EW}}$, $g_{\text{eff}}$ decreases after the temperature drops below $v_{\text{EW}}$ due to the fact that masses are generated in the SM and the number of particles such as the Higgs and the top-quark will exponentially be suppressed when the temperature becomes smaller than their mass, reducing the total value of $g_{\text{eff}}$. Another major drop at $T\lesssim 1$ MeV in this plot corresponds to the same effect happening when the temperature drops below the electron mass $m_e$. Around $T\sim 200~\text{MeV}=\Lambda_{\text{QCD}}$, $g_{\text{eff}}$ are drastically reduced, caused by the QCD phase transition binding quarks and gluons together. Below $\Lambda_{\text{QCD}}$ the propagating states are color-neutral composite states (baryons and mesons) and all the color degrees of freedoms of the Standard Model become inaccessible. At low temperatures $T \ll m_e$, only photons and neutrinos contribute to the effective number of degrees of freedom which is of the order $g_{\text{eff}}(T\ll 1~\text{MeV})\sim 3-4$.

\subsubsection{Kinetic decoupling}
\label{sec:kindec}
The kinetic decoupling time of a species in equilibrium with a thermal bath can be defined as the time for which the interaction rate of this species becomes insufficient compared to the Hubble expansion rate to ensure kinetic equilibrium. From this time, the considered species would \textit{decouple} from the thermal bath, evolving freely with the expansion of the universe afterwards. This concept is particularly important for early universe cosmology and for Dark Matter phenomenology therefore we investigate the neutrino decoupling in the Standard Model as an illustration of this effect. Neutrinos interact with leptons and quarks in the Standard Model only via weak interactions with processes such as
\begin{equation}
\bar{\nu}_\ell + \nu_\ell \leftrightarrow \bar{\ell}+\ell \quad \text{and} \quad
\nu_\ell+\ell \leftrightarrow \nu_\ell+\ell~,
\end{equation}
for which the typical interaction cross section is given by $\sigma \sim (G_F T)^2$ where $G_F \sim 10^{-5}~\text{GeV}^{-2}$ is the Fermi constant. An order of magnitude of the interaction rate is thus given by
\begin{equation}
\Gamma \sim G_F^2 T^5~.
\end{equation} 
Using the kinetic equilibrium condition stated in Eq.~(\ref{eq:kineqcondition}), one can estimate the neutrino decoupling temperature $T_{\text{d}}$ as 
\begin{equation}
T_{\text{d}}\simeq \Big( G_F^2 M_{\text{Pl}} \Big)^{-1/3} \simeq 1~\text{MeV}~.
\end{equation}

\begin{figure}[h!]
\begin{center}
\includegraphics[width=0.6\linewidth]{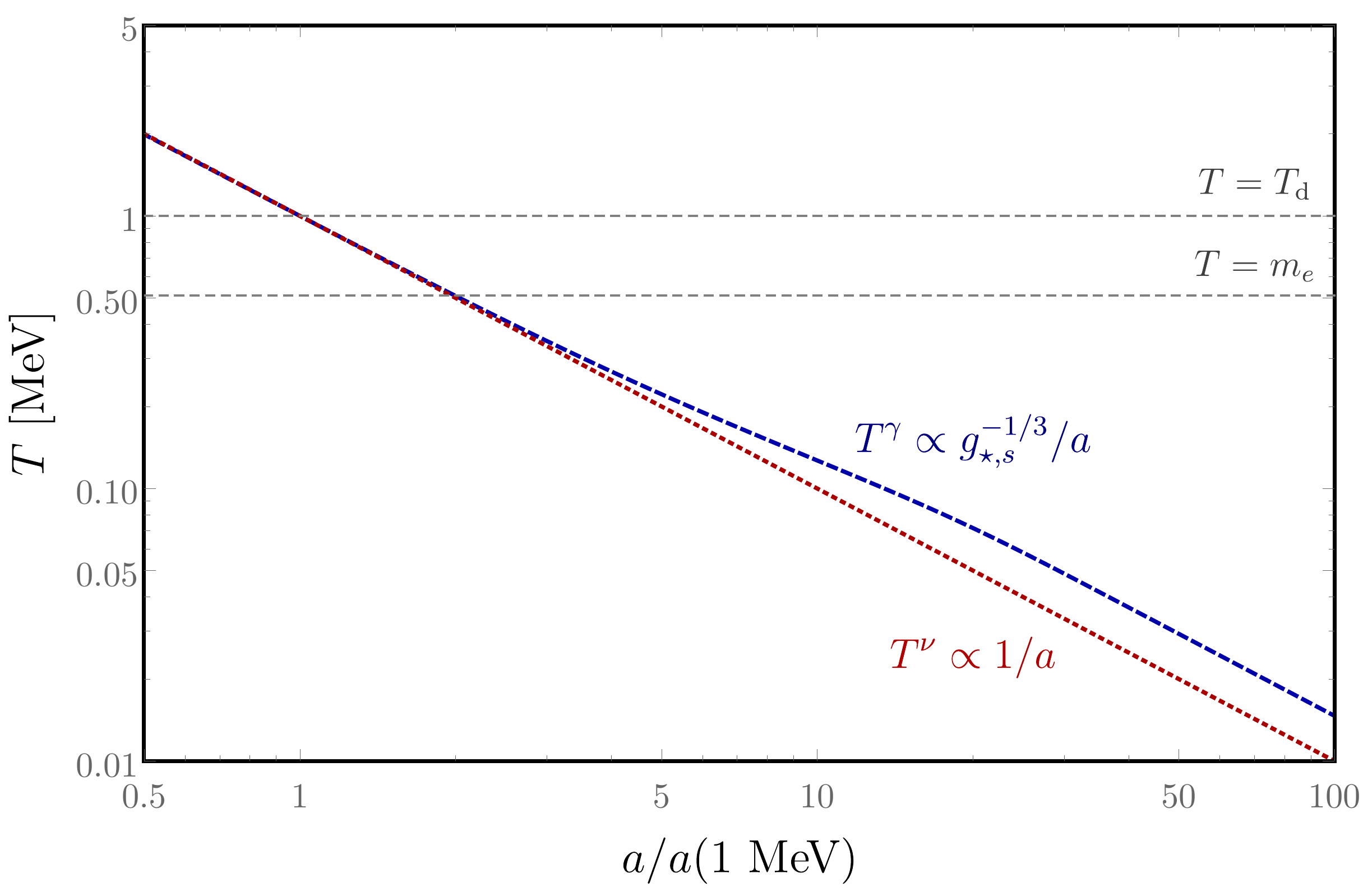}
\caption{Scale factor dependence of the temperature of the photons $T^\gamma$ and neutrinos $T^\nu$ after the neutrino decouling temperature $T_{\text{d}}\simeq~\text{MeV}$} 
\label{fig:TVSa}
\end{center}
\end{figure}

As a result, neutrinos decoupled from the SM particle content while being still relativistic as their masses are sub-eV, therefore after $T_{\text{d}}$ their distributions were affected only by the expansion until the present time keeping the shape of a Fermi-Dirac distribution even though neutrinos are non-relativistic nowadays. After the neutrino decoupling temperature, the SM thermal bath is composed only of photons and electrons. When the temperature drops below the electron mass $T\lesssim m_e$, the annihilation $\gamma \gamma \rightarrow e^+e^-$ becomes exponentially suppressed compared to the reverse process. This results in a \textit{reheating} of the photon temperature compared to the neutrino temperature. One can show this effect using conservation of entropy between $T_{\text{d}}$ and some temperature $T_{\text{f}}\ll m_e$ in the $\gamma$ system comprised of $e^- e^+$ as well and conservation of entropy in a system $\nu$ composed of the three neutrino families
\begin{align}
g_{\star,s}^\gamma(T_{\text{d}})T_{\text{d}}^3= g_{\star,s}^\gamma(T_{\text{f}}^\gamma)(T_{\text{f}}^\gamma)^3~,\\
g_{\star,s}^\nu(T_{\text{d}})T_{\text{d}}^3=g_{\star,s}^\gamma(T_{\text{f}}^\nu)(T_{\text{f}}^\nu)^3~,
\end{align}
where $g_{\star,s}^{\gamma}$ and $g_{\star,s}^{\nu}$ are the effective degrees of freedom of the $\gamma$ and $\nu$ systems respectively. Notice the difference in the temperatures $T_{\text{f}}^\nu$ and $T_{\text{f}}^\gamma$. The degrees of freedom evaluated at $T_{\text{d}}$ are
\begin{equation}
g_{\star,s}^\gamma(T_{\text{d}})=\dfrac{11}{2} \quad \text{and} \quad g_{\star,s}^\nu(T_{\text{d}})=\dfrac{21}{4}~.
\end{equation}
The effective number of degrees of freedom is conserved in the $\nu$ system and only photons contribute in the $\gamma$ system at $T_{\text{f}} \ll m_e$, giving
\begin{equation}
g_{\star,s}^\gamma(T_{\text{f}}^\gamma)=2 \quad \text{and} \quad g_{\star,s}^\nu(T_{\text{f}}^\nu)=\dfrac{21}{4}~.
\end{equation}
The relation between photon and neutrino temperatures can be derived straightforwardly 
\begin{equation} 
T_{\text{f}}^\gamma=\left( \dfrac{g_{\star,s}^\gamma(T_{\text{d}})}{g_{\star,s}^\gamma(T_{\text{f}}^\gamma)} \right)^{1/3} T_{\text{f}}^\nu = \left( \frac{11}{4} \right)^{1/3} T_{\text{f}}^\nu~,
\end{equation}
Therefore we expect the neutrino temperature to be slightly smaller than the photon one as illustrated in Fig.~\ref{fig:TVSa} where the temperature evolution of the $\gamma$ and $\nu$ systems are shown as a function of the scale factor based on entropy conservation. This figure shows that photons inherit some energy density from the electrons compared to the neutrinos.\\
\subsubsection{Conclusion}
After describing the various stages of the universe, we presented some important thermodynamics quantities relevant for describing the universe in the radiation domination era and we emphasized the notion of kinetic decoupling which is essential to be understood in the context of Dark Matter phenomenology, as discussed in the following chapters.

%% file: parts/evidences.tex
\vspace{0.3cm}

\noindent
The Dark Matter conundrum is actually a long standing issue that  has puzzled physicists for almost a century and has been accepted only in the few past decades. After reviewing the initial hints of a missing mass problem that has emerged in the first part of the twentieth century, we expose in this chapter the various probes and observations that lead to the conclusion that a Dark Matter component is present in our universe based on modern measurements and theory developments.

\section{Pioneers of the dark universe : historical approach}
More than one hundred years ago, after working on the mathematical formulation of special relativity, Henri Poincaré was impressed by Lord Kelvin's idea of applying the recently elaborated kinetic theory of gases to astrophysical systems. He applied this theory to the Milky Way by computing the number of stars that our galaxy should host in order to reproduce the sun velocity with respect to the galactic center and he compared with an estimation based on observations from this epoch~\cite{1906PA.....14..475P}. His conclusion was he following: "\textit{then there is no Dark Matter, or at least not so much as there is of shining matter}"\footnote{For futher reading regarding historical aspects, a detailed review can be found in~\cite{Bertone:2016nfn}}. There was no Dark Matter problem at that time but obviously he had absolutely no chance to guess that one of the most puzzling problem in physics one hundred years later would be related to his erroneous conclusion. However his pioneering work opened the way to improved studies of this kind by the Dutch physicist Jacobus Kapteyn who was the first to propose a theory of arrangement and motion of stars in the galaxy~\cite{1922CMWCI.230....1K}. He had the idea of decomposing the galaxy in spherical shells and applied the kinetic theory of gases relating the distribution of mass to the velocity dispersion of the stars. He understood that one could determine the Dark Matter density by comparing the effective mass of stars to the luminosity-curve which is summarized in the sentence: "\textit{We therefore have a means of estimating the Dark Matter mass in the universe}".
The work of Kapteyn led his former student, Jan Oort, to study stellar kinematics in order to estimate the total matter density in the vicinity of the Sun~\cite{1932BAN.....6..249O}. He reached the conclusion that some non-negligible amount of Dark Matter must be present to account for the observed stellar dynamics: "\textit{There is an indication that the invisible mass is more strongly concentrated to the galactic plane that of the visible stars}".\\
Later in the 1920's, Edwin Hubble studied the relation between redshift of galaxies and their distances to the earth. The redshift was measured from spectral lines displacement and distances were determined using various methods such as Cepheid for instance. Cepheids are periodically changing luminosity stars whose period is tightly related to the maximum luminosity, therefore measuring their period and luminosity provide informations regarding their distances to the observer. Edwin Hubble deduced the famous linear relation between redshift and distances~\cite{1929PNAS...15..168H}. A couple of years later Fritz Zwicky, aware of the recent results by Hubble, studied particular velocities of galaxies inside galaxy clusters, in particular the Coma cluster located $\sim 10^{10}$ light years away from our galaxy. Zwicky, assuming the Coma cluster in a stationary state, used the Virial theorem to relate the velocity dispersion $\bar{v^2}$ of the galaxies to the average density of the cluster~\cite{1937ApJ....86..217Z}.
\begin{equation}
\bar{v^2}\simeq \dfrac{3G_N M}{5R}~.
\end{equation}
He derived the conclusion that the Coma system must be at least 400 times larger than the value inferred from observation of luminous matter in order to explain such a large velocity dispersion. He suspected that a Dark Matter halo was present in a much larger amount than the luminous matter and popularized the \textit{Dark Matter} denomination.
\section{Rotation curves}
\begin{center}
\begin{figure}[h!]
  \begin{minipage}[c]{0.5\textwidth}
\includegraphics[width=0.85\linewidth]{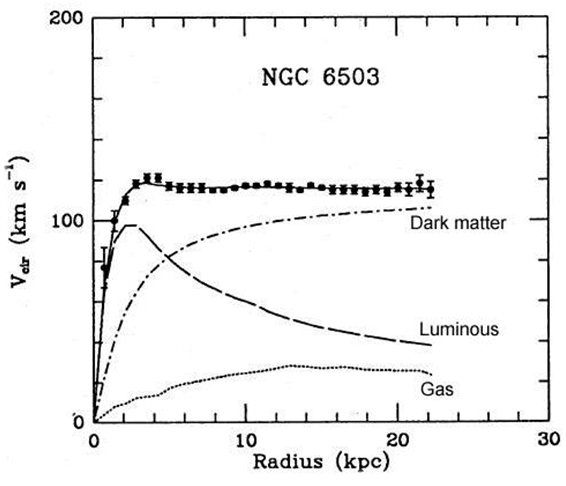}
   \end{minipage}\hfill
   \begin{minipage}[c]{0.5\textwidth}   
\includegraphics[width=\linewidth]{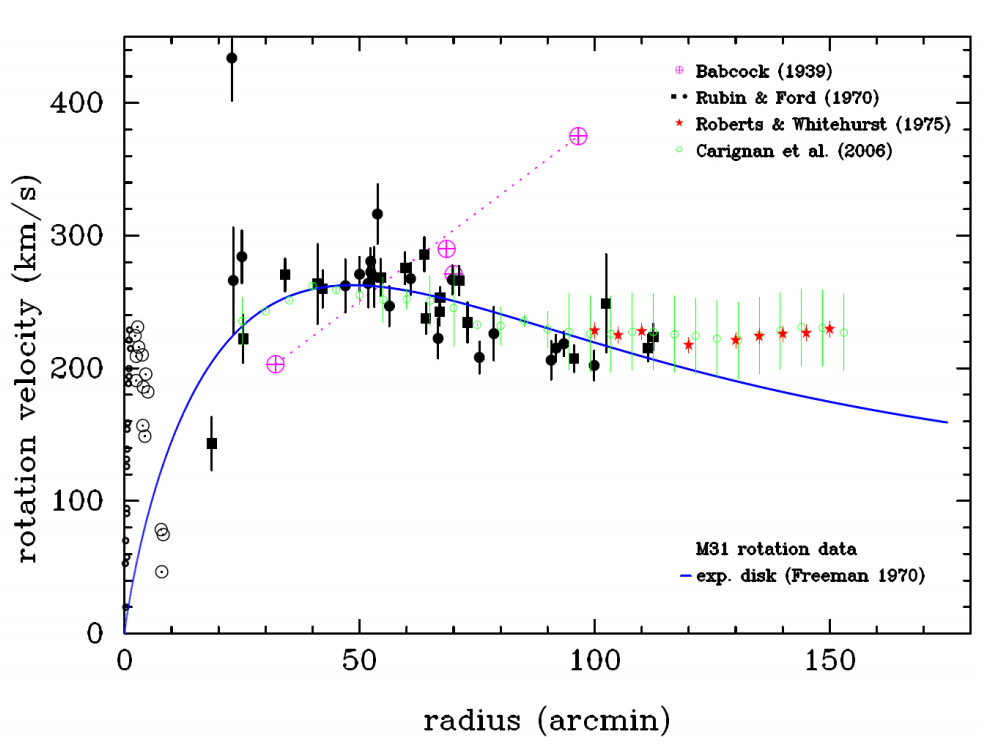}
   \end{minipage}
   \caption{\textbf{Left:} Rotation curve of the NGC6503 galaxy, showing contributions of the gas, luminous matter (stars) and a Dark Matter component fitting the experimental data. Taken from~\cite{1991MNRAS.249..523B}. \textbf{Right:} Rotation curve of M31 showing measurements from Babcock~\cite{1939LicOB..19...41B}, Rubin and Ford~\cite{1970ApJ...159..379R}, Roberts and Whitehurst~\cite{1975ApJ...201..327R} and Carigan et al.~\cite{2006ApJ...641L.109C}. The blue line correspond to Freeman prediction assuming a exponential disk~\cite{1970ApJ...160..811F}. Figure taken from~\cite{Bertone:2016nfn}}
\label{fig:rotationcurve}
\end{figure}
\end{center}

The study of rotation curves has been playing a very important role in the acceptance of the existence of a missing mass problem by the scientific community. Horace Babcock was the first to extend the study of the rotation curve of the Andromeda (M31) galaxy to the most outer regions~\cite{1939LicOB..19...41B}. The rotation curve of a galaxy represents the evolution of the radial velocity $v(r)$  of stars with respect to their distances from the center of the galaxy. Assuming a spherical matter distribution in the galaxy, according to Newtonian dynamics, one can relate the gravitational potential $\Phi(r)$ to the matter density $\rho(r)$ through the Poisson equation  
\begin{equation}
\Delta \Phi(r)=4\pi G_N \rho(r)~,
\end{equation}
and therefore deduce the expression of the radial velocity at a distance $r$ from the center
\begin{equation}
v(r)=\sqrt{\dfrac{G_N M(r)}{r}}~,
\label{eq:vofr}
\end{equation}
where $M(r)$ is the mass of a sphere of radius $r$ such that
\begin{equation}
M(r)=\int_0^r \rho(r') \diff^3 r'~.
\label{Mofr}
\end{equation}
Assuming that a large proportion of the galaxy mass is comprised in a sphere of radius $R$, one should expect $M(r\gg R)\simeq M(R)$, therefore on large radii the velocity should decrease with $r$ as 
\begin{equation}
v(r\gg R)\simeq \sqrt{\dfrac{G_N M(R)}{r}} \propto \dfrac{1}{\sqrt{r}}~.
\end{equation}
However Babcock realized that radial velocities were not decreasing on large radii as expected: "\textit{The approach to constant angular velocity discovered for the outer spiral arms is hardly to be anticipated from current theories of galactic rotation}". Even though Babcock's observations are not exactly in agreement with modern measurements, he raised the problem that rotation curves tend to \textit{flatten} on large radii, whose explanation is attributed to the existence of a Dark Matter halo.\\
One can deduce the shape of the Dark Matter density profile allowing for an explanation of the flattening of rotation curves by noticing from Eq.~(\ref{eq:vofr}) that we expect $M(r) \propto r$ for a constant velocity on large radii. Therefore according to Eq.~(\ref{Mofr}) the Dark Matter density distribution $\rho_{\text{DM}}(r)$ should behave on large radii as
\begin{equation}
\rho_{\text{DM}}(r) \propto \dfrac{1}{r^2}~.
\end{equation}
In Fig.~\ref{fig:rotationcurve} the velocity curve of the NGC6503 spiral galaxy is depicted, showing explicitely the impact of a Dark Matter component whose density behaves as $\rho_{\text{DM}}(r) \propto r^{-2}$ on large radii.\\
In 1951 the (HI) $21$cm spectral line of the hydrogen hyperfine structure\footnote{When the spin configurations of the electron and protons in a hydrogen atom flip from a parallel excited configuration ($\uparrow\uparrow$) to an anti-parallel fundamental arrangement ($\uparrow\downarrow$), a photon is emitted with a wavelength $\lambda \simeq 21$ cm.} was detected for the first time by Harold Ewen and Edward Purcell. This line was suggested as a new way of observing the universe and in particular, to measure rotation curves, which have been widely used in the last fifty years.\\
Later on in the 1970's, using a recently developped image tube-spectrograph Vera Rubin and Kent Ford observed HII regions of the (M31) Andromedra galaxy. From the H$\alpha$ line ($\sim 656.3$ nm), they extended previous measurements of the M31 rotation curve to the most outer regions~\cite{1970ApJ...159..379R}. Their results were compatible with radio observations of the same galaxy, implying the presence of an invisible component. A similar conclusion was reached by Ken Freeman while observing peaks of rotation curves of galaxies such as M$33$ and NGC$300$~\cite{1970ApJ...160..811F}. 
Two important papers were published in 1974, one by Einasto, Kaasik and Saar~\cite{1974Natur.250..309E}, the other by Peebles, Ostriker and Yahil~\cite{1974ApJ...193L...1O}. In both paper, they raised the important point that the Dark Matter halo introduced in order to explain galactic rotation curves could address at the same time similar discreprancies on galaxy cluster scales. They both estimated the total matter energy density present in the universe based on their galaxy cluster masses to be of the order of $\sim 20 \%$, quite close to the value measured nowdays. The important message of these two papers is that the Dark Matter problem is present on a wide variety of scales:  galactic, galaxy clusters and cosmological scales.\\ 
Later in the 1970's, more rotation curve studies were pursued, in particular Albert Bosma confirmed the flatness of rotation curves of 25 galaxies~\cite{1981AJ.....86.1825B} beyond luminous matter using radio waves. More recently, the authors of~\cite{2006ApJ...641L.109C} confirmed Horace Babcock and Vera Rubin results regarding the rotation curve of M31 as shown in Fig.~\ref{fig:rotationcurve} on the right pannel. Rotation curves are still widely used to infer mass distributions of astrophysical system such as galaxies, clusters of galaxies and dwarf satellites.

\section{The Cosmic Microwave Background}
\label{sec:CMB}
The Cosmic Microwave Background (CMB) is a radiation emitted around 300.000 years after the Big Bang, propagating in every direction in the universe. It is a very strong prediction of Big Bang cosmology and an extremely useful tool to scrutinize early universe cosmology, predicted by Ralph A. Alpher and Robert C. Herman~\cite{1948Natur.162..774A} based on work by George Gamow~\cite{1948Natur.162..680G} in the late 1940's. A fraction of second after the Big Bang, according to our current understanding of cosmology, the universe was extremely hot, thermalized and dense but was cooling down due to the Hubble expansion of the universe. Photons could not propagate freely and the universe was opaque due to Compton scattering with baryonic matter. When the temperature of the universe reached a value close to the binding energy of an electron with a hydrogen nucleus $T \lesssim 1$ eV (i.e. $z \simeq 1090$), atoms started to form and photons decoupled from baryons causing the universe to become transparent, known as the \textit{recombination} time. Photons emitted at this time could freely stream in the universe and were detected accidentally on earth for the first time by Pensias and Wilson in 1964 at Bell Labs while realizing an isotropic irreducible background was present in the radio-wave spectrum~\cite{1965ApJ...142..419P}. Although affected by the expansion redshift, the spectrum of this almost isotropic signal is still currently described today with a high degree of precision by a black body spectrum of intensity
\begin{equation}
I_\nu=\dfrac{4 \pi \hbar \nu^3}{c^2}\dfrac{1}{e^{2\pi \hbar \nu /k_B T_{\text{\tiny{CMB}}}}-1}~,
\end{equation}
with a value of the temperature measured today $T_{\text{CMB}}=2.72547 \pm 0.00057~$K~\cite{2009ApJ...707..916F}. 
%
Since the solar system is moving within our galaxy, photons coming from this specific direction of motion in the sky are redshifed by Doppler effect, also causing a blueshift of photons coming from the opposite direction. Therefore this dipole effect is responsible for anisotropies in the CMB temperature map of the order of $\delta T/T\sim10^{-3}$. Astrophysical processes such as the Sunyaev-Zeldovich effect, corresponding to inverse-Compton scattering of CMB photons on hot gas present in galaxy clusters for instance, are the cause of additional contibutions to the anisotropy map of the CMB. After removing the dipole effect and several other astrophysical contributions, the CMB temperature map reveals anisotropies of the order of $\delta T/T\sim10^{-5}$ which correspond to a non-uniform matter distribution at the time of recombination.

\begin{figure}[h!]
\begin{center}
\includegraphics[width=\linewidth]{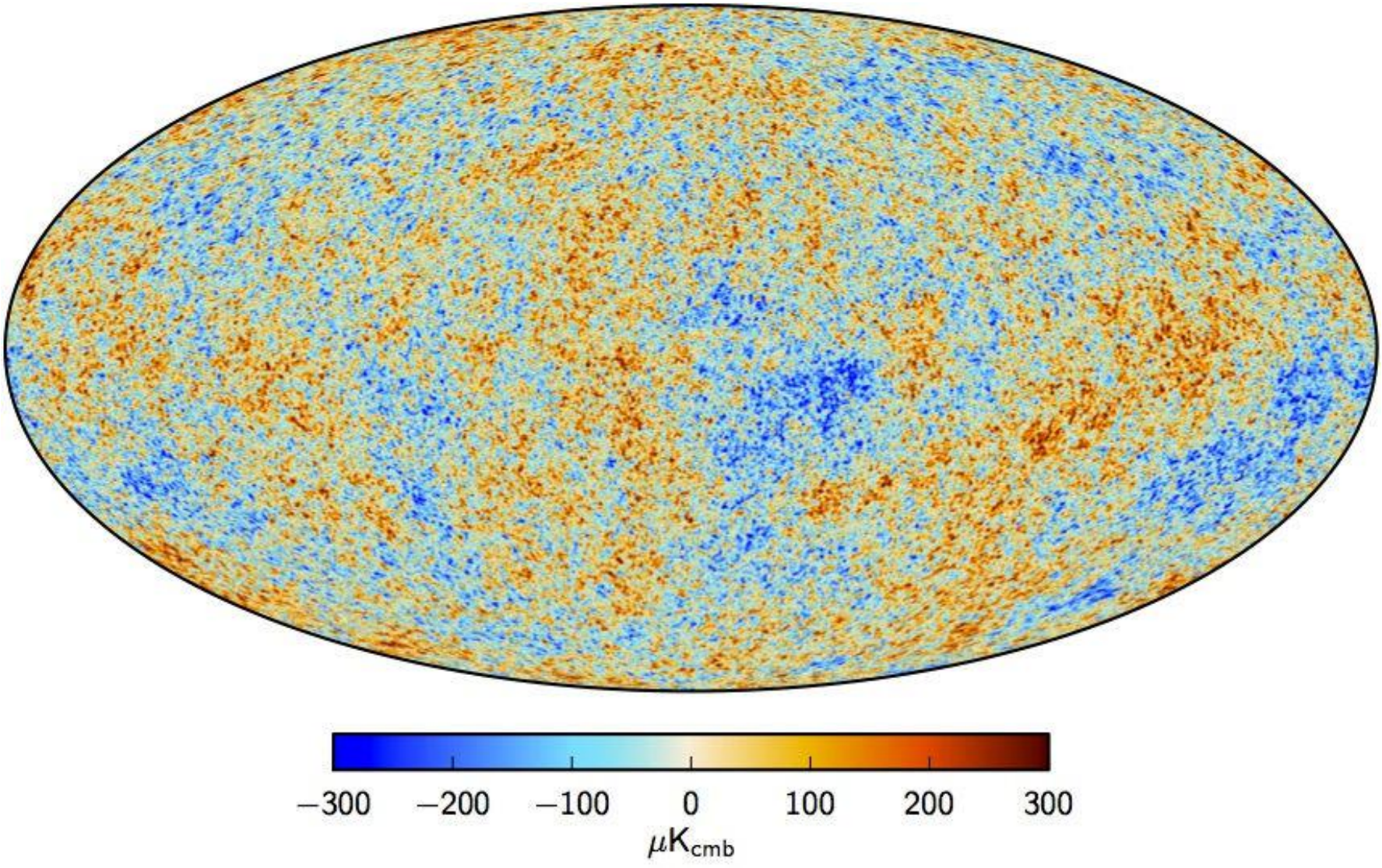}
\caption{CMB temperature anisotropy map as measured by the Planck collaboration. Figure taken from~\cite{Adam:2015rua}.} 
\label{fig:CMBmap}
\end{center}
\end{figure}
\begin{figure}[h]
\begin{center}
\includegraphics[width=\linewidth]{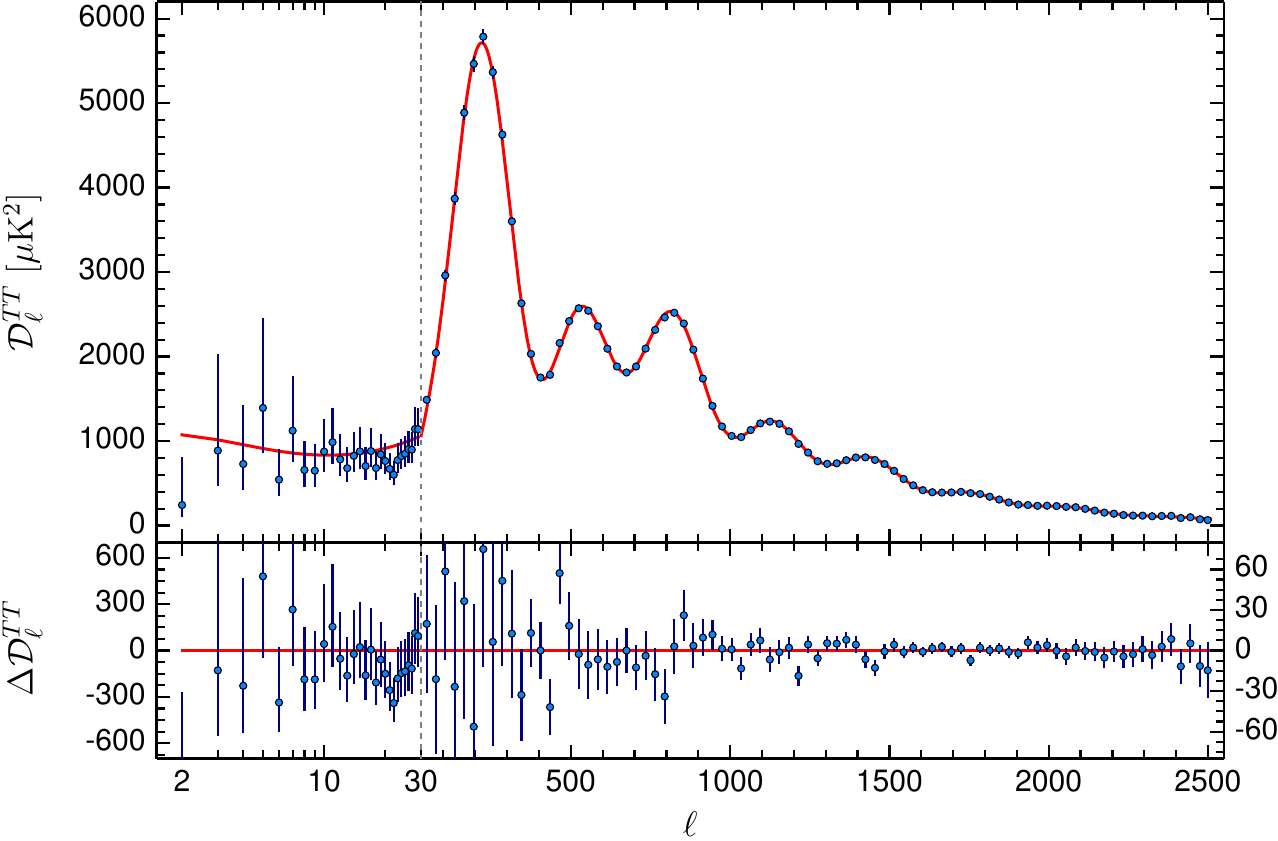}
\caption{Power spectrum of the CMB temperature anisotropy map. The top pannel shows experimental data from the Planck collaboration, featuring also in red solid line the best fit to the spectrum. The bottom pannel shows residual of the fit. Figure taken from~\cite{Adam:2015rua}.} 
\label{fig:CMBspectrum}
\end{center}
\end{figure}

The map of these anisotropies was measured recently by the Planck collaboration~\cite{Ade:2013kta} as shown in Fig.~\ref{fig:CMBmap}. It was historically detected by the Far InfraRed Absolute Spectrophotometer (FIRAS) on board the Cosmic Background Explorer (COBE)~\cite{Fixsen:1996nj} and later on measured by the ballon experiments BOOMERANG~\cite{Netterfield:2001yq} and MAXIMA~\cite{Hanany:2000qf}, and the Wilkinson Microwave Anisotropy Probe (WMAP) satellite~\cite{2013ApJS..208...20B}.
It is common to decompose the temperature anisotropies in one specific direction $\hat{n}$ of the sky using the basis of spherical harmonics $Y_{\ell m}(\hat{n})$:
\begin{equation}
\Theta(\hat{n})\equiv\frac{\delta T}{T}(\hat{n})=\sum_{\ell=2}^\infty \sum_{m=-\ell}^\ell a_{\ell m} Y_{\ell m}(\hat{n}) ~.
\end{equation}
In the early tighly-coupled baryon-photon plasma, quantum fluctuations gave rise to baryonic overdensity fluctuations that were amplified by gravitational attraction but were repelled by radiation pressure. The balance of these two processes generated acoustic oscillations in the baryonic fluid, propagating at the speed of sound $c_{\text{s}}$ related to the mean bayon and photon densities $\bar{\rho}_{b,\gamma}$:
\begin{equation}
c_{\text{s}}^2=\dfrac{1}{3(1+R)}, \qquad R \equiv \dfrac{4\bar{\rho}_{\text{b}}}{3\bar{\rho}_\gamma}~.
\end{equation}
These acoustic waves propagated in the baryon-photon plasma until the end of the Compton drag epoch $\eta_{\text{drag}}$\footnote{$\eta$ is the conformal time defined in Eq.~(\ref{eq:conformaltime}) and $z_{\text{drag}}\simeq 1060$.} when baryons started to fall freely without feeling radiation pressure anymore, that decreased sharply after the photon-baryon decoupling $\eta_{\text{dec}}$.
The Dark Matter distribution, not affected by radiation pressure would evolve with the baryon distribution under the effect of gravity. Photons would then propagate without experiencing any Compton scattering, while keeping the imprint of the size of the baryon spherical shell, the \textit{last scattering surface}, whose radius is given by the sound horizon at $\eta_{\text{drag}}$
\begin{equation}
d_{\text{s}}(\eta_{\text{drag}})=a \int_{\eta_{0}}^{\eta_{\text{drag}}} c_{\text{s}} \diff \eta^\prime~.
\end{equation}
Therefore the CMB anisotropy map two-point correlation function $\langle \Theta(\hat{n}) \Theta (\hat{n}')\rangle$  between two directions $\hat{n}$ and $\hat{n}'$ features peaks separated by a typical length corresponding to the angular scales $\theta \sim d_{\text{s}}(\eta_{\text{dec}})/d_{\text{A}}(\eta_{\text{dec}})$\footnote{$d_A$ is the angular distance as defined in Eq.~(\ref{eq:dA}).} as shown in Fig.~\ref{fig:CMBspectrum}. The quantity $\langle \Theta(\hat{n}) \Theta (\hat{n}')\rangle$ is defined as
\begin{equation}
\langle \Theta(\hat{n}) \Theta (\hat{n}')\rangle = \sum_{\ell,m}\sum_{\ell',m'} \langle a_{\ell m} a^*_{\ell' m'} \rangle Y_{\ell m}(\hat{n}) Y^*_{\ell' m'}(\hat{n}') =\dfrac{1}{4\pi}\sum_{\ell}(2\ell+1)C_\ell P_\ell(\cos \theta)~,
\end{equation}
with $\cos \theta = \hat{n} \cdot \hat{n}'$ and $P_\ell$ are the Legendre polynomial functions. The coefficients $C_\ell$ represent the power spectrum related to the quantity $\mathcal{D}_\ell$ defined as follow:
\begin{equation}
\mathcal{D}_\ell\equiv \dfrac{\ell(\ell+1)}{2\pi}C_\ell \quad \text{and} \quad C_\ell \equiv \langle |a_{\ell,m}|^2 \rangle= \dfrac{1}{2\ell+1}\sum_{m=-\ell}^{m=+\ell} |a_{\ell m}|^2~.
\end{equation}

The value of the typical angular separation $\theta$ between two points in the CMB anisotropy map is related to $\ell \simeq \pi/\theta$. Therefore small (large) values of $\ell$ correspond to large (small) physical distances in the last scattering surface. Even though the oscillation pattern corresponding to these acoustic waves is observable in the spectrum today, numerous physical effects alter its shape from emission to observation. Therefore, making precision measurements allows to efficiently probe cosmological models. Some of the most important of these effects on the spectrum are:

\begin{itemize}
\item The Sachs-Wolfe effect is responsible for a plateau at small  $\ell$, corresponding to the redshift of photons escaping gravitational potential wells which generates fluctuations $\delta T/T \sim - 1/3 \Phi$ where $\Phi$ is the gravitational potential. The typical physical scales associated to the modes corresponding to small values of $\ell$ were super-Hubble at the time of photon decoupling and they could not experience the effects of the acoustic oscillations, the corresponding spectral shape remains scale-invariant as suggested by inflation scenarios. 
\item An overall damping of the spectrum, known as \textit{Silk-damping} is due to photon diffusion on small distances in the baryon-photon plasma before being Compton-scattered by electrons. This effect is responsible for washing out overdensities on short distances corresponding to an exponential damping of the spectrum for large values of $\ell$. 
\item Integrated Sachs-Wolfe effects: photons propagating in a time-variating gravitational potential would experience a redshift when leaving a potential well not compensating the blueshift caused when entering, due to time dependent Hubble expansion. This process would affect scales the most sensitive to the expansion, i.e. the largest ones, creating a small tilt in the spectrum for very small values of $\ell$.
\end{itemize}
The CMB spectrum allows for an estimate of the free parameters of the $\Lambda$CDM model which includes dark energy in the form of a cosmological constant $\Lambda$ and a cold Dark Matter component whose density is $\Omega_{\text{c}}$. Currently the most precise determination of the matter energy density $\Omega_{\text{m}}$ at the present time, deduced from the CMB anisotropy map, is provided by the Planck collaboration~\cite{Ade:2015xua}:
\begin{align}
\Omega_{\text{m}}& = 0.3156\pm 0.0091~,\\
\Omega_{\text{b}}h^2& = 0.02225\pm 0.00016~,\\
\Omega_{\text{c}}h^2& = 0.1198\pm 0.0015~,
\end{align}
where $h \equiv H_0 / (100 \text{ km s}^{-1}\text{ Mpc}^{-1})$ with $H_0 = 67.27 \pm 0.66 \text{ km s}^{-1}\text{ Mpc}^{-1}$. From these measurements on can infer the value of the curvature density $\Omega_{\text{k}}=0.000~\pm~0.005$ which is compatible with a flat universe. These precise measurements indicate the presence of a non-negligible Dark Matter component in the matter energy budget of the universe of the order of $85\%$ and around $25\%$ of its total energy density. The impact of a Dark Matter density component on the CMB spectrum is indirectly related to several features of the spectrum including:
\begin{itemize}
\item The typical peak scale depends on the sound horizon at the decoupling time and thus is related to the baryon density at that time. It would also depend on the expansion history and in particular the time of matter-radiation equality, linked to the total matter density.
\item A change in the duration between the time of matter-radiation equality and the decoupling would impact the efficiency of the baryonic damping of the oscillations controlled by the value of $\Omega_{\text{m}}$ and leads to a vertical shift of the peaks in the spectrum.
\end{itemize}

As a summary, the CMB observation is undoubtedly one of the strongest evidence of the presence of Dark Matter in our universe and the spectrum allows for a rigorous test of the $\Lambda$CDM model while strongly constraining other alternative scenarios.

\section{Gravitational lensing}
\label{sec:lensing}
\begin{center}
\begin{figure}[h!]
  \begin{minipage}[c]{0.5\textwidth}
  \begin{center}
\includegraphics[height=5.5cm]{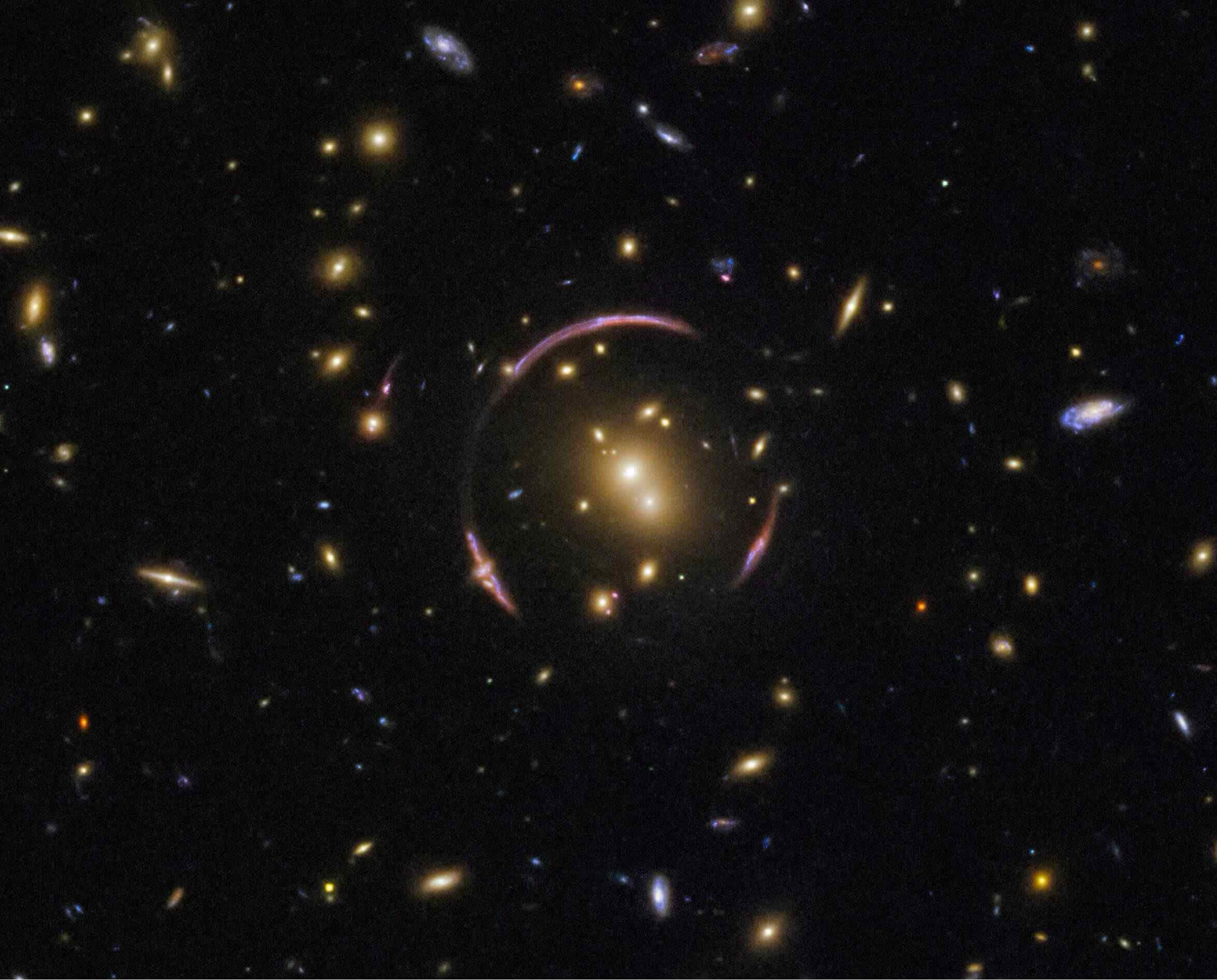}  
  \end{center}
   \end{minipage}\hfill
   \begin{minipage}[c]{0.5\textwidth}
   \begin{center}
\includegraphics[height=5.5cm]{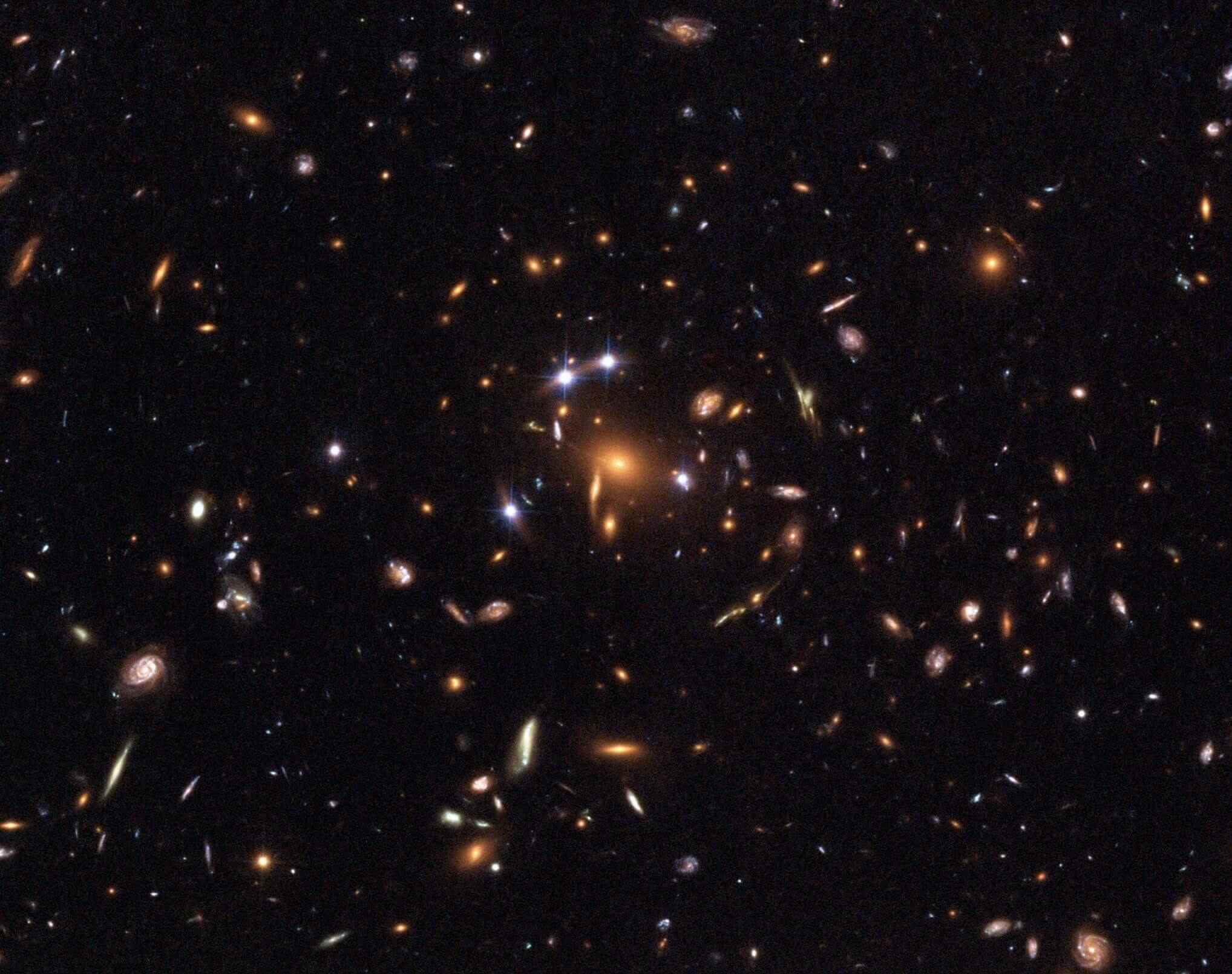}   
   \end{center}
   \end{minipage}
   \caption{\textbf{Left:} Image of the galaxy cluster  SDSS J0146-0929 from the Hubble Space Telescope showing Einstein's ring due to strong gravitational lensing. \textbf{Right:} Image of the galaxy cluster SDSS J1004+4112 showing multiples images of the same quasar around the center. Credit: ESA/Hubble and NASA}
\label{fig:stronglensing}
\end{figure}
\end{center}

One very important consequence of General Relativity is that light can be deflected by a gravitational potential and therefore does not propagate in a straight line around a massive object as one would expect in a flat space. Using Schwarzschild metric and Einstein equations one can estimate the deflection angle $\delta \phi$ of a light-ray passing nearby an object of mass $M$ as
\begin{equation}
\delta \phi \simeq \dfrac{4 G_N M}{b}~,
\end{equation}
where $b$ is the impact parameter of the incident photon. As expected, in the limit where the mass $M$ vanishes, photons follow a straight line. As a result, one can deduce the mass of a massive object located between a source of photons and an observer, by measuring the deflection angle of the incoming photons~\footnote{For futher reading, a review of gravitational lensing can be found in~\cite{Massey:2010hh}.}. 
\paragraph{Strong lensing}
In the regime where a very dense region is present between a source and an observer, light emitted from the source could follow several geodesics to reach the observer, resulting in mutiple images of the same physical source in the field of view as illustrated in Fig.~\ref{fig:stronglensing} on the right pannel. In the case where a spherically symmetric object is located exactly in the source-observer axis, on should expect to observe an \textit{Einstein ring} due to the cylindrical symmetry of the system. The \textit{Einstein radius} of this ring is given by
\begin{equation}
\theta_{\text{E}}\simeq \sqrt{\dfrac{4 G_N M  d_{\text{OS}}}{d_{\text{O}} d_{\text{S}}}}~,
\end{equation} 
where $d_{\text{OS}}, d_{\text{O}}$ and $d_{\text{S}}$ are respectively angular distances, as defined in Eq.~(\ref{eq:dA}) between the massive object and the source, between the observer and the object and between the source and the observer. In case of a massive object slightly shifted from the observer-source axis, instead of observing rings, one would observe a series of arcs as shown in Fig.~\ref{fig:stronglensing} on the left pannel. Strong lensing have been used since the 1980's in order to measure masses of galaxies and more recently the Sloan Digital Sky Survey
(SDSS) Quasar Lens Search (QLS) published an estimate of the cosmological constant and the matter density based on lensing measurements~\cite{Oguri:2007sv}. Their results are in good agreement with several other cosmological probes.

\paragraph{Weak lensing}
The \textit{weak lensing} regime corresponds to distortions of the apparent shape of luminous objects by the gravitational potential of some massive structure located near to the line of sight, between the source and the observer, resulting in a possible magnification or shear of the source image. Even though the average shape of galaxies is circular, galaxies along adjacent lines of sight are expected to be sheared in a similar way by the effect of weak lensing, resulting in a ellipsoid shape on average. Therefore combining observations over a large sample of galaxies allows for the reconstruction of the gravitational potential along a line of sight. This method has been used by the Sloan Lens ACS Survey to infer the average galactic baryonic and Dark Matter fraction from large sample galaxies in~\cite{Gavazzi:2007vw}. They found that in a sphere of radius $\sim8~\text{kpc}$ around the center of galaxies, the Dark Matter mass-fraction represents $\sim 27\%$ showing that the inner core of galaxies is mostly dominated by baryons. \\
The weak gravitational lensing produces an effect observable in background galaxies known as \textit{cosmic shear} allowing to trace the total mass distribution induced by the foreground large-scale structure in the universe. Results from the Dark Energy Survey (DES) collaboration regarding the estimate of the total matter density, combining galaxy clustering and weak lensing studies, were shown to be consistent with CMB data~\cite{Abbott:2017wau}.
\begin{figure}[h!]
\begin{center}
\includegraphics[width=0.5\linewidth]{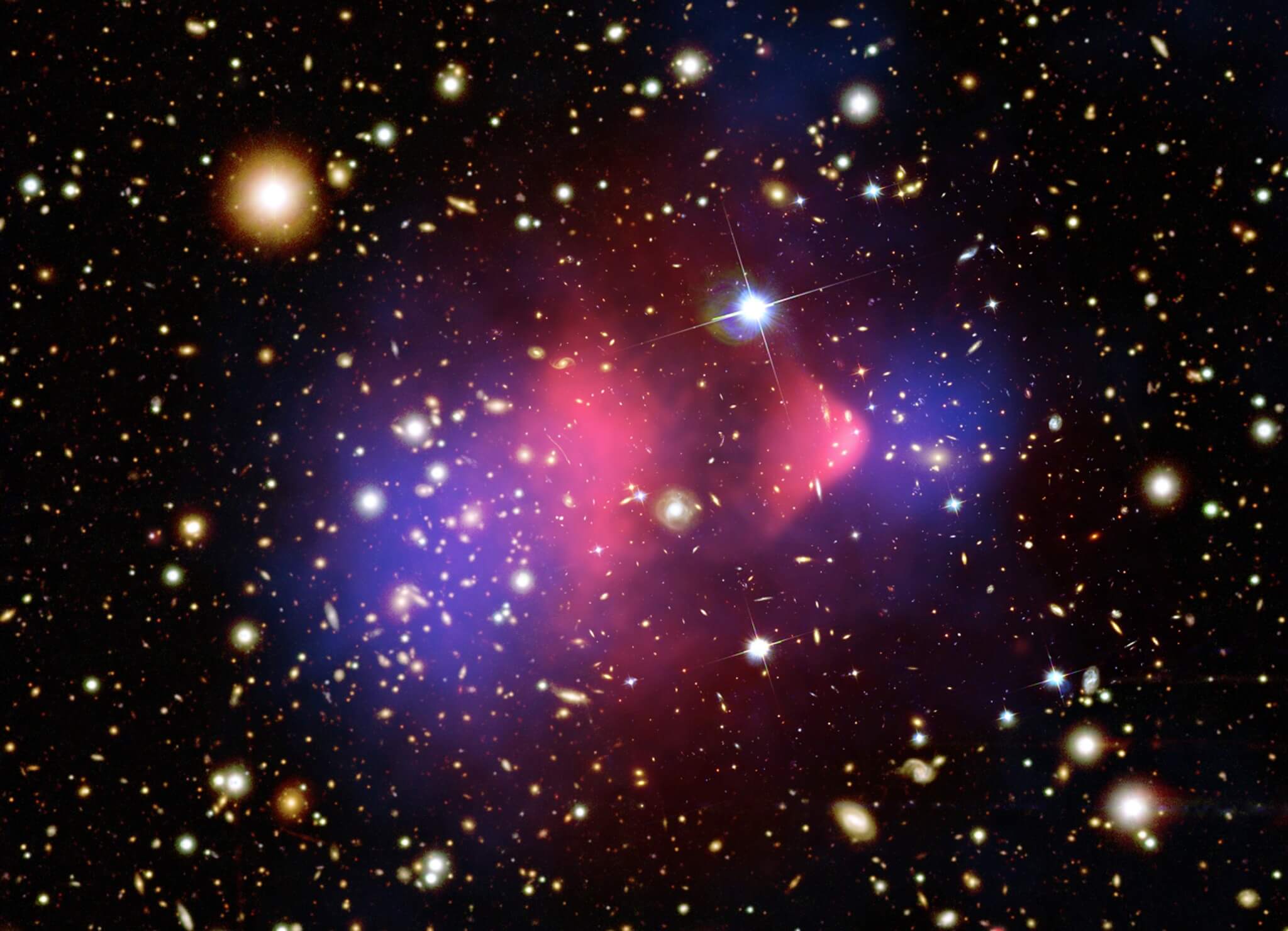}
\caption{The bullet cluster (1E0657-56) composite image: the background represents the optical image of the system, the blue region shows the mass distribution infered from weak lensing and the pink region shows the X-ray emission from hot gas. Credit: X-ray map: NASA/CXC/CfA~\cite{Markevitch:2005vi}; Weak lensing map and optical image: NASA/STScI; ESO WFI; Magellan/U.Arizona~\cite{Clowe:2006eq}.}
\label{fig:bulletcluster}
\end{center}
\end{figure}

One of the most convincing argument regarding the presence of Dark Matter is derived from the weak lensing mass-contour of the so-called \textit{Bullet cluster}. The bullet cluster (1E0657-56) is actually a system of two merging clusters, as depicted in Fig.~\ref{fig:bulletcluster}, where the total mass distribution have been infered using weak lensing and the luminous mass distribution is deduced using X-ray emission from the hot intra-cluster gas. The particular shape of the luminous matter contour shows that the two cluster gases interacted while passing through each other. However the total cluster mass-distributions coincide with the location of the galaxies, displaying spherical shapes therefore indicating that as the stars within galaxies, most of the mass contributions did not interact during the collision of the clusters showing the \textit{collisionless} property of the Dark Matter. Such an observation is a strong argument in favor of the particle interpretation of Dark Matter contrary to modified gravity theories for which providing a simultaneous interpretation of the Bullet Cluster event and accomodating the observed rotation curves is challenging, as discussed further on in Sec.~\ref{sec:alternative}.

\section{Baryon Acoustic Oscillations and the matter power spectrum}
\label{sec:BAO}
Primordial baryon acoustic waves propagating in the tighly coupled baryon-photon plasma and giving rise to anistropies in the CMB became the seed of overdensities in the matter distribution after recombination. These overdensities have been affected by gravity, which tends to link baryon and Dark Matter densities together, allowing for this pattern to be still observable nowdays in the matter power spectrum $P(k)$, which can be expressed
\begin{equation}
\la \delta_\rho (t,\vec{k}) \delta_\rho (t,\vec{k}^\prime)\ra \equiv P(k) \delta^{(3)}(\vec{k}-\vec{k}^\prime)~,
\end{equation}
where $\la \delta_\rho (t,\vec{k}) \delta_\rho (t,\vec{k}^\prime)\ra$ is the Fourier transform of the overdensity two-point correlation function
\begin{equation}
\la \delta_\rho (t,\vec{k}) \delta_\rho (t,\vec{k}^\prime)\ra = \int \dfrac{\diff^3 \vec{r}}{(2\pi)^3} e^{i \vec{k}\cdot \vec{r}} \langle \delta_\rho(t,\vec{x}) \delta_\rho(t,\vec{x}+\vec{r}) \rangle \delta^{(3)}(\vec{k}-\vec{k}^\prime)~,
\end{equation}
with
\begin{equation}
\delta_\rho(t,\vec{x}) \equiv \dfrac{\bar{\rho}(t)-\rho(t,\vec{x})}{\bar{\rho}(t)}~,
\end{equation}
where $\rho(t,\vec{x})$ is the matter density distribution and $\bar{\rho}(t)$ the mean matter density.\\
One technique used to obtain the matter power spectrum is to measure the absorption spectrum corresponding to the Lyman-$\alpha$ (Ly-$\alpha$) emission line\footnote{Line corresponding to an electron transiting from the $n=2$ orbital to the $n=1$ level in a hydrogen atom. The wavelength corresponding to this transition is $\lambda_{\text{Ly-}\alpha} \sim 121,6~\text{nm}$ (ultraviolet).} from distant quasars. Quasars are extremely luminous sources in particular near the Ly-$\alpha$ emission line. Light emitted from distant quasars propagating in the intergalactic medium while being redshifted by the Hubble expansion. Photons with wavelengths $\lambda < \lambda_{\text{Ly-}\alpha}$  at the time of emission will at some point reach a region were $\lambda \simeq \lambda_{\text{Ly-}\alpha}$ because of the redshift. Neutral hydrogen atoms present in the intergalactic medium would absorb photons with $\lambda \simeq \lambda_{\text{Ly-}\alpha}$ causing the initial emitted spectra to show absorption lines at different redshifts corresponding to photons crossing overdense regions along their way. Such a spectral shape is known as the \textit{Lyman-$\alpha$ forest}. As a result, measuring the absorption spectrum of a distant quasar allows to derive a 1D density-map along the line of sight of the intergalactic medium, and combining informations from multiples quasars provides informations regarding the matter density in the universe.\\
\begin{figure}[h!]
\begin{center}
\includegraphics[width=0.5\linewidth]{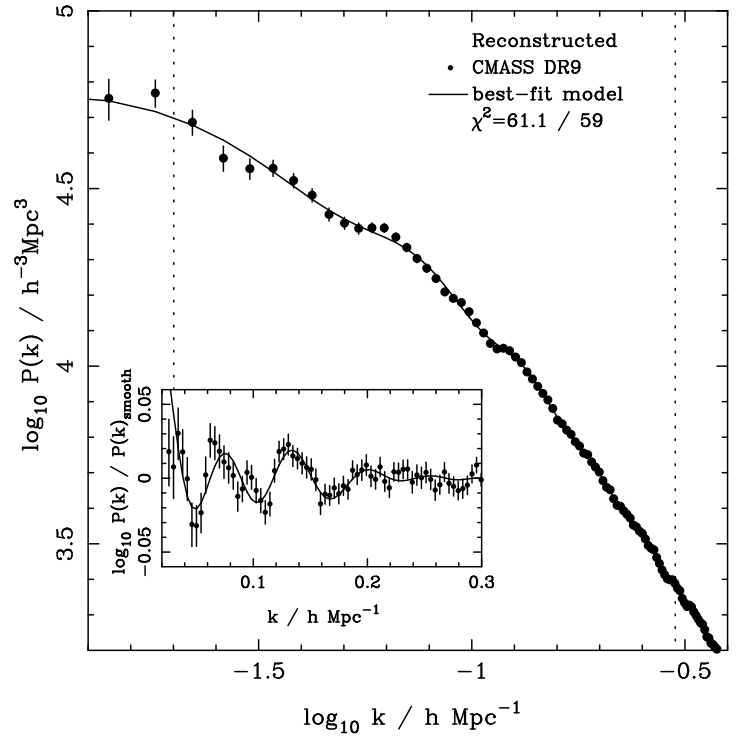}
\caption{Matter power spectrum as measured by the SDSS-III BOSS collaboration showing the best-fit model in solid line. The inset displays the same spectrum normalized by the best-fit model with no BAO. Figure taken from~\cite{2012MNRAS.427.3435A}.}
\label{fig:matterpowerspectrum}
\end{center}
\end{figure}
A complementary method to access the power spectrum is to map galaxies in a very large volume of the universe and derive the correlation function. This analysis was performed by the 2dFGRS collaboration~\cite{Cole:2005sx} and more recently by the Baryon Oscillation Spectroscopic Survey (BOSS) as part of the the Sloan Digital Sky Survey III (SDSS-III)~\cite{2012MNRAS.427.3435A} as shown in Fig.~\ref{fig:matterpowerspectrum} which combined data from more than $200.000$ galaxies corresponding to a volume of $\sim2~\text{Gpc}^3$. Fig.~\ref{fig:matterpowerspectrum} shows an oscillation pattern in the power spectrum corresponding to a physical size of $\sim 150~\text{Mpc}$ and can be predicted from the CMB spectrum using gravitational perturbation theory. However it cannot be described by considering only baryons in the matter content of the universe, a Dark Matter component is essential to explain the features present in the spectrum. One can infer the cosmological parameters from the power spectrum and a recent analysis from the BOSS collaboration shows a good agreement with previous measurements~\cite{Alam:2016hwk}, confirming the $\Lambda$CDM model as the main cosmological paradigm.

\section{Simulations and Dark Matter distribution}
\label{sec:simdistrib}

Although a Dark Matter component in our universe seems well established at the present time, beside measuring its global abundance on cosmological scales, quantizing the Dark Matter density in smaller environments such as galaxies and the Milky-Way in particular, remains a challenging task. Nowdays, numerical simulations have shown to be a particularly efficient way of testing our understanding of structure formation and provide a way to estimate velocity and matter density distribution. Peebles and Ostriker in their seminal paper of 1963~\cite{1973ApJ...186..467O} argued, based on N-body simulations (including $\mathcal{O}(100)$ particles), that galactic disks such as the one of the Mily-Way might be actually unstable. However, adding a spherical dark halo much larger than the disk would improve the stability. In fifty years, simulations have drastically evolved and today they are an essential tool to infer Dark Matter properties such as density distribution. In this section we discuss our current understanding regarding the Dark Matter distribution and local density in the solar system based on numerical simulation results and local measurements of the Dark Matter density.

\subsubsection{The isothermal Dark Matter model}

One could try to estimate the shape of the Dark Matter density distribution $\rho(\vec{x})$ in a galaxy by assuming a thermal origin for the Dark Matter, whose phase space distribution can be approximated by $f(\vec{x},\vec{v})\propto \exp \left( -E/T \right)$ and related to $\rho(\vec{x})=\int \diff^3 \vec{v}f(\vec{x},\vec{v})$ where $E=(1/2)m_{\text{DM}}v^2+m_{\text{DM}}\phi(\vec{x})$ with $\phi(\vec{x})$ the gravitational potential. Assuming that this distribution is not time-dependent, according to the Poisson equation and assuming a Maxwell-Boltzmann distribution, the Dark Matter density at the present time would behave as $\rho(r)\propto 1/r^2$. As the baryon density is dominating the mass distribution in the inner part of the halo, the distribution for radii below a typical scale $r<r_s$ can be truncated giving the (pseudo-)isothermal density profile
\begin{equation}
\rho_{\text{iso}}(r)=\dfrac{\rho_s}{1+ \left(\dfrac{r}{r_s}\right)^2}~,
\end{equation}
where $\rho_s$ and $r_s$ are normalization factors. Based on the same argument, one can deduce the velocity distribution of the Dark Matter particles in the galaxy as a truncated Maxwellian distribution, also known as the \textit{Standard Halo Model}:
\begin{equation}
f(\vec{v})=\left\{ \begin{split}
&\dfrac{1}{N_{\text{esc}}}\dfrac{1}{\sqrt{2\pi}\sigma}e^{-\frac{|\vec{v}|^2}{2\sigma^2}} , &&|\vec{v}|<v_{\text{esc}} \\ 
&0, &&|\vec{v}|\geq v_{\text{esc}}
\end{split}  \right.
\end{equation}
where $v_{\text{esc}}$ corresponds to the velocity above which particles are no longer gravitationally bound to the galaxy and would eventually escape. $\sigma$ is the velocity dispersion related to the most probable velocity $v_0 \simeq 230~\text{km s}^{-1}$ and $N_{\text{esc}}=\text{erf}(v/v_0)-2\pi^{-1/2}(v/v_0)e^{-v^2/v_0^2}$. Although the Dark Matter velocity and density distribution of galaxies cannot easily be determined, it turns out N-body simulations have been playing a significant role in shedding light on these aspects over the past decades.

\subsubsection{N-body simulations}

\begin{center}
\begin{figure}[h!]
  \begin{minipage}[c]{0.5\textwidth}
  \begin{center}
\includegraphics[height=5cm]{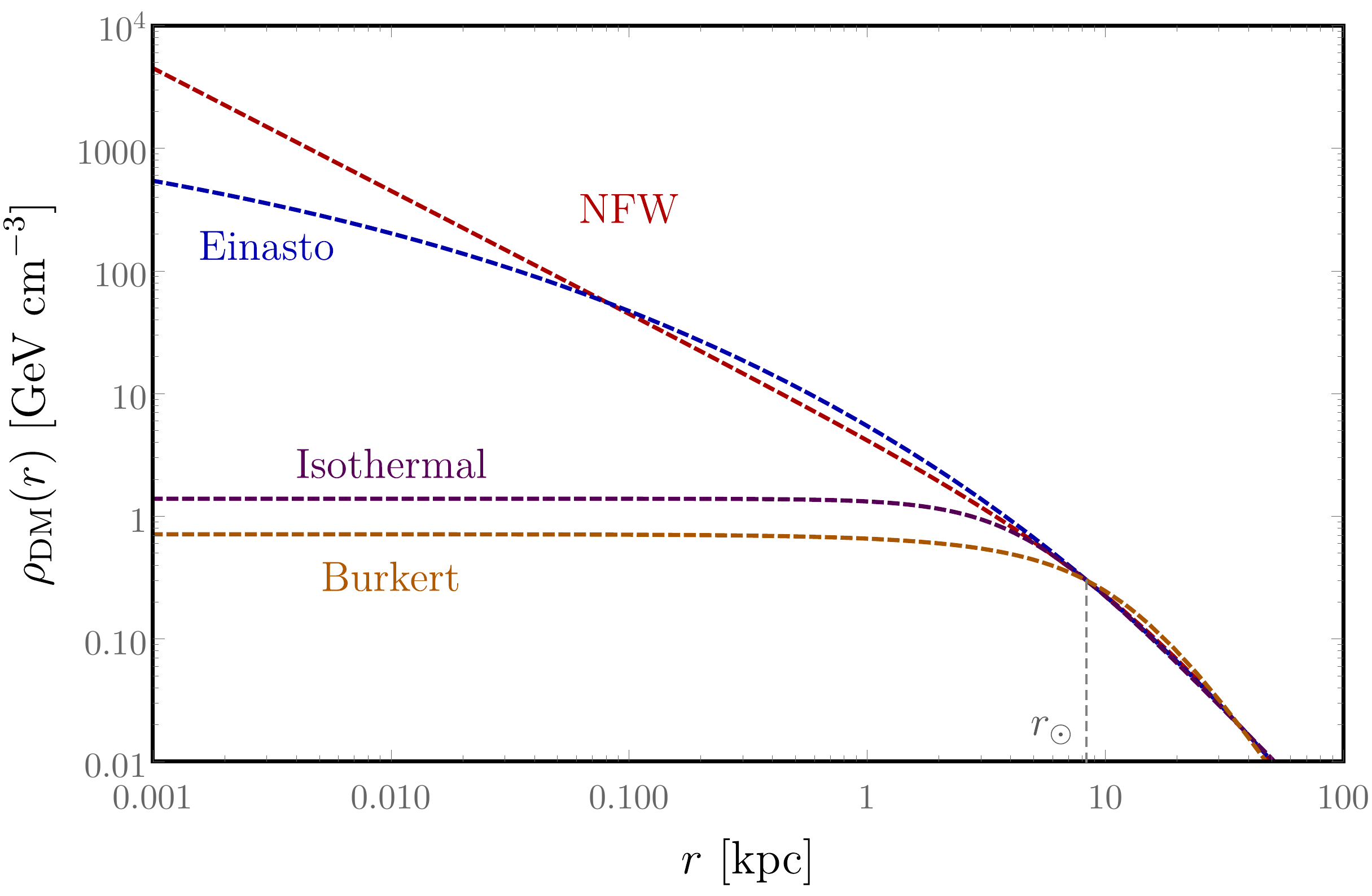}  
  \end{center}
   \end{minipage}\hfill
   \begin{minipage}[c]{0.5\textwidth}
   \begin{center}
\includegraphics[height=5.3cm]{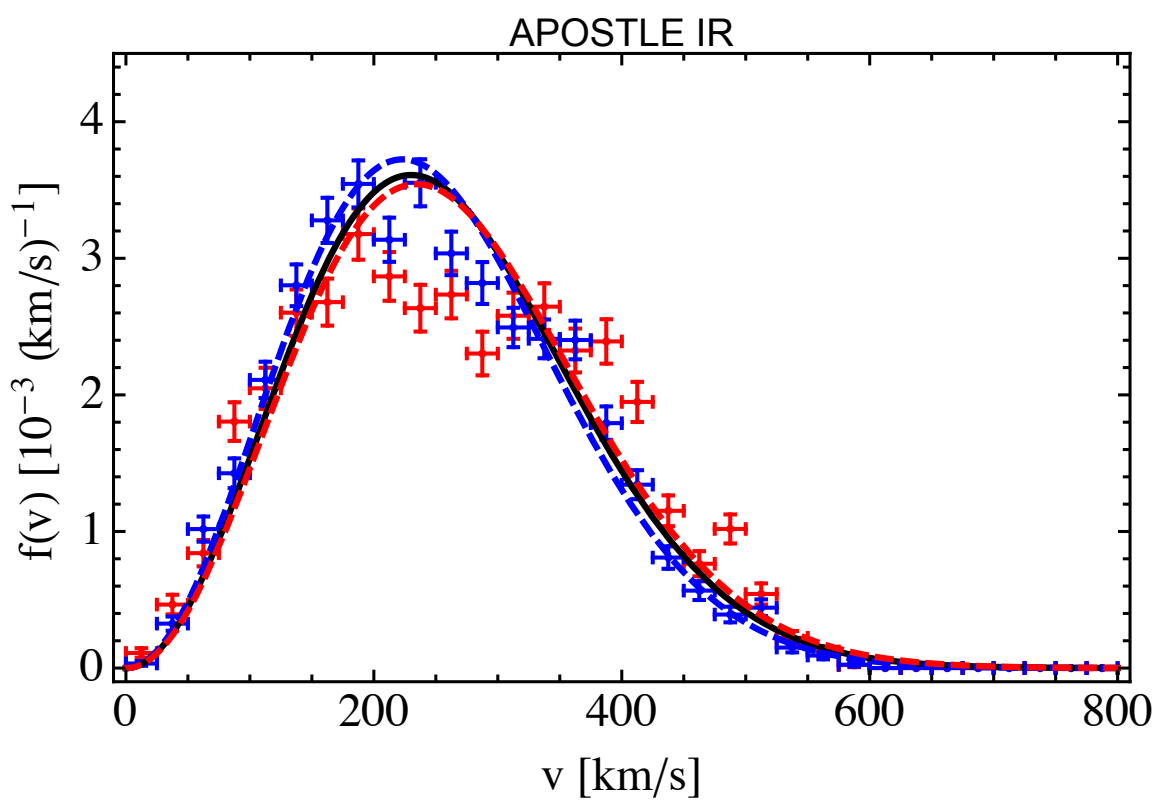}   
   \end{center}
   \end{minipage}
   \caption{\textbf{Left:} NFW, Einasto, Isothermal and Burkert galactic Dark Matter density profile using $\alpha=0.17$ and $r_s=\{24.4, 28.4, 4.3, 12.6\}~\text{kpc}$ respectively. The profiles are normalized for the Milky-Way such that $\rho_\odot=0.3~\text{GeV cm}^{-3}$ and $r_\odot=8.33~\text{kpc}$.  \textbf{Right:} The black line shows the standard halo model distribution with $v_0=230~\text{km s}^{-1}$ and the colored dashed lines the best-fit Maxwellian distributions corresponding to the values generated for two different MW like galaxies from the APOSTLE IR simulations~\cite{Sawala:2015cdf}. Figure taken from~\cite{Bozorgnia:2016ogo}.}
\label{fig:DMdensityprofile}
\end{figure}
\end{center}

The first N-body simulations on cosmological scales were considering only a component of collisionless Dark Matter particles interacting via gravity~\cite{1985ApJ...292..371D}, however they have shown to be determinant to impose the cold Dark Matter as an essential ingredient for the understanding of structure formation. Computational capabilities have improved exponentially over the past decades, allowing to include semi-analytical models of galaxy formation and to perform a complete hydrodynamical simulations including baryonic effects such as supernovae explosion and Active Galactic Nucleus (AGN) modelization. Recent simulations such as Illustris~\cite{Vogelsberger:2014kha} or EAGLE~\cite{Schaye:2014tpa} have shown to reproduce numerically the cosmic web observed in the universe, the distribution of galaxies in clusters, the star formation rate as well as element abundances on small scales in a $\Lambda$CDM universe.
In the 1990's, based on the first improvements of N-body cosmological simulations, Navarro, Frenk and White realized~\cite{Navarro:1995iw} that halos with sizes covering several orders of magnitude could be described by rescaling a single function\footnote{Actually, it is a spherically averaged radial density profile as Dark Matter halos are found to be triaxial~\cite{Dubinski:1991bm}.}, the so-called NFW profile:
\begin{equation}
\rho_{\text{NFW}}(r)=\dfrac{\rho_s}{\left( \dfrac{r}{r_s}\right)\left[1+ \left(\dfrac{r}{r_s}\right)^2\right]}~,
\end{equation}
which diverges in the limit $r \rightarrow 0$ and decreases for large radii as $\rho(r\gg r_s)\propto r^{-3}$ which is suppressed compared to the isothermal profile. More recent simulations~\cite{Navarro:2008kc} have shown that halo densities could be described more accurately using the function introduced by Einasto~\cite{1965TrAlm...5...87E}, the so-called Einasto profile
\begin{equation}
\rho_{\text{Einasto}}=\rho_s \exp \left( -\dfrac{2}{\alpha} \left[ \left( \dfrac{r}{r_s}\right)^\alpha -1 \right]   \right)~,
\end{equation} 
However, this function contains one more free parameter $\alpha$ which has to be adjusted to account for the diversity in the simulated halos. In addition, a comonly used profile, the so-called Burkert profile~\cite{Burkert:1995yz}, is motivated by rotation curves measurements of dwarf galaxies and has the following shape
\begin{equation}
\rho_{\text{Burkert}}(r)=\dfrac{\rho_s}{\left(1+\dfrac{r}{r_s} \right)\left[ 1+ \left(\dfrac{r}{r_s} \right)^{2} \right]}~.
\end{equation}
This profile reproduces a flat profile in the inner part of the halo but decreases more sharply at high radii $\rho(r\gg r_s)\propto r^{-3}$. The various density profiles discussed in this section are shown in Fig.~\ref{fig:DMdensityprofile} on the left pannel in the Milky-Way case, showing in particular the behavior at low radius. Additional features should be present in a more realistic halo model for instance clumps, as remnants of galaxy merging processes, resulting in localized Dark Matter overdensities. The velocity distribution can be inferred from simulations and compared to the expected Standard Halo Model distribution as shown in Fig.~\ref{fig:DMdensityprofile}. It turns out the overall distribution of Milky-Way like galaxies are close to a Maxwellian distribution. However deviations can be observed. In particular numerical simulations have shown to generate distribution tails which are less suppressed compared to the Maxwellian case. This can be interpreted as non-equilibrium processes interfering with galaxy formation, such as galaxy merging effects or tidal disruption of the galactic halo, slightly affecting the expected Maxwellian distribution.

\subsubsection{The local Dark Matter density}

Local Dark Matter density measurements present a strong challenge as direct detection experiments sensitivity depends on its value, as discussed in Sec.~\ref{sec:DD}. One method used to derive the Dark Matter density in the solar system is to use the Milky Way's rotation curve as performed in~\cite{Bovy:2012tw,Iocco:2011jz,Salucci:2010qr}\footnote{For futher reading, a detailed review can be found in~\cite{Read:2014qva}.}. Even though the results depend slightly on the assumptions made in the derivation, recent measurements seem to converge toward a value of the order of $\rho_\odot \sim 0.4 ~\text{GeV cm}^{-3}$. A Bayesian analysis based on several dynamical observables for the Milky-Way~\cite{Catena:2009mf} provided an estimation of the local density assuming a NFW profile:
\begin{equation}
\rho_{\odot}=0.39 \pm{0.025}~\text{GeV cm}^{-3}~.
\end{equation}
In a more recent analysis~\cite{Sivertsson:2017rkp}, measurements of the local DM density are found to be:
\begin{equation}
\rho_{\odot}=0.46^{+0.07}_{-0.09}~\text{GeV cm}^{-3}~.
\end{equation}
The Gaia satellite~\cite{Perryman:2001sp} is expected to catalogue the positions and velocities of billions of stars in the galaxy and therefore should permit the local Dark Matter density to be measured with increased precision in the following years. 

\subsubsection*{Conclusion}

In this chapter we discussed the several evidences and observations that have driven (most of) the scientific community to admit the presence of Dark Matter. Notably, we discussed the presence of Dark Matter on different physical scales based on gravitational lensing observations, measurements of the CMB anisotropy map, the matter power spectrum and rotation curves of galaxies. In the last section we discussed the role of N-body simulations in establishing the Dark Matter as an essential ingredient of structure formation.
In the next chapter we discuss elements related to the particle interpretation of the dark component of our universe.

%% file: parts/particleDM.tex
\vspace{0.3cm}

\noindent
The aim of this chapter is to present an overview of a class of solutions to the missing mass issue presented in the previous chapters, which assumes that the Dark Matter component of our universe is made of particles. First, we start by describing the current landscape of the established Standard Model of particle physics before motivating the need to go beyond. In the following sections, we present one of the most motivated mechanism to produce the Dark Matter density as well as some of the most popular particle physics models. For completeness, we also present the most common alternative solutions as well as constraints and arguments against the particle hypothesis.

\section{The Standard Model of Particle Physics}
The Standard Model of particle physics provides a modern description of the strong, weak and electromagnetic interactions. It describes in an elegant way how interactions between quantum fields, mediated by gauge bosons, emerge from the concept of local gauge invariance and Lorentz invariance. The Standard Model is a renormalizable Quantum Field Theory (QFT) implying that the experimental determination of a finite set of parameters is sufficient to theoretically predict observables up to any arbitrary large order in perturbation theory. However, the Standard Model is not valid up to infinite energy as it does not embed a proper quantum description of gravity. Therefore a natural energy cutoff of this theory is expected to be around the Planck scale, where gravitational corrections are expected to become significant. In the Standard Model, elementary matter constitutents are described by chiral fermionic fields charged under the local gauge group:
\begin{equation}
\mathcal{G}_{\text{SM}}=SU(3)_c \otimes SU(2)_L \otimes U(1)_Y ~,
\end{equation}
where $SU(3)_c$ is associated to Quantum Chromodynamics, $SU(2)_L$ to the weak isospin and $U(1)_Y$ to the hypercharge. In addition, it relies on the Higgs mechanism in order to describe the spontaneous breaking of the electroweak symmetry and the generation of masses of most of the particle content of the Standard Model. 
The full Lagrangian of the Standard Model can be expressed as a sum of four terms:
\begin{equation}
\mathcal{L}_{\text{SM}}=\mathcal{L}_{\text{gauge}}+\mathcal{L}_{\text{fermion}}+\mathcal{L}_{\text{scalar}}+\mathcal{L}_{\text{Yukawa}} ~,
\end{equation}
where each term will be explicited in the following.

\subsection{The Standard Model Langrangian}

\subsubsection{The Gauge sector}
Twelve gauge fields are present in the SM which are spin-1 bosons mediating the Standard Model interactions: eight gluons $G^a_\mu(a=1,...,8)$ and three electroweak fields $W^a_\mu(a=1,...,3)$ transforming respectively under the adjoint representation of $SU(3)_c$ and $SU(2)_L$ and one hypercharge field $B^\mu$. The $SU(N)$ non-abelian gauge group generators $T_N$ obey the following Lie algebra relations: 
\begin{equation}
[T_2^a,T_2^b]=i\varepsilon^{a b c}T_2^c ~, \qquad \text{and} \qquad [T_3^a,T_3^b]=if^{a b c}T_3^c ~,
\end{equation}
where $\varepsilon^{a b c}$ and $f^{a b c}$ are called the structure constants of the $SU(2)_L$ and $SU(3)_c$ groups. The $SU(2)_L$ generators are given by $T_2^a=(1/2)\sigma^a$ where $\sigma^a$ are the three Pauli matrices:
\begin{equation}
\sigma^1=\left(\begin{matrix}
0 & 1\\
1 & 0
\end{matrix} \right), \qquad \sigma^2=\left(\begin{matrix}
0 & -i\\
i & 0
\end{matrix} \right), \qquad \sigma^3=\left(\begin{matrix}
1 & 0\\
0 & -1
\end{matrix} \right)~.
\end{equation}
The generators $T_3^a$ can be represented as $3 \times 3$ matrices with similar properties and are called Gell-Mann matrices.
Field strength tensors of the gauge fields are defined by the following relations:
\begin{align}
B_{\mu \nu}&=\partial_\mu B_\nu-\partial_\nu B_\mu ~,\\
W^a_{\mu \nu}&=\partial_\mu W^a_\nu-\partial_\nu W^a_\mu+g \varepsilon^{a b c} W_\mu^b W_\nu^c ~, \\
G^a_{\mu \nu}&=\partial_\mu G^a_\nu-\partial_\nu G^a_\mu+g_s f^{a b c} G_\mu^b G_\nu^c~,
\end{align}
where $g$ and $g_s$ are the electroweak and strong gauge couplings.
From these definitions, one can build a gauge invariant Lagrangian for the gauge fields including kinetic terms and interaction terms, between non-abelian gauge fields, in the compact form:
\begin{equation}
\mathcal{L}_{\text{gauge}}=-\dfrac{1}{4}B_{\mu \nu}B^{\mu \nu}-\dfrac{1}{4}W^a_{\mu \nu}W^{\mu \nu,a}-\dfrac{1}{4}G^a_{\mu \nu}G^{\mu \nu,a}~.
\end{equation}
Any mass term for the gauge fields is forbidden by the gauge symmetry structure. In the following sections we will see how the Spontaneous Symmetry Breaking (SSB) of the $SU(2)_L \times U(1)_Y$ gauge symmetry via the Higgs-Brout-Englert mechanism can lead to the mass generation of the $W^{\pm}$ and $Z$ gauge boson.


\subsubsection{Quarks and leptons}
The fermionic content of the Standard Model is devided intro three families of quarks and leptons with identical quantum numbers as summarized in Table~\ref{charges_SM}. Each generation contains $SU(2)_L$ singlets that are right-handed fields: $u_R, d_R$ and $\ell_R$ and two $SU(2)_L$ doublets:
\begin{equation}
L_i=\left( \begin{array}{c}
\nu_{\ell_i} \\ 
\ell_i^-
\end{array} \right)_L,  \qquad Q_i=\left( \begin{array}{c}
u_{i} \\ 
d_i
\end{array} \right)_L~.
\end{equation}
\begin{table}[t]
\begin{center}
\begin{tabular}{|c||c c c|}
\hline 
 & $U(1)_Y$ & $SU(2)_L$ & $SU(3)_c$ \\ 
\hline 
\hline 
$Q^i$ & 1/6 & $\mathbf{2}$ & $\mathbf{3}$ \\ 
\hline 
$u_R^i$ & 2/3 & $\mathbf{1}$ & $\mathbf{3}$ \\ 
\hline 
$d_R^i$ & -1/3 & $\mathbf{1}$ & $\mathbf{3}$ \\ 
\hline 
$L^i$ & 1/2 & $\mathbf{2}$ & $\mathbf{1}$ \\ 
\hline 
$\ell_R^i$ & -1 & $\mathbf{1}$ & $\mathbf{1}$ \\ 
\hline
\end{tabular}
\end{center}
\caption{\label{charges_SM} Charges of the Standard Model fermionic content. The index $i=1,...,3$ refers to the three generations. The bold notation used for the non-abelian gauge group denotes the dimension of the representation.}
\end{table}
The gauge invariant kinetic and gauge interaction terms are given by the Dirac Lagrangian:
\begin{equation}
\mathcal{L}_{\text{fermion}}= \sum_{i=e,\mu,\tau} i \xbar{L}_i \slashed{D}L_i + \sum_{i=e,\mu,\tau} i \xbar{\ell}_{R i} \slashed{D}\ell_{Ri}+\sum_{\text{quarks}} i \xbar{q}_{R i} \slashed{D}q_{Ri}+\sum_{\text{quarks}} i \xbar{Q}_{L i} \slashed{D}Q_{Li}~,
\end{equation}
where we used the Feynman notation $\slashed{D}\equiv D_\mu \gamma^\mu$ with $D_\mu$ the covariant derivative that can be written after electroweak spontaneous symmetry breaking as:
\begin{equation}
D_\mu=\partial_\mu-\dfrac{ig}{\sqrt{2}}(\sigma^+ W^+_\mu+\sigma^- W_\mu^-)-\dfrac{ig}{2\cos \theta_W}(\sigma_3-2Q\sin^2\theta_W)Z_\mu-ie Q A_\mu~,
\end{equation}
where $Y$ is the hypercharge, $\sigma^{\pm}=\dfrac{1}{2}(\sigma^1 \pm i\sigma^2)$ and the electric charge $Q=T^3 + Y$. A mass term for any fermion $\Psi$ should take the form:
\begin{equation}
\mathcal{L} \supset m_\Psi\xbar{\Psi}\Psi=m_\Psi(\xbar{\Psi}_L \Psi_R+\xbar{\Psi}_R \Psi_L)~.
\end{equation}
However because of the chiral $SU(2)_L$ charges assignment, gauge symmetry forbids any mass term in the fermionic sector of the Standard Model. In the following section we describe how the Higgs mechanism provides a theoretical explanation for the presence of massive fermions in the Standard Model while respecting the gauge invariance condition.

\subsubsection{The scalar sector}
The price to pay to introduce masses for the weak bosons and fermions of the Standard Model is to allow for the breaking of the $SU(2)_L \times U(1)_Y$ gauge group. This can be achieved, while ensuring the renormalizability of the theory, by invoking the mechanism introduced independently by Higgs~\cite{Higgs:1964ia,Higgs:1964pj}, Brout and Englert~\cite{Englert:1964et}. In this mechanism, the symmetry transformation properties of the Lagrangian is maintained but it is not respected by the ground state. This is achieved in the Standard Model by introducing a complex scalar $SU(2)_L$ Higgs doublet $\Phi$ with hypercharge $Y=1/2$ that can be written as a function of four real degrees of freedom $\phi_i (i=1,...,4)$:
\begin{equation}
\Phi=\dfrac{1}{\sqrt{2}}\left( \begin{array}{c}
\phi_1+i\phi_2 \\ 
\phi_3+i\phi_4
\end{array} \right)~,
\end{equation}
The most general renormalizable kinetic and potential terms for this complex scalar respecting the gauge invariance reads
\begin{equation}
\mathcal{L}_{\text{scalar}}=(D^\mu \Phi)^\dagger (D_\mu \Phi) -V(\Phi)~,
\end{equation}
where $D_\mu \Phi$ is the covariant derivative of the Higgs doublet with
\begin{equation}
D_\mu = \partial_\mu - \dfrac{ig}{2}\sigma^a W^a_\mu - i g^\prime Y B_\mu~,
\end{equation}
where $ g^\prime$ is the gauge coupling associated to the hypercharge gauge group. $V(\Phi)$ is the Higgs potential that can be parametrized by two parameters $\mu$ and $\lambda$, which are dimension-1 and dimensionless respectively
\begin{equation}
V(\Phi)=-\mu^2|\Phi|^2+\lambda|\Phi|^4~.
\end{equation}
The parameter $\lambda$ has to be positive to ensure that the potential is bounded from below. However if the parameter $\mu^2$ is positive, the minimum of the potential $V(\Phi)$ occurs for a non-vanishing vacuum expectation value $v$
\begin{equation}
 \langle 0 |\Phi^\dagger \Phi|0\rangle  = \dfrac{v^2}{2}~,
\end{equation}
where $v$ can be computed by minimizing the Higgs potential $v=\mu/\sqrt{\lambda}$. We can choose a specific vacuum configuration by performing a gauge transformation and write the vacuum expectation value of Higgs doublet as
\begin{equation}
\langle \Phi \rangle= \dfrac{1}{\sqrt{2}}\left( \begin{array}{c}
0 \\ 
v
\end{array} \right)~.
\end{equation}
We can rewrite the four degrees of freedom that we introduced as fluctuations around the electroweak vacuum, three degrees of freedom as Goldstone bosons of the theory, will be absorbed by gauge fields while generating their masses and one is called the Higgs boson $h$ used to parametrize the Higgs doublet in the unitary gauge
\begin{equation}
\Phi = \dfrac{1}{\sqrt{2}}\left( \begin{array}{c}
0 \\ 
v+h
\end{array} \right)~.
\end{equation}
Making explicit the covariant derivative from the kinetic terms of the Higgs doublet in this parametrization leads to mass terms for 3 gauge fields $W^\pm_\mu$ and $Z_\mu$ redefined in the following way
\begin{equation}
W_\mu^\pm = \dfrac{W^1_\mu\mp i W^2_\mu}{\sqrt{2}}~,
\end{equation}
and
\begin{equation}
\left( \begin{array}{c}
Z_\mu \\ 
A_\mu
\end{array} \right)=\left( \begin{array}{c c}
\cos \theta_W & -\sin \theta_W \\ 
\sin \theta_W  & \cos \theta_W 
\end{array} \right) \left( \begin{array}{c}
W^3_\mu \\ 
B_\mu
\end{array} \right)~,
\end{equation}
where we used the Weinberg angle $\theta_W$ defined as follow
\begin{equation}
\tan \theta_W \equiv \dfrac{g^\prime}{g}~.
\end{equation}
The masses of the elecroweak bosons can be related to the vacuum expectation value and the Weinberg angle
\begin{equation}
m_W=\dfrac{gv}{2} \qquad  m_Z = \dfrac{gv}{2 \cos \theta_W} \qquad \text{and} \qquad m_A=0~.
\end{equation}
A mass term for the Higgs boson is also generated after the electroweak SSB and reads:
\begin{equation}
m_h=\sqrt{2 \lambda} v~.
\end{equation}
Therefore all the physical masses of Standard Model bosons depends only on very few parameters and they have been determined experimentally quite precisely over the past decades~\cite{Olive:2016xmw} as:
\begin{equation}
m_Z=91.1876(21)~\text{GeV}~,\quad m_W=80.385(15)~\text{GeV}~,\quad m_h=125.09(24)~\text{GeV}~,
\end{equation}
\subsubsection{The Yukawa structure}
In the Standard Model the only gauge invariant Yukawa terms that one can write are the following
\begin{equation}
\mathcal{L}_{\text{Yukawa}}=-(y_\ell)_{ij} \xbar{L_i}\Phi
\ell_{R j} - (y_u)_{ij} \xbar{Q_i}\widetilde{\Phi}
u_{R j} -(y_d)_{ij} \xbar{L_i}\Phi
d_{R j}+\text{h.c.}
\end{equation}
where we introduced the conjugate $SU(2)_L$ doublet $\widetilde{\Phi}\equiv i\sigma^2 \Phi^*$ with an hypercharge $Y_{\widetilde{\Phi}}=-1$ and couplings to up-type quarks $y_{ij} $ are $3 \times 3$ complex matrices. After the electroweak spontaneous symmetry breaking, in unitary gauge the Yukawa Lagrangian reads:

\begin{equation}
\mathcal{L}_{\text{Yukawa}}=-\left(\dfrac{v +h}{\sqrt{2}}\right)\left[(y_\ell)_{ij}  \xbar{\ell}_{L i}
\ell_{R j} + (y_u)_{ij} \xbar{u}_{Ri}
d_{L j} +(y_d)_{ij} \xbar{d}_{Li} d_{R j}+\text{h.c.} \right]~.
\end{equation}
A coupling between the Higgs boson and fermions is generated from the previous term. The non-vanishing vacuum expectation values introduced flavor non-diagonal mass terms that can be diagonalized by perfoming the following unitarity redefinition of the fields

\begin{equation}
u_{L i} \rightarrow V_u^{ij}u_{L j} \quad \text{and} \quad d_{L i} \rightarrow V_d^{ij}d_{L j}~.
\end{equation}
From these unitarity transformations we can rewrite the charge current of the quarks as
\begin{equation}
J^{\mu +}=\dfrac{1}{\sqrt{2}}\xbar{u}_{Li} \gamma^\mu (V^\dagger_u V_d)_{ij}d_{L j} + \text{h.c.}=\dfrac{1}{\sqrt{2}}\xbar{u}_{Li} \gamma^\mu (V_{\text{CKM}})_{ij}d_{L j}~,
\end{equation}
where we introducted the Cabibbo-Kobayashi-Maskawa (CKM) matrix $(V_{\text{CKM}})_{ij}$ which is flavor non-diagonal. The CKM matrix is almost diagonal and can be expressed as powers of some small parameter $\lambda\simeq 0.22$ as
\begin{equation}
V_{\rm CKM}\sim \left( \begin{array}{ccc}
1 & \lambda & \lambda^3 \\ 
\lambda & 1 & \lambda^2 \\ 
\lambda^3 & \lambda^2 & 1
\end{array} \right)~.
\end{equation}
Coming back to the Yukawa Lagrangian after the diagonalization of the mass matrix, it can be written in the following form:
\begin{equation}
\mathcal{L}_{\text{Yukawa}}\supset -\sum_{\Psi=u,d,\ell} \dfrac{y_\Psi v}{\sqrt{2}} \xbar{\Psi}_L \Psi_R +\text{h.c}
\end{equation}
where $y_\Psi$ denotes the Yukawa couplings in the diagonal basis. The generated fermion masses are then
\begin{equation}
m_\Psi = \dfrac{y_\Psi v}{\sqrt{2}}~.
\end{equation}
After diagonalization of the mass terms, the interaction term with the Higgs boson becomes
\begin{equation}
\mathcal{L}_{\text{Yukawa}}\supset -\sum_{\Psi=u,d,\ell} h \dfrac{m_\Psi}{v} \Big( \xbar{\Psi}_L \Psi_R + \text{h.c.} \Big)~,
\end{equation}
The previous formula is a strong prediction of the Standard Model showing that couplings between the Higgs boson and fermionic states are proportional to their masses. Often considered as the cornerstone of the Standard Model, the discovery of the Higgs bosons in 2012 highlights the concistency and establishment of the Standard Model as an accurate description of microscopic interactions.

\subsection{The need to go beyond the Standard Model}
We presented the Standard Model as being a remarkable theory allowing for a description of microscopic processes occuring in our universe. However, the Standard Model is far from being a complete theory and it is established that supplementary degrees of freedom have to be introduced in order to solve inconvenient features discussed in the following.

\subsubsection{The hierarchy problem}
Altough being a renormalizable theory, the Standard Model is not valid up to some infinite scale as it does not embed a proper quantum description of gravitational interactions. Therefore the Standard Model must be an effective theory valid up to the Planck scale $M_{\text{Pl}}\sim10^{19}~\text{GeV}$ where we expect the degrees of freedom of some quantum version of gravity to become relevant. In the Standard Model, masses of gauge bosons and fermions $m_F$ are generated via the Higgs mechanism and can be written is the schematic way:
\begin{equation}
m_F^{\text{R}}=m_F^{\text{bare}}+\delta m_F~,
\end{equation}
where $m_F^{\text{bare}}$ denotes a bare mass term present in the Lagrangian, $\delta m_F$ the one loop correction and $m_F^{\text{R}}$ the renormalized value. Because a mass term is forbidden by the imposed symmetries of the Standard Model and is generated via the Higgs mechanism, its quantum corrections are proportional to the vacuum expectation value $v$ of the Higgs field and vanishes in the limit where the symmetry is restored $v \rightarrow 0$, which prevents quadratic or linear divergences such that $ \delta  m_F \propto v \log \left(\Lambda/m_F \right)$. Considering the cutoff of the order of the Planck scale gives quantum correction $\delta m_F\sim \mathcal{O}(m_F^{\text{bare}})$. However a mass term for the Higgs doublet is not forbidden by any symmetry, and as a result the quantum corrections $\delta m_h$ of the Higgs mass are quadratically dependent on the cutoff scale $\Lambda$. One of the leading contribution is given by:
\begin{equation}
\delta m_h^2 \supset-\dfrac{3y_t^2}{8 \pi^2} \Lambda^2~,
\end{equation}
with $y_t$ being the Yukawa coupling of the top-quark. In this case if we consider the Planck mass as the cutoff scale of the theory, one would expect relative quantum corrections to be roughly:
\begin{equation}
\dfrac{\delta m_h}{m_h}  \sim 10^{15}~,
\end{equation}
which corresponds to an extreme fine tuning of 15 decimals between the bare mass term and quantum corrections, required for the theoretical prediction to match the experimentally measured value of the Higgs mass $m_h \simeq 125~\text{GeV}$. Actually, taking a cutoff of the order of the TeV scale already raises a fine tuning issue. This problem is known as \textit{hiearchy problem} and is a motivation to search for beyond-the-Standard-Model physics at the TeV scale or above.

\subsubsection{Neutrino masses}
One important missing piece of the Standard Model is related to the fact that right-handed neutrinos are not included in this theory. Indeed, in the SM, neutrinos only interact with left-handed currents through weak interactions. However it has been established over the past few years that neutrinos oscillate and therefore, they have to be massive. A global analysis of several experiments~\cite{Wang:2015rma}, including solar, accelerator, atmospheric and reactor neutrino experiment, allowed to estimate the difference squared mass $\Delta m_{ij}^2 \equiv m_{\nu_i}^2-m_{\nu_j}^2$, where $i,j=1,2,3$ denote the mass eigenstate, as
\begin{equation}
\begin{split}
\Delta m_{21}^2 = 7.54 \times 10^{-5}~\text{eV}^2 \qquad \Delta m_{31}^2 = 2.47 \times 10^{-3}~\text{eV}^2 \qquad & \text{(normal)}\\
\Delta m_{21}^2 = 7.54 \times 10^{-5}~\text{eV}^2 \qquad \Delta m_{13}^2 = 2.42 \times 10^{-3}~\text{eV}^2 \qquad & \text{(inverted)}\\
\end{split}
\end{equation}
which depends on the hierarchy between masses $m_3>m_2>m_1$ (normal) and $m_2>m_1>m_3$ (inverted). The unitary transformation performed to change from the eigenstate to the mass basis is parametrized by the PMNS matrix similarly to the CKM matrix in the quark sector. One of the easiest solution to explain neutrino masses is to introduce three right-handed neutrinos $\nu_R$ singlet under the SM gauge groups such that one can write a term similar to the quark sector
\begin{equation}
\mathcal{L}\supset \sum_{i=e,\mu,\tau}y^\nu_{ij} \xbar{L}_i \Phi \nu_{R j}+~\text{h.c.}  \supset \sum_{i=1,2,3} - m_{\nu_i} \xbar{\nu_{Ri}} \nu_{Li}~.
\end{equation}
After electroweak spontaneous symmetry breaking, masses are generated for the three neutrino generations. However, in order to generate such a large hiearchy between the Higgs mass and the neutrino masses, the Yukawa couplings have to be extremely small $y^\nu_{ij}\sim m_\nu/v \lesssim 10^{-10}$ which could seem \textit{unatural}. Therefore, in order to avoid invoking such small numbers, one could consider large Majorana masses $M_{\text{M}}$ for the three right-handed neutrinos and Dirac masses $m_{\text{D}}$ such as\footnote{We remove flavor indices for clarity but the generalization is straightforward.}
\begin{equation}
\mathcal{L}\supset -\dfrac{1}{2} M_{\text{M}} \xbar{\nu_R^c} \nu_R - m_{\text{D}}\xbar{\nu_R} \nu_L +~\text{h.c.}~.
\end{equation}
This term is allowed by gauge symmetries but would imply a lepton-number violation by generating a low energy gauge invariant dimension-5 effective operator, known as \textit{Weinberg operator}~\cite{Weinberg:1979sa}
\begin{equation}
\mathcal{O}^5=\dfrac{1}{M_{\text{M}}}\left( \xbar{L_L^c} \Phi^* \right) \left( \tilde{\Phi}^\dagger L_L  \right) \supset \dfrac{v^2}{M_{\text{M}}} \xbar{\nu_L^c} \nu_L~.
\end{equation}
which generates a Majorana mass term for the left-handed neutrinos after spontaneous symmetry breaking. This can be shown considering the mass matrix
\begin{equation}
\mathcal{L}=\dfrac{1}{2}\left(    \begin{array}{cc}
\xbar{\nu_L} & \xbar{\nu_R^c}
\end{array}\right) \left( \begin{array}{cc}
0 & m_{\text{D}} \\ 
m_{\text{D}} & M_{\text{M}}
\end{array} \right) \left( \begin{array}{c}
\nu_L \\ 
\nu_R^c
\end{array} \right)  +~\text{h.c.}~,
\end{equation}
whose eigenvectors are Majorana fields corresponding to the eigenvalues $m\simeq m_{\text{D}}^2/M_{\text{M}}$ and $M\simeq M_{\text{M}}$ in the limit where $M_{\text{M}}\gg m_{\text{D}}$. Thefore the large Majorana mass term of the right-handed neutrinos would explain the hierachy between the Higgs mass and the neutrino masses without introducing very small Yukawa couplings, also known as the~\textit{see-saw mechanism}. This framework has widely been studied~\cite{Yanagida:1979as,Minkowski:1977sc} and many extensions~\cite{Mohapatra:1986bd} have been considered invoking a higher number of degrees of freedom and usually new physics at the TeV scale or higher.

\subsubsection{The gauge coupling unification}
\begin{figure}[h!]
\begin{center}
\includegraphics[width=0.7\linewidth]{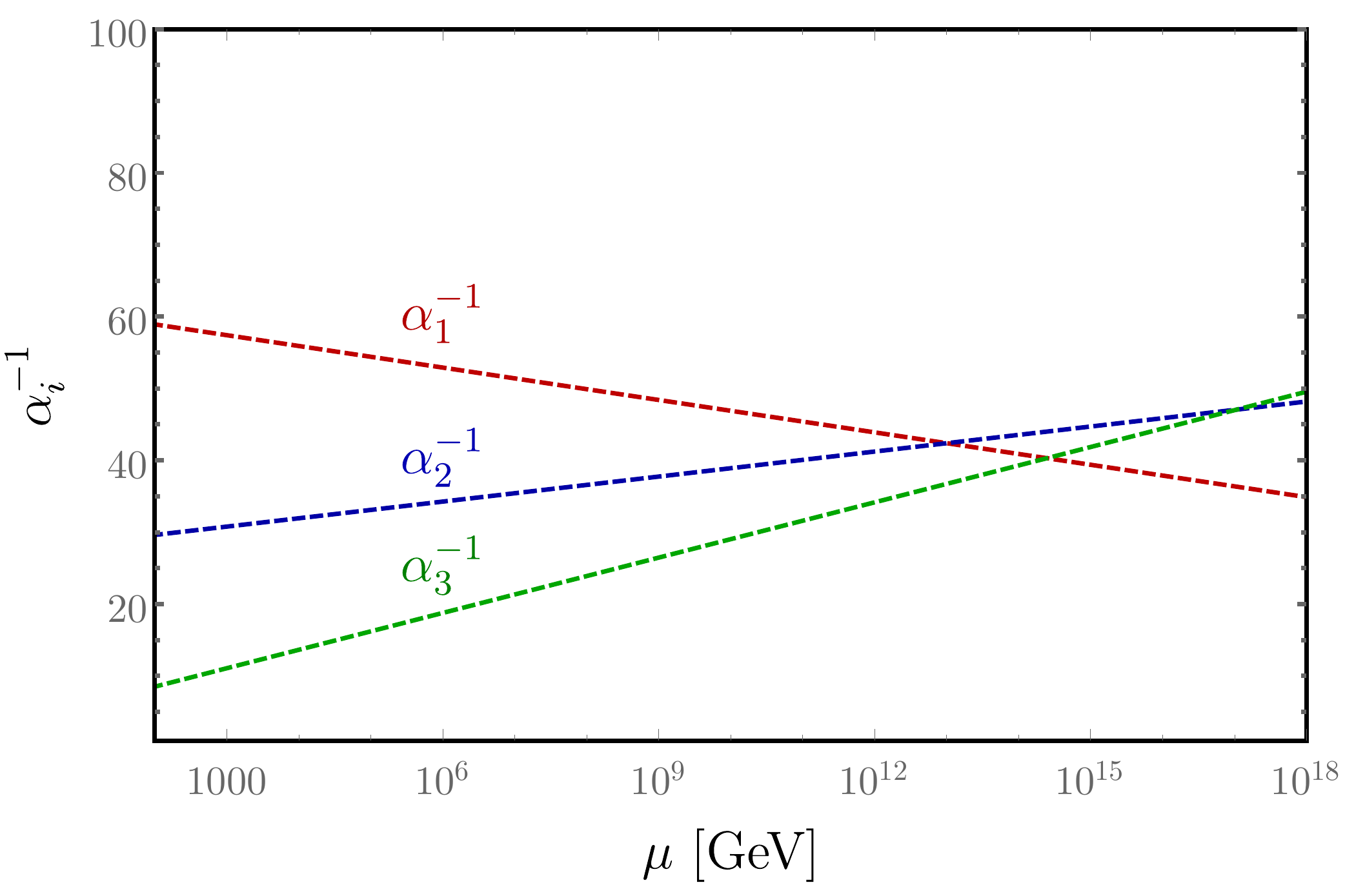}
\caption{One loop Renormalization Group Equation evolution of the gauge coupling constants of the Standard Model with the energy scale $\mu$. The parameter $\alpha_i$ is related to the gauge coupling $g_i$ through $\alpha_i=g_i^2/4\pi$} 
\label{fig:RGESM}
\end{center}
\end{figure}

One argument to look for supplementary degrees of freedom at high energy is based on the fact that when looking at the running of the Standard Model gauge couplings, as depicted in Fig.~\ref{fig:RGESM} at the one loop level, one can realize that the three gauge couplings seem to converge toward the same value. 

This motivates beyond-the-Standard-Model constructions such as Grand Unification Theories for instance which are theories considering an extended gauge structure at high energies which contains the SM gauge group and therefore, unifying the known microscopic interactions.

\section{Dark Matter thermal production : the WIMP miracle}
\label{sec:WIMP}
\begin{figure}[h!]
\begin{center}
\includegraphics[width=0.8\linewidth]{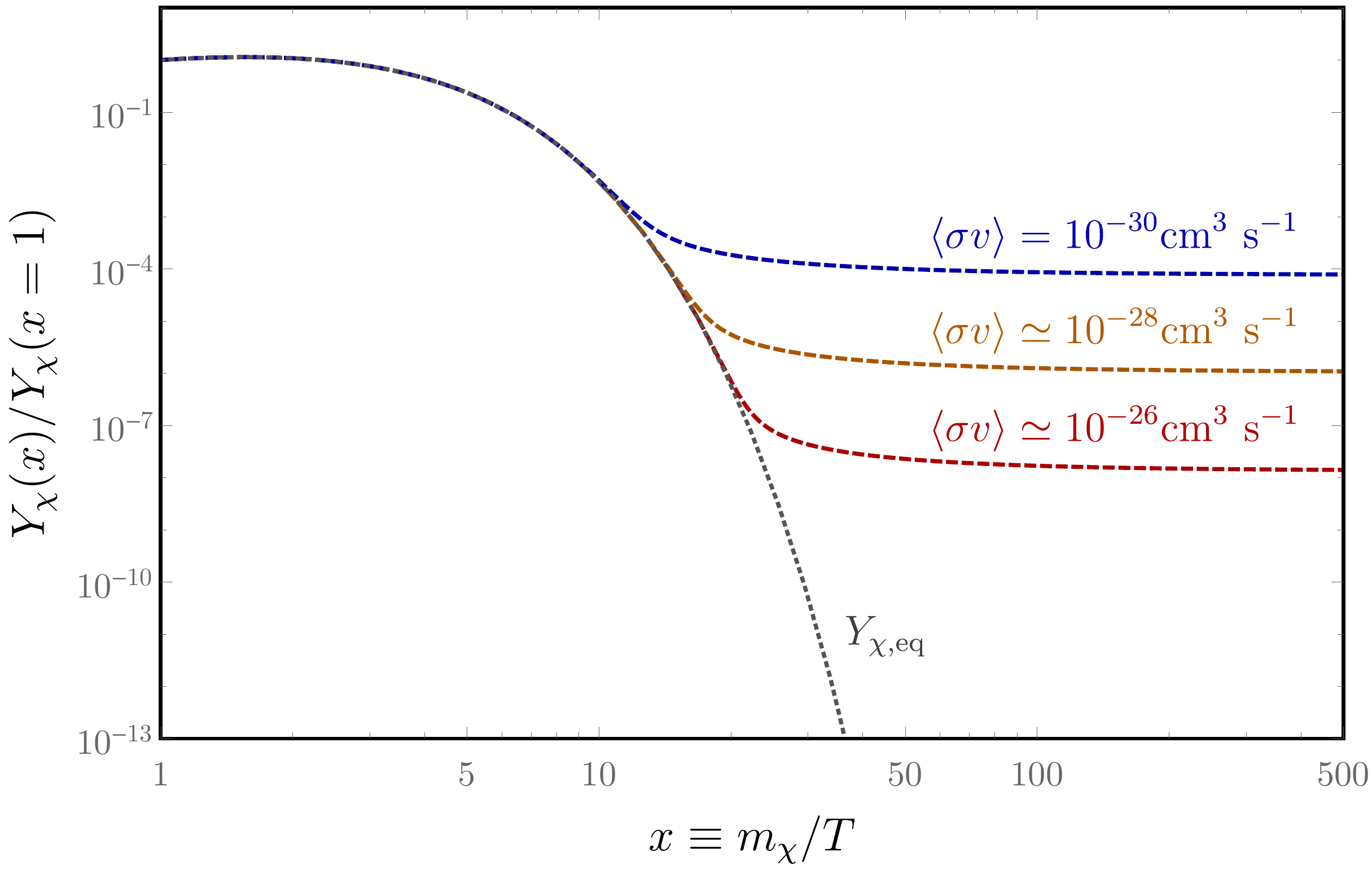}
\caption{Evolution of the Dark Matter yield $Y_\chi$ as a function of the variable $x$, growing with time. The dashed blue, orange and red curves are numerical solution of Eq.~(\ref{eq:BoltzmannYield}) for several values of $\la \sigma v \ra$ and the dotted black line is the thermal yield distribution.} 
\label{fig:BoltzmannWIMP}
\end{center}
\end{figure}

In the previous section we exposed some arguments motivating the need to look beyond the Standard Model and to introduce new degrees of freedom. In the case where some of these supplementary degrees of freedom would play the role of the Dark Matter, one could ask the questions: Why the Dark Matter density is the one observed nowdays? How can it be generated? One possibility to account for the Dark Matter relic abundance is to assume that the evolution of the Dark Matter particles has followed a story similar to the Standard Model particle content. Because of the weak interactions with Standard Model particles that the Dark Matter must manifest according to experimental suggestions, one may assume that the Dark Matter is made of \textit{Weakly Interacting Massive Particles} (WIMPs) formely in thermal equibrium with the SM bath, that decoupled at some early stage of the universe. This idea emerged in the 1980's and remains at the present time the most common paradigm to explain the Dark Matter relic abundance\footnote{For further reading, reviews can be found in~\cite{
Jungman:1995df,Baer:2014eja,Arcadi:2017kky,Bertone2005a}.}. Assuming a primordial thermal contact between SM and DM particles, the DM number density evolution with time depends on the Hubble expansion and possible annihilation/creation processes involving DM particles. Considering for instance a process $\chi+  \chi \leftrightarrow \psi +\psi$ where $\chi$ is a DM candidate whose mass $m_\chi \sim~\text{GeV-TeV}$ and $\psi$ is a SM field, the Dark Matter density evolution is given by the Boltzmann equation\footnote{Details regarding the derivation and resolution of this equation are given in Sec.~\ref{sec:Boltzmann}.}
\begin{equation}
\dfrac{\diff n_\chi}{\diff t}+3Hn_\chi=-\langle \sigma v \rangle (n_\chi^2-n_{\chi,\text{eq}}^2)~,
\end{equation}
where $\la \sigma v \ra$ is the \textit{velocity averaged annihilation cross section} of the process $\chi+  \chi \rightarrow \psi +\psi$ and $n_{\chi,\text{eq}}$ is the expected DM density in case of thermal equilibrium with $\psi$, i.e. if DM pair annihilation and creation occur at the same rate.
The left-hand side of this equation represents the evolution of DM density 
in the case where interaction processes are negligible, i.e. the right-hand side vanishes, then the Dark Matter density would evolve according to the Hubble expansion $n_\chi \propto a^{-3}$. The right-hand side terms tend to ensure thermal equilibrium between DM and SM particles, and lead the DM density to evolve according to its thermal distribution. However, since $n_{\chi,\text{eq}}\propto T^{3/2}e^{-m_\chi/T}$ in the non-relativistic regime, the interaction rate would become suppressed at $T\sim m_{\rm DM}$ compared to the Hubble expansion rate, resulting in a \textit{freeze-out} of the Dark Matter density.\\
The Boltzmann equation can more suitedly be expressed by defining the \textit{yield} $Y_\chi \equiv n_\chi/s$ as the ratio of the DM number density over the SM entropy density, which is a quantity proportional to the comoving number in the radiation domination era, and using the variable $x\equiv m_\chi/T$ yields
\begin{equation}
\dfrac{\d Y_\chi}{\d x}  = - \frac{\la \sigma v \ra s(x) \xi(x)}{ H(x) x}(Y_\chi^2-Y^{2}_{\chi,\text{eq}} )~,
\end{equation}
where the function $\xi(x) \sim 1$ corresponds to the temperature variation of the effective number of relativistic degrees of freedom:
\begin{equation}
\xi(x)\equiv 1-\dfrac{1}{3}\dfrac{\diff \log g_{\star,s}}{\diff \log x}~.
\end{equation}
Neglecting temperature variation of $g_{\star(,s)}$ allows to write the Boltzmann equation in the compact form
\begin{equation}
\dfrac{\d Y_\chi}{\d x}  = - \la \sigma v \ra \frac{\kappa}{ x^2}(Y_\chi^2-Y^{2}_{\chi,\text{eq}} ) \quad \text{with} \quad \kappa \equiv \dfrac{g_{\star,s}}{g_\star^{1/2}}\dfrac{2 \sqrt{2}\pi}{45} M_{\text{Pl}} m_\chi~.
\label{eq:BoltzmannYield}
\end{equation}
Assuming that $\la \sigma v \ra$ does not depend on the temperature\footnote{This assumption is not always satisfied, a more detailed treatment can be found in Sec.~\ref{sec:Boltzmann}.}, this equation can be integrated numerically as shown in Fig.~\ref{fig:BoltzmannWIMP} where several solutions corresponding to different values of $\la \sigma v\ra$ are depicted. This figure shows that the yield follows at first its thermal expected value when the DM becomes non-relativistic and after $x\sim10-20$ the yield $Y_\chi$ remains constant over time. A large annihilation cross section implies smaller values of the yield at the present time because the annihilation process would be more efficient.
In order to estimate the DM relic density expected from the Boltzmann equation at the present time, we can integrate from the freeze-out time $x_{\text{F}}$ until the present time in the approximation of instantaneous freeze-out where we assume the term $Y_{\chi,\text{eq}}$ exponentially suppressed to be negligible with respect to $Y_{\chi}$ for $x > x_{\text{F}}$ which yields:
\begin{equation}
\dfrac{1}{Y_{\chi,\infty}}-\dfrac{1}{Y_\chi(x_{\text{F}})}=\dfrac{\la \sigma v \ra \kappa}{x_{\text{F}}}~,
\end{equation}
where $Y_{\chi,\infty}$ denotes the DM yield at the present time. Considering that the DM yield dropped substantially since the freeze-out time, we can neglect the second term in the left-hand side and we obtain the following expression:
\begin{equation}
Y_{\chi,\infty}\simeq\dfrac{x_{\text{F}}}{\kappa \la \sigma v \ra }~.
\end{equation}
Then we can deduce the DM relic density at the present time
\begin{equation}
\Omega_\chi h^2=\dfrac{m_\chi s_0 Y_{\chi,\infty} h^2}{\rho_c^0}=\dfrac{x_{\text{F}}}{\kappa \la \sigma v \ra}\frac{m_\chi s_0 h^2}{\rho_c^0}~,
\end{equation}
where $s_0$ and $\rho_c^0$ are the entropy density and the critical density at present time. The ratio of these quantities is $s_0/\rho_c^0 \simeq 2.5 \times 10^8\text{ GeV}^{-1} h^{-2}$. In order to derive a numerical result for the relic density we can estimate the value of $x_{\text{F}}$ by considering the freeze-out time as the instant where the expansion rate becomes of the order of the interaction rate
\begin{equation}
H(x_{\text{F}})=\la \sigma v \ra n_{\text{eq}}(x_{\text{F}})~,
\end{equation}
which gives $x_{\text{F}}$ as the solution of the transcendental equation:
\begin{equation}
x_{\text{F}} \simeq \log \left( \dfrac{3 \sqrt{5} M_{\text{Pl}} m_\chi  \la \sigma v \ra \sqrt{x_{\text{F}}}}{2 \pi ^{5/2} \sqrt{g_\star}} \right)~.
\end{equation}
This equation shows that the freeze-out time depends only mildly on the particle physics input and an approximated solution is given by $x_{\text{F}}\sim 20$, confirming our hypothesis of non-relativistic decoupling. Using this value, an estimation of the relic density at the present time is given by
\begin{align}
\Omega_\chi h^2 \simeq 0.1 \left(  \dfrac{m_\chi}{100~\text{GeV}}  \right) \left(  \dfrac{3 \times 10^{-26}~\text{cm}^3 \text{s}^{-1}}{\la \sigma v \ra}  \right)~.
\end{align}
This expression shows the relation between the particle physics input, i.e. the annihilation cross section and Dark Matter mass, and the value of the energy density on cosmological scales. The fact that a DM mass of the order of the electroweak scale $m_\chi \sim 100~\text{GeV}$ and a typical electroweak cross section $\sigma \sim g^{4}m_{\rm DM}^2/m_Z^4 \sim 10^{-9}~\text{GeV}^{-2} \simeq 10^{-26}~\text{cm}^3~\text{s}^{-1}$ lead to the observed DM relic density is known as the \textit{WIMP miracle} and is considered as a strong motivation to study electroweak-scale based DM models. Such a typical electroweak cross section would confirm our initial hypothesis of thermal equilibrium and ensure the theoretical consistency of the freeze-out mechanism. However the cross section value required to match the computed Dark Matter density to its observed value would be different in a couple of cases, for instance if some particle participates in the Dark Matter annihilation process, known as \textit{co-annihilation}~\cite{Edsjo:1997bg}. Public codes such as \texttt{MicrOMEGAs}~\cite{Belanger:2018ccd}, \texttt{DarkSUSY}~\cite{Bringmann:2018lay} or \texttt{Gambit}~\cite{Athron:2017ard} allow for numerical treatment of the Boltzmann equation and relic density computation for a given model. An important point about the freeze-out mechanism is that it does not rely on any specific assumption regarding the prior Dark Matter history before thermalization with the Standard Model particle content. As long as the thermalization condition is satisfied at some stage in the radiation domination era, this mechanism can be applied independently of the post-inflation history of the universe.

\section{Beyond the Standard Model candidates}

In the previous section we discussed about the necessity to enlarge the Standard Model particle content and how the Dark Matter could be produced thermally via the freeze-out mechanism. In this section we present some concrete realizations of these ideas by exposing beyond-the-Standard-Model famous theoretical constructions adressing the Dark Matter puzzle.

\subsection{Neutrinos as Dark Matter}

In the Standard Model, the only particles possessing all the required properties to be considered as a viable Dark Matter candidate are neutrinos as they are stable, neutral and massive. As discussed in Sec.~\ref{sec:kindec}, the neutrino temperature is of the order of the photon temperature and as neutrinos are non-relativistic nowdays, one can estimate their current density as:
\begin{equation}
\Omega_\nu h^2=\dfrac{\sum_i m_{\nu_i}}{94~\text{eV}}~.
\end{equation}
However experimental results suggest that neutrino masses have to be smaller than the eV scale implying that they can contribute to the Dark Matter density up to $10\%$ at most. Since neutrinos decoupled while being relatistic, even though they are non-relativistic at the present time, their phase space density still has the shape of a Fermi-Dirac distribution meaning that neutrinos are contributing as a \textit{hot} component to the Dark Matter density.\\
Assuming that some extra heavy singlet state $N$ mixes with Standard Model neutrinos with an angle $\theta_\alpha$ ($\alpha=e,\mu,\tau$), known as \textit{sterile neutrino}, the state $N$ can play the role of the Dark Matter providing that its lifetime is larger than the age of the universe as $N$ can decay to $N \rightarrow \nu + \gamma$ via loop induced processes. Sterile neutrinos are viable DM candidates and typically have masses of the order of the keV scale $m_N \lesssim 10~\text{keV}$ and a mixing angle $\theta_\alpha \lesssim 10^{-9}$ in order to evade bounds from a plethora of constraints~\cite{Adhikari:2016bei}, depending on the exact production mechanism.

\subsection{Supersymmetric candidates}

Supersymmetry (SUSY) has been initially introduced as a solution to the hierachy problem. The main idea is to introduce a new symmetry protecting the loop corrections of the Higgs mass to quadratically diverge with the energy cutoff of the theory. A supersymmetry transformation turns a bosonic states into a fermionic state and vice versa. For instance in the Minimal Supersymmetric Standard Model (MSSM), each of the fermionic (bosonic) Standard Model particle possesses a bosonic (fermionic) \textit{superpartner} with identical charge assignment, with a particle content augmented by a second Higgs doublet and associated superpartner. However if supersymmetry were an exact symmetry we should expect particles and their superpartner to be degenerate in mass but since such particles have not been discovered yet, supersymmetry must be broken in some way. If supersymmetry is broken softly at the TeV scale, one could still naturally solve the hierarchy problem by generating quantum corrections to the Higgs mass depending logarithmically on the cutoff scale of the theory. \\
In order to forbid fast proton decay mediated by some superpartner, highly unfavoured by experiments, an additional discrete symmetry called \textit{R-parity} is introduced whose conserved number can be expressed for each particle as:
\begin{equation}
R_P \equiv (-1)^{3(\rm{B}-\rm{L})+2s}~,
\end{equation}
where B and L are the baryon and lepton number and $s$ is the spin of the particle. As a result, particles of the Standard Model possess a charge $R_P=+1$ and their superpartner $R_P=-1$ implying that every interaction vertex contains only an even number of superparners allowing for the Lightest Supersymmetric Particle (LSP) of the theory to be stable and thefore to be a viable Dark Matter candidate. In the MSSM, the four fermionic neutral superpartner $\tilde{B}^0$, $\tilde{W}^0$, $\tilde{H}_u^0$ and $\tilde{H}_d^0$ associated to the electroweak neutral gauge fields and the Higgs doublets mix to give the \textit{neutralinos} $\tilde{\chi}^0_1$, $\tilde{\chi}^0_2$, $\tilde{\chi}^0_3$ and $\tilde{\chi}^0_4$. As the lightest neutralino is stable and features electroweak interactions, with a mass expected of the order of the TeV scale, it possesses all the required properties to be a viable WIMP candidate. Direct detection experiments have already excluded some part of the available parameter space in phenomenological versions of the MSSM but there is still some parameter space compatible with the requirement of generating the correct relic density~\cite{Munoz:2017ezd,Jungman:1995df}.\\
Another category of potential supersymmetric Dark Matter candidate is the superpartner of the left-handed neutrinos, the so-called \textit{sneutrinos}. However this candidate has sizable interactions with the $Z$-boson implying a large expected direct detection cross section in contradiction with experimental results, therefore ruling out this kind of particle as a solution to the Dark Matter problem.\\
Supersymmetry can be promoted to a \textit{local} symmetry and in this case a spin-3/2 superpartner of the graviton called \textit{gravitino} is present in the theory featuring Planck suppressed interactions. Gravitinos with very large masses could be produced in high-scale SUSY models from the SM thermal bath via gluon fusion  at temperature close to the reheating temperature~\cite{Benakli:2017whb}. In TeV SUSY frameworks, LSP gravitinos with masses typically of the order of the GeV scale could be produced via gluon fusion or from the decay of the Next-to-LSP (NLSP)~\cite{Rychkov:2007uq}. However gravitinos cannot be arbitrary light as NLSP decays to gravitinos might significantly alter BBN. 

\subsection{Axions and Axion-like particles}
\subsubsection{The QCD axion}
The QCD axion was introduced for the first time by Peccei and Quinn in 1977~\cite{Peccei:1977hh} as a solution to the strong CP problem. In the Standard Model the only source of CP violation is expected as a phase in the CKM matrix, however it was realized in the 70's that a non-perturbative CP violating term must be included in the QCD sector of the SM Lagrangian generated by non-trivial topological configurations of the vacuum 
\begin{equation}
\mathcal{L} \supset  \theta_{\text{QCD}}\dfrac{\alpha_s}{8\pi} G^{\mu \nu,a} \widetilde{G}_{\mu \nu}^a~,
\end{equation}
where $\theta_{\text{QCD}}$ is a parameter and $\widetilde{G}_{\mu \nu}^a=(1/2)\varepsilon^{\mu \nu \rho \sigma}G_{\rho \sigma}^a$ is the dual gluon field strength tensor. However strong experimental bounds from the measurement of the neutron electric dipole moment $|d_n| ~\lesssim~3 \times 10^{-26}~e\text{ cm}$~\cite{Afach:2015sja} are pushing the value of $\theta_{\text{QCD}}$ towards unaturally small values $\theta_{\text{QCD}}\lesssim 10^{-10}$, known as the \textit{strong CP problem}.
Therefore Peccei and Quinn suggested an elegant method to cancel this term in a dynamical way. They considered a global $U(1)_{\text{PQ}}$ symmetry and the would-have-been Nambu-Goldstone boson associated to the global spontaneous symmetry breaking is called the \textit{axion} $a$. However since this symmetry is broken explicitely by chiral anomalies generated by triangle diagrams, $a$ becomes a pseudo Nambu-Goldstone boson and a additional term has to be considered in the Lagrangian

\begin{equation}
\mathcal{L} \supset \dfrac{a}{f_{\text{PQ}}}\dfrac{\alpha_s}{8\pi}G^{\mu \nu,a} \widetilde{G}_{\mu \nu}^a~,
\end{equation}
where $f_{\text{PQ}}$ is the $U(1)_{\text{PQ}}$ breaking scale. This term corresponds to a potential term generated by instanton effects for $a$ which is minimized by the vacuum expectation value
\begin{equation}
\langle a \rangle = -\theta_{\textrm{QCD}} \cdot f_{\text{PQ}}~.
\end{equation}
For this vacuum configuration the $\theta_{\text{QCD}}$ is cancelled dynamically and the strong CP problem is solved. The mass of this light pseudo-scalar can be related to its coupling to matter $\propto 1/f_{\text{PQ}}$ as
\begin{equation}
m_a \simeq 6~\text{meV} \left(\dfrac{10^9 \text{GeV}}{f_{\text{PQ}}}\right)~.
\end{equation}
Due to mixing with neutral pion, the QCD axion would couple to a pair of photons
\begin{equation}
\mathcal{L} \supset -\dfrac{g_{a \gamma}}{4}a F_{\mu \nu} \widetilde{F}^{\mu \nu}, \quad \text{with} \quad |g_{a \gamma}|~=10^{-12} ~\text{GeV}^{-1} \left( \dfrac{10^9\text{ GeV}}{f_{\text{PQ}}} \right)~.
\end{equation}
The QCD axion has been considered as a Dark Matter candidate whose relic density can be achieved with a thermal or non-thermal production. The thermal QCD axion production would be essentially similar to the standard freeze-out mechanism and would require a sizable coupling to the SM particles in order to ensure thermal equilibrium in the early universe, leading to an axion mass too large to be compatible with the large scale structure formation. Another possibility for considering QCD axion as Dark Matter would be to produce it through the vacuum misalignment mechanism (MIS). Assuming that the breaking scale $f_{\text{PQ}}$ is large, when the temperature of the universe cooled down and reached $T \sim f_{\text{PQ}}$, the axion potential adopted the form of a mexican-hat similar to the Higgs potential in the SM. However, when $T \sim \Lambda_{\text{QCD}} \sim 200$MeV, instanton effects would break the $U(1)_{\text{PQ}}$ symmetry and the mexican-hat potential would tilt in such a way that only one vacuum configuration is stable, i.e. the configuration providing for a cancellation of the $\theta_{\text{QCD}}$ term. The QCD axion field would coherently oscillate around the minimum of its potential forming a Bose-Einstein condensate while its kinetic energy would contribute to the energy density of the universe, whose value at the present time is given by
\begin{equation}
\Omega_a h^2|_{\text{MIS}} \simeq 0.11 \left( \dfrac{40~\mu \text{eV}}{m_a} \right)^{1.19}~.
\end{equation}
Even though the mass of the axion is very light compared to other common Dark Matter candidates, it would still behave as a cold Dark Matter component. 
\subsubsection{Axion-like particles}
Similar axion-like particle (ALPs) constructions can be formulated as generalizations of the Peccei-Quinn idea where the axion mass and coupling are independent. Strong constraints are set by cosmological and astrophysical arguments as well a dedicated experiments~\cite{Essig:2013lka}. In this context ALPs could be a viable cold Dark Matter candidate while evading current bounds for couplings $\lesssim 10^{-13}$ GeV and masses $\lesssim 100$ eV.\\
One specific interesting scenario dubbed \textit{fuzzy cold Dark Matter}~\cite{Hu:2000ke} considers an extremely light ALP particle with a mass $m_a \sim 10^{-22}~\text{eV}$. This was initially motivated by a discrepancy existing between $\Lambda$CDM predictions and N-body simulations as discussed in Sec.~\ref{sec:smallscalescontroversies}. These very light particles would form a Bose-Einstein condensate on galactic scales while being stable due to the Heisenberg uncertainty principle. Therefore, free oscillations are leading to suppression of small-scale structures present in galaxies below the Jeans scale $r_{\text{J}}\sim 50~\text{kpc}$ for $m_a \sim 10^{-22}~\text{eV}$ while behaving as a cold Dark Matter component on larger scales.

\subsection{Extra dimensions}
In some theories, the possibility of having more than four dimensions is  considered. In particular, in the Universal Extra Dimensions (UED) framework all the particles of the Standard Model are free to propagate in all the dimensions. Assuming only one single extra dimension compactified on a physical scale $R$ implies a quantization of the momentum in the fifth dimension that can be seen as an apparent mass term in four dimensions. Therefore, a \textit{KK-tower} of particles with identical quantum numbers but different masses is expected for each particle present in the 4-dimensional theory, whose masses are given by
\begin{equation}
m_n=\sqrt{(n/R)^2+m_0^2}~,
\end{equation}
where $m_0$ is the mass of the zero mode and $n$ an integer. A conserved quantity called \textit{KK-parity} is the consequence of momentum conservation in the extra dimension and as $R-$parity is supersymmetry, KK-parity would prevent the lightest Kaluza-Klein mode from decaying. In this framework, the first massive KK mode of the hypercharge field $B^1$ is a good Dark Matter candidate and could be produced via the freeze-out mechanism for a mass of the order of the TeV scale~\cite{Servant:2002aq}.\\
Another possibility is to consider the first massive KK mode associated to the graviton as a viable Dark Matter candidate. In this case the DM would only interact via gravity with the Standard Model and its density can be generated via non-thermal processes.

%% file: parts/alternative.tex
\section{Alternative solutions to collisionless cold dark matter}
\label{sec:alternative}
In the previous section we discussed particle physics interpretation of the dark matter problem, in the following section we expose the current status of the most popular alternative solutions.
\subsection{Modified gravity}
\subsubsection{Modified Newtonian Dynamics (MOND)}
One of the first idea that one could think of when looking at the evidences in favour of the exitence of dark matter is that our current theory of gravity is not correct or at least must be different on some scales. This idea was exploited by Milgrom in the 1980's~\cite{1983ApJ...270..365M} by considering a modified version of Newton's theory of gravity in the low-acceleration regime :
\begin{equation}
m\mu\left( \dfrac{a}{a_0} \right) \vec{a}=\vec{F}
\end{equation}
where $a_0$ is a constant, the function $\mu(x) \rightarrow 1$ when $a\gg a_0$ and $\mu(x)\rightarrow x$ when $a \ll a_0$. $a_0$ was determined later on by studying rotation of galaxies~\cite{1983ApJ...270..371M} as one expect the assymptotic circular velocity of stars to be $V_\infty^4 = a_0 G_N M$ in a host galaxy of mass $M$ which corresponds to the relation expected according to the empirical Baryonic Tully-Fisher Relation (BTFR) as discussed in Sec.~\ref{sec:BTFR}. Milgrom estimated $a_0 \simeq 10^{-8}~\text{cm}^2~\text{s}^{-1}$ by comparing its theory to the mass-luminosity ratio in a sample of galaxies.
Recently, velocity dispersion studies of the NGC1052–DF2 galaxy have been perfomed showing a value compatible with a ratio of total matter over luminous matter to be of the order of one, implying that no dark matter halo is present in this galaxy~\cite{2018Natur.555..629V}. This isolated observation challenges modified gravity scenarios which are expected to yield the same effect in galaxies with similar characteristics. However a galaxy only made of baryons is possible in the cold dark matter context depending on its astrophysical history. Recently, it was shown that even if this observation is in tension with MOND, the discrepancy remains reasonable and does not exclude the MOND framework~\cite{Famaey:2018yif}.
One convincing argument usually employed against modified gravity extensions is the Bullet Cluster as discussed in Sec.~\ref{sec:lensing} to assert the existence of dark matter. Indeed, one striking argument regarding this merging cluster event is that even though most of the mass is in the form of hot intra-cluster gas whose shape does not seem spherially symmetric, the lensing map seems spherically distributed and seems to match the location of the galaxies in the cluster. One might argue that a lensing map analysis should be performed in a specific theory in order to conclude. A gravitational wave analysis in similar events have been performed in the MOND context and a dark halo is still necessary in order to explain observations~\cite{Takahashi:2007nj}.
\subsubsection{Tensor-Vector-Scalar theories}
The Tensor-Vector-Scalar (TeVeS) was proposed by Bekenstein as a relativistic generalization of the MOND theory~\cite{Bekenstein:2004ne}. One of the main aspect of this theory is that the geometrical part of the Lagrangian is composed of the Einstein-Hibert Lagrangian with a metric $g_{\mu \nu}$ as General Relativity but the matter lagrangian is constructed with a physical metric $\tilde{g}_{\mu \nu}$ such that
\begin{equation}
\tilde{g}_{\mu \nu}=e^{2 \phi} g_{\mu \nu}-2 A_\mu A_\nu \sinh (2\phi)
\end{equation}
where $\phi$ and $A^\mu$ are respectively scalar and vectors fields. Varying the action with respect to $g_{\mu \nu}$ gives some relation between the Einstein tensor and the energy-momentum tensor similar to Einstein equations except some terms that depends on the scalar and vectors degrees of freedom introduced. Therefore in the low gravitational potential regime, one expects Newtonian dynamics to be modified according to those extra terms. However, it was shown that cosmological probes such as the CMB spectrum could not be explained in this framework without adding new degrees of freedom, rendering this theory much less appealing~\cite{Skordis:2005xk}. The recent observations a gravitational wave signal GW170817 from the coalescence of binary neutron stars by the LIGO collaboration~\cite{TheLIGOScientific:2017qsa} were used to constrain modified-gravity based interpretation of dark matter~\cite{Boran:2017rdn} by computing the expected Shapiro time delay\footnote{time delay experienced by following a geodesic with or without a massive object nearby.} caused by the dark matter density along the line of sight. This time was compared to the delay between the gravitational wave signal and $\gamma$-ray signal in order to place constraints on the violation of the weak equivalence principle which is expected to be the case in some modified gravity such as TeVeS. It turns out General Relativity still provides an accurate description of gravity, therefore this constraint has been shown to exclude TeVeS as a possible interpretation of the dark matter.
\subsubsection{Emergent gravity}
Emergent gravity has been proposed by Eric Verlinde as a theoretical idea that spacetime and gravity emerge from an underlying microscopic theory~\cite{Verlinde:2016toy}. Based on entropy considerations, he argued that a new dark gravity force must be present and could explain the issues currently
attributed to the dark matter. In this setup one would expect the total centripetal acceleration $g_{\text{t}}$ in a spherically symmetric system to behave as~\cite{Lelli:2017sul}
\begin{equation}
g_{\text{tot}}=g_{\text{b}}\left( 1+\sqrt{\dfrac{a_0}{g_{\text{b}}}} \sqrt{1+\dfrac{G_N}{g_{\text{\tiny{b}}}r}\dfrac{\partial M_{\text{\tiny b}}}{\partial r}} \right)
\end{equation}
where $M_{\text{b}}$ is the baryonic mass, $g_{\text{b}}$ the acceleration due to the baryonic contribution, $a_0$ some acceleration scale $a_0 \sim c H_0$ and $r$ the radial distance from the center of the system. However it was shown in~\cite{Lelli:2017sul} that predicitions from emergent gravity are in tension with expectations from the empirical Mass-Acceleration Discrepancy Relation as discussed in Sec.~\ref{sec:MDAR}. Therefore this theory is unlikely to account for the dark matter presence in our universe.

\subsection{Baryonic DM candidates}
One simple possibility to explain the dark matter in our universe is to suppose that it is actually made of interstellar medium gas that is not emitting a substantial amount of light but whose abundance is underestimated. In this case, one would expect cosmic rays to interact with the gas, emitting pions which will eventually decay to $\gamma$-rays that should be observed by current experiments. However, as shown in~\cite{1996A&A...313....1S}, the fraction of interstellar gas is severely constrained from $\gamma$-ray flux observations and cannot account for the dark matter. 

\subsubsection{Massive astrophysical compact objects}
Massive astrophysical compact objects (MACHOs) have been suggested as a potential category of dark matter candidates composed of heavy astrophysical dim objects such as brown dwarfs, remnants of early stars, neutron stars or black holes, as discussed further on. In order to observe this kind of objects, \textit{microlensing} techniques have been employed. Microlensing effects of MACHOs consist of a magnification of the observed flux of some luminous object caused by a MACHO of mass $M$ crossing the line of sight for a typical time  
\begin{equation}
t\sim100~\text{days}\left( \dfrac{M}{M_\odot}\right)^{1/2}
\end{equation}
allowing to potentially observe microlensing effects~\cite{1991ApJ...366..412G,1986ApJ...304....1P} for objects of masses $M \in [10^{-7},10^2 ] M_\odot$. This strategy was employed successfully by the EROS, MACHO, OGLE, MOA and SuperMACHO colaborations. EROS-2 excluded~\cite{Tisserand:2006zx}  MACHOs in the following mass range
\begin{equation}
0.6 \times 10^{-7} M_\odot <M < 15M_\odot
\end{equation}
However, a more recent analysis~\cite{Calcino:2018mwh} in light of updated Milky Way halo data from rotation curves, has shown looser constraints which actually depends on the mass distribution of these objects such that masses $M\sim M_\odot$ are still allowed as possible explanation of dark matter but monochromatic interpretations seem in tension with these limits.
\subsubsection{Primordial Black Holes}
Primordial Black Holes (PBHs) are a particularly interesting subcategory of MACHOs. The possibility of forming PBH from gravitational collapse initiated by overdensities in the early universe was suggested by Carr and Hawking in 1974~\cite{1974MNRAS.168..399C}. Hawking published his famous 1974 paper~\cite{Hawking:1974sw} on particle production by black holes, the so-called \textit{Hawking radiation}, implying that a black hole would totally evaporate on a time scale
\begin{equation}
t \sim 10^{64}~\text{years}\left( \dfrac{M}{M_\odot}\right)^3
\end{equation}
Therefore black holes with masses $M \lesssim 10^{-18} M_\odot$ which formed in the early universe must have been evaporated entirely by the present time. For large PBHs masses, distorsion effects are expected in the CMB~\cite{Blum:2016cjs}, thefore masses $M\gtrsim 5 M_\odot$ are disfavored. For black hole masses $M\gtrsim 10^{-18}M_\odot$, PBH capture by white dwarfs or neutron star are expected to destroy their host structure in a short amount of time, therefore a large abundance of PBH would not allow for the observation of these star remnants which constrains PBH as dark matter interpretation~\cite{Capela:2012jz}. As a result, the only remaining window lies for masses $M\sim M_\odot$ in order to explain the total dark matter density with PBH, which interestingly is the mass range corresponding to those of the black hole merger observed by the LIGO collaboration recently~\cite{TheLIGOScientific:2017qsa}. A recent study~\cite{Raidal:2017mfl} have shown that the non-observation of gravitational waves with LIGO implies constraints on the PBH abundance $\Omega_{\text{PBH}}/\Omega_{\text{DM}}\lesssim 10\%$ for $M\in[0.5M_\odot,50M_\odot]$.
As a summary, PBHs are perhaps the best-motivated MACHO candidate in spite of strong constraints over a large mass range. PBHs are still possible dark matter candidates but the viable mass range remains quite narrow and should be accessible by gravitational waves experiments in the future.

\section{Controversies within $\Lambda$CDM}

$\Lambda$CDM has shown to be an appealing and accurate model, facing numerous experimental constraints. However, some experimental results are still not completely understood in the context of $\Lambda$CDM or based on simulations. In this section we review some of the dark matter related controversies within the $\Lambda$CDM landscape.

\subsection{Small scales controversies}
\label{sec:smallscalescontroversies}
The so-called \textit{small-scales controversies} denote a series of observed discrepancies between astrophysical measurements and $\Lambda$CDM-based N-body simulations that appear on galactic scales\footnote{For futher reading, detailed reviews can be found in~\cite{Weinberg:2013aya,DelPopolo:2016emo,Tulin:2017ara}}.
\begin{itemize}
\item \underline{The cusp-core problem}~\cite{Moore:1999gc} is related to the fact that in some galaxies, the inner part of rotation curves can be better fitted with a cored profile such as isothermal than by the NFW profile as suggested by simulations, which is steeper at small radius and therefore implies a larger predicted amount of dark matter in the inner region of galaxies.
\item \underline{The too-big-to-fail problem}~\cite{2011MNRAS.415L..40B}: subhalos predicted by simulations in Milky-Way like galaxies are too dense and massive to host the brightest Milky Way satellites and they shouldn't have failed forming more stars. It can be related to the cusp-core problem, $\Lambda$CDM-based simulations seem to predict a larger amount of dark matter in halo and subhalos.
\item \underline{The missing satellite problem}~\cite{2012MNRAS.422.1203B}: Simulations predict a larger number of satellite galaxies that is actually observed in the Wilky Way. However he discrepancy have been reduced over the past year by discovering more satellites to $\sim 50$ observed and $\sim 100$ expected satellites from simulations. It was shown in a recent study~\cite{Kim:2017iwr} that this problem might be solved if the Mily Way were populated with lower mass satellites $M\sim 10^8 M_\odot$ which are not efficient in forming a luminous component and therefore more difficult to observe.
\end{itemize}

\subsection{Solutions to the small-scale controversies}

In this subsection we discuss potential solutions to the small-scale controversies within the $\Lambda$CDM cosmology.

\subsubsection{Warm dark matter}
\label{sec:waarmdarkmatter}
One possibility in order to solve at least the Missing Satellite problem is to consider Warm Dark Matter. If dark matter particles are relatistic when the photon temperature is $T\sim1~\text{keV}$, the DM is considered as \textit{hot} and in this case the formation of the large scale structures of the universe would follow a top-down pattern, i.e. the large clusters would form before galaxies. Otherwise if the DM is non-relativistic at $T\sim1~\text{keV}$, it is considered as cold and formation of structures would take place in the opposite order (bottom-up). Cold dark matter has been shown to be favourized by N-body simulations to reproduce our observable universe. Warm dark matter candidates have typically masses of the order of $\sim1~\text{keV}$ and decouple from the SM thermal bath while being still relativistic whereas they are non-relativistic at the time of matter-radiation equality. The main effect of WDM is that density perturbations with large modes $k\gtrsim k_{\text{FS}}$ are damped relative to the CDM case due to relativistic free streaming effects where $k_{\text{FS}}$ can be estimated as
\begin{equation}
k_{\text{FS}}\simeq 50~\text{Mpc}^{-1}\left( \dfrac{m_{\text{DM}}}{2~\text{keV}} \right) \left( \dfrac{T_{\text{DM}}/T_\nu}{0.2} \right)^{-1}
\end{equation}
which is expressed as a function of the neutrino and DM temperatures. Small WDM masses can have a sizable impact on the Lyman-$\alpha$ forest~\cite{Viel:2013apy} measurements and therefore masses below $m_{\text{DM}}\lesssim 2~\text{keV}$ are excluded in the case of thermally produced DM. However it was shown~\cite{Schneider:2013wwa} that WDM cannot concistently solve the small-scale crisis and evade bounds from the Lyman-$\alpha$ forest measurements.

\subsubsection{Self interacting dark matter}
\label{sec:selfinteractingDM}
Another possibility to explain cored dark matter profiles is to consider Self Interacting Dark Matter (SIDM). In the inner region of halos, where the DM is the more present, a sizable self interaction cross section might wash out overdensities, resulting in a cored profile. This effects has been shown by considering a self interaction cross sections of the order of $\sigma_{\text{self}}/m_{\text{DM}}\lesssim 1~\text{cm}^2~\text{g}^{-1}$ in simulations~\cite{Rocha:2012jg}. However based on the observation of cluster merging event such as the Bullet Cluster, if the DM particles possess a sizable self interaction cross section, one could expect an offset between the center of the star and dark matter distributions as the the stars behave as a non-interacting gas whereas the dark matter particles would feel a drag force pushing them back. This argument has been used in~\cite{Randall:2007ph,Kahlhoefer:2015vua} and to derive bounds on $\sigma_{\text{self}}/m_{\text{DM}}\lesssim 1~\text{cm}^2~\text{g}^{-1}$. More recent analyses~\cite{Robertson:2016xjh,Kim:2016ujt} based on a cluster merging simulations showed that the bound on $\sigma_{\text{self}}$ might not be as strong. However the authors of~\cite{Kim:2016ujt} discussed the fact that after dark halo coallescence in galaxy collisions, the remnant of the collision should oscillate around the center with a large orbit, that could eventually constrain $\sigma_{\text{self}}/m_{\text{DM}}\lesssim 0.1~\text{cm}^2~\text{g}^{-1}$.

\subsubsection{Baryonic effects}

One important effect included in recent simulations is the baryonic feedback on the dark matter particles. As the inner part of halos are supposed to be mostly populated by baryons, their effect on the dark matter distribution can be sizable. In particular Active Galactic Nuclei (AGN) effects such as hot gas streaming from supernovae explosion or adiabatic contractions can substantially heat the dark matter up and flatten the inner part of the density core~\cite{2012MNRAS.421.3464P,2017MNRAS.472.2153P}.

\subsection{Empirical relations}
The following empirical relations have been shown to face some inconsistencies when compared to N-body simulations. However, they are still being discussed in the litterature and their status as being an affirmed issue within $\Lambda$CDM is not established yet.

\subsubsection{The Baryonic Tully-Fisher Relation}
\label{sec:BTFR}
The Baryonic Tully-Fisher Relation (BTFR) expresses the correlation between the 21cm line width of spiral galaxies to their rotation velocities. In the case where a galaxy of baryonic mass $M_{\text{b}}=M_{\text{gas}}+M_*$ (the sum of the gas and star masses) is related to the velocity $V_\infty$ at large radius\footnote{valid when the rotation curve is indeed flat}, the BTFR relation takes the simple form
\begin{equation}
M_{\text{b}} \propto V_\infty^4~.
\end{equation}
This relation is satisfied for masses over several orders of magnitude and actually naturally emerges in the MOND framework~\cite{2012AJ....143...40M}. In a combination of simulated galaxies~\cite{Sales:2016dmm}, the BTFR was recovered over a large range of masses, however low mass galaxies seem to deviate from this prediction. In addition, this relation is expected to present a large scatter in the $\Lambda$CDM context~\cite{Lelli:2015wst} larger than what is actually observed.

\subsubsection{Mass-Discrepancy Acceleration Relation}
\label{sec:MDAR}
The Mass-Discrepancy Acceleration Relation (MDAR) expresses a correlation between the effective acceleration $g_{\text{eff}}$ of stars in galaxies and the baryonic acceleration $g_{\text{b}}$ which can be related via the empirical relation~\cite{2016PhRvL.117t1101M,2017ApJ...836..152L}
\begin{equation}
g_{\text{eff}}=\dfrac{g_{\text{b}}}{1-e^{-\sqrt{g_{\text{\tiny b}}/g^\dagger}}}
\end{equation}
where $g^\dagger \sim 10^{-10}~\text{m}~\text{s}^{-2}$ corresponds to the minimal acceleration scale for which the relation $g_{\text{eff}}/g_{\text{b}}\simeq 1$ holds. A remarkable fact about this empirical relation is that it is satisfied for a large variety of galaxies while the scatter for high quality data remains small~\cite{Li:2018tdo}. It was argued that this tight scattering could not be reproduced in a $\Lambda$CDM context but naturally emerges from MOND~\cite{2017ApJ...836..152L}. In a recent analysis based on $\Lambda$CDM simulations\cite{Keller:2016gmw}, this relation has been shown to be satisfied and attributed to the role of baryons. However it was argued in~\cite{Li:2018tdo} that the sampling of the simulated galaxies was not large enough to claim recovering the MDAR over the large variety of observed galaxies.
\section{Constraints on dark matter particles}
The particle physics interpretation of dark matter seems the most appealing solution up to this day as physics beyond-the-Standard-Model is expected and no complete alternative solution has been proposed even though the situation is not clear about the fact that $\Lambda$CDM can adress all the issues mentioned in the previous sections. In the following we expose some of the constraints related to particle physics models of dark matter.

\label{section:properties}
\begin{itemize}
\item \textbf{Dark matter interaction with photons:} Assuming that dark matter particles can scatter significantly with photons, one should expect a damping of the peak amplitudes of the CMB spectrum for large $\ell$ as overdensities on small scales would be washed out by rapid DM-photon scatterings. Facing the great precision measurements of the CMB spectrum, the constraint on the DM-photon scattering cross section to this day is given by~\cite{Stadler:2018jin}: $\sigma_{\gamma-\text{DM}} \lesssim 2 \times 10^6~\sigma_{\rm Th}(m_{\rm DM}/\text{GeV})$ where $\sigma_{\rm Th}\sim 10^3~\text{GeV}^{-2}$ is the Thomson cross section. Another constraint on DM-photon interaction can be derived by considering the effect of magnetic fields that could significantly alter DM density profiles of galaxy clusters in contradiction with observations. The authors of~\cite{Kadota:2016tqq} derived a bound on the DM electric charge $\epsilon_{\rm DM} \lesssim 10^{-14}(m_{\rm DM}/\text{GeV})$.
\item \textbf{Theoretical constraint on the dark matter mass:} In order to achieve the correct relic density in the WIMP paradigm, the value required for the velocity averaged annihilation cross section must be $\la \sigma v \ra \sim 10^{-26}~\text{cm}^3~\text{s}^{-1}$. Using the partial wave decomposition of the unpolarized cross section $\sigma=\sum_J \sigma_J$ and based on the unitarity of the $S$ matrix, one can derive a bound on $\sigma v$ as:
\begin{equation}
\sigma v \lesssim \dfrac{4\pi(2J+1)}{m_{\rm}^2 v}~,
\end{equation}
implying for the $J=0$ partial wave, a constraint on $m_\chi \lesssim 340~\text{TeV}$ assuming that the correct relic density is achieved~\cite{Griest:1989wd}. Initially presented as a lower bound on the mass of a stable neutral heavy lepton and by assuming a $Z$-boson mediated interaction, the Lee-Weinberg bound is derived from the condition of not overclosing the universe in the freeze-out process of such heavy particle. Therefore, the relic density of a particle with a mass $m_{\rm DM}\lesssim 3~\text{GeV}$ is expected to overclose the universe if the depletion process is mediated by the $Z$ boson~\cite{Lee:1977ua}. Even though these limits are model dependent, they can be rescaled easily allowing to picture the parameter space compatible with a standard freeze-out scenario in a given model.
\item \textbf{Dark matter lifetime:} The dark matter should be stable, or at least meta-stable. Indeed, assuming a very long lifetime the dark matter could still be present in the early days of the universe and at the present time to account for the large scale structure formation. A naïve constraint on the dark matter lifetime $\tau_{\rm DM}$ can be derived by simply considering that $\tau_{\rm DM}$ has to be larger than the age of the universe $\tau_{\rm universe}\sim 10^{17}~\text{s}$. Stronger constraint such as energy injection in the dark ages from CMB measurements allows to set the bound $\tau_{\rm DM} \lesssim 10^{25}~\text{s}$ which can be stronger or looser depending on the DM mass and decay channels~\cite{Slatyer:2016qyl}.
\item \textbf{Dark matter self-interaction:} The dark matter cannot be (strongly) self-interacting. As discussed in Sec.~\ref{sec:selfinteractingDM}, the observation of the Bullet Cluster allows to set the bound $\sigma_{\text{self}}/m_{\text{DM}}\lesssim 0.1~\text{cm}^2~\text{g}^{-1}$.
\item \textbf{Dark Matter cannot be hot:} As discussed in Sec.~\ref{sec:waarmdarkmatter}, based on their phase space distribution, Dark Matter particles have to be cold or at least warm, as hot Dark Matter would cause overdensities to be washed out in the early universe, constraining the Dark Matter mass $m_{\rm DM}\lesssim 2~\text{keV}$.
\end{itemize}

\subsubsection*{Conclusion}
The Standard Model of particle physics provides a description of three of the four fundamental interactions with an unprecedented accuracy. However this theory is not complete and in this chapter we motivated the fact that new degrees of freedom have to be introduced. Such degrees of freedom could constitue the totality of the Dark Matter component in the universe and the Dark Matter density could be generated by the freeze-out mechanism in the WIMP paradigm. Alternative theories such as MOND and relativistic extensions are appealing however up to this day they cannot explain all the features of the missing mass issue on all scales. The particle hypothesis remains the most complete solution even though the properties of such particles are strongly constrained. In the following chapter, we discuss the current status of Dark Matter searches.

%% file: parts/DMsearches.tex
\vspace{0.3cm}

\noindent
Motivated by the theoretical arguments exposed in the previous chapter, we know that the Standard Model is lacking of some ingredients. Beyond-the-Standard-Model constructions might include dark matter candidates that potentially present interactions with the Standard Model particle content, in some other way than purely gravitationally, which is the key assumption on which the WIMP paradigm relies.
\begin{figure}[h!]%
\begin{center}
\includegraphics[width=0.9\linewidth]{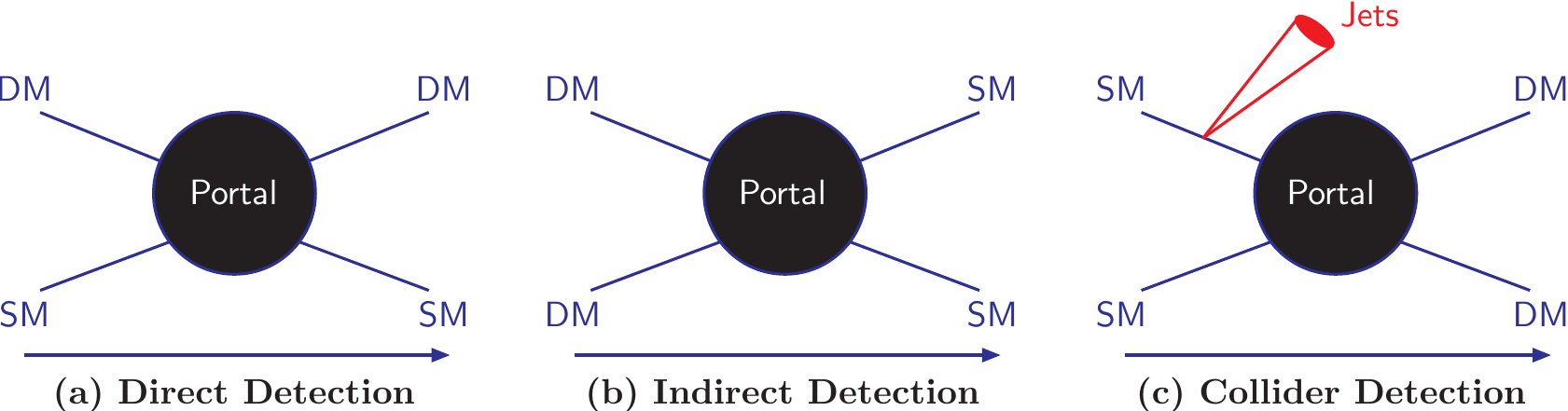}
\caption{Illustration of the three main DM detection strategies presented in this chapter.\label{fig:DD_ID_collider_picture}
}
\end{center}
\end{figure}
Assuming that such a coupling exists, as depicted in Fig.~\ref{fig:DD_ID_collider_picture}, the possible Dark Matter detection strategies can be sorted in three large categories based on the following processes:
\begin{itemize}
\item Dark Matter scattering off a SM particle $\text{DM}+\text{SM}\rightarrow \text{DM} + \text{SM}$: Direct Detection
\item Dark Matter annihilation $\text{DM}+\text{DM}\rightarrow \text{SM} + \text{SM}$: Indirect Detection
\item Dark Matter production $\text{SM}+\text{SM}\rightarrow \text{DM} + \text{SM}$: Collider Searches
\end{itemize}
In this chapter we present an overview of the current status of direct, indirect and collider searches.

\section{Direct detection}
\label{sec:DD}
The basic idea of direct detection is to aim at measuring the recoil energy of a nucleus scattering off a DM particle that is present in the galactic halo. This idea was at first suggested by M. Goodman and E. Witten in the 1980's~\cite{Goodman:1984dc}. The sensitivity of direct detection experiments have been increased by several order of magnitudes every decade ever since. The expected number of events in this kind of experiment is typically quite low therefore most of the detectors are located deeply underground to avoid atmospheric background. In this section we review the formalism associated to the computation of the event rate and review the current status of direct detection experiments.

\subsection{Event rate}
The expected event rate of a specific experiment depends on several inputs from astrophysics such as the DM phase space distribution and the local DM density as well as inputs from nuclear physics making it hard to estimate with high accuracy. The differential event rate expressed in terms of number of events per kilogram and per day is given by:
\begin{equation}
\frac{\diff R}{\diff E_{R}}=\frac{ N_T\rho_\odot}{m_{\text{DM}}} \int_{v_{\text{min}}}^{v_{\text{max}}} vf_E(\vec{v},t)\frac{\diff \sigma}{\diff E_{R}}(v,E_{R})\diff \vec{v}~,
\label{eq:eventrate}
\end{equation}
where $N_T$ is the number of target nuclei, $\rho_\odot \sim 0.3~\text{GeV cm}^{-3}$ is the DM density in the solar system, $\diff \sigma/\diff E_{R}$ is the differential DM-nucleus scattering cross section, $\vec{v}$ the WIMP velocity relative to Earth and $f_E(\vec{v},t)$ is the velocity distribution of the WIMP in the frame of the Earth. $v_{\text{min}}$ is the minimal velocity producing a recoil energy $E_R$ such that
\begin{equation}
v_{\text{min}}=\sqrt{(m_{N}E_{R})/(2\mu_N^{2})}~,
\end{equation}
with $\mu_N=m_{\text{DM}}m_{N}/(m_{\text{DM}}+m_{N})$ being the DM-nucleus reduced mass. $v_{\text{esc}}\simeq~500-600~\text{km s}^{-1}$ is the escape velocity, i.e. the velocity for which the WIMP are no longer gravitationnally bound to the Milky Way. As discussed in Sec.~\ref{sec:simdistrib}, the WIMP velocity distribution in the galactic halo frame $f_G(\vec{v},t)$ can be approximated by a Maxwellian distribution. However one has to take into account the rotation of the Earth around the Sun with a velocity along the Sun's direction $v_\circledcirc\simeq \tilde{v}_\circledcirc \cos[\omega(t-t_0)]$ where $\tilde{v}_\circledcirc\sim 15~\text{km s}^{-1}$ and $\omega=2\pi/\text{year}$, and the velocity of the Solar System in the galactic halo $v_\odot \sim 230~\text{km s}^{-1}$. The total observed velocity can be expressed as the sum of these contributions
\begin{equation}
\vec{v}_{\text{obs}}=\vec{v}+\vec{v}_\odot+\vec{v}_{\circledcirc}~.
\end{equation}
The modulation effect of the velocity $\vec{v}_{\circledcirc}$ due to the rotation of the Earth around the Sun is the observation strategy of some experiments as discussed further on. The typical WIMP velocity in the Earth reference frame is $v\sim10^{-3}$ implying that collisions between WIMPs and atomic nuclei can be treated in the non-relativistic approximation. Therefore the recoil energy $E_{R}$ can be expressed as:
\begin{equation}
E_{R}\simeq\dfrac{q^2}{2 m_N}=\frac{\mu_N^{2}v^{2}(1-\cos\theta^{\star})}{m_{N}}~,
\end{equation}
where $\theta^{\star}$ is the scattering angle in the centre-of-mass frame. For a typical WIMP mass $m_{\text{DM}}\sim 100~\text{GeV}$ the expected recoil energy is 
\begin{equation}
E_R\simeq 50~\text{keV} \left( \dfrac{m_{\text{DM}}}{100~\text{GeV}}\right)^2 \left( \dfrac{100~\text{GeV}}{m_N}\right)~.
\end{equation}
In order to account for the unknown interactions between a dark sector and the SM particles, the differential cross section is usually parametrized as a sum of two terms known as Spin Dependant (SD) and Spin Independent (SI)~\cite{Jungman:1995df,Bednyakov:2006ux}. It is common to express the differential cross section as a function of the cross section in the limit of vanishing momentum tranfert $\sigma_{\text{SI,SD}}$ and as a function of the form factors $F^{2}_{\text{SI,SD}}(E_{\text{R}})$ which carry the momentum dependency:
\begin{equation}
\frac{\diff\sigma}{\diff E_{R}}=\frac{m_N}{2\mu_N^{2}v^{2}}[\sigma_{\text{SI}}F^{2}_{\text{SI}}(E_{\text{R}})+\sigma_{\text{SD}}F^{2}_{\text{SD}}(E_{\text{R}})]~.
\end{equation}
The most important aspects affecting experimental sensitivity to a WIMP signal is a combination of
\begin{itemize}
\item \underline{Energy threshold:} drives the sensitivity to low WIMP masses, and consequently the sharpening of the direct detection limits on the scattering cross section at low masses.
\item \underline{Control over the background and exposure:} determine the overall sensitivity of the experiment pushing the limits to lower scattering cross sections.
\item \underline{Target:} has an impact on the experiment sensitivity to low and heavy WIMP masses, as well as on capability to probe spin-dependent scatterings.
\end{itemize}

\subsection{Spin (in)dependent interactions}

\subsubsection{Effective Operator analysis}
Scattering between WIMPs and nucleons are expected to occur in the non-relativistic regime, where the typical exchanged momentum is well below the QCD Landau pole and the nucleon masses. In this regime it is possible to use a model-independent effective approach in order to understand the kind of coupling that can lead to spin dependent or independent contributions. In order to describe interactions between a fermionic DM candidate $\chi$\footnote{the same kind of analysis can be performed for scalar~\cite{Fitzpatrick:2012ix,DelNobile:2013sia} or vector~\cite{Hisano:2010yh} Dark Matter.} and a nucleon $\text{n}$, one can write in the most general way a set of 10 Lorentz-invariant operators as follow:
\begin{equation}
\mathcal{L}_{\text{eff}}=\sum_{i=1}^{10} \sum_{\text{n}=n,p}c_i^{\text{n}} \mathcal{O}_i^{\text{n}}~,
\end{equation}
where $c_i^{\text{n}}$ are Wilson coefficients and n denotes either a neutron $(n)$ or a proton $(p)$. The operators $\mathcal{O}_i^{\text{n}}$ can be constructed from the basis of $4\times4$ matrices $\Gamma \in \{1,\gamma_5,\gamma_\mu,\gamma_\mu\gamma_5,\sigma_{\mu \nu} \}$\footnote{$\sigma_{\mu \nu}\equiv \dfrac{1}{2} [ \gamma_\mu , \gamma_\nu ]$.} in the symbolic form
\begin{equation}
\mathcal{O}_i^{\text{n}}=\bar{\chi} \Gamma_\chi \chi \bar{\text{n}} \Gamma_{\text{n}} \text{n}~,
\end{equation}
by writing all the possible Lorentz-invariant combinations.
These operators can be matched into a set of 12 non-relativistic operators as discribed in~\cite{Fitzpatrick:2012ix,DelNobile:2013sia}. For instance, the non-relativistic limit of the following operators 
\begin{equation}
\mathcal{O}_1^{\text{n}}=\bar{\chi}\chi \bar{{\text{n}}}{\text{n}} , \qquad \mathcal{O}_8^{\text{n}}=\bar{\chi}\gamma^\mu \gamma_5\chi \bar{{\text{n}}} \gamma_\mu \gamma_5 {\text{n}}~,
\end{equation}
can be derived using the low-energy expansion of the four-component spinor 
\begin{equation}
u^{s}(p)=\begin{pmatrix}
\sqrt{p^{\mu}\sigma_{\mu}}\xi^{s}\\
\sqrt{p^{\mu}\bar{\sigma}_{\mu}}\xi^{s}\\
\end{pmatrix}\simeq\frac{1}{\sqrt{4m}}\begin{pmatrix}
(2m-\vec{p}\cdot  \vec{\sigma})\xi^{s}\\
(2m+\vec{p}\cdot \vec{\sigma})\xi^{s}\\
\end{pmatrix}+\mathcal{O}(p^{2})
\end{equation}
where $\sigma^{\mu}=(1,\vec{\sigma})$, $\bar{\sigma}^{\mu}=(1,-\vec{\sigma})$ and $\xi$ is a 2-component Weyl spinor, implying that in the process $\chi(p)+{\text{n}}(k)\rightarrow \chi(p^\prime)+{\text{n}}(k^\prime)$, at the leading order in $p$
\begin{equation}
\bar{u}(p^\prime)u(p)\simeq  2m \xi^{\prime \dagger}\xi ~, \qquad \bar{u}(p^\prime)\gamma^{\mu}\gamma^{5}u(p) \simeq \left( \begin{array}{cc}
2 \vec{P}\cdot\vec{s} & 2 m \vec{s}
\end{array}  \right)~,
\end{equation}
where $\vec{P}=\vec{p}+\vec{p}^\prime$ and the spin operator is defined as $\vec{s}\equiv\xi^{\prime \dagger}\frac{\vec{\sigma}}{2}\xi$. The non-relativistic-limit of $\mathcal{O}_1^{\text{n}}$ and $\mathcal{O}_5^{\text{n}}$ can be written as
\begin{equation}
\mathcal{O}_1^{\text{n}}\simeq 4 m_\chi m_{\text{n}}~, \qquad \mathcal{O}_8^{\text{n}}\simeq -16 m_\chi m_{\text{n}} \vec{s}_\chi \cdot \vec{s}_{\text{n}}~.
\end{equation}
Therefore the $\mathcal{O}_1^{\text{n}}$ and $\mathcal{O}_8^{\text{n}}$ operators are responsible for SI and SD interactions repectively\footnote{As discussed in~\cite{Fitzpatrick:2012ix} angular-momentum dependent as well as spin and angular-momentum dependent interactions can also be considered depending on the remaining non-relatisitic operator in a specific theory.}.

\subsubsection{Spin Independent}

Assuming only Spin Independent interactions, for a typical WIMP mass $m_{\text{DM}}\sim~\text{GeV}$ the associated De Broglie wavelength would be of the order of the typical nucleus size $\lambda_{\text{DM}}\sim 10^{-15}~\text{m}$ implying that nucleons will contribute coherently to the scattering process of a DM particle with a nucleus. In the Born approximation, the amplitude corresponding to a DM-nucleus scattering can be written as:
\begin{equation}
\mathcal{M}_{\rm fi} \propto \int \diff^{3}r\chi_{\text{f}}^{\ast}(r)V(r)\chi_{\text{i}}(r) \quad \text{with} \quad V(r) \propto \rho(r)~,
\end{equation}
where $\chi_{\rm i,f}=e^{-i\vec{p}_{\rm i,f}\cdot\vec{r}}$ are the DM wave functions in the initial and final states and $V(r)$ is the potential generated by a nucleus with a mass distribution $\rho(r)$. Therefore the matrix element can be expressed as:
\begin{equation}
\mathcal{M}_{\rm fi}\propto [Zf_{p}+(A-Z)f_{n}]F_{\text{SI}}(q)~, \quad \text{with} \quad F_{\text{SI}}(q) \propto \int \diff ^{3}r\rho(\vec{r})e^{-i\vec{q}\cdot\vec{r}}~,
\end{equation} 
where $\vec{q}=\vec{p}_{\rm f}-\vec{p}_{\rm i}$, $Z$ is the atomic number, $A$ is the mass number, $f_{p,n}$ are the coupling to protons and neutrons respectively. The form factor $F_{\text{SI}}(q)$ can be understood at the Fourier transform of the nucleus mass distribution, which is usually described by the Helm form factor~\cite{Duda:2006uk} :
\begin{equation}
|F_{\text{SI}}(q)|^2=\left( \dfrac{3 j_1(qR)}{qR} \right)^2e^{-q^2s^2}~, \quad \text{with} \quad j_1(x)=\dfrac{\sin x}{x^2}-\dfrac{\cos x}{x}~,
\end{equation}
where $R$ is the effective nuclear radius satisfying $R^2=c^2+(7/3)\pi^2a^2-5s^2$ with $s\simeq 0.9~\text{fm}$ being the nuclear skin thickness, $a\simeq 0.52~\text{fm}$ and $c\simeq1.23A^{1/3}-0.6~\text{fm}$. The Helm form factor is a rapidly decreasing function accounting for the lost of nucleus coherence for large exchanged momenta. In the limit where $f_n \simeq f_p$, the total nucleus scattering cross section scales as:
\begin{equation}
\sigma_{\text{SI}}=\sigma_{\text{SI}}^{\text{n}}\dfrac{\mu_N^2}{\mu_{\text{n}}^2}A^2~,
\label{eq:SI}
\end{equation}
where $\sigma_{\text{SI}}^{\text{n}}$ is the nucleon scattering cross section and $\mu_{\text{n}}$ the DM-nucleon reduced mass. The expression~(\ref{eq:SI}) illustrates the fact that target material composed of heavy nuclei are typically the most sensitive to spin independent interactions and therefore are used for this purpose.

\subsubsection{Spin Dependent}
In the case of a Spin Dependent interaction term, the cross section does not depend on the number of nucleons but on the spin structure of the nucleus
\begin{equation}
\sigma_{\text{SD}}=\sigma_{\text{SD}}^{\text{n}}\dfrac{\mu_N^2}{\mu_{\text{n}}^2}\dfrac{4}{3}\dfrac{J+1}{J}\dfrac{[a_p \la S_p\ra+a_n \la S_n\ra]^2}{a^2_{\text{n}}}~,
\end{equation}
where $J$ is the total spin of the nucleus, $a_{\text{n}}$ is the DM-nucleon coupling and $\la S_{\text{n}}\ra$ is the total contribution of the nucleon n to the total spin of the nucleus. The SD form factors $F_{\text{SD}}(q)$ as well as the contributions $\la S_{\text{n}}\ra$ can be estimated from detailed nuclear calculations or by using simple modelization. 
In order to reach a large sensitivity, SD target detectors are usually composed of unpaired proton and neutrons which cannot be arbitrary large therefore the senstivity achieved by SD detectors is typically lower than for SI-sensitive experiments.

\subsection{Current status of direct detection}
In this section we give an overview of the present status of direct detection. It is important to highlight that in the computation of the expected scattering rate in Eq.~(\ref{eq:eventrate}), some assumptions have been made about the velocity distribution, nuclear form factor, type of DM-nucleon scattering, and local DM density that suffers from large uncertainties as discussed in Sec.~\ref{sec:simdistrib}. In particular, the common assumptions are that the DM phase space is described by a Maxwellian velocity distribution, that the nucleus can be treated as a hard sphere as indicated by the Helm form factor, and that the DM-nucleon scattering is elastic. Based on these assumptions, facing the fact that none of the direct detection experiments has observed a significant event that was reliably attributed to a DM scattering, it is common for experimental collaborations to derive constraints in the $\{\sigma_{\text{SI,SD}}^{\text{n}},m_{\text{DM}} \}$ plane as depicted in Fig.~\ref{fig:sigmaSI}. \footnote{For further reading, a review can be found in~\cite{Undagoitia:2015gya}.}

\subsubsection{Recent experimental results}
\begin{figure}[h!]
\begin{center}
\includegraphics[width=0.9\linewidth]{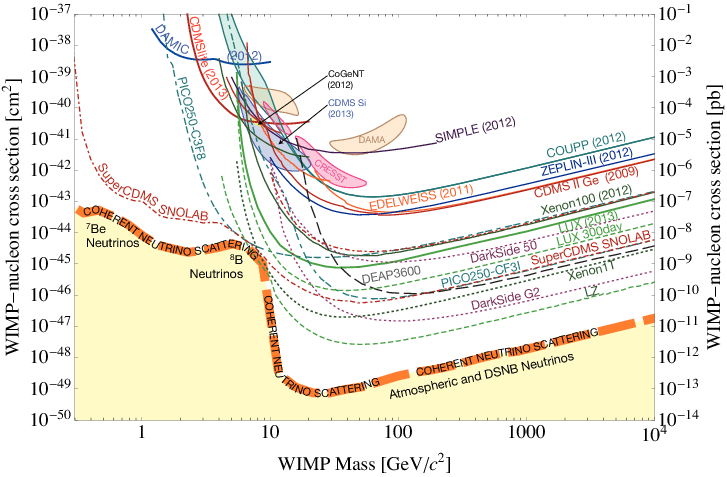}
\caption{Current constraints on DM-nucleon scattering cross sections from various experiments as well as expected sensitivity for the experiments mentioned in the text. The orange dashed line shows the neutrino floor. Figure taken from~\cite{Cooley:2014aya}} 
\label{fig:sigmaSI}
\end{center}
\end{figure}

In order to observe the nucleus-recoil due to a DM particle scattering off it, several kind of experiments can be designed based on three physical effects : production of heat, ionization and scintillation. They can be sorted into three large categories :
\begin{itemize}
\item \textbf{Phonon/heat detector:} experiments such as COUPP~\cite{Behnke:2012ys}, PICASSO~\cite{Archambault:2012pm}, PICO~\cite{Amole:2017dex} and SIMPLE~\cite{Felizardo:2014awa} are composed of supearheated detector in a metastable state, aiming at observing a DM particle scattering off a nuclei while depositing some energy, resulting in a phase transition and formation of bubbles. These experiments are typically composed of fluor-based molecules which contains a high number of unpaired protons and neutrons, therefore making them mostly sensitive to SD interactions. The strongest constraints to this day on the SD DM-proton cross section is set by PICO which excludes $\sigma_{\text{SD}}^p \sim 4\times10^{-41}~\text{cm}^2$ for $m_{\text{DM}}\sim40~\text{GeV}$~\cite{Amole:2017dex}.
\item \textbf{Liquid noble gases:} These experiments are composed of two-phases time projection chamber using both ionization and scintillation signals in order to discriminate a DM signal from background events. They are typically made of heavy nuclei such as xenon for XENON100~\cite{Aprile:2016swn}, LUX~\cite{Akerib:2016vxi}, ZEPLIN~\cite{Akimov:2011tj} and PandaX~\cite{Cui:2017nnn} in order to be highly sensitive to SI interactions. This material have very efficient self-shielding capacities and can easily be scaled lo larger masses, making them the most constraining direct detection experiments on SI cross section. Recently LUX~\cite{Akerib:2016vxi} and PandaX~\cite{Cui:2017nnn} have reached limits on SI cross section of $\sigma_{\text{SI}}^{\text{n}} \lesssim 10^{-46}~\text{cm}^2 $ for $m_{\text{DM}}\sim 50~\text{GeV}$ but currently the strongest bound is set by the XENON1T collaboration~\cite{Aprile:2017iyp} constraining $\sigma_{\text{SI}}^{\text{n}} \lesssim 8\times 10^{-47}~\text{cm}^2 $. As depicted in Fig.~\ref{fig:sigmaSI}, the sensitivity of these experiments is limited at low DM masses $m_{\text{DM}}\lesssim 10~\text{GeV}$ because of their energy thresholds and at high DM masses the sensitivity decreases because for the considered DM density in the solar system, a larger DM mass implies a smaller number density, therefore less numerous events.

\item \textbf{Solid state cryogenic detectors:} The experiments of this category can reach typically sub-Kelvin temperatures and are either based on bolometer-type detectors such as CRESST~\cite{Petricca:2017zdp}, EDELWEISS~\cite{Armengaud:2017rzu} or SuperCDMS~\cite{Agnese:2017njq} with a low energy threshold $\lesssim~\text{keV}$ or semiconductor with high-purity germanium detectors such as CoGeNT~\cite{Aalseth:2012if}. Due to their low energy thresholds, these experiments are still sensitive to low Dark Matter masses $m_{\text{DM}}\lesssim 10~\text{GeV}$. Some of these experiments have reported excesses over the past few year~\cite{Angloher:2011uu,Aalseth:2012if,Agnese:2013rvf} which have not been confirmed by upgraded versions of the same experiments. However the DAMA/LIBRA have observed a long-standing $\sim 9\sigma$ annual modulation excess ~\cite{Bernabei:2010mq} compatible with a DM interpretation but this result is not compatible with recent xenon-based experiments which exclude the value of the cross section required for this interpretation by several orders of magnitudes as depicted in Fig.~\ref{fig:sigmaSI}. This excess still remains unexplained nowdays.

\end{itemize}

\subsubsection{Detection prospects}
\begin{figure}[h!]
\begin{center}
	\includegraphics[width=12cm]{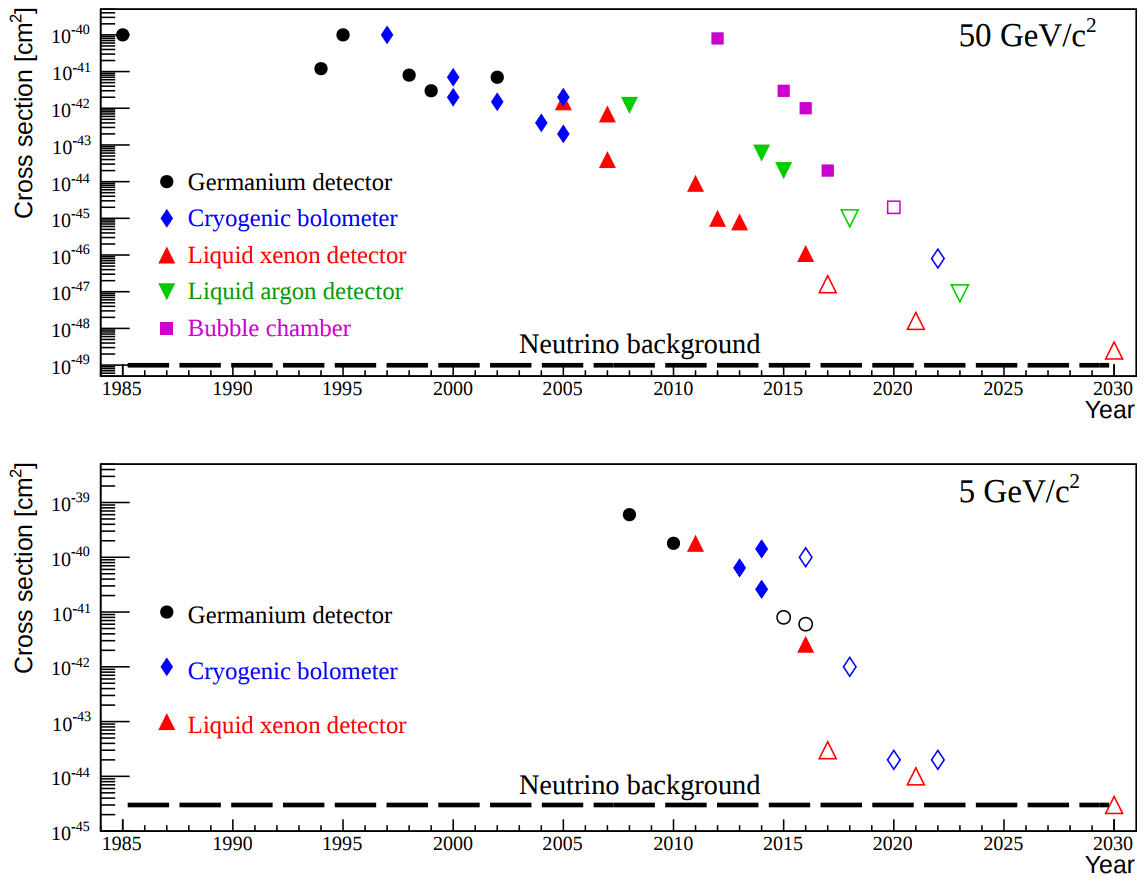}
\caption{Sensitivity evolution of SI WIMP-nucleon cross section for the cases $m_{\rm DM}=50~\text{GeV}$ and $m_{\rm DM}=5~\text{GeV}$. The neutrino floor is represented with solid dashed line. Figure taken from~\cite{Undagoitia:2015gy}.
\label{fig:DDfuture}}
\end{center}
\end{figure}

Several possibilities are considered in order to aim at observing a scattering event beyond the current sensitivity. The two main options at the present time are focusing of the low DM mass regime and improving the sensitivity for standard DM masses $m_{\text{DM}}\sim50~\text{GeV}$. \\
Sensitivity improvement is expected by SuperCDMS which aims to operate few hundreds of kg of target material in order to cover the $m_{\text{DM}}\sim~\text{GeV}$ parameter space~\cite{Cushman:2013zza}. Some efforts are also realized in order to explore the sub-GeV DM mass regime with specific detector materials which should improve the current sensitivity at low masses~\cite{Hochberg:2017wce}.\\
As represented in Fig.~\ref{fig:DDfuture}, several upcoming xenon-based experiments are expected to improve the current limits by several orders of magnitude in the future years by increasing the total mass of detector above the ton-scale such as LZ~\cite{Akerib:2015cja}, DEAP~\cite{Amaudruz:2014nsa}, DarkSide~\cite{Aalseth:2015mba}, XMASS2~\cite{Hiraide:2015cba}, XENONnT~\cite{Aprile:2014zvw} and DARWIN~\cite{Aalbers:2016jon}. The expected sensitivity of these experiments should  constrain $\sigma_{\text{SI}}^{\text{n}}\lesssim 10^{-48}~\text{cm}^2$ as depicted in Fig.~\ref{fig:sigmaSI}. However the sensitivity of these kind of experiments cannot allow to reach arbitrary small values of the scattering cross section as at some point neutrinos emitted by supernovae or ${}^{7}\text{Be}$ and ${}^{8}\text{B}$ as well as atmospheric and solar neutrinos will be an irreducible background, the so-called \textit{neutrino floor}~\cite{Billard:2013qya} as illustrated in Fig.~\ref{fig:sigmaSI}. In order to explore the region in the $\{\sigma_{\text{SI}},m_{\text{DM}}\}$ plane covered by the neutrino floor, one possibility to discriminate a DM from a neutrino scattering event is to measure the direction of the incident particle. The basic idea is that the DM flux depends on the motion of the earth around the Sun, which is expected to be annually modulated therefore one could predict the DM flux at a given epoch. For this purpose, collaborations such as DRIFT~\cite{Battat:2014oqa} or MIMAC~\cite{Santos:2013hpa} are developping the required experimental techniques for directional detection, a promising next-generation of Dark Matter direct detection experiments~\cite{Mayet:2016zxu}.

\section{Indirect Detection}
In the standard WIMP freeze-out mechanism, a sizable Dark Matter annihilation cross section is required in order to achieve the relic density observed at the present time. Therefore, a complementary approach to direct detection in order to detect WIMPs is to look for the products of Dark Matter annihilations or decays that could occur inside large astrophysical structures. These searches are not limited to WIMPs and extend to any framework where DM interactions with the SM particle content are assumed. Dark Matter ($\chi$) annihilations or decays into SM particles could eventually produce quarks ($q$), leptons ($\ell$), gauge bosons ($W^\pm,Z,\gamma$) or Higgs bosons $h$ which will eventually decay to electrons ($e^-$), protons ($p$), nuclei ($N$), gamma-rays or neutrinos ($\nu$) as
\begin{equation}
\bar{\chi} \chi ~ \longrightarrow ~ \bar{q}q, \ell^- \ell^+, W^+ W^-, ZZ, \gamma \gamma, hh ~ \longrightarrow ~ e^+e^-, \bar{p}p, \bar{N}N, \gamma \gamma, \bar{\nu}\nu~.
\end{equation}
In order to claim the observation of some Dark Matter related event, an excess has to be detected over the astrophysical background which is not well understood because of large uncertainties regarding the nature of the sources and propagation models. In the following we review briefly the most common signatures and associated detection stategies for indirect detection. For further reading some review can be found in~\cite{Cirelli:2015gux,Gaskins:2016cha}

\subsection{Gamma-rays}
Gamma-rays are potentially one of the most prominent DM indirect observation channel as $\gamma$-rays propagate in a straight line without being affected between the source and the observer. Observation of DM annihilations or decays to $\gamma$-rays has been attempted by using the Fermi-LAT~\cite{Atwood:2009ez} satellite and also several ground-based Imaging Atmospheric Cerenkov Telescopes (IACTs) such as VERITAS~\cite{Holder:2011fg}, HESS~\cite{Abdallah:2016ygi} and MAGIC~\cite{Aleksic:2013xea}. The prompt Dark Matter annihilation to $\gamma$-rays could occur directly or in two step: by annihilating to some heavy SM particles first, which will enventually decay to $\pi_0$ decaying as well to a photon pair in a second step.
The prompt $\gamma$-ray spectrum should exhibit a continuous shape and a hard cut for energies $E_\gamma \sim m_{\text{DM}}$ as the annihilation to photons with $E_\gamma > m_{\text{DM}}$ becomes kinematically forbidden. This continuous shape can be complicated to disentangle from the background. However some specific spectral features could allow the $\gamma$-ray flux to typically exceed the background. For instance the direct production $\bar{\chi} \chi \rightarrow 2 \gamma$ generates a $\gamma$-ray spectral line around $E_\gamma \sim m_{\text{DM}}$ and the two-steps production $\bar{\chi} \chi \rightarrow 2 \Phi \rightarrow 4 \gamma$, where $\Phi$ is some particle decaying to a photon pair, could lead to a typical spectral "box" shape centered around $\Delta E_\gamma \simeq \sqrt{m_\chi^2-m_\Phi^2}$.

\subsubsection{The gamma-ray flux}

The differential gamma-ray intensity (photons per area per time per solid angle per energy) from annihilation of two DM particles $\chi$ present in the galactic center (GC) for instance is given by
\begin{equation}
\frac{\diff \Phi_{\rm ann}}{\diff\Omega\,\diff E} =\frac{\langle \sigma v \rangle}{8\pi m_{\chi}^{2}} \frac{\diff N_{\gamma}}{\diff E} \underbrace{\int_{\rm los}{\rho_{\chi}^{2}(r) \diff \ell}}_{J_{\rm ann}}~,
\end{equation}
where $\langle \sigma v \rangle$ is the averaged annihilation cross section times relative velocity, $m_{\chi}$ is the mass of the DM particle, and $\diff N_{\gamma}/\diff E$ is the energy spectrum of photons emitted per annihilation.  The function $\rho_{\chi}(r)$ is the DM density as a function of the distance $r$ from the GC. The coordinate $\ell$ runs along the line-of-sight (los), and $r(\ell,\psi)=\sqrt{r_{\odot}^{2}+\ell^{2}-2r_{\odot}\ell \cos(\psi)}$ where $r_{\odot}$ is the distance between the Sun and the GC, and $\psi$ the angle between the line-of-sight and the direction of the GC.  The line-of-sight integral of the DM density squared is often referred to as the "astrophysical factor" or "$J$-factor" and is denoted $J_{\rm ann}$, defined here as differential in solid angle. In the case where the DM particles are unstable and decaying, the differential gamma-ray intensity is given by
\begin{equation}
\frac{\diff \Phi_{\rm dec}}{\diff \Omega\,\diff E} =\frac{1}{4\pi \tau m_{\chi}} \frac{\diff N_{\gamma}}{\diff E} \underbrace{\int_{\rm los}{\rho_{\chi}(r) \diff \ell}}_{J_{\rm dec}}~,
\end{equation}
where $\tau$ is the lifetime of the DM particles, and here $\diff N_{\gamma}/\diff E$ is the energy spectrum of photons emitted per decay.  The "astrophysical factor" for decay $J_{\rm dec}$ is given by the line-of-sight integral over the DM density. The energy spectrum of the photons produced by DM annihilation or decay can be written as a sum over all possible final states
\begin{equation}
\frac{\diff N_{\gamma}}{\diff E}=\sum_{f} B_{f}\frac{\diff N_{f}}{\diff E}~,
\end{equation}
where $B_{f}$ is the branching fraction of final state $f$, and $\diff N_{f}/\diff E$ is the photon spectrum from annihilation or decay to the final state $f$. \\
 However no signal attributed to DM annihilation has been observed to this day, therefore several collaborations were able to constrain the value of $\la \sigma v\ra$ based on the absence of any observed signal.

\subsubsection{Constraining Dark Matter properties with IACTs}

\begin{figure}[h!]
\begin{center}
	\includegraphics[width=6cm]{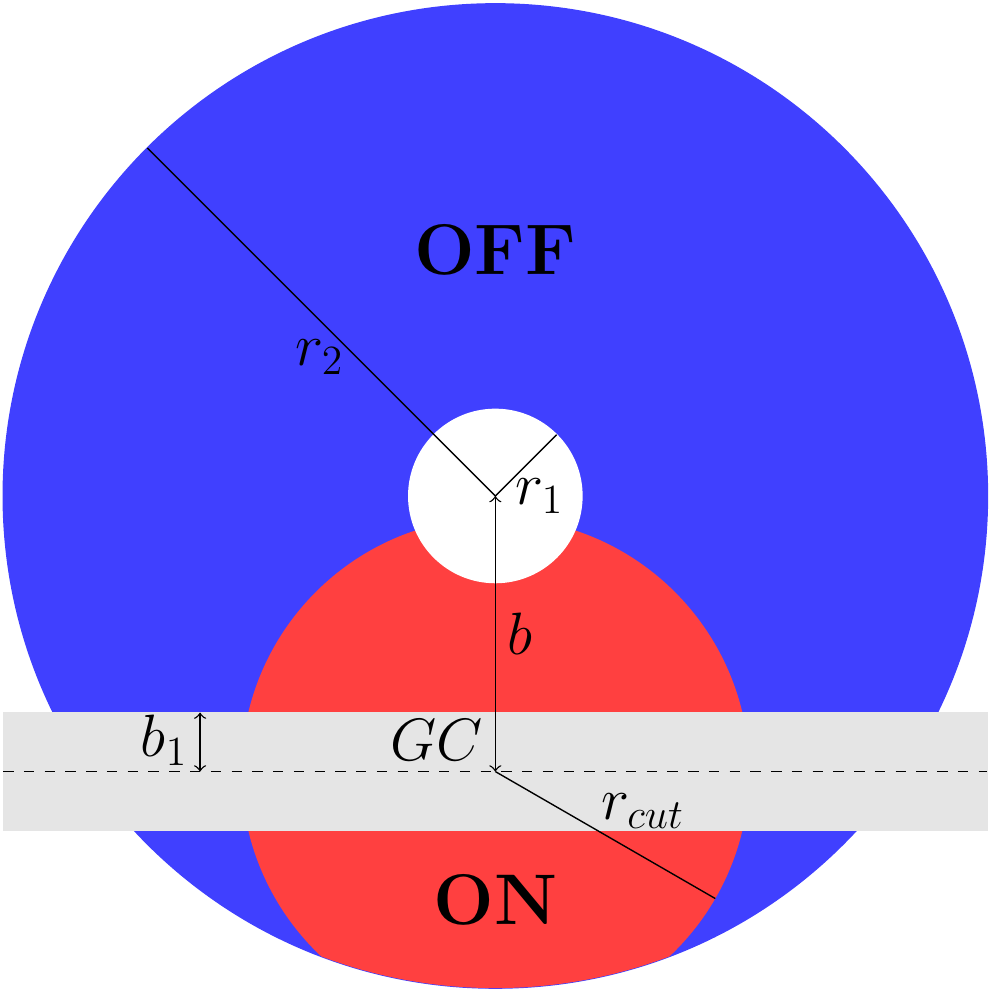}
\caption{Illustration of the choice of ON and OFF regions.  The ON and OFF regions are chosen within a ring centered on the FOV with inner radius $r_1$ and outer radius $r_2$.  For the GC observation considered here, the center of the FOV is offset by $b$ degrees in latitude from the GC\@.  The ON region is shown in red, defined by the intersection of a circle of radius $r_{\rm cut}$ centered on the GC and the ring with inner radius $r_1$ and outer radius $r_2$.  The OFF region, shown in blue, is defined by the remainder of the ring outside of the ON region.  The Galactic plane is excluded by a latitude cut of $b_1$ degrees (shown by the gray rectangle) from both the ON and OFF regions.  
\label{fig:ring}}
\end{center}
\end{figure}

In order to derive a constrain on $\la \sigma v\ra$ by using IACTs for instance, one can perform a likelihood analysis based on the so-called \textit{Ring Method}~\cite{Doro:2012xx} as described in the following~\footnote{the same analysis can be done in the decaying Dark Matter case by substituting $\la \sigma v \ra \rightarrow \tau$ in the derivation.}. The most important background for IACTs is the total cosmic ray electron (CRE)~\footnote{refers actually to electron plus positron.} spectrum as CRE-induced atmospheric showers cannot be distinguished from gamma-induced showers. The CRE spectrum has been measured by the Fermi LAT from $\sim20$~GeV up to $\sim1$~TeV~\cite{Abdo:2009zk}, and is well approximated by a power law $\propto E^{-3}$. In general IACTs can reject hadronic showers with high efficiency therefore can be neglicted as a possible source of background. To search for a DM signal one can define a signal region (denoted ON) and background region (denoted OFF) within the field of view (FOV) using the Ring Method. The ON and OFF regions are illustrated in Fig.~\ref{fig:ring}, and are chosen to lie within a ring centered on the FOV. The geometry is chosen and optimized to reduce systematics associated with variation of the effective area across the FOV\@. We define the geometrical parameter $\alpha=\Delta \Omega_{\rm ON}/\Delta \Omega_{\rm OFF}$, which is the ratio of the solid angles of the ON and OFF regions. The number of photons observed from a specified region of the sky from DM annihilation is
\begin{equation}
\label{eq:nann}
N_{\rm ann}=t_{\rm obs}\frac{\langle \sigma v \rangle}{8\pi m_{\chi}^{2}} N_{\gamma,{\rm obs}} \int_{\Delta \Omega}J_{\rm ann}(\psi)\diff\Omega~,
\end{equation}
where $t_{\rm obs}$ is the observation time, and
\begin{equation}  
N_{\gamma,{\rm obs}}=
\int_{\Delta E}
{\int_{-\infty}^{+\infty}\frac{\diff N_{\gamma}(\bar{E})}{\diff E}A_{\rm eff}(\bar{E})\frac{e^{-\frac{(E-\bar{E})^{2}}{2\sigma^{2}}}}{\sqrt{2\pi\sigma^{2}}}\diff\bar{E}
\diff E}~,
\end{equation}
where $A_{\rm eff}(E)$ is the energy-dependent effective area. 
The quantity $N_{\gamma,{\rm obs}}$ is the energy spectrum per annihilation multiplied by the effective area of the IACT and convolved with its energy resolution, integrated over the energy range considered ($\Delta E$). Here we have modeled the energy resolution of the considered IACT by convolving the source energy spectra with a Gaussian with energy-dependent width $\sigma(E)$. The Galactic plane is excluded within $|b_1|<0.3^\circ$ to avoid non-DM astrophysical gamma-ray emission. The number of background events observed from a specified region of the sky over an energy window $\Delta E$ is then 
\begin{equation}
\label{eq:nbg}
N_{\rm bg} = t_{\rm obs} \Delta \Omega \int_{\Delta E}{ \int_{-\infty}^{+\infty}{ \frac{\diff N_{\rm CRE}(\bar{E})}{\diff E\,\diff A\,\diff t\,\diff \Omega} A_{\rm eff}(\bar E) \frac{e^{-\frac{(E-\bar{E})^{2}}{2\sigma^{2}}}}{\sqrt{2\pi\sigma^{2}}} \diff \bar{E}}\, \diff E}~,
\end{equation}
where $\Delta \Omega$ is the solid angle of the region and $\diff N_{\rm CRE}/\diff E\,\diff A\,\diff t\,\diff \Omega$ is the differential intensity spectrum of the CRE events. Analyses using the Ring Method search for an excess of counts in the ON region compared to the OFF region. After rescaling the observed counts in the OFF region by the factor $\alpha$,
the excess of counts between the ON and the rescaled OFF regions is defined as  $\theta_{\rm diff}=\theta_{\rm ON}-\alpha \theta_{\rm OFF}$, where $\theta_{\rm ON}$ and $\theta_{\rm OFF}$ are the total numbers of events (the sum of signal photons and background events) in the ON and OFF regions, respectively. We assume that the likelihood of observing $\theta$ counts in a given region is
Poisson-distributed with mean value $N$ therefore assuming that no excess is observed ($\theta=0$) the likelihood can be expressed as
\begin{equation}
\mathcal{L}(m_{\chi},\langle \sigma v \rangle)= e^{-(N_{\rm ON}+\alpha N_{\rm OFF})} I_{0}(2\sqrt{\alpha N_{\rm ON} N_{\rm OFF}})~,
\end{equation}
where $I_{0}$ is a Bessel function of the first kind and $N_{\rm ON,OFF}$ are the expected number of events in the ON and OFF regions. We can take advantage of the spectral information by calculating the likelihood over small energy bins, and define the total likelihood as the product of the likelihoods over each energy bin:
\begin{equation}
\mathcal{L}(m_{\chi},\langle \sigma v \rangle)=\prod_{j}\mathcal{L}_{j}(m_{\chi},\langle \sigma v \rangle)~,
\end{equation}
where $j$ indexes the energy bins. In order to derive a contraint on $\la \sigma v \ra$ one can calculate the likelihood statistic test ratio (TS) expressed as
\begin{equation}
\text{TS}=-2\ln\left(    \dfrac{\mathcal{L}(m_{\chi},\langle \sigma v \rangle)}{\max\limits_{\langle \sigma v \rangle}[\mathcal{L}(m_{\chi},\langle \sigma v \rangle)]} \right)~,
\end{equation}
which is $\chi^{2}$-distributed with one degree of freedom and can be well approximated by a Gaussian distribution by the central limit theorem. The likelihood ratio is maximized for $\langle \sigma v \rangle=0$ (i.e., no signal events), so this ratio can be compared to a Gaussian distribution and find the value of $\langle \sigma v \rangle$, for a certain $m_{\chi}$, which constrains some model at a certain confidence level.

\subsubsection{Overview of gamma-ray searches}
\begin{figure}[h!]
\begin{center}
	\includegraphics[width=12cm]{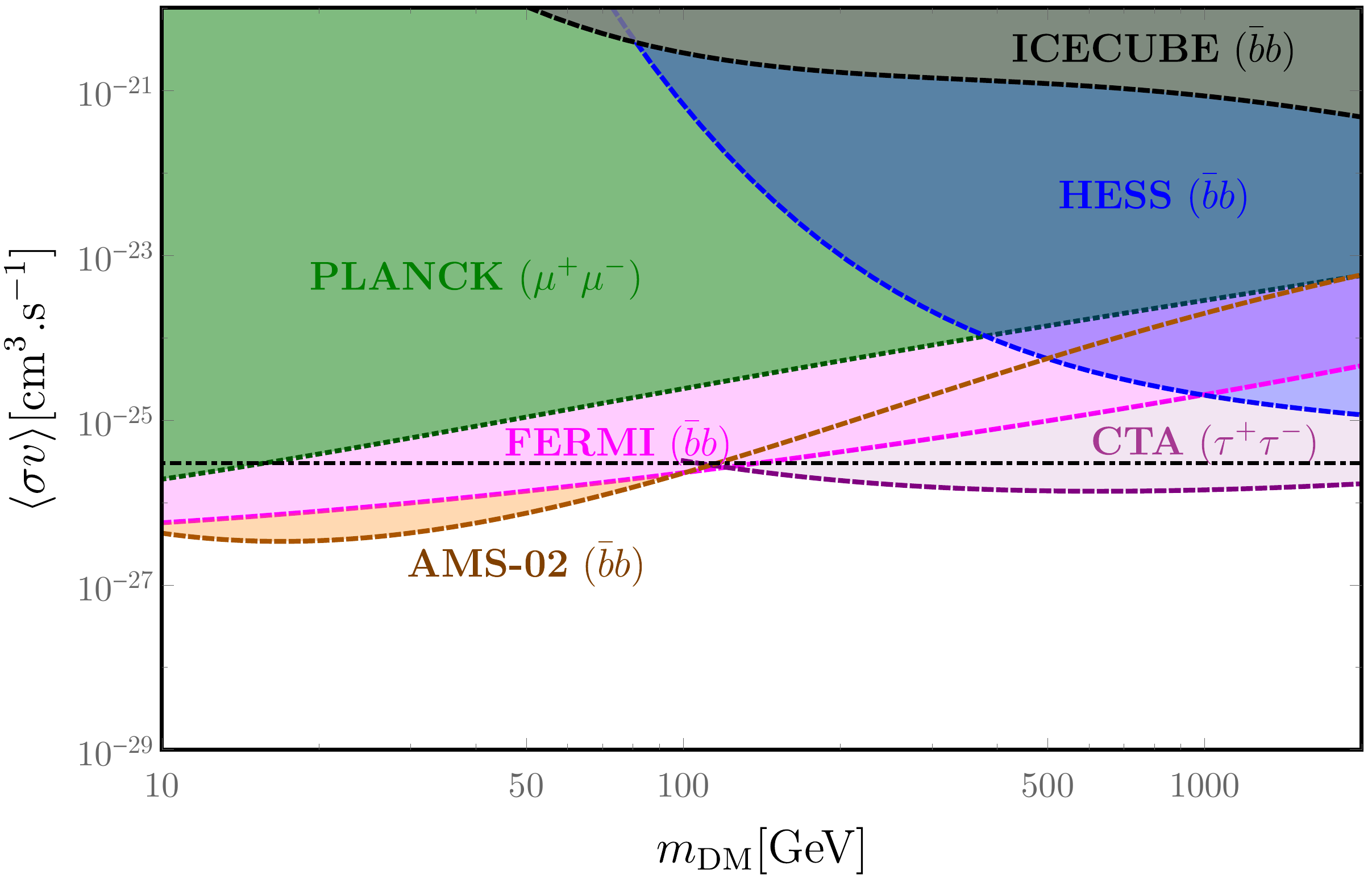}
\caption{Indirect detection constraints for several channels denoted in parentheses: from IceCube~\cite{Aartsen:2014hva}, measurement of the $\bar{p}/p$ ratio with AMS-02~\cite{Lin:2015taa}, H.E.S.S~\cite{Abdallah:2016ygi}, observation of late DM annihilations in the CMB with Planck~\cite{Slatyer:2015jla}, CTA sensitivity estimation~\cite{Pierre:2014tra} and joined analysis from Fermi and MAGIC collaborations~\cite{Ahnen:2016qkx}. The canonical value of $\la \sigma v \ra =3\times 10^{-26}~\text{cm}^3~\text{s}^{-1}$ is represented with a black dotted-dashed line.
\label{fig:IDconstraints}}
\end{center}
\end{figure}

Some of the most promising target in order to attempt at observing DM annihilations are the Galactic Center, the galactic halo and the Dwarf Spheroidal Satellites (dSphs) of the Milky Way which are objects typically dominated by a dark component. From the observation of the inner part of the galactic halo the H.E.S.S. collaboration strongly constrain a thermal DM production $\la \sigma v \ra \sim 10^{-27}~\text{cm}^3~\text{s}^{-1}$ for DM annihilations to $\gamma$-ray lines~\cite{Abdallah:2018qtu} and for DM masses above $m_{\text{DM}} \gtrsim 300~\text{GeV}$. At lower masses $m_{\text{DM}}\lesssim 100~\text{GeV}$, a joined analysis based on Fermi-LAT and MAGIC data on dSphs excludes the canonical value $\la \sigma v \ra \sim 10^{-26}~\text{cm}^3~\text{s}^{-1}$ for DM annihiliations to a $\bar{b}b$ pair~\cite{Ahnen:2016qkx}. The future Cerenkov Telescope Array (CTA) might reach the sensitivity required to probe the thermal expected value of $\la \sigma v \ra$ for DM masses up to several TeV depending on the annihilation channel~\cite{Silverwood:2014yza,Pierre:2014tra}. Some of the stronger constraints from indirect detection are depicted in the $\{ \la \sigma v \ra,m_{\text{DM}}\}$ plane in Fig.~\ref{fig:IDconstraints}.

\subsection{Antimatter}
Astrophysical antimatter particles are promising targets for Dark Matter annihilation as only few processes are responsible for their production. They are mostly produced by spallation of primary cosmic ray and are strongly affected by propagation in the galaxy through inverse Compton scattering and synchrotron radiation as well as influenced by solar magnetic field. The ratio of $\bar{p}/p$ have been measured by the PAMELA~\cite{Adriani:2012paa} and AMS-02 collaborations quite precisely with a fine agreement. A discrepancy at large energies $E\gtrsim 10~\text{GeV}$ was reported by these collaborations. However it was shown in~\cite{Giesen:2015ufa} that uncertainties regarding the source and propagation of anti-protons could explain this excess. The precise measurement of the $\bar{p}/p$ ratio allows to constrain DM annihilations to $\bar{b}b$~\cite{Lin:2015taa} as depicted in Fig.~\ref{fig:IDconstraints}. Positrons and anti-nuclei are also considered as potential annihilation channels but are not discussed in this thesis.

\subsection{Neutrinos}
Neutrinos have the interesting property to not being affected over large distances during the propagation in the galaxy as they interact only weakly, therefore they preserve the spectral information of the source. They are detected by Cerenkov light produced by some detector material as neutrinos pass through it. An interesting scenario quite studied in the litterature is to consider the effect of DM being trapped in the gravitational potential of the Sun in the case where the DM-baryon scattering cross section is large enough. The only way to detect annihilations of these trapped particles would be via the neutrino channel as only neutrinos can escape efficiently the Sun. The non-observation of a neutrino excess from the Sun allowed the collaborations IceCube~\cite{Aartsen:2012kia} and ANTARES~\cite{Adrian-Martinez:2013ayv} so set constraints on $\la \sigma v \ra_{\bar{\nu}\nu}$. Constraints on different annihilation channels, such as $\bar{b}b$, are also derived from neutrino experiments as depicted in Fig.~\ref{fig:IDconstraints}. However these experiments cannot contrain sufficiently $\la \sigma v \ra$ at the present time to probe its thermal expected value.

\section{Collider Searches}
Another possibility to attempt at observing Dark Matter is to produce it at high energy colliders such as the Large Hadron Collider (LHC). LHC proton-proton collisions  might result in the production of WIMPS in association with one or more QCD jets, photons as well as other detectable SM debris. Since WIMPs are electrically neutral 
and cosmologically stable massive particles, they manifest at colliders as missing transverse energy $\slashed{E}_T$. For this reason searches for DM are based on the observation of the visible counterpart as trigger of the event such as charged leptons, jets or 
a photon, generally referred to as mono-X searches. By selecting events with large missing energy one can reduce the SM background and potentially disentangle a DM signal. However, colliders can allow to identify only missing energy, and therefore they cannot uniquely ascertain  the presence of DM in a signal event. They can simply confirm the presence of a neutral and "stable" particle, that might have even decayed outside the detector. Anyhow, colliders offer an exciting and complementary search strategy to identify WIMPs. Indeed, assuming that the production of WIMPs at colliders is uniquely connected to WIMP-nucleon scatterings at underground laboratories, one can use the non-observation signals with large missing transverse momentum to derive limits on the WIMP-nucleon scattering cross-section \cite{Goodman:2010yf,Bai:2010hh,Goodman:2010ku,Rajaraman:2011wf,Busoni:2013lha,Alves:2014yha,
Busoni:2014haa,DeSimone:2016fbz}.\\
 However an EFT model independent approach is often limited as the effective approach might attain its  validity limit at the high energies reached by modern colliders. For instance it is possible that the centre-of-mass energy of a process producing a pair of DM particles reaches the mass of the mediator that has been integrated out in the EFT to generate the contact operator involving the DM and some SM field. In this case the EFT is no longer valid and the underlying degrees of freedom of some UV complete theory might manifest such as Breit-Wigner resonance effect. Therefore one would need the formulation of the complete UV theory in order to extract constraints on the corresponding parameter space.\\
Even though colliders are extremely useful at constraining DM models, one would often have to perform a phenomenological analysis in a specific theory rather than rely on model independent approaches which are limited for collider studies for the previously stated reasons.

\subsection{Mono-X Searches}
Mono-X searches stands for the search for WIMPs ($\chi$) produced in association with one or more QCD jets or 
potentially other SM particles as:
\begin{equation}
pp~\rightarrow ~\chi \chi +\text{X}~,
\end{equation}
where X denotes QCD jets, $\gamma$, $h$, $Z$ etc. The idea is to search for 
events with a jet of high transverse momentum $p_T$ within an event with large missing transverse 
momentum. In particular, the most recent studies performed at the LHC include up to four jets and require 
the leading jet to have $p_T > 250$~GeV \cite{Ratti:2016pwi,Tolley:2016lbg}, while others do not limit the 
number of jets while selecting events with at least one jet with $p_T > 100$~GeV \cite{CMS:2016pod}. While 
being more inclusive, these recent searches have become more challenging due to the number of jets analyzed, 
requiring a substantial improvement on the background coming from $Z~+~\text{jet}$ and $W^\pm~+~\text{jet}$ channels.
Important detector effects, such as fake jets, and QCD backgrounds weaken the LHC sensitivity 
to WIMPs, and for these reasons mono-jet searches are subject to large systematics. Nevertheless, an enormous effort has been carried out in this direction with data driven background and optimized event  selections, which combined with the increase in luminosity has led to an overall improvement on the LHC 
sensitivity to WIMPs. The latest results from CMS and ATLAS collaborations in the 
search for DM based on mono-X searches are given in~\cite{Aaboud:2016tnv,CMS:2016pod}.


\subsection{Invisible decays}

\subsubsection{Higgs decays}
In the case where WIMPs are lighter than $m_h/2\simeq 62.5$~GeV, the Higgs boson might invisibly decay into WIMP pairs. Therefore, one can use bounds from LHC on the invisible branching ratio of the Higgs, ${\text{Br} (h\rightarrow \text{inv})} \leq 0.25$ at 95\% C.L. \cite{Aad:2015pla,Khachatryan:2016whc}, to set constraints on WIMP models. 

\subsubsection{$Z$-boson decays}

The decay width of the $Z$-boson has been precisely measured and therefore stringent limits can be derived on any  extra possible decay mode of the Z boson. In particular, one can use only direct measurements of the invisible partial width using the single photon channel to obtain an average bound which is derived by computing the difference between the total and the observed partial widths assuming lepton universality. The current limit is
$\Gamma {  (Z\rightarrow \rm{inv})} \leq 499 \pm 1.5$~MeV \cite{Olive:2016xmw}.

\section*{Summary}

In this chapter we reviewed some theoretical elements of modern cosmology in order to explain how the Dark Matter problem has emerged in the last century. Based on modern observations, we discussed the various probes and evidences of the presence of a non-relativistic Dark Matter component representing $\sim 25\%$ of the energy budget of the universe and its impact on structure formation. We discussed various possibilities to account for this missing mass issue, in particular in the context of the WIMP freeze-out scenario. In the last chapter of this first part, we discussed the current status of standard searches for Dark Matter which aim at observing DM-SM scatterings, annihilations or direct production of DM. In the following part we propose some phenomenological solutions to the Dark Matter problem based on the WIMP paradigm, from the studies of simplified Dark Matter models to more ellaborate constructions motivated by particle physics considerations. In the last part we explore more complex scenarios involving alternative Dark Matter production mechanisms.

%% file: parts/simplifiedmodels.tex
\label{sec:simpmodels}
\section{Introduction}
Little is, however, as of yet known about DM as a particle; any candidate for (most of) the DM must 
nevertheless be consistent with the following five observationally-motivated constraints: 
\begin{itemize}
\item The relic abundance of DM needs to account for the observed CDM abundance.
\item The DM particle should be non-relativistic at matter-radiation equality to form structures in the 
early Universe in agreement with the observation. As a result, if the DM was produced as a thermal relic 
in the early Universe, its mass cannot be arbitrarily light. Specifically, cosmological simulations rule 
out DM masses below a few keV \cite{Benson:2012su,Lovell:2013ola,Kennedy:2013uta}. 
\item The DM should be  electromagnetically neutral, as a result of null searches for stable 
charged particles \cite{SanchezSalcedo:2010ev,McDermott:2010pa} as well as Direct Detection (DD) experiments, 
which we will review subsequently.
\item The DM particle must be cosmologically stable since its presence is ascertained today, implying that its 
lifetime is larger than the age of the Universe. Under certain assumptions, much stronger limits are applicable 
conservatively requiring a lifetime order of magnitude larger can be derived \cite{Queiroz:2014yna,Audren:2014bca,
Giesen:2015ufa,Mambrini:2015sia,Baring:2015sza,Lu:2015pta,Slatyer:2016qyl,Jin:2017iwg}.
\item Cluster collisions, such as the Bullet Cluster \cite{Clowe:2006eq}, constrain the level of self-interactions 
that DM particles can have (see however~\cite{Hochberg:2014dra,Hansen:2015yaa} for alternative scenarios).
\end{itemize}

Within the generous parameter space outlined by the observational requirements listed above, we will argue below 
that the paradigm of WIMPs~\cite{Steigman:1984ac} is one of the most compelling options for DM as a particle. In order to maximally profit of the information from the different kind of experimental searches we need an 
efficient interface between the experimental outcome and theoretical models. The processes responsible for the 
DM relic density and its eventual detection can be described by simple extensions of the SM in which a DM 
candidate interacts with the SM states (typically the interactions are limited to the SM fermions) through a 
mediator state (dubbed portal). This idea is at the base of the so-called "Simplified Models"~\cite{DiFranzo:2013vra,
Berlin:2014tja,Abdallah:2014hon,Buckley:2014fba,Godbole:2015gma,Abdallah:2015ter,Duerr:2015wfa,Baek:2015lna,
Carpenter:2016thc,Bauer:2016gys,Sandick:2016zut,Bell:2016uhg,Bell:2016ekl,
Khoze:2017ixx,ElHedri:2017nny} which are customarily adopted especially in the context of 
collider studies, see e.g., Refs.~\cite{Jacques:2015zha,Xiang:2015lfa,Backovic:2015soa,Bell:2015rdw,Brennan:2016xjh,
Boveia:2016mrp,Englert:2016joy,Goncalves:2016iyg,DeSimone:2016fbz,Liew:2016oon,Kraml:2017atm,Bauer:2017ota,Albert:2017onk}. 

The first class of models which will be the object of study are the SM Dark Portals\footnote{An analogous 
study has been performed in Ref.~\cite{Escudero:2016gzx}. Our results are in substantial agreement with the 
ones reported in this reference.}, i.e., models in which the DM interacts with the SM state through the Higgs 
or the $Z$-boson. In the case when the DM is a pure SM singlet, gauge invariant renormalizable operator connecting 
the DM with the $Z$ or the SM-Higgs boson can be build only in the latter case and only for scalar and vectorial 
DM. In the other cases one should rely either on higher dimensional operators, or on the case that the 
coupling with the Higgs and/or the $Z$ is originated by their mixing with new neutral mediators. The latter 
case can imply the presence of additional states relevant for the DM phenomenology and will be then 
discussed later on in the text. We will instead quote below some example of higher dimensional operator 
but we will not refer to any specific construction for our analysis. 

The simplied models presented in the previous cases for the SM-Higgs
and $Z$-boson portals will be generalized and discussed in
more details in the case of generic BSM spin-0
mediator interacting with a pair of scalar, fermion or vector
DM fields. Contrary to the case of SM portals, interactions of
the mediator with the gauge bosons are not mandatory. We
will thus stick, in this chapter to the case, analogous to the so called
simplified models, in which the DM is coupled only to
the SM fermions.
%

\section{Higgs portal}

The most economical way to connect a SM singlet DM candidate with the SM Higgs doublet $H$ is through four 
field operators built to connect the Higgs bilinear $H^\dagger H$, which is a Lorentz and gauge invariant 
quantity, with a DM bilinear. Assuming CP conservation, the possible~\footnote{We limit, for simplicity, 
to the lowest dimensional operators. Higher dimensional operators are discussed, for example, in Ref.~\cite{Greljo:2013wja}.} 
operators connecting the SM-Higgs doublet with scalar, fermion and vector DM are given by \cite{Silveira:1985rk,
McDonald:1993ex,Burgess:2000yq,Kim:2006af,Andreas:2010dz,Kanemura:2010sh,Lebedev:2011iq,Mambrini:2011ik,
Djouadi:2011aa,LopezHonorez:2012kv,
Djouadi:2012zc}: 

\begin{equation}
\label{eq:HpLagrangian}
\xi \lambda^H_{\chi} \chi^{*} \chi H^\dagger H, \,\,\,
 \xi \frac{\lambda^H_{\psi}}{\Lambda} \bar{\psi} \psi H^{\dagger} H
~~{\rm and~~} \xi \lambda^H_V V^\mu V_\mu H^\dagger H,
\end{equation}
where, in the unitary gauge, $H ={\left(0\,\,\,\frac{v_h+h}{\sqrt{2}}\right)}^T$ with $h,\,v_h$ denoting the 
physical SM Higgs boson, Vacuum Expectation Value (VEV) and $\xi=1/2 (1)$ in case the DM is (not) its own antiparticle. We have indicated a
scalar, fermionic and vectorial DM as $\chi$, $\psi$ and $V$, respectively.
From Eq.~(\ref{eq:HpLagrangian}) note that stability of the DM is protected either by a discrete $\mathbb{Z}_2~({\rm for}~ 
\psi,\,V_\mu$ and when $\chi=\chi^*$) or by a $U(1)~({\rm for}~\chi\neq \chi^*)$ symmetry. In the case of a scalar and vectorial DM it is possible to rely on a dimension-4 renormalizable 
operator. On the contrary, a fermionic DM requires at least a dimension-5 operator which depends on an unknown 
Ultra-Violet (UV) scale $\Lambda$. After EW symmetry breaking (EWSB), trilinear couplings between the Higgs field $h$ and DM pairs are induced. 
In the case of fermionic DM it is possible to absorb the explicit $\Lambda$ dependence by a redefinition of 
the associated coupling, i.e., $\lambda_\psi^H \frac{v_h}{\Lambda}$ as $\lambda_\psi^H $, so that it does not 
appear explicitly in computations.

The models defined by Lagrangians of Eq.~(\ref{eq:HpLagrangian}) have only two free parameters, the DM masses 
$m_{\chi,\psi,V}$ and couplings $\lambda_{\chi,\psi,V}^H$
with the SM-Higgs. The constraints on these models can be then 
easily summarized in bi-dimensional 
(i.e., DM mass vs its coupling with the SM-Higgs) planes. 
\begin{center}
\begin{figure}[h!]
  \begin{minipage}[l]{0.33\textwidth}
\includegraphics[width=\linewidth]{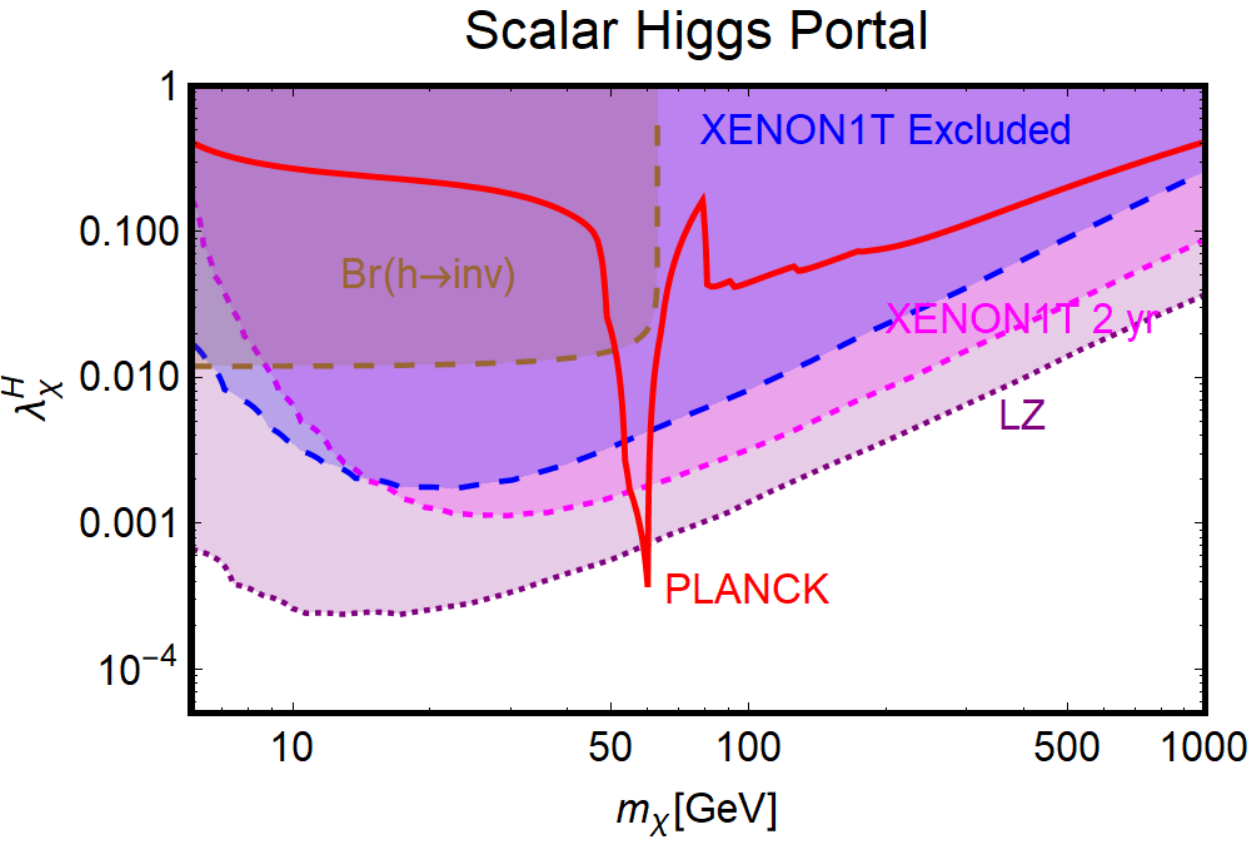}
   \end{minipage}\hfill
   \begin{minipage}[c]{0.323\textwidth}   
\includegraphics[width=\linewidth]{figures/simplifiedmodels/FermionHp.pdf}
   \end{minipage}
      \begin{minipage}[r]{0.33\textwidth}   
\includegraphics[width=\linewidth]{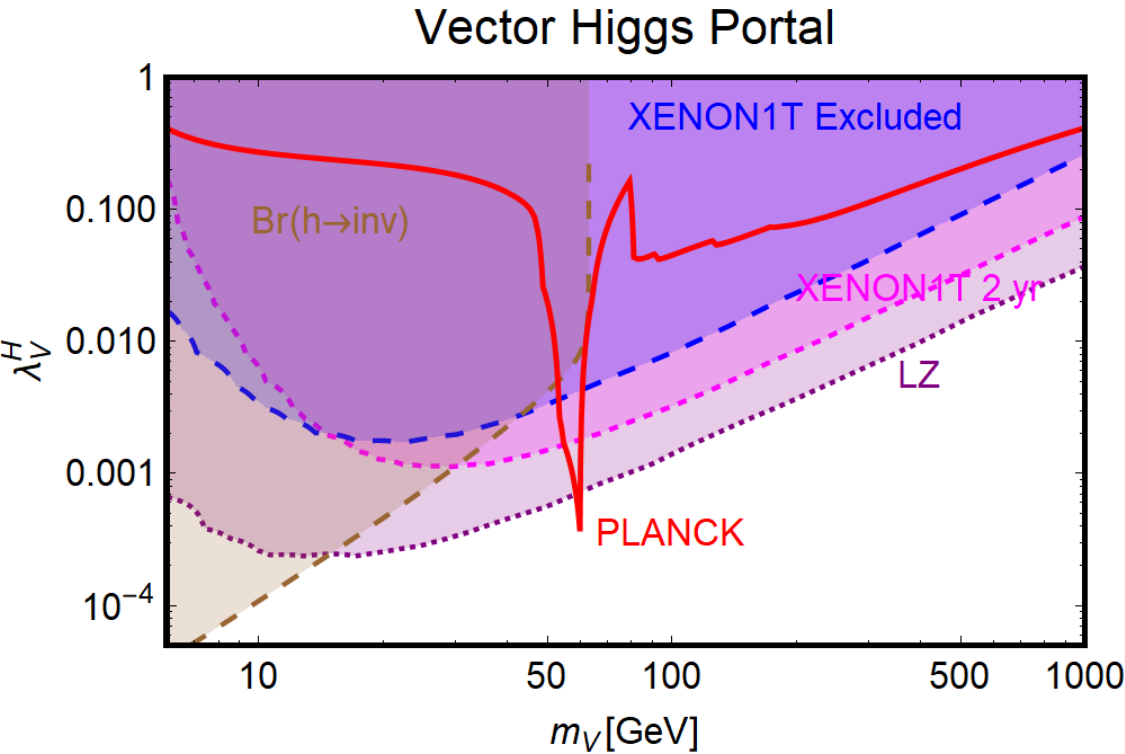}
   \end{minipage}
   \caption{Illustration of the SM-Higgs portal
in the relevant bi-dimensional planes for a
scalar (left panel), fermionic (middle panel) 
and vectorial (right panel) DM. In each plot, the red line represents the model points featuring the correct DM relic 
density. The blue region is excluded by the current SI DD limits. The magenta region would be excluded in 
case of absence of signals in XENON1T after two years of exposure time while the purple region is within the reach of 
future LZ limits. Finally, the brown region is excluded because of a experimentally disfavored invisible decay branching 
fraction of the SM-Higgs boson.}
\label{fig:ScalarHp}
\end{figure}
\end{center}
In Fig.~\ref{fig:ScalarHp} we summarize our results for scalar, fermionic and vectorial DM, respectively. All the plots 
report basically three set of constraints~\footnote{We will report in the main text just the results of the analysis.
Analytical expressions of the relevant rates are extensively reported in the appendix.}. The first one
(red contours) is represented by the achievement of the correct DM relic density~\footnote{In the mentioned figure 
and throughout all this thesis we will, as convention, use the label "Planck", as the corresponding experiment, for 
iso-countours corresponding to the correct DM relic density.}. The DM annihilates into SM
fermions and gauge bosons, through s-channel exchange of the SM-Higgs boson, and, for higher masses, also into
Higgs pairs through both s- and t-channel diagrams (in this last case a DM particle is exchanged). Since the
coupling of the SM-Higgs with SM fermions and gauge bosons depends on the masses of the particles themselves, the
DM annihilation cross-section is suppressed, at the exception of the pole region $m_\chi \sim m_h /2$,
until the $W^+W^-$, $ZZ$ and $\bar{t} t$ final states are kinematically accessible. Even in this last case, the
cosmologically allowed values for the couplings are in strong tension with the constraints from DD of the 
DM, which for all the considered spin assignation of the DM, arise from SI
interactions of the DM with the SM quarks originated by t-channel exchange of the SM-Higgs boson. As can be easily
seen that the entire parameter space corresponding to thermal DM is already ruled out, at the exception,
possibly, of the pole region, for DM masses at least below 1 TeV. Eventual surviving resonance regions will
be ruled-out in case of absence of signals at the forthcoming XENON1T, assuming a two years of exposure time. 
As expected, the most constrained
scenario is the fermionic DM one because of the further suppression of the p-wave suppression of its annihilation 
cross-section. 

Notice that, scalar and vectorial DM, due to the s-wave annihilation cross-section, might also be probed through 
ID. The corresponding limits are nevertheless largely overpower by the ones from DD and hence, have been then omitted 
for simplicity. The limits from DD experiments are complemented at low DM masses, i.e.,
$m_{\chi,\psi,V} < m_h/2$, by the one from invisible decay width of the Higgs. Indeed this constraint would
exclude DM masses below the energy threshold of DD experiments. Our findings are in agreement with 
the other recent studies in the 
topic \cite{Cline:2013gha,Queiroz:2014pra,Buckley:2014fba,Dutra:2015vca,Freitas:2015hsa,Anchordoqui:2015fra,
Fedderke:2015txa,Duch:2015jta,Abdallah:2015ter,Chen:2015dea,Abercrombie:2015wmb,Kumar:2015wya,Han:2016gyy,
He:2016mls,Bambhaniya:2016cpr}.

%

\section{$Z$ portal}

An interaction between the $Z$-boson and a SM singlet DM candidate is not gauge invariant for any dimension-4 
operators. In the case of scalar and fermionic DM models, the simplest option is to consider a dimension 6 
operator~\footnote{Similar to the case of the Higgs portal we just quote, as an example, the lowest dimensional 
operator. This is however, not the only possible option.}~\cite{Cotta:2012nj,deSimone:2014pda,Kearney:2016rng}. 
In the case of scalar DM it is of the form: 
\bea
\mathcal{L}=  \lambda_\chi \frac{H^\dagger \overleftrightarrow{D^{\mu}} H}{\Lambda^2}
\chi^* \overleftrightarrow{\partial_{\mu}} \chi,
\eea 
which give rise to a trilinear interaction between the $Z$-boson and a DM pair once the SM-Higgs field in the Lagrangian is 
replaced by its VEV, so that $H \overleftrightarrow{D^\mu} H \rightarrow \frac{g v_h^2}{4 \cos \theta_W}Z^\mu$. $\Lambda$ 
is again the relevant cutoff scale of the effective theory. Similar to the case of a fermionic 
DM in Higgs portal we 
can absorb it in the definition of a dimensionless coupling as $\lambda^Z_\chi\equiv \lambda_\chi v^2_h/\Lambda^2$. In addition, after EWSB, an effective dimension-4 interaction like $(g^2/16\cos^2\theta_W)\lambda^{ZZ}_{\chi\chi}{|\chi|}^2
Z^\mu Z_\mu$ can emerge from the dimension-6 SM gauge invariant operator $\lambda_{\chi\chi}$ $(D^\mu H)^\dagger$ $D_\mu 
H {|\chi|}^2/\Lambda^2$ such that $\lambda^{ZZ}_{\chi\chi}=\lambda_{\chi\chi}v^2_h/\Lambda^2$. For simplicity 
we maintain a rescaling with powers of the $SU(2)_L$
gauge coupling $g$. The interaction Lagrangian for the DM, along with the relevant SM parts, can thus be written as: 
\bea
\mathcal{L}=i \frac{g}{4 c_W} \lambda_\chi^Z 
\chi^* \overleftrightarrow{\partial_{\mu}} \chi Z^\mu
+\frac{g}{4 c_W} \sum_f \overline{f} \gamma^\mu \left(V_f^Z-A_f^Z \gamma_5\right) f Z_\mu + \frac{g^2}{16 c^2_W} \lambda^{ZZ}_{\chi \chi} {|\chi|}^2 Z^\mu Z_\mu,
\eea
where $c_W=\cos\theta_W$ and $\theta_W$ is Weinberg angle \cite{Amsler:2008zzb} and $f$ generically refers to a SM fermion. Note that we have used a 
normalization of $g/4\cos\theta_W$ throughout in analogy to the SM $\bar{f} f Z$ couplings. The interaction Lagrangian for fermion DM is built in a similar fashion as the scalar case. In the case of 
Dirac DM the starting operator is:
\begin{equation}
\mathcal{L}= \frac{H^\dagger \overleftrightarrow{D_\mu} H}{\Lambda^2}\left( \overline{\psi}\gamma^\mu 
\left(v_\psi^Z-a^Z_\psi \gamma^5\right) \psi\right),
\end{equation}which, after the EWSB, together with the apposite SM part leads to:
\bea
\label{eq:Zlagrangian}
\mathcal{L}=\frac{g}{4 \cos\theta_W} \overline{\psi}\gamma^\mu \left(V_\psi^Z-A^Z_\psi \gamma^5\right) 
\psi Z_\mu + \frac{g}{4 \cos\theta_W}
\sum_f \overline{f}\gamma^\mu \left(V^Z_f-A^Z_f \gamma^5\right) f Z_\mu,
\eea
with ${V}_\psi^Z={v}_\psi^Z \frac{v_h^2}{\Lambda^2}$ and ${A}_\psi^Z={a}_\psi^Z \frac{v_h^2}{\Lambda^2}$. 
In the case of Majorana DM $V_\psi^Z=0$ and we rescale the remaining DM coupling by a factor of $1/2$. In the case of spin-1 DM we will consider two possible kind of interactions, namely, self- (Abelian) 
and not self-conjugated (non-Abelian) DM, respectively. For the latter, along with the necessary SM parts, 
we can write the following Lorentz invariant interaction:
\bea
\label{eq:ZVlagrangian}
\mathcal{L}=\frac{g}{4 \cos\theta_W} \eta^Z_V [[VVZ]]\nonumber+ \frac{g}{4 \cos\theta_W} \sum_f \overline{f}\gamma^\mu \left(V^Z_f-A^Z_f \gamma^5\right) f Z_\mu~,\nonumber \\
{\rm with~~}[[VVZ]] \equiv i \left[V_{\mu\nu} V^{\dagger\, \mu}Z^\nu-V^\dagger_{\mu\nu} V^{\mu}Z^\nu 
\right. \left.+\frac{1}{2}Z_{\mu\nu} \left(V^\mu V^{\dagger\,\nu}-V^\nu V^{\dagger\,\mu}\right)\right],
\eea where $V_{\mu\nu},V^\dagger_{\mu\nu},Z_{\mu\nu}$ represent the respective field strengths. In 
Eq.~(\ref{eq:ZVlagrangian}) 
the $[[VVZ]]$ coupling is normalized as $g/4\cos\theta_W$ while the model specific information are 
parametrized as $\eta^Z_V$. In the case of self-conjugate spin-1 DM, an interaction with the gauge boson can be built through the Levi-Civita symbol 
as Ref.~\cite{Mambrini:2009ad}:
\bea
\label{eq:abelianZ}
\mathcal{L}=\frac{g}{4 \cos\theta_W} \eta^{Z}_V \epsilon^{\mu \nu \rho \sigma} 
V_{\mu} Z_{\nu} {V}_{\rho \sigma} +\frac{g}{4 \cos\theta_W}\sum_f\overline{f}\gamma^\mu \left(V^{Z}_f-A^{Z}_f \gamma^5\right) f Z_\mu~.
\eea

\begin{center}
\begin{figure}[t!]
\begin{minipage}[t]{\textwidth}
\begin{center}
\includegraphics[width=0.50\textwidth]{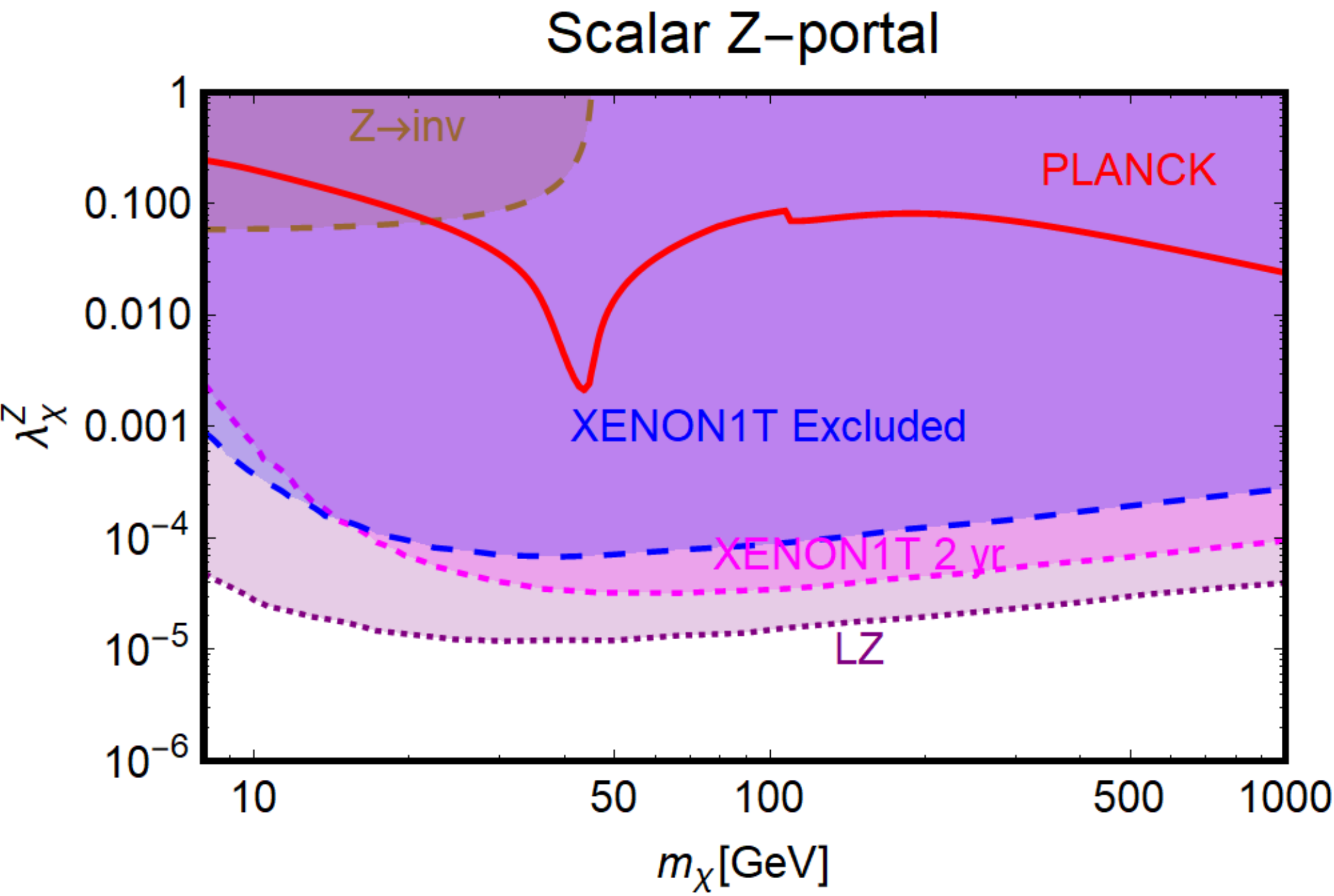}
\end{center}
\end{minipage}
 \begin{minipage}[l]{0.49\textwidth}
\includegraphics[width=\linewidth]{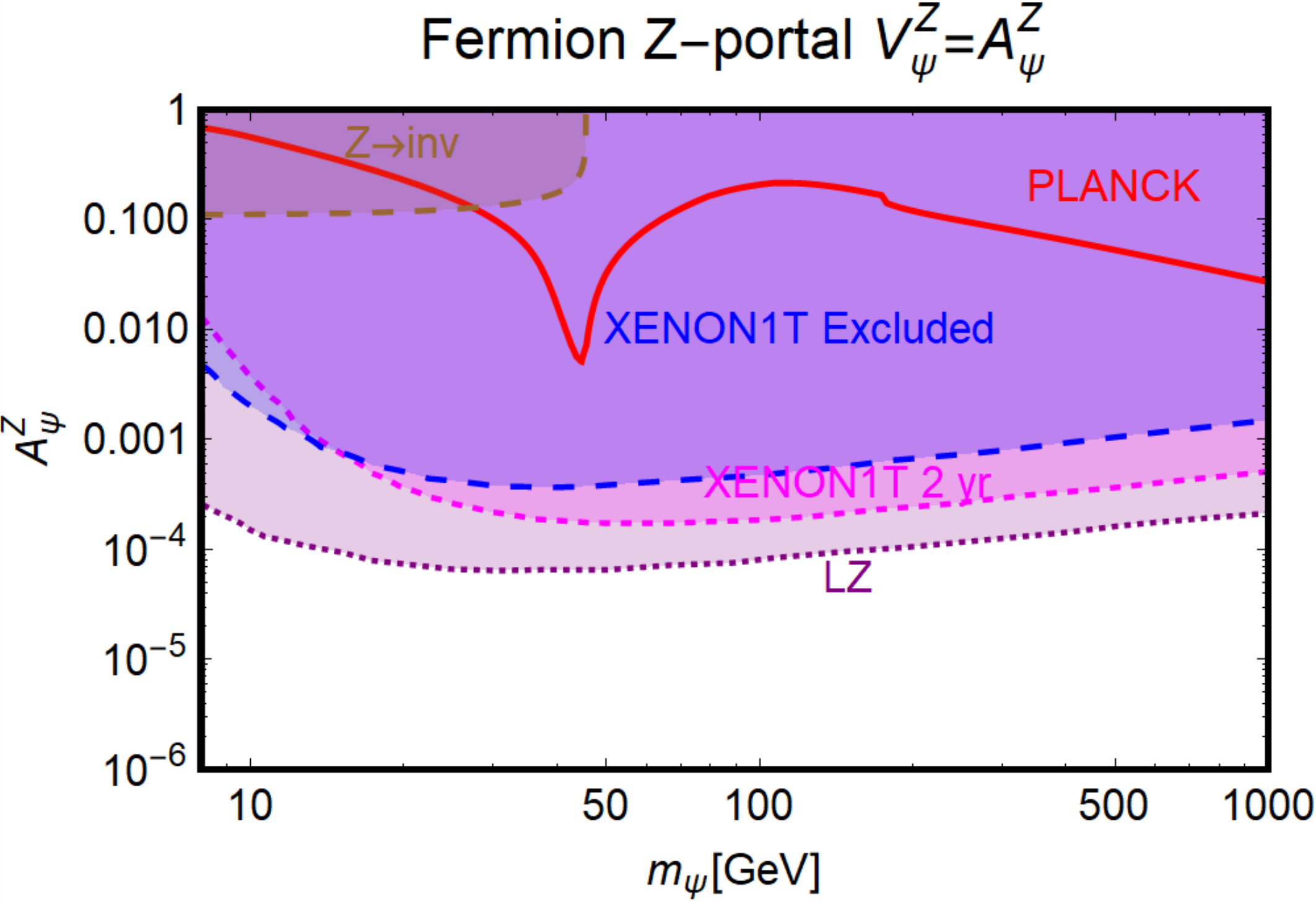}
   \end{minipage}\hfill
   \begin{minipage}[r]{0.49\textwidth}   
\includegraphics[width=\linewidth]{figures/simplifiedmodels/fermionZpA.pdf}
   \end{minipage}
     \begin{minipage}[l]{0.49\textwidth}
\includegraphics[width=\linewidth]{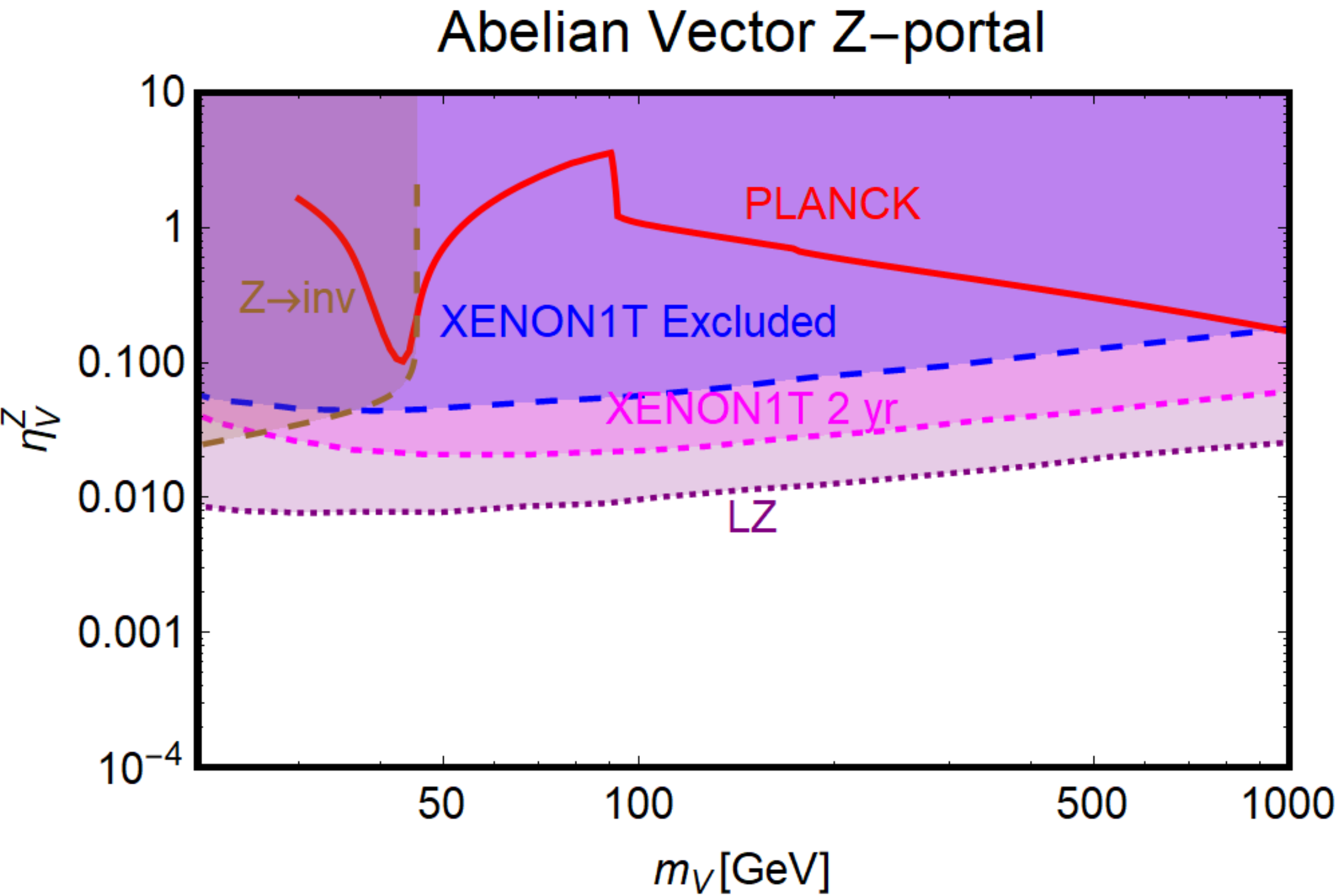}
   \end{minipage}\hfill
   \begin{minipage}[r]{0.49\textwidth}   
\includegraphics[width=\linewidth]{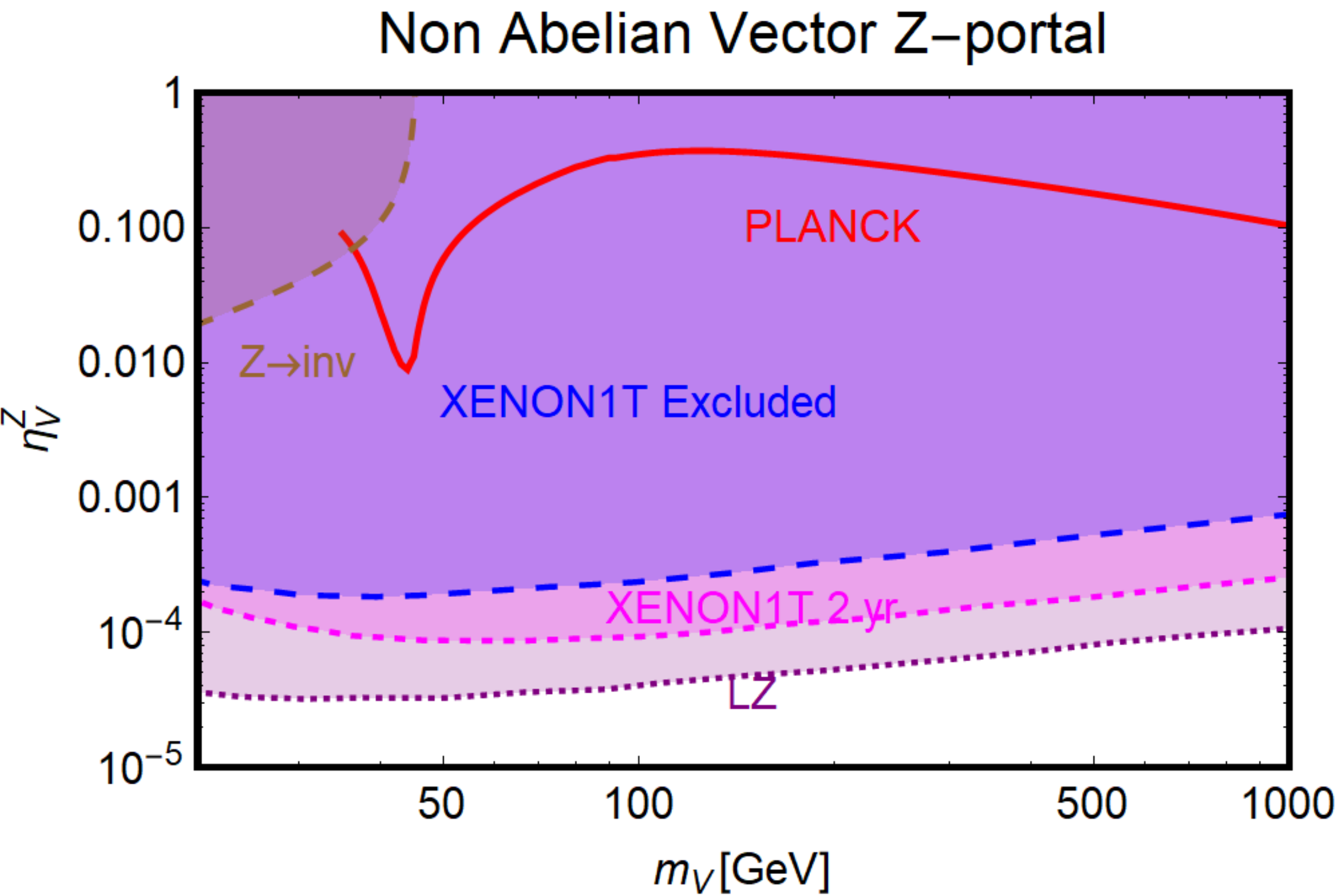}
   \end{minipage}
      \caption{Combined constraints for $Z$-portal with scalar, fermion and vector DM in the $m_\chi$-$\lambda^Z_\chi$ bi-dimensional
plane. Color specifications are the same as
Fig.~\ref{fig:ScalarHp}, except the fact that now 
the brown region represents experimentally
excluded invisible decay width of $Z$-boson.\label{fig:Zportal}}
\end{figure}
\end{center}

Similar to the previous cases the coupling $\eta_V^Z$ in Eq.~(\ref{eq:abelianZ}) encodes a cut-off scale 
(see e.g., Refs.~\cite{Anastasopoulos:2006cz,Antoniadis:2009ze,Dudas:2009uq,Dudas:2013sia} for the construction 
of effective theories). The 
theoretical derivation of Eq.~(\ref{eq:ZVlagrangian})
is however, more contrived. Further, similar to the Higgs portal, the $Z$-portal models are fully defined by two parameters so that one can repeat the 
same kind of analysis performed in the previous subsection. The results are summarized in
Figs.~\ref{fig:Zportal}~\footnote{Similar to the SM-Higgs portal case we will report in the 
main text only the main results while discussing the computation in more detail in the appendix.}.

As evident, in all but the Majorana $Z$-portal case, thermal DM is already excluded, even for masses above 
the TeV scale, by current constraints from XENON1T. These constraints are even stronger with respect to the case of 
the SM-Higgs portal. This is because, apart from the lighter mediator, the scattering cross-section 
on Xenon nuclei is 
enhanced by the isospin violating interactions of the $Z$ with light quarks. Low DM masses, possibly out of 
the reach of DD experiments, are instead excluded by the limit on the invisible decay width of the $Z$-boson. As already 
pointed out, the only exception to this picture is represented by the case of Majorana DM where the SI component 
of the DM scattering cross-section is largely suppressed due to the absence of a vectorial coupling of the 
DM with the $Z$. This scenario is nevertheless already (partially) within the reach of current searches for 
a SD component of the scattering cross-section. The increased sensitivity of XENON1T will 
allow to exclude DM masses below 300 GeV, except the ``pole'' region. 
The latter, however, will meet the same fate from a projected
future LZ sensitivity.

\section{Scalar portal}

\subsection{Scalar Dark Matter}

\begin{center}
\begin{figure}[t!]
  \begin{minipage}[l]{0.33\textwidth}
\includegraphics[width=\linewidth]{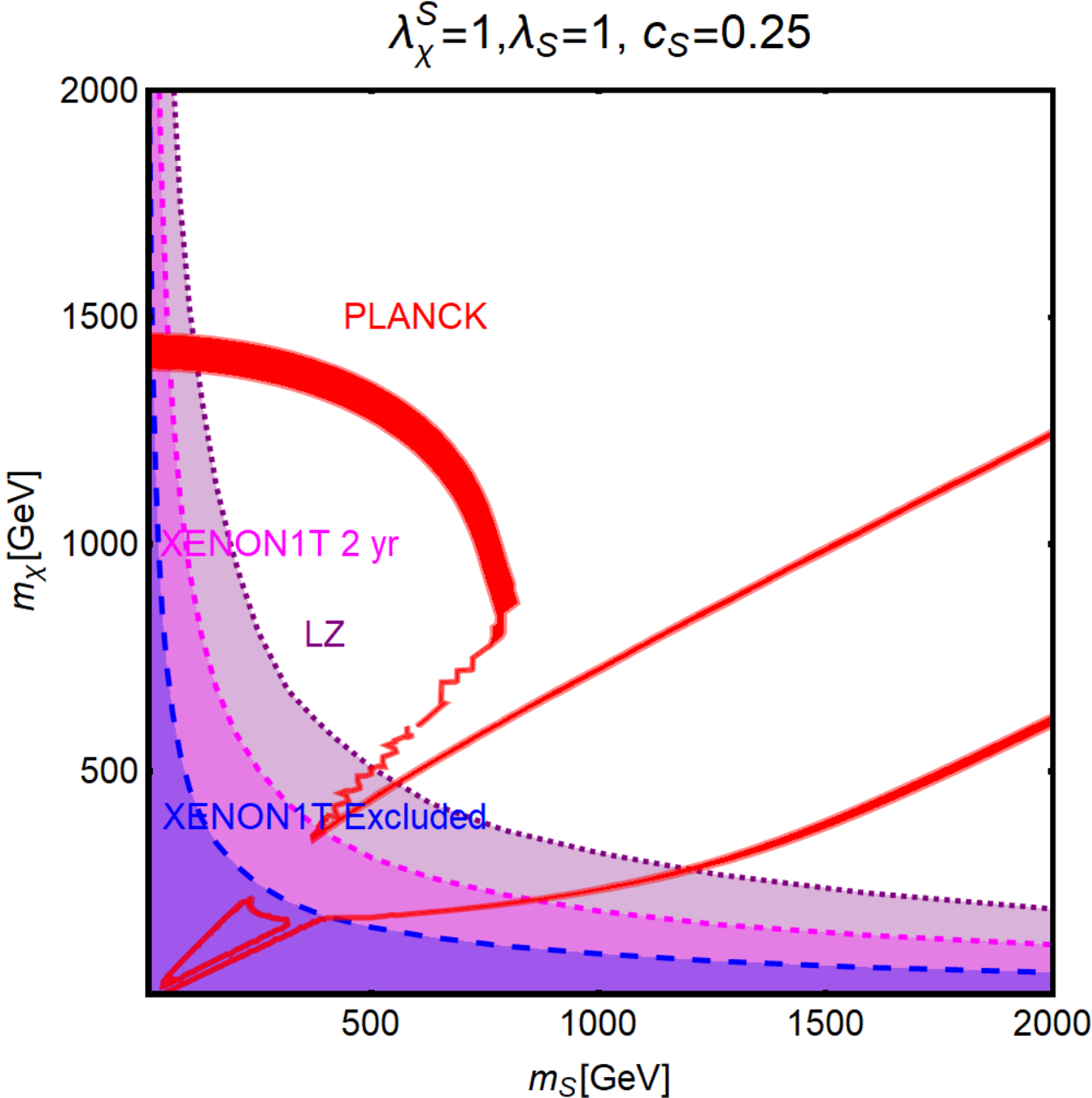}
   \end{minipage}\hfill
   \begin{minipage}[c]{0.323\textwidth}   
\includegraphics[width=\linewidth]{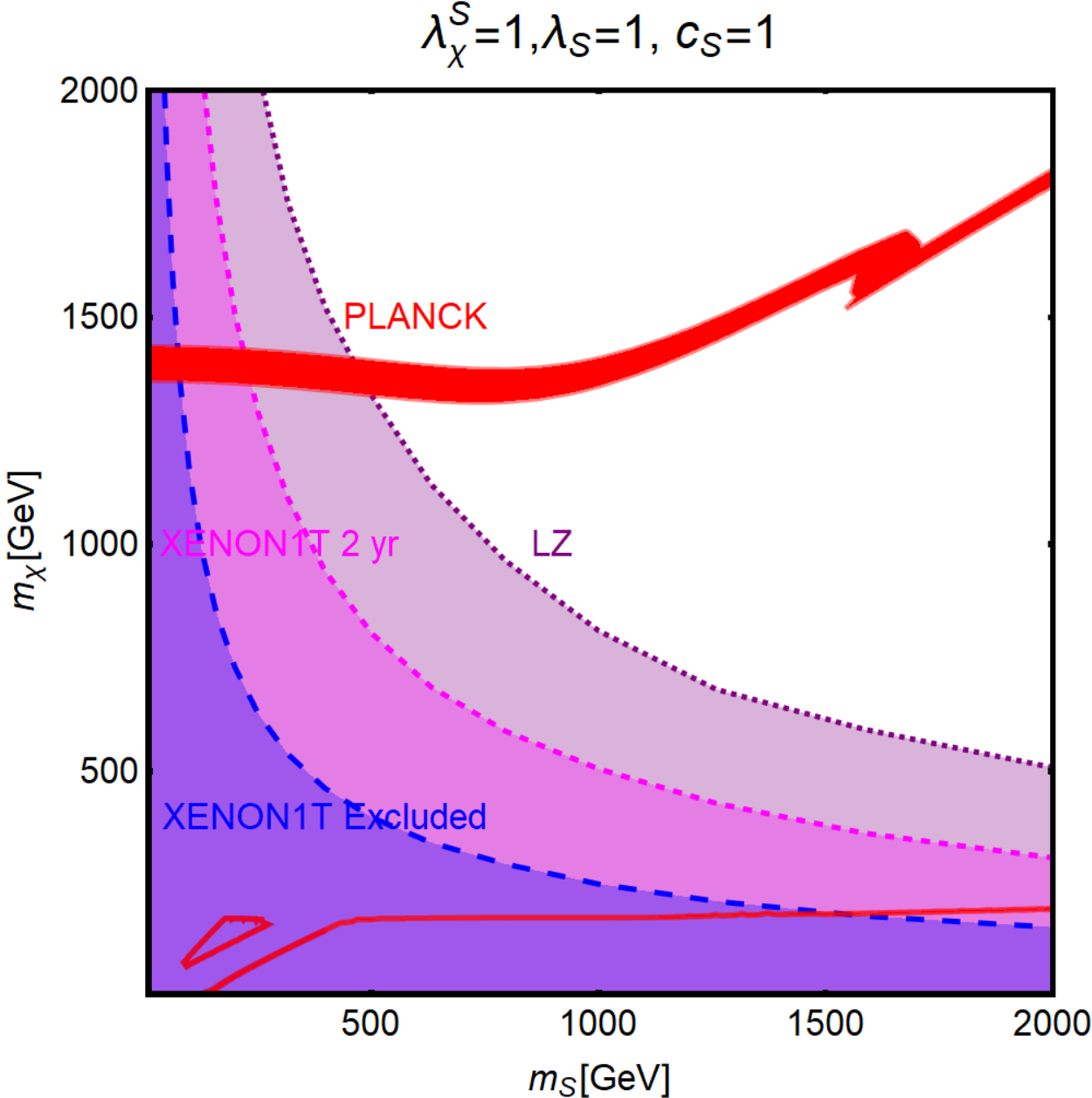}
   \end{minipage}
      \begin{minipage}[r]{0.33\textwidth}   
\includegraphics[width=\linewidth]{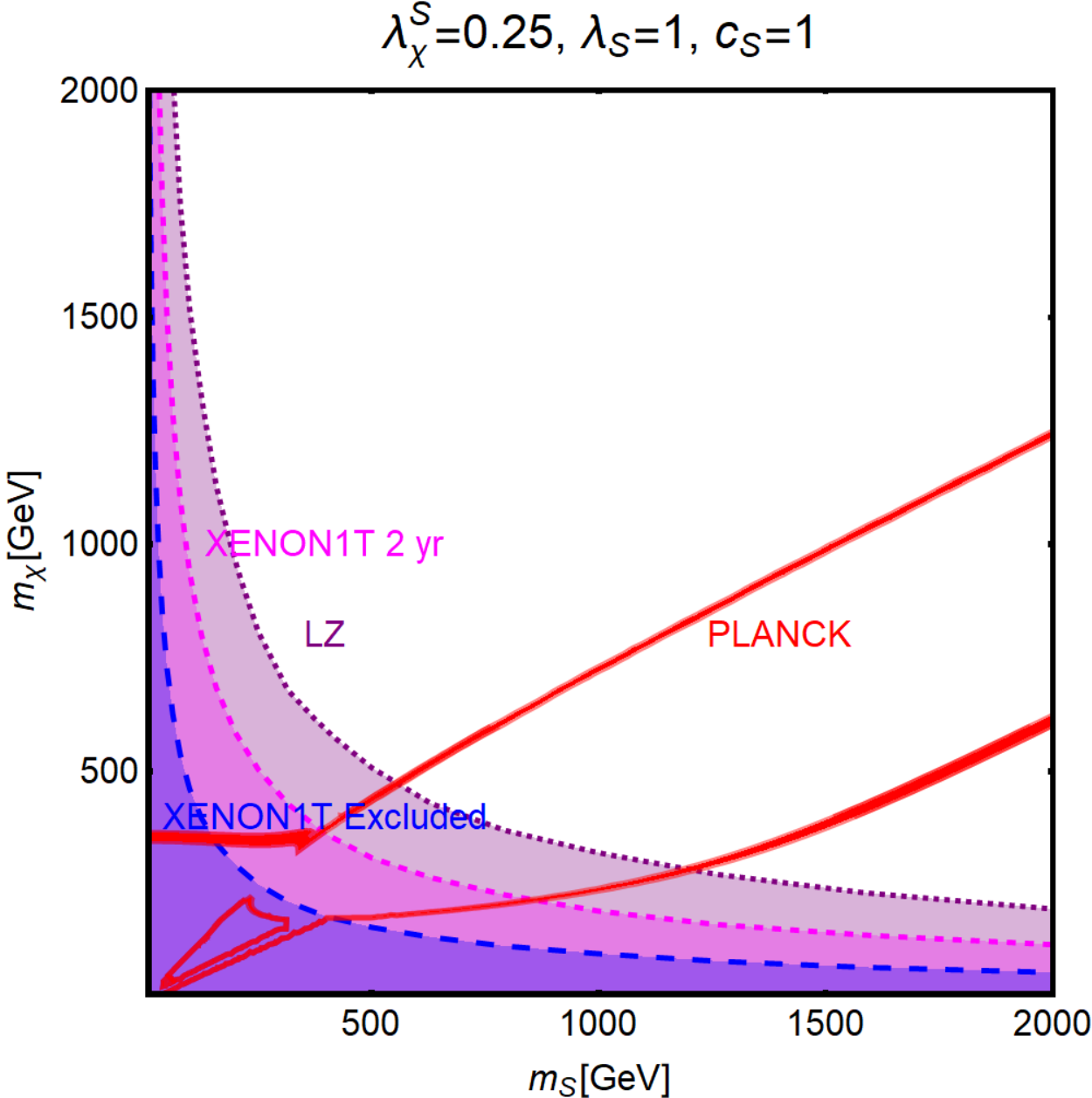}
   \end{minipage}
   \caption{Combined constrains for a scalar DM with scalar mediator scenario in the bi-dimensional 
plane $(m_S,m_\chi)$ for three assignations of the relevant couplings, i.e., $(\lambda_\chi^S,\,\lambda_S,\,c_S)
=(1,\,1,\,0.25)$ (left), $(1,\,1,\,1)$ (middle) 
and $(0.25,\,1,\,1)$ (right). Here the iso-contours of the 
correct DM relic density are represented by red bands. The blue, magenta and purple regions 
represent the current exclusion and the projected sensitivity of XENON1T (assuming 2 years of exposure 
time in the latter case) and LZ, respectively.}
\label{fig:Sportal}
\end{figure}
\end{center}

We will consider the following Lagrangian:
\begin{equation}
\label{eq:bsms0sdm1}
\mathcal{L}=-\xi \mu_\chi^S {|\chi|}^2 S- \xi \lambda_\chi^{S} {|\chi|}^2 S^2 
- \frac{c_S}{\sqrt{2}} \frac{m_f}{v_h}\bar{f} f S,
\end{equation}
where $S$ is a real scalar field and $\xi$ denotes the normalization factor, accounting, similar to the 
previous section, for the case when the DM  coincides (or not) with its own antiparticle. In the case of SM 
fermions we have assumed a Yukawa-like structure of the couplings with the mediator while for the scalar 
DM $(\chi)$ we have 
parametrized all the information, including possible normalization factors (e.g., factor of $1/2$ in the 
second term of Eq.~(\ref{eq:bsms0sdm1})), in the respective couplings. Note that $\mu_\chi^S$ parameter has
the dimension of mass. Unless differently stated we will assume $\mu_\chi^S=\lambda_\chi^S m_S$ with 
$\lambda_\chi^S$ being a dimensionless coupling and $m_S$ as the mass of $S$. We will also add self 
interaction term for the scalar field given by:
\begin{equation}
\label{eq:scalar_self}
\mathcal{L}_S=-\frac{1}{3!}m_S \lambda_S S^3.
\end{equation}
The assignation for the dimensional couplings, as well as the introduction of the Lagrangian term in 
Eq.~(\ref{eq:scalar_self}), are inspired to scenarios in which the scalar field $S$ acquires a VEV. In 
this setup the Lagrangian of Eq.~(\ref{eq:scalar_self}) originates from the quartic term in the scalar potential 
whose presence cannot be forbidden by any symmetry argument. In the same fashion a quartic interaction term 
$S^2 H^\dagger H$ with the SM-Higgs doublet, responsible for a mixing of the $S$ and 
$h$ states,  should also be included. For simplicity we will assume 
here that the coupling of this last operator is negligible\footnote{The most general case is dicussed in Sec.~\ref{ssec:higgs}}. \\
Contrary to the case of the SM portals, which have only the DM 
mass and its coupling as free parameters, 
we have expressed, as reported on Fig.~\ref{fig:Sportal}, our main results in the bi-dimensional plane 
$(m_\chi,m_S)$ for the three free coupling assignations $(\lambda_\chi^S,\lambda_S,c_S)=(1,1,0.25)$, $(1,1,1)$, $(0.25,1,1)$. 
Figure~\ref{fig:Sportal} hence shows the comparison between current DD limits, as well as the projected 
sensitivities from XENON1T and LZ, and the requirement of the correct DM relic density.
The results reported in Fig.~\ref{fig:Sportal} can be explained as follows. A t-channel exchange of the scalar 
mediator induces SI interactions of the DM, which are written, in the case of the proton as: 
\begin{align}
\sigma_{\chi p}^{\rm SI} & =\frac{\mu_{\chi p}^2}{4 \pi}\frac{{(\lambda_\chi^S)}^2 c_S^2}{m_\chi^2 m_S^2} 
\frac{m_{p}^2}{v_h^2}{\left[f_p \frac{Z}{A}+f_n \left(1-\frac{Z}{A}\right)\right]}^2 \nonumber \\ & \approx 10^{-45}{\mbox{cm}^2} {(\lambda_\chi^S)}^2 c_S^2
{\left(\frac{400~\mbox{GeV}}{m_S}\right)}^2 {\left(\frac{400~\mbox{GeV}}{m_\chi}\right)}^2~.
\end{align}
Here $A$, $Z$ represent, respectively, the atomic and proton number of the material constituting the detector,  
$\mu_{\chi p}=m_\chi m_p/(m_\chi+m_p)$ denotes reduced mass of the WIMP-proton system with $m_p$ representing 
the mass of the latter while $f_p$ and $f_n$ represent the effective couplings of the DM with protons and neutrons. 
In the case of a scalar mediator we have:
\begin{align}
\label{eq:formfac1}
f_N=\sum_{q=u,d,s} f_q^N+\frac{6}{27}f^N_{\rm TG}, \quad ~\text{with} \quad 
 f^N_{\rm TG}=1-\sum_{q=u,d,s} f_q^N,\,\,\,\,N=p,n,
\end{align}
with $f_N\,(f_q^N)$ being the form factor whose physical meaning is associated to the contribution from all the six quark
flavours (up, down and strange quark flavours)
to the mass of the proton and the neutron. The contribution of the heavy quarks in $f_N$ is described by a unique 
form factor $f_{c}^N=f_{b}^N=f^N_{t}=\frac{2}{27}f_{TG}^N$ (in Eq.~(\ref{eq:formfac1}) we have implicitly summed over 
the three heavy quark flavours). For their numerical values we have adopted the default assignations by micrOMEGAs. 
Notice that the factor ${\left[f_p \frac{Z}{A}
+f_n \left(1-\frac{Z}{A}\right)\right]}$ is actually a rescaling factor which is introduced for a consistent 
comparison with the experimental limits which customarily assume $f_p=f_n$~\cite{Feng:2013fyw}. This assumption is 
justified in the case of the spin-0 mediator since $f_p$ and $f_n$ differ only by the contributions of up and 
down quarks which are sub-dominant with respect to the contribution from the strange quark (and then from the 
heavy quarks, see Eq.~(\ref{eq:formfac1})), which is the same for proton and neutron. In the 
following numerical estimates we will then 
automatically set $\left[f_p \frac{Z}{A}+f_n \left(1-\frac{Z}{A}\right)\right]$ $\rightarrow f_p \sim 0.3$. For spin-1 mediators one expects in general $f_p \neq f_n$ and this often 
translates into an enhancement of the cross-section and, hence, stronger limits on the model parameters. 
This can be already noticed by comparing the limits in the case of the SM-Higgs and $Z$-boson portals. Current limits exclude then low values for both the mass of the DM and the one of the mediator. These 
limits will become, of course, progressively stronger, in case of absence of signals at XENON1T and/or LZ.
Concerning the DM relic density for $m_\chi < m_S,m_t$ the DM annihilation cross-section is suppressed by 
Yukawa structure of the couplings so that the correct relic density is obtained only around the resonance 
region $m_\chi \sim m_S/2$. This corresponds to the wide region between the two lines in Fig.~\ref{fig:Sportal}. 
This region is of course determined by $\lambda_\chi^S \times c_{S} $ and is then the same as shown in the 
left and right plots of Fig. \ref{fig:Sportal}.
The difference between these two figures is the disappearance of the region corresponding to the $SS$ 
final state (see right plot) where, annihilation cross-section is proportional to $(\lambda_\chi^S)^4$ 
(see Eq.~(\ref{Eq:chichiss})). 
This channel is then not sufficient to avoid an over-density of the Universe for
$\lambda_\chi^S=0.25$. In the middle plot, the pole region is enlarged, to the point of covering almost all the 
parameter space, even joining the $SS$ final state at $m_S \simeq 1$ TeV. For higher DM masses, instead, the correct 
relic density is achieved also far from s-channel resonances through either the $\bar{t} t$ channel or the $SS$ channel, 
whether kinematically open. In such a case, the following analytical estimates, through the conventional velocity 
expansion, of the DM annihilation cross-section, can be obtained in the case where $m_t < m_\chi < m_S$:
\begin{align}
\langle \sigma v \rangle (\chi \chi \rightarrow \bar{t} t)
\approx \frac{3}{16 \pi} {(\lambda_\chi^S)}^2 c_S^2 
\frac{m_t^2}{v_h^2}\frac{1}{m_S^2}  \approx 3.4 \times 10^{-25}{\mbox{cm}}^3 {\mbox{s}}^{-1}{(\lambda_\chi^S)}^2 c_S^2  \times  {\left(\frac{1~\mbox{TeV}}{m_S}\right)}^2~,
\end{align}
and in the case where $m_t < m_S < m_\chi$:
\begin{align}
\langle \sigma v \rangle (\chi \chi \rightarrow \bar{t} t) \approx \frac{3}{64 \pi} {(\lambda_\chi^S)}^2 c_S^2 
 \frac{m_t^2}{v_h^2}\frac{m_S^2}{m_\chi^4} \approx 10^{-26}{\mbox{cm}}^3 {\mbox{s}}^{-1}
 {(\lambda_\chi^S)}^2 c_S^2  \times {\left(\frac{2~\mbox{TeV}}{m_\chi}\right)}^4 {\left(\frac{m_S}{1.5~\mbox{TeV}}\right)}^2~.
\end{align}
For $ m_S < m_\chi$ :
\begin{align} 
\langle \sigma v \rangle (\chi \chi \rightarrow S S) 
\approx \frac{{(\lambda_\chi^S)}^4}{64 \pi m_\chi^2} \approx 5.8 \times 10^{-26}{\mbox{cm}}^3 {\mbox{s}}^{-1}{(\lambda_\chi^S)}^4  \times {\left(\frac{1~\mbox{TeV}}{m_\chi}\right)}^2.
\label{Eq:chichiss}
\end{align}
As evident that both the $\bar{t} t$ and $SS$ cross-sections are s-wave dominated and thus, velocity independent. 
As a consequence residual annihilation would occur at present times which can be probed by DM ID strategies. 
Similar to the case of the SM-Higgs portal, DD limits are much more competitive with respect 
to the ones from ID, hence the latter have not been explicitly exhibited on Fig.~\ref{fig:Sportal}. 
We also notice that the dominant contribution of the annihilation cross-section into $SS$ depends only on the 
$\lambda_\chi^S$ coupling; as a consequence the scalar self-coupling $\lambda_S$ does not play a relevant role 
for DM phenomenology~\footnote{This statement is strictly valid in the case, considered here, of $\lambda_\chi^S=1$. 
In the case $\lambda_\chi^S \ll \lambda_S$ the contribution of the trilinear couplings $\lambda_S$ to the 
annihilation cross-section into $SS$ might be sizable and event dominant. However, we expect in such a case, 
since the annihilation cross-section would scale at least as $(\lambda_\chi^{S})^2$, the DM to be in general 
overabundant.}.

\subsection{Fermionic Dark Matter}

\begin{center}
\begin{figure}[h!]
  \begin{minipage}[l]{0.33\textwidth}
\includegraphics[width=\linewidth]{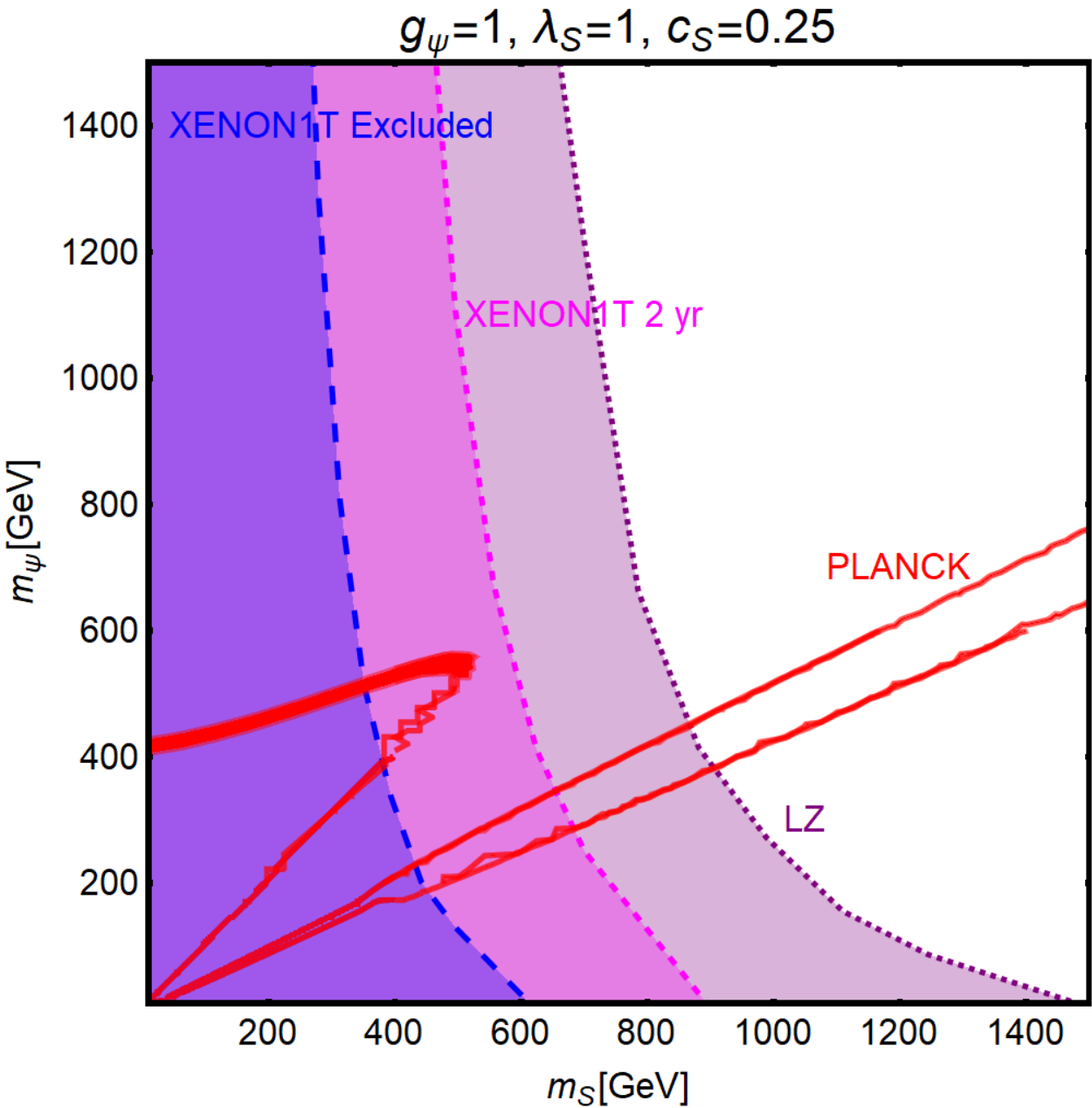}
   \end{minipage}\hfill
   \begin{minipage}[c]{0.323\textwidth}   
\includegraphics[width=\linewidth]{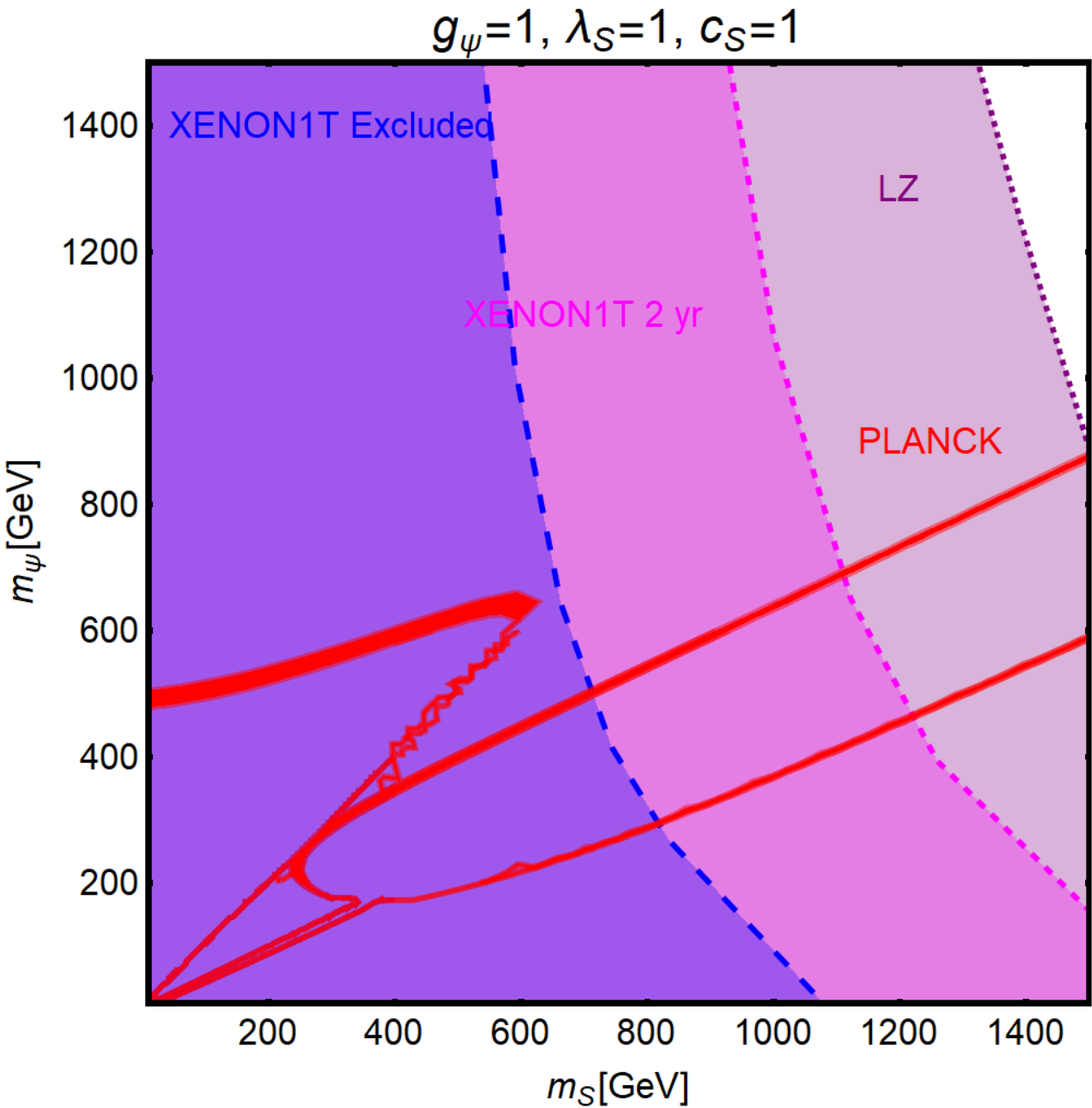}
   \end{minipage}
      \begin{minipage}[r]{0.33\textwidth}   
\includegraphics[width=\linewidth]{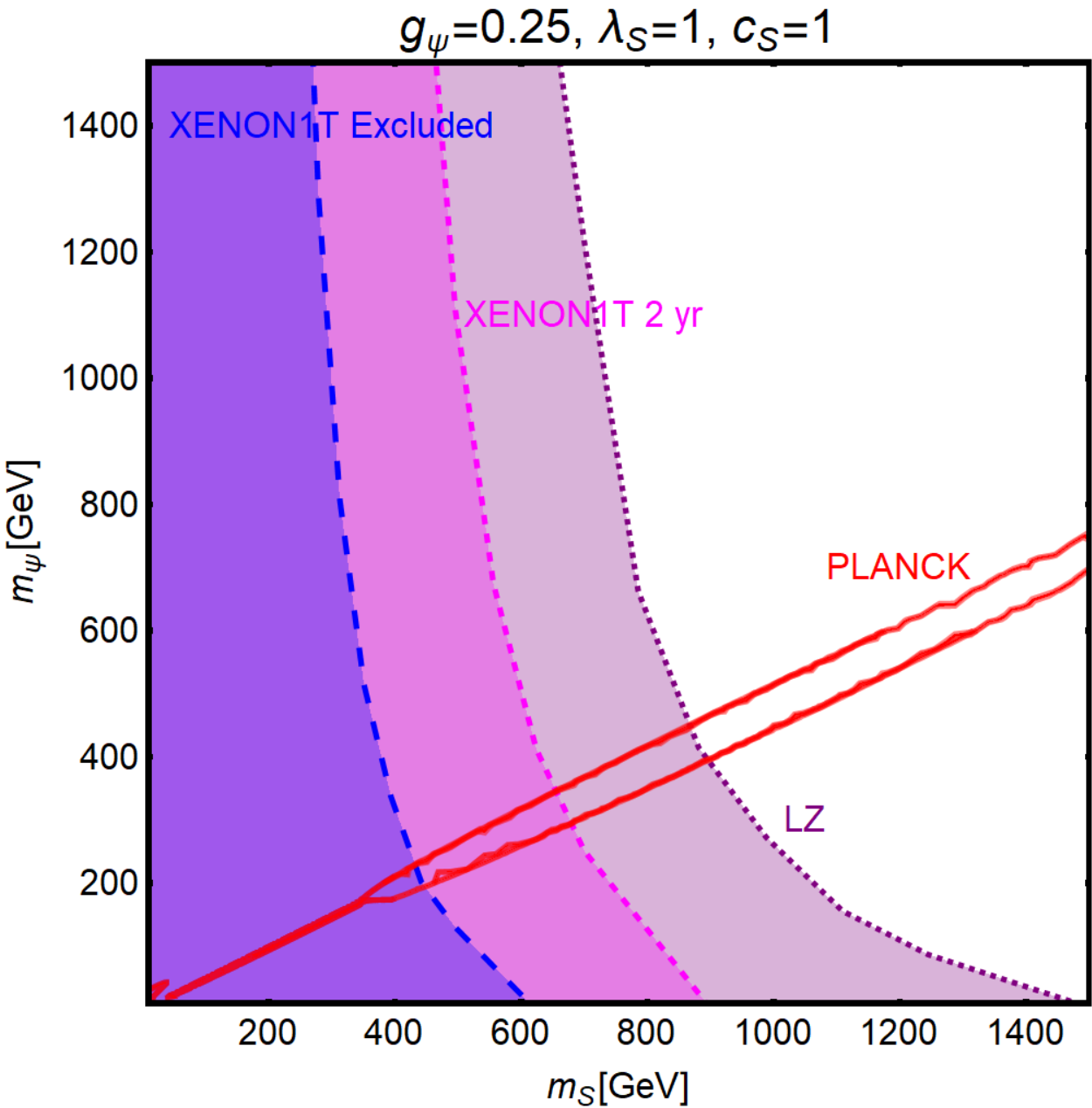}
   \end{minipage}
   \caption{The same as Fig.~\ref{fig:Sportal} but for a Dirac fermion DM, i.e., replacing $\lambda^S_\chi,
\,m_\chi$ by $g_\psi,\,m_\psi$, respectively.}
\label{fig:Fportal}
\end{figure}
\end{center}

The interaction of a fermionic DM and a scalar s-channel mediator can be described by the following 
phenomenological Lagrangian:
\begin{equation}
\mathcal{L}=-\xi g_\psi \bar{\psi} \psi S-\frac{c_S}{\sqrt{2}} \frac{m_f}{v_h} \bar{f} f S+\mathcal{L}_S,
\end{equation}
where $\mathcal{L}_S$ is defined in Eq.~(\ref{eq:scalar_self}). Contrary to the case of a scalar DM, 
the operator $\bar{\psi} \psi S$ is of dimension 4, so that $g_\psi$ is already a dimensionless parameter. 
Note that similar to Eq.~(\ref{eq:bsms0sdm1}) we have parametrized $g_\psi$ to contain all the information 
of the $\bar{\psi} \psi S$ vertex including a normalization factor. One could think that an eventual VEV of the 
scalar mediator $S$ can be the origin of the DM mass, so that $g_\psi \sim m_\psi/v_S$, with $v_S$ being the 
VEV of $S$~\cite{Kim:2008pp,Kim:2009ke,Kim:2016csm}. We won't make this assumption in this work and regard 
$g_\psi$ as a generic  dimensionless constant. The main results of our analysis have been summarized in Fig.~\ref{fig:Fportal}. We have once again considered 
the DM and scalar masses as free parameters and an analogous assignation of the couplings as in the previous subsection. 

%

The results shown in the figure can be described analytically as follows: the DD of the DM is again 
principally determined by SI interactions whose cross-section is given by:
\begin{align}
\sigma^{\rm SI}_{\psi p}=\frac{\mu_{\psi p}^2}{\pi}g_\psi^2 c_S^2 \frac{m_p^2}{v_h^2}f_N^2 \frac{1}{m_S^4}\approx 
\frac{1}{\pi}g_\psi^2 c_S^2 \frac{m_p^{4}}{v_h^2}f_N^2\frac{1}{m_S^4}\approx 2.9 \times 10^{-45}\, {\mbox{cm}}^{2} g_\psi^2 c_S^2 {\left(\frac{500~\mbox{GeV}}{m_S}\right)}^4~,
\end{align}
where $\mu_{\psi p}=m_\psi m_p/(m_\psi+m_p)$ denotes reduced mass
of the associated WIMP-proton system. As evidenced from Fig.~\ref{fig:Fportal}, DM masses even above the TeV scale, are excluded by current DD limits for $m_S \lesssim 400-500\,\mbox{GeV}$. Values below the TeV scale for both the DM and mediator 
masses will be excluded in the absence of signals from the next generation  experiments. The correct DM relic density can be achieved, without relying on s-channel resonances, only when at 
least one between the $\bar{t} t$ and $SS$ final states is kinematically accessible. In such a case the 
DM pair annihilation cross-section can be approximated in the case  where $m_t < m_\psi < m_S$ as:
%
\begin{align}
\langle \sigma v \rangle(\bar{\psi} \psi \rightarrow \bar{t} t)~
\approx \frac{3}{4\pi}g_\psi^2 c_S^2 \frac{m_t^2}{v_h^2}\frac{m_\psi^2}{m_S^4}v^2 \approx 1.5 \times 10^{-26} {\mbox{cm}^3} {\mbox{s}}^{-1} g_\psi^2 c_S^2  {\left(\frac{m_\psi}{300~\mbox{GeV}}\right)}^2 {\left(\frac{1~\mbox{TeV}}{m_S}\right)}^4 ~, 
\end{align}
and in the case where $m_S, m_t < m_\psi $ as:
\begin{align}
 \langle \sigma v \rangle(\bar{\psi} \psi \rightarrow \bar{t} t)  & \approx \frac{3}{64\pi}g_\psi^2 c_S^2 
\frac{m_t^2}{v_h^2}\frac{1}{m_\psi^2}v^2 \approx 2.8 \times 10^{-26} {\mbox{cm}^3} {\mbox{s}}^{-1} g_\psi^2 
c_S^2 {\left(\frac{600~\mbox{GeV}}{m_\psi}\right)}^2 ,\nonumber\\
 \langle \sigma v \rangle(\bar{\psi} \psi \rightarrow S S)& \approx 
 \frac{3}{64\pi}g_\psi^4 \frac{1}{m_\psi^2}v^2 \approx 2.0 \times 10^{-26} {\mbox{cm}^3} {\mbox{s}}^{-1} 
g_\psi^4 {\left(\frac{1~\mbox{TeV}}{m_\psi}\right)}^2  ~.
\end{align}
Here $v^2 \sim 0.23$. We notice again that in the limit $m_\psi \gg m_S$, the scalar self-coupling $\lambda_S$ does 
not influence the DM relic density. The dependence on the couplings between the three plots of Fig.~\ref{fig:Fportal} is the same as of the scalar case.
However contrary to the case of scalar DM, all the annihilation channels are now velocity suppressed, hence cannot account 
for a sizable ID signals.

\subsection{Vector Dark Matter}

For the description of the vectorial DM case we consider the following
Lagrangian:
\bea
\label{eq:bsmscap1}
\mathcal{L}=\frac{1}{2} m_V \eta_V^S V^\mu V_\mu S+ \frac{1}{8}{(\eta_V^{S})}^2 V^\mu V_\mu S S - \frac{c_S}{\sqrt{2}}
\frac{m_f}{v_h}S \bar{f} f+\mathcal{L}_S,
\eea
which is inspired by the construction proposed in Refs.~\cite{Gross:2015cwa,Arcadi:2016kmk}. Note that the
first three terms of Eq.~(\ref{eq:bsmscap1}) appear after the spontaneous symmetry breaking, once the portal 
field is expanded as $(S+v_S)/\sqrt{2}$ with $v_S$ as the concerned VEV. The quantity $m_V$ is expressed 
as $\eta^S_V v_S/2$. A similar construction is also possible from a gauge invariant $D^\mu {\bf S} {(D_\mu {\bf S})}^*$ operator
for a complex scalar field ${\bf S}$ with $D_\mu=\partial_\mu-i \frac{1}{2} \eta^S_V V_\mu$. 
However, in this scenario the third term of Eq.~(\ref{eq:bsmscap1}) would require new BSM charges
for the SM fermions.


\begin{center}
\begin{figure}[h!]
  \begin{minipage}[l]{0.33\textwidth}
\includegraphics[width=\linewidth]{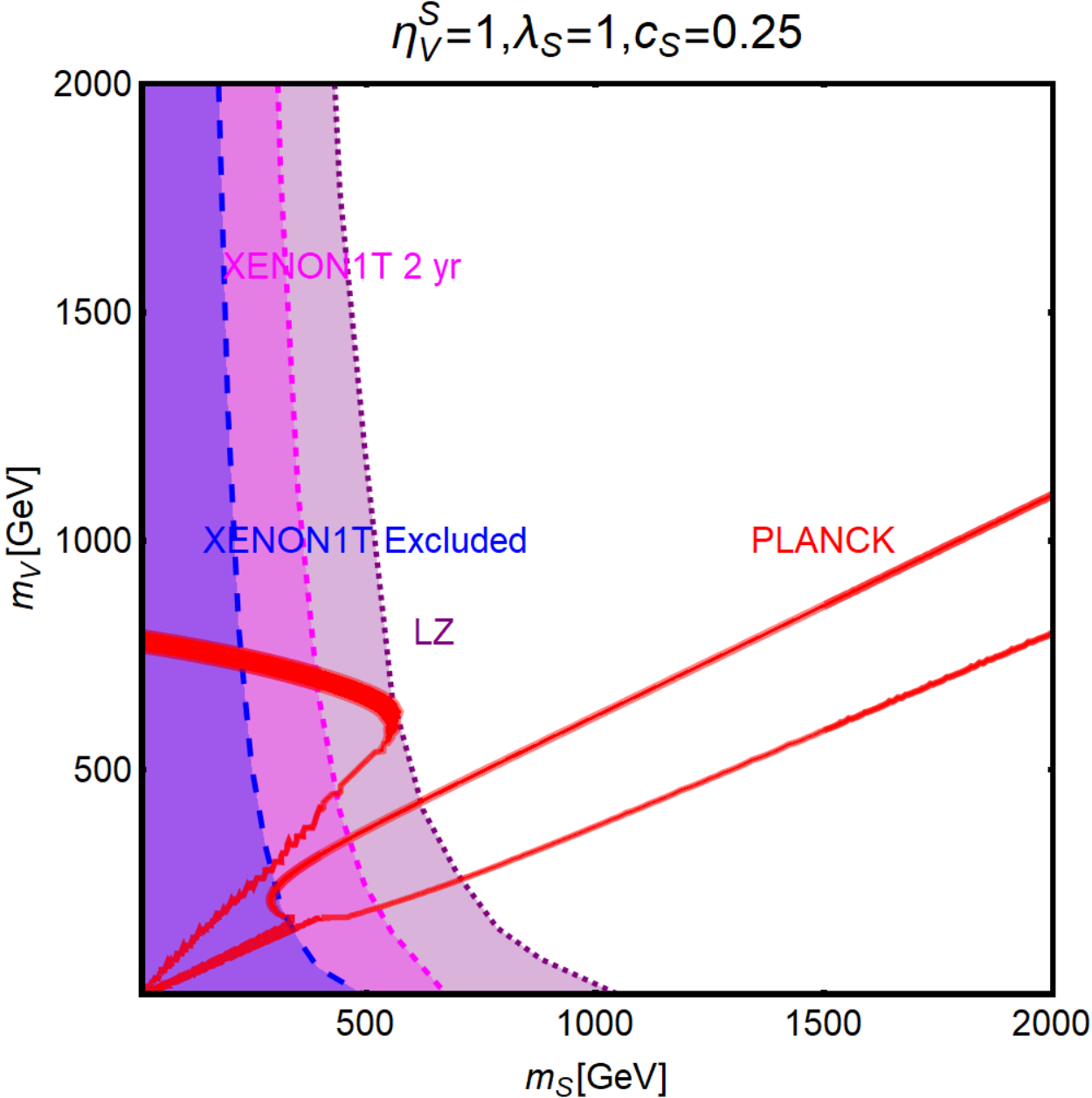}
   \end{minipage}\hfill
   \begin{minipage}[c]{0.323\textwidth}   
\includegraphics[width=\linewidth]{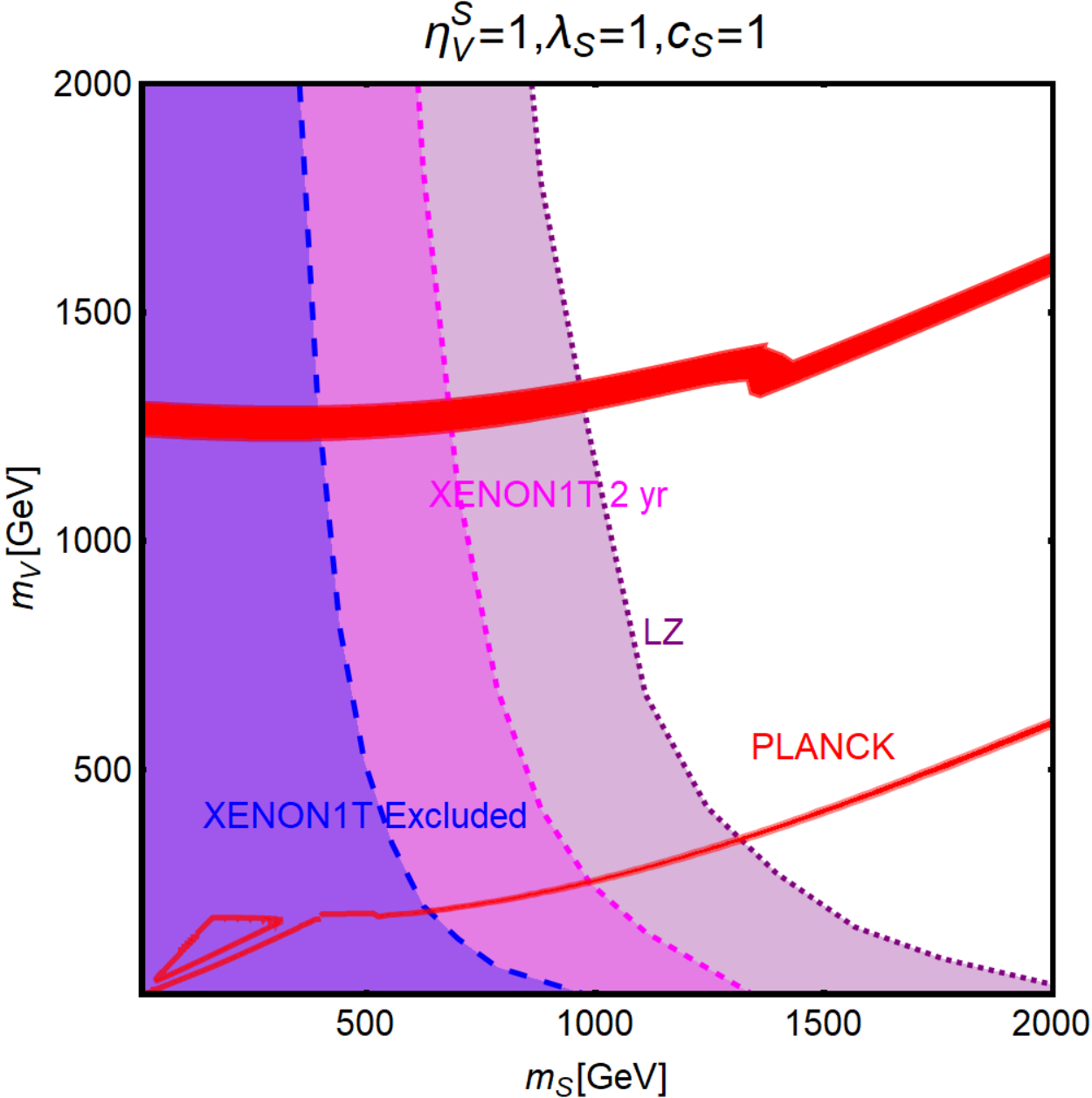}
   \end{minipage}
      \begin{minipage}[r]{0.33\textwidth}   
\includegraphics[width=\linewidth]{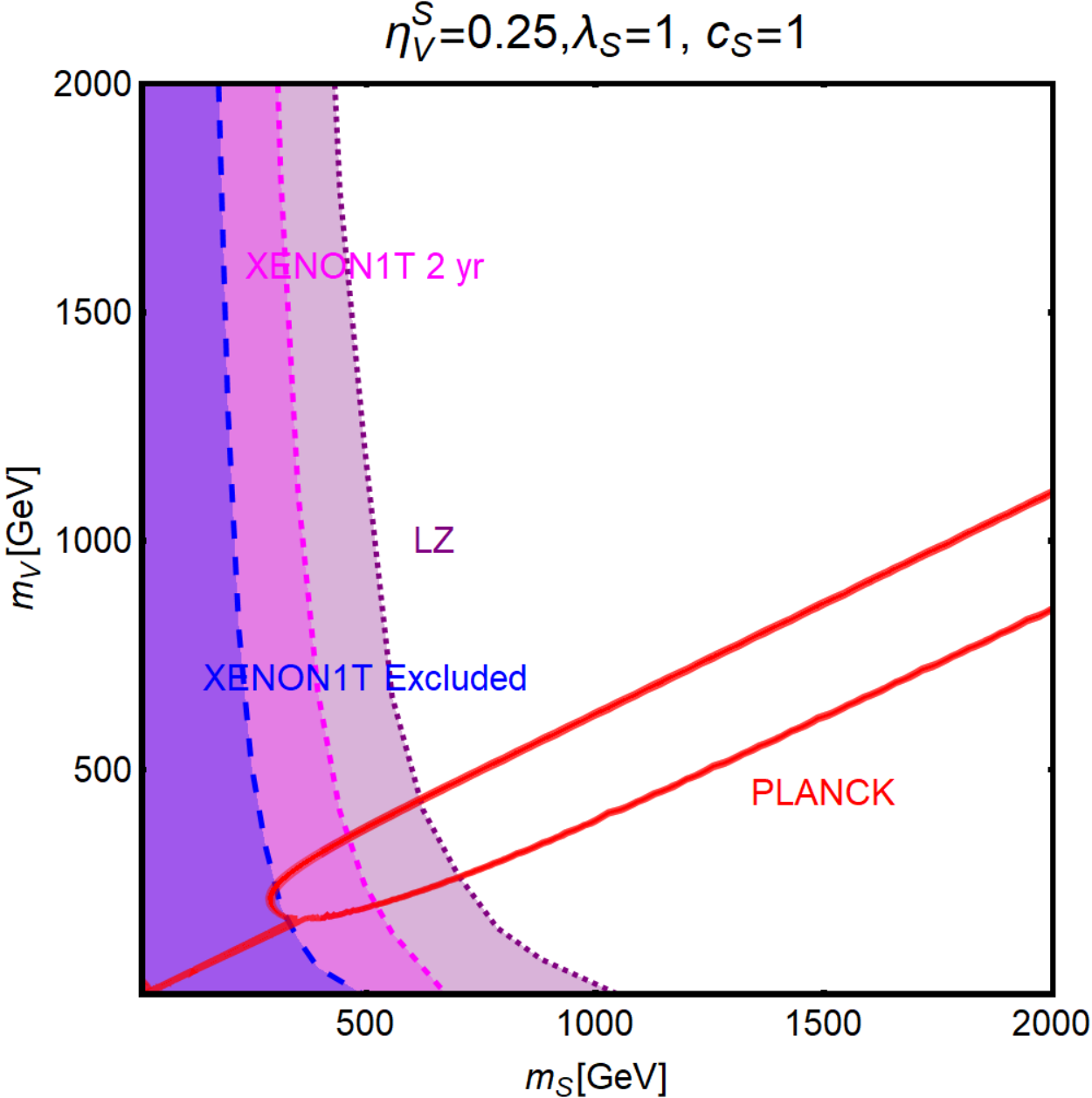}
   \end{minipage}
   \caption{The same as Fig.~\ref{fig:Sportal} for a vector DM with scalar mediator, i.e., 
trading $\lambda^S_\chi,\,m_\chi$ with $\eta^S_V,\,m_V$, respectively.}
\label{fig:Vportal}
\end{figure}
\end{center}


This scenario has been analyzed with the same procedure as the scalar and fermionic DM cases. The results, 
reported in Fig.~\ref{fig:Vportal} appear not to be very different from what obtained in the case of 
a scalar DM in Fig.~\ref{fig:Sportal}. This can be explained by the fact that a massive vectorial DM can be viewed 
as three scalar degrees of freedom. The DM scattering rate on protons and its most relevant annihilation 
channels are described by the following analytical expressions:
%
\bea
\sigma_{Vp}^{\rm SI}=\frac{\mu_{Vp}^2}{4\pi} {(\eta_V^S)}^2 c_S^2 \frac{m_p^2}{v_h^2}f_p^2 \frac{1}{m_S^4} \approx 8.2 \times 10^{-45}{\mbox{cm}}^2~{(\eta_V^S)}^2 c_S^2 {\left(\frac{1~\mbox{TeV}}{m_S}\right)}^4.
\eea

The parameter $\mu_{V p}$ $=m_V m_p$/$(m_V+m_p)$ as usual represents reduced mass
of the relevant WIMP-proton system and in the case where $m_t <m_V < m_S$
%
\begin{align}
\langle \sigma v \rangle (V V \rightarrow \bar{t} t) & \approx
\frac{1}{4\pi} {(\eta_{V}^S)}^2 c_S^2 \frac{m_t^2}{v_h^2}\frac{m_V^2}{m_S^4} \nonumber \\ & \approx 4.1 \times 10^{-26} {\mbox{cm}^3} {\mbox{s}}^{-1} {(\eta_V^S)}^2 c_S^2  {\left(\frac{m_V}{300~\mbox{GeV}}\right)}^2 
{\left(\frac{1~\mbox{TeV}}{m_S}\right)}^4 ~,
\end{align}
and for $m_S<m_V$
\begin{align}
\langle \sigma v \rangle (V V \rightarrow \bar{t} t) \approx \frac{1}{64\pi}{(\eta_V^S)}^2 c_S^2 \frac{m_t^2}{v_h^2}\frac{1}{m_V^2} \approx 2.8 \times 10^{-26} {\mbox{cm}^3} {\mbox{s}}^{-1} {(\eta_V^S)}^2 c_S^2  {\left(\frac{1~\mbox{TeV}}{m_V}\right)}^2 ~,
\end{align}
and
%
\begin{align}
\langle \sigma v \rangle(V V \rightarrow S S) \approx \frac{11}{2304\pi}{(\eta_V^S)}^4 \frac{1}{m_V^2}  \approx 1.7 \times 10^{-26} {\mbox{cm}^3} {\mbox{s}}^{-1} {(\eta_V^S)}^4 {\left(\frac{1~\mbox{TeV}}{m_V}\right)}^2.
\end{align}

\section{Conclusion}
In the light of extensive programme of direct DM searches, we assessed the status of the WIMP
paradigm in the context of simplified models, accounting for
the current and projected limits. In the minimal simplified models, the particle spectrum of the SM
should be complemented by just a new state, i.e., the DM
candidate, since portal interactions can be mediated either by
the SM-Higgs or by the $Z$-boson, although in the last case a theoretically consistent construction is more contrived. In the
case of SM-Higgs portal, for all the DM spin assignations, SI
interactions with nucleons are induced. The consequent very
strong limits, due to the light mediator, are incompatible with
the thermal relic density ad exception of DM masses above
the TeV scale or the “pole”, i.e., $m_{\rm DM} \sim  m_h/2$, region. This
last scenario would nevertheless be ruled out in the absence
of signals at XENON1T (assuming a 2 years of exposure
time) and LZ. In the $Z$-portal scenario current limits on the
SI cross-section already exclude the pole region. These strong
limits can nevertheless be partially overcome in the case where fermionic DM possesses only axial couplings with the $Z$ boson, as
naturally realized in the case of a Majorana fermion DM. In this particular case, the thermal DM with mass of a few hundreds GeV would remain viable even in the absence of signals at the next generation detectors.

The SM-Higgs and $Z$-portal setups are easily extended to
the cases of BSM scalar mediator. Despite of the different velocity dependencies of the annihilation
cross-sections, the regions of the correct DM relic
density are then mostly determined by Yukawa structure of
the couplings between the mediator and the SM fermions.
The correct DM relic density is indeed obtained, far from the
resonance regions, only when the $\bar{t}t$ and/or $SS$ annihilation
channels are kinematically open. Given
the several free parameters, for clarity of the picture, we have
focused our investigation on the masses of the new particle states and fixed the couplings to be close to $\mathcal{O}(1)$. The current limits
still allow masses of a few hundreds GeV for both the DM
and the mediator while XENON1T, in the absence of signals
after 2 years of exposure, will exclude mediator masses up to
approximately 1 TeV and DM masses up to a few TeV.
As a consequence, in this setup, the thermal DM is achieved
in the most part of the parameter space, in the pole region which requires particular fine tuning
because of the typical small decay width of the scalar
mediator.

In conclusion, simplified models are quite appealing and extremely predictive as WIMP constructions
as only a few number of extra degrees of freedom need to be invoked and because of the posibility to confront these models with existing constraints from direct detection. However, we have shown that upcomming constraints from future experiments will completely exclude or push the viable values of some parameters of the most simplified model in a corner of the parameter space, resulting in the need to introduce extra degrees of freedom to evade the future bounds from DD experiments. As new degrees of freedom have to be introduced, the simplicity of such constructions might become questionable and one would need more elaborated and motivated models to explain the presence of Dark Matter in our universe in the WIMP paradigm, while facing all the current experimental constraints and theoretical consistency of these models.

%% file: parts/chernsimons.tex
\label{ch:CS}

\section{Introduction}

As discussed in Chapter~\ref{sec:simpmodels}, in the simplest WIMP realization, a Dark Matter candidate is introduced, typically a scalar of fermion, and is assumed to be singlet 
under the SM gauge group, featuring interactions with the SM states mediated by the $Z$ or the SM-Higgs 
boson. Similar theoretical frameworks have also been addressed considering a  vectorial DM~\cite{Hambye:2008bq,DiazCruz:2010dc,Mizukoshi:2010ky,Bhattacharya:2011tr,Farzan:2012hh,
Baek:2012se,Carone:2013wla,Chen:2014cbt,Graham:2015rva,Chen:2015dea,DiFranzo:2015nli,Bambhaniya:2016cpr,
Barman:2017yzr}. 

Unfortunately, such simple models, except for a very few exceptions, are critically challenged by the existing and expected upcoming DM searches from the direct, indirect and collider probes. However, to ensure elucidate predictions, it is nevertheless necessary to investigate 
theoretical competence of these simplified setups~\cite{Kahlhoefer:2015bea,Englert:2016joy,Bell:2016uhg,Goncalves:2016iyg,Bell:2016ekl}, i.e., for example, whether they have consistent unitarity behaviours, reasonable Ultra-Violet (UV) completion and 
how they can be embedded into unified theory frameworks.

In the simplified model framework, the easiest solution to account for the DM stability is to assume an additional discrete symmetry. From the Standard Model, we know that the typical structure of such global symmetries is intimately connected to gauge symmetries as in some cases, global symmetries can be remnants of spontaneously broken gauge symmetries for instance. In such cases, the Dark Matter stability could naturally emerge as a consequence of a more complex broken gauge structure which would naturally embed supplementary degrees of freedom that could mediate interactions between the dark sector and the SM particle content. 
The case of spin-1 mediators deserves serious attentions as phenomenological 
study of these frameworks reveals intricate complementary aspects of DM searches with 
relevant collider observations (see Ref.~\cite{Arcadi:2017kky} for a thorough discussion). An intriguing origin of such spin-1 mediator(s)
can arise as gauge boson(s) of some beyond the SM (BSM) gauge group(s),
Abelian or non-Abelian, that simultaneously assigns non-zero
gauge charge(s) also for the DM candidate. A spin-1 mediator,
maintaining gauge invariance and renormalizability,
can couple to the SM Electro-Weak (EW) gauge boson, and thus, subsequently to other SM particles, in a few different ways,
\footnote{A BSM spin-1 mediator can couple
to the SM EW gauge bosons also via the well-known spontaneous symmetry breaking and Higgs mechanism, provided that the SM-Higgs
doublet has non-zero charges under the BSM gauge groups. Spontaneous symmetry breaking in the BSM sector is triggered 
with new SM singlet scalar which may or may not mix
with the SM-Higgs. We do not consider this possibility in our analysis.} e.g., via a kinetic mixing\footnote{A kinetic mixing, from the principle of gauge invariance, is allowed only between the vector bosons of Abelian groups.}
\cite{Holdom:1985ag,delAguila:1988jz,delAguila:1995rb,Foot:1991kb} or using a Chern-Simons (CS) interaction~\cite{Dudas:2009uq,Mambrini:2009ad,Dudas:2012pb,Dudas:2013sia}. The former can either appear naturally
in Lagrangian preserving the gauge invariance and renormalizability of the SM \cite{Holdom:1985ag,delAguila:1995rb,Foot:1991kb} or can be generated after integrating
out the heavy fermionic degrees of freedom \cite{Holdom:1985ag,delAguila:1988jz} charged under both the SM and BSM gauge groups. The latter can also 
arise in an analogous way after integrating out such heavy degrees of freedom,
as extensively studied in Ref.~\cite{Anastasopoulos:2006cz}. The presence of these new fermionic degrees
of freedom, if chiral under some representations, introduces new challenges to construct an anomaly free model framework~\footnote{One can always 
consider these new fermions to transform vector-like with respect to the SM gauge groups such that no new chiral anomalies appear in the SM.}.
This goal is customarily achieved by arranging anomaly cancellation in the chiral sectors of the theory, by assigning specific couplings/charges
for the involved particle species, e.g., by considering distinct couplings between the SM chiral fermions, 
and possibly also the DM, with the BSM spin-1 mediator as discussed in Ref.~\cite{Arcadi:2017atc} in the 
context of unified theories. Some classes of anomalies, like the triangle ones
involving Abelian, non-Abelian or a mixture of the two gauge groups can, alternatively be cured through the 
Green-Schwarz mechanism~\cite{Green:1984sg,Green:1984ed}.

In this chapter, we consider specific frameworks that are inspired by the Green-Schwarz mechanism~(see also Refs.~\cite{Dudas:2009uq,Dudas:2012pb}), i.e., the connection between the DM and the $Z'$ is mediated by a CS interaction.
The vectorial DM can be identified as the cosmologically stable gauge boson of a new BSM symmetry group while the $Z'$ can arise from another
BSM Abelian or non-Abelian gauge groups. Confining within the framework of Abelian theory, we will explore two possibilities of how a $Z'$
can coordinate with the SM:
(1) gauge invariant renormalizable
kinetic mixing between the field strength of $Z'$ with $B_{\mu\nu}$, the SM hypercharge field strength
and, (2) a second CS interaction involving $Z'$
and the SM hypercharge vector $B$.
For both these scenarios we extensively investigate the impact of measured relic density \cite{Ade:2015xua} as well as the existing and anticipated sensitivity reaches
from Direct Detection (DD) and Indirect Detection (ID) experiments
on the associated model parameter spaces.
We also explore relevant theoretical constraints like EW Precision Tests (EWPTs), UV completion etc. for these setups. Finally, 
for completeness, we also discuss the possible pertinent collider aspects of these models such as 
invisible $Z$-decay width, mono-{\bf X}, dijets or dilepton searches.

\section{Chern-Simons couplings}
In this section we briefly review some relevant aspects of anomalies in gauge theories and the role of Chern-Simons terms in anomaly cancellation mechanisms. Sec.~\ref{sec:UVcompletion} and~\ref{sec:appendixC} are devoted to the construction of a specific framework giving rise to the Chern-Simons couplings considered in this chapter. 
\subsection{Anomalies in gauge theories}
\begin{center}
\begin{figure}[t!]
  \begin{minipage}[l]{0.49\textwidth}
  \centering
  \hspace{0.7cm}
\includegraphics[width=0.6\linewidth]{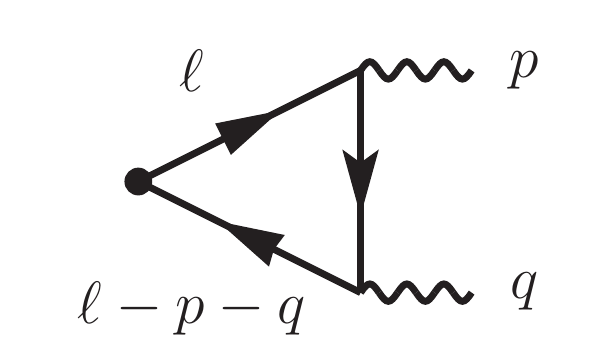}
   \end{minipage}\hfill
   \begin{minipage}[c]{0.49\textwidth}
   \centering  
\includegraphics[width=0.6\linewidth]{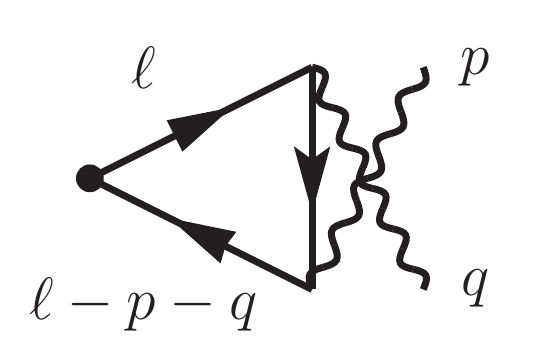}
  \hspace{0.7cm}
   \end{minipage}
   \caption{Diagrams generating potential divergences of the vector and axial currents.}
\label{fig:anomalies}
\end{figure}
\end{center}

Anomalies are non-conservation of a symmetry of the classical action that can arise from quantum corrections in theories comprising chiral fermionic content. As gauge invariance is an essential ingredient of the Standard Model by ensuring its renormalizability and unitarity, the presence of anomalies could have severe consequences and jeopardize the gauge invariance, rendering the theory completely inconsistent. In order to understand how anomalies can arise in gauge theories, consider the following QED Lagrangian:
\begin{equation}
\mathcal{L}\supset -\dfrac{1}{4}F^{\mu \nu}F_{\mu \nu} + \bar{\psi}(i \cancel{\partial}-m)\psi-e\bar{\psi}\cancel{A}\psi~.
\end{equation}
This Lagrangian is invariant under a global \textit{vector} $U(1)_V$ and non-invariant under a global \textit{axial} $U(1)_A$ transformations:
\begin{equation}
\psi \rightarrow e^{i\alpha} \psi~,\qquad \psi \rightarrow e^{i\beta \gamma_5} \psi~.
\end{equation}
We can associate a conserved Noether current $J_V^\mu$ to the vector symmetry
\begin{equation}
J_V^\mu=\bar{\psi}\gamma^\mu \psi~, \qquad \partial_\mu J_V^\mu=0~.
\end{equation}
Axial symmetry is not a symmetry of the QED Lagrangian but is restored in the limit where the mass is sent to $0$ as only a mass term involves two different chiralities. The corresponding current $J_A^\mu$ can be written as:
\begin{equation}
J_A^\mu=\bar{\psi}\gamma^\mu \gamma_5 \psi~, \qquad \partial_\mu J_A^\mu=2im\bar{\psi}\gamma_5 \psi~.
\end{equation}
However as discussed further on, even in the limit where $m \rightarrow 0$, divergences of the axial and vector current can be generated from loop diagrams involving three external gauge bosons and a triangular loop of fermions. For instance consider the 3-point correlation function, as depicted in Fig.~\ref{fig:anomalies}, involving a loop of fermions in presence of background fields:
\begin{equation}
\Gamma^{\mu \alpha \beta}(p,q)=\la J_A^\mu J_V^\alpha J_V^\beta \ra 
\end{equation}
The amplitude can be written as a sum of two terms corresponding to the diagrams represented in Fig.~\ref{fig:anomalies} as:
\begin{align}
\Gamma_{\mu \alpha \beta}(p,q)=&e^2 \int \dfrac{\diff^4 \ell}{(2 \pi)^4}\text{Tr}\Big[ \dfrac{i}{\cancel{\ell}-m+i\epsilon} \gamma_\mu \gamma_5  \dfrac{i}{\cancel{\ell}-\cancel{p}-\cancel{q}-m+i\epsilon}\gamma_\beta \dfrac{i}{\cancel{\ell}-\cancel{p}-m+i\epsilon}\gamma_\alpha  \Big] \nonumber \\&+\left( \begin{array}{c}
p \leftrightarrow q \\ 
\alpha \leftrightarrow \beta
\end{array} \right) ~.
\end{align}
In the limit of vanishing external momenta and mass the amplitude reads:
\begin{equation}
\Gamma^{\mu \alpha \beta}(p,q) \propto \int_0^\Lambda \dfrac{\ell^3 \diff \ell }{\ell^3}\gamma^\mu \gamma_5 \gamma^\alpha \gamma^\beta \propto \Lambda
\label{eq:anomalousamplitude}
\end{equation}
The high energy behavior of the integrand shows that each diagram present a divergence linear in the energy cutoff $\Lambda$. However, the value of linearly divergent integrals depends explicitely on the shift of the momentum to be integrated over. For instance, shifting the momentum $\ell^\mu \rightarrow \ell^\mu + a^\mu$ of a linearly divergent integral by a vector $a^\mu$ would result in a different value for the integral of a factor $\Delta^\alpha(a^\mu)$  which can be schematized as:
\begin{equation}
\Delta^\alpha(a^\mu) = \int \dfrac{\diff^4 \ell}{(2 \pi)^4}(F^\alpha[\ell+\alpha]-F^\alpha[\alpha]) \propto \dfrac{i a^\alpha}{32\pi^2}~,
\end{equation}
where $F^\alpha$ is the integrand function. Shifting the integrand would change the value of the linearly divergent integral independently of fermion masses. Therefore we can compute the amplitude in Eq.~(\ref{eq:anomalousamplitude}) by setting $m=0$ and by shifting the momentum $\ell^\mu \rightarrow \ell^\mu + \beta_1 p^\mu+\beta_2 q^\mu $ in the first diagram and shifting $\ell^\mu \rightarrow \ell^\mu + \beta_1 q^\mu+\beta_2 p^\mu $ in the second diagram, by requiring Bose symmetry. The divergences of the currents are thus given by:
\begin{align}
(p+q)_\mu \Gamma^{\mu \alpha \beta }(p,q)=\dfrac{e^2}{4\pi^2}\epsilon^{\alpha \beta \rho \sigma}(\beta_1-\beta_2)p_\rho q_\sigma\nonumber ~,\\
p_\alpha \Gamma^{\mu \alpha \beta }(p,q)=\dfrac{e^2}{4\pi^2}\epsilon^{\mu \beta \rho \sigma}(1-\beta_1+\beta_2)p_\rho q_\sigma~.
\end{align}
As a result, there is no choice of the parameters $\beta_1$ and $\beta_2$ that allows for a simultaneous conservation of the axial and vector currents. In order to preserve the gauge invariance, and since the axial symmetry is already broken by fermion masses, we choose $\beta_1-\beta_2=1$, in which case:
\begin{equation}
(p+q)_\mu \Gamma^{\mu \alpha \beta }(p,q)=\dfrac{e^2}{4\pi^2}\epsilon^{\alpha \beta \rho \sigma}p_\rho q_\sigma~, \quad \text{and} \quad p_\alpha \Gamma^{\mu \alpha \beta }(p,q)=0~,
\label{eq:divergenceaxial}
\end{equation}
showing explicitely the non-invariance under an axial transformation which is the prize to pay to restore the gauge symmetry of QED at least to the one loop level. Another way to understand the non-conservation of the axial symmetry is to look at the effect of a chiral transformation on the path integral measure in QED
\begin{equation}
\int \mathcal{D}\psi  \mathcal{D}\bar \psi  \mathcal{D}A \exp \Big[ i \int \diff^4 x  \mathcal{L}_{\rm QED}\Big]~,
\end{equation}
which becomes after the transformation $\psi \rightarrow e^{i \beta \gamma_5} \psi$:
\begin{align}
\int \mathcal{D}\psi  \mathcal{D}\bar \psi  \mathcal{D}A \exp \Big[ i \int \diff^4 x \Big( & \mathcal{L}_{\rm QED}  -J_A^\mu \partial_\mu \beta +\beta \dfrac{e^2}{16 \pi^2} \epsilon^{\mu \nu \alpha \beta} F_{\mu \nu} F_{\alpha \beta}\Big)\Big]~,
\end{align}
which gives after integrating by parts, dropping the surface terms and performing an expansion in $\beta$:
\begin{equation}
\partial_\mu J_A^\mu=-\dfrac{e^2}{16\pi^2}\epsilon^{\mu \nu \rho \sigma} F_{\mu \nu} F_{\rho \sigma}~.
\end{equation}
This formula matches the expression found in Eq.~(\ref{eq:divergenceaxial}) and it can be shown that higher orders in perturbation theory does not contribute to the axial or vector anomalies implying an exact determination at the one loop level. This unique effect is due to the very specific topology of triangle diagrams involving fermions in four dimensions. Therefore, controlling divergences of the currents at the one loop level suffices to ensure the validity of the corresponding symmetries at all orders in perturbation theory.

\subsection{Anomaly cancellation and Chern-Simons couplings}
In QED it is always possible to choose a shift vector to ensure the conservation of the vector current. However in chiral theories, the left- and right-handed fermionic components do not interact identically with gauge fields, therefore the previous statement is not always verified. For instance, in the Standard Model, fermions charged under the electroweak gauge group $U(1)_Y \times SU(2)_L$ possesses chiral couplings, thus one could expect anomalies arising from diagrams with $SU(2)_L^2 U(1)_Y $\footnote{Notation corresponding to diagrams with two $SU(2)_L$ and one $U(1)_Y$ gauge fields as external legs.} gauge fields as external legs. One can generalize anomalies to non-abelian gauge theories with the following currents:
\begin{equation}
J_\mu^a=\sum_\psi \bar{\psi}_i T^a_{ij}\gamma_\mu \psi_j~,
\end{equation}
where $T^a_{ij}$ are the group generators. In this case, the divergence of the current is generated by triangle diagrams, as depicted in Fig~\ref{fig:anomalies}, connected to non-abelian gauge fields $A^b$ and $A^c$:
\begin{equation}
\partial_\alpha J^{\alpha,a}=\sum_{\rm left, right} 
\text{Tr}[T^a_R\{T_R^b,T_R^c \}]\dfrac{g^2}{128\pi^2}\epsilon^{\mu \nu \rho \sigma} F_{\mu \nu}^b F_{\rho \sigma}^c~,
\label{eq:anomalousterm}
\end{equation}
where $R$ denotes the representation of the chiral fermion running in the loop and the sum as to be performed over all possible left- and right-handed states. It can be shown that all the possible anomalous terms $\text{Tr}[T^a_R\{T_R^b,T_R^c \}]$ in the Standard Model cancel out. For instance the lepton contribution $\ell$ in the $SU(2)_L^2 U(1)_Y$ anomalous term is cancelled by the quark contribution $q$:
\begin{equation}
Y_\ell+3Y_q=0~,
\end{equation}
where $Y_{\ell,q}$ denote the hypercharges of leptons and quarks. These fine-tuned cancellations might seems mysterious and actually set strong constraints on theories involving more complex gauge structures. Another possibility to ensure gauge invariance if these cancellations does not occur naturally in a theory, is to rely on the Green-Schwarz mechanism~\cite{Green:1984sg}. Consider a theory respecting several abelian gauge symmetries indexed by $i=1,2,3$ with corresponding gauge fields $A^\mu_{i}$. Assuming that these symmetries are realized a la Stueckelberg with axions named $a_i$. Below the symmetry breaking scales of the $U(1)$ gauge groups considered, if the charge assignment of the fermions does not allow the anomalous term in Eq.~(\ref{eq:anomalousterm}) to vanish, the set fermions is said to be \textit{anomalous}\footnote{Defined as a set of fermions for which the value of $\partial_\mu J^\mu$ computed using Eq.~(\ref{eq:anomalousterm})} and as a result the gauge variation of the effective Lagrangian generated by triangle diagrams involving those fermions would not vanish
\begin{equation}
\delta \mathcal{L}_{\rm triangle} \neq 0~.
\end{equation}
However, in this setup, one would expect effective couplings between axions and gauge fields in the form
\begin{equation}
\mathcal{L}_{\rm axion} \propto \dfrac{a_i}{v_i}\epsilon^{\mu \nu \rho \sigma} F^{\mu \nu}_j F^{\rho \sigma}_k~,
\label{eq:anomalousaxion}
\end{equation}
where $v_i$ is the breaking scale of $U(1)_i$. As the gauge variation of the axion $a_i$ under a $U(1)_i$ transformation $A^\mu_i \rightarrow A^\mu_i+\partial^\mu \alpha_i$ is $a_i \rightarrow a_i +m_i \alpha_i$, where $m_i$ is the mass of $A^\mu_i$, the gauge variation of $\mathcal{L}_{\rm axion}$ does not vanish. In order to restore the gauge symmetry and ensure the validity of the theory, one can consider additional so-called Generalized Chern-Simons (GCS)~\cite{Anastasopoulos:2006cz} terms\footnote{abusively called Chern-Simons in the following for simplicity.}:
\begin{equation}
\mathcal{L}_{\rm GCS} \propto \epsilon_{\mu \nu \rho \sigma} A_i^{\mu} A_j^{\nu} F^{\rho \sigma}~.
\end{equation}
These local terms are not gauge-invariant but the idea behind the Green-Schwarz mechanism is that the total gauge variation of the triangle, axion and GCS contributions cancels out:
\begin{equation}
\delta \Big( \mathcal{L}_{\rm triangle} +\mathcal{L}_{\rm axion} +\mathcal{L}_{\rm GCS} \Big)=0~,
\end{equation}
rendering the low-energy theory anomaly free and consistent. GCS terms can arise from string constructions or can be generated in quantum field theory, as a low energy remnant term generated by a set of heavy anomalous fermions. In the former case, one interesting point regarding GCS terms is that their effective couplings are expected to depend on the heavy fermion masses. However, gauge variations of the GCS terms do not depend on any energy scale allowing to cancel anomalies of light fermions in spite of several orders of magnitude of difference between the heavy and light fermion mass scales. More details regarding GCS terms can be found in~\cite{Anastasopoulos:2006cz} and in Sec.~\ref{sec:UVcompletion} and~\ref{appendix:CS}.

\section{Scenario-I: $Z'$-$Z$ interaction via Kinetic mixing}
\label{sec:scenario1}

In this section we study the aforesaid type-I scenario when
the "dark sector", comprised of a vectorial DM $X_\mu$ and a spin-1
vector boson $\wt V_{\mu}$, is "secluded" from the visible sector,
i.e., there exists no direct coupling between this dark sector and the SM 
fermions~\footnote{The other possibility, i.e., the dark sector has direct 
couplings with the SM fermions is reviewed 
recently in Ref.~\cite{Arcadi:2017kky}.}. These $X_{\mu}$
and $\wt V_\mu$, for example, can appear as the gauge bosons of 
some BSM $U(1)_X$ and $U(1)_V$ groups and we consider a
CS interaction to connect them together.
A bridge between the dark sector and the SM now appears via a kinetic mixing of $\wt V_{\mu\nu}$ and
$B_{\mu\nu}$, the field strengths associated with BSM $U(1)_V$
and the SM $U(1)_Y$ gauge group, respectively. A similar kinetic mixing between $X_{\mu \nu} $ and $B_{\mu \nu}$,
being renormalizable and allowed by the SM gauge invariance,
should also be included in a general Lagrangian. However, we do not consider this possibility for the stability of the DM and 
postpone further discussion in this direction till Sec.~\ref{sec:UVcompletion}. The relevant phenomenology of the said model can be described by the following low-energy effective Lagrangian:
\begin{align}
\label{eq:starting_lagrangian}
 \mathcal{L}\supset &-\dfrac{1}{4} {B}^{\mu \nu} {B}_{\mu \nu}-\dfrac{1}{4} X^{\mu \nu} X_{\mu \nu} -\dfrac{1}{4} \wt{V}^{\mu \nu} \wt{V}_{\mu \nu}  - \dfrac{\sin \delta}{2}\widetilde{V}^{\mu \nu}{B}_{\mu \nu} \nonumber\\
& + \alpha_{\rm CS} \epsilon^{\mu \nu \rho \sigma}  X_{\mu} \wt{V}_{\nu} X_{\rho \sigma} +\dfrac{m_{V}^2}{2}\wt{V}^\mu \wt{V}_\mu+\dfrac{m_{X}^2}{2}X^\mu X_\mu,
\end{align}
here $\delta$ is the kinetic mixing parameter and $\alpha_{\rm CS}$
represents the effective coupling of CS operator.
$X_{\mu\nu}$ gives the field strength of $U(1)_X$ group
and $m_V,\,m_X$ represent mass terms of the mediator and the DM.
Thus, one gets a set of four free inputs, namely,
$\delta,\, \alpha_{\rm CS}$, $m_V$ and $m_X$ whose ranges
will be tested subsequently imposing a series of theoretical
and experimental constraints. The presence of kinetic mixing in eq.~(\ref{eq:starting_lagrangian}) implies non-canonical kinetic
term for $B_{\mu\nu}$ and also for $\wt V_{\mu\nu}$. In order 
to generate diagonal kinetic terms in the physical or mass
basis one should invoke three different rotations
~\cite{Babu:1997st,Chun:2010ve,Mambrini:2010dq,Mambrini:2011dw}. 
The first 
rotation, involving then angle $\delta$, takes $B_{\mu},\,
\wt V_{\mu}$ to a basis (say $B^{\text{int}}_{\mu},\,
\wt V^{\text{int}}_{\mu}$) with diagonal kinetic terms. The
second rotation, after EW symmetry breaking (EWSB), via angle $\theta_{\wt W}$, takes this $B^{\text{int}}_{\mu}$
together with $W^3_{\mu}$ to the intermediate $A_\mu,\,Z^{\text{int}}_\mu$ basis. Finally,
the third rotation through another angle $\phi$, leaving
the massless photon aside, takes $Z^{\text{int}}_\mu,\,\wt V^{\text{int}}_\mu$
to the $Z_\mu,\,Z'_\mu$ basis where $Z_\mu,\,Z'_\mu$
are associated with the physical $Z$ and $Z'$ boson. 
In summary, the initial $B_\mu,\, W^3_\mu,\, \wt V_\mu$
basis can be related to the physical $A_\mu,\,Z_\mu,\,Z'_\mu$
basis in the following way:

\begin{equation}
\label{eq:transformation}
\begin{pmatrix}
{B}_\mu \\ 
W_{3 \mu} \\
\wt{V}_{\mu}
\end{pmatrix}=\begin{pmatrix}
{c}_{\wt W} & -{s}_{\wt W} c_\phi + t_\delta s_\phi &-{s}_{\wt W} s_\phi - t_\delta c_\phi \\ 
{s}_{\wt W} & {c}_{\wt W} c_\phi & {c}_{\wt W} s_\phi \\ 
0 & -\dfrac{s_\phi}{c_\delta} & \dfrac{c_\phi}{c_\delta}
\end{pmatrix}
\begin{pmatrix}
A_\mu \\ 
Z_\mu \\
Z^\prime_\mu
\end{pmatrix},
\end{equation}
where $t_\delta,\,c_\delta,\,c_{\wt W},\,s_{\wt W} ,\,s_\phi,\,c_\phi \equiv \tan\delta,\,\cos\delta,\, \cos\theta_{\wt W},\,\sin\theta_{\wt W},\,\sin\phi,\,\cos\phi$ with:
\begin{equation}
\tan2 \phi=\dfrac{\wt{m}_Z^2 {s}_{\wt W} \sin 2\delta}{m_V^2-\wt{m}_{Z}^2(c_\delta^2 -s^2_{\wt W} s_\delta^2)},
\end{equation}
here $s_\delta\equiv \sin\delta$. The quantities $\theta_{\wt W},\,
\wt m_Z$ do not represent the measured values of Weinberg angle
and $Z$-boson mass \cite{Olive:2016xmw} but are related to
them as will be explained later. 
The masses of $Z^\prime$ and $Z$ are written as:
\beq
\label{eq:mzzpmass}
m_{Z^\prime,Z}^2=\dfrac{1}{2}\left[ \wt{m}_Z^2(1+s^2_{\wt W} t_\delta^2)+\dfrac{m_V^2}{c_\delta^2}    \pm \sqrt{(\wt{m}_Z^2(1+s^2_{\wt W} t_\delta^2)+\dfrac{m_V^2}{c_\delta^2})^2 - \dfrac{4}{c_\delta^2}\wt{m}_Z^2 m_V^2 } \right],
\eeq
which gives $m_Z \simeq \wt m_Z$ in the experimentally favoured limit $\delta \ll 1$, along with $m_{Z'} \simeq m_V$. Notice that the transformation used in Eq.~(\ref{eq:transformation}) is valid only if one of these two conditions is met:
\begin{align}
\label{eq:masses}
 \frac{m_V^2}{\wt{m}_Z^2} &\geq 1+2 s_{\wt{W}} \tan^2 \delta +2 \sqrt{s_{\wt{W}}^2 \tan^2 \delta \left(1+s_{\wt{W}}^2 \tan^2 \delta\right)}, \nonumber\\
 \frac{m_V^2}{\wt{m}_Z^2} &\leq 1+2 s_{\wt{W}} \tan^2 \delta -2 \sqrt{s_{\wt{W}}^2 
 \tan^2 \delta \left(1+s_{\wt{W}}^2 \tan^2 \delta\right)}.
\end{align}
One should note that the transformation of Eq.~(\ref{eq:transformation})
does not change the photon coupling~\cite{Babu:1997st}, implying the following identity:
\begin{equation}
\label{eq:stwswrelation}
s_{\wt W} c_{\wt W} \wt{m}_Z^2=s_W c_W m_Z^2=\frac{\pi \alpha(m_Z)}{\sqrt{2} G_F},
\end{equation}
where $s_W,\,c_W\equiv \sin\theta_W,\,\cos\theta_W$ are associated with the measured value of Weinberg angle $\theta_W$ \cite{Olive:2016xmw}. $\alpha(m_Z)$ is the fine structure constant at the energy scale $m_Z$ and $G_F$ represents the Fermi constant \cite{Olive:2016xmw}. One should also consider the invariance of $W$-boson mass
under the transformation of Eq.~(\ref{eq:transformation}), i.e., $m_W^2=m^2_Z c_W^2=\wt{m}_Z^2 c^2_{\wt W}$ which allows us to express the $\rho$ 
parameter \cite{Olive:2016xmw} as:
\begin{equation}
\label{eq:rhoparam}
\rho = \frac{\wt{m}_Z^2 c^2_{\wt W}}{m^2_Z c^2_W},
\end{equation} 
with the experimental measured value given by $\rho-1=4^{+8}_{-4} \times 10^{-4}$~\cite{Olive:2016xmw}.
Further, from the EWPT one can consider a simple and conservative limit on 
$\delta$ as~\cite{Kumar:2006gm,Chun:2010ve}:
\begin{equation}
\label{eq:EWPT}
\delta \lesssim \arctan \left[0.4 \left( \dfrac{m_{Z'}}{\text{TeV}} \right) \right].
\end{equation}

It is now apparent that one can use Eq.~(\ref{eq:stwswrelation}),
Eq.~(\ref{eq:rhoparam}) and Eq.~(\ref{eq:EWPT})
to discard experimentally disfavoured values of the kinetic mixing parameter $\delta$.
Further constraints on $\delta$ can emerge from various other experimental observations.
The mixing among $B_\mu,\,W^3_\mu $ and $\wt V_\mu$ (see Eq.~(\ref{eq:transformation})), 
couples the SM fermions and the DM with the $Z$ boson. The latter coupling implies
an enhancement of the invisible $Z$ decay width for $2 m_X < m_Z$.  Hence, the parameter $\delta$, along with $m_V$, will receive constraints from a plethora of different collider searches like dileptons $(pp\to Z'$ $\to e^+e^-$, $\mu^+\mu^-)$ \cite{
Aaboud:2016cth,Khachatryan:2016zqb,ATLAS:2017wce}, dijets $(pp\to Z'\to \bar{q} q)$ 
\cite{Khachatryan:2015dcf,ATLAS:2015nsi,Sirunyan:2016iap,Aaboud:2017yvp}, mono-{\bf X} $(pp\to Z'+{\mathbf{X}},\, Z'\to {\rm DM~pairs})$ with ${\mathbf{X}}=q/g$
\cite{Sirunyan:2017hci}, $W$ \cite{Aaboud:2016qgg}, $Z$ \cite{Aaboud:2016qgg,CMS:2016hmx}, $\gamma$ \cite{Aaboud:2016uro,CMS:2016fnh,Aaboud:2017dor},
SM-Higgs \cite{ATLAS:2017uwx,ATLAS:2017pqx,Sirunyan:2017hnk}, etc., 
invisible $Z$ decay width~\cite{Olive:2016xmw} and a few others. The 
dijets and mono-{\bf X} searches also restrict the parameters $\alpha_{\rm CS},\,m_X$. 
Finally, the DM phenomenology, i.e., the correct relic density, DD and ID results will also put limits
on the parameters $\delta,\,\alpha_{\rm CS},\,m_V$ and $m_X$.
We note in passing that for numerical analyses we have traded the parameter 
$m_V$ with the physical mass $m_{Z'}$ using Eq.~(\ref{eq:mzzpmass}). In the mass basis, the Lagrangian relevant for our subsequent analysis is given by: 
\begin{align}
\label{eq:mass_basis_lagrangian}
 \mathcal{L}_{Z/Z',\text{SM}}&=\bar{f} \gamma^\mu \left(g_{f_L}^{Z}P_L+g_{f_R}^{Z}P_R\right) f Z_\mu
 +\bar{f} \gamma^\mu \left(g_{f_L}^{Z'}P_L+g_{f_R}^{Z'}P_R\right) f Z'_\mu + g_W^Z [[W^+ W^-Z]] \nonumber\\
& +g_W^{Z'} [[W^+ W^-Z']] +\dfrac{g_{hZZ}}{2} Z^\mu Z_\mu h + g_{hZZ'}  Z'_\mu Z^\mu h+\dfrac{g_{hZ'Z'}}{2}  Z'_\mu Z^{\prime \mu} h\nonumber\\
&- \alpha_{\rm CS}\,\frac{s_\phi}{c_\delta}\,\epsilon^{\mu \nu \rho \sigma}\, X_\mu X_{\rho \sigma} Z_\nu+ \alpha_{\rm CS}\,\frac{c_\phi}{c_\delta}\,\epsilon^{\mu \nu \rho \sigma}\, X_\mu X_{\rho \sigma} Z^{'}_\nu,
\end{align}
here $P_{L(R)}=(1\pm\gamma_5)/2$ and
%
\begin{align}
\label{eq:zzpsmfermion}
g^{Z}_{f_L}&&= g_Y Y_L (- s_W c_{\phi }+t_{\delta }s_{\phi })+g_W I_z (c_W c_\phi),\,\,\,\,
g^{Z}_{f_R}= g_Y Y_R (- s_W c_{\phi }+t_{\delta }s_{\phi }),\nonumber\\
g^{Z^\prime}_{f_L}&&= g_Y Y_L (- s_W s_{\phi }-t_{\delta }c_\phi)+g_W I_z (c_W s_\phi),\,\,\,\,
g^{Z^\prime}_{f_R}= g_Y Y_R (- s_W s_{\phi }-t_{\delta }c_\phi),
\end{align}
where $g_Y$ is the gauge coupling of $U(1)_Y$,
$I_z$ is the $3^{rd}$ component of the weak isospin, $Y_{L,R}$ are hypercharge of the left- and 
right- chiral fermions, $g_W$ is the $SU(2)_L$ gauge coupling.
\begin{align}
\label{eq:zzpww}
g^{Z}_{W}=g_W c_W c_\phi,\,\,\, \qquad
g^{Z^\prime }_{W}=g_W c_W s_\phi,
\end{align}
%
\begin{align}
\label{eq:zzph}
g_{hZZ}&=2 v_h \Big[ \frac{g_Y}{2} (- s_W c_{\phi }+t_{\delta }s_{\phi }) -\frac{g_W}{2}(c_W c_\phi) \Big]^2,\nonumber \\
g_{hZZ^\prime }&=2  v_h \Big[ \frac{g_Y}{2} (- s_W s_{\phi }-t_{\delta }c_{\phi }) -\frac{g_W}{2}(c_W s_\phi) \Big]\Big[ \frac{g_Y}{2} (- s_W c_{\phi }+t_{\delta }s_{\phi }) -\frac{g_W}{2}(c_W c_\phi) \Big],\nonumber\\
g_{hZ^\prime Z^\prime }&=2 v_h \Big[ \frac{g_Y}{2} (- s_W s_{\phi }-t_{\delta }c_{\phi }) -\frac{g_W}{2}(c_W s_\phi) \Big]^2,
\end{align}
here $v_h$ is the vacuum expectation value (VEV) of the SM-Higgs
doublet and a factor `$2$' appears in $ZZh,\,Z'Z'h$ vertices
due to the presence of identical particles.
For the 
$\alpha_{\rm CS}\,\frac{-s_\phi(c_\phi)}{c_\delta}\,\epsilon^{\mu \nu \rho \sigma}\, X_\mu X_{\rho \sigma} Z_\nu(Z^{'}_\nu)$ term, considering the following momentum assignments
$X_{\mu}(p_a) X_\nu (p_b) Z_{\rho}/Z'_\rho$ one can use
\begin{align}
\label{eq:zzpdm}
g_{X}^Z =-2 \alpha_{\rm CS}\, 
\frac{s_\phi}{c_\delta}, \qquad
g^{Z^\prime}_X =2 \alpha_{\rm CS}\,
\frac{c_\phi}{c_\delta},
\end{align}
for subsequent relevant analytical formulae like DM pair annihilation
cross-section etc. Once again the factor `$2$' appears
due to identical particles. Lastly
\begin{align}
[[W^+ W^- Z^{(')}]] \equiv i \left[W^+_{\mu\nu} W^{-\, \mu}Z^{(')\nu}-W^+_{\,\mu\nu} W^{-\,\mu} Z^{(') \nu}
+\frac{1}{2}Z^{(') \mu\nu} \left(W^{+}_\mu W^{-}_\nu-W^{+}_\nu W^{-}_\mu\right)\right].
\end{align}
With these analytical expressions now we will discuss the DM phenomenology of this framework in the next subsection.

\subsection{Dark Matter Phenomenology}

In this subsection we discuss the DM phenomenology in the light
of various constraints coming from the requirement of correct relic density and consistency with the existing and/or upcoming DD and ID results. The DM is produced in the early Universe according to the 
WIMP paradigm. We will subsequently discuss
how these observations can affect the accessibility of the chosen set-up at the present or in the near future DM detection experiments, assuming projected search sensitivities.
Nevertheless, for the sake of completeness, we will also qualitatively discuss the 
complementary limits arising
from the EWPT, collider searches, etc. 
We start our discussion in the context of accommodating correct relic density
and successively will address the restrictions coming from direct
and indirect DM searches.

\subsubsection{Relic density}
\label{sssec:rdsc1}

In the chosen framework, the DM pair can annihilate into the SM fermions $(\bar{f} f)$ and $W^+W^-$, as well as in $Zh$ and $Z'h$ final states, through s-channel exchange of the $Z/Z'$ boson, and into $ZZ,\,ZZ'$ and $Z'Z'$ final states, through t-channel exchange of a DM state. Approximate analytical expressions of the corresponding annihilation cross-sections are obtained through the expansion $\langle \sigma v \rangle \simeq a +b x^{-1}+...$ with $x=\dfrac{m_X}{T}$~\cite{Gondolo:1990dk,Jungman:1995df}, $T$
being the temperature. We remind, however, that the aforesaid approximation is not reliable over the entire parameter space~\cite{Griest1991c}, e.g., the pole regions $m_X \sim \dfrac{m_{Z,\,Z'}}{2}$. Hence, all results presented 
in this chapter are obtained through full numerical computations using the package micrOMEGAs~\cite{Belanger:2006is,Belanger:2008sj,Belanger:2013oya},
after implementing the model in FeynRules \cite{Christensen:2008py,Alloul:2013bka}.
The DM annihilation cross-sections in the possible final states are
given as:
\begin{itemize}
\item $XX \to \bar{f}f:$
\begin{align}
\label{eq:XXff}
\langle \sigma v\rangle_{\bar{f}f} &\simeq  \frac{n_c\,m_f^2}{3 \pi} \sqrt{1-\frac{m_f^2}{m_X^2}} {\left(\frac{\left(g_{f_L}^Z-g_{f_R}^Z\right) g_X^Z}{m_Z^2}+\frac{\left(g_{f_L}^{Z'}-g_{f_R}^{Z'}\right) g_X^{Z'}}{m_{Z'}^2}\right)}^2\dfrac{1}{x}\nonumber\\
&+\frac{n_c}{108 \pi}\left(\frac{(g_{X}^Z)^2 \left((g_{f_L}^Z)^2+(g_{f_R}^Z)^2\right)}{m_Z^4}+\frac{(g_{X}^{Z'})^2 \left((g_{f_L}^{Z'})^2+(g_{f_R}^{Z'})^2\right)}{m_{Z'}^4}\right. \nonumber\\
&\left. \hspace{1.6cm} + \frac{2 g_{X}^{Z'} g_{X}^{Z} \left(g_{f_L}^{Z'}g_{f_L}^{Z}+g_{f_R}^{Z'} g_{f_R}^{Z}\right)}{m_{Z'}^2 m_Z^2}\right)\dfrac{1}{x^2},
\end{align}
where we have written not only the leading $1/x$ (p-wave) term but also
the second order $1/x^2$ (d-wave) one which appears to be
the dominant one since the first one is suppressed by the square
of SM fermion mass $m^2_f$. The parameter $n_c$ denotes colour factor
with a value of $3\,(1)$ for quarks (leptons).
\item $XX\to W^+W^-:$
\begin{align}
\label{eq:XXWW}
\langle \sigma v\rangle_{W^+W^-}\simeq \frac{5}{9 \pi m_W^4}{\left(\frac{g_X^Z g_W^Z m_X}{4}-\frac{g_X^{Z'} g_W^{Z'} m_X^3}{m_{Z'}^2}\right)}^2\dfrac{1}{x^2},
~~{\rm for~} m_{W,\,Z} \ll m_X \ll m_{Z^\prime}.
\end{align}
%
\item $XX\to Zh:$
\begin{align}
\label{eq:XXZh}
\langle \sigma v\rangle_{Zh}\simeq \frac{2}{3\pi}\frac{m_X^6}{m_Z^6}{\left(\frac{g_X^Z g_{hZZ}}{4 m_X^2}-\frac{g_X^{Z'} g_{hZZ'}}{m_{Z'}^2}\right)}^2\dfrac{1}{x},
~~{\rm for~}m_Z \sim m_h \ll m_X \ll m_{Z^\prime}.
\end{align}
%
\item $XX\to Z^\prime h:$
\begin{align}
\label{eq:XXzph}
\langle \sigma v\rangle_{Z^\prime h}\simeq \frac{1}{24 \pi}\frac{m_X^2}{m_{Z'}^6}{\left(g_{hZZ'} g_X^Z+g_{hZ'Z'} g_X^{Z'}\right)}^2 \dfrac{1}{x},
~~{\rm for~}m_Z \sim m_h  \ll m_{Z^\prime} \ll m_X.
\end{align}
%
\item $XX\to ZZ:$
%
\begin{align}
\label{eq:XXZZ}
\langle \sigma v\rangle_{ZZ}\simeq\frac{8 \alpha_{\rm CS} ^4 \delta ^4 m_X^2 m_Z^4 s_W^4}{9 \pi  m_{Z'}^8},
~~{\rm for~} m_Z \ll m_X  \ll m_{Z^\prime}.
\end{align}
%
\item $XX\to Z^\prime Z^\prime:$ 
%
\begin{align}
\label{eq:XXzpzp}
\langle \sigma v\rangle_{Z^\prime Z^\prime}\simeq \frac{8 \alpha_{\rm CS} ^4 m_X^2}{9 \pi  m_{Z'}^4},
~~{\rm for~} m_Z   \ll m_{Z^\prime}  \ll m_X.
\end{align}
%
\item $XX\to Z^\prime Z:$ 
\begin{align}
\label{eq:XXzpz}
\langle \sigma v\rangle_{Z^\prime Z}\simeq \frac{16 \alpha_{\rm CS} ^4 \delta ^2 m_X^2 m_Z^2 s_W^2}{9 \pi  m_{Z'}^6},
~~{\rm for~} m_Z   \ll m_{Z^\prime}  \ll m_X.
\end{align}
\end{itemize}

As evident from the above expressions that annihilation channels originated by s-channel exchange of the $Z/Z'$ feature a suppressed annihilation cross-section, at least p-wave or even d-wave in the cases of $W^+ W^-$ and $\bar f f$ final states  (see Eq.~(\ref{eq:XXWW})
and Eq.~(\ref{eq:XXff})). For the latter a p-wave contribution is also present but it is helicity suppressed. The t-channel induced annihilations feature, instead, a s-wave cross-section. The cross-section into $ZZ$ pairs (see Eq.~(\ref{eq:XXZZ})) is anyway suppressed by an higher power of the kinetic mixing parameter $\delta$, with respect to the other annihilation channels.     

\begin{figure}[t!]
\includegraphics[width=\linewidth /2]{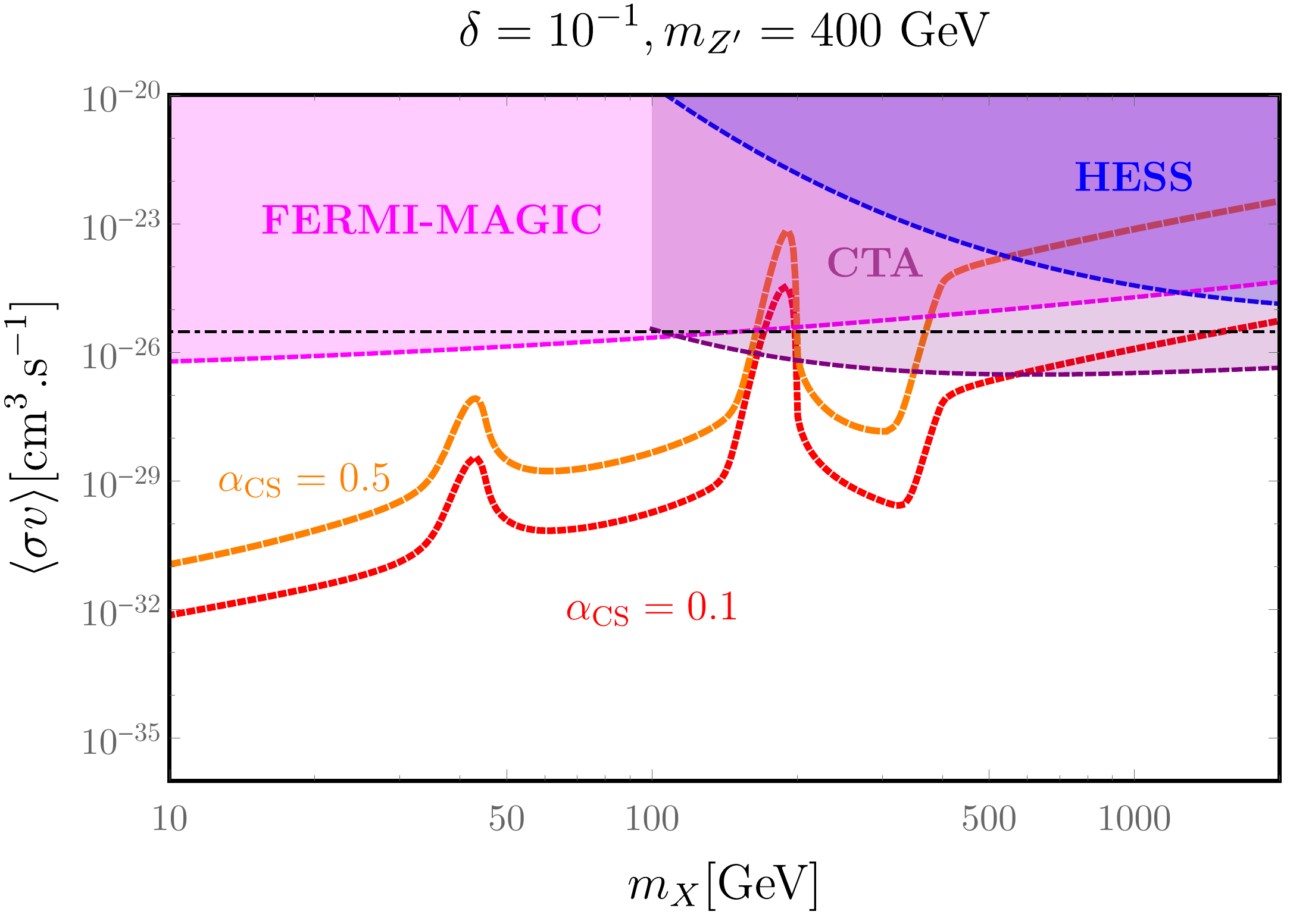}
\includegraphics[width=\linewidth /2]{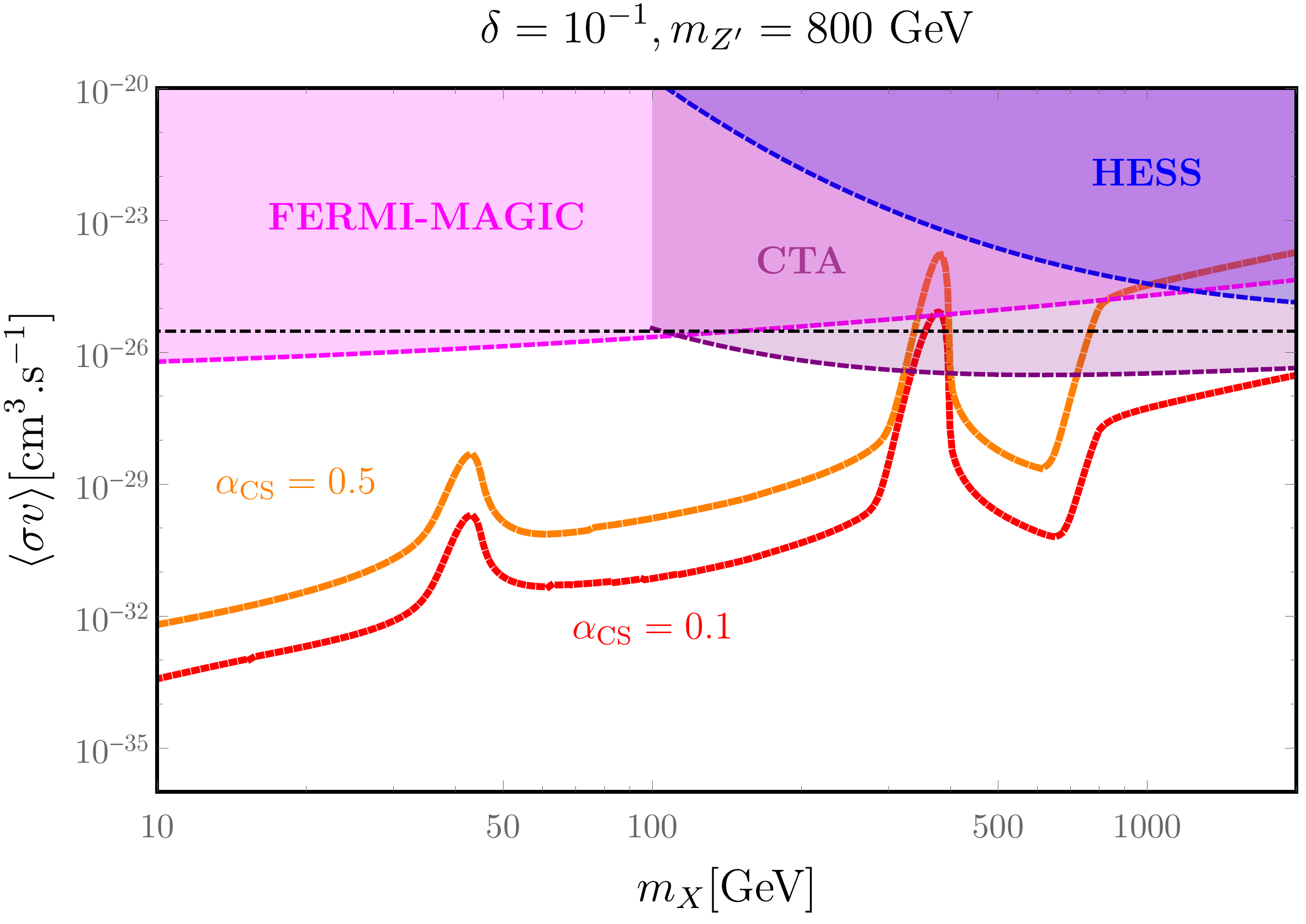}
\caption{Variation of the thermally averaged DM pair 
annihilation cross-section $\langle \sigma v\rangle$, at thermal freeze-out, 
as function of the DM mass $m_X$ for  
$m_{Z'}=400$ GeV (left) and $800$ GeV (right) with two values of $\alpha_{\rm CS}=0.1$
(red coloured solid line) and $0.5$ (orange coloured solid line)
keeping kinetic mixing parameter $\delta=0.1$. The black coloured dashed line represents the thermally 
favoured value of $\langle \sigma v\rangle=$  
$3 \times 10^{-26}\,{\mbox{cm}}^3\, {\mbox{s}}^{-1}$. The magenta 
coloured region with dashed outline represents the resultant exclusion from the combined FERMI-LAT and MAGIC observations 
(abbreviated as FERMI-MAGIC in all the successive relevant figures) while the blue coloured region
with dashed boundary represents the region already 
excluded by HESS for a Einasto density profile of the DM. Finally, the purple colour region
with dashed outline represents
the expected future exclusion from CTA.}
\label{fig:sigmav}
\end{figure}

The DM pair annihilation cross-section, as depicted in Fig.~\ref{fig:sigmav},
typically lies much below the thermally favoured value,
i.e., $10^{-{26}}\,{\rm cm}^3\,{\rm s}^{-1}$, ad exception of the pole regions 
$m_{X}\sim m_Z/2, m_{Z'}/2$, or DM masses above several hundreds of GeV so that at 
least annihilations into $ZZ'$ final state appears kinematically accessible.

\subsubsection{Indirect Detection}

At the moment, the strongest limits come from searches of the gamma-rays produced in 
the DM annihilations. For DM masses below a few hundred GeVs, 
these limits are set by the FERMI 
satellite \cite{Ackermann:2015zua} and exclude the thermally favoured values of $\langle \sigma v\rangle$ for $m_X < 100$ GeV. For higher values of $m_X$ the best sensitivity is achieved by 
HESS \cite{Abramowski:2011hc} which has put a limit $\langle \sigma v \rangle \lesssim 
10^{-25}\text{cm}^3\,\text{s}^{-1}$ for $m_X \sim 1\,{\mbox{TeV}}$,
considering a Einasto density profile for the DM.

As shown earlier while discussing the relic density that most of the DM annihilation processes have a p-wave 
(or even d-wave) annihilation cross-section, i.e., they are velocity dependent. 
As a consequence the 
DM annihilation at present times is suppressed by several orders of magnitude with respect to 
its value at the thermal freeze-out and thus, limits from DM ID are actually not 
effective. ID can probe the WIMP paradigm for the DM relic density only 
when the latter is mostly determined by processes with an s-wave (i.e., velocity independent) 
annihilation cross-section. In our setup this requirement is fulfilled by annihilation into $Z'Z$ and $Z'Z'$ states (see Eq.~(\ref{eq:XXzpz})
and Eq.~(\ref{eq:XXzpzp})). The other s-wave 
dominated annihilation into $ZZ$ final state (see Eq.~(\ref{eq:XXZZ})),
as already addressed, is insignificant
as it is suppressed by the fourth power of kinetic mixing parameter $\delta$.

We have thus compared the DM annihilation cross-section into $Z'Z$ and $Z'Z'$ 
final states with the limits derived from combined analysis of the MAGIC and FERMI-LAT observations
(abbreviated as FERMI-MAGIC) of the dwarf spheroidal galaxies (dSphs)~\cite{Ahnen:2016qkx}. We also consider limits from $10$ years of 
observations towards the inner galactic halo by HESS~\cite{Abdallah:2016ygi}. In the absence of a dedicated analysis for the considered final states we have applied the limit for gamma-rays 
originating from $W^+W^-$ pairs. This choice is reasonable since the $Z^\prime$ decays efficiently into hadrons as the SM gauge bosons. Any mild change in the limits as a result of this assumption is beyond the scope of our current work.

Results concerning changes of the DM pair annihilation cross-section at thermal freeze-out
with the DM mass $m_X$ are reported in Fig.~\ref{fig:sigmav} for specific assignations of the other relevant inputs. As evident from this figure that the impact of ID limits 
from FERMI-MAGIC (magenta coloured region) and HESS 
(blue coloured region) appears to be rather limited. This can 
be understood by looking at this simple analytical estimate for $\langle \sigma v \rangle_{Z'Z'}$:
%
\begin{equation}
\langle \sigma v\rangle_{Z^\prime Z^\prime}\simeq  \dfrac{8 \alpha_{\text{CS}}^4 m_X^2}{9 \pi  m_{Z^\prime}^4} \simeq 3\times 10^{-28}~\text{cm}^3\,\text{s}^{-1}  \Big( \dfrac{\alpha_{\text{CS}}}{0.1} \Big)^{4} \Big( \dfrac{m_X}{1~\text{TeV}}  \Big)^{2} \Big( \dfrac{m_{Z'}}{1~\text{TeV}}\Big)^{-4},
\end{equation}
which shows that a value of the cross-section equal or bigger than the thermal expectation
i.e., $\sim\mathcal{O} (10^{-26}\, \text{cm}^3\,\text{s}^{-1})$,
can be achieved only for masses of the $Z'$ not exceeding a few hundreds of GeV and/or 
$\alpha_{\rm CS} \sim 0.5$. Such value of $\alpha_{\rm CS}$, however, is somewhat extreme since this coupling is expected to have a radiative origin as will be discussed later.
The impact of possible limits from ID will be anyway increased in the near 
future by Cherenkov Telescope Array a.k.a. CTA~\cite{Doro:2012xx,Pierre:2014tra,Wood:2013taa,Silverwood:2014yza,
Lefranc:2015pza,Lefranc:2016dgx} which is reported
(purple coloured region) in Fig.~\ref{fig:sigmav},
assuming a projected limits from $500$~h of observation towards the galactic center.
  
\subsubsection{Direct detection}

In the case of Spin Independent (SI) interactions of the 
DM with nuclei, a cross-section
$\sim\mathcal{O}(10^{-46}\text{ cm}^2)$, 
for DM mass of $50$ GeV, has been excluded by the LUX~\cite{Akerib:2016vxi}, 
PandaX~\cite{Tan:2016zwf} and Xenon1T~\cite{Angle:2007uj} experiments. 
In the chosen setup (see Eq.~(\ref{eq:starting_lagrangian})), scattering of \rm{the} DM with nucleons is originated, at the microscopic level, 
by an interaction between the DM and the SM quarks mediated via t-channel 
exchange of the $Z/Z'$ boson. An interaction of this kind, in the non-relativistic limit,
is described by the following effective operator~\cite{Belanger:2008sj}:
\begin{equation}
\mathcal{L}_{\rm DM~scattering}=\left(\frac{g_X^Z a_q^Z}{m_Z^2}
+\frac{g_X^{Z'} a_q^{Z'}}{m_{Z'}^2}\right) 
\left(\partial_\alpha X_\beta X_\nu-X_\beta \partial_\alpha X_\nu\right) 
\epsilon^{\alpha \beta \nu \mu}\bar q \gamma_\mu \gamma_5 q,
\end{equation}
where $a_q^{Z'}=\dfrac{g^{Z^\prime}_{q_R}-g^{Z^\prime}_{q_L}}{2}$ 
and $a_q^Z=\dfrac{g^{Z}_{q_R}-g^{Z}_{q_L}}{2}$, which corresponds to a SD 
interaction with squared amplitude:
\begin{equation}
\overline{|\mathcal{M}|^2}=\dfrac{32 m^2_X m^2_N}{m^4_Z m^4_{Z^\prime}}
\Big( \sum_q a^{Z}_q \Delta_{q}^N g_{X}^{Z}m_{Z^\prime}^2+ 
\sum_q a^{Z'}_q \Delta_{q}^N g_{X}^{Z^\prime} m_{Z}^2 \Big)^2,
\end{equation}
with $m_N$ denoting the mass of a nucleon $N=p,\,n$ while $\Delta_q^N$ represents the contribution of the quark $q$ to the spin of the nucleon $N$. 
The SD scattering cross-section can be straightforwardly derived, taking into account the 
multiple isotopes present in the detector material as:
\begin{equation}
\sigma_{Xp}^{\text{SD}}=\dfrac{2}{\pi}\dfrac{\mu^2_{Xp}}{m_Z^4m_{Z^\prime}^4} 
\dfrac{ \sum\limits_A \eta_A \Big( S^A_{Z} g_{X}^{Z} m_{Z^\prime}^2+ 
S^A_{Z^\prime} g_{X}^{Z^\prime} m_{Z}^2 \Big)^2 }{\sum\limits_{A} \eta_A \Big( S^A_n + S^A_p \Big)^2},
\end{equation}
where $\mu_{Xp}=m_X m_p/(m_X+m_p)$ is the reduced mass for 
the DM-proton system and $S^A_{Z^{(\prime)}}=a^{Z^{(\prime)}}_u(\Delta^p_u S^A_p+\Delta^p_d S^A_n)+
a^{Z^{(\prime)}}_d[(\Delta^p_d+\Delta^p_s)S^A_p+(\Delta^p_u+\Delta^p_s)S^A_n]$. 
Here $S^A_{p,n}$ represents the contribution of protons and neutrons to the spin of 
a nucleus with atomic number $A$ while $\eta_A$ represents the relative abundance 
of a given isotope of the element constituting the target material. Notice that the 
result is almost independent of the DM mass since the only dependence 
through $\mu_{Xp}$ will vanish for $m_X \gg m_p$ giving
$\mu_{Xp} \sim m_p$. A simple estimate 
of $\sigma_{Xp}^{\text{SD}}$ can be 
performed assuming $\delta,\alpha_{\text{CS}} \ll 1$ as:
\begin{equation}
\sigma_{Xp}^{\text{SD}}\simeq 6 \times 10^{-50} \text{cm}^2 \Big( \dfrac{\delta}{0.1} 
\Big)^2 \Big( \dfrac{\alpha_{\text{CS}}}{0.1}  \Big)^2 
\Big( \dfrac{m_{Z^\prime}}{1 \text{ TeV}}  \Big)^{-4},
\end{equation}
which gives values of $\sigma_{Xp}^{\text{SD}}$ well below the 
current maximal sensitivity ($10^{-41} \text{cm}^2$),  and remains also beyond the reach of next generation detectors
for the chosen set of parameter values, consistent with 
other existing constraints.

\subsection{Collider phenomenology}

Phenomenological models, where
the DM candidate interacts with the SM fields through a spin-1 mediator 
typically posses a rich collider phenomenology. In the case when 
on-shell production of a $Z'$ is kinematically accessible
in proton-proton collision, detectable signals can appear from the 
decays of $Z'$ into SM fermions, showing a new resonance 
in the invariant mass distribution of dijets~\cite{Khachatryan:2015dcf,ATLAS:2015nsi,Sirunyan:2016iap,Aaboud:2017yvp} or 
dileptons~\cite{Aaboud:2016cth,Khachatryan:2016zqb,ATLAS:2017wce} as well as from possible 
decay of a $Z'$ into DM pairs which can be probed through mono-{\bf{X}} searches 
({\bf{X}}= {hadronic jets}, photon, weak gauge bosons, SM-Higgs)~\cite{
Sirunyan:2017hci,Aaboud:2016qgg,Aaboud:2016qgg,CMS:2016hmx,Aaboud:2016uro,CMS:2016fnh,Aaboud:2017dor,
ATLAS:2017uwx,ATLAS:2017pqx,Sirunyan:2017hnk} accompanied by moderate/large missing transverse
energy/momentum. Interestingly, the relative relevance of these two 
kinds of searches, i.e., resonances and mono-{\bf X} events, is mainly set by the invisible decay branching 
fraction (Br) of the $Z'$ which, in turn, is constrained by the DM observables 
that primarily appears from the requirement of correct relic density for the chosen framework.

\begin{figure}[t!]
\includegraphics[width=7.5 cm]{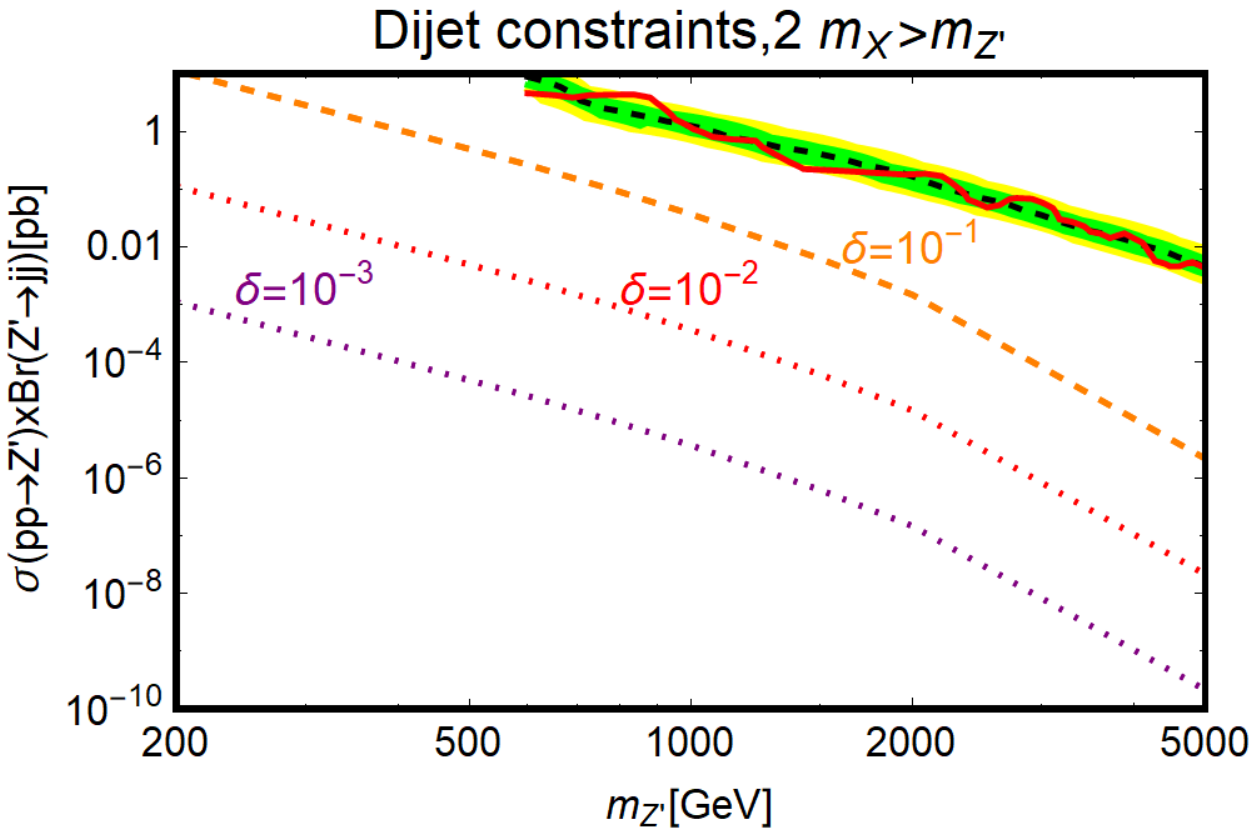}
\includegraphics[width=7.5 cm]{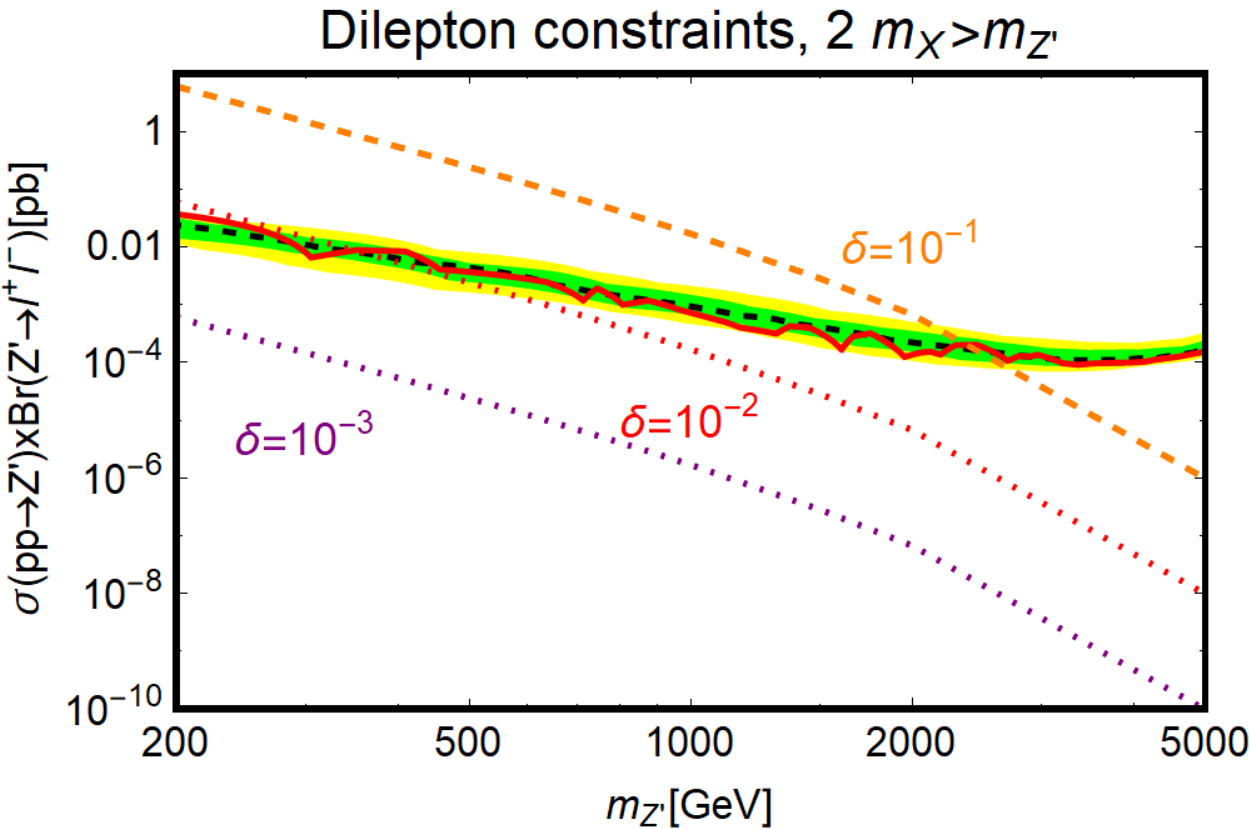}
\caption{Comparison of the production cross-sections for 
$pp\to Z' \to jj$ (left panel) and  $pp\to Z' \to l^+ l^-$ (right panel)
processes for an on-shell $Z'$ in the chosen theory framework
with the relevant experimental limits~\cite{Khachatryan:2015dcf} 
and~\cite{ATLAS:2017wce}, respectively, as a function
of $m_{Z'}$ for three fixed
values of $\delta=10^{-3},\,10^{-2}$ and $10^{-1}$.
Here the solid red coloured
curve represents the observed experimental limit while 
dashed black coloured line, green and yellow coloured bands represent
the expected limit and its $1\sigma,\,2\sigma$ ranges. The dotted
purple, red and dashed orange coloured lines are used for 
$\delta=0.001,\,0.01$ and $0.1$ configurations, respectively.
For simplicity, we have considered 
the case of $2 m_X > m_{Z'}$ so that $Z'$ has only visible decays.
In our analysis we consider $j=u,\,d,\,c,\,s,\,b$ and 
$t$ for $m_{Z'}> 2m_t$ and $l=e,\,\mu$.}
\label{fig:collider1}
\end{figure}

The impact of experimental limits from resonance searches is
depicted in Fig.~\ref{fig:collider1} where, for simplicity, we consider
$2m_X > m_{Z'}$ to forbid $Z'\to$ DM pairs process.
For these analyses we consider all possible quark flavours
including top for $m_{Z'}> 2m_t$ regime while $l=e$ and $\mu$ only.
Here we have compared the results of numerical
simulations for the chosen setup with the experimental ones
including $1\sigma,\,2\sigma$ variations of the production
cross-sections as observed in Refs.~\cite{Khachatryan:2015dcf} 
and~\cite{ATLAS:2017wce}, respectively. Since the $Z'$-SM fermions mixing (see Eq.~(\ref{eq:zzpsmfermion})) appears from an effective
$Z$-$Z'$ mixing, both $\sigma(pp\to Z')$ and Br($Z\to jj/l^+l^-$) are sensitive to the parameter $\delta$. The resultant $\delta^2$ dependence thus, hints diminishing $\sigma(pp\to Z')\times {\rm Br} (Z'\to jj/l^+l^-)$ values
for decreasing $\delta$ values,
as also reflected in Fig.~\ref{fig:collider1}. It is evident from Fig.~\ref{fig:collider1} that
once values of the kinetic mixing parameter $\delta$ smaller 
than $\mathcal{O}(1)$ are considered to comply with the theoretical and EWPT 
(see Eq.~(\ref{eq:EWPT})) constraints, only the limits 
from dileptons resonance searches remain effective. 
For $\delta=0.1$, values of $m_{Z'}$ below 2 TeV 
are ruled out while for $\delta=0.01$, a much weaker lower bound of approximately 300 GeV is obtained
on $m_{Z'}$. For further lower values of $\delta$, very small $\delta^2$
dependence~\footnote{Holds true for a resonant production with no/suppressed invisible decay of $Z'$.} makes $m_{Z'}$ unconstrained from the aforesaid searches.

In the presence of a non-zero and sizable invisible branching fraction for $Z'$, i.e., Br($Z'\to$ DM pairs),
the production cross-sections of dijets/dileptons get suppressed by 
the enhanced decay width of $Z'$. In such a scenario, the 
cross-section corresponding to mono-{\bf X} signals might become sizable, possibly providing complementary 
constraints. We preserve the discussion of such complementary signals for the next subsection.

\subsection{Results}

\begin{figure}[h!]
\begin{center}
\includegraphics[width=9 cm]{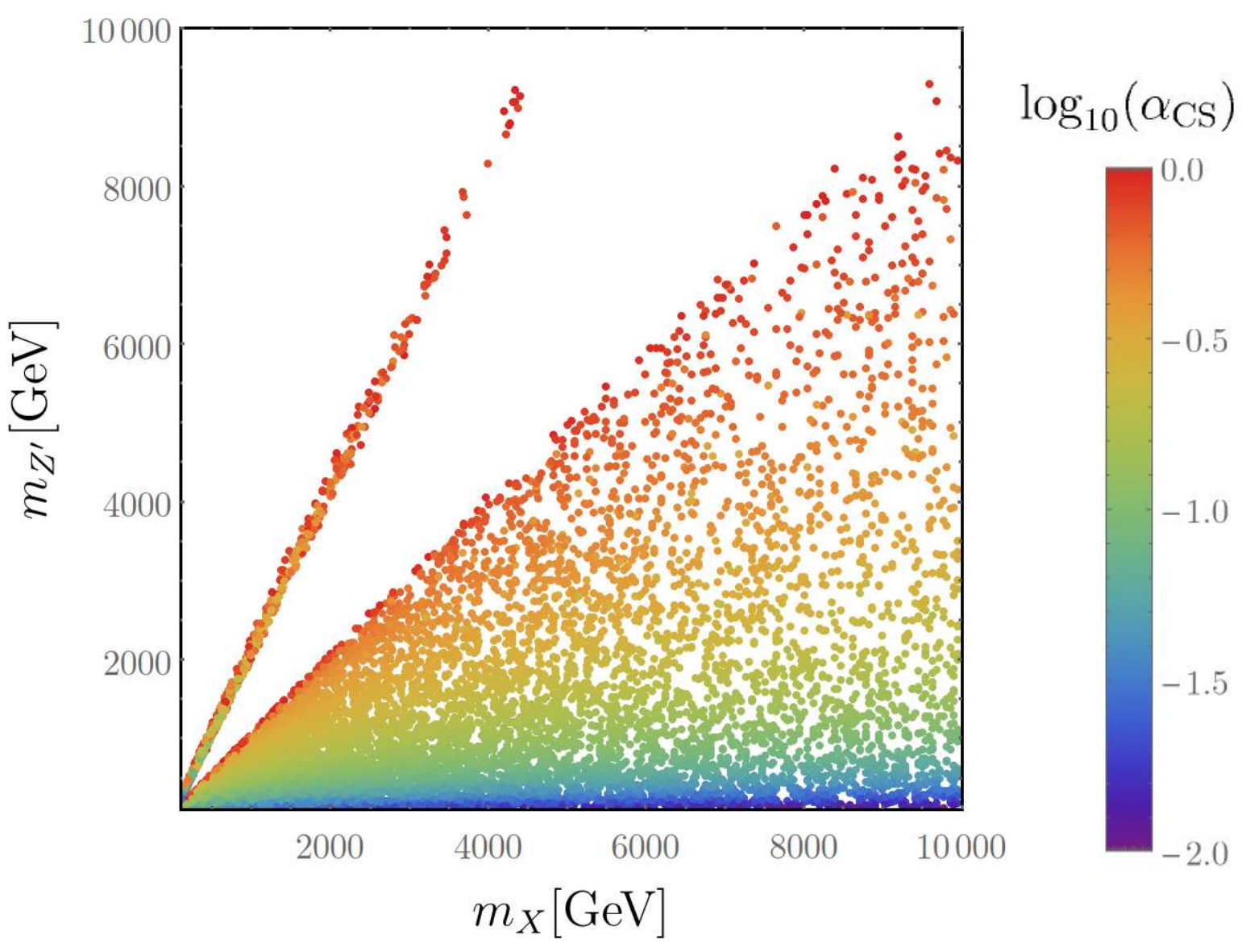}
\caption{Summary of results in the bi-dimensional plane $m_X,m_{Z'}$
for a vectorial DM with $Z'$-portal in the presence
of CS interaction and kinetic mixing terms.
The colour code corresponds to the variation in $\alpha_{\rm CS}$ values.
The plot reports model points, generated through a parameter scan, 
illustrated in the main text, passing constraints from the DM phenomenology 
and general constraints on the kinetic mixing parameter.}
\label{fig:scan}
\end{center}
\end{figure}

In this subsection we report the impact of various constraints, as mentioned in the previous subsections, on the parameter space of the model.
As already pointed out the effect of DD and ID constraints on our
model framework is substantially negligible. Concerning the DM phenomenology, 
the only relevant constraint comes from the requirement of the correct relic density. We 
have then conducted a more extensive analysis, with respect to the one presented in Fig.~\ref{fig:sigmav} 
by performing a scan over the four free input parameters in the following ranges:
\begin{align}
\label{eq:scan1}
\delta \in \left[10^{-3}, 1\right], \,\,
\alpha_{\rm CS} \in \left[10^{-2}, 1\right],\,\,
m_X \in \left[100\,\mbox{GeV},10\,\mbox{TeV}\right],\,\,
m_{Z^\prime} \in \left[100\,\mbox{GeV},10\,\mbox{TeV}\right],
\end{align}
and retaining the model points featuring the correct DM relic density~\footnote{We consider points 
corresponding to $\Omega_X h^2 = 0.12 \pm 10\%$ variation.} and respecting, at the same time, 
constraints from the EWPT, SM $\rho$-parameter measurement as well as reproducing experimentally viable mass 
and width for the $Z$-boson.
The ensemble of points respecting these aforementioned constraints has been reported in Fig.~\ref{fig:scan}, 
in the $m_X$-$m_{Z'}$ bi-dimensional plane with a colour code showing variations in $\alpha_{\rm CS}$ values.

As evident from Fig.~\ref{fig:scan}, in agreement with the general discussion of Sec.~\ref{sssec:rdsc1}, that the correct relic density can be achieved 
either in the "pole" region $m_X \sim m_{Z'}/2$ or for $m_X \gtrsim m_{Z'}$, when the process 
$XX \rightarrow Z'Z'$ is kinematically allowed. 

\begin{figure}[h!]
\begin{center}
\includegraphics[width=5.0 cm]{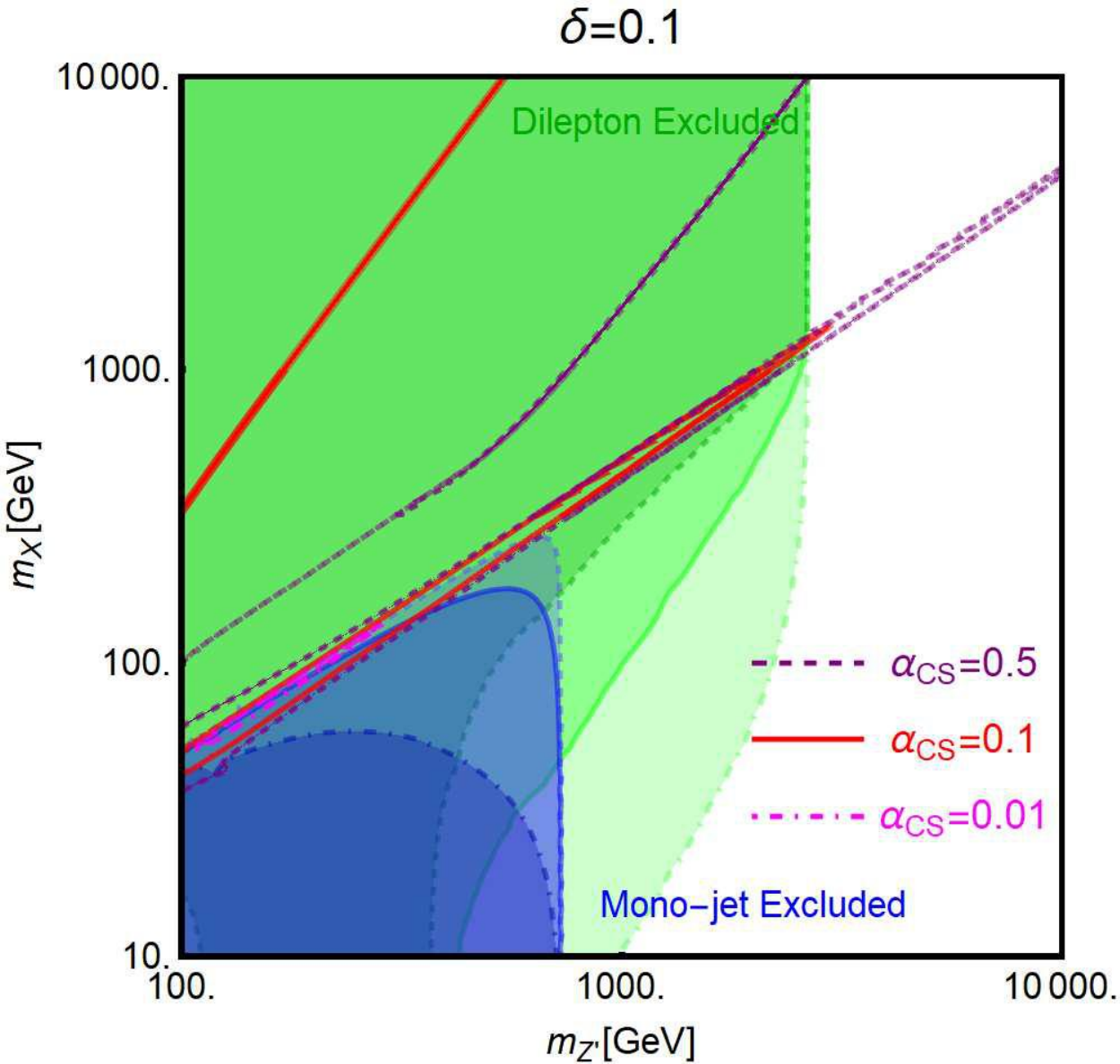}
\includegraphics[width=5.0 cm]{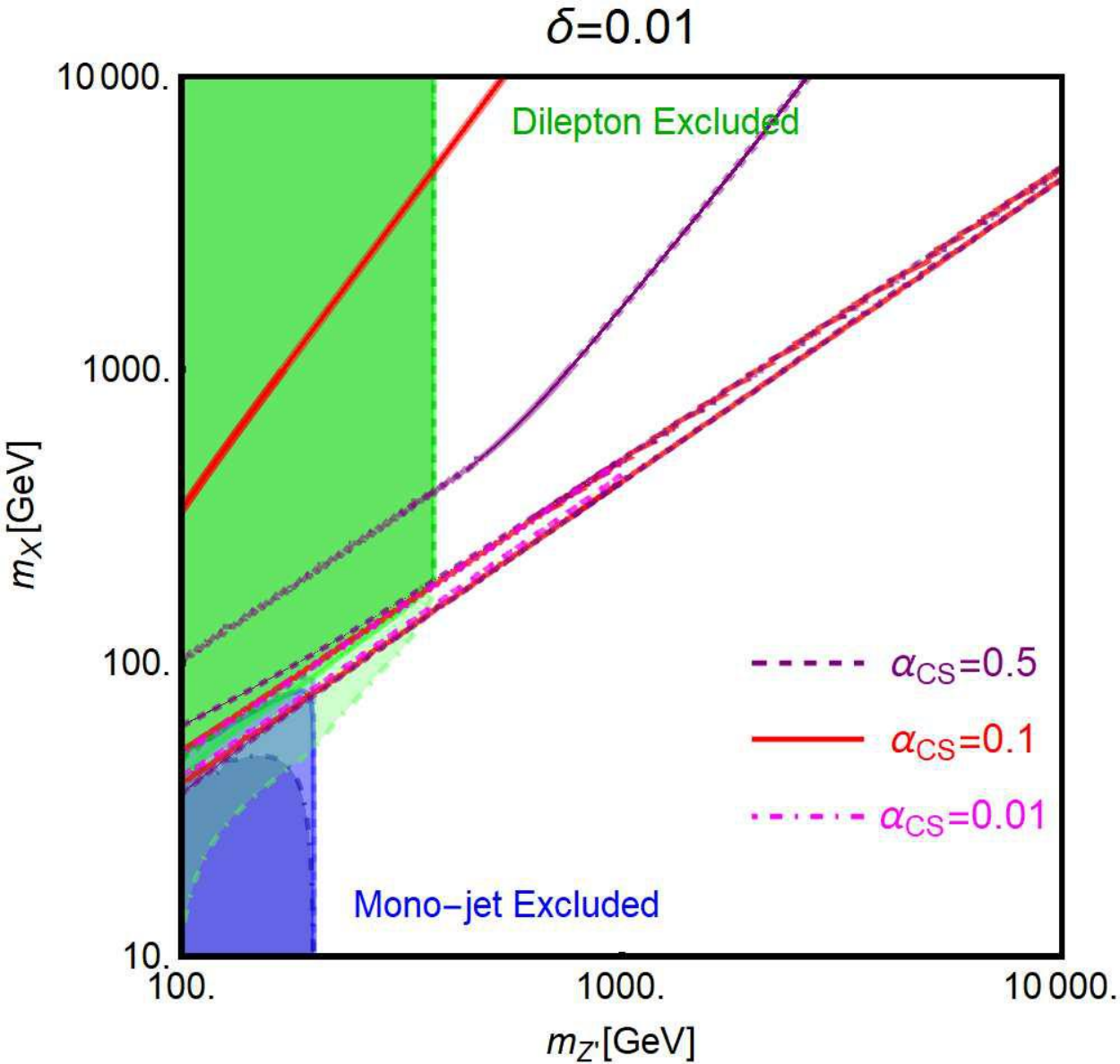}
\includegraphics[width=5.0 cm]{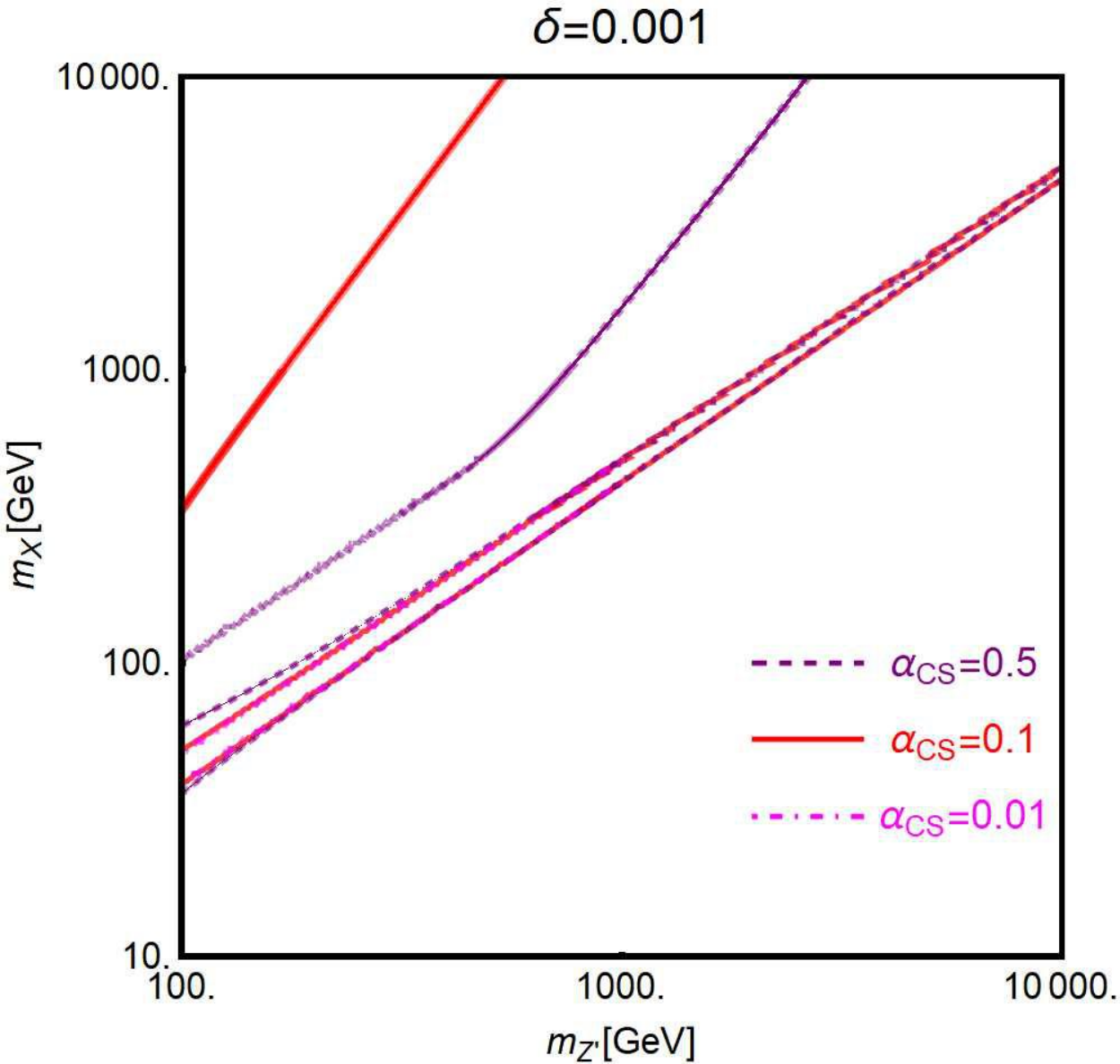}
\caption{Summary of the relic density and the collider constraints in the $(m_{Z'},m_X)$ bi-dimensional plane 
for the three values of $\delta$, $0.1$ (top-left), $0.01$ (top-right) and $0.001$ (bottom) choosing
three assignations of $\alpha_{\rm CS}$, namely $0.5$, $0.1$, and $0.01$. 
In each plots the dashed magenta, solid red and dot-dashed purple coloured curves 
represent the correct DM relic density for $\alpha_{\rm CS}=0.5,\,0.1$ and $0.01$, respectively. 
The regions covered by the light blue colour with dashed boundary, blue colour with solid outline 
and dark blue colour with dot-dashed boundary
are excluded by mono-jet searches when one considers
the value of $\alpha_{\rm CS}=0.5,\,0.1$ and $0.01$, respectively}
\label{fig:collider_summary}
\end{center}
\end{figure}

We have successively examined the impact of collider constraints on our construction 
in Fig.~\ref{fig:collider_summary}. For a better and elucidate illustration
of our findings, we have considered three fixed assignations of $\delta$, namely $0.001,\,0.01$ and 
$0.1$ and the same for $\alpha_{\rm CS}$, $0.01,\,0.1$ and $0.5$ keeping $m_X$ and $m_{Z'}$ as 
the two free varying input parameters. The relevant results are
reported in the three panels of Fig.~\ref{fig:collider_summary} corresponding to the three different 
values of $\delta$, namely $0.1$ (top-left), $0.01$ (top-right)
and $0.001$ (bottom) where isocontours of the correct DM relic density 
are shown for the three assignations of $\alpha_{\rm CS}$ parameter using three different
representations (purple coloured dot-dashed line for $\alpha_{\rm CS}=0.01$, 
solid red coloured line for $\alpha_{\rm CS}=0.1$ and magenta coloured dashed line for $\alpha_{\rm CS}=0.5$).
These isocontours have been compared with the limits from  
dilepton resonances and mono-jet searches. The 
former is evaluated in a similar fashion 
as of Fig.~\ref{fig:collider1} by computing the associated cross-section, 
as a function of the input parameters $\delta,\alpha_{\rm CS}, m_X$ and $m_{Z'}$ using the package 
MadGraph5$\_$aMC$@$NLO~\cite{Alwall:2014hca} and compared with the relevant
experimental observations. Contrary 
to what we did earlier, now we have also accounted 
for the possibility of a sizable invisible branching fraction 
of $Z'$ by suitably rescaling the experimental limit according to the procedure illustrated in
Ref.~\cite{Arcadi:2013qia}.

As evidenced from the top-left panel of Fig.~\ref{fig:collider_summary} 
that the previously quoted limit of approximately 2 TeV (see Fig.~\ref{fig:collider1}, right panel plot) 
for $\delta=0.1$ is actually effective only when $m_{Z'}< 2 m_X$. When  
decay of a $Z'$ into DM pairs is kinematically accessible, i.e., $m_{Z'}> 2 m_X$,
and $\alpha_{\rm CS} \gtrsim \mathcal{O}(1)$, the lower limit on $m_{Z'}$ 
can be reduced even to a few hundreds 
of GeV. The mono-jet limits have been derived by evaluating the production cross-section times the detector efficiency 
and acceptance in the selection of the final state by generating and analyzing events, corresponding to the process 
$p p \rightarrow XXj$, through the combination of MadGraph5$\_$aMC$@$NLO~\cite{Alwall:2014hca} 
(matrix element calculation) PYTHIA 
8~\cite{Sjostrand:2014zea} (event generation and hadronization) and 
DELPHES 3 (fast detector simulations)~\cite{deFavereau:2013fsa}. 
The selection acceptance for the generated events has been determined by 
imposing a minimal value of 500 GeV 
for the missing transverse energy. 
The mono-jet limits have been determined by imposing 
an upper bound of $\sim 6$ fb on the product of production cross-section,
signal acceptance and the detection efficiency \cite{Sirunyan:2016iap}
Our procedure has been validated by reproducing the 
excluded region for the benchmark model adopted in Ref.~\cite{Sirunyan:2016iap}.

It is apparent from Fig.~\ref{fig:collider_summary} that
both kinds of the collider limits, i.e., dileptons and mono-jet, are essentially effective for large and
moderate values of $\delta$, i.e., significant
for $\delta=0.1$ (top-left panel) and moderate for 
$\delta=0.01$ (top-right panel).
These behaviours, as already stated,
are expected since the parameter $\delta$ determines the production vertex of a 
$Z'$ as well as its decay rate into the SM fermions. Hence, 
collider production cross-section suppresses very fast 
as the kinetic mixing parameter decreases. On the 
contrary, the requirement of the correct relic density is more moderately affected by the decrease 
in the value of the kinetic 
mixing parameter. This happens because the correct relic density 
is achieved mostly through the annihilation into $Z'Z'$ final states, whose rate does 
not depend on $\delta$, or in the ``pole'' region, where variations of the couplings can be compensated by slight 
changes of $|2m_X-m_{Z'}|$. 

Among the different collider constraints the most effective ones are the ones 
which emerge from searches of dilepton resonances, 
even if the invisible decay channel for a $Z'$ is taken into account. Using the invariant mass distribution of 
the heavy dilepton resonance, peaked at the $Z'$ mass, one can discriminate the signal 
from the background nicely and for this reason the dilepton channel is 
considered to be a very good probe for these kinds of models.

As evidenced from the two top-row plots of 
Fig.~\ref{fig:collider_summary} that these searches
exclude most of the viable thermal DM region, leaving just a small portion 
of the parameter space around the $Z'$ pole region. On the contrary, the impact of the mono-jet 
constraints is much more moderate and exists only 
for the light DM masses. The collider constraints, as already discussed, 
disappear very fast as the 
value of the kinetic mixing parameter $\delta$ decreases. For $\delta=0.01$,
(top-right panel of Fig.~\ref{fig:collider_summary}),
compared to the $\delta=0.1$ scenario,
a much smaller portion of the parameter space
remains excluded from the dileptons and mono-jet
constraints.
All the collider constraints disappear for $\delta=0.001$ 
(bottom panel of Fig.~\ref{fig:collider_summary}) where the only constraint on 
the model parameter space comes from the requirement of the correct DM relic density.

\section{Scenario-II: $Z'$-$Z$ interaction via a second Chern-Simons term}
\label{sec:scenario2}

In this section, just like Sec.~\ref{sec:scenario1},
we consider a CS term to connect a vectorial DM
with a $Z'$. However, unlike  Sec.~\ref{sec:scenario1}
the coupling between the $Z'$ mediator and the neutral 
EW gauge bosons arises via a second independent CS term.
Being guided by our previous approach, we will address the phenomenological implications of this 
setup as the result of a numerical scan over the free inputs after a brief illustration of the model followed
by concise discussions on the constraints of correct relic density
and different DM searches.

\subsection{The Lagrangian}

The Lagrangian describing the low-energy phenomenology of the 
aforementioned setup can be written as:
\begin{equation}
\mathcal{L}\supset \alpha_{\rm CS} \epsilon^{\mu \nu \rho \sigma}  X_{\mu} Z^\prime_{\nu} 
X_{\rho \sigma} + \beta_{\text{CS}} \epsilon^{\mu \nu \rho \sigma}  
Z_{\mu} Z^\prime_{\nu} B_{\rho \sigma}  +\frac{m_{Z^\prime}^2}{2}Z^{\prime \mu} 
Z^\prime_\mu+\frac{m_{X}^2}{2}X^\mu X_\mu,
\label{eq:lagdoubleCS}
\end{equation}
where $\alpha_{\text{CS}}$ and $\beta_{\text{CS}}$ are the $XXZ^\prime$ and $ZZZ',\,Z\gamma Z'$ coupling constants, 
respectively.  $B_{\rho \sigma}$ denotes the field strength of the SM hypercharge gauge field. The origin 
of the coupling $\alpha_{\text{CS}}$ is the same as of section \ref{sec:scenario1} which will be discussed 
later in appendix \ref{sec:appendixC}. The second CS coupling $\beta_{\text{CS}}$ is non-invariant under the SM gauge group transformation. Nevertheless, as pointed out in Ref.~\cite{Dudas:2012pb}, it could be generated by considering the following gauge invariant effective operator obtained after integrating out some heavy degrees of freedom:
\begin{equation}
\label{eq:twoCSLorigin}
\mathcal{L}\propto i\, \epsilon^{\mu \nu \rho \sigma} D_\mu 
\theta_{Z^\prime}\Big( (D_\nu H)^\dagger H - H^\dagger D_\nu H \Big)B_{\rho \sigma},
\end{equation}
where $D_\mu\theta_{Z^\prime}=\partial_\mu \theta_{Z^\prime}-v_{Z^\prime}q_{Z^\prime}g_{Z^\prime} Z^\prime_\mu$ \
represents the covariant derivative of Stueckelberg axion $\theta_{Z^\prime}$ while $D_\nu H$ denotes the usual covariant derivative of the SM-Higgs doublet. $q_{Z^\prime},\,v_{Z^\prime}$ are the charge and VEV of the associated
complex scalar field and $Z'_\nu$ is the gauge boson of the concerned
$U(1)_{Z^\prime}$ group with $g_{Z^\prime}$ as the gauge coupling.
After the EWSB and choosing unitary gauge for the $U(1)_{Z'}$ group
(such that $\theta_{Z^\prime}$, connected to the phase of an associated heavy Higgs 
field, gets ``eaten" by the longitudinal component of $Z'_\mu$), we recover the operator considered in Eq.~(\ref{eq:lagdoubleCS}).

The structure of Eq.~(\ref{eq:lagdoubleCS})
contains two CS couplings, namely $\alpha_{\rm CS}$ and $\beta_{\rm CS}$. The former is the same as of Eq.~(\ref{eq:starting_lagrangian}) whose origin is explained in the appendix \ref{sec:appendixC} while the latter
appears from Eq.~(\ref{eq:twoCSLorigin}). Given that
the relevant charges are $\sim\mathcal{O}(1)$ (see Eq.~(\ref{eq:effctivealphaCS}), the parameter
$\alpha_{\rm CS}$ from the associated pre-factor goes
as $\sim\mathcal{O}(10^{-3})$. On the other hand, from Eq.~(\ref{eq:twoCSLorigin}) the pre-factor for $\beta_{\rm CS}$ varies as $v_{Z^\prime} v_h^2/M^3$ 
with $M$ as the cut-off scale of the theory,
related to the mass of the associated heavy BSM fermions. Now assuming other relevant parameters as $\sim \mathcal{O}(1)$, even when $v_{Z^\prime}\sim M$, one gets
$\beta_{\rm CS}\sim v_h^2/M^2$. Hence, for $v_h \sim \mathcal{O}(10^2$ GeV) and $M\sim \mathcal{O}(10$ TeV)
(this conservative limit is consistent with
the hitherto undetected evidence of BSM physics at the
13 TeV LHC operation), $\beta_{\rm CS}\sim \mathcal{O}(10^{-4})$ or $0.1 \times \alpha_{\rm CS}$. It is
thus apparent that in general a hierarchy between the values of two CS couplings is rather natural as they have two different theory origins. In our analysis
we, however, also consider the possibility of $\alpha_{\rm CS}=\beta_{\rm CS}$. 

\subsection{Relic density}
\begin{figure}[h!]
\begin{center}
\includegraphics[width=8 cm]{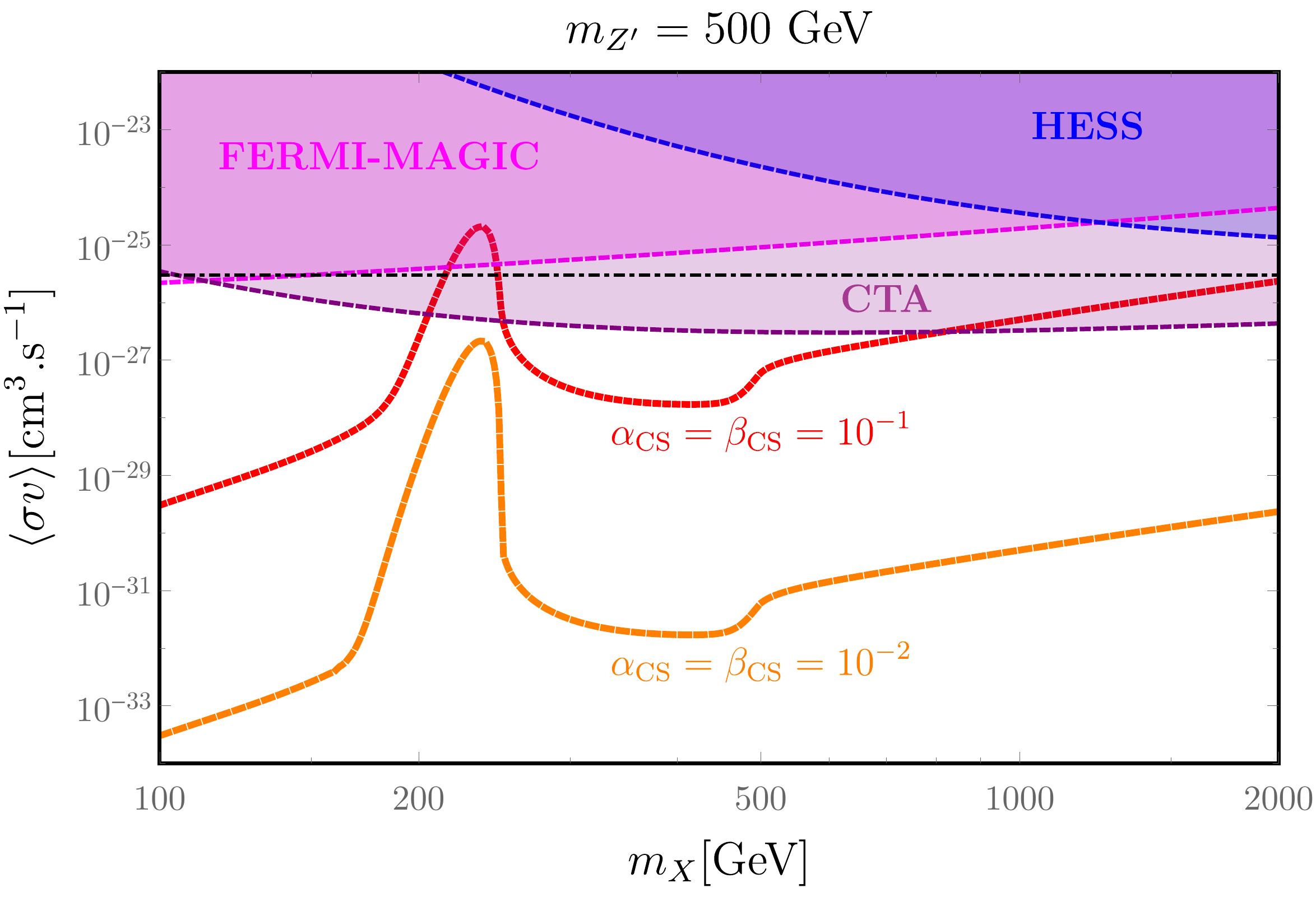}
\caption{Thermally averaged DM pair annihilation cross-section at the time of thermal freeze-out, 
in the presence of double CS terms, as function of the DM mass $m_X$, for $m_{Z'}=500\,\mbox{GeV}$ 
and two assignations for $\alpha_{\rm CS}$ and $\beta_{\rm CS}$, namely $\alpha_{\rm CS}=\beta_{\rm CS}=0.1$ 
(red coloured curve) and $\alpha_{\rm CS}=\beta_{\rm CS}=0.01$ (orange coloured curve). 
The black coloured dashed line represents the thermally 
favoured value of $\langle \sigma v\rangle=$  
$3 \times 10^{-26}\,{\mbox{cm}}^3\, {\mbox{s}}^{-1}$.
Different exclusion regions are the same as of Fig.~\ref{fig:sigmav}.}
\label{fig:sigmav_dCS}
\end{center}
\end{figure}

In the case of double CS terms, unlike the kinetic
mixing scenario, the number of accessible DM pair annihilation channels 
is limited just to three options, i.e., $Z\gamma$, $ZZ$ and $Z'Z'$. The first
two are induced by s-channel exchange of the $Z'$ while the third one is 
induced by t/u channel exchange of a DM state. Simple analytical approximations of 
the corresponding DM pair annihilation cross-sections can be
derived, as usual, through the customary velocity expansion:

\begin{itemize}
\item $XX\to Z\gamma:$
\begin{equation}
\label{eq:XXZgam}
\langle \sigma v\rangle_{Z\gamma}\simeq \frac{\alpha_{\text{CS}} ^2 
\beta_{\text{CS}} ^2 c_W^2 \left(4 m_X^2-m_Z^2\right)^3}{48 \pi  m_X^4 m_{Z^\prime}^4}\dfrac{1}{x},
~~{\rm for~} m_Z \ll m_X .
\end{equation}
%
\item $XX\to ZZ:$
\beq
\label{eq:XXZZ2}
\langle \sigma v\rangle_{ZZ}\simeq \frac{8 \alpha_{\text{CS}}^2 
\beta_{\text{CS}} ^2 s_W^2 \left(m_X^2-m_Z^2\right)^{3/2} }{3 \pi m_X  m_{Z^\prime}^4}\dfrac{1}{x},
~~{\rm for~} m_Z \ll m_X .
\eeq
%
\item $XX\to Z^\prime Z^\prime:$
\end{itemize}
\begin{align}
\label{eq:XXzpzp2}
\langle \sigma v\rangle_{Z^\prime Z^\prime}\simeq \frac{\alpha_{\text{CS}} ^4  
\left(32 m_X^8+14 m_{Z^\prime}^8-56 m_X^6 m_{Z^\prime}^2+69 m_X^4 m_{Z^\prime}^4-50 m_X^2 m_{Z^\prime}^6\right)}{9 \pi  m_X^2 m_{Z^\prime}^4 \left(m_{Z^\prime}^2-2 m_X^2\right)^2} \sqrt{1-\frac{m_{Z^\prime}^2}{m_X^2}}~.
\end{align}
The behavior of the DM pair annihilation cross-section, as function of its mass, 
is reported in Fig.~\ref{fig:sigmav_dCS}. Similar to the 
scenario discussed in Sec.~\ref{sec:scenario1}, 
the DM pair annihilation cross-sections are typically velocity suppressed except the 
s-wave annihilation channel into $Z'Z'$ final states.
%
%
As evidenced from Fig.~\ref{fig:sigmav_dCS} that value of the thermally averaged DM pair 
annihilation cross-section can match the experimentally favoured value only at  
$m_X \sim \frac{m_{Z'}}{2}$, (i.e., pole region), and for $m_X>m_{Z'}$, so that 
the DM pair annihilation into $Z'$ pairs is allowed. Furthermore, to avoid 
overproduction of the DM, $\alpha_{\rm CS}$ values 
at least $\sim \mathcal{O}(0.1)$ are needed.

\subsection{Indirect detection}

Possible prospects of ID rely, as the kinetic mixing scenario, 
mostly on the detection of gamma-rays produced after the DM pair annihilation into $Z'$ pairs which 
subsequently decay into hadrons. Indeed the other two annihilation channels, i.e., $ZZ$ and $Z\gamma$ have p-wave (velocity dependent) annihilation cross-sections  lying several 
orders of magnitude below the present and the near future experimental sensitivities 
\cite{Hooper:2012sr,Gomez-Vargas:2013bea,Ando:2013ff,Gonzalez-Morales:2014eaa,
Queiroz:2014yna,Li:2015kag,Mambrini:2015sia,Massari:2015xea,Queiroz:2016zwd,
Profumo:2016idl,Adams:2016alz,Archambault:2017wyh,Khatun:2017adx,Campos:2017odj}. That said, similar to the previous case, we assumed that the annihilations into $Z'$ pairs lead to a gamma-ray yield comparable to the $W^+W^-$ final state. Therefore, one can use CTA sensitivity to DM annihilations into the $W^+W^-$ channel to constrain the model as can be seen in Fig.~\ref{fig:sigmav_dCS}. It is clear that only CTA is expected to mildly probe this setup for DM masses above $1$~TeV.

\subsection{Direct Detection}

In the case of double CS interactions, the $Z'$ has no direct/tree-level 
couplings with the SM-quarks (or the gluons) and hence, no 
operators relevant for DD is induced at the tree-level.

\subsection{Collider phenomenology}

In the scenario with double CS interactions,
the $Z'$ is directly coupled only with the $Z$-boson and the photon. 
Tree level production of the $Z'$ at the LHC, nevertheless, is possible
through vector boson fusion (VBF) in association with 
two hadronic jets. The production cross-section, however, is more suppressed
compared to the case of a single CS interaction with kinetic mixing
(see Sec.~\ref{sec:scenario1})
which has direct couplings with quarks at the tree level~\footnote{A richer collider phenomenology could
appear in extensions of the proposed scenario in which the $Z'$ 
has direct coupling with 
the $W$-boson and the gluons
as studied in Refs.
~\cite{Antoniadis:2009ze,Bramante:2011qc,Kumar:2012ba,Dudas:2013sia,Ducu:2015fda}.}. 

\begin{figure}[h!]
\begin{center}
\includegraphics[width=8.5 cm]{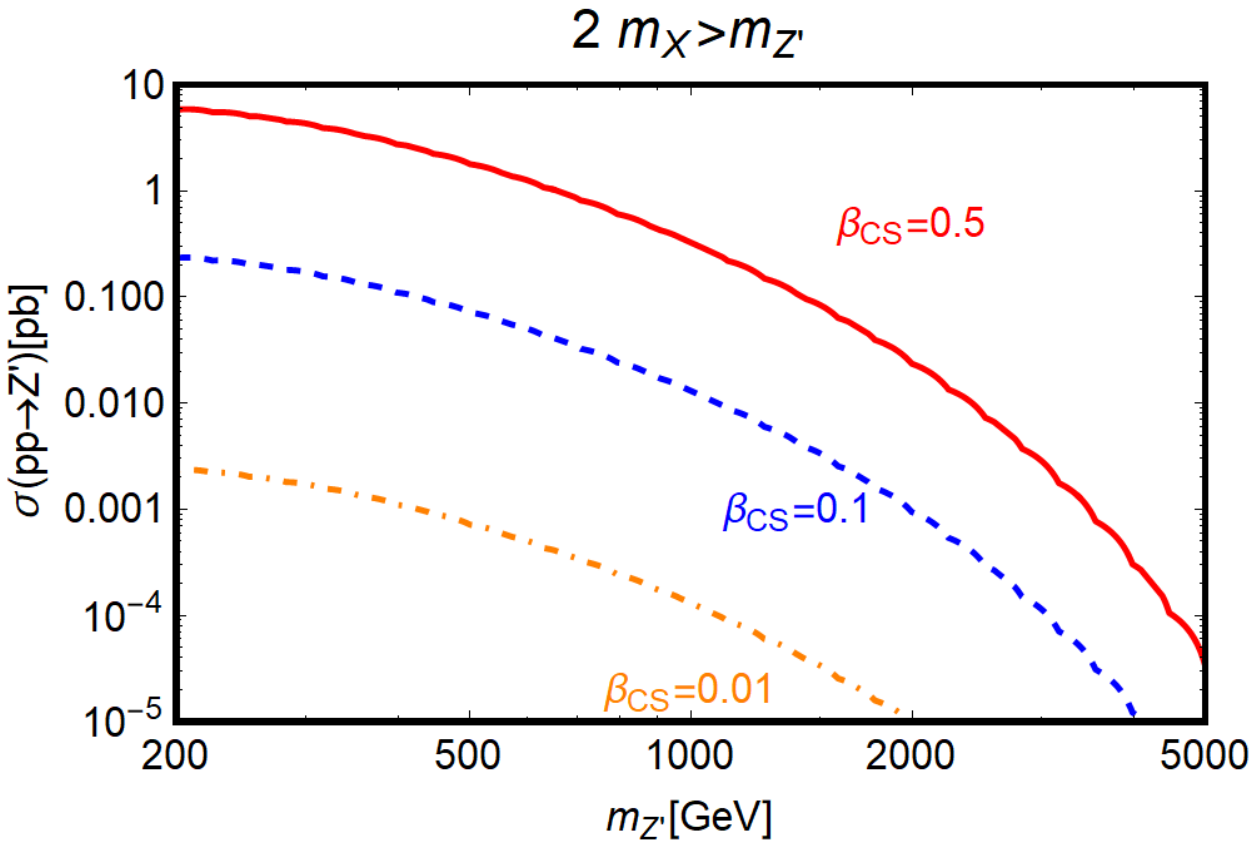}
\end{center}
\caption{Production cross-section of the $Z'$ at the LHC through VBF as 
a function of $m_{Z'}$ for 13 TeV centre-of-mass
energy for three different values of $\beta_{CS}$ parameter, namely, $0.5$, $0.1$ and $0.01$
that are represented with solid red,
dashed blue and dot-dashed orange coloured lines, respectively. 
The choice of $2 m_X > m_{Z'}$ forbids invisible decay of the $Z'$.}
\label{fig:VBF}
\end{figure}

We have reported in Fig.~\ref{fig:VBF} the expected $Z'$ production cross-section at 
the LHC for 13 TeV centre-of-mass 
energy as a function of the mass of $Z'$ $(m_{Z'})$ for three values of
parameter $\beta_{\rm CS}=0.5,\,0.1$ and $0.01$, depicted with solid red coloured line,
dashed blue coloured line and dot-dashed orange coloured line, respectively. The 
cross-section has been computed using the procedure illustrated in Ref.~\cite{Kumar:2007zza}
where discovery prospects of a $Z'$, produced via $Z$-fusion and decaying to four leptons final states,
i.e., $Z' \rightarrow ZZ \rightarrow 4l$, have been investigated.  
These kinds of signals could be probed at the LHC with 14 TeV 
centre-of-mass energy provided that 
$\sigma (pp \rightarrow Z') \simeq 1\,\mbox{pb}$. 
It is now apparent from Fig.~\ref{fig:VBF} that a future $Z'$ discovery
would appear feasible only for $\beta_{\rm CS}\simeq 0.5$
with $m_{Z'} \lesssim 1$ TeV. For 
lower values of $\beta_{\rm CS}$ the production 
cross-section, goes as $\beta^2_{\rm CS}$, would 
appear very suppressed to escape detection at the LHC unless
(possibly) one considers higher luminosity. One should,
however, remain careful about $\sim \mathcal{O}(1)$ value 
of the parameter $\beta_{\rm CS}$ as this would
indicate a scale for the associated heavy fermion mass $M$ 
(see discussions after Eq.~(\ref{eq:twoCSLorigin})) well within the 
reach of ongoing LHC operation where, unfortunately, no evidence
of BSM physics has been confirmed till date.

Notice that in the above discussion 
we have implicitly assumed that invisible decay of the $Z'$ is kinematically forbidden,
i.e., $2\,m_X> m_{Z'}$. 
If this was not the case, the production cross-section of $4l+2j$ final states would be suppressed further by the non-zero invisible branching fraction of the $Z'$. On the other hand, a sizable 
invisible branching fraction for $Z'$ might offer meaningful detection
prospects for $p p \rightarrow Z'\rightarrow XX+2j$ process. Investigation 
of such signature would require a dedicated study which 
is beyond the scope of this work.

\subsection{Results}
\begin{figure}[h!]
\includegraphics[width=8cm]{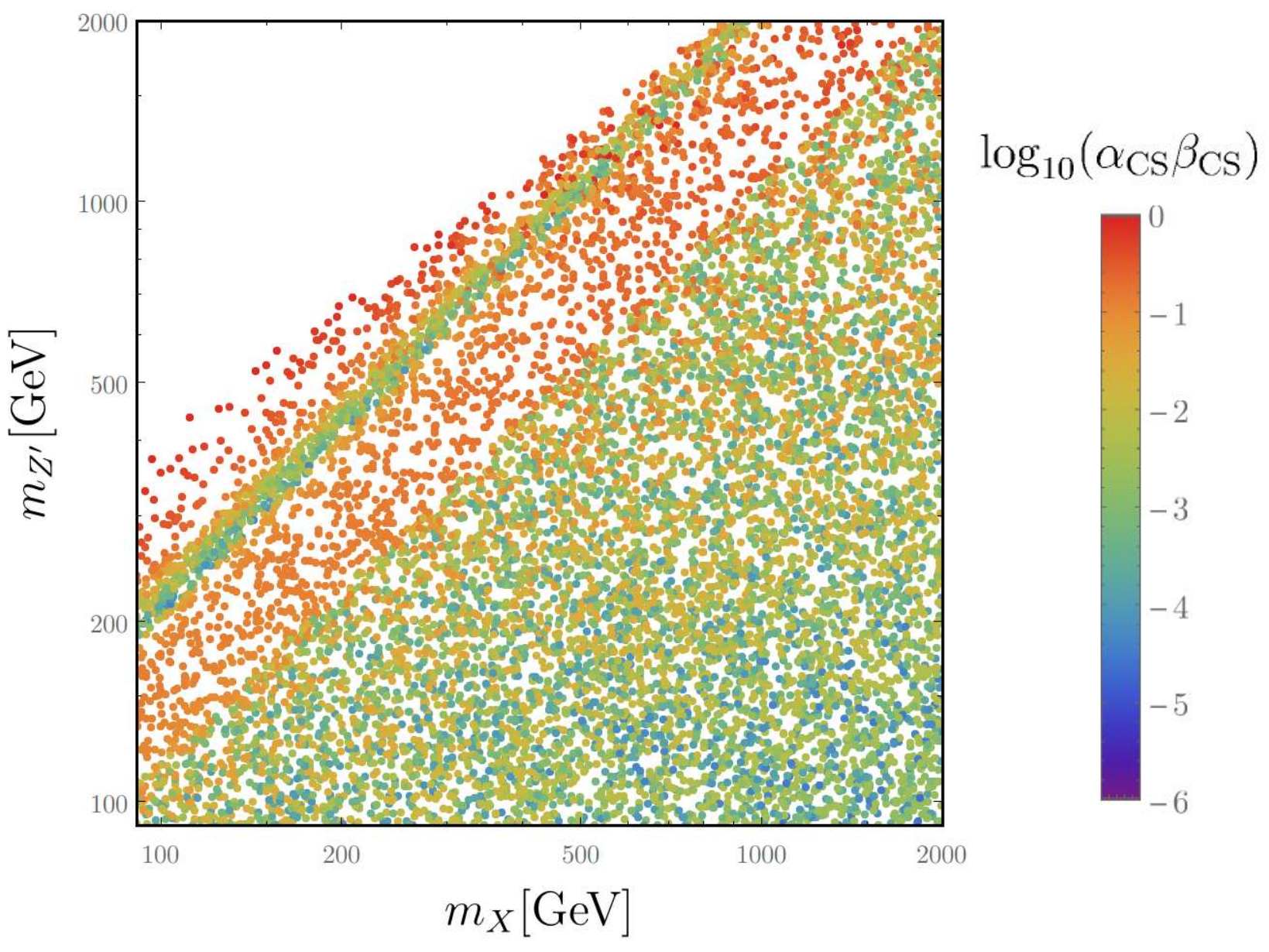}
\includegraphics[width=7.4cm]{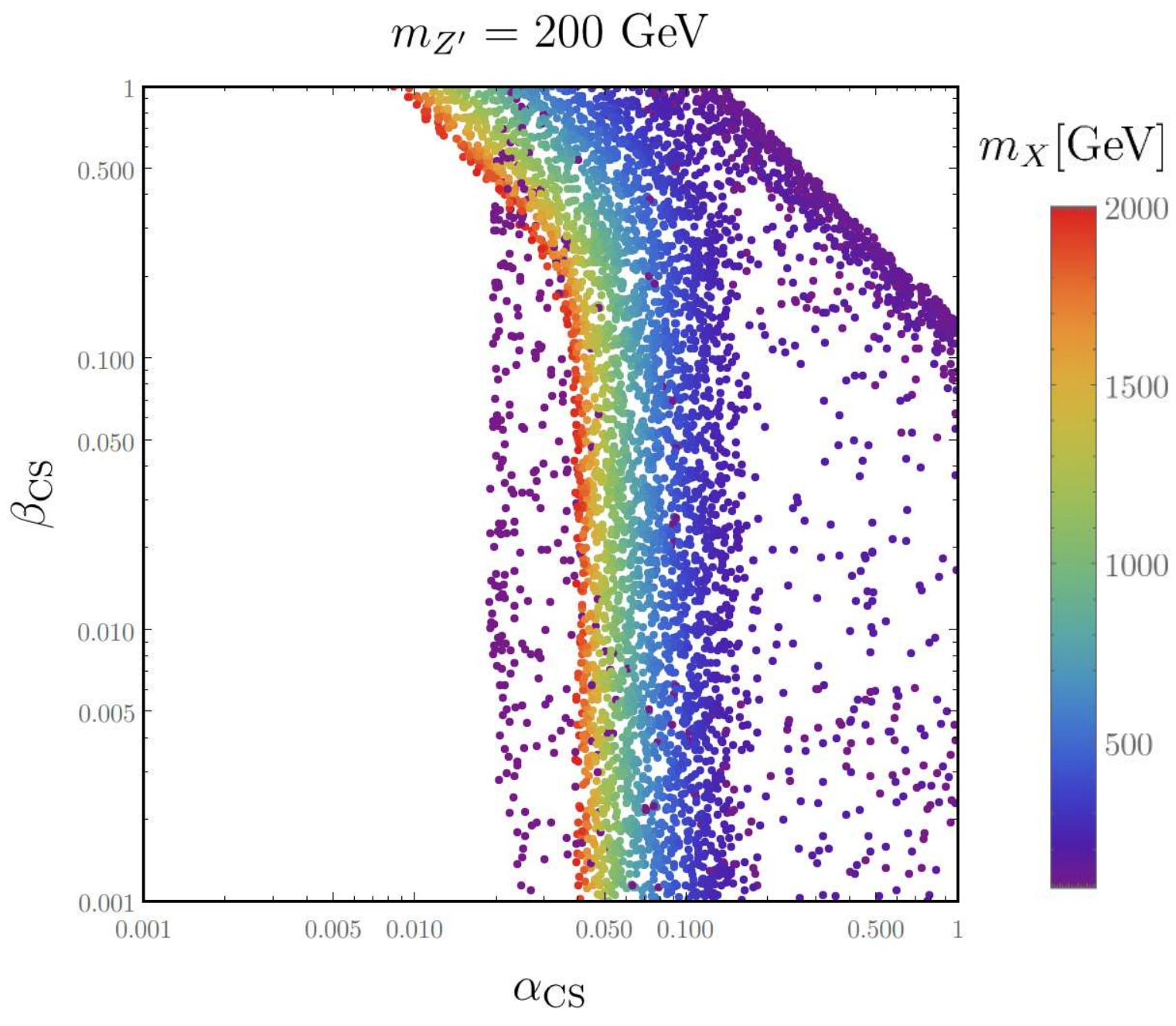}
\caption{Results of a parameter scan for the setup with two CS terms
showing points compatible with the DM phenomenology.
The left plot is in $m_X,\,m_{Z'}$ bi-dimensional plane with
a colour coding for the product of two CS couplings
$\alpha_{\rm CS}\beta_{\rm CS}$ while the right plot is in 
$\alpha_{\rm CS},\,\beta_{\rm CS}$ plane for $m_{Z'}=200$ GeV with a 
colour coding for various $m_X$ values.}
\label{fig:scandoubleCS}
\end{figure}

Similar to the first model considered in this work, we perform a scan in 
the parameter space over $\alpha_{\text{CS}},\,\beta_{\text{CS}},\,m_{Z^\prime}$ and $m_X$ 
and represent our findings in Fig.~\ref{fig:scandoubleCS} showing the phenomenologically viable model points.
One should note that, contrary to the kinetic mixing
scenario, the absence of tree-level $Z'$ couplings
with the SM fermions appears useful to efface a set of constraints
coming from the precision $Z$-physics and collider observations.
The scan is performed in the following ranges:

\begin{align}
\label{eq:scan2}
\alpha_{\rm CS} \in \left[10^{-3}, 1\right],\,\,\,\,
\beta_{\rm CS} \in \left[10^{-3}, 1\right],\,\,\,\,
m_X \in \left[90 \,\mbox{GeV},2\,\mbox{TeV}\right],\,\,\,\,
m_{Z^\prime} \in \left[90\,\mbox{GeV},2\,\mbox{TeV}\right].
\end{align}
%

As expected, the left panel of Fig.~\ref{fig:scandoubleCS}, representing the viable model 
points in the bi-dimensional plane $(m_X,m_{Z'})$, shows, similar to the kinetic mixing scenario, 
a sensitive preference for configurations with $m_X > m_{Z'}$, for which the correct relic density 
can be more easily achieved through the velocity unsuppressed annihilation rate into $Z'$ pairs. This is 
further evidenced by the right panel of Fig.~\ref{fig:scandoubleCS}, investigating the bi-dimensional 
plane $(\alpha_{\rm CS},\,\beta_{\rm CS}$) for a fixed $m_{Z'}$. Indeed, most of the points approximately trace isocontours 
with a shape independent of $\beta_{\rm CS}$, unless this coupling 
is much bigger than $\alpha_{\rm CS}$. 
This behavior would be exactly expected in the case when the DM relic density is mostly accounted by the 
annihilation into the $Z'Z'$ final state since the corresponding rate depends only on $\alpha_{\rm CS}$ (see Eq.~(\ref{eq:XXzpzp2})). 
One must note that the demand of $\beta_{\rm CS} \gg \alpha_{\rm CS}$ is theoretically
challenging since normally one would expect $\beta_{\rm CS} < \alpha_{\rm CS}$ as
already discussed in the context of Eq.~(\ref{eq:twoCSLorigin}).

It is curious to note the apparent conflicts among the DM detection prospects
in DD, ID and collider experiments with $Z'$ searches at the collider and
theoretical consistency of the studied framework.
For example, from Fig.~\ref{fig:VBF} it is evident
that a $Z'$ discovery at the LHC with 13 TeV centre-of-mass
energy would require $\beta_{\rm CS}\sim \mathcal{O}(1)$ and $m_{Z'}\lesssim 1$ TeV.
Such high value of $\beta_{\rm CS}$, as already mentioned,
is hard to accommodate phenomenologically. Pushing $\beta_{\rm CS}$
towards its natural regime, i.e., $\sim \mathcal{O}(10^{-4})$, on the other hand, would predict a 
$\alpha_{\rm CS}\sim \mathcal{O}[10-10^4]$ to retain some of the viable 
points, notably for log$_{10}(\alpha_{\rm CS}\beta_{\rm CS})\gtrsim {-3}$, as
shown in the left plot of Fig.~\ref{fig:scandoubleCS}
to respect the DD, ID and relic density constraints. Now keeping
in mind the radiative origin of $\alpha_{\rm CS}$, as explained
in the appendix \ref{sec:appendixC}, such high values of $\alpha_{\rm CS}$
would require either a pathologically behaved strongly
coupled theory or unnatural high values for the associated charges.
Lower and rather natural $\alpha_{\rm CS}$ values, e.g., $\sim 10^{-3}$,
can ameliorate such theoretical shortcomings and one can still
get viable points consistent with the DM observables
in the region of log$_{10}(\alpha_{\rm CS}\beta_{\rm CS})\lesssim {-5}$.
The associated low $\beta_{\rm CS}$ values, however, failed
to produce detectable $Z'$ signals at the LHC as already shown in
Fig.~\ref{fig:VBF}. Similar contradictions also appear
for the right plot of Fig.~\ref{fig:scandoubleCS} which,
keeping in mind the collider detection prospect of a $Z'$, 
predicts $0.01 \lesssim \alpha_{\rm CS}\lesssim 0.3$ for $m_{Z'}=200$ GeV.
This range of the parameter $\alpha_{\rm CS}$ once again
either requires moderate values of the involved charges
or asks for a strongly coupled model frameworks.
In a nutshell, keeping in mind the detection possibilities
of the BSM physics, either in DM or in collider searches,
together with an elegant theoretical construction, the scenario with two
CS interactions is less appealing compared
to a scenario with one CS interaction and a kinetic mixing.
Such simple conclusion, however, is not obvious when
a scenario with two CS interactions involves
non-Abelian gauge groups.

\section{Generation of a Chern-Simons coupling from a UV complete model}
\label{sec:UVcompletion}
In this section we propose a UV complete model that 
can address the origin of CS and  kinetic mixing terms.
The generation of a generalized CS interaction term between the 
gauge boson of a BSM $U(1)$ group and the SM gauge bosons has already been
proposed in Refs.~\cite{Anastasopoulos:2006cz,Dudas:2009uq,Antoniadis:2009ze}. 
The effective CS coupling is originated from the triangle loops 
involving BSM heavy fermions, charged under 
both the SM $U(1)_Y$ and a BSM $U(1)$ symmetry groups,
after integrating out the heavy degrees of freedom.
In this work we consider a similar construction where a CS 
interaction term is generated between the gauge bosons of two new BSM symmetry groups 
labeled as $U(1)_X$ (associated to the DM) and $U(1)_V$ (associated with $\wt{V}_\mu$), respectively~\footnote{For the convenience of
reading note that in scenario-I, $\wt{V}_\mu$ denotes the gauge field of $U(1)_V$ which, after diagonalization of the flavour basis Lagrangian, is written as $Z'_\mu$ while
in scenario-II the kinetic and mass terms are already diagonalized
and hence, $\wt{V}_\mu=Z'_\mu$.}.

Concerning the origin of kinetic mixing, in the most general scenario, i.e., when the new BSM 
fermions are charged~\footnote{One could expect the BSM fermions to have an impact on the 
DM decoupling process from the SM thermal bath, however, since the freeze out occurs only 
when the DM particles are non-relativistic, the
BSM fermions would have already decoupled and their density would be exponentially 
suppressed due to Boltzmann factor.} under the BSM gauge groups $U(1)_X,\, U(1)_V$ as well as
with respect to $U(1)_Y$ of the SM, one can radiatively generate three possible kinetic mixing
terms like $X^{\mu\nu}\wt V_{\mu\nu}, \, X^{\mu\nu} B_{\mu\nu}$ and
$\wt V^{\mu\nu} B_{\mu\nu}$ with~\footnote{Such terms, being Lorentz invariant, renormalizable 
and invariant under the SM gauge groups, can directly appear in  
Lagrangian. We, however, do not consider such possibility
and confine our discussion on scenarios where such interactions are radiatively generated.} 
appropriate field strengths of the involved
Abelian groups. The coefficients for these terms emerge as a result
of integrating out the aforesaid heavy BSM fermionic degrees of freedom 
and also include associated gauge charges. One should remain
careful at this stage as a kinetic mixing like $ X^{\mu\nu} B_{\mu\nu}$
triggers a mixing between the DM $X_\mu$ and $Z_\mu$ which
would subsequently allow the DM to decay into the SM fermions
and thereby, spoiling its stability. A similar situation can also
appear for $X^{\mu\nu}\wt V_{\mu\nu}$, given that $\wt V_\mu$
is somehow, e.g., via a second kinetic mixing $\wt V^{\mu\nu} B_{\mu\nu}$, 
decaying into the SM particles. It is thus important to assign
gauge charges for the BSM fermions in an elegant way such that
the stability of the DM remains preserved.

We denote new BSM fermions as $\chi$ ($\psi$), chiral with respect to $U(1)_X$ ($U(1)_V$)
while vector like compared to $U(1)_V$ ($U(1)_X$) as well as to the SM \cite{Antoniadis:2009ze}.
This way no new chiral anomalies are introduced in the SM while one can find 
suitable charge assignments for these BSM fermions to efface new anomalies.
Regarding the mass generation of these new fermions the most plausible option is represented via 
Yukawa interactions with two BSM complex scalar fields $\phi_{X,V}$. The 
relevant Lagrangian is written as:
%
\begin{align}
\label{eq:bsmfm}
\mathcal{L}^{\text{fermions}}=- y_F \phi_V \bar{\psi}_{1L} \psi_{1R}- y_F \phi_V^* \bar{\psi}_{2L} \psi_{2R} -  y_F \phi_X \bar{\chi}_{1L} \chi_{1R}- y_F \phi_X^* \bar{\chi}_{2L} \chi_{2R} + \text{h.c.},
\end{align}
where $y_F$ represents generic Yukawa couplings~\footnote{We, without
the loss of generality, consider identical Yukawa couplings for all the BSM 
fermions for simplicity since the final results are independent of them.}.
The complex scalar fields $\phi_X$, $\phi_V$ can be written using the 
following parametrization as:
\begin{equation}
\phi_X=(v_X+h_X)e^{i\theta_X/v_X} \qquad \text{and} \qquad \phi_V=(v_V+h_V)e^{i\theta_V/v_V},
\end{equation}
with $v_X,\,v_V$ and $h_X,\,h_V$ denoting the corresponding VEVs
and the Higgs fields. Here $\theta_X,\,\theta_V$
are Stueckelberg axions \cite{Stueckelberg:1938zz,Kors:2004dx}. Investigating the kinetic
terms for these scalars with suitable BSM covariant derivatives,
one gets, for example, for $\phi_X$:

\begin{equation}
|D_\mu \phi_X|^2 = |\partial_\mu \phi_X -i q_X g_X \phi_X X_\mu|^2\supset 
(\partial_\mu \theta_X - q_X g_X v_X X_\mu)^2, 
\end{equation}
with $g_{X}$ as the gauge coupling of the $U(1)_X$ gauge group
and $q_X$ as the charge of $\phi_X$ with respect to this group. 
The DM mass thus, is generated via the 
Stueckelberg mechanism \cite{Stueckelberg:1938zz,Kors:2004dx}. The 
masses for the new BSM fermions, $\chi_1,\,\chi_2$, are also generated after the spontaneous symmetry
breaking (SSB) in the $U(1)_X$ sector.
In fact, after SSB one ends up with the following tree-level
realizations for various masses:
\begin{equation}
\label{eq:XVgenmass}
m_{h_X}\sim \sqrt{\lambda_X} v_X, \qquad m_{F} \sim y_F v_X, \qquad m_X \sim g_X q_X v_X,
\end{equation}
where $\lambda_{X}$ is the quartic coupling of 
the $\phi_{X}$ potential and the notation $m_F$ is used to represent
generic BSM fermion masses. A similar construction holds for 
$U(1)_V$ with $\lambda_V,\,g_V,\,q_V,\,m_{h_V}$ and $m_V$ as the appropriate replacements.
We further consider the following hierarchy among the various couplings: 
$\lambda_X,\,\lambda_V \gg y_F \gg g_X,\,g_V$. Such choice implies
$m_{h_X},m_{h_V} \gg m_{F} \gg m_X,\,m_V$ for $\lambda_X,g_X\sim \lambda_V,\,g_V$
with $v_X\sim v_V$.
Hence, at an energy scale $E\sim m_F \ll m_{h_X},\,m_{h_V}$, the new scalars
$h_X,\,h_V$ are nearly decoupled from the theory given that 
$v_X,\,v_V$ are reasonably high. In this limit one can use the following
approximations $\phi_X \simeq v_X e^{i\theta_X/v_X}\simeq v_X+i\theta_X$ and 
$\phi_V \simeq v_V e^{i\theta_V/v_V}\simeq v_V+i\theta_V$ which help
to recast Eq.~({\ref{eq:bsmfm}}) as:
%
\begin{equation}
\mathcal{L}^{\text{fermions}} \supset - i y_F \theta_V \bar{\psi}_1 \gamma_5 \psi_1 + i y_F \theta_V \bar{\psi}_2 \gamma_5 \psi_2 - i y_F \theta_X \bar{\chi}_1 \gamma_5 \chi_1 + i y_F \theta_X \bar{\chi}_2 \gamma_5 \chi_2.
\end{equation}

\begin{table}[t]
\begin{center}
\begin{tabular}{|c||c|c|c|c||c|c|c|c|}
\hline 
 & $\psi_{1L}$ & $\psi_{1R}$ & $\psi_{2L}$ & $\psi_{2R}$ & $\chi_{1L}$ & $\chi_{1R}$ & $\chi_{2L}$ & $\chi_{2R}$ \\ 
\hline 
\hline
$U(1)_X$ & $e_1$ & $e_1$ & $e_2$ & $e_2$ & $e_4$ & $e_3$ & $e_3$ & $e_4$ \\ 
\hline 
$U(1)_V$ & $q_1$ & $-q_1$ & $-q_1$ & $q_1$ & $q_2$ & $q_2$ & $-q_2$ & $-q_2$ \\ 
\hline 
\end{tabular} 
\end{center}
\caption{Charge assignments of the left- and right- chiral components of 
the BSM fermions which belong to gauge group $U(1)_{X}\times U(1)_V$.}
\label{tab:charges}
\end{table}

The structure of Yukawa couplings of the new fermions,
as well as 
the requirement of anomaly cancellations, restrict the possible charge assignments 
under the BSM $U(1)$ groups. 
A set of assignations complying with these two requirements is reported in 
Table~\ref{tab:charges}. 
As evident from Table~\ref{tab:charges} that this kind of charge assignations 
allows natural cancellations of the $U(1)_{V,X}^3$ 
anomalies, independently for the $\psi$ and $\chi$ sectors, irrespective of the values 
of $q_i,e_i$ charges\footnote{One could see from Table~\ref{tab:charges} that the sum
of all left- and right-chiral charges for $\psi$ and $\chi$
vanishes for $U(1)_V$ while for $U(1)_X$ it is $2\sum\limits^4_{i=1}e_i$.
If we also set this sum to be zero, as expected from the requirement
of gauge-gravity anomaly cancellation for an Abelian group, we can
recast Eq.~(\ref{eq:anorel}) as $q_2=\frac{q_1(e_2-e_1)}{e_3-e_4}$.}.
The cancellation of the mixed anomalies, e.g.,
$U(1)_V \, U(1)^2_X$, is instead achieved in a non-trivial way. 
This, indeed, requires the following relation between the charges:
%
\begin{equation}
\label{eq:anorel}
q_2=\frac{q_1 (e_1^2-e_2^2)}{ (e_3^2-e_4^2)}.
\end{equation}

Further, the charge assignments of Table~\ref{tab:charges} 
predicts a vanishing kinetic mixing between $X$ and $V$ as 
\begin{equation}
\label{eq:XVkinmix}
\sum\limits^{i=1,\,2}_{\xi=\psi_i,\,\chi_i} 
c^X_{\xi_L} c^V_{\xi_L}+ c^X_{\xi_R} c^V_{\xi_R}=0,
\end{equation}
with $c^{X(V)}_{\xi_{L(R)}}$ representing appropriate charges
shown in the Table~\ref{tab:charges}. This, as discussed already, is crucial since
a mixing between $X$ and $V$, in the presence
of a kinetic mixing between $V,\,Z$, can trigger
a subsequent mixing between $X$ and $Z$, such that
$X$ can decay into the SM fermions and thereby, 
the stability of the DM gets spoiled.

We show in detail later in appendix \ref{sec:appendixC} that when Eq.~(\ref{eq:anorel})
is satisfied, it appears feasible to construct an anomaly
free theory where the following effective operator emerges
after integrating out heavy fermionic degrees of freedom
from the triangular loops:
%
\begin{equation}
\label{eq:sasaX}
\epsilon^{\mu \nu \rho \sigma} D_\mu \theta_X D_\nu \theta_V X_{\rho \sigma},
\end{equation}
where $\theta_X,\,\theta_V$ are Stueckelberg axions
of the $U(1)_X,\,U(1)_V$ groups and $D_\mu\theta_X = \partial_\mu \theta_X-g_X q_X v_X X_\mu$, and $D_\nu\theta_V = \partial_\nu \theta_V-g_V q_V v_V \wt V_\nu$ with 
$X_\mu,\,\wt V_\nu$ as gauge bosons of the concerned
$U(1)_X,\, U(1)_V$ groups, respectively. Eq.~(\ref{eq:sasaX}) is invariant under the following gauge transformations:
\begin{equation}
 X_\mu \rightarrow X_\mu + \partial_\mu \alpha_X, \,\,\, \wt V_\mu \rightarrow \wt V_\mu + \partial_\mu \alpha_V, \,\,\, \theta_X \rightarrow \theta_X + g_X q_X v_X \alpha_X, \,\,\,
 \theta_V \rightarrow \theta_V + g_V q_V v_V \alpha_V,
\end{equation}
with $\alpha_X$ and $\alpha_V$ as the transformation parameters. The same equation, after considering unitary gauge, leads to the following operator
as introduced earlier in Eq.~(\ref{eq:starting_lagrangian}):
\begin{equation}
\mathcal{L} = \alpha_{\text{CS}} \epsilon^{\mu \nu \rho \sigma} X_\mu \wt V_\nu X_{\rho \sigma}, 
\end{equation}
with
\begin{equation}
\label{eq:alphaCSwork}
\alpha_{\text{CS}} \equiv \frac{q_1 \left(e_2^2-e_1^2\right)}{8 \pi ^2}.
\end{equation}

One can define an effective charge $\wt{Q}^3\equiv q_1(e_2^2-e_1^2)$ 
to get a simple relation:
%
\begin{equation}
\alpha_{\rm CS} =\dfrac{\wt{Q}^3}{8\pi^2}.
\end{equation}

Clearly $\alpha_{\rm CS}$ $\sim \mathcal{O}$ $[ 10^{-2},1]$
(see Eq.~(\ref{eq:scan1})) or $\sim\mathcal{O}$ $[ 10^{-3},1]$ 
(see Eq.~(\ref{eq:scan2})) corresponds to a $\sim \mathcal{O}(1)$ value of the effective charge $\wt Q$ for one generation of the BSM fermions.

Note that the CS coupling $\alpha_{\rm CS}$ has no explicit dependence
on the BSM fermion mass, i.e., it seems to remain finite as
$m_F$, the relevant heavy fermion mass, $\longrightarrow \infty$ and thereby, 
resembles a non-decoupling effect. This is a consequence of the assumption
$\lambda_{X},\,\lambda_{V} \gg y_F \gg g_X,\,g_V$, as considered
earlier, which makes $\alpha_{\rm CS}$ independent of $m_F$
as long as $m_F \gg m_X,\,m_{Z'}$ such that the adopted effective
approach remains justified for an energy scale $E$ below 
the mass of the ``lightest'' BSM fermion of the theory.
The parameter $\beta_{\rm CS}$ (see Eq.~(\ref{eq:twoCSLorigin})),
on the contrary, vanishes as the associated BSM fermion masses
$\longrightarrow \infty$, as expected according to the decoupling effect.

One should further note that as the effective charge $\wt Q$ includes the gauge couplings in its definition,
$\wt{Q}\sim \mathcal{O}(1)$ implies either $g_X\sim g_V\sim \mathcal{O}(1)$ or a large multiplicity of the BSM 
fermions having gauge charges $\sim \mathcal{O}(1)$.
However, as mentioned  previously, the aforementioned
theoretical construction relies on the assumption of $\lambda_{X},\,\lambda_{V} \gg g_X,\,g_V$ which,
for $g_X,\,g_V\sim \mathcal{O}(1)$,
hints towards a strongly coupled theory. In this regime one would encounter several
theoretical issues like the vacuum instability, etc. which might spoil viability of the
effective approach. However, from the view point of a radiative origin, 
$\alpha_{\rm CS} \sim \mathcal{O}(1)$ is unnatural.

The kinetic mixing parameter $\delta$ (see Eq.~(\ref{eq:starting_lagrangian})), as already stated 
in the beginning of this section, can get generated at the loop level
from two sets of the BSM fermions (preferably vector-like to avoid new anomalies in a trivial way)
charged under the SM $U(1)_Y$ and BSM $U(1)_V$ groups and having masses
$m$ and $M$, respectively. The parameter $\delta$ is then estimated
as $\delta \simeq (q_Y g_Y q_V g_V/16\pi^2)\times {\rm log}(m/M)$ \cite{Holdom:1985ag} with 
$q_Y,\,q_V,\,g_Y,\,g_V$ as the relevant combination of gauge charges and gauge
couplings of the associated gauge groups. Assuming these gauge charges
and couplings, as well as log$(m/M)$, to be $\sim \mathcal{O}(1)$, 
one would expect natural range of $\delta$ as $\sim \mathcal{O}(10^{-3}-10^{-2})$.
This range, as evident from Eq.~(\ref{eq:alphaCSwork}) and Eq.~(\ref{eq:effctivealphaCS}), 
is almost the same as of $\alpha_{\rm CS}$, considering
$\sim \mathcal{O}(1)$ values of the involved gauge charges. Both
these natural ranges of CS coupling $\alpha_{\rm CS}$ and 
kinetic mixing parameter $\delta$ are connected with their
radiative origins.

\section{Conclusion}

In this chapter, we discussed the theoretical framework related to CS couplings and we scrutinized experimental viability and theoretical consistency of the WIMP
DM models comprised of an Abelian vectorial DM $X_\mu$ and an Abelian $Z'$ portal, coupled
through a CS interaction.
Regarding the DM phenomenologies we investigated the detection
prospects in the light of accommodating the correct relic density
and sensitivity reaches of the various existing as well as
anticipated upcoming DD and ID experiments. Concerning collider probes we examined
the observational aspects of these models from the view point
of dijet, dilepton resonances and mono-{\bf X} searches
using the 13 TeV LHC data. Further, we also studied the 
viable ranges of the associated parameters focusing on
the possible theoretical and/or ``well-measured'' experimental constraints,
mainly for the kinetic mixing scenario, arising from the 
EWPT, $\rho$-parameter, $Z$-mass, total and invisible $Z$-decay widths, etc.
Finally, we also explored possible origins of a kinetic mixing term
and a CS interaction term, arising via a set of heavy BSM fermions running
in the triangle loops, from the standpoint of an UV complete
theory.
A radiative origin for CS coupling $\alpha_{\rm CS}$,
from the perspective of an UV complete construction, predicts
a natural range for $\alpha_{\rm CS}$ as $\lesssim \mathcal{O}(10^{-3})$
whereas discovery/exclusion prospects, with the existing and 
near future experimental setups, favour $\alpha_{\rm CS}\sim \mathcal{O}(1)$.
Such $\alpha_{\rm CS}$ values, along with a $\delta$ of similar order,
can accommodate the correct relic density rather easily and can be probed/excluded
from DD, ID and collider (via mono-{\bf X}) searches.
An $\mathcal{O}(1)$ value of $\alpha_{\rm CS}$, just like 
the kinetic mixing parameter $\delta$, is hard to explain 
with a radiative origin unless more families of the BSM fermions
are included. Any such non-minimal constructions, i.e., large 
number of BSM fermions to increase $\delta$ and/or
$\alpha_{\rm CS}$ value(s) or a multi-component
DM to account for the correct relic density with ``natural'' $\delta$
values would reduce the model predictivity. 

Experimental attainments of the second case study with two 
CS couplings are more contrived due to the absence of a tree-level
mixing between the $Z'$ and the SM fermions, unlike the first
case study with one CS coupling and a kinetic mixing.
Missing tree-level couplings between $Z'$ and 
the SM fermions conceal this framework from constraints
like the EWPT, $\rho$-parameter, precision $Z$ physics, etc.
which offer notable effects on the model parameter space for scenario-I.
In this framework, DD prospects are missing at the tree-level
and ID sensitivities remain orders of magnitude below
the ongoing and upcoming experimental reaches. Further, by construction, in general one expects $\alpha_{\rm CS} > \beta_{\rm CS}$
which indicates natural range of $\beta_{\rm CS}$ in the ballpark of 
$10^{-4}$. Thus, this scenario remains practically hidden 
from the collider searches, even considering the high-luminosity LHC
or a $100$ TeV proton-proton collider. The trick of pushing $\beta_{\rm CS}$
values upwards by adding more 
BSM fermions is rather intricate compared to the kinetic mixing
scenario.

%% file: parts/flavor.tex
\section{Introduction}
\label{sec:intro}

Recently, the phenomenological motivation for considering non-universal $Z'$ models has increased due to mounting evidence for semi-leptonic $B$ decays whose rates and differential distributions are inconsistent with those predicted by the Standard Model~\cite{Descotes-Genon:2013wba,Altmannshofer:2013foa,Ghosh:2014awa}. In this chapter we investigate the phenomenology of such a $Z^\prime$ playing the role of mediator between the Standard Model and a Dark Matter candidate.
Over the past years, the LHCb Collaboration has reported a number of deviations from $\mu$-$e$ universality  in  $B\rightarrow K^{(*)}\ell^+\ell^-$  decays. Since Flavour-Changing Neutral Current (FCNC) processes are forbidden in the Standard Model at tree level, such decays have to occur through loop induced processes, causing the precise value of these decay rates to be extremely sensitive to contributions from beyond-the-standard-model physics featuring tree-level FCNC interactions. The ratios of $\mu^+ \mu^-$ to $e^+ e^-$ final states, $R_K$ and $R_{K^*}$, defined as
\begin{equation}
R_{K^{(*)}}\equiv\dfrac{\int_{q_{\rm min}^2}^{q_{\rm max}^2} \dfrac{\diff \Gamma(B\rightarrow K^{(*)} \mu^+ \mu^-)}{\diff q^2}\diff q^2}{\int_{q_{\rm min}^2}^{q_{\rm max}^2} \dfrac{\diff \Gamma(B\rightarrow K^{(*)} e^+ e^-)}{\diff q^2}\diff q^2}~, 
\end{equation}
have the advantage that some hadronic uncertainties cancel out rendering the theoretical computations of these ratios sufficiently reliable to probe deviations from the Standard Model predictions. In order to reduce contamination from $J/\Psi$ resonance $\sim 9~\text{GeV}^2$, these ratios have been measured in the energy bins $q^2\in [1, 6 ] ~\text{GeV}^2$ for $R_K$~\cite{Aaij:2014ora} and $q^2\in [0.045, 6 ] ~\text{GeV}^2$ for $R_{K^{*}}$~\cite{Aaij:2017vbb} to be about $70\%$ of their expected values:
\begin{align}
R_{K^{*}}^{[0.045,1.1]}&=0.66^{+0.113}_{-0.074} ~, \quad  R_{K^{*}}^{[1.1,6]}=0.685^{+0.122}_{-0.083} ~, \quad R_K^{[1,6]}=0.745^{+0.097}_{-0.082} ~,
\end{align}
each displaying a $\sim 2.5\sigma$ deviation from the SM prediction.  
Combining that with the input from other  $b\rightarrow s \ell^+ \ell^-$ processes, the SM is disfavored by $4$ to $5$ standard deviations~\cite{Capdevila:2017bsm,Altmannshofer:2017yso}. The $R_K$ and $R_{K^*}$ anomalies could be the first evidence of New Physics (NP). The relevant Hamiltonian describing a NP contribution to $R_{K^{(*)}}$, providing an effective approach holds, can be parametrized in a model-independent way as~\cite{Altmannshofer:2017yso}: 
  
\begin{equation}
\mathcal{H}_{\rm eff}=-\dfrac{4G_F}{\sqrt{2}}V_{tb} V_{ts}^* \sum_{i,\ell}\Big( C_{i \ell}^{\rm NP}(\mu) \mathcal{O}_i(\mu)+C_{i \ell}^{\prime \rm NP}(\mu) \mathcal{O}_i^\prime(\mu) \Big)+\text{h.c.}~, 
\end{equation} 
where $\mathcal{O}^{(\prime)}_i(\mu)$ are a set of effective operators defined at a scale $\mu$, including relevant lepton-universality violation four-fermion contact interactions:
\begin{align}
\mathcal{O  \ell}_9^{(\prime)}=(\bar{s}\gamma_\mu P_{L (R)} b)(\bar{\ell}\gamma^\mu \ell), \qquad
\mathcal{O}_{10  \ell}^{(\prime)}=(\bar{s}\gamma_\mu P_{L (R)} b)(\bar{\ell}\gamma^\mu\gamma_5 \ell)~.
\end{align}
The corresponding Wilson coefficients are denoted $C_{i \ell}^{(\prime) \rm NP}$. 
A number of recent phenomenological analyses, 
see e.g. \cite{Capdevila:2017bsm,Altmannshofer:2017yso,Ciuchini:2017mik,Hiller:2017bzc,Geng:2017svp,Ghosh:2017ber,Bardhan:2017xcc,DAmico:2017mtc,Alok:2017sui}, conclude that these data can be well fit  
when the low-energy Lagrangian below the weak scale contains a new physics operator of the $C^{\text{NP}}_{9\mu}=-C^{\text{NP}}_{10\mu}$ form:  
\begin{equation}
\Delta \mathcal{L}_{\text{eff}} \supset 
G_{b s\mu} (\bar{b}_L \gamma^\mu s_L) (\bar{\mu}_L \gamma_\mu \mu_L) + {\rm h.c.}, 
\qquad G_{bs\mu} \sim \frac{1}{(30\text{ TeV})^2}. 
\label{eq:c9mc10}
\end{equation}

\begin{figure}[h!]
\includegraphics[width=0.49\linewidth]{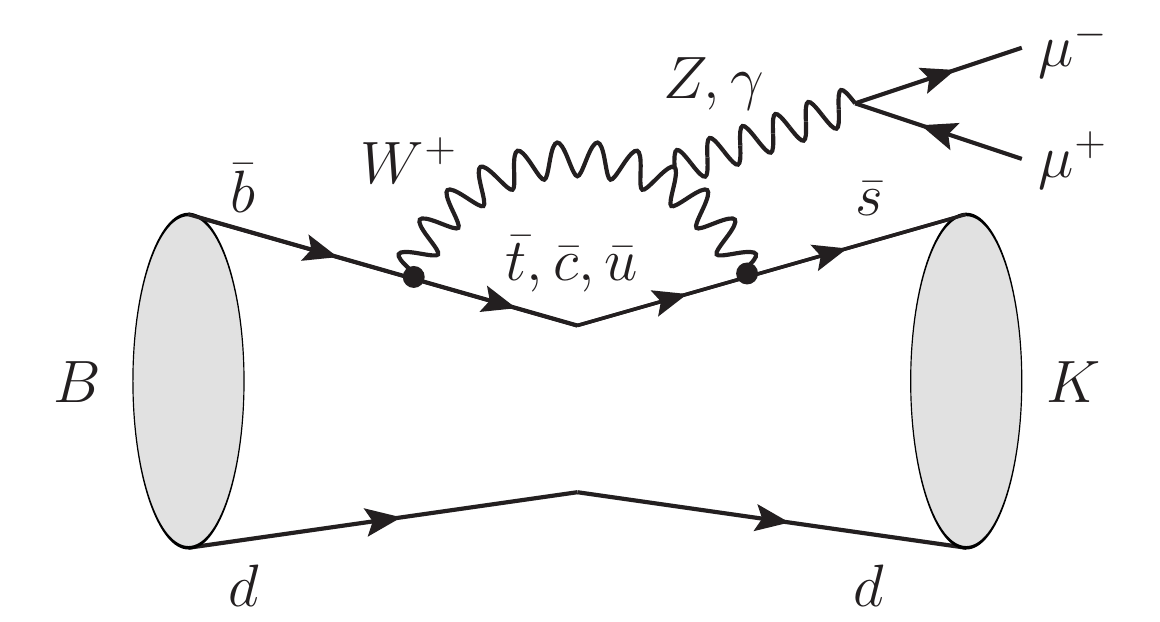}
\includegraphics[width=0.49\linewidth ]{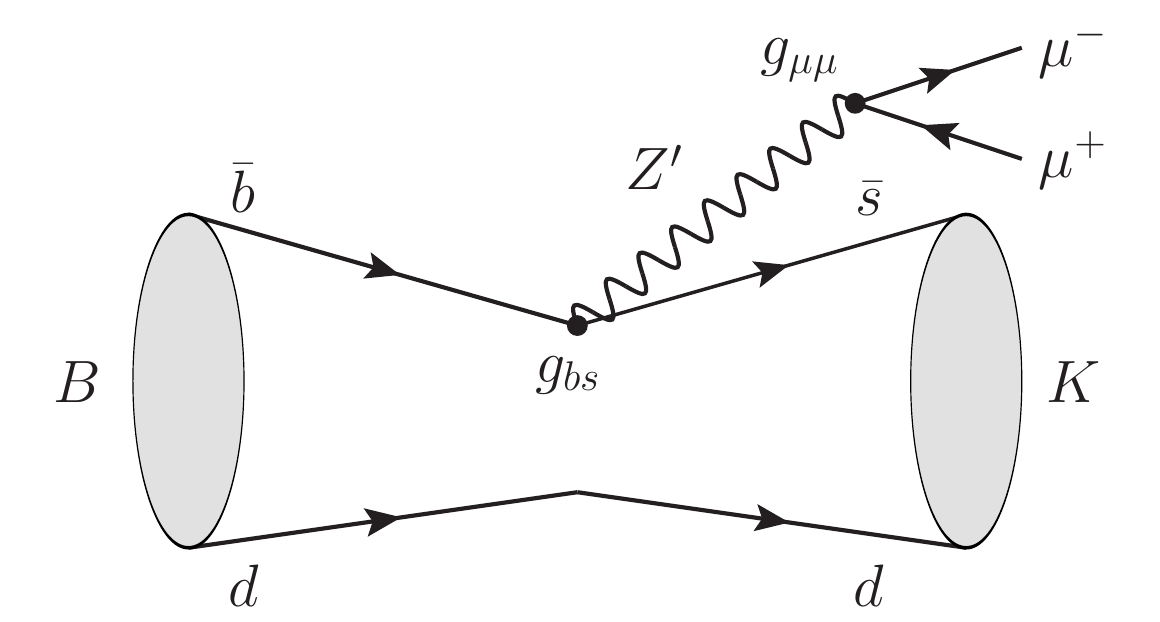}
\caption{\textbf{Left:} One of the diagrams responsible for the $B-$meson decay in the Standard Model. \textbf{Right:} Diagram responsible for the $R_{K^{(*)}}$ discrepancy in a new physics model involving a $Z^\prime$ with flavour changing neutral current at tree-level.}
\label{fig:diagramsRK}
\end{figure}

In a flavourful $Z'$ model, the new physics operator in Eq.~(\ref{eq:c9mc10}) will arise from tree-level $Z'$ exchange: $G_{b s\mu} = -g_{bs} g_{\mu\mu}/M_{Z^\prime}^2$, where 
$g_{bs}$ is the flavour-violating $Z'$ coupling to left-handed b- and s-quarks  and $g_{\mu \mu}$ is the couplings to left-handed muons, as represented on the right pannel of Fig.~\ref{fig:diagramsRK}.  
There is already a vast literature discussing the $Z'$ explanation of the B-anomalies and phenomenological constraints on the parameter space of such models, see e.g.~\cite{Gauld:2013qba,Buras:2013qja,Altmannshofer:2014cfa,Crivellin:2015mga,Crivellin:2015lwa,Niehoff:2015bfa,Celis:2015ara,Greljo:2015mma,Niehoff:2015iaa,Altmannshofer:2015mqa,Falkowski:2015zwa,Carmona:2015ena,GarciaGarcia:2016nvr,Megias:2016bde,Chiang:2016qov,Altmannshofer:2016oaq,Boucenna:2016qad,Foldenauer:2016rpi,Kamenik:2017tnu,Chivukula:2017qsi,Faisel:2017glo,Ellis:2017nrp,Alonso:2017uky,Carmona:2017fsn,Dalchenko:2017shg,Raby:2017igl,Bian:2017rpg,Bian:2017xzg,Alok:2017jgr,Fox:2018ldq,Chala:2018igk}.
In realistic models of this kind, the coupling  $g_{bs}$ is strongly constrained by precision measurements of the $B_s$ meson mass difference.
Taking that into account, one can derive the constraint $M_{Z'} \lesssim 1.2 g_{\mu \mu}$~TeV, implying that $M_{Z'}$ must be close to the weak scale in weakly coupled models. 
The corollary is that the $Z'$ is in the correct mass range to act as mediator between the SM and  thermally produced Dark Matter~\cite{Sierra:2015fma,Belanger:2015nma,Celis:2016ayl,Altmannshofer:2016jzy,Baek:2017sew,Cline:2017qqu,Fuyuto:2017sys}. 
In this chapter we further pursue this direction, and discuss a $Z'$ model that can account for the B-anomalies and, simultaneously, explain the observed relic abundance via a weakly interacting massive particle (WIMP) communicating with the SM through the same $Z'$.

We follow Ref.~\cite{King:2017anf},  
which introduces a fourth vector-like family with non-universal gauged $U(1)'$ charges.
The idea is that the $Z'$ couples universally to the three chiral families, which then mix with the non-universal fourth family to induce effective non-universal couplings in the physical light mixed quarks and leptons. 
Such a mechanism has wide applicability, for example it was recently discussed in the context of 
F-theory models with non-universal gauginos \cite{Romao:2017qnu}.
Two explicit examples were discussed in \cite{King:2017anf}. 
Firstly an $SO(10)\rightarrow SU(5)\times U(1)_X$  model, where we identified $U(1)'\equiv U(1)_X$,
which however was subsequently shown to be not consistent with both explaining $R_{K^*}$ and respecting the $B_s$ mass difference \cite{Antusch:2017tud}.
Ref.~\cite{King:2017anf} also  discussed a fermiophobic model where the gauged $U(1)'$ charges are not carried by the three chiral families, only by fourth vector-like family.
In the absence of mixing, the $Z'$ is fermiophobic, having no couplings to the three chiral families, but does couple to a fourth vector-like family. 
Due to mixing effects, we shall suppose that the $Z'$ gets induced couplings to second family left-handed lepton doublets (containing the left-handed muon and its neutrino) and third family left-handed quark doublets (containing the left-handed top and bottom quarks). 
Including only such couplings is enough to address the B-anomalies, in analogy to related scenarios where new  vector-like fermions mix with the SM ones~\cite{Niehoff:2015bfa,Niehoff:2015iaa,Carmona:2015ena,Boucenna:2016qad,Megias:2016bde,Carmona:2017fsn,Raby:2017igl,Chala:2018igk}. 
In addition, this set-up  provides a natural WIMP Dark Matter candidate: the neutrino residing in the fourth family. 
We  are interested in the parameter space of this model where both B-anomalies and the relic abundance of Dark Matter are simultaneously explained. 
We show that this can be achieved without conflicting a myriad of direct and indirect Dark Matter constraints as well as experimental constraints such as $B_s$ mixing, LHC searches, neutrino trident, and so on. 
The requirement to satisfy all these constraints in a natural way  points to a specific corner of the parameter space, with $300~{\rm GeV} \lesssim m_{Z'} \lesssim 1$~TeV, Dark Matter heavier than a TeV, and a narrow range of possible $Z'$ couplings. 

In Sec.~\ref{sec:model} we define the gauged $U(1)'$ model with a vector-like fourth family.
The $Z'$ couplings relevant for the subsequent analysis are summarized in Eq.~(\ref{eq:Zp_Rk_couplings}) . 
In Sec.~\ref{constraints} we discuss the constraints these parameters need to satisfy in order to address the B-anomalies without conflicting other experimental results. 
In Sec.~\ref{darkmatter} we turn to the Dark Matter sector, and identify the masses and  couplings of the vector-like fourth family singlet Dirac neutrino which lead to a correct relic density, while evading all indirect and direct searches so far. 
The main results are contained in Sec.~\ref{conclusion}, where we put together the requirements imposed by the B-anomalies and by the relic density, and identify the viable parameter space where both are satisfied.  

\section{The model}
\label{sec:model}
We consider a model in which, in addition to the SM with the usual three chiral families of left-handed quarks and leptons, including the right-handed neutrinos, we add a dark $U(1)^\prime$ gauge symmetry and a fourth vector-like family of fermions. The idea is to have the SM quarks and leptons neutral under the  $U(1)^\prime$ while the vector-like family has the SM quantum numbers and is charged under the  $U(1)^\prime$, leading to a Dark Matter candidate and flavour-changing $Z^\prime$ operators after the vector-like fermion mass term mix with the SM fermions. 

Table~\ref{tab:model} shows all the particle content and their corresponding representations and charges. The non-universal $U(1)^\prime$ charges forbid mixing between the fourth family and the chiral families via the usual Higgs Yukawa couplings. Therefore, we need to add new singlet scalars, with appropriate $U(1)^\prime$ charges, to generate mass mixing of quarks and leptons with the vector-like family. The $U(1)^\prime$ is broken by the VEVs of the new Higgs singlets $\phi_\psi$ to yield a massive $Z^\prime$.
\begin{table}[h!]
		\centering
		\begin{tabular}[t]{| c | c c c c|}
			\hline
			\multirow{2}{*}{\rule{0pt}{4ex}Field}	& \multicolumn{4}{c |}{Representation/charge} \\
			\cline{2-5}
			\rule{0pt}{3ex}			& $SU(3)_c$ & $SU(2)_L$ &  $ U(1)_Y$ & $ U(1)^\prime $  \\ [0.75ex]
			\hline \hline
			\rule{0pt}{3ex}%
			$Q_{Li}$ & $\bf{3}$ &$\bf{2}$   &$1/6$ & $ 0 $ \\
			$u_{Ri}$ & $\bf{3}$ &$\bf{1}$   &$2/3$ & $ 0 $ \\
			$d_{Ri}$ & $\bf{3}$ &$\bf{1}$   &$-1/3$\phantom{+} & $ 0 $ \\
			$L_{Li}$ & $\bf{1}$ &$\bf{2}$   &$-1/2$\phantom{+} & $ 0 $ \\
			$e_{Ri}$ & $\bf{1}$ &$\bf{1}$   &$-1$\phantom{+} & $ 0 $ \\
			$\nu_{Ri}$ & $\bf{1}$ &$\bf{1}$   &$0$ & $ 0 $ \\
			\hline
			\hline
			$H$ & $\bf{1}$ &$\bf{2}$   &$1/2$ & $ 0 $ \\
			\hline
			\hline 
			$Q_{L4},\tilde{Q}_{R4} $ & $\bf{3}$ &$\bf{2}$   &$1/6$ & $ q_{Q4} $ \\
			$u_{R4}, \tilde{u}_{L4}$ & $\bf{3}$ &$\bf{1}$   &$2/3$ & $ q_{u4} $ \\
			$d_{R4}, \tilde{d}_{L4}$ & $\bf{3}$ &$\bf{1}$   &$-1/3$\phantom{+} & $ q_{d4} $ \\
			$L_{L4}, \tilde{L}_{R4}$ & $\bf{1}$ &$\bf{2}$   &$-1/2$ \phantom{+}& $ q_{L4} $ \\
			$e_{R4}, \tilde{e}_{L4}$ & $\bf{1}$ &$\bf{1}$   &$-1$\phantom{+} & $ q_{e4} $ \\
			$\nu_{R4}, \tilde{\nu}_{L4}$ & $\bf{1}$ &$\bf{1}$   &$0$ & $ q_{\nu4} $ \\
			\hline
			\hline
				$\phi_{Q,u,d,L,e}$ & $\bf{1}$ &$\bf{1}$   &$0$ & $ -q_{Q_4,u_4,d_4,L_4,e_4} $ \\
				\hline
		\end{tabular}
		\caption{The model consists of the usual three chiral families of quarks and leptons $\psi_i$ $(i=1,2,3)$, including the right-handed neutrino, a Higgs doublet $H$, plus a fourth vector-like family of fermions $\psi_4, \tilde{\psi}_4$ and new Higgs singlets $\phi_\psi$ which mix fourth family fermions with the three chiral families. Note that we exclude $\phi_\nu$ so that $\nu_{R4}, \tilde{\nu}_{L4}$ do not mix and are stable.}
		\label{tab:model}
	\end{table}
    
The Higgs Yukawa couplings of the first three chiral families can be written in a $4\times 4$ matrix notation
\begin{equation}
\mathcal{L}^{\text{Yukawa}}= y^u \bar{Q}_{L} \tilde{H} u_{R}+ y^d \bar{Q}_{L} H  d_{R}+y^e  \bar{L}_{L} H e_{R}  +y^\nu \bar{L}_{L}  \tilde{H} \nu_{R} +\text{h.c.}~,
\label{eq:yuk1}
\end{equation}
where $\tilde{H}=i\sigma_2 H^*$ and $y^u,$ $y^d$, $y^e$, $y^\nu$ are $4\times 4 $ matrices with the fourth row and columns consisting of all zeros, since the fourth family does not couple to the Higgs doublets. The $U(1)^\prime$ charges allow Yukawa couplings between the singlet fields $\phi$, the fourth family $\tilde{\psi}_4 $ and the first three chiral families $\psi_i$. Furthermore, there is an explicit mass term between the opposite chirality fourth family fields $\psi_4$ and $\tilde{\psi}_4$,
\begin{equation}
\begin{split}
\mathcal{L}^{\text{mass}}&=x_i^Q \phi_Q \bar{Q}_{Li}\tilde{Q}_{R4}+x_i^u \phi_u \bar{\tilde{u}}_{L4}u_{Ri}+x_i^d \phi_d \bar{\tilde{d}}_{L4}d_{Ri}+x_i^L \phi_L \bar{L}_{Li}\tilde{L}_{R4}+x_i^e \phi_e \bar{\tilde{e}}_{L4}e_{Ri} \\
 &+ M_4^Q \bar{Q}_{L4}\tilde{Q}_{R4}+ M_4^u \bar{\tilde{u}}_{L4}u_{R4} + M_4^d \bar{\tilde{d}}_{L4}d_{R4} +M_4^L \bar{L}_{L4}\tilde{L}_{R4}+M_4^e \bar{\tilde{e}}_{L4}e_{R4} \\
 &+M_4^\nu \bar{\tilde{\nu}}_{L4}\nu_{R4} 
+\text{h.c.}~, 
\end{split}
\end{equation}
where $i=1,...,3$. 

The fourth-family vector-like singlet neutrinos 
$\nu_{R4},\tilde{\nu}_{L4}$ 
are special since we don't have a singlet field $\phi_\nu$ that couples them to the other families, which is why such terms are absent in the above equation. This implies that $\nu_{R4},\tilde{\nu}_{L4}$ are absolutely stable, with their stability guaranteed by an unbroken global $U(1)_{\nu_{R4}}$ and, since they do not carry any Standard Model quantum numbers, they may play the role of Dark Matter.  Note that we also impose lepton number conservation $U(1)_L$ for all four families of leptons which forbids Majorana mass terms. Hence
all neutrinos (including those in the fourth vector-like family) will have purely Dirac masses.\footnote{Alternatively it is possible to introduce various seesaw mechanisms into this kind of model, leading to Majorana masses, as recently discussed \cite{Antusch:2017tud}. However in this work we only consider Dirac neutrinos.}

After the singlet scalar fields $\phi$ obtain a non-zero vacuum expectation value (VEV), we may rewrite the Lagrangian in terms of new mass parameters $M_i^Q=x_i^Q\left<\phi_Q\right>$, similarly for the other mass parameters, such that
\begin{equation}
\begin{split}
\mathcal{L}^{\text{mass}}&=M_\alpha^Q \bar{Q}_{L\alpha}\tilde{Q}_{R4}+ M_\alpha^u \bar{\tilde{u}}_{L4}u_{R\alpha} + M_\alpha^d \bar{\tilde{d}}_{L4}d_{R\alpha} +M_\alpha^L \bar{L}_{L\alpha}\tilde{L}_{R4}+M_\alpha^e \bar{\tilde{e}}_{L4}e_{R\alpha} \\
 &+M_4^\nu \bar{\tilde{\nu}}_{L4}\nu_{R4} 
+\text{h.c.}~, 
\end{split}
\end{equation}
where $\alpha=1,...,4$. We may diagonalize the mass matrix before electroweak symmetry breaking, when only the fourth family is massive 
\begin{equation}
\begin{split}
\mathcal{L}^{\text{mass}}&=\tilde{M}_4^Q \bar{Q'}_{L4}\tilde{Q}_{R4}+ \tilde{M}_4^u \bar{\tilde{u}}_{L4}u'_{R4} + \tilde{M}_4^d \bar{\tilde{d}}_{L4}d'_{R4} +\tilde{M}_4^L \bar{L'}_{L4}\tilde{L}_{R4}+\tilde{M}_4^e \bar{\tilde{e}}_{L4}e'_{R4} \\
 &+M_4^\nu \bar{\tilde{\nu}}_{L4}\nu_{R4} 
+\text{h.c.}~.
\end{split}
\end{equation}
The prime states for the heavy mass basis where only the fourth family has explicit vector-like Dirac mass terms and it's related to the original charge basis by unitary mixing matrices, 
\begin{equation}
Q'_L=V_{Q_L}Q_L, \quad u'_R=V_{u_R}u_R, \quad d'_R=V_{d_R}d_R, \quad L'_L=V_{L_L}L_L, \quad e'_R=V_{e_R}e_R,
\label{eq:transf_heavybasis}
\end{equation}
while for the neutrino states $\tilde{\nu}_{L4}$ and $\nu_{R4}$ the original and the mass basis coincides. In this basis, the Yukawa couplings in Eq.~(\ref{eq:yuk1}) become
\begin{equation}
\mathcal{L}^{\text{Yukawa}}= y^{\prime u} \bar{Q'}_{L} \tilde{H} u'_{R}+  y^{\prime d} \bar{Q'}_{L} H d'_{R}+y^{\prime e} \bar{L'}_{L} H  e'_{R} + y^{\prime \nu} \bar{L'}_{L} \tilde{H} \nu_{R}+ \text{h.c.}~, 
\label{eq:yuk2}
\end{equation}
where
\begin{equation}
y^{\prime u}=V_{Q_L}y^u V^\dagger_{u_R}, \quad y^{\prime d}=V_{Q_L}y^d V^\dagger_{d_R}, \quad y^{\prime e}=V_{L_L}y^e V^\dagger_{e_R} \quad y^{\prime \nu}=V_{L_L}y^\nu.
\end{equation}
This shows that there is a coupling between the heavy fourth family and the Higgs due to their mixing with the first three chiral families. However, this coupling will be small since the original  $y^u,$ $y^d$, $y^e$, $y^\nu$ contain zeroes in the fourth row and column and they are mixing suppressed. Therefore, we can integrate out the fourth family and look at the low energy effective theory by simply removing the fourth rows and columns of the primed Yukawa matrices in Eq.~(\ref{eq:yuk2}). The three massless families, below the heavy mass scale, are described by 
\begin{equation}
\mathcal{L}^{\text{Yukawa}}_{\text{light}}= y^{\prime u}_{ij} \bar{Q'}_{Li} \tilde{H} u'_{Rj}+ y^{\prime d}_{ij} \bar{Q'}_{Li}H d'_{Rj}+y^{\prime e}_{ij} \bar{L'}_{Li} H  e'_{Rj}+y^{\prime \nu}_{ij} \bar{L'}_{Li} \tilde{H}  \nu_{Rj} + \text{h.c.}~, 
\label{eq:yuk_light}
\end{equation}
where 
\begin{align}
y^{\prime u}_{ij}=(V_{Q_L}y^u V^\dagger_{u_R})_{ij}, \quad y^{\prime d}_{ij}=(V_{Q_L}y^d V^\dagger_{d_R})_{ij}, \quad y^{\prime e}_{ij}=(V_{L_L}y^e V^\dagger_{e_R})_{ij}, \quad y^{\prime \nu}_{ij}=(V_{L_L}y^\nu )_{ij}~,
\end{align}
and $i,j=1,...,3$. The Yukawa matrices for the quarks and charged leptons can be now diagonalized 
\begin{align}
V'_{uL}y^{\prime u} V^{\prime \dagger}_{uR}=\text{diag}(y_u,y_c, y_t), ~ V^{\prime}_{dL}y^{\prime d} V^{\prime \dagger}_{dR}=\text{diag}(y_d,y_s,y_b), ~ V'_{eL}y^{\prime e} V^{\prime \dagger}_{eR}=\text{diag}(y_e,y_\mu,y_\tau).
\label{eq:diag_Yuk_mat}
\end{align}
The unitary CKM matrix is then given by
\begin{equation}
V_{\text{CKM}}=V'_{uL}V^{\prime \dagger}_{dL}.
\end{equation}
In the case of neutrinos, since we are forbidding Majorana masses, the light physical neutrinos have Dirac mass eigenvalues given by,
\begin{equation}
v V_{\nu L}^{\prime \nu} V^{\prime \dagger}_{\nu R}=\text{diag}(m_1, m_2, m_3).
\label{eq:nu_diag}
\end{equation}
The lepton mixing matrix or PMNS matrix can be constructed from the transformations in Eqs.~(\ref{eq:diag_Yuk_mat}) and~(\ref{eq:nu_diag})
\begin{equation}
V_{\text{PMNS}}=V'_{eL} V^{\prime \dagger}_{\nu L}.
\end{equation}
To look at the Lagrangian involving the SM gauge couplings, we emphasize that all the four families have the same charges under the SM. The unitary transformations in Eq.~(\ref{eq:transf_heavybasis}) cancel as in the usual GIM mechanism and the gauge couplings in the heavy mass basis remains the same as in the SM. After integrating out the fourth family and electroweak symmetry is broken, and the light Yukawa matrices are diagonalised, the couplings to the $W^\pm$ gauge bosons are 
\begin{align}
\mathcal{L}^{\text{int}}_W \supset \frac{g_2}{\sqrt{2}}\pmatr{\bar{u}_L  & \bar{c}_L  &\bar{t}_L }V_{\text{CKM}}W^+_\mu \gamma^\mu \pmatr{d_L \\ s_L \\ b_L}  
+ \frac{g_2}{\sqrt{2}}\pmatr{\bar{e}_L  & \bar{\mu}_L  &\bar{\tau}_L }V_{\text{PMNS}} W^+_\mu \gamma^\mu \pmatr{\nu_{1L} \\ \nu_{2L} \\ \nu_{3L}}~,
\end{align}
where $g_2$ is the usual $SU(2)_L$ gauge coupling. For the couplings to the $Z$ gauge boson, the same happens, the charges are the same for the fourth families and the transformations in Eq.~(\ref{eq:transf_heavybasis}) cancel, such that in the heavy mass basis, after electroweak symmetry breaking, we are left with
\begin{equation}
\mathcal{L}^{\text{int}}_Z=\frac{e}{2 s_W c_W}\bar{\psi}'_\alpha Z_\mu \gamma^\mu (C^\psi_V-C^\psi_A \gamma_5) \psi'_\alpha
\label{eq:Z_couplings}
\end{equation}
where
\begin{equation}
\psi'_\alpha=u'_\alpha, d'_\alpha, e'_\alpha, \nu'_\alpha \quad \alpha=1,...,4
\end{equation}
and
\begin{equation}
C^\psi_A=t_3, \quad C^\psi_V=t_3-2s^2_W Q.
\end{equation}
The electric charge of the fermions is denoted by $Q$ and $t_3$ are the eigenvalues of $\sigma_3/2$. The couplings to the $Z$ boson are flavour diagonal, even after diagonalization of the light fermion mass matrices, due to the unitary transformations cancelling. The interactions will be the same as in Eq.~(\ref{eq:Z_couplings}), replacing the fields $\psi'_\alpha$ by their three family mass eigenstates. 

In the case of the couplings to the $Z'$ gauge bosons, we have non-universal couplings that lead to flavour changing. In the original basis, after the $U(1)'$ symmetry is broken, we have diagonal gauge couplings between the massive $Z'$ gauge boson and the four families
\begin{equation}
\mathcal{L}^{\text{gauge}}_{Z'}=g' Z^\prime_\mu (\bar{Q}_L D_Q \gamma^\mu Q_L+\bar{u}_R D_u \gamma^\mu u_R+\bar{d}_R D_d \gamma^\mu d_R 
+  \bar{L}_L D_L \gamma^\mu L_L+\bar{e}_R D_e \gamma^\mu e_R)
\end{equation}
where, 
\begin{equation}
\begin{split}
D_Q=\text{diag}(0,0,0,q_{Q4}), \quad D_u=\text{diag}(0,0,0,q_{u4}), \quad D_d=\text{diag}(0,0,0,q_{d4}) \\
D_L=\text{diag}(0,0,0,q_{L4}), \quad D_e=\text{diag}(0,0,0,q_{e4}), \quad D_\nu=\text{diag}(0,0,0,q_{d4}).
\end{split}
\end{equation}
In addition there are the fourth family couplings involving the opposite chirality states $\tilde{\psi}_4$. Using the transformations in Eq.~(\ref{eq:transf_heavybasis}), we get the $Z'$ couplings in the diagonal heavy mass basis
\begin{equation}
\mathcal{L}^{\text{gauge}}_{Z'}=g' Z^\prime_\mu (\bar{Q'}_L D'_Q \gamma^\mu Q'_L+\bar{u'}_R D'_u \gamma^\mu u'_R+\bar{d'}_R D'_d \gamma^\mu d'_R 
+  \bar{L'}_L D'_L \gamma^\mu L'_L+\bar{e'}_R D'_e \gamma^\mu e'_R)
\end{equation}
where $D'_Q=V_{Q_L}D_QV^\dagger_{Q_L}$, and similarly with $Q\rightarrow L$, etc. Ignoring phases, these matrices can be parametrized as
\begin{equation}
D'_Q=q_{Q_4} \left(
\begin{array}{cccc}
 s_{14}^2 & c_{14} s_{14} s_{24} & c_{14} c_{24} s_{14} s_{34} & c_{14} c_{24} c_{34} s_{14} \\
 c_{14} s_{14} s_{24} & c_{14}^2 s_{24}^2 & c_{14}^2 c_{24} s_{24} s_{34} & c_{14}^2 c_{24} c_{34} s_{24} \\
 c_{14} c_{24} s_{14} s_{34} & c_{14}^2 c_{24} s_{24} s_{34} & c_{14}^2 c_{24}^2 s_{34}^2 & c_{14}^2 c_{24}^2 c_{34} s_{34} \\
 c_{14} c_{24} c_{34} s_{14} & c_{14}^2 c_{24} c_{34} s_{24} & c_{14}^2 c_{24}^2 c_{34} s_{34} & c_{14}^2 c_{24}^2 c_{34}^2 \\
\end{array}
\right)
\end{equation}
where $s_{ij}$ and $c_{ij}$ refer to $\sin\theta_{ij}$ and $\cos\theta_{ij}$ (we have also suppressed the superscript in the angles  $ s^Q_{14} \rightarrow  s_{14}$ for simplicity). Since the $U(1)'$ charges differ for the fourth family, the unitary transformations do not cancel and the matrices $D'_Q$, etc., are not generally diagonal. Therefore, $Z'$ exchange can couple to light families of different flavour.

We are interested in the $\bar{s}bZ'$ and $\bar{\mu}\mu Z'$ couplings, needed for the $R_K$ anomaly. Assuming that only the mixing angles $\theta^{Q_L}_{34}$ and $\theta^{L_L}_{24}$ are different from zero\footnote{A more natural possibility would be to assume that the new vector-like fermions have a large mixing only with the 3rd generation of the SM doublet, that is with taus instead of muons. 
Then the coupling to muons could arise due to a mixing between the SM charged leptons, as in \cite{King:2017anf}. 
However,  explaining the B-meson anomalies in such a set-up  runs in conflict with the strong bounds from  non-observation of $\tau \rightarrow 3 \mu$. 
}
the mixing mass matrices become
\begin{equation}
D'_Q=q_{Q_4}\left(
\begin{array}{cccc}
 0 & 0 & 0 & 0 \\
 0 & 0 & 0 & 0 \\
 0 & 0 & (s^Q_{34})^2 & c^Q_{34} s^Q_{34} \\
 0 & 0 & c^Q_{34} s^Q_{34} & (c^Q_{34})^2 \\
\end{array}
\right),
\quad
D'_L=q_{L_4} \left(
\begin{array}{cccc}
 0 & 0 & 0 & 0 \\
 0 & (s^L_{24})^2 & 0 & c^L_{24} s^L_{24} \\
 0 & 0 & 0 & 0 \\
 0 & c^L_{24} s^L_{24} & 0 & (c^L_{24})^2 \\
\end{array}
\right)
\end{equation}
while the rest of them being zero. In the low energy effective theory, after integrating out the fourth heavy family, the $Z'$ couplings to the the three massless families of quarks and leptons are
\begin{equation}
\mathcal{L}_{Z'}^{\text{gauge}}= g'Z'_{\mu} \left(q_{Q_4}(s^Q_{34})^2\bar{Q'}_{L_3} \gamma^\mu Q'_{L_3}+ q_{L_4}(s^L_{24})^2\bar{L'}_{L_2} \gamma^\mu L'_{L_2} \right),
\label{eq:Z'_general_couplings}
\end{equation}
where $Q'_{L3}=(t'_L,b'_L)$ and $L'_{L2}=(\nu'_{\mu L}, \mu'_L)$. Using now the diagonalization of the Yukawa matrices in Eq.~(\ref{eq:diag_Yuk_mat}), we can expand the primed fields in terms of the mass eigenstates, 
\begin{eqnarray}
b'_L&=&(V^{\prime \dagger}_{dL})_{31} d_L +(V^{\prime \dagger}_{dL})_{32}s_L+(V^{\prime \dagger}_{dL})_{33}b_L \nonumber \\
t'_L&=&(V^{\prime \dagger}_{uL})_{31} u_L +(V^{\prime \dagger}_{uL})_{32}c_L+(V^{\prime \dagger}_{uL})_{33}t_L \nonumber \\
\nu'_{\mu L}&=&(V^{\prime \dagger}_{\nu L})_{21} \nu_{1 L} +(V^{\prime \dagger}_{\nu L})_{22}\nu_{2 L}+(V^{\prime \dagger}_{\nu L})_{23} \nu_{3 L} \\
\mu'_L&=&(V^{\prime \dagger}_{eL})_{21} e_L +(V^{\prime \dagger}_{eL})_{22}\mu_L+(V^{\prime \dagger}_{eL})_{23}\tau_L. \nonumber
\end{eqnarray}
For simplicity, we assume that the charged lepton mass matrix is diagonal so that we may drop the primes 
on the muon field so that $\mu'_L = \mu_L$. Under this assumption, in the lepton sector,
the $Z'$ only couples to muon mass eigenstates $\mu_L$ and muon neutrinos $\nu_{\mu L}$,
where the latter are related to neutrino mass eigenstates by the PMNS matrix,
\begin{equation}
\nu'_{\mu L} = (V_{\text{PMNS}})_{21} \nu_{1 L} +(V_{\text{PMNS}})_{22}\nu_{2 L}+(V_{\text{PMNS}})_{23} \nu_{3 L} 
\end{equation}
Given the hierarchies of the CKM matrix, we will assume similar hierarchies of the rotation matrix elements:
\begin{equation}
\lvert (V'_{(d,u)L})_{31} \rvert^2 \ll \lvert (V'_{(d,u)L})_{32}  \rvert^2 \ll \lvert(V'_{(d,u)L})_{33}  \rvert^2 \approx 1 
\label{eq:mixing_supp}
\end{equation}

The vector-like neutrino $\nu_4$ is not charged under the SM and it is considered as a Dark Matter candidate. The portal that allows it to annihilate into ordinary matter is the $Z'$ mediator. 
The explicit coupling between the $Z'$ and the Dark Matter candidate $\nu_4$ is
\begin{equation}
\mathcal{L}_{Z'}^{\nu_4} = g'q_{\nu_4} Z'_\mu \overbar{\nu}_4 \gamma^\mu \nu_4,
\end{equation}
where the Dirac Dark Matter field is given by $\nu_4=\tilde{\nu}_{4L}+\nu_{4R}$ with a Dirac mass 
$m_{\nu}\overbar{\nu}_4  \nu_4$ where we have defined $m_{\nu}\equiv M^{\nu}_4$.

We finish this section by summarizing all non-SM interactions that will later be relevant for the phenomenological analysis, introducing the notation that we shall subsequently use: 
\begin{equation}
{\cal L} \supset Z'_\mu \left(
g_{bb} \bar{q}_{L} \gamma^\mu q_{L} 
+ g_{bs}\bar{b}_{L} \gamma^\mu s_{L} 
+ g_{\mu \mu}\bar{\ell}_{L} \gamma^\mu \ell_{L} 
+ g_{\nu \nu} \overbar{\nu}_4 \gamma^\mu \nu_4
\right),
\label{eq:Zp_Rk_couplings}
\end{equation}
where $q_L = (t_L,b_L)^T$, $\ell_L = (\nu_{\mu \, L}, \mu_L)^T$, 
$g_{bb}=g'q_{Q4}(s^Q_{34})^2$, $g_{bs}=g_{bb} (V^{\prime \dagger}_{dL})_{32}$, $g_{\mu \mu}=g'q_{L_4}(s^L_{24})^2$, $g_{\nu \nu} =  g'q_{\nu_4}$.
We expect $|(V^{\prime \dagger}_{dL})_{32}| \lesssim |V_{ts}|$, where $|V_{ts}|\approx 0.04$ is the 3-2 entry of the CKM matrix, as otherwise unnatural cancellations would be required.  
It follows that $|g_{bs}| \lesssim  |V_{ts}g_{bb}|$;  
in the following for simplicity we assume  $g_{bs} = V_{ts} g_{bb}$, and that $g_{bb}$ and $g_{\mu \mu}$ have the same sign. 
Thus, the relevant parameter space is 5-dimensional: 3 couplings ($g_{bb}$, $g_{\mu \mu}$, $g_{\nu \nu}$) and 2 masses ($M_{Z'}$ and the Dark Matter mass $m_{\nu}$).  
From the theory point of view these are all essentially free parameters, 
although one naturally expects $g_{\nu \nu} \gg  g_{bb}, g_{\mu \mu}$ in the absence of large mixings or large  hierarchies of $U(1)'$ charges.   
These parameters are then constrained by flavour physics, multiple low-energy precision measurements, colliders, and Dark Matter detection experiments.
In the following sections we work out these constraints, and identify the regions of the parameter space where both the B-anomalies and the Dark Matter relic abundance can be explained without conflicting any existing experimental data.  
We note that $Z^{\prime}$ models simultaneously addressing the B-anomalies and Dark Matter have been previously discussed in Refs.~\cite{Sierra:2015fma,Belanger:2015nma,Celis:2016ayl,Altmannshofer:2016jzy,Baek:2017sew,Cline:2017qqu,Fuyuto:2017sys}.   
In particular, Ref.~\cite{Altmannshofer:2016jzy} performed a detailed analysis of collider, precision, Dark Matter constraints in a similar model based on gauged $L_\mu-L_\tau$ symmetry. 
The main practical difference between the considered setup and that model is the presence of $Z^{\prime}$ couplings to b-quarks in Eq.~(\ref{eq:Zp_Rk_couplings}), which affects the LHC phenomenology as well as direct and indirect detection signals.  

\section{$R_{K^{(*)}}$ anomalies and flavour constraints}
\label{constraints}

In this section we review and update the constraints on the parameter space of $Z'$ models motivated by the current B-meson anomalies.      
One possible explanation of the $R_K$ and $R_{K^*}$ measurements in LHCb is that the low-energy Lagrangian below the weak scale contains an additional contribution to the effective 4-fermion operator with left-handed muon, $b$-quark, and $s$-quark fields:  
\begin{equation}
\Delta \mathcal{L}_{\text{eff}} \supset G_{b s\mu} ( \bar{b}_L \gamma^\mu s_L) (\bar{\mu}_L \gamma_\mu \mu_L )+ {\rm h.c.}, \qquad G_{b s\mu} \approx \frac{1}{(31.5\text{ TeV})^2}. 
\label{eq:bsmu_fit}
\end{equation}
Above, the numerical value of the effective coefficient corresponds to the best fit quoted in Ref.~\cite{Altmannshofer:2017yso}. 
In the considered model, this operator arises from tree-level $Z'$ exchange  and it dominates over the analogous operator with $\mu_L$ replaced by $e_L$ according to Eq.~(\ref{eq:mixing_supp}).  
We can express the coefficient $G_{b s\mu}$ as function of the couplings in Eq.~(\ref{eq:Zp_Rk_couplings}), 
\begin{equation}
G_{b s\mu}= - \frac{g_{bs}g_{\mu\mu}}{M^2_{Z'} }= - \frac{V_{ts} g_{bb}g_{\mu\mu}}{M^2_{Z'}}. 
\label{eq:bsmu_model}
\end{equation}
Together, Eqs.~(\ref{eq:bsmu_fit}) and (\ref{eq:bsmu_model}) imply the constraint on the parameters 
$g_{bb}$, $g_{\mu\mu}$ and $M_{Z'}$:  
\begin{equation}
\frac{g_{bb}g_{\mu\mu}}{M^2_{Z'}} \approx \frac{1}{(6.4\text{ TeV})^2}.
\label{eq:coeff_RK}
\end{equation} 
There are additional constraints on these parameters coming from flavour physics and low-energy precision measurements. 
In the following we  determine the region of the parameter space where the $R_{K^{(*)}}$ anomalies can be explained without conflicting other experimental data.
\newline

\begin{figure}[h!]
\begin{center}
\includegraphics[width=12cm]{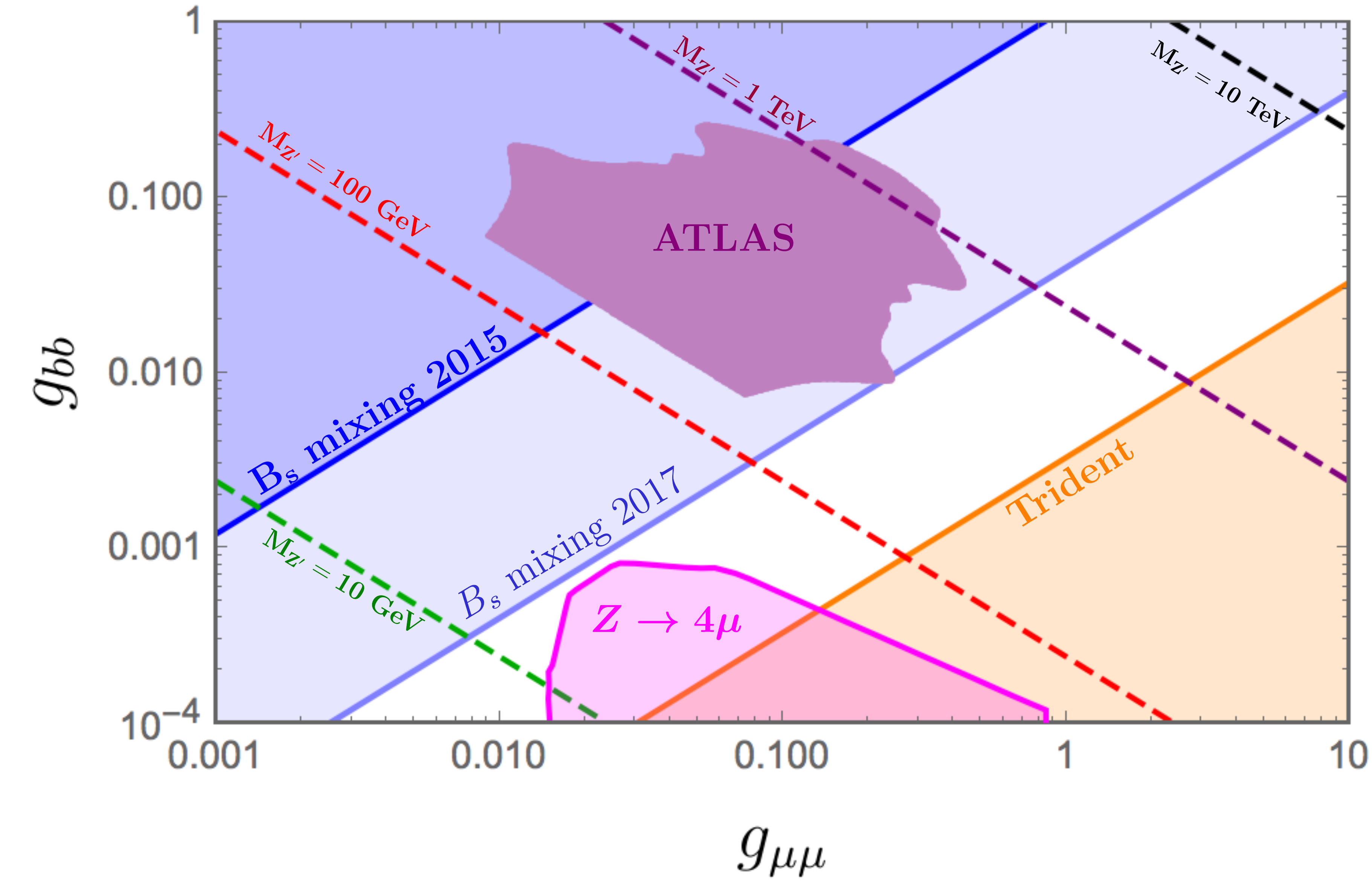}
\caption{The parameter space in the $(g_{\mu\mu},g_{bb})$ plane compatible with $R_K$ anomalies and flavour constraints (white). 
The $Z'$ mass varies over the plane, with a unique $Z'$ mass for each point in the plane as determined by Eq.~(\ref{eq:coeff_RK}). 
We show the recent $B_s$ mixing constraints (light blue), and the trident bounds (orange);   
for reference we also display the previous weaker $B_s$ mixing bounds (dark blue). 
The green, red, purple and black lines correspond to $M_{Z'}=10, 100, 1000, 10000 \text{ GeV}$ respectively.\label{fig:constrained_couplings}} 
\end{center}
\end{figure}

\paragraph{\bf \underline{$B_s-\bar{B}_s$ mixing:}} The $Z^{\prime}$ coupling to $bs$ leads to an additional tree-level contribution to $B_s-\bar{B}_s$ mixing. Low-energy observables are affected by the effective operator arising from integrating out the $Z'$ at tree level: 
\begin{equation}
\Delta \mathcal{L}_{\text{eff}} \supset - {G_{b s}\over 2}  (\bar{s}_L \gamma^\mu b_L)^2 + {\rm h.c}, 
\qquad G_{b s} = \frac{g_{bs}^2}{M^2_{Z'}} = \frac{g_{bb}^2 V^2_{ts}}{M_{Z'}^2}.
\label{eq:effective_bs}
\end{equation}
Such a new contribution is highly constrained by the measurements of the mass difference $\Delta M_s$ of neutral $B_s$ mesons.
In this work we follow the recent analysis of Ref.~\cite{DiLuzio:2017fdq} which, using updated lattice results, obtains a stronger bound on $G_{b s}$: 
\begin{equation}
- \frac{1}{(180\text{ TeV})^2}  \lesssim G_{b s} \lesssim \frac{1}{(770\text{ TeV})^2}, 
\qquad @ \, 95\% {\rm CL}.
\end{equation}
The resulting constraints in the $(g_{\mu\mu},g_{bb})$ plane are shown as the light blue region in Fig.~\ref{fig:constrained_couplings}. 
The updated constraint is particularly strong for the models that generate a strictly positive $G_{b s}$ \cite{DiLuzio:2017fdq} (as is the case in $Z'$ models) due to the $\sim 1.8\sigma$ discrepancy between the measured $\Delta M_s$ and the updated SM predictions which favors $G_{b s} < 0$.   
As a consequence, $Z'$ models explaining the B-meson anomalies required $M_{Z'} \lesssim 1$~TeV, 
assuming weak coupling $g_{\mu\mu} \lesssim 1$. 
For easy reference, we also show the $B_s$ mixing constraints based on the previous SM determination of $\Delta M_s$ \cite{Artuso:2015swg},
$- (160\textrm{ TeV})^{-2}  \lesssim G_{b s} \lesssim (140 \textrm{ TeV})^{-2}$, 
see the dark blue region in Fig.~\ref{fig:constrained_couplings} labeled ``$B_s$ mixing 2015''.  
\newline

\paragraph{\bf \underline{Neutrino trident:}} The $Z'$ coupling to left-handed muons leads to a new tree-level contribution to the effective 4-lepton interaction  
\begin{equation}
\Delta \mathcal{L}_{\text{eff}} \supset - {G_{\mu}\over 2}  (\bar \ell_L \gamma^\mu \ell_L)^2, 
\qquad G_{\mu} = \frac{g_{\mu\mu}^2}{M^2_{Z'}}.
\label{eq:effective_mu}
\end{equation}
This operator is constrained by the trident production $\nu_\mu \gamma^* \rightarrow \nu_\mu \mu^+ \mu^-$ \cite{Geiregat:1990gz,Mishra:1991bv,Altmannshofer:2014pba}. 
Using the results of the global fit in Ref.~\cite{Falkowski:2017pss}, the bound on the effective coefficient is given by    
\begin{equation}
-  \frac{1}{(390\text{ GeV})^2} \lesssim G_{\mu} \lesssim \frac{1}{(370\text{ GeV})^2}, 
\qquad @ \, 95\% {\rm CL}.
\end{equation}
The limits in the $(g_{\mu\mu},g_{bb})$ plane are shown as the orange region in Fig.~\ref{fig:constrained_couplings}.
Since the trident constraints probe much lower scales than the $B_s$ mixing, a much larger $Z'$ coupling to muons is allowed, $g_{\mu \mu} \gtrsim 1$ for a heavy enough $Z'$. 
Nevertheless, together with the $B_s$ mixing constraints, the trident leaves only a narrow sliver of the parameter space that could address the $B$ meson anomalies.
  
\begin{figure}[h!]
\begin{minipage}[b]{0.5\textwidth}
\includegraphics[width=7.6cm]{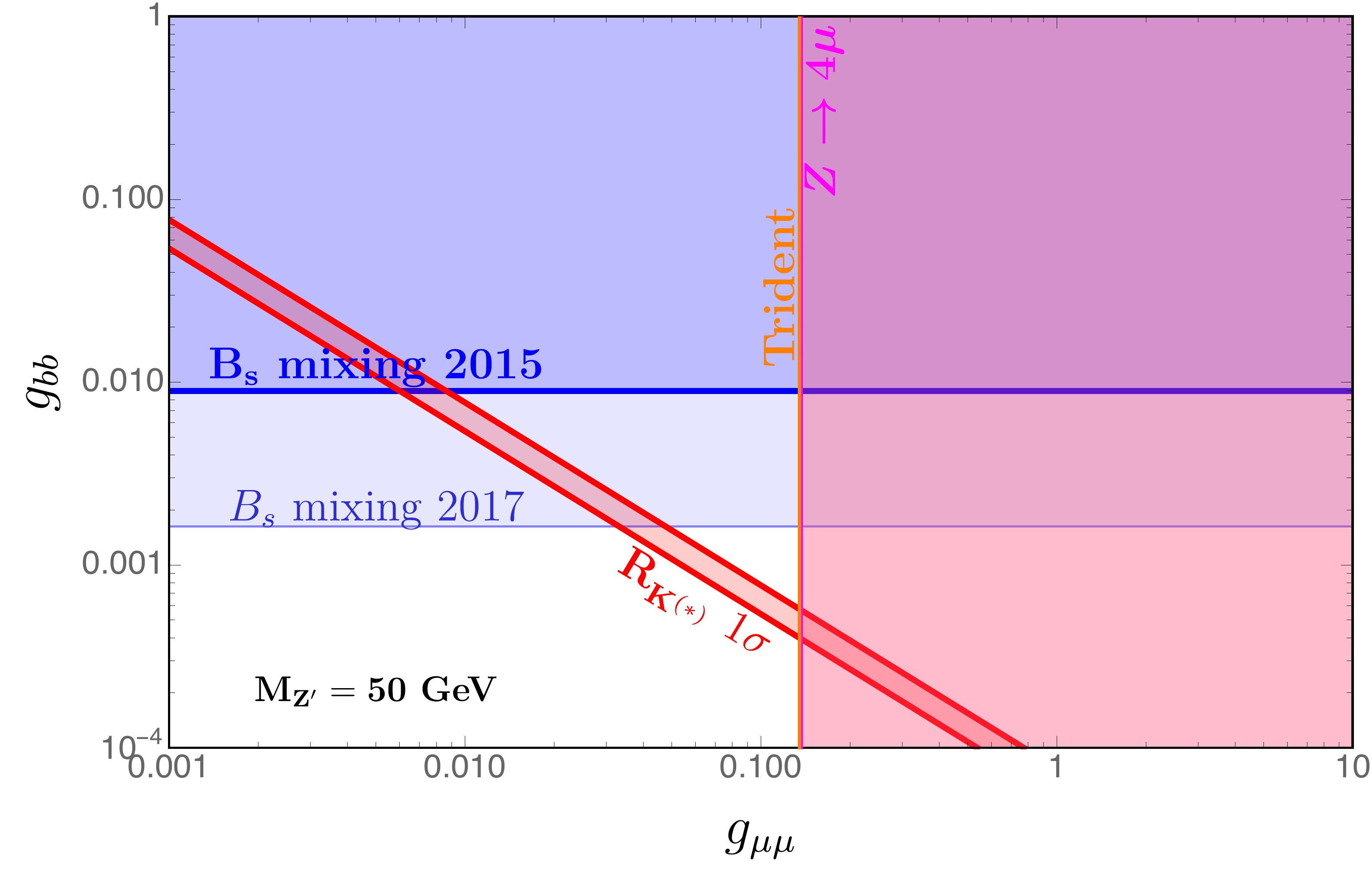} 
\end{minipage}
\begin{minipage}[b]{0.5\textwidth}
\includegraphics[width=7.6cm]{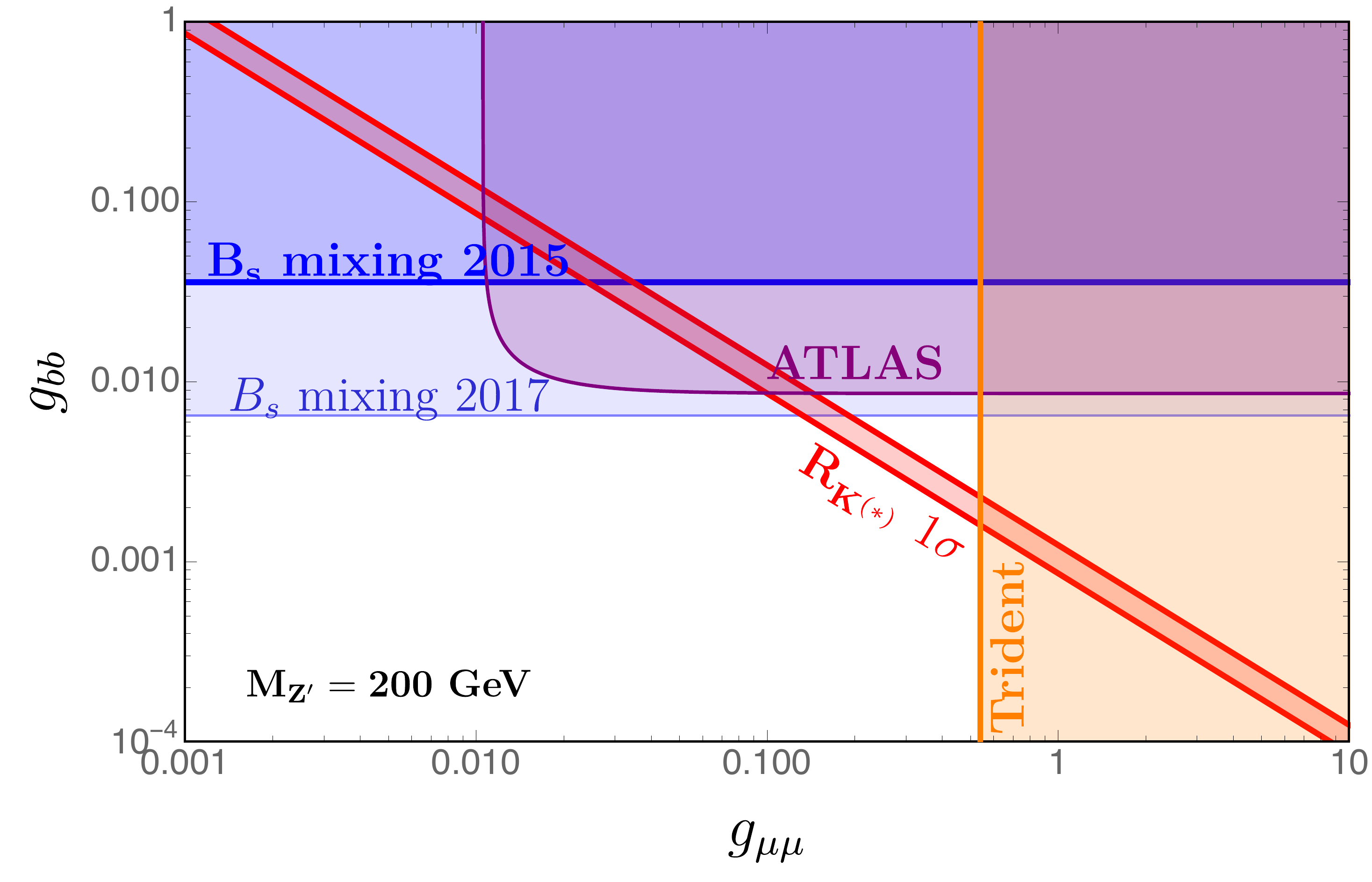}
\end{minipage}
\begin{minipage}[b]{0.5\textwidth}
\includegraphics[width=7.6cm]{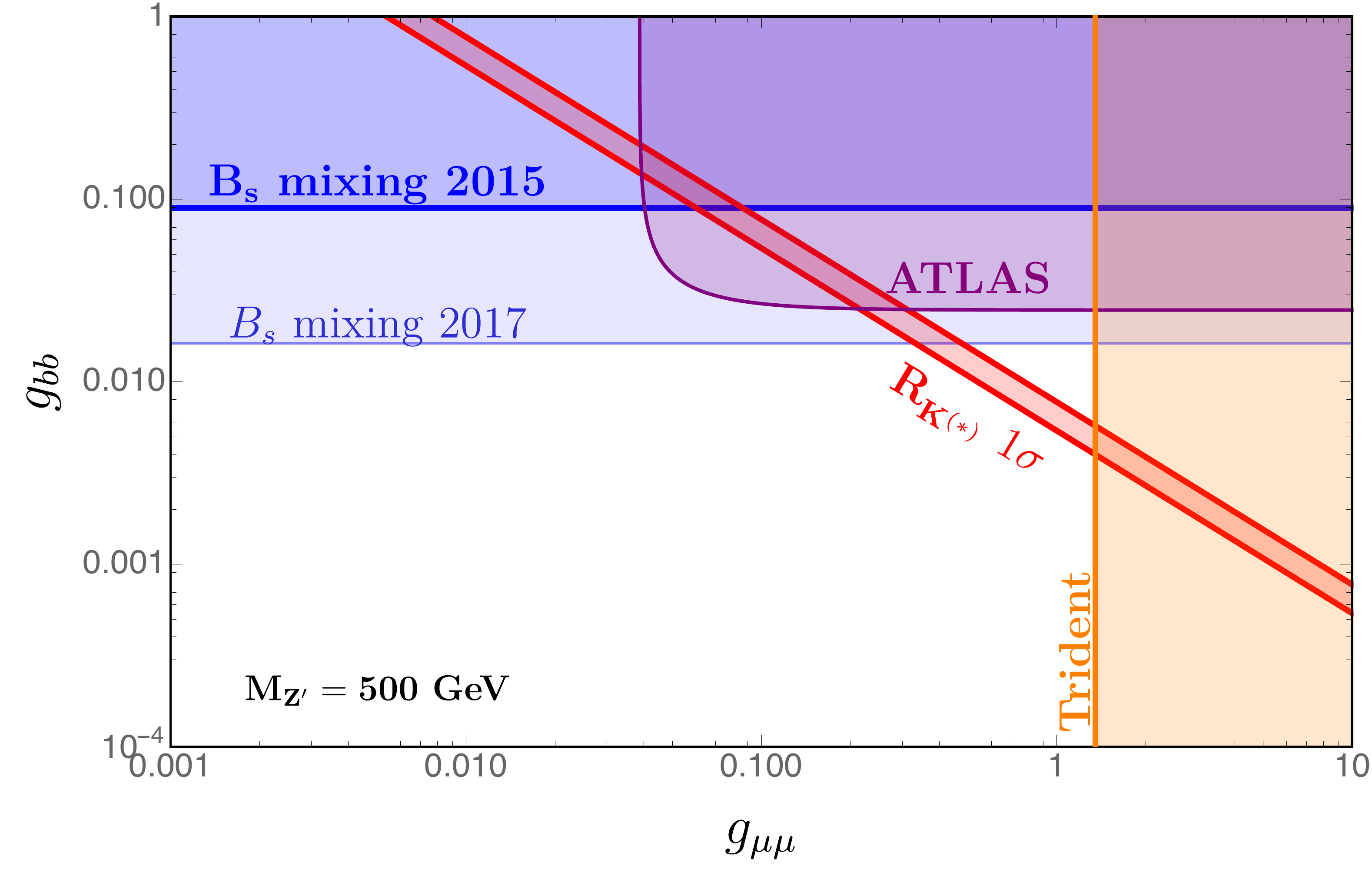}
\end{minipage}
\begin{minipage}[b]{0.5\textwidth}
\includegraphics[width=7.6cm]{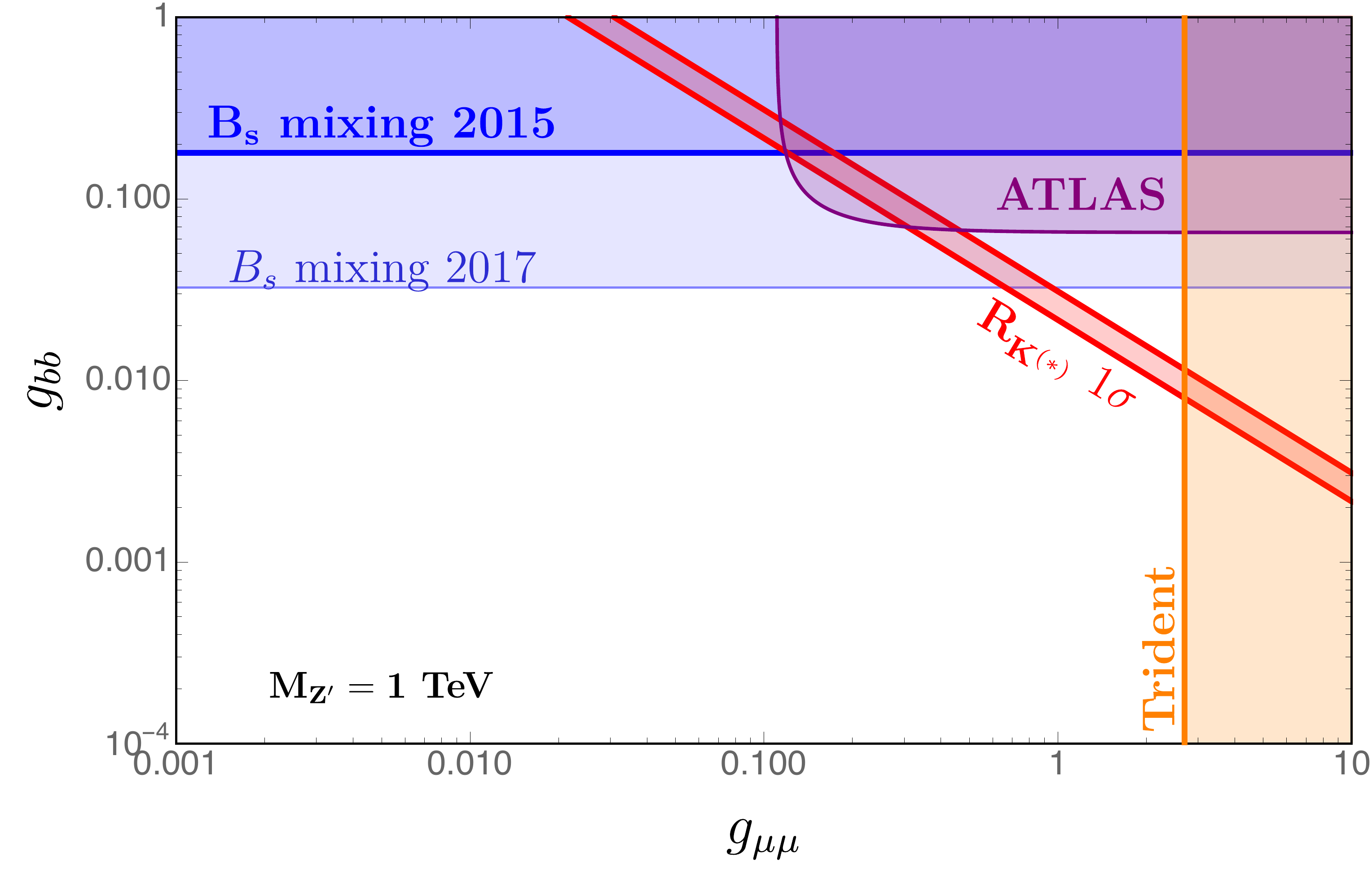}
\end{minipage}
\caption{Bounds on the parameter space in the $(g_{\mu\mu},g_{bb})$ plane for fixed $Z'$ masses: 50, 200, 500 and 1000 GeV, as indicated on each panel. The red bands explain $R_K$ at $1\sigma$. The blue and orange areas show the $B_s-\overbar{B}_s$ mixing~\cite{DiLuzio:2017fdq} and neutrino trident~\cite{Falkowski:2017pss} $2\sigma$ exclusions, respectively. For low $Z'$ masses we have additional constraints from $Z\rightarrow4\mu$ as shown in pink. The ATLAS limits~\cite{Aaboud:2017buh} from dimuon resonance searches for $36 \text{ fb}^{-1}$luminosity  are given in purple for larger $Z'$ masses.\label{fig:ATLASlimits}}
\end{figure}

\paragraph{\bf \underline{LHC searches:}} Further constraints on this model come from collider searches. 
For light $Z'$ masses, the LHC measurements of the Z decays to four muons, with the second muon pair produced in the SM via a virtual photon~\cite{CMS:2012bw,Aad:2014wra}, $pp \rightarrow Z \rightarrow 4 \mu$, sets relevant constraints in the low mass region of $Z'$ models, $5\lesssim M_{Z'} \lesssim 70$ GeV. 
The $Z \rightarrow 4 \mu$  constraints on the magnitude of the  $Z'$ coupling to muons were analyzed in Refs.~\cite{Altmannshofer:2014cfa,Altmannshofer:2014pba,Altmannshofer:2016jzy}. 
Projecting these results onto this model, the excluded parameter space is marked as the pink regions in Fig.~\ref{fig:constrained_couplings} and in the upper-left panel of Fig.~\ref{fig:ATLASlimits}. 
All in all, the $Z \rightarrow 4 \mu$ constraint is non-trivial but for any $Z'$ mass it always leaves some available parameter space to explain the B-meson anomalies. 

For a heavier $Z'$, the strongest constraints comes from LHC dimuon resonance searches, $pp\rightarrow Z'\rightarrow \mu^+ \mu^-$, see also  \cite{Dalchenko:2017shg}. 
In this model the $Z'$ is dominantly produced at the LHC through its couplings to bottom quarks, $b \bar b \to Z'$. 
The cross section $\sigma (p p \to Z')$ from $b \bar b$ collisions is taken from Fig.~3 of Ref.~\cite{Faroughy:2016osc}.
The contribution of bottom-strange collisions, which is subleading in this model, is estimated using {\tt Madgraph}~\cite{Alwall:2014hca}. 
The $Z'$ boson can subsequently decay into muons, muon neutrinos, bottom or strange quarks, and also into top quarks and Dark Matter when kinematically allowed. 
The partial decay widths are given by 
\begin{align}
 & \Gamma_{Z'\rightarrow \mu \bar{\mu}}=\frac{1}{24\pi}g_{\mu \mu}^2 M_{Z'}  
=  \Gamma_{Z'\rightarrow \nu_\mu \bar{\nu}_\mu}~, \nonumber \\  & \Gamma_{Z'\rightarrow b \bar{b}}=\frac{1}{8\pi}g_{bb}^2 M_{Z'}~, \nonumber \\ & \Gamma_{Z'\rightarrow b \bar{s}}=\frac{1}{8\pi}g_{bb}^2V_{ts}^2 M_{Z'}~, \nonumber \\
 & \Gamma_{Z'\rightarrow t \bar{t}}=\frac{1}{8\pi}g_{bb}^2 M_{Z'} \left( 1-\frac{m_t^2}{M_{Z'}^2} \right) \sqrt{1-\frac{4 m_t^2}{M_{Z'}^2}}~, \nonumber \\
 & \Gamma_{Z'\rightarrow \nu_4 \bar{\nu}_4}=\frac{1}{24\pi}g_{\nu\nu}^2 M_{Z'} \left( 1-\frac{m_\nu^2}{M_{Z'}^2} \right) \sqrt{1-\frac{4 m_\nu^2}{M_{Z'}^2}}~, 
\end{align}
from which we  calculate ${\rm Br}(Z' \to \mu \mu)$ analytically. 
Then $\sigma (p p \to Z' \to \mu \mu)$ is estimated using the narrow-width approximation, and compared with the limits from the recent dimuon resonance search  by ATLAS~\cite{Aaboud:2017buh}, which allows us to constrain $Z'$ mases between $150$ GeV and $5$ TeV. 
We verified that the analogous  Tevatron analyses give weaker constraints, also in the low mass regime. 
Fig.~\ref{fig:ATLASlimits} shows the ATLAS constraints for specific $Z'$ masses (200, 500 and 1000 GeV) with Dark Matter couplings set to zero and arbitrary $(g_{\mu \mu}, g_{bb})$ couplings.   
Fig.~\ref{fig:constrained_couplings} shows the same limits for the $Z'$ mass fixed in function of  $(g_{\mu \mu}, g_{bb})$ by the condition in Eq.~(\ref{eq:coeff_RK}). 
We conclude that in the  parameter space of the model relevant for explaining the B-meson anomalies   
the ATLAS dimuon limits are always weaker that the new $B_s$ mixing constraints. 
\newline

\paragraph{\bf \underline{Other constraints}:} Finally we comment on other precision observables which yield subleading constraints on the model.

The  contribution of $Z'$ to the muon magnetic moment is given by 
\begin{equation}
\Delta^\mu_{g-2} = {1 \over 12 \pi^2} m_\mu^2 \left (g_{\mu \mu} \over M_{Z'} \right )^2 . 
\end{equation}
The measured discrepancy of the muon magnetic moment is 
$\Delta^\mu_{g-2} = (290 \pm 90)\times 10^{-11}$ \cite{Jegerlehner:2009ry}. 
This sets weaker limits on the ratio $g_{\mu \mu}/M_{Z'}$ than the trident production. 

Next, $Z'$ exchange generates the effective interaction between b-quarks and muons: 
\begin{equation}
\mathcal{L}_{\text{eff}} \supset  G_{b\mu} (\bar{b}_L \gamma^\mu b_L)(\bar{\mu}_L\gamma_\mu \mu_L),
 \qquad G_{b\mu} = - \frac{g_{bb} g_{\mu \mu}}{M^2_{Z'}} =  -  \frac{1}{(6.4\text{ TeV})^2},
\label{eq:op_upsilon}
\end{equation}
where we used Eq.~(\ref{eq:coeff_RK}). 
The operator in Eq.~(\ref{eq:op_upsilon}) is constrained by lepton flavour universality of upsilon meson decays \cite{Aloni:2017eny}. 
Focusing on the $\Upsilon_{1s}$ state, given the measured ratio~\cite{Olive:2016xmw} 
\begin{equation}
R^{\tau/\mu}_{1s}=\frac{\Gamma(\Upsilon_{1s}\rightarrow \tau^+ \tau^-)}{\Gamma(\Upsilon_{1s}\rightarrow \mu^+\mu^-)}=1.008 \pm 0.023,
\end{equation}
and the SM prediction is $R^{\tau/\mu}_{1s}=0.9924$, one finds the constraint 
\begin{equation}
-\frac{1}{(150\text{ GeV})^2} < G_{b\mu} < \frac{1}{(190\text{ GeV})^2}
\qquad @ \, 95\% {\rm CL}.
\end{equation}
This is automatically satisfied in this model in the parameter space where the $R_K$ anomalies are explained.

\section{Dark Matter}
\label{darkmatter}

This model  comprises a fourth neutrino ($\nu_4$) which  possesses all the properties of a viable Dark Matter candidate. Indeed, $\nu_4$ is an electrically neutral particle interacting weakly with the SM sector through an exchange of $Z'$. 
Furthermore, charge assignments under the local symmetries forbid any mixing with other fields such as the SM neutrinos. 
Therefore, conservation of fermion number in the dark sector can be effectively seen as a $\mathbb{Z}_2$ symmetry forbidding the Dark Matter from decaying and as a consequence ensuring its stability. 
In this section we discuss in some detail the generation of the Dark Matter relic density and show the constraints  from indirect and direct Dark Matter searches.
For brevity, in the following the Dark Matter candidate is simply denoted as $\nu$.

\subsection{Relic abundance} 
\begin{figure}[h!]
\begin{center}
\includegraphics[width=12 cm]{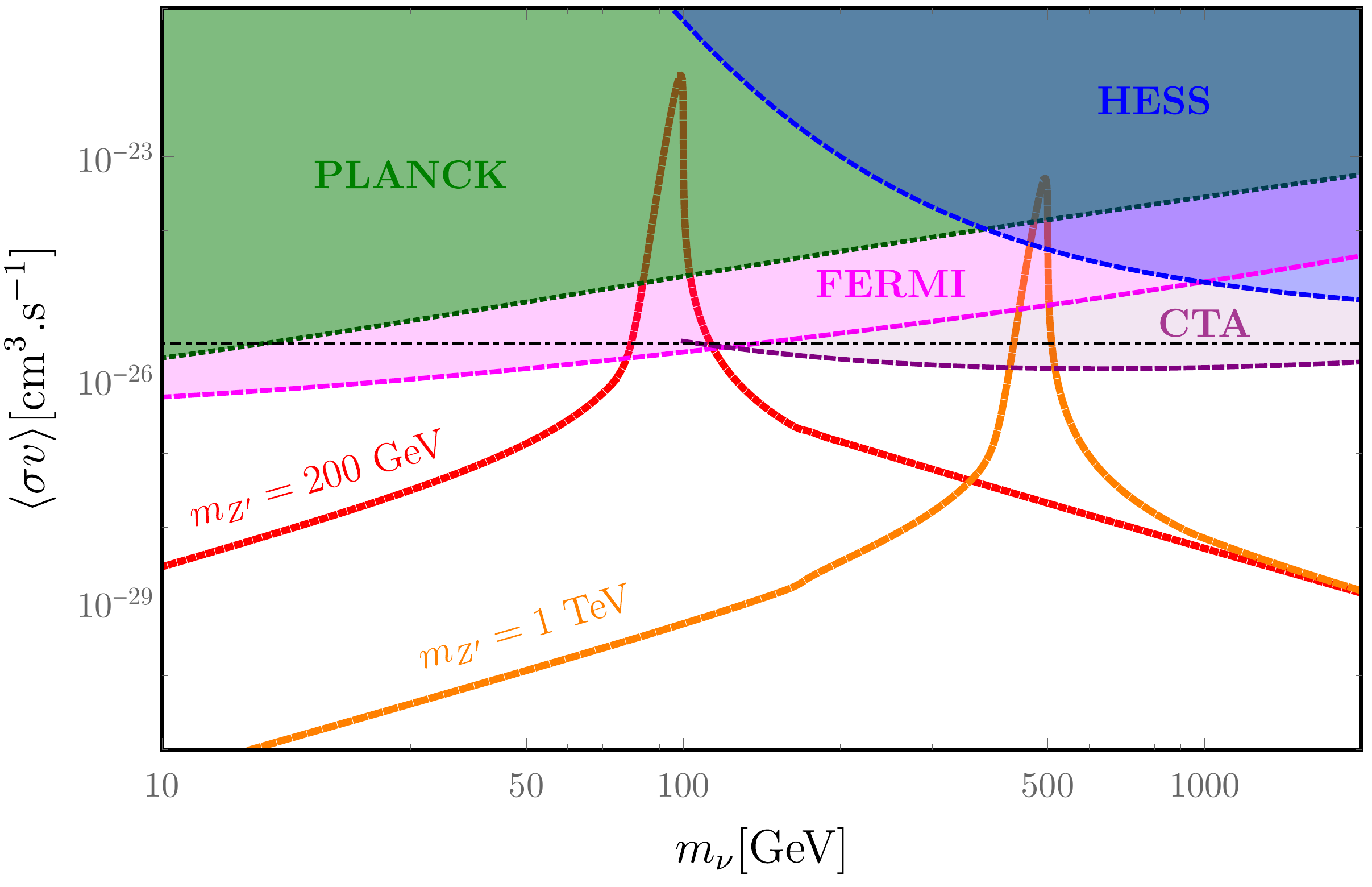}
\caption{Dark Matter velocity averaged annihilation cross section for $M_{Z^\prime}=200(1000)$ GeV in red (orange) assuming $g_{bb}=g_{\nu \nu}=g_{\mu \mu}=0.1$ and indirect detection limits assuming $\bar{b}b$ as final state from HESS~\cite{Abdallah:2016ygi} in blue, Fermi~\cite{Ackermann:2015zua} in pink and predictions for the upcoming CTA~\cite{Wood:2013taa} assuming 500h of observation toward the Galactic Center in purple. Limits from the Planck collaboration~\cite{Slatyer:2015jla} are shown in green, assuming Dark Matter annihilation to $\bar{\mu} \mu$. The dotted-dashed black line represents the canonical value of the cross section $\langle \sigma v \rangle =3 \times 10^{-26} \text{cm}^3~\text{s}^{-1}$.} 
\label{fig:relic_density_ID}
\end{center}
\end{figure}

The Dark Matter candidate is a weakly interacting massive particle (WIMP) whose relic abundance can be generated via the well studied freeze-out scenario~\cite{Bertone2005a,Arcadi:2017kky}. 
We will fit the parameters to reproduce the present Dark Matter density measured by the Planck collaboration: $\Omega_{\text{DM}} h^2 = 0.1198 \pm{0.0015}$~\cite{Ade:2015xua}. 
In this model the Dark Matter particles can annihilate to\footnote{%
Since $g_{bs} \ll g_{bb}$, we can safely ignore annihilation to $\bar{b}s$ and $\bar{s}b$.
}
$\bar{\mu}\mu, \bar{\nu}_\mu \nu_\mu, \bar{b}b$, and possibly to $\bar{t}t,Z^\prime Z^\prime$, if kinematically accessible:   
\begin{equation}
\langle \sigma v \rangle = \sum_{\psi=b,t,\mu,\nu_\mu} \langle \sigma v \rangle_{\bar{\nu} \nu \rightarrow \bar{\psi} \psi} + \langle \sigma v \rangle_{\bar{\nu} \nu \rightarrow Z^\prime Z^\prime} . 
\end{equation}
One can derive an analytical approximation of $\langle \sigma v \rangle$ by expanding it in powers of $x^{-1}$ around the typical freeze-out temperature $x_{\text{F}} \sim 23$.
Away from the pole and thresholds, each component of $\langle \sigma v \rangle$ can be approximated by the s-wave expression:   
\begin{align}
\label{eq:sigmavapp}
\langle \sigma v \rangle_{\bar{\nu} \nu \rightarrow \bar{\psi} \psi}  \simeq &
\left \{ \begin{array}{ccr}
c_\psi \frac{g_{\nu \nu}^2 g_{\psi \psi}^2 }{4\pi} \frac{  m_{\nu}^2 }{M_{Z'}^4} & \hspace{2cm} &[M_{Z^\prime} \gg m_{\nu} \gg m_\psi ] ~
\\
c_\psi \frac{g_{\nu \nu}^2 g_{\psi \psi}^2 }{64\pi m_{\nu }^2}  & \hspace{2cm}   &[ m_{\nu} \gg M_{Z^\prime} \gg m_\psi ] ~ 
\end{array} \right . ,
\nonumber \\
\langle \sigma v \rangle_{\bar{\nu} \nu \rightarrow Z^\prime Z^\prime}  \simeq & \frac{ g_{\nu \nu}^4}{32 \pi  m_{\nu}^2} \hspace{4.2cm}   [ m_{\nu} \gg M_{Z^\prime} ] ~,
\end{align}
where $c_\psi$ is a color factor. 
One can see that the annihilation cross section grows as $m_{\nu}^2$ for small Dark Matter masses, and evolves as $m_{\nu}^{-2}$ for large Dark Matter masses.  
Therefore, for fixed couplings and $M_{Z'}$, there are typically two possible values of $m_{\nu}$ reproducing $\langle \sigma v \rangle_{\rm thermal}$, as illustrated in Fig.~\ref{fig:relic_density_ID}. 
For small couplings, $g \lesssim 0.1$, the annihilation cross section is substantially lower than the thermal one  except in the pole region, 
and the two solutions approach $m_{\nu} \sim M_{Z^\prime}/2$. 
As demonstrated in~\cite{Griest1991c}, the presence of a pole in the annihilation cross section may invalidate the $1/x$ expansion. 
In such a case one cannot use  Eqs.~(\ref{eq:sigmavapp}) and instead one has to rely on numerical evaluations using Eq.~(\ref{eq:sigmavfullintegral}).
In order to explore the complete available parameter space, we compute the relic density and $\langle \sigma v \rangle$ numerically using the package { \tt micrOMEGAs}~\cite{Belanger:2013oya} after implementing the model in FeynRules~\cite{Alloul:2013bka}.
For higher values of the couplings, $g \gtrsim 1$, the correct relic density can be achieved away from the pole region where Eqs.~(\ref{eq:sigmavapp}) are adequate.

\subsection{Indirect detection constraints}
In the WIMP framework, Dark Matter annihilations to SM states occurring inside large astrophysical structures such as the galactic center, dwarf spheroidal (dSphs) galaxies or galaxy clusters might be relatively frequent at the present time. This could lead to indirect Dark Matter observation by detecting the by-products of these annihilations in high-energy cosmic rays~\cite{Gaskins:2016cha}. 
In this model  $\langle \sigma v \rangle$ is approximately velocity independent in the non-relativistic limit, therefore the same value of order $\langle \sigma v \rangle_{\rm thermal}$ required to match the relic density is also relevant for indirect detection. 
The two annihilation channels most relevant for indirect detection are $\nu \bar \nu \to b \bar b$ and $\nu \bar \nu \to \mu^+ \mu^-$. 
In the parameter space where $\nu \bar \nu \to b \bar b$ (and possibly to $t \bar t$) dominates, the best current limits on $\langle \sigma v \rangle$ are derived by the Fermi-LAT collaboration from a combined analysis of 15 Milky Way dSphs and excludes Dark Matter masses $m_{\text{DM}} \lesssim 100$ GeV~\cite{Ackermann:2015zua}, 
assuming the Navarro-Frenk-White profile~\cite{Navarro:1995iw}. 
For larger Dark Matter masses stronger constraints on  the same annihilation channel come from the HESS experiment~\cite{Abdallah:2016ygi}, however the typical limits are  $\langle \sigma v \rangle \lesssim 10^{-25}\text{cm}^3~\text{s}^{-1}$ and therefore cross sections of order $\langle \sigma v \rangle_{\rm thermal}$ are not probed. 
In the future, sensitivity of the Cherenkov Telescope Array (CTA) might be sufficient to probe annihilation the thermal cross section for 
$m_{\text{DM}} \gtrsim 100$~GeV~\cite{Wood:2013taa,Pierre:2014tra,Silverwood:2014yza,Lefranc:2016dgx,Lefranc:2015pza}.
The current and future constraints in the $b \bar b$ annihilation channel are illustrated in Fig.~\ref{fig:relic_density_ID}, where we also show predictions of this model for two particular points in the parameter space.   

As can be seen in Fig.~\ref{fig:constrained_couplings}, given the newer $B_s$ mixing constraints the allowed parameter space has $g_{\mu \mu} \gg g_{bb}$, and therefore annihilation into $\mu^+\mu^-$ (and the corresponding neutrinos) dominates. 
In such a case, the indirect detection limits on $\langle \sigma v \rangle$ are substantially weaker, such that the thermal annihilation cross section is allowed for Dark Matter masses above a few GeV \cite{Ackermann:2015zua}.  
For this reason, the indirect limits are not relevant in most of the interesting parameter space of this model. 
However, for small  Dark Matter masses $m_\nu \sim \text{GeV}$ annihilation  into leptons  at redshift $z \sim 1000$ can be constrained by CMB spectrum observations, as it  could modify the ionization history. 
For the thermal annihilation cross section, the Planck collaboration constraints on CMB spectrum distortions exclude Dark Matter masses below $m_\nu \lesssim 10 $ GeV~\cite{Slatyer:2015jla},  as illustrated in Fig.~\ref{fig:relic_density_ID}. 
We note that annihilation into leptonic final states can be relevant for experiments such as AMS-02  measuring cosmic-ray positrons and electrons, 
from which several studies have obtained strong constraints  on $\langle \sigma v \rangle$~\cite{Kopp:2013eka,Bergstrom:2013jra,Ibarra:2013zia,Lu:2015pta}. 
However these constraints are subject to strong dependence on the propagation model and uncertainties regarding cosmic-ray propagation in the interstellar medium, 
and for this reason we do not include them in the following.  
All in all, in this model Dark Matter masses $\lesssim 10$ GeV are excluded by the Planck collaboration results. Moreover, in the parameter space where annihilation into $b \bar b $ dominates, Dark Matter masses below 100 GeV are excluded by the Fermi-LAT results, although that parameter space is also disfavored by the recent $B_s$ mixing constraints.

\subsection{Direct Searches}

Sensitivity of Direct Detection (DD) experiments has improved by several orders of magnitude during the past decade, 
and currently the xenon-based experiments LUX~\cite{Akerib:2016vxi}, PandaX~\cite{Tan:2016zwf} and Xenon1T~\cite{Aprile:2017iyp} probe Dark Matter spin-independent (SI) scattering cross section of the order of $\sigma_{\text{SI}} \gtrsim 10^{-45}{\rm cm}^2$ for Dark Matter masses of the order of 100~GeV. 

\begin{figure}
\begin{minipage}[h!]{0.49\textwidth}
\includegraphics[height=5.3cm]{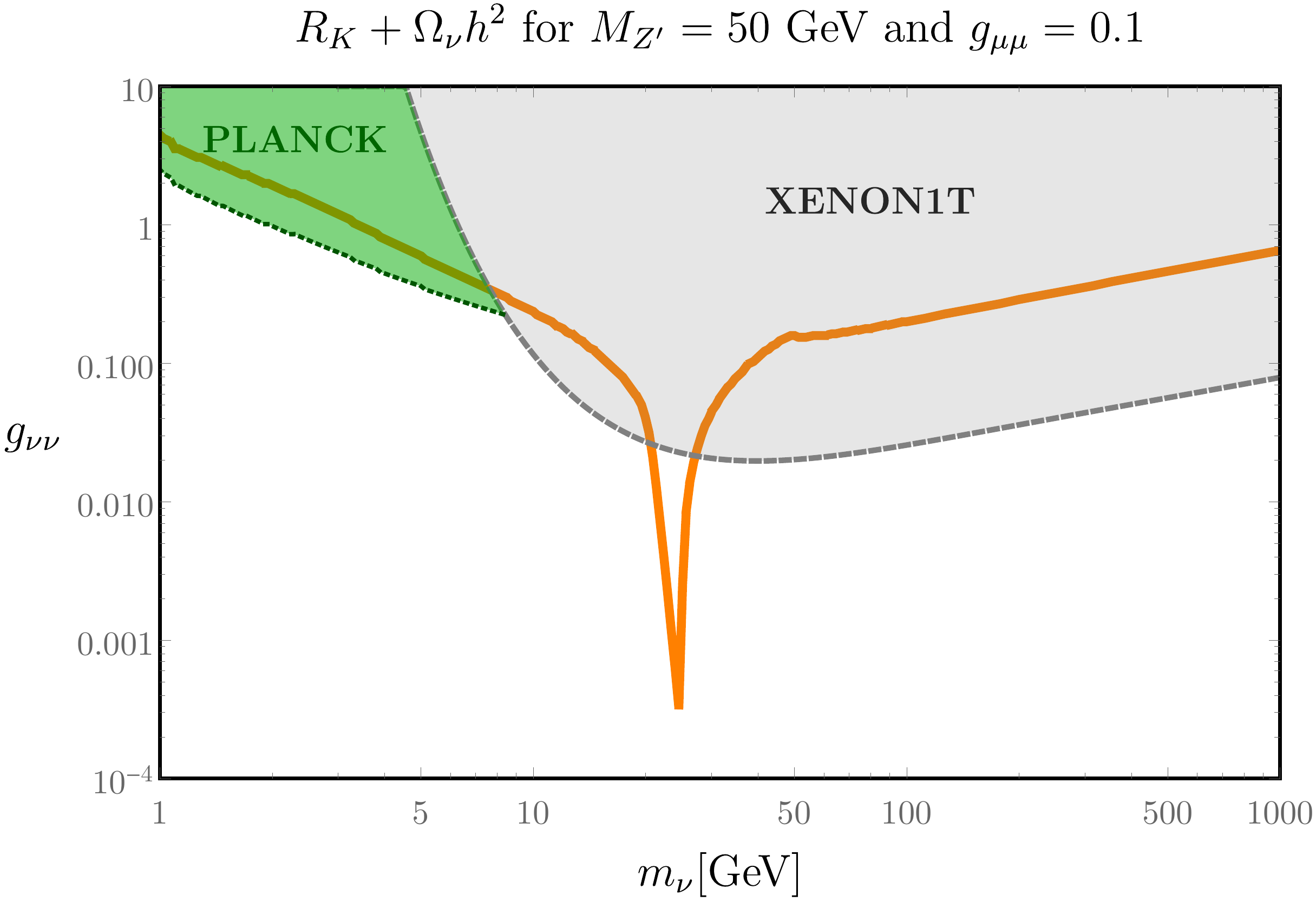}
\end{minipage}
\hfill
\begin{minipage}[h!]{0.49\textwidth}
\includegraphics[height=5.3cm]{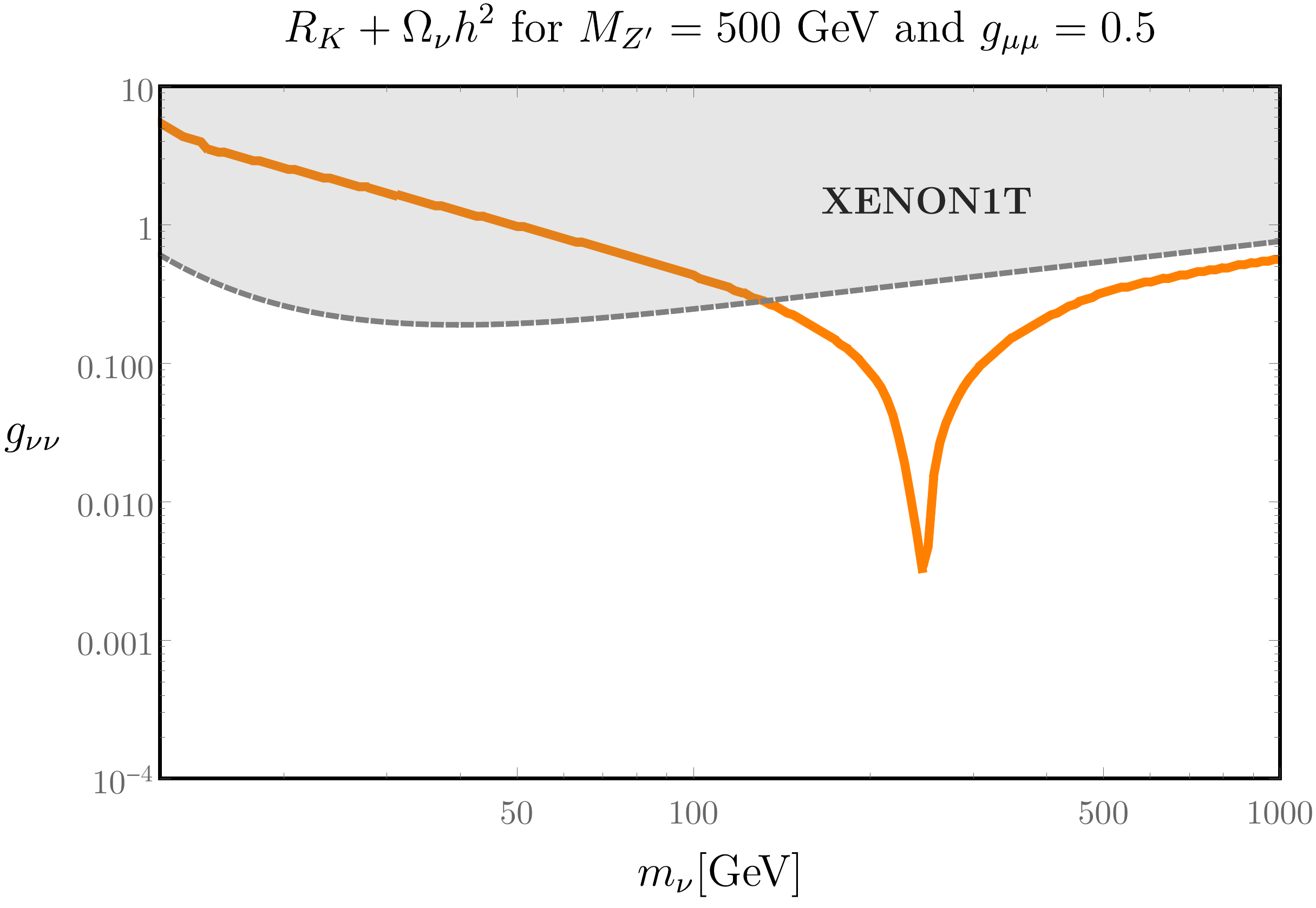}
\end{minipage}
\caption{Direct and indirect detection constraints on the parameter space : the orange line represents the model points featuring the correct Dark Matter  relic density and the appropriate Wilson coefficient explaining the $R_K$ discrepancy. The gray region shows the parameter space excluded by the Xenon1T experiment~\cite{Aprile:2017iyp} and the green region represents the parameter space not excluded by direct detection experiments but in tension with the Planck collaboration~\cite{Slatyer:2015jla} results.}
\label{fig:DD}
\end{figure}

In this set-up, integrating out the tree-level $Z^\prime$ exchange between Dark Matter and the SM leads to the following effective operators at the scale $\mu \simeq M_{Z'}$:  
\begin{equation}
\mathcal{L}_{\text{eff}} \supset - \sum_{f=\mu,b} \dfrac{g_{\nu \nu} g_{f f}}{M_{Z^\prime}^2} \bar{f}_L \gamma^\alpha f_L  \bar{\nu} \gamma_\alpha \nu ~, 
\end{equation}
where again one can neglect the effective coupling to $b s$.  
Below the scale $M_{Z'}$, vector-like Dark Matter couplings to light quarks are induced via renormalization group (RG) running:  
\begin{equation}
\label{eq:Leffchiq}
\mathcal{L}_{\text{eff}} \supset \sum_{q = u,d} C_{1,f}^{(6)} (\mu) \bar{q} \gamma^\alpha q  \bar{\nu} \gamma_\alpha \nu ~ .  
\end{equation}
The complete RG equations can be found e.g. in \cite{DEramo:2014nmf}; schematically, one has $C_{1,f}^{(6)}(\mu) \sim {\alpha \over 4\pi} \frac{g_{\nu \nu} g_{ff}}{M_{Z^\prime}^2} \log\left (M_{Z\prime}  \over \mu \right )$.  
Other tensor structures beyond that in Eq.~(\ref{eq:Leffchiq}) also appear but they give subleading effects in direct detection.   
Finally, at $\mu \simeq 2$~GeV the couplings in Eq.~(\ref{eq:Leffchiq}) can be mapped to momentum- and spin-independent non-relativistic interactions of Dark Matter with protons and neutrons:  
\begin{equation}
\mathcal{L}_{\text{eff,NR}} \supset \sum_{N= p,n} c_1^N\bar{\nu} \nu \bar{N} N 
\end{equation}
where $c_1^p = 2 C_{1,u} + C_{1,d\chi}|_{\mu \simeq 2{\rm GeV}}$ and $c_1^n = C_{1,u} + 2 C_{1,d}|_{\mu \simeq 2{\rm GeV}}$. 
We evaluate numerically the one-loop RG evolution of effective couplings.  
To this end, above $m_Z$ we use the {\tt RunDM} package~\cite{DEramo:2016gos,DEramo:2014nmf,Crivellin:2014qxa}, while running below $m_Z$ and the coefficients $c_1^N$ are obtained by {\tt DirectDM}~\cite{Bishara:2017nnn}. 
For example, for $M_{Z'} = m_Z$ one finds  
\begin{equation}
c_1^p \simeq 3.1 \times 10^{-3}\left(\frac{g_{\mu \mu } g_{\nu \nu}}{M_{Z'}^2} \right) 
+ 2.5 \times 10^{-3}\left(\frac{ g_{bb} g_{\nu \nu}}{M_{Z'}^2}\right)~,
\qquad c_1^n = 0~, 
\end{equation}
while for $M_{Z'} = 1$~TeV: 
\begin{equation}
c_1^p \simeq 5.6 \times 10^{-3}\left(\frac{g_{\mu \mu } g_{\nu \nu}}{M_{Z'}^2} \right) 
+ 2.3 \times 10^{-3}\left(\frac{ g_{bb} g_{\nu \nu}}{M_{Z'}^2}\right)~,
\quad c_1^n \simeq 
4.5 \times 10^{-2}\left(\frac{ g_{bb} g_{\nu \nu}}{M_{Z'}^2}\right)~. 
\end{equation}
The coupling to neutrons vanishes within the approximation $M_{Z'} \leq m_Z$.
For $M_{Z'} > m_Z$ a non-zero $c_1^n$ can be  generated, and is dominated by the top Yukawa contributions to  the RG running. 
The Dark Matter-nucleon spin-independent cross section can be straightforwardly derived from $\mathcal{L}_{\text{eff,NR}}$: 
\begin{equation}
\sigma^{\text{N}}_{\text{DD}} = \frac{ (c_1^N)^2 m_p^2 m_\nu^2}{\pi (m_p + m_\nu)^2}.   
\end{equation}
To compare with experimental bounds, which typically assume equal cross section on protons and neutrons, for a target nucleus with $Z$ protons and $A-Z$ neutrons we introduce the averaged cross section
\begin{equation}
\sigma_{\text{DD}} \simeq \frac{m_p^2 m_\nu^2}{\pi (m_p + m_\nu)^2}
{(Z c_1^p  + (A-Z)  c_1^n)^2 \over A^2}. 
\end{equation}
In the allowed parameter space relevant for the B-meson anomalies we have $g_{bb} \ll g_{\mu \mu}$. 
Assuming that  hierarchy, and also $m_{\text{p}} \ll m_\chi$, for xenon targets an approximate expression for the averaged cross section reads 
\begin{equation}
\sigma_{\text{DD}} \sim  
\Big( \dfrac{g_{\nu\nu}}{0.2} \Big)^2  
\Big( \dfrac{g_{\mu \mu}}{0.1} \Big)^2 
 \Big( \dfrac{m_Z}{M_{Z^\prime}} \Big)^{4} 
 10^{-45} ~\text{cm}^2 .
\end{equation}
In Fig.~\ref{fig:DD} we depicted the values of the $g_{\nu \nu}$ coupling satisfying the requirement of having the observed Dark Matter density as well at the correct value of the couplings $g_{\mu \mu}$ and $g_{bb}$ explaining the $R_K$ discrepancy. The left panel of that figure illustrates that, for low $M_{Z'}$, the Xenon1T collaboration excludes Dark Matter masses away from the $Z^\prime$ pole but still allows for low Dark Matter masses $m_\nu \lesssim 10$~GeV. However, as discussed in the previous subsection, such low masses are excluded by the indirect Planck constraints, therefore the complementarity of direct and indirect detection searches indicates that the Dark Matter mass has to be close to the pole $m_\nu \sim M_{Z'}/2$. 
For larger $M_{Z'}$, Dark Matter masses away from the pole region are allowed, see the right panel of Fig.~\ref{fig:DD}.

\section{Discussion and Conclusion}
\label{conclusion}

\begin{figure}
\begin{minipage}[h!]{0.47\textwidth}
\includegraphics[height=4.5cm]{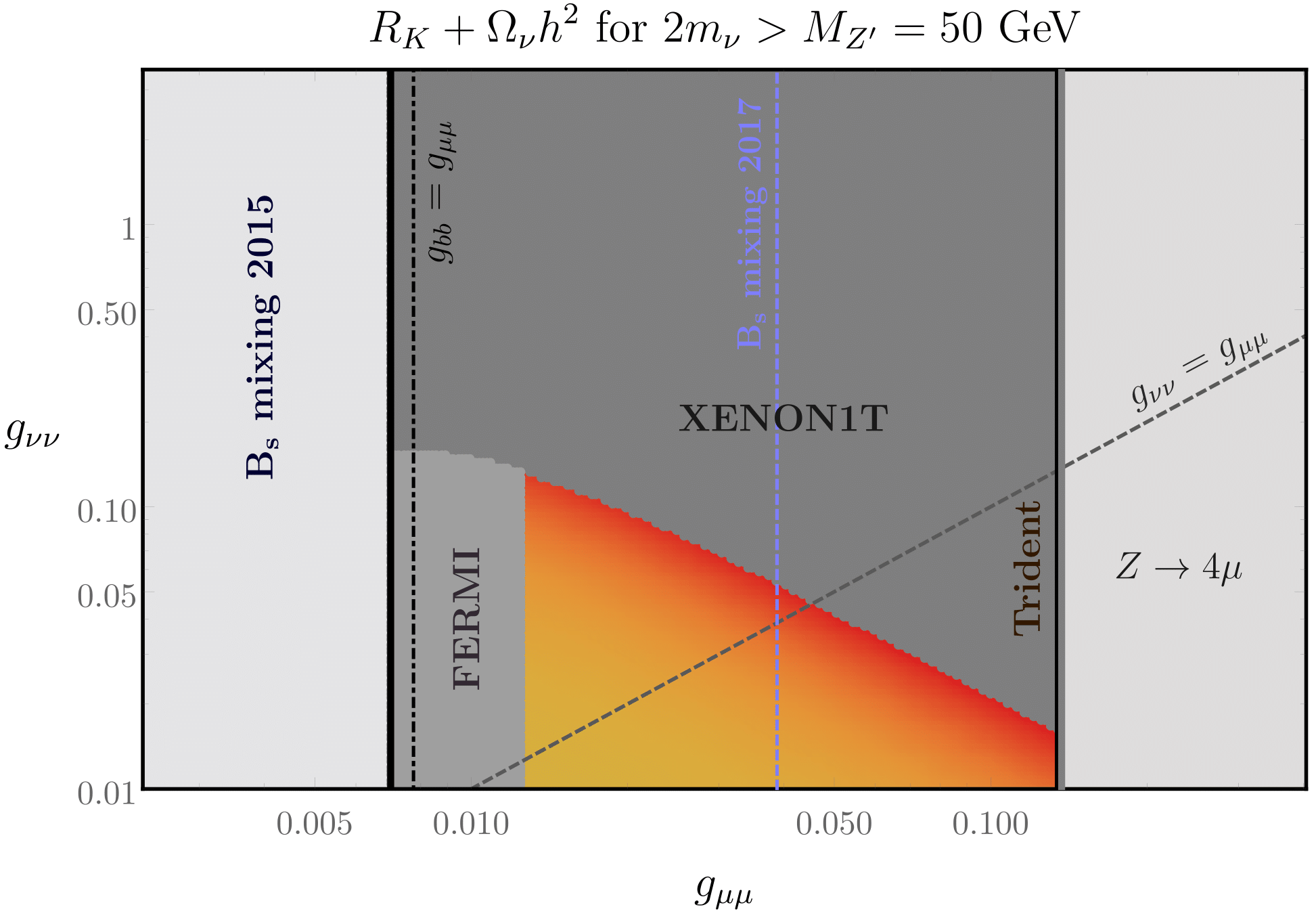}
\end{minipage}
\hfill
\begin{minipage}[h!]{0.52\textwidth}
\includegraphics[height=4.5cm]{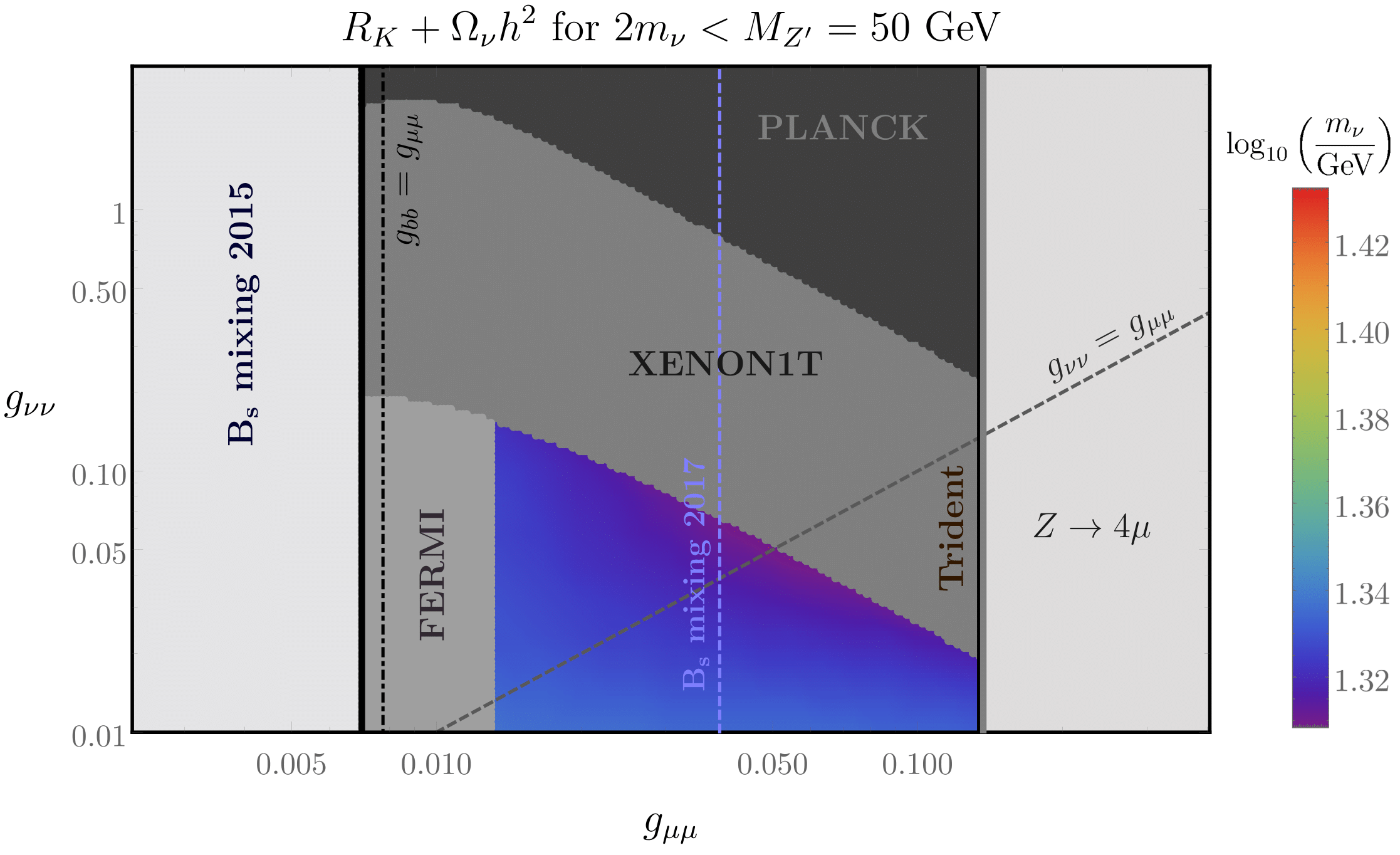}
\end{minipage}
\caption{
\label{fig:mzp50}
Summary of the constraints for $M_{Z^\prime}=50$ GeV.
See text in Sec.~\ref{conclusion} for details. }
\begin{minipage}[h!]{0.47\textwidth}
\includegraphics[height=4.5cm]{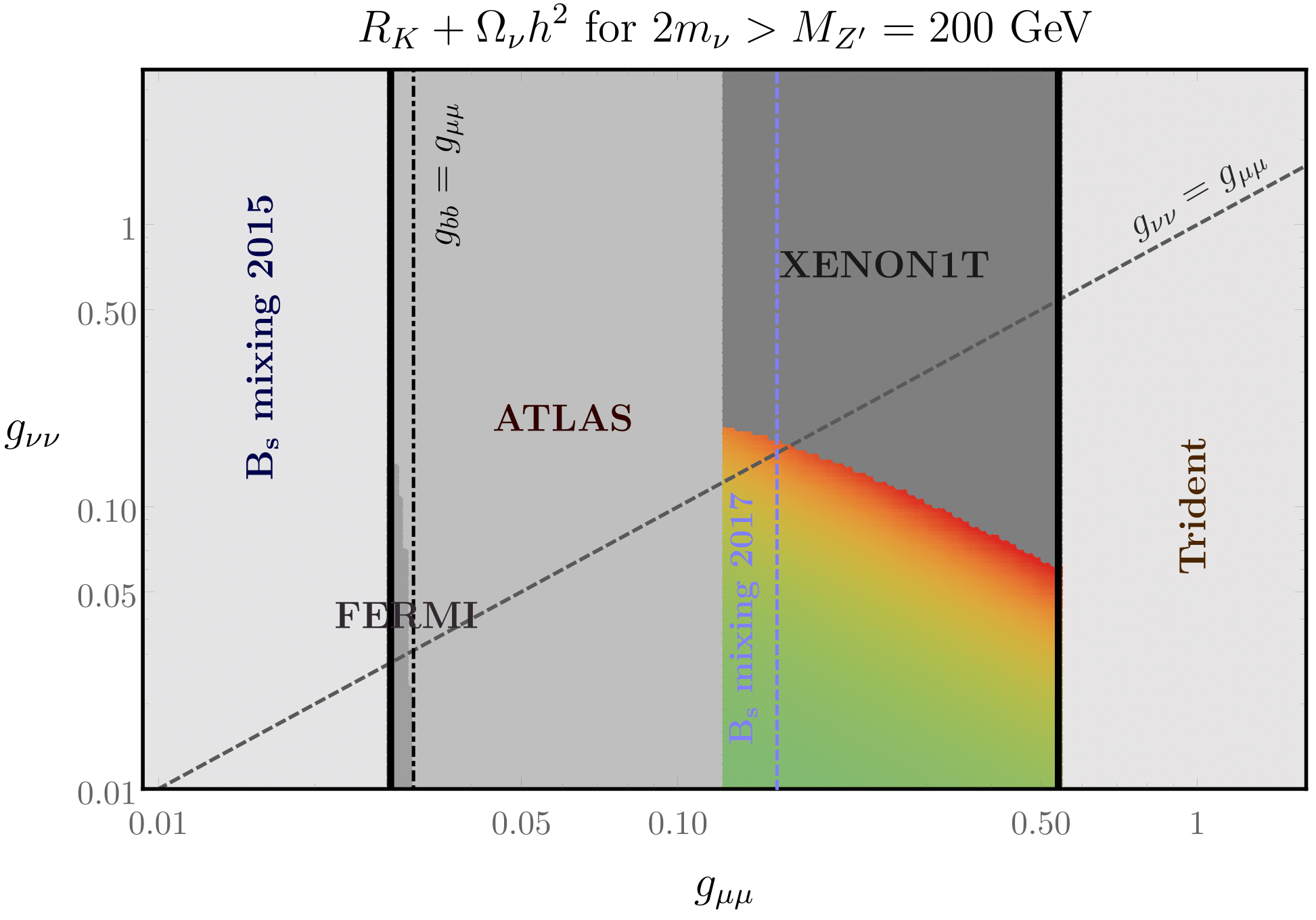}
\end{minipage}
\hfill
\begin{minipage}[h!]{0.52\textwidth}
\includegraphics[height=4.5cm]{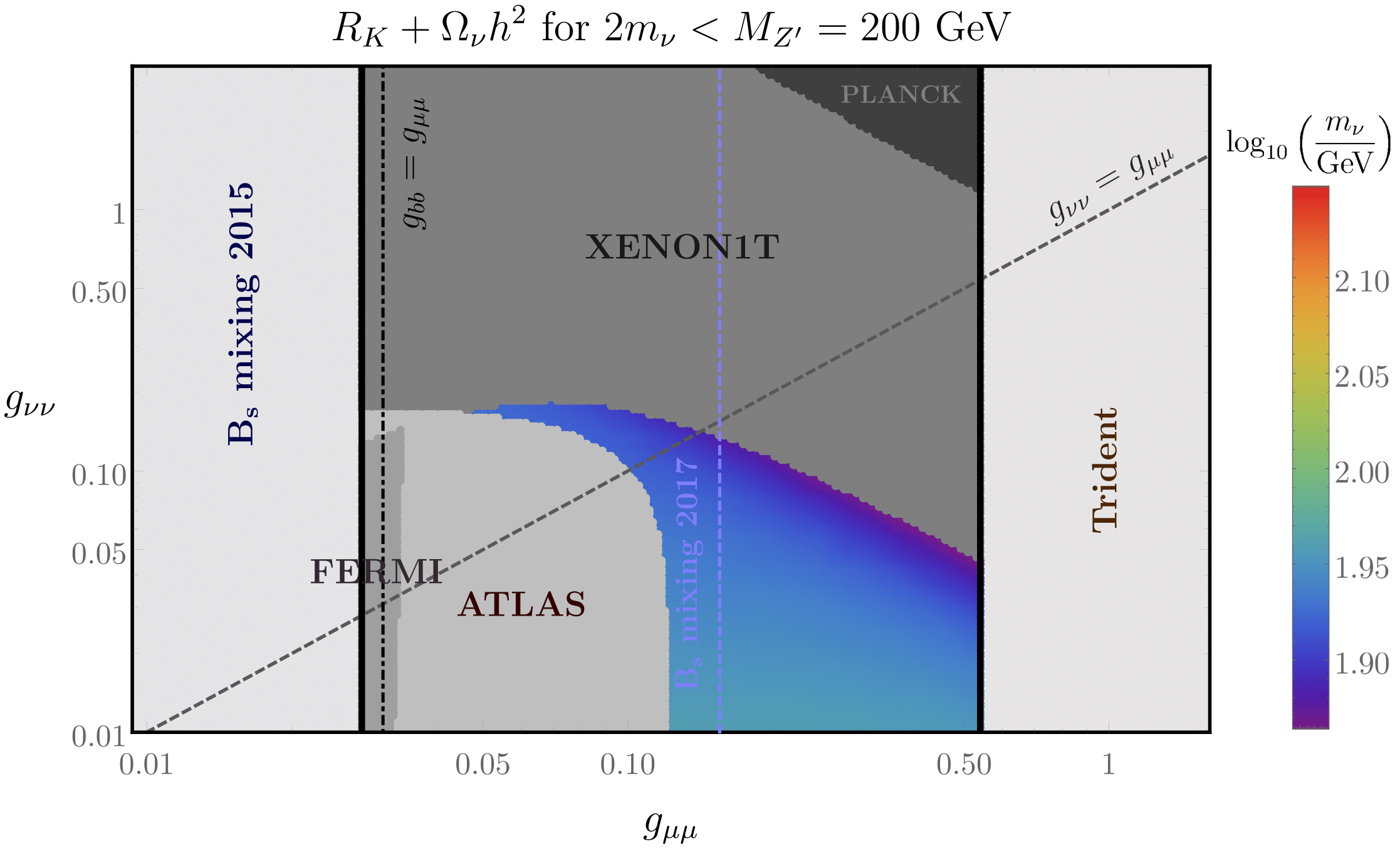}
\end{minipage}
\caption{Summary of the constraints for $M_{Z^\prime}=200$ GeV. See text in Sec.~\ref{conclusion} for details.}
\begin{minipage}[h!]{0.47\textwidth}
\includegraphics[height=4.5cm]{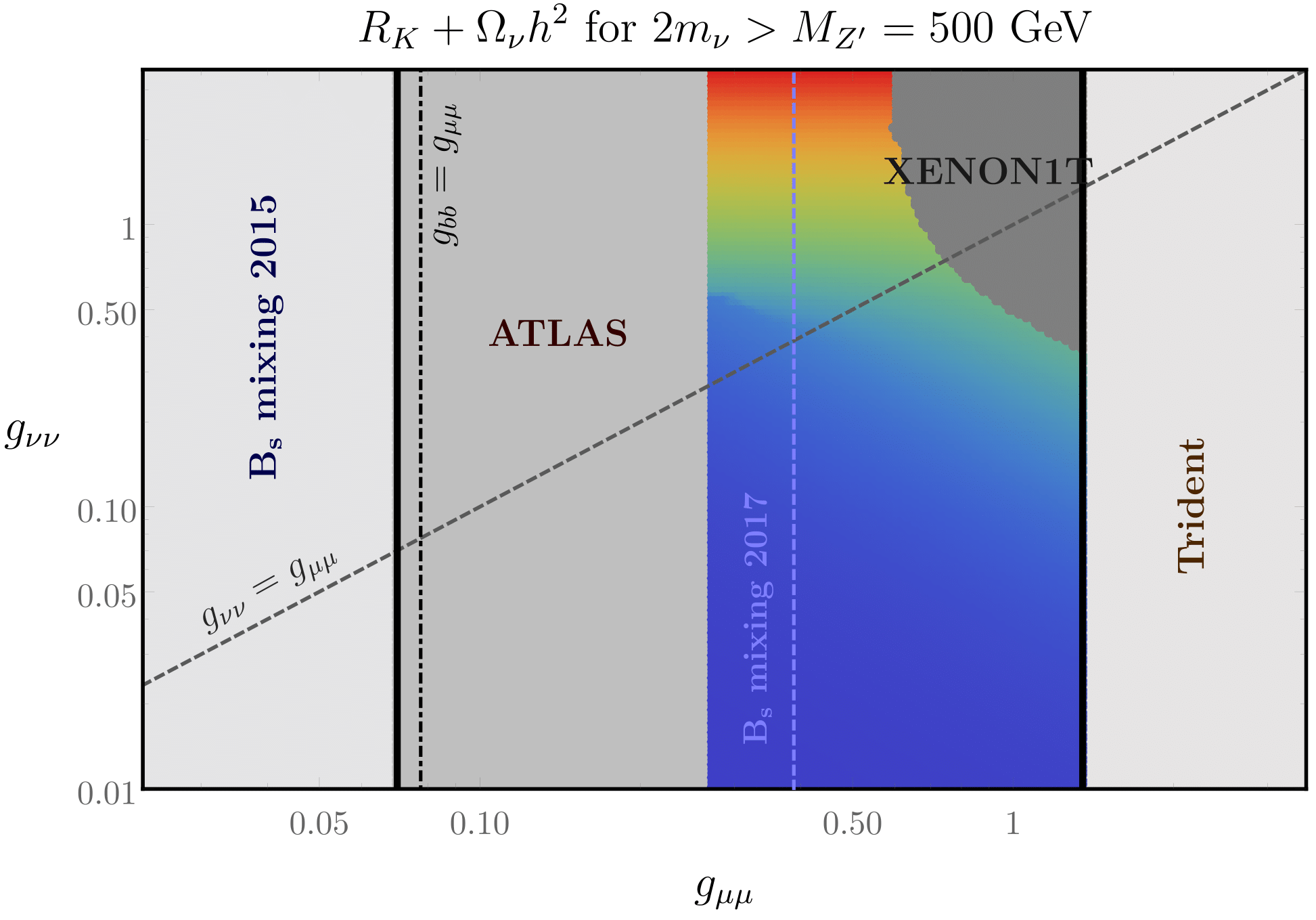}
\end{minipage}
\hfill
\begin{minipage}[h!]{0.52\textwidth}
\includegraphics[height=4.5cm]{figures/flavor/MZp500_firsthalf.png}
\end{minipage}
\caption{Summary of the constraints for $M_{Z^\prime}=500$ GeV. See text in Sec.~\ref{conclusion} for details.}
\begin{minipage}[h!]{0.47\textwidth}
\includegraphics[height=4.5cm]{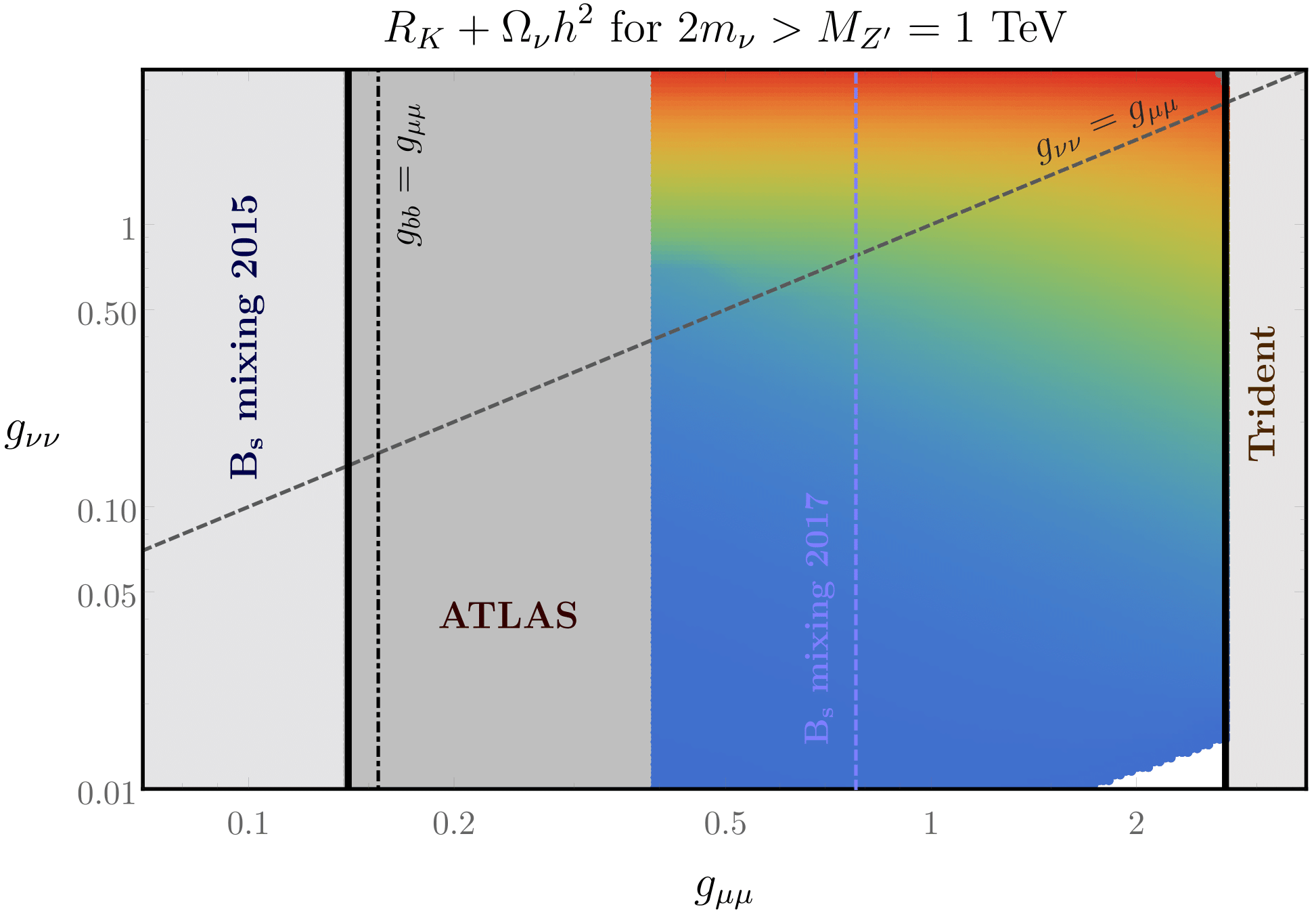}
\end{minipage}
\hfill
\begin{minipage}[h!]{0.52\textwidth}
\includegraphics[height=4.5cm]{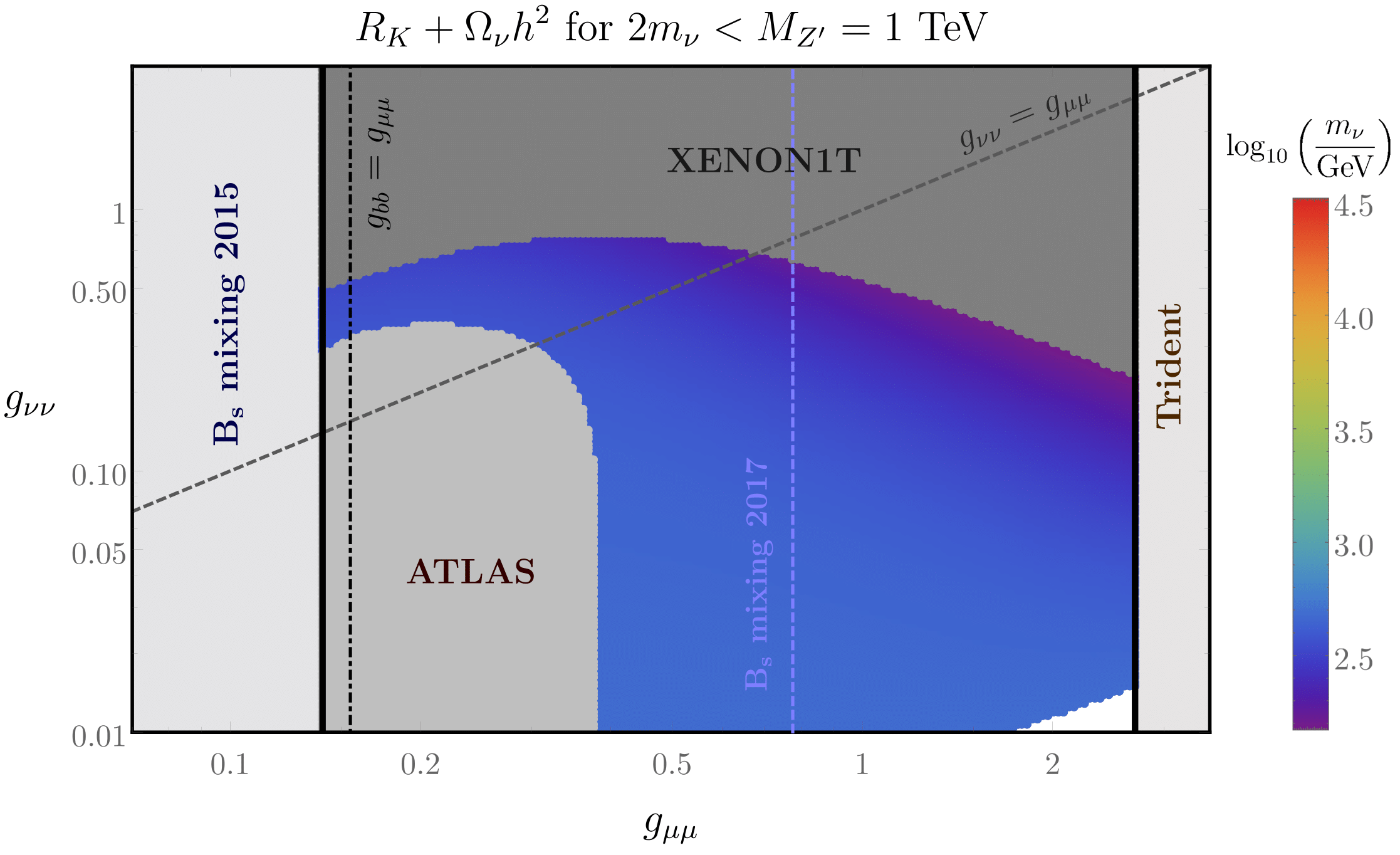}
\end{minipage}
\caption{Summary of the constraints for $M_{Z^\prime}=1$~TeV. See text in Sec.~\ref{conclusion} for details. }
\end{figure}

\begin{figure}
\begin{minipage}[h!]{0.47\textwidth}
\includegraphics[height=4.5cm]{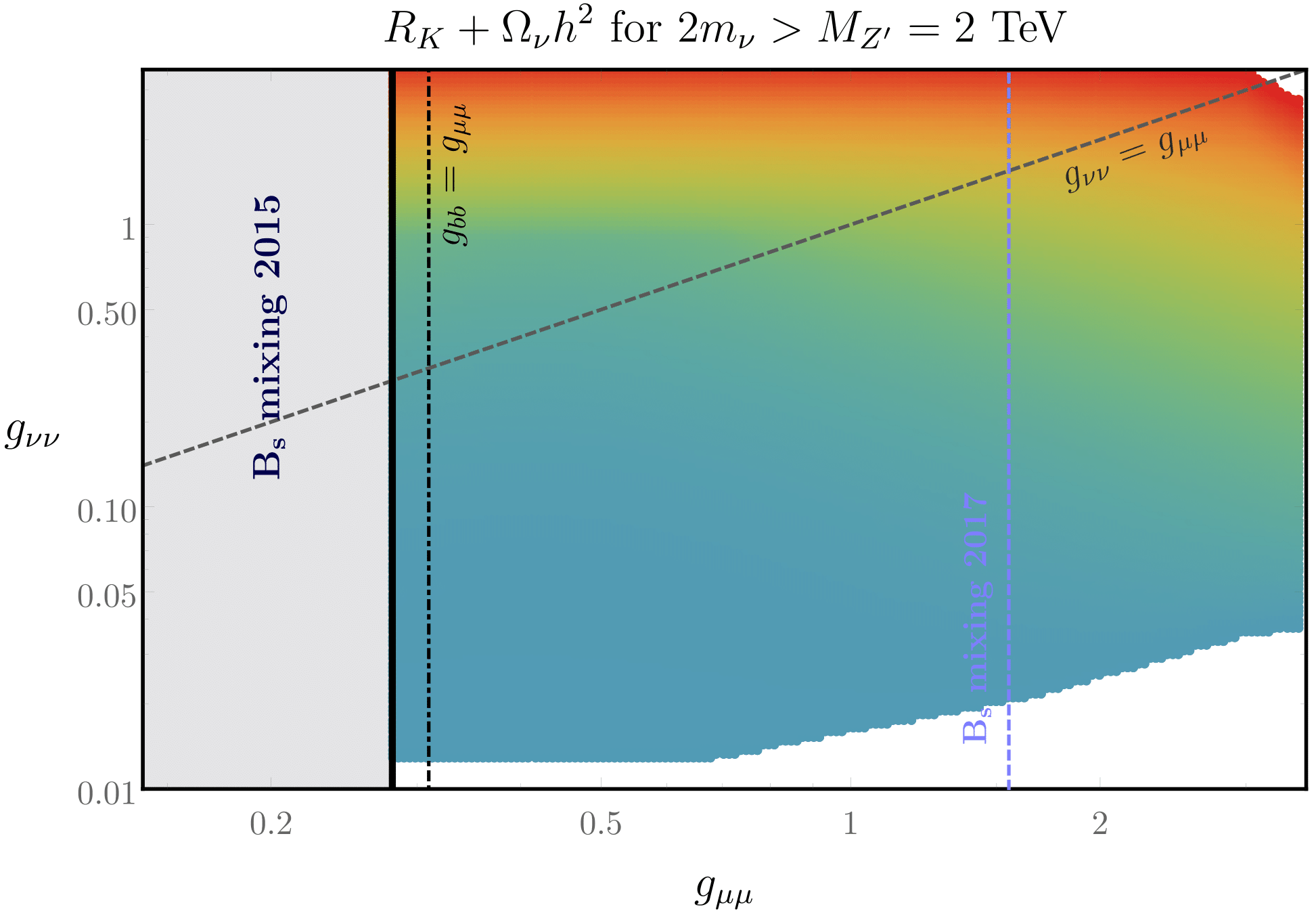}
\end{minipage}
\hfill
\begin{minipage}[h!]{0.52\textwidth}
\includegraphics[height=4.5cm]{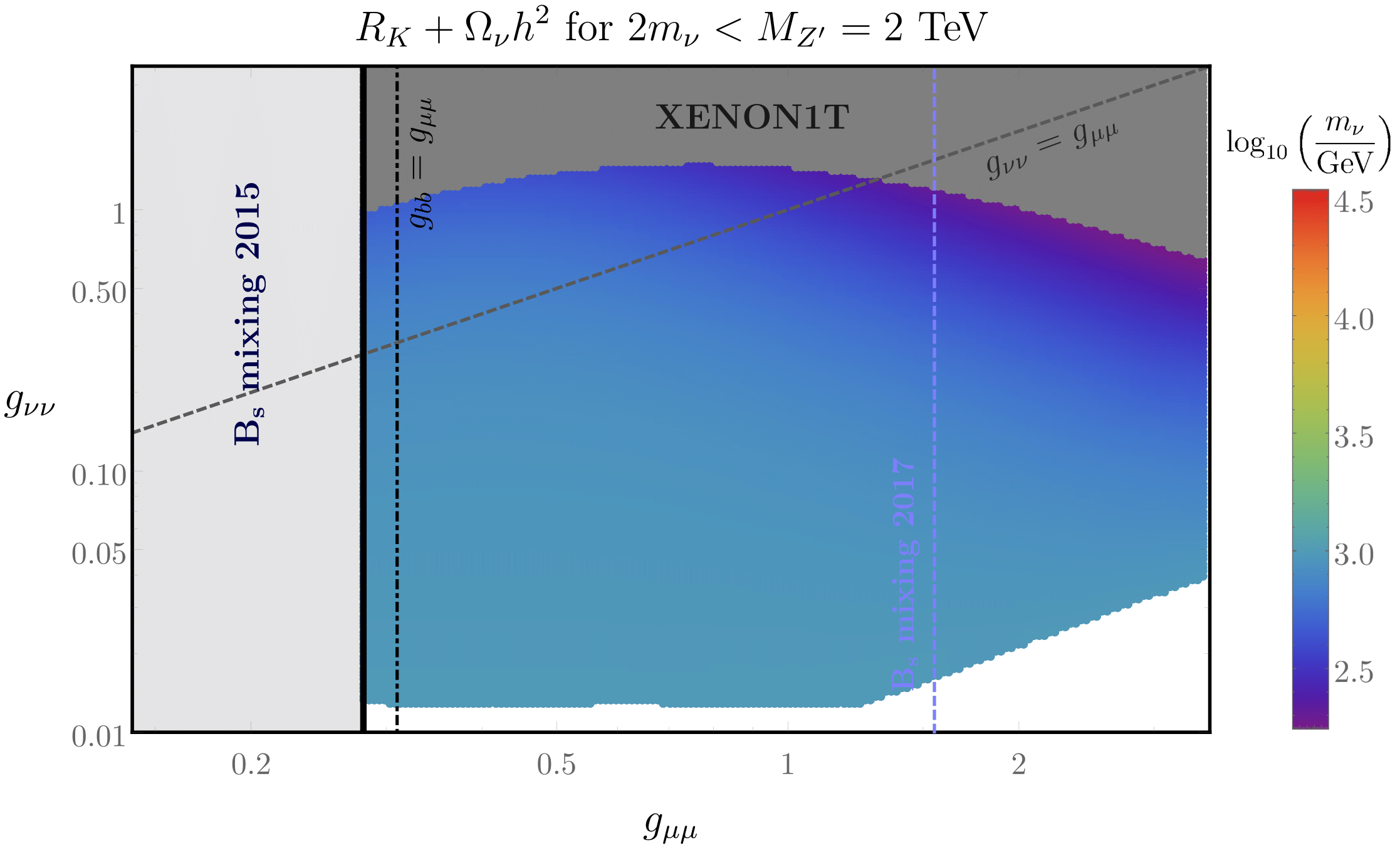}
\end{minipage}
\caption{
\label{fig:mzp2000}
Summary of the constraints for $M_{Z^\prime}=2$~TeV. See text in Sec.~\ref{conclusion} for details. }
\end{figure}

The main results are shown in Figs.~\ref{fig:mzp50}-\ref{fig:mzp2000} which show for which parameters  this model can address the B-meson  anomalies while satisfying all experimental and cosmological constraints. 
As discussed below Eq.~(\ref{eq:Zp_Rk_couplings}), the relevant parameter space is effectively five-dimensional, and spanned by the $Z^\prime$ couplings to Dark Matter ($g_{\nu \nu}$), muons ($g_{\mu \mu}$), and $b$ quarks ($g_{bb}$), and by the masses of Dark Matter ($m_{\nu}$) and the $Z^\prime$ vector messenger ($M_{Z'}$). 
We display it in the $\{g_{\mu \mu}$,$g_{\nu \nu}\}$ plane for several representative values of $M_{Z'}$. 
For each  $g_{\mu \mu}$ and $M_{Z'}$, $g_{bb}$ is fixed according to Eq.~(\ref{eq:coeff_RK}) to the best fit value reproducing the $R_{K^{(*)}}$ measurements. 

Then $m_{\nu}$ is fixed by the requirement of reproducing the correct relic abundance of Dark Matter.
There are typically two distinct solutions for $m_{\nu}$ satisfying 
$\langle\sigma v \rangle = \langle\sigma v \rangle_{\rm thermal}$, therefore for each $M_{Z'}$ in the left (right) panel we display the solutions with $m_{\nu} > M_{Z^\prime}/2$ ($m_{\nu} < M_{Z^\prime}/2$).
These solutions are color coded in Figs.~\ref{fig:mzp50}-\ref{fig:mzp2000}, from smaller (blue) to larger (red) $m_{\nu}$. 
The white regions are where we find no parameters choice to tune the annihilation cross section to the thermal value. 
The continuously gray-shaded regions are excluded by direct detection, indirect detection, $B_s$ mixing, dimuon searches at the LHC, $Z$ decay to four muons and/or muon trident constraints.
However, we choose not to shade the region excluded by the recent update of the $B_s$ mixing constraints in \cite{DiLuzio:2017fdq}, and instead represent those by a dashed blue line labeled ``$B_s$ mixing 2017''. The region represented on the left of this line is excluded by these constraints.
For any value of the $Z^\prime$ mass in the considered range there exists a range of parameters reproducing the $R_{K^{(*)}}$ anomalies and the relic abundance, and passing all experimental constraints to date. 
However, for lower $M_{Z^\prime}$ the allowed region corresponds to $g_{\nu \nu} \lesssim g_{\mu \mu}$, once the direct (Xenon1T) and indirect (Planck) detection constraints together with updated $B_s$ mixing constraints are taken into account. 
In this model the $Z^\prime$ coupling to muons is suppressed by a mixing angle between the SM 2nd generation lepton doublet and the 4th generation vector-like lepton doublet, and thus we expect  $g_{\nu \nu} \gg g_{\mu \mu}$. 
Conversely, $g_{\nu \nu} \lesssim g_{\mu \mu}$ is unnatural and would require a large hierarchy between the corresponding $U(1)^\prime$ charges, $q_{\nu_4} \ll q_{L_4}$. 
On the other hand, for $300~{\rm GeV} \lesssim M_{Z'} \lesssim 1$~TeV we find some allowed parameter space where $g_{\nu \nu}$ is a factor of few larger than $g_{\mu \mu}$, which is plausible.  
Further increasing $M_{Z'}$ requires a sizable $Z^\prime$ coupling to muons in order to address the B-meson anomalies, $g_{\mu \mu} \gtrsim 1$. 
Then we are forced back into the unnatural $g_{\nu \nu} \sim g_{\mu \mu}$ region, simply  due to perturbativity constraints on $g_{\nu \nu}$ rather than some experimental bounds.   
To summarize, assuming this model is indeed the correct explanation of the observed $R_{K^{(*)}}$ anomalies and Dark Matter relic abundance, this analysis hints at a particular corner of the parameter space where $300~{\rm GeV} \lesssim M_{Z'} \lesssim 1$~TeV, $m_{\nu} \gtrsim 1$~TeV, $g_{\nu \nu} \gtrsim 1$, $g_{bb} \sim 0.1 g_{\mu \mu}$ and $0.1 \lesssim g_{\mu \mu} \lesssim 1$. 
Incidentally, that parameter space can be probed by several distinct methods. 
First of all, the allowed  window can be further squeezed by better precision measurements of the trident $\nu_\mu N \to \mu^+ \mu^- \nu_\mu N$ process, and by improving the theoretical precision of the SM prediction for the $B_s$ meson mass difference. 
The above statement is in fact valid for all models where  the B-anomalies are addressed by a tree-level $Z'$ exchange.
What is more specific to models where the  $Z'$ interactions with the SM fermions originates from mixing of the latter with vector-like fermions is a non-vanishing $Z'$ coupling not only to muons but also to b-quarks.   
This results in a non-negligible rate of the  partonic process $b \bar b \to Z' \to \mu^+ \mu^-$ which  can be probed by dimuon resonance searches at the LHC. 
In fact, the preferred $M_{Z'}$ range is where the LHC sensitivity is optimal. 
Targeted searches for b-quark-collision initiated process (rather than recast of generic dimuon searches) could lead to a discovery signal in the near future, or to better constraints that are more stringent than the $B_s$ mixing one. 
\begin{figure}
\begin{minipage}[h!]{0.47\textwidth}
\includegraphics[height=4.5cm]{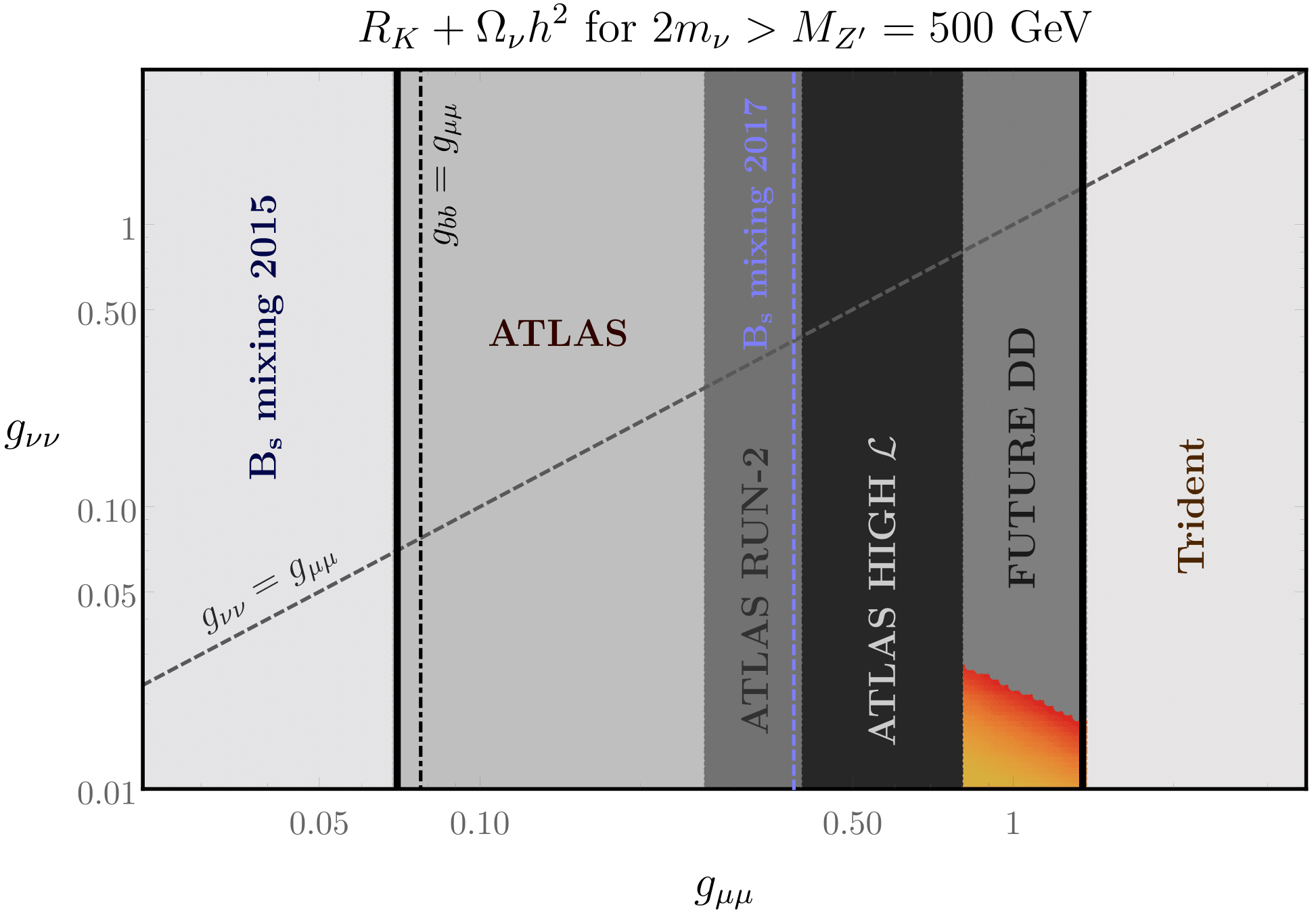}
\end{minipage}
\hfill
\begin{minipage}[h!]{0.52\textwidth}
\includegraphics[height=4.5cm]{figures/flavor/MZp500_firsthalf_prospects.png}
\end{minipage}
\caption{
\label{fig:mzp500prospects}
Projection of future constraints on the parameter space of the model for $M_{Z^\prime}=500$~GeV. 
The current ATLAS dimuon limits \cite{Aaboud:2017buh} are scaled with integrated luminosity to ${\cal L} = 200$~fb${}^{-1}$ (ATLAS RUN-2) and ${\cal L} = 3000$~fb${}^{-1}$(ATLAS HIGH ${\cal L}$).
Future direct detection limits (FUTURE DD) assume that the current Xenon1T~\cite{Aprile:2017iyp} constraints on the DM-nucleon scattering cross section are improved by two orders of magnitude.}
\end{figure}
Finally, the preferred range of Dark Matter masses and couplings can be probed by direct detection experiments, such that the improvements of one or two orders of magnitude in sensitivity in the next years, which is expected to be achieved by the LZ~\cite{Akerib:2018lyp}, DARWIN~\cite{Aalbers:2016jon} and DarkSide-20k~\cite{Aalseth:2017fik} experiments. 
As illustrated in Fig.~\ref{fig:mzp500prospects}, these future improvements should exclude the remaining most natural parameter space of the model.

%% file: parts/VSIMP.tex
\section{Introduction}\label{sec:intro}
The WIMP mechanism is potentially one of the most natural and simplest way to explain the Dark Matter abundance in our universe. However, the absence of experimental signals in direct~\cite{Arcadi:2017kky,Aprile:2016swn,Akerib:2016vxi,Tan:2016zwf} and indirect~\cite{Ackermann:2015tah,Abdallah:2016ygi} detection has strongly constrained the simplest WIMP models and the naturalness of the parameters invoked to achieve the correct relic density has become questionable. The possibility to solve some discrepancies between astrophysical observations and simulations based on $\Lambda$CDM\footnote{known as the small scale controversies, discussed in Sec.~\ref{sec:smallscalescontroversies}.} by considering a Dark Matter candidate with a sizable self-interaction has led researchers to focus attention in recent years on 
alternative thermal scenarios embedding such a DM candidate. For these reasons, alternative production mechanisms such as Strongly Interacting Massive Particles (SIMP)~\cite{Hochberg:2014dra} and ELastic DEcoupling Relic (ELDER)~\cite{Kuflik:2015isi,Kuflik:2017iqs}, with a slightly different thermal history than WIMPs, has been developed recently and typically sub-GeV Dark Matter candidates are considered in both frameworks. Thermal production of such light Dark Matter is possible if, for instance, standard $\text{DM + DM}\to \text{SM + SM}$ annihilations proceed with small couplings~\cite{Feng:2008ya} or if new annihilation mechanisms are present, such as $3~\text{DM}\to2~\text{DM}$ annihilations~\cite{Carlson:1992fn,Hochberg:2014dra,Hochberg:2014kqa} or forbidden $2\to2$ channels~\cite{Griest1991c,DAgnolo:2015ujb}.

The thermal production of SIMPs is based on freezeout of $3\rightarrow 2$ self-annihilation of Dark Matter, with  coupling between SIMPs and light Standard Model (SM) particles, which maintain kinetic equilibrium between the two sectors until  freeze-out occurs.  Various realizations of SIMP Dark Matter have been proposed in the literature, which often contain (pseudo)scalar Dark Matter particles with dark abelian or non-abelian gauge symmetries~\cite{Hochberg:2014dra,Lee:2015gsa,Choi:2015bya,Hochberg:2015vrg,Choi:2016hid,Choi:2016tkj,Choi:2017mkk}.
Massive dark vector bosons can also be SIMP candidates when stemming from non-abelian dark gauge bosons~\cite{Hambye:2009fg,Hambye:2008bq,Karam:2015jta,Karam:2016rsz,Bernal:2015ova,Heikinheimo:2017ofk}, as can be dark fermions or scalars when accompanied with a light dark photon or another scalar ~\cite{Cline:2017tka,Dey:2016qgf}.
Vector SIMP models are particularly predictive since the cubic and quartic self-interactions of Dark Matter are determined by a single gauge coupling.
If the non-abelian dark gauge symmetry is spontaneously broken by the Higgs mechanism, the resulting massive dark Higgs can equilibrate the vector SIMPs and the SM via a Higgs portal coupling~\cite{Hambye:2009fg,Hambye:2008bq,Karam:2015jta,Bernal:2015ova,Kamada:2016ois}. The spin information of the Dark Matter could be then be inferred from the invisible Higgs decay, as is the case for the WIMP~\cite{Lebedev:2011iq}.

In this work, we consider vector SIMP Dark Matter in an $SU(2)_X$ dark gauge theory, where the three massive (degenerate) $SU(2)_X$ gauge bosons play the role of vector SIMPs.
Equilibration between the dark and visible sectors can be achieved by elastic scattering between the Dark Matter and the $SU(2)_X$ dark Higgs, provided that the latter is light enough to be thermalized with the SM via the Higgs portal until freeze-out occurs. As we will see, the dark Higgs can successfully thermalize the two sectors only when it is close in mass to the Dark Matter, in which case additional forbidden $2\to2$ annihilations are important as well.
Alternatively, the dark $U(1)_{Z'}$ photon can thermalize the dark and visible sectors via its kinetic mixing with the SM hypercharge alongside its coupling to the DM, which proceed through generalized Chern-Simons (CS) terms~\cite{Anastasopoulos:2006cz,Antoniadis:2009ze,Mambrini:2009ad,Dudas:2009uq,Kim:2015vba,Arcadi:2017jqd}. In both cases of the Higgs and vector portals, we find parameter space consistent with all existing constraints. 
Our results indicate that the framework can be probed via Higgs/$Z$-boson invisible decays as well as dark Higgs/dark photon searches in current and future collider and beam dump experiments.

This chapter is organized as follows. First, we start by presenting the SIMP and ELDERs mechanisms which possess common features. Then, we present the model considered in this chapter based on a $SU(2)_X$ dark gauge theory in Sec.~\ref{sec:model}, including the relevant Higgs and gauge-mixing vector portals to the SM.
Section~\ref{sec:relic} discusses the $3\rightarrow 2$ annihilation processes setting the DM abundance, the self-scattering cross sections, and the effects of forbidden channels on the relic density.
Methods for achieving kinetic equilibrium between the dark and visible sectors via Higgs mixing and/or gauge mixing are addressed in Sec.~\ref{sec:portal}. We conclude in Sec.~\ref{sec:conc}.

\section{The SIMP and ELDER mechanisms}\label{sec:model}
In the WIMP paradigm, one of the key assumption to generate the correct relic density via the freeze-out mechanism is the primordial thermal equilibrium between the dark sector and the Standard Model particle content. This equilibrium can be ensured through rapid scatterings between a DM candidate $\chi$ and some SM particle: $\chi + \text{SM} \rightarrow \chi + \text{SM}$ and in the simplest WIMP cases, the DM particles annihilate while becoming non-relativistic $\chi + \chi \rightarrow \text{SM} + \text{SM}$. However the story would be different if interactions among the dark sector were also present in the theory, such as self-annihilations which might affect the DM density. In the former case, the fate of the DM density evolution would depend on the relative importance of the rates of these processes:
\begin{itemize}
\item Scattering rate $\Gamma_{\rm scat}$ of processes such as $\chi + \text{SM} \leftrightarrow \chi + \text{SM}$
\item Annihilation rate $\Gamma_{\rm ann}$ of the process $\chi + \chi \rightarrow \text{SM} + \text{SM}$
\item Self-annihilation rate  $\Gamma_{\rm self}$ among the dark sector : $\chi + \chi + \chi \rightarrow \chi + \chi  $.
\end{itemize}
Assuming that $\Gamma_{\rm scat}$ is large enough to ensure a primordial thermal equilibrium between the dark sector and the SM particles, three distinct regimes are possible:

\paragraph{WIMP:} In the WIMP case, $\Gamma_{\rm scat}$ is large enough to ensure kinetic equilibrium until the freeze-out time and the DM density undergo a depletion through annihilations into SM particles: $\Gamma_{\rm ann} \gg \Gamma_{\rm self}$. Part II of this thesis is completely devoted to the study of specific realizations of this mechanism. 

\paragraph{Strongly Interacting Massive Particles (SIMP)~\cite{Hochberg:2014dra}:} The SIMP case correponds to the regime where scattering are still efficient enough to ensure a kinetic equilibrium between the dark sector and the SM bath until the freeze-out temperature. However in this case the freeze-out occurs through annihilations in the dark sector, implying $\Gamma_{\rm ann} \ll \Gamma_{\rm self}$. In order for the kinetic equilibrium state to be achieved until the freeze-out time, the typical kinetic energy injected by self annihilation processes per unit of time $\dot{K}_{\rm self}$ must be smaller than the typical energy transferred per unit of time $\dot{K}_{\rm el}$ in elastic scatterings between the dark sector and the SM bath. The freeze-out temperature $T_{\rm F}$ can be defined as the temperature for which these two processes become comparable:
\begin{equation}
\dot{K}_{\rm self} (T_{\rm F}) \sim \dot{K}_{\rm el} (T_{\rm F})~.
\end{equation}
Assuming that the reaction $3 \chi \rightarrow 2 \chi$ is the dominant self-annihilation process, as shown in Sec.~\ref{sec:SIMPappendix}, the Boltzmann equation for the Dark Matter density $n_{\chi}$ typically takes the form: 
\begin{equation}
\dfrac{\diff n_\chi}{\diff t}+3Hn_\chi=-\la \sigma v^2 \ra (n_\chi^3-n_\chi^{\text{eq}}n_\chi^2)
\end{equation}
where $n_\chi^{\text{eq}}$ is the expected DM thermal equilibrium distribution and the quantity $\la \sigma v^2 \ra$ is the analogous of the velocity averaged annihilation cross section for the $3\rightarrow 2$ process. As detailed in Sec.~\ref{sec:SIMPappendix}, the relic density can be expressed as:
\begin{equation}
\Omega_\chi h^2=\dfrac{m_\chi s_0 h^2}{\rho_c^0}\dfrac{\sqrt{2 H(m_\chi)}}{s(m_\chi)}x_{\rm F}^2 \la \sigma v^2 \ra^{-1/2}~,
\end{equation}
where $x_{\rm F} \equiv m_\chi / T_{\rm F}$ and $T_{\rm F}$ is the freeze-out temperature. $T_{\rm F}$ can be estimated as the temperature for which the RHS of the Boltzmann equation becomes comparable to the Hubble expansion dependent term on the LHS. In this regime, typically the correct relic density can be achieved for DM masses $m_{\rm DM}\sim 50~\text{MeV}$, assuming sizable couplings in the dark sector of $\mathcal{O}(1)$ for a gauge interaction for instance. Thus, in such a construction, the self-interaction cross section is typically much larger than in the WIMP context.

\begin{figure}[h!]
\begin{center}
\includegraphics[width=0.6\linewidth]{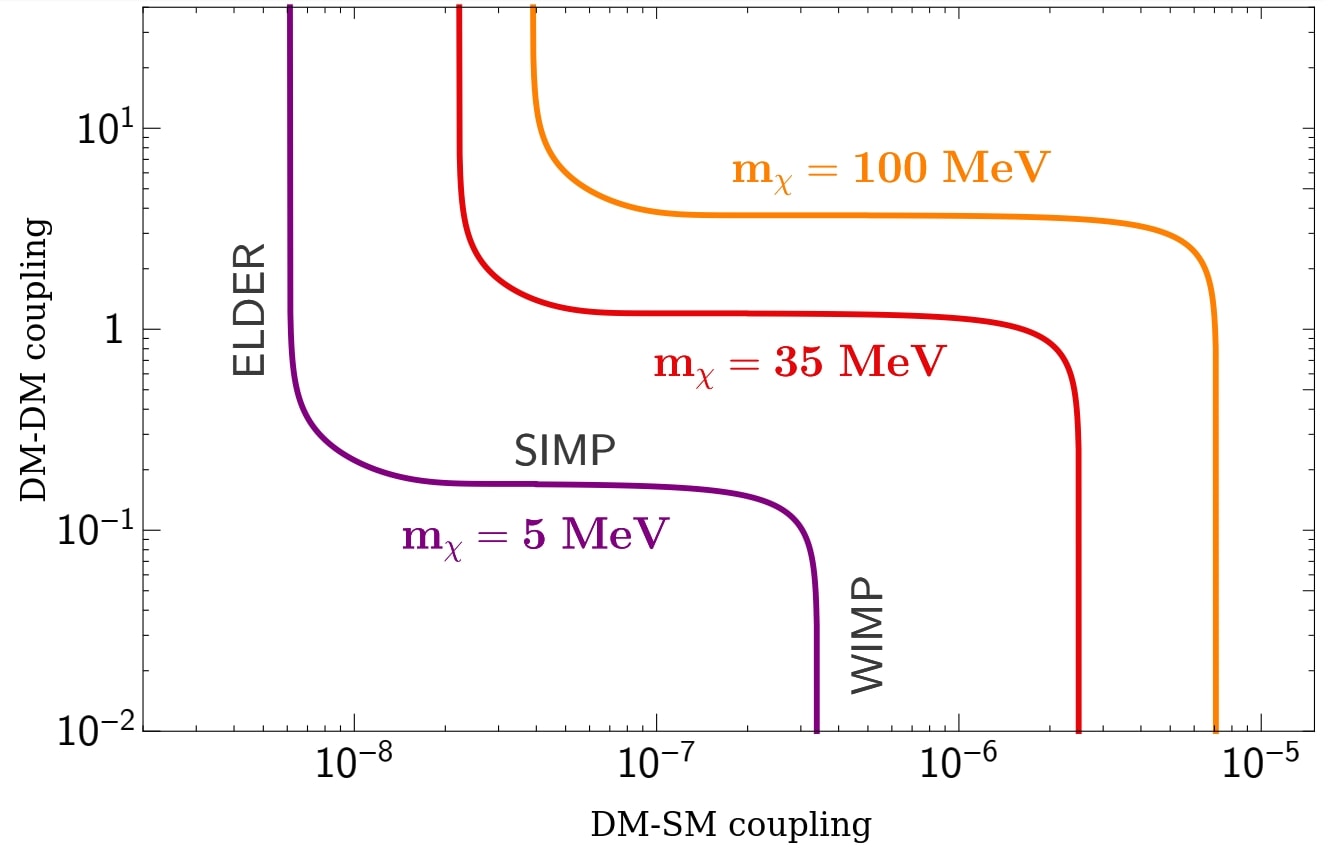}
\caption{Illustration of the correct relic density curve in the 3 regimes discussed in this section showing the influence of the dark sector coupling ("DM-DM coupling") and the coupling between the SM bath and the dark sector ("DM-SM coupling"). Taken and adapted from~\cite{Kuflik:2017iqs}.} 
\label{fig:ELDER-SIMP-WIMP}
\end{center}
\end{figure}

\paragraph{Elastic Decoupling Relic (ELDER)~\cite{Kuflik:2015isi}:} In this case the following hierachy is present among the rates $ \Gamma_{\rm ann} \lesssim \Gamma_{\rm scat} \lesssim \Gamma_{\rm self}$ therefore the dark sector and the SM bath would decouple before freeze-out, while maintaining a thermal equilibrium in their own seperate baths. The thermal decoupling typically occurs while the Dark Matter become non-relativistic and by entropy conservation in both sectors:
\begin{equation}
T_\chi^{1/2}e^{-m_\chi/T_\chi}\propto T^3~,
\end{equation}
where $T_\chi$ and $T$ are the temperatures of the dark sector and the SM respectively. The previous relation implies the following approximate behavior of $T_\chi$:
\begin{equation}
T_\chi \simeq \dfrac{T_{\rm d}}{1+3 \dfrac{T_{\rm d}}{m_\chi} \log \Big( \dfrac{T_{\rm d}}{T}
\Big)}~,
\end{equation}
where $T_{\rm d}$ is the decoupling temperature. As discussed in Sec.~\ref{sec:SIMPappendix} the DM temperature evolution, assuming the dark sector in thermal equilibrium, is given by
\begin{equation}
\dfrac{\partial T_\chi}{\partial T}=\dfrac{3 T_\chi^2}{m_\chi T}+\dfrac{4 T_\chi^2}{3m_\chi^2}\dfrac{\gamma(T)}{HT}\Big( T_\chi - T \Big)~.
\end{equation}
The first term on the RHS of this equation drives the DM temperature to evolve grossly as $T_\chi \appropto 1/(-\log T)$ while the second term tends to moderate the temperature difference between the two sectors with an efficiency proportional to the momentum relaxation rate $\gamma(T)$ defined in Sec.~\ref{sec:SIMPappendix}. The decoupling temperature can be estimated as the temperature for which the second term becomes of the order of one:
\begin{equation}
\dfrac{4 T_{\rm d}^2}{3 m_\chi^2} \dfrac{\gamma( T_{\rm d})}{H} \sim 1~.
\end{equation}
When the temperature drops below $T_{\rm d}$, the dark sector and SM bath decouple. Thermal equilibrium is maintained in the dark sector because of a large $\Gamma_{\rm self}$ rate, imposing the DM chemical potential to vanish. Dark Matter particles then undergo a canibalization period where the process $3 \chi \rightarrow 2 \chi$, kinematically favoured, is highly dominant over the reverse process. Therefore the freeze-out will occur when the $3 \rightarrow 2$ process is not frequent enough to ensure a chemical equilibrium in the dark sector. Thus, we can define the freeze-out temperature $T_{\rm F}$ via the condition:
\begin{equation}
\Gamma_{\rm self} (T_{\rm F}) \sim H(T_{\rm F})
\end{equation}
The relic density can be expressed as a function of the freeze-out and decoupling temperatures as~\cite{Kuflik:2017iqs}:
\begin{equation}
\Omega_\chi = \dfrac{45}{2^{5/2} \pi^{3/2}}\left( \dfrac{m_\chi s_0}{\rho_c^0} \right)\left( \dfrac{g_\chi}{g_{\star s}(x_{\rm d})} \right) \dfrac{x_{\rm d}^{5/2} e^{-x_{\rm d}}}{x_{\rm F}}
\end{equation}
where $x_{\rm d, F} \equiv m_\chi/T_{\rm d, F}$, $g_\chi$ is the number of degrees of freedom of the DM states and $g_{\star s}(x_{\rm d})$ is the effective number of relastivistic degrees of freedom of the SM bath at the decoupling temperature. In order to effectively realize the ELDER mechanism, one typically needs to consider larger couplings than in the SIMP case, pushing them towards the perturbativity limit as represented in Fig.~\ref{fig:ELDER-SIMP-WIMP}.

\section{The model}\label{sec:model}

Here we present the framework for vector SIMPs: We start with the dark gauge theory, and then describe the Higgs interactions as well as kinetic gauge mixing and couplings between the dark photon and the Dark Matter.

\subsection{The dark sector}\label{ssec:darksector}

We consider as a model for non-abelian SIMP Dark Matter an $SU(2)_X$ gauge theory in the dark sector, broken completely due to the VEVs of a dark Higgs doublet $\Phi$.  The massive gauge bosons of $SU(2)_X$, denoted by $X^i_\mu$ $(i=1,2,3)$, are degenerate and stable due to a dark custodial isospin symmetry, and are a Dark Matter candidate~\cite{Hambye:2009fg,Hambye:2008bq,Bernal:2015ova}. The accidental custodial symmetry persists in the presence of the Higgs portal and $Z'$ portal with the generalized Chern-Simons term which we discuss later, maintaining the stability of the Dark Matter.

The Lagrangian for the dark sector is given by
\be
{\cal L}= -\frac{1}{4} {\vec X}_{\mu\nu}\cdot {\vec X}^{\mu\nu}+ \mathcal{L}_{\rm scalar}\,,
\ee
where the field strength tensors are ${\vec X}_{\mu\nu}=\partial_\mu {\vec X}_\nu-\partial_\nu {\vec X}_\mu+g_X ({\vec X}_\mu\times {\vec X}_\nu) $. The scalar potential is given by
\bea\label{eq:Vhiggs}
\mathcal{L}_{\rm scalar}&=&   |D_\mu \Phi|^2 +m^2_\Phi |\Phi|^2 -\lambda_\Phi |\Phi|^4\,,
\eea
with the covariant derivatives for the dark Higgs doublet is $D_\mu \Phi=(\partial_\mu-\frac{1}{2} i g_X {\vec\tau}\cdot {\vec X}_\mu)\Phi$.

After expanding the dark Higgs fields around the VEV as $\Phi=\frac{1}{\sqrt{2}}(0,v_X+ \phi)^T$ in unitary gauge, one obtains gauge boson mass of $m_X=\frac{1}{2} g_X v_X$. The self-interactions of the vector Dark Matter and its interactions with the dark Higgs $\phi$ are given by
\bea
{\cal L} &\supset&-\frac{1}{2}g_X (\partial_\mu {\vec X}_\nu -\partial_\nu {\vec X}_\mu)\cdot ({\vec X}^\mu\times {\vec X}^\nu) -\frac{1}{4} g^2_X ({\vec X}_\mu\cdot {\vec X}^\mu)^2 \nonumber \\
&&+\frac{1}{4}g^2_X({\vec X}_\nu\cdot {\vec X}^\mu)({\vec X}_\mu\cdot {\vec X}^\nu)+\frac{1}{2} m^2_X {\vec X}_\mu \cdot {\vec X}^\mu \left(\frac{2\phi}{v_X}+\frac{\phi^2}{v^2_X} \right)\,. \label{darkHiggs}
\eea

The non-abelian interactions among the vector bosons $X$ allow for $3\rightarrow 2$ annihilations as SIMPs.  This idea is actually much more general than we discussed above.  This symmetry breaking can also be considered as dynamical, as a result of chiral symmetry breaking $SU(2)_L \times SU(2)_R \rightarrow SU(2)_V$ in an $SU(N_c)$ gauge theory.  This corresponds to the limit where $m_\phi \rightarrow \infty$ at low energies, while resonances can play an important role at higher energies.  In this case, the coupling $g_X$ is still considered perturbative.

Alternatively, we can consider the theory with a Higgs doublet in the strongly coupled regime $g_X \gg 1$.  As pointed out by 't Hooft~\cite{tHooft:1979yoe}, an $SU(2)$ gauge theory with a doublet scalar does not have an order parameter to distinguish the broken and confining phases, and hence the two phases are continuously connected, akin to liquid and gas phases of water at high pressures.  In the strong coupling case, the vector SIMP is described by the interpolating field $\Phi^\dagger i\mathop{D}\limits^{\leftrightarrow}{}_\mu \Phi$, while the dark Higgs by $\Phi^\dagger \Phi$.  Given enough parameters in the model $(g_X, m_\Phi^2, \lambda_\Phi)$, one can most likely have the dark Higgs heavier than the vector SIMP as required (see below); such a discussion requires numerical simulations and is beyond the scope of this work.

\subsection{Higgs portal}\label{ssec:higgs}
The dark Higgs provides a portal between the dark sector and the visible sector, since the dark and SM scalars may interact at the renormalizable level,
\beq
{\cal L}_{\rm higgs} =  \lambda_{\Phi H}|\Phi|^2|H|^2+\lambda_{SH}|S|^2 |H|^2+ \lambda_{\Phi S}|\Phi|^2|S|^2\,.
\eeq
Here, a complex scalar field $S$ is introduced for giving mass to $Z’$ gauge boson by Higgs mechanism in the later discussion on $Z’$ portal in Sec.~\ref{ssec:gauge}. Since $Z’$ is assumed to be heavier than Dark Matter in our model, we assumed that the radial mode of $S$ has no significant mixing with the dark Higgs $\phi$ and the SM Higgs.

The SM and dark Higgs bosons are then mixed by
\be\label{eq:hmix}
\left( \begin{array}{c} h_1 \\ h_2 \end{array} \right)=  \left( \begin{array}{cc} \cos\theta & -\sin\theta \\  \sin\theta & \cos\theta \end{array} \right) \left( \begin{array}{c} \phi \\ h \end{array} \right)\,,
\ee
where $h_1,h_2$ are mass eigenstates of mass
\be
m^2_{h_1,h_2}=\lambda_\Phi v^{ 2}_X+\lambda_H v^2 \mp \sqrt{(\lambda_\Phi v^{ 2}_X-\lambda_H v^2)^2+ \lambda^2_{\Phi H}v^{2}_X v^2 }\,,
\ee
and the mixing angle is given by
\be
\tan 2\theta = \frac{\lambda_{\Phi H} v_X v}{\lambda_H v^2-\lambda_\phi v^2_X}.
\ee
Here, we assume that the additional Higgs field $s$ for $U(1)_{Z'}$ is heavy enough so that its mixing effects with the above Higgs fields is negligible.
The Higgs mixing yields interactions between the vector DM and the SM particles,
\bea
{\cal L}&\supset&\frac{m^2_X}{v_X}\, {\vec X}_\mu \cdot {\vec X}^\mu (\cos\theta\, h_1+\sin\theta\, h_2)+\frac{m^2_X}{2v^2_X}\, {\vec X}_\mu \cdot {\vec X}^\mu (\cos\theta\, h_1+\sin\theta\, h_2)^2 \nonumber \\
&&-\frac{m_f}{v}\, {\bar f} f (-\sin\theta\, h_1+\cos\theta\, h_2)\,,
\eea
enabling communication between the two sectors.

In the presence of such Higgs-portal couplings, the SM Higgs can decay invisibly into a pair of dark gauge bosons or dark higgses, with decay rates
\bea\label{eq:h2decay}
\Gamma(h_2\rightarrow X X)&=& \frac{3\sin^2\theta m^3_{h_2}}{32 \pi v^2_X}\,\bigg(1-\frac{4m^2_X}{m^2_{h_2}}+\frac{12 m^4_X}{m^4_{h_2}} \bigg) \sqrt{1-\frac{4m^2_X}{m^2_{h_2}}}\,,\nonumber \\
\Gamma(h_2\rightarrow h_1 h_1)&=& \frac{\lambda^2_{\Phi H} v^2}{32\pi m_{h_2}}\sqrt{1-\frac{4m^2_{h_1}}{m^2_{h_2}}}\,.
\eea
The visible decays of the SM Higgs are scaled down universally by $\cos^2\theta$ due to the Higgs mixing.  As we will see in Sec.~\ref{ssec:higgsportal}, the bound on invisible Higgs decays places a strong constraint on the allowed mixing, and hence on the possibility that the Higgs portal maintains kinetic equilibrium between the two sectors.

\subsection{Vector portal}\label{ssec:gauge}

In addition to the Higgs portal, we can gauge a $U(1)_{Z'}$ symmetry acting on the complex scalar $S$, with the covariant derivative $D_\mu S=(\partial_\mu -ig_{Z'}Z'_\mu )S$.  The $U(1)_{Z'}$ massive gauge boson $Z'$ can connect the dark and visible sectors, in the presence of gauge kinetic mixing with the SM hypercharge as well as DM-$Z'$ interactions:
\beq
\mathcal{L}_{\rm vector} = -\frac{1}{2} \,\sin\xi\, Z'_{\mu\nu} B^{\mu\nu}+{\cal L}_{XXZ^\prime}\,.
 \eeq
Here ${\cal L}_{XXZ^\prime}$ generates a 3-pt interaction between $XXZ^\prime$; it may be generated by a non-abelian Chern-Simons (CS) term, as will be discussed below.

The kinetic and mass terms for the $Z$ and $Z'$ gauge bosons~\cite{Choi:2015bya} is diagonalized
\be
\left(\begin{array}{c} B_\mu \\ W^3_\mu \\ Z'_\mu \end{array}\right)=\left( \begin{array}{ccc} c_W & -s_W c_\zeta +t_\xi s_\zeta  & -s_W s_\zeta-t_\xi c_\zeta  \\   s_W & c_W c_\zeta & c_W s_\zeta   \\  0 & -s_\zeta/c_\xi & c_\zeta/ c_\xi   \end{array} \right)  \left(\begin{array}{c} A_\mu \\ Z_{1\mu} \\ Z_{2\mu} \end{array}\right)
\ee
where $(B_\mu, W^3_\mu,Z'_\mu)$ are hypercharge, neutral-weak and dark gauge fields, $(A_\mu, Z_{1\mu},Z_{2\mu})$ are mass eigenstates, and $s_W\equiv \sin\theta_W, c_W\equiv \cos\theta_W$, etc.  Here, $Z_{1}$ is $Z$-boson-like and $Z_{2}$ is $Z'$-boson-like, with masses
\begin{align}
m^2_{1,2}= \frac{1}{2}\left[m^2_Z (1+s^2_W t^2_\xi)+m^2_{Z'}/c^2_\xi\pm \sqrt{(m^2_Z(1+s^2_W t^2_\xi)+m^2_{Z'}/c^2_\xi)^2- 4m^2_Z m^2_{Z'} /c^2_\xi} \,\right]\,,
\end{align}
where the mixing angle is
\be
\tan 2\zeta =\frac{m^2_Z s_W \sin 2\xi}{m^2_{Z'}-m^2_Z(c^2_\xi-s^2_W s^2_\xi )}\,.
\ee
The electromagnetic and neutral-current interactions are then
\bea
{\cal L}_{\rm EM/NC}&=& e A_\mu J^\mu_{\rm EM} + Z_{1\mu} \bigg[ e \varepsilon J^\mu_{\rm EM}+\frac{e}{2s_Wc_W} (c_\zeta-t_W \varepsilon/t_\zeta  )J^\mu_Z -g_{Z'} \frac{s_\zeta}{c_\xi} J^\mu_{Z'} \bigg] \nonumber\\
&&+Z_{2\mu}  \bigg[- e \varepsilon J^\mu_{\rm EM}+\frac{e}{2s_Wc_W} (s_\zeta+t_W\varepsilon)J^\mu_Z +g_{Z'} \frac{c_\zeta}{c_\xi} J^\mu_{Z'} \bigg]\,,
\eea
 where $\varepsilon\equiv c_W t_\xi c_\zeta\simeq c_W \xi$ for $|\xi|\ll 1$, and $J^\mu_{\rm EM}$, $J^\mu_Z$ and $J^\mu_{Z'}$ are electromagnetic, neutral and dark currents, respectively.
 For $m_{Z'}\ll m_Z$, one has $\zeta\simeq -s_W \xi=-t_W \varepsilon$, so the neutral current interaction of the dark photon is negligible due to $s_\zeta+t_W\varepsilon \simeq \zeta+ s_W\xi\simeq 0$.

There are no direct couplings between the SM and the non-abelian vector Dark Matter at the renormalizable level, because of the non-abelian gauge symmetry. Likewise, there are no direct renormalizable interactions between the $Z^\prime$ and the $X$-boson, since the dark Higgs are not charged under both symmetries (in other words, the dark Weinberg angle vanishes).

 If heavy fermions charged under both $SU(2)_X$ and $U(1)_{Z'}$ are present in the theory, they may generate low-energy effective $XXZ^\prime$ interactions via triangle diagrams. From the effective theory point of view these may manifest as generalized non-abelian Chern-Simons terms~\cite{Dudas:2009uq,Kim:2015vba},
\be
{\cal L}_{\rm CS,EFT} \supset c_1 \epsilon^{\mu\nu\rho\sigma} Z'_\mu {\vec X}_\nu\cdot (\partial_\rho {\vec X}_\sigma - \partial_\sigma {\vec X}_\rho)  \label{CS1}\,.
\ee
Although the coefficient $c_1$ is dimensionless, these are non-renormalizable operators and arise from gauge dimension-8 operators, known as D'Hoker-Farhi terms~\cite{DHoker:1984izu},
\bea
{\cal L}_{\rm CS}&\supset & \frac{i}{M^4} S^\dagger D^\mu S  (D^\nu \Phi)^\dagger {\tilde X}_{\mu\nu}\Phi+{\rm c.c.}
\eea
Likewise, an effective 3-pt interaction can be generated by the gauge invariant dimension-8 operator of the form
\bea
{\cal L}_{\rm D8}= \frac{1}{M^4} \, \epsilon^{\mu\nu\rho\sigma} (\Phi^\dagger X_{\mu\nu} D_\lambda \Phi) \partial^\lambda Z'_{\rho\sigma}\,.
\eea

In this work we will consider the phenomenology of the effective operator Eq.~(\ref{CS1}). Sec.~\ref{sec:CSnonabelianappendix} contains a concrete example of generating the effective Chern-Simons term.

We remark on the invisible decays of $Z$ and $Z'$ bosons in our setup.
The $Z$ boson can decay invisibly into a pair of vector Dark Matter particles through the generalized CS terms in the presence of a gauge kinetic mixing between $Z'$ and $Z$ bosons. But, if $N_f$ heavy fermions $f$ running in triangle diagrams are relatively light  for a sizable CS term (but heavy enough not to affect our discussion on vector SIMPs in the later sections) as discussed in Sec.~\ref{sec:CSnonabelianappendix}, the $Z$-boson preferentially decays directly into a pair of heavy fermions at tree level. Then, the corresponding $Z$-boson invisible decay width is given by
\bea
\Gamma(Z_1\rightarrow f{\bar f})=\frac{N_f \alpha_{Z'} \varepsilon^2m_{Z}}{3 c^2_W}\left(1+\frac{2m^2_f}{m^2_{Z'}}\right)  \left(1-\frac{4m^2_f}{m^2_Z} \right)^{1/2}  \label{Z-inv}
\eea
with $\alpha_{Z'}\equiv g^2_{Z'}/(4\pi)$.
On the other hand, if the heavy fermions are heavier than $m_{Z'}/2$, the  $Z'$ boson decays into a pair of vector Dark Matter particles via the CS term, with the width
\bea
\Gamma(Z_2\rightarrow XX)= \frac{c^2_1 m^3_{Z'}}{8\pi m^2_X} \left(1-\frac{4m^2_X}{m^2_{Z'}} \right)^{5/2}. \label{Zp-inv}
\eea

\section{Vector SIMP Dark Matter}\label{sec:relic}

Having established the interactions of the framework, we now address the cross section for the Dark Matter relic abundance and self-scatterings.
We first determine the relic density of Dark Matter from $3\to2$ processes in Sec.~\ref{ssec:SIMP}, and discuss the role of additional forbidden annihilation channels in Sec.~\ref{ssec:forb}.

\subsection{SIMP channels}
\label{ssec:SIMP}

Here we compute the relic density assuming the $3\to2 $ annihilation processes are the dominant number-changing processes. In the presence of an isospin symmetry for the vector Dark Matter, all components of Dark Matter have the same mass, and can be treated as identical particles.
Assuming the Dark Matter remains in kinetic equilibrium with the SM until the time of freeze-out,
the Boltzmann equation for the vector Dark Matter is given by~\cite{Hochberg:2014dra}
\begin{align}
\frac{\diff n_{\rm DM}}{\diff t}+ 3 H n_{\rm DM}=& -\Big(\langle\sigma
v^2\rangle_{3\rightarrow 2}-\langle\sigma v^2\rangle^h_{3\rightarrow 2}\Big)
\Big(n^3_{\rm DM}- n^2_{\rm DM}n^{\rm eq}_{\rm DM}\Big) \nonumber \\
&-\langle\sigma v^2\rangle^h_{3\rightarrow 2}\Big(n^3_{\rm DM}- n_{\rm DM}(n^{\rm
eq}_{\rm DM})^2 \Big)\,.
\end{align}
Here, the thermally averaged $3\rightarrow 2$ annihilation cross-section (away from a
resonance) is given by
\begin{align}
\langle\sigma v^2\rangle_{3\rightarrow 2}&=&\frac{25\sqrt{5}g_X^6}{23887872\pi
m_X^5}\frac{1}{(m_{h_1}^2-4m_X^2)^2(m_{h_1}^2+m_X^2)^2} \bigg(14681m_{h_1}
^8-87520m_{h_1}^6m_X^2 \nonumber \\
&&+21004m_{h_1}^4m_X^4+327580m_{h_1}^2m_X^6 +290775m_X^8\bigg) +
\langle\sigma v^2\rangle^h_{3\rightarrow 2}
\label{3to2}
\end{align}
with
\begin{align}
\langle\sigma v^2\rangle^h_{3\rightarrow 2}= \frac{\sqrt{5}g_X^6m^{16}_{h_1}}{80621568\pi m_X^{10}} \frac{(1-m^2_{h_1}/(16 m^2_X) )^{1/2} }{(m_{h_1}
^2-4m_X^2)^{7/2}(m_{h_1}^2+2m_X^2)^2}\bigg(C_1+\frac{2C_2 m^4_{h_1}}
{(m^2_{h_1}-7m^2_X)^2} \bigg)
\end{align}
where $C_1$ and $C_2$ are dimensionless quantities given in Eqs.~(\ref{C1}) and
(\ref{C2}), respectively. We note that the first term in $\langle\sigma
v^2\rangle_{3\rightarrow 2}$ stems from $XXX\rightarrow XX$ channels and $
\langle\sigma v^2\rangle^h_{3\rightarrow 2}$ due to $XXX\rightarrow X h_1$ channels
contributes only for $m_{h_1}<2m_X$, becoming dominant near the resonance at $m_{h_1}
=2m_X$.
where $C_1$ and $C_2$ are dimensionless quantities given in Eqs.~(\ref{C1}) and (\ref{C2}), respectively.  We note that the first two lines stem from $XXX\rightarrow XX$ channels and the last line due to $XXX\rightarrow X h_1$ channels contributes only for $m_{h_1}<2m_X$, becoming dominant near the resonance at $m_{h_1}=2m_X$. On the other hand, $XXX\rightarrow h_1 h_1$ channels are $p$-wave suppressed so they are not included here.  
Additional terms that give an approximate resonance when $m_{h_1}= 3 m_X$ are present, but as they are $p$-wave suppressed they are always subdominant and hence can be neglected. Further details of the $3\rightarrow 2$ cross section and discussion of the Boltzmann equation can be found in Appendix~\ref{sec:VSIMP3to2}.

In the instantaneous freeze-out approximation, the relic abundance for $3\rightarrow 2$ annihilation is found to be
\begin{equation}
\Omega_{\rm DM}  \simeq \frac{m_X s_0/\rho_c}{s({m_X})^2 / H(m_X)} \frac{x_f^2}{\sqrt{\langle\sigma v^2\rangle_{3\rightarrow 2}}}\,,
\end{equation}
where ${s_0/\rho_c \simeq 6\cdot 10^8 /\rm GeV}$ is the ratio of the entropy density today to the critical density, $s(m_X)$ is the entropy density at $T= m_X$, and $H(m_X)$ is the Hubble rate at $T= m_X$. Here $x_f=m_X/T_f$ indicates the freezeout temperature, which is typically $x_f \in [15,20]$ for $3\to 2 $ freezeout.
For $m_{h_1} \gtrsim 3 m_X$, the Higgs contributions to the cross-section effectively decouples, and we have
\begin{equation}
\Omega_{\rm DM} \simeq 0.33 \left( \frac{x_f}{20}\right)^2 \left( \frac{10.75}{g_*}\right)^{3/4} \left(\frac{m_X/\alpha_X }{100~ \rm MeV} \right)^{3/2}\,.
\end{equation}

\begin{figure}[t!]
  \begin{center}
      \includegraphics[height=0.45\textwidth]{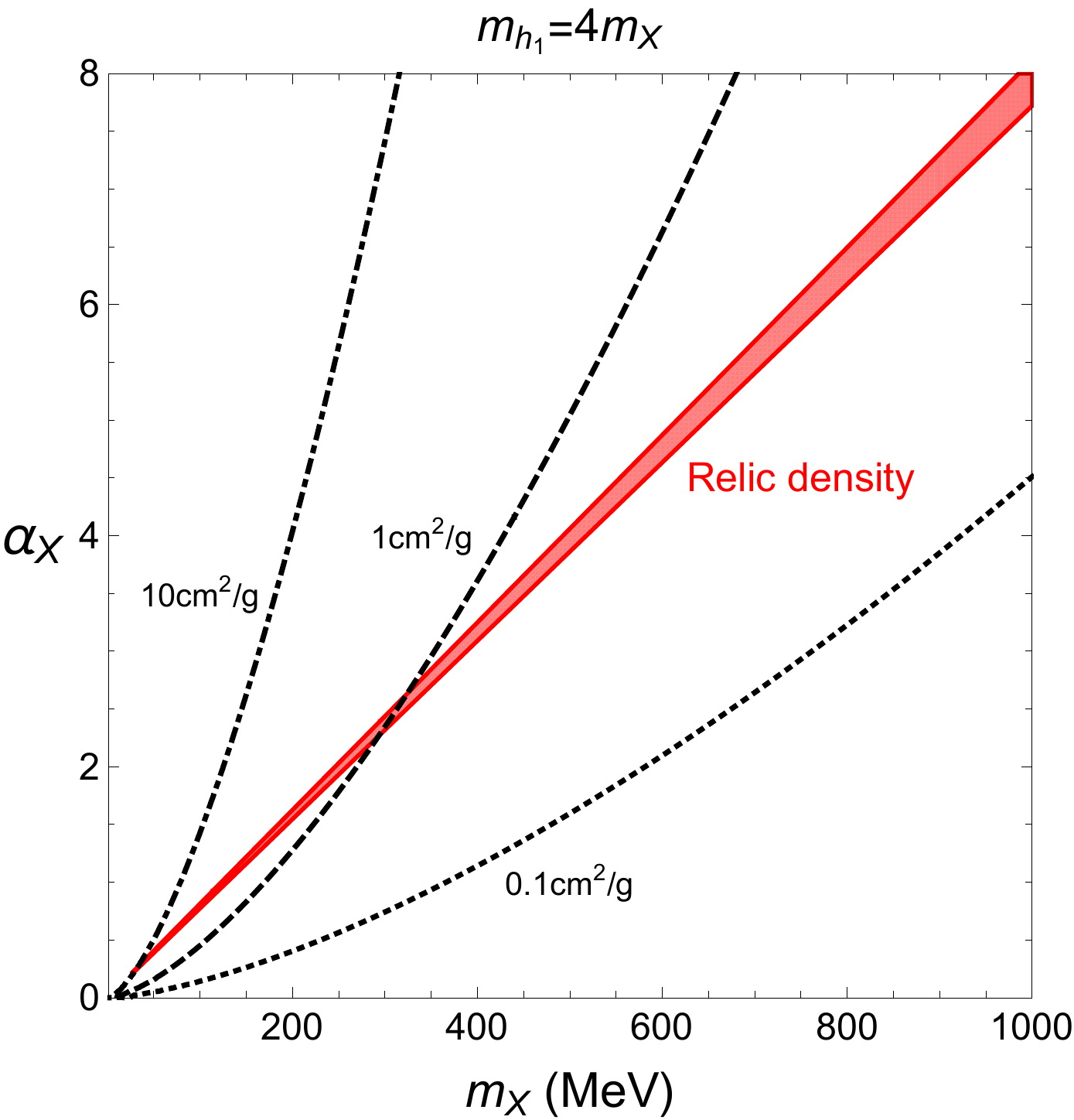}\\
             \includegraphics[height=0.45\textwidth]{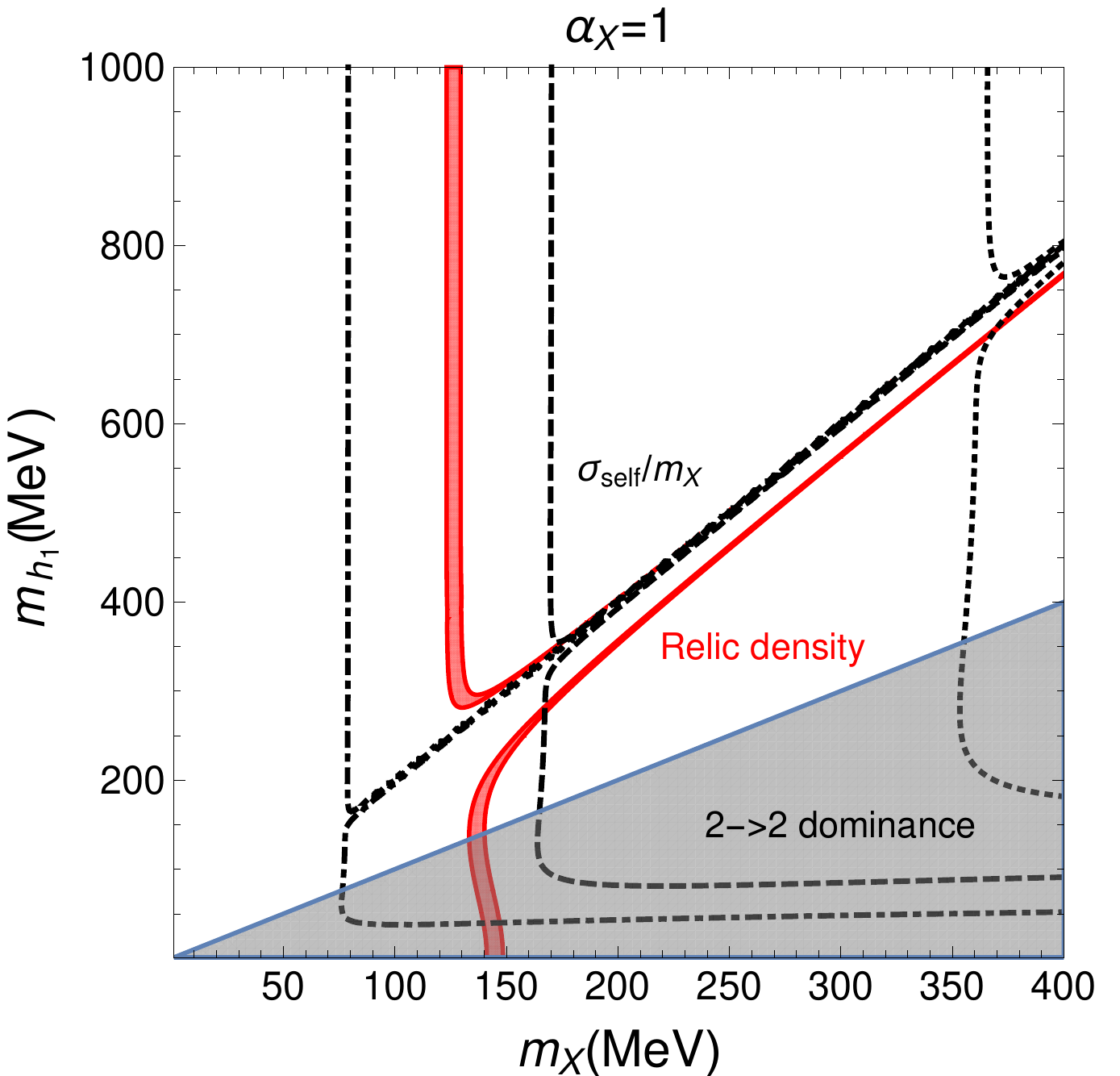}
      \includegraphics[height=0.45\textwidth]{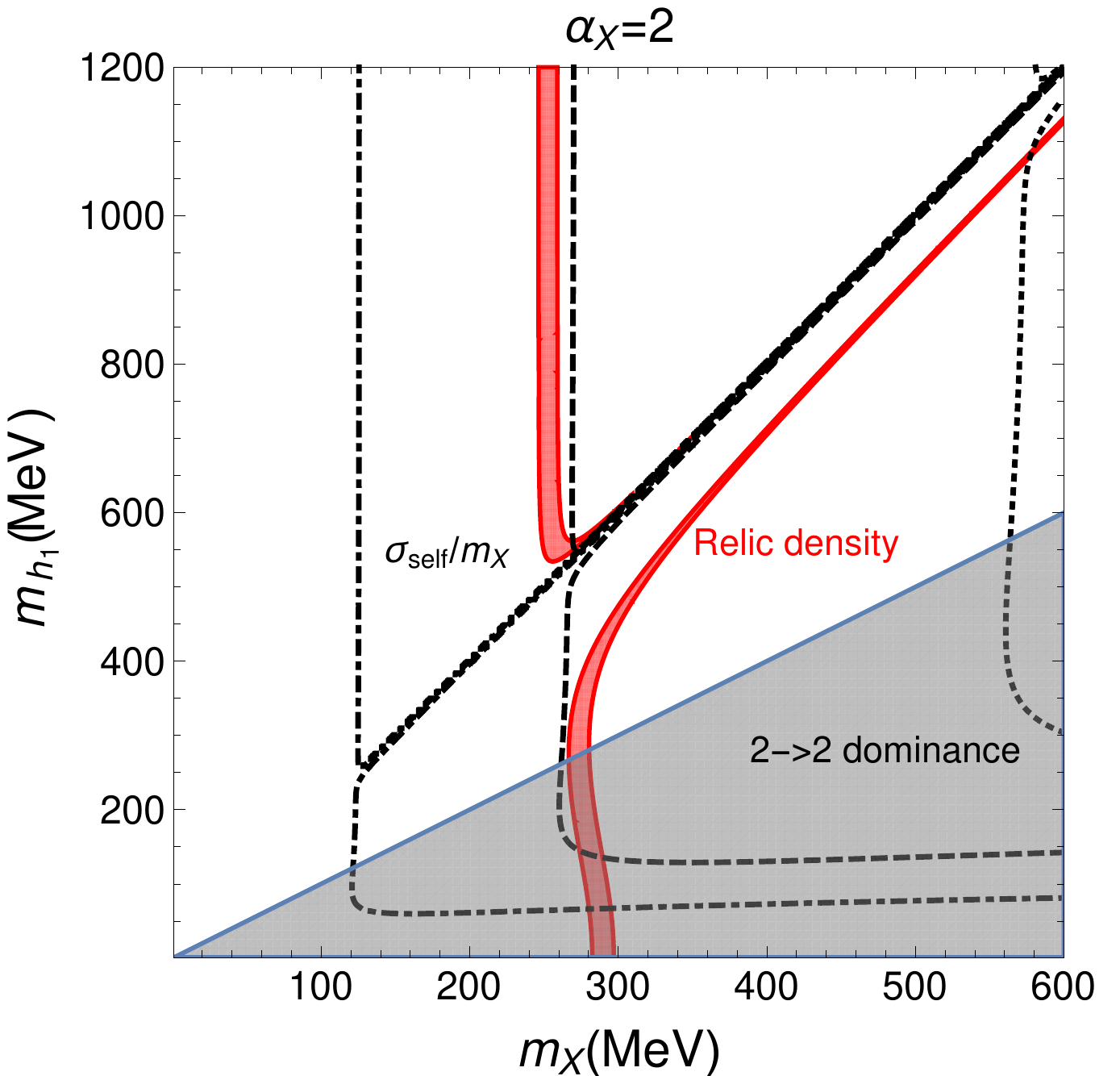}
   \end{center}
  \caption{The parameter space of vector SIMP Dark Matter in the $m_X$ vs. $\alpha_X\equiv g^2_X/(4\pi)$ (top) or $m_{h_1}$ (bottom), when considering $3\to2$ annihilation channels only. The Planck $3\sigma$ measurement of the relic density is show in red in all panels. Contours of the self-scattering cross section of $\sigma_{\rm self}/m_X=0.1,1,10\,{\rm cm^2/g}$ are shown in the dotted, dashed and dot-dashed curves, respectively.  We have chosen $m_{h_1}=4m_X$ on top and $\alpha_X=1, 2$ on bottom. The shaded gray regions in the lower panels are where other $2\rightarrow 2$ channels dominate over $3\to2$ processes. }
  \label{relic}
\end{figure}

In Fig.~\ref{relic} we depict the parameter space in which the measured Dark Matter relic density is obtained within $3\sigma$ (red region) for $\alpha_X\equiv g^2_X/(4\pi)$ ($m_{h_1}$) and $m_X$ in the upper (lower) panel. For illustration, the top panel shows the results for dark Higgs mass of $m_{h_1}=4m_X$ where no resonance enhancement is present, while in the bottom panel we fix $\alpha_X=1, 2$ and vary $m_X$ and $m_{h_1}$.

In addition to $3\to2$ annihilations, the vector SIMP Dark Matter undergoes self-scattering processes, which are constrained by the bullet cluster~\cite{Clowe:2003tk,Markevitch:2003at,Randall:2007ph} and by elliptical halo shapes~\cite{Rocha:2012jg,Peter:2012jh}. Away from a resonance, the self interacting cross-section is
\bea
\sigma_{\rm self}
&=&\frac{g^4_X}{1152\pi m_{h_1}^4m_X^2(m_{h_1}^2-4m_X^2)^2}
\Big(520m_{h_1}^8-4208m_{h_1}^6m_X^2+8801m_{h_1}^4m_X^4 \nonumber \\
&&-1200m_{h_1}^2m_X^6+320m_X^8\Big)\,.
\eea
A simple approximation can be derived in the limit $m_{h_1} \gg m_X$ : \begin{equation}
\dfrac{\sigma_{\text{self}}}{m_X}\simeq \frac{65 \pi  \alpha _X^2}{9 m_X^3} \simeq 5 \alpha_X^2 \Big( \dfrac{m_X}{100\text{ MeV}} \Big)^{-3}  \text{ cm}^2 / \text{g} \qquad \qquad [m_{h_1} \gg m_X]
\end{equation}
Contours of the self-scattering cross section obeying $\sigma_{\rm self}/m_X=0.1,1,10\,{\rm cm^2/g}$ are shown in Fig.~\ref{relic} in dotted, dashed and dot-dashed lines, respectively.

We learn that away from a resonance region, vector SIMP $3\to2$ Dark Matter consistent with self-scattering constraints points to Dark Matter masses of $m_X\gtrsim {\cal O}(100~{\rm MeV})$ and strong couplings of $\alpha_X\gtrsim 1$. Indeed, strong coupling is a frequent common feature in SIMP Dark Matter models~\cite{Hochberg:2014dra,Lee:2015gsa,Choi:2015bya,Hochberg:2015vrg,Choi:2016tkj,Hambye:2009fg,Hambye:2008bq,Bernal:2015ova,Cline:2017tka}, though exceptions can arise ({\it e.g.} on resonance~\cite{Choi:2016hid,Choi:2017mkk}).  
Close to the resonance region, the relic density is sensitive to the dark Higgs mass, and the viable parameter space is broadened further to include larger DM masses at fixed dark gauge coupling, or smaller dark gauge couplings for fixed DM masses.

We comment that the strong gauge coupling leads to a question on the potential breakdown of perturbativity in relic density calculation. In our case, however, the $SU(2)_X$ gauge symmetry is completely broken by the VEV of the dark Higgs, and there are no light particles below the confinement or symmetry breaking scale ({\it i.e.}\/, vector SIMP mass). Therefore, given that there is no phase transition separating the Higgs phase and confining phase, namely, the complementarity between the Higgs and confining phases~ \cite{Fradkin:1978dv,Banks:1979fi,tHooft:1979yoe,Susskind:1979up,Raby:1979my,Georgi:2016qbt}, the Higgsed theory can be pushed into regions where perturbativity is questionable. Closer inspection of the issue of complementarity may be worthwhile, though is beyond the scope of this work. 

As the dark Higgs mass approaches the DM mass, when $m_X< m_{h_1}\lesssim 1.5 m_X$, forbidden $2\to2$ annihilation channels contribute significantly to the relic density and must be included as well; we study this in the next subsection. (The regions in which $2\to2$ processes dominate the relic density are shown in shaded gray in Fig.~\ref{relic}.) As we will see, the self-scattering rate is reduced in this case, allowing smaller Dark Matter masses consistent with observational constraints.

\subsection{Forbidden channels}
\label{ssec:forb}

When the dark Higgs is slightly heavier than the Dark Matter, forbidden $2 \rightarrow 2$ channels such as $X_iX_i\rightarrow h_1 h_1$ and $X_iX_j\rightarrow X_k h_1$---although kinematically inaccessible at zero temperature---can be important in determining the relic density at the time of freeze-out~\cite{Griest1991c,DAgnolo:2015ujb}.  For $m_X\lesssim m_{h_1}\lesssim2(1.5)m_X$, new $3\rightarrow 2$ channels such as $XXX\rightarrow X h_1 (h_1 h_1)$ open up as well so they have been already included in Fig.~\ref{relic}.  Here we discuss the effects of the forbidden channels on the relic abundance and identify the parameter space of vector SIMP Dark Matter that is consistent with the observed relic density when including these effects.
(This will be particularly relevant when kinetic equilibrium between the SIMP and SM sectors is obtained via the Higgs portal, as will become evident in Sec.~\ref{ssec:higgsportal}.)

Assuming that the forbidden channels are dominant,  the approximate Boltzmann equation is given by
\bea
\frac{\diff n_{\rm DM}}{\diff t}+3Hn_{\rm DM}&\approx &-\frac{2}{3}\langle\sigma v\rangle_{ii\rightarrow h_1 h_1} n^2_{\rm DM}+6\langle\sigma v\rangle_{h_1 h_1\rightarrow ii} (n^{\rm eq}_{h_1})^2  \nonumber \\
&&- \frac{1}{3}\langle\sigma v\rangle_{ij\rightarrow k h_1} n^2_{\rm DM} +\langle\sigma v\rangle_{k h_1\rightarrow ij}n^{\rm eq}_{h_1} n_{\rm DM}\,,
\eea
where we have assumed that $h_1$ maintains chemical and thermal equilibrium with the SM bath throughout freezeout. Detailed balance conditions at high temperature determine the annihilation cross sections for the forbidden channels in terms of the unforbidden channels,
\bea
\langle\sigma v\rangle_{ii\rightarrow h_1 h_1} &=& \frac{9(n^{\rm eq}_{h_1})^2}{(n^{\rm eq}_{\rm DM})^2} \,\langle\sigma v\rangle_{h_1 h_1\rightarrow ii}
= (1+\Delta_{h_1})^3 e^{-2\Delta_{h_1}x}  \,\langle\sigma v\rangle_{h_1 h_1\rightarrow ii}\,, \label{dbal1} \\
\langle\sigma v\rangle_{ij\rightarrow k h_1} &=& \frac{ 3n^{\rm eq}_{h_1}}{n_{\rm DM}^{\rm eq}}\, \langle\sigma v\rangle_{k h_1\rightarrow ij} = (1+\Delta_{h_1})^{3/2} e^{-\Delta_{h_1}x} \,  \langle\sigma v\rangle_{k h_1\rightarrow ij}\,,  \label{dbal2}
\eea
with $\Delta_{h_1}\equiv (m_{h_1}-m_\chi)/m_\chi$.   The cross section formulas for the allowed $2\rightarrow 2$ channels in the RHS above are given in Sec.~\ref{sec:VSIMPforbidden}.

Denoting the allowed $2\to2$ cross sections in the RHS above by $ \langle\sigma v\rangle_{k h_1\rightarrow ij} =a$ and $ \langle\sigma v\rangle_{h_1 h_1\rightarrow ii}=b$, the DM abundance is found to be~\cite{DAgnolo:2015ujb}
\be
Y_{\rm DM}(\infty)\approx \frac{x_f}{\lambda}\, e^{\Delta_{h_1}x_f} \, f(\Delta_{h_1},x_f)
\ee
with 
\begin{align}
f(\Delta_{h_1},x_f)= \bigg[& \frac{1}{3} a (1+\Delta_{h_1})^{3/2}\Big(1-(\Delta_{h_1} x_f) \, e^{\Delta_{h_1}x_f} \int^\infty_{\Delta_{h_1} x_f} \diff t\, t^{-1} e^{-t} \Big) \nonumber \\
&+ \frac{2}{3} b(1+\Delta_{h_1})^3 e^{-\Delta_{h_1}x_f}\Big(1-2(\Delta_{h_1}x_f)\,e^{2\Delta_{h_1} x_F} \int^\infty_{2\Delta_{h_1} x_f} \diff t\, t^{-1} e^{-t} \Big)  \bigg]^{-1}\,,
\end{align}
resulting in the relic density
\be
\Omega_{\rm DM}h^2 = 5.20\times 10^{-10}\,{\rm GeV}^{-2} \Big(\frac{g_*}{10.75} \Big)^{-1/2} \Big(\frac{x_f}{20} \Big)\, e^{\Delta_{h_1} x_f} f(\Delta_{h_1},x_f)\,.
\ee
In general, however, one must account simultaneously for both the $3\rightarrow 2$ processes and the $2\rightarrow 2$ forbidden channels in determining the Dark Matter relic abundance.

\begin{figure}
  \begin{center}
  \includegraphics[height=0.45\textwidth]{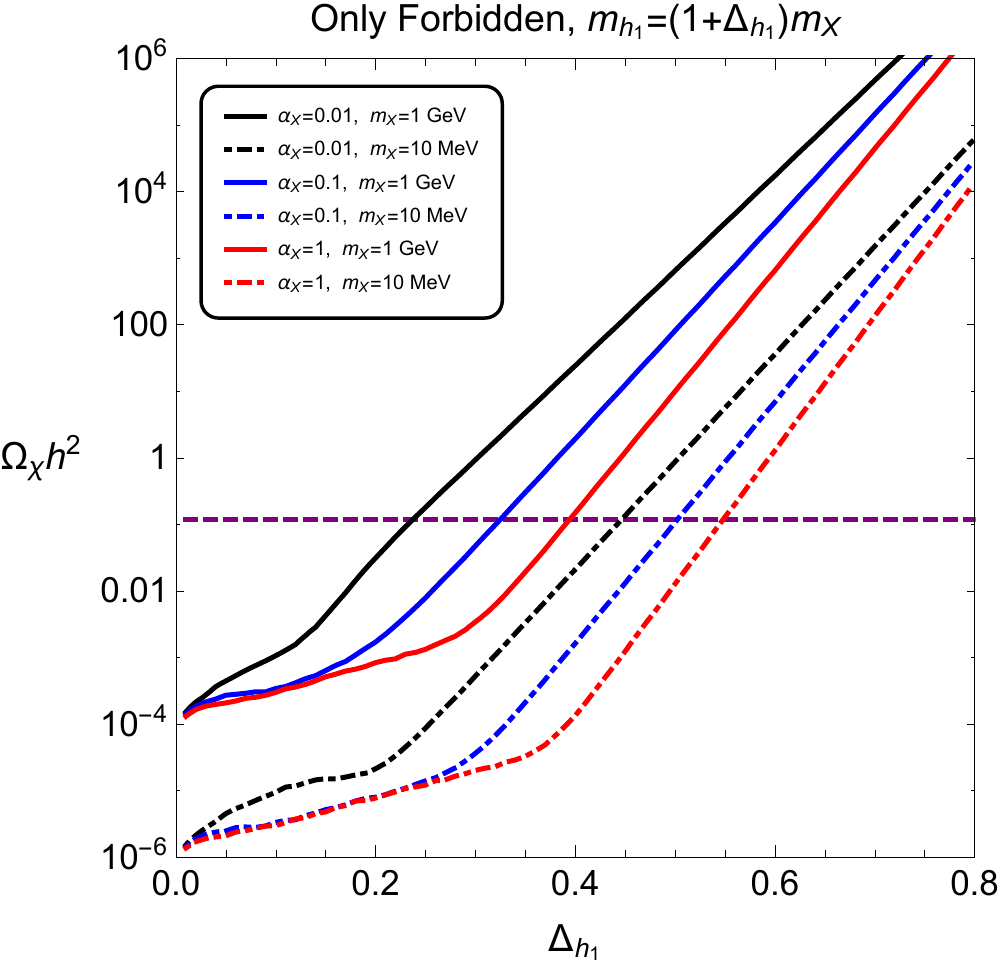}
      \includegraphics[height=0.45\textwidth]{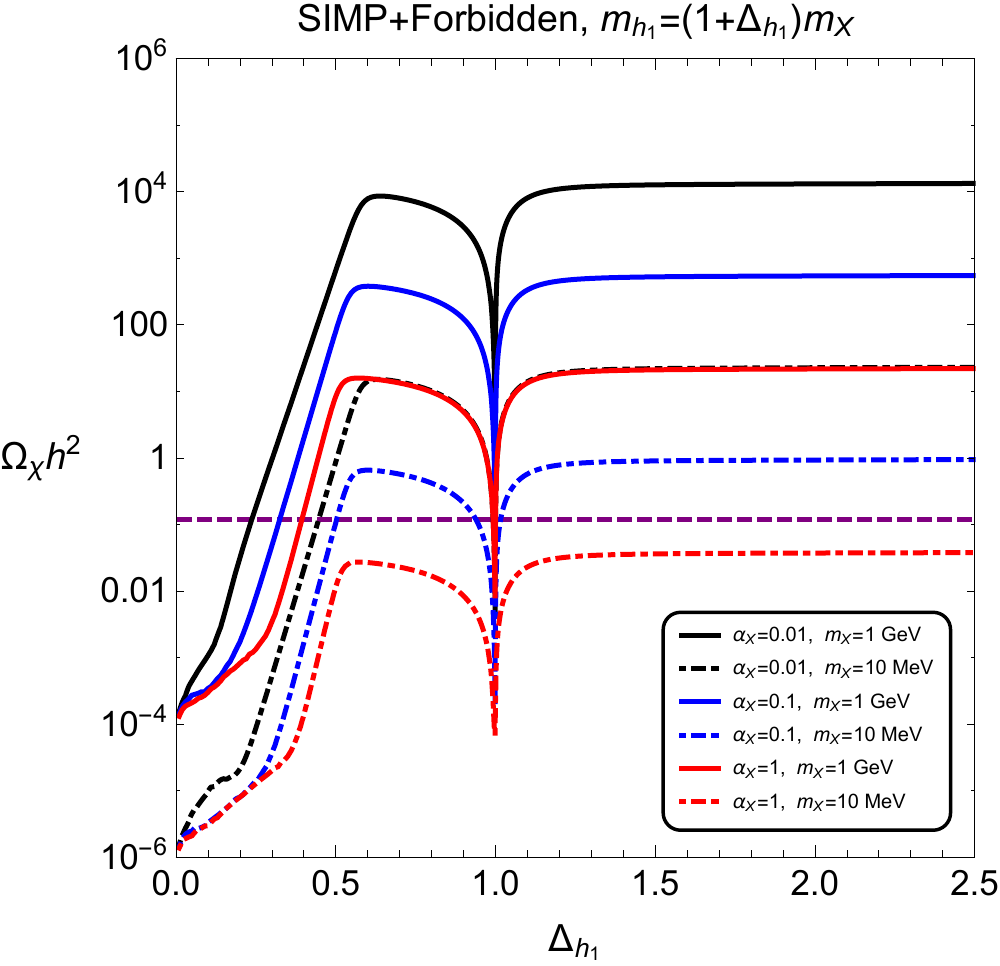}
   \end{center}
  \caption{Dark Matter relic density as a function of $\Delta_{h_1}=(m_{h_1}-m_X)/m_X$, for forbidden channels only ({\bf left}) and both forbidden and SIMP channels ({\bf right}). The measured relic density is shown by the purple curve. We show the results for various illustrative values of coupling and mass: $\alpha_X=0.01,0.1,1$ and $m_X=0.1\,{\rm MeV}, 1\,{\rm GeV}$.}
  \label{fdm2}
\end{figure}

In Fig.~\ref{fdm2}, we show the Dark Matter relic density as a function of $\Delta_{h_1}=(m_{h_1}-m_X)/m_X$, first when including only forbidden channels ({\bf left panel}) and then when taking both forbidden and SIMP channels into account ({\bf right panel}). We have varied $\alpha_X$ and $m_X$ between $0.01-1$ and $10\;{\rm MeV}-1\,{\rm GeV}$, respectively.
We learn that forbidden channels play an important role for $\Delta_{h_1}\lesssim 0.5$, where the observed relic density can be achieved over a broad range of couplings $\alpha_X$ and masses $m_X$. As the mass difference increases, $3\to2$ SIMP annihilations begin dominating the relic abundance as a saturated value for mass differences $\Delta_{h_1}\gtrsim 0.5$. We note that the importance of the forbidden semi-annihilation channels for $\Delta_{h_1}\lesssim 0.5$, in contrast to the naive expectation from the Boltzmann suppression factors of $\Delta_{h_1}\lesssim 1$, is due to a large numerical factor in the SIMP $3\to2$ annihilation cross section.

\section{Kinetic equilibrium}\label{sec:portal}

In order for the SIMP mechanism to be viable, we require that the SIMP sector efficiently dumps entropy into the SM bath. In the proposed framework, this can be achieved either by a Higgs or $Z'$ portal between the vector SIMPs and SM particles. After a general discussion of the relevant Boltzmann equation in terms of the dark sector temperature and the requirement of equilibration in Sec.~\ref{ssec:boltz}, we study the Higgs portal in Sec.~\ref{ssec:higgsportal} and the $Z'$ portal in Sec.~\ref{ssec:Zprimeportal}.

\subsection{Equilibration conditions}\label{ssec:boltz}

Following Ref.~\cite{Kuflik:2015isi}, we find the decoupling temperature by comparing the rate of change in kinetic energy injected by the $3\to 2$ annihilations compared to the kinetic energy lost due to elastic scattering.  When the $3\to 2$ occurs, the mass of one dark particle is converted to the kinetic energy of the 2 outgoing particles. These  particles quickly scatter off the Dark Matter particles, and distribute the energy to the dark bath. Thus, the $3\to 2$  annihilations maintain chemical equilibrium in the DM gas, while releasing kinetic energy per particle
\beq
\dot{K}_{3\to 2} = m_{\rm DM}\, \frac{ \dot{n}_{\rm DM}}{n_{\rm DM}}\simeq  -m_{\rm DM}^2 H T^{-1}~.\label{hubbleloss}
\eeq

Elastic scattering processes transfer this excess kinetic energy to the SM gas at a rate
\begin{align}
\dot{K}_{\rm el} = \frac{1}{2 E_{p}} \sum_i \frac{g_i \diff^3 k_i}{(2\pi)^3 2 E_i} \frac{ \diff^3 k^\prime_i}{(2\pi)^3 2 k^\prime_i}  \frac{ \diff ^3 p^\prime}{(2\pi)^3 2 p^\prime} \delta^4(p+k_i - p^\prime - k^\prime_i) \overline{|\mathcal{M}|^2} \left(E_{p} -E_{p^\prime} \right).
\end{align}
Here, the sum is taken over the species $i$ in the relativistic plasma with initial(final) momentum $k(k^\prime)$, $p(p^\prime)$ is the Dark Matter initial(final) momentum. 
The decoupling occurs when the DM-to-SM energy transfer can no longer keep up with the kinetic energy production. The quantity $\dot{K}_{\rm el}$ can be related to the momentum relaxation rate $\gamma(T)$ defined in Sec.~\ref{sec:SIMPappendix} as $\dot{K}_{\rm el} \simeq T \gamma(T)$. Equating Eq.~(\ref{hubbleloss}) with Eq.~(\ref{elasticloss}) leads to the following conditions satisfied at the kinetic decoupling temperature $T_{\rm KD}$:
\beq
\gamma(T_{\rm KD}) \simeq H(T_{\rm KD}) \frac{m_{\rm DM}^2}{T_{\rm KD}^2} \label{eq:eqcond}~,
\eeq
where $H=0.33 g^{1/2}_* T^2/M_{\rm Pl}$ with $g_*=10.75$ the effective relativistic number of species for $1\,{\rm MeV}\lesssim T\lesssim 100\,{\rm MeV}$ and $M_{\rm Pl}=2\times 10^{18}$~GeV the Planck mass. In what follows we use Eq.~\eqref{eq:eqcond}, evaluated at $T_{\rm KD} = m_{\rm DM}/20$, to place a lower bound on the interactions between the vector SIMPs and the SM particles, needed to achieve the correct DM abundance. The ELDER DM curve~\cite{Kuflik:2015isi,Kuflik:2017iqs}, corresponds to $T_{\rm KD} \simeq m_{\rm DM}/15$, where the relic abundance is determined by the elastic scattering rate.

The Dark Matter can also thermalize with the SM, if the Dark Matter maintains equilibrium with the dark Higgs, while the dark Higgs maintains equilibrium with the SM bath via decay and inverse decays into SM fermions. The dark Higgs should be heavier than the dark $X$-bosons, or else the Dark Matter will efficiently annihilate into dark Higgs, effectively becoming a WIMP-like scenario. However, if the dark Higgs is much heavier than the Dark Matter, then the dark Higgs abundance will have been sufficiently depleted and it will not be able to maintain equilibrium between the two sectors.
This pushes the spectrum to a forbidden regime, $m_X< m_{h_1} \lesssim 1.5 m_X$, where the Dark Matter can annihilate into dark Higgses, but with a large Boltzmann suppression. At the time right before freezeout, both the semi-annihilation $XX \to X h_1 $ and self-annihilation $XXX \to XX$ processes will be active for large gauge coupling. The dark sector will be in thermal equilibrium with vanishing chemical potential. Thus in order for freezeout to occur one just needs to check that the dark Higgs can deplete the density in the dark sector fast enough up until freezeout,
\beq
n^{\rm eq}_{h_1}(T_{\rm FO}) \Gamma_{h_1 \to \rm SM} >  H(T_{\rm FO}) \left[n_{X }^{\rm eq}(T_{\rm FO})+ n_{h_1}^{\rm eq}(T_{\rm FO}) \right]\,. \label{eq:eqcond2}
\eeq
We use the above condition on the dark Higgs decay rate in the case that vector SIMPs are in kinetic equilibrium through the scattering with the dark Higgs.

\subsection{Higgs portal}\label{ssec:higgsportal}

The coupling $\lambda_{\Phi H}$ present in Eq.~\eqref{eq:Vhiggs} leads to mixing between the SM and dark Higgs, which enables a Higgs portal between the dark and visible sectors.

In the presence of Higgs-portal induced mixing between the SM and dark Higgs, the SM Higgs can decay invisibly into a pair of Dark Matter particles, with decay rate given by Eq.~\eqref{eq:h2decay}:
\bea
\Gamma(h_2\rightarrow X X)= \frac{3\sin^2\theta m^3_{h_2}}{32 \pi v^2_X}\,\bigg(1-\frac{4m^2_X}{m^2_{h_2}}+\frac{12 m^4_X}{m^4_{h_2}} \bigg) \sqrt{1-\frac{4m^2_X}{m^2_{h_2}}}.
\eea
The combined VBF, $ZH$ and gluon fusion production of Higgs bosons at CMS leads to ${\rm BR}(h_2\rightarrow XX)<0.24$ at $95\%$ CL~\cite{Khachatryan:2016whc}, while the ATLAS bounds from the VBF~\cite{Aad:2015txa} and $ZH$~\cite{Aad:2014iia} modes give ${\rm BR}(h_2\rightarrow XX)<0.29$ and ${\rm BR}(h_2\rightarrow XX)<0.75$, respectively. These decays provide a strong constraint on the mixing: $\sin\theta \lesssim 10^{-5}$ for $\alpha_X\sim \mathcal{O}(1)$.

The mixing also induces direct couplings of the darks Higgs to the SM electron and muons, which in turn induces tree-level scattering of the SM of the leptons. However, the smallness of the electron Yukawa coupling and the Boltzmann-suppression of the muons at the time of freezeout combined with constraints on the Higgs invisible decay result in the elastic scattering being inefficient for thermalization.

\begin{figure}
  \begin{center}
      \includegraphics[height=0.46\textwidth]{figures/VSIMP/dHiggs4}
       \includegraphics[height=0.44\textwidth]{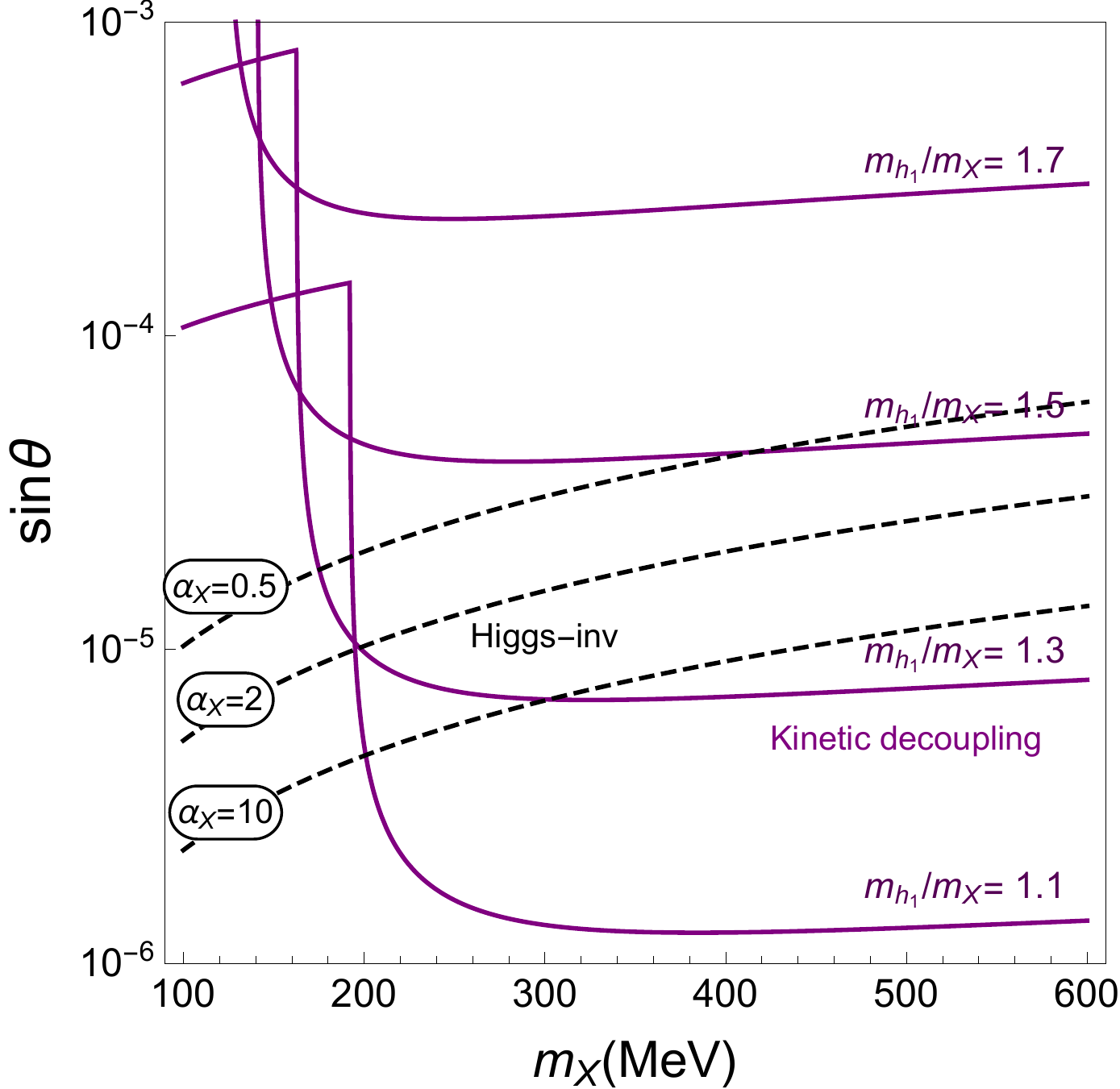}
   \end{center}
  \caption{Vector SIMPs through the Higgs portal, with DM-dark Higgs scattering and dark Higgs-SM decays. {\bf Left:} Parameter space of $m_{h_1}$ vs. $\sin\theta$ for DM-dark Higgs scattering. The shaded purple regions indicate where kinetic equilibrium between the DM and dark Higgs fails.  {\bf Right:} Parameter space of $m_X$ vs. $\sin\theta$. The purple lines are the lower bounds on $\sin\theta$ from kinetic equilibrium for fixed ratios $m_{h_1}/m_X$.  {\bf In both panels:} the dashed black curves are the upper bounds on $\sin\theta$ from Higgs invisible decays.
  }
  \label{darkH}
\end{figure}

Alternatively, if the dark Higgs is fairly light, scattering between the vector Dark Matter and the dark Higgs can equilibrate the dark sector, with decays and inverse decays of the dark Higgs into SM particles completing the equilibration requirement between the SIMP and SM sectors.
The momentum relaxation rate from the elastic scattering of Dark Matter off of dark Higgs, $X_i h_1\rightarrow X_i h_1$, is given by
\bea\label{eq:eqHDM}
\gamma(T)_{h_1}= \frac{g_{h_1} g^4_X m^2_{h_1}}{12\pi^3 m_X (m_X+m_{h_1})^2}
\bigg(\frac{m^2_{h_1}-6m^2_X}{m^2_{h_1}- 4m^2_X} \bigg)^2 T^2\,e^{-m_{h_1}/T}
\eea
where $g_{h_1}=1$.  We note that the above result is valid for $m_{h_1}(m_{h_1}-2m_X)\gtrsim p^2_{\rm DM}\sim m^2_{\rm DM} v^2_{\rm DM}$.
Plugging this into the kinetic equilibrium condition Eq.~\eqref{eq:eqcond}, we find that equilibrium between the dark Higgs and the DM is effective in most of parameter space satisfying the Dark Matter relic abundance.

Simultaneously, kinetic equilibrium between the dark Higgs and the SM is maintained by decays and the inverse decays of the Higgs into a pair of SM fermions,
\bea\label{eq:eqHSM}
\Gamma(h_1\to f \bar f)=\frac{m^2_f m_{h_1} \sin^2\theta}{8\pi v^2} \left(1-\frac{4m^2_f}{m^2_{h_1}}\right)^{3/2}\,.
\eea
In Fig.~\ref{darkH}, we illustrate this second requirement of equilibration between the dark Higgs and the SM, as a function of $\sin\theta$ and $m_{h_1}$ ({\bf left}) or $m_X$ ({\bf right}) for fixed $m_X$ and $\alpha_X$ ({\bf left}) or fixed ratio $m_{h_1}/m_X$ ({\bf right}). The upper bound on the mixing angle from invisible Higgs decays is indicated by the dashed black curves in both panels. Here the active thermalization process comes primarily from decays into muons when kinematically accessible, and from electrons for smaller masses.

We learn that the Higgs portal is a viable mediator between vector SIMPs and the SM when the dark Higgs is close in mass to the DM. In this regime, $2\rightarrow 2$ forbidden (semi)-annihilations channels of DM and the dark Higgs, $X_i X_j\rightarrow X_k h_1 (h_1 h_1)$, can be active and are then important contributors in determining the Dark Matter relic density, as discussed in Sec.~\ref{ssec:forb}. In this case, the semi-annihilations are also active thermalization processes within the dark sector.

We note that current limits on Higgs mixing from rare kaon- and $B$-meson decays are weaker than the bound we impose from the Higgs invisible decay. However future beam dump or fixed target experiments, such as SHiP at CERN SPS, have the potential to probe the Higgs mixing angle further down~\cite{Alekhin:2015byh}. The allowed parameter space for the Higgs portal to vector SIMPs could then be further probed as the invisible Higgs decay constraint improves.

Before ending this subsection, we remark that a Higgs portal coupling could allow in principle for the elastic scattering of relic vector SIMP Dark Matter with electrons in direct-detection experiments~\cite{Essig:2011nj,Graham:2012su,Lee:2015qva,Essig:2015cda,Hochberg:2015pha,Hochberg:2015fth,Hochberg:2016ntt,Essig:2016crl,
Derenzo:2016fse,Tiffenberg:2017aac,Essig:2017kqs}.  For $m_e, m_X,m_{Z'}\gg p_{\rm DM}\simeq m_X v_{\rm DM}$, the DM-electron direct detection scattering cross section via the Higgs portal is given by
\bea
\sigma_{\rm DD}&=&\frac{\alpha_X\sin^2\theta\cos^2\theta m^4_e m^2_X}{v^2 (m_e+m_X)^2}\left(\frac{1}{m^2_{h_1}}-\frac{1}{m^2_{h_2}}\right)^2\nonumber  \\
&\approx & 4\times 10^{-50}\,{\rm cm}^2 \left(\frac{\alpha_X}{2}\right)\left(\frac{\sin\theta}{10^{-4}}\right)^2\left(\frac{1.2}{m_{h_1}/m_X}\right)^4 \left(\frac{100\,{\rm MeV}}{m_X}\right)^4\,.
\eea
The small electron Yukawa coupling suppresses the cross section substantially, yielding a currently unconstrained spin-independent direct detection cross section.

\subsection{$Z'$ portal}\label{ssec:Zprimeportal}

\begin{figure}
  \begin{center}
   \includegraphics[height=0.42\textwidth]{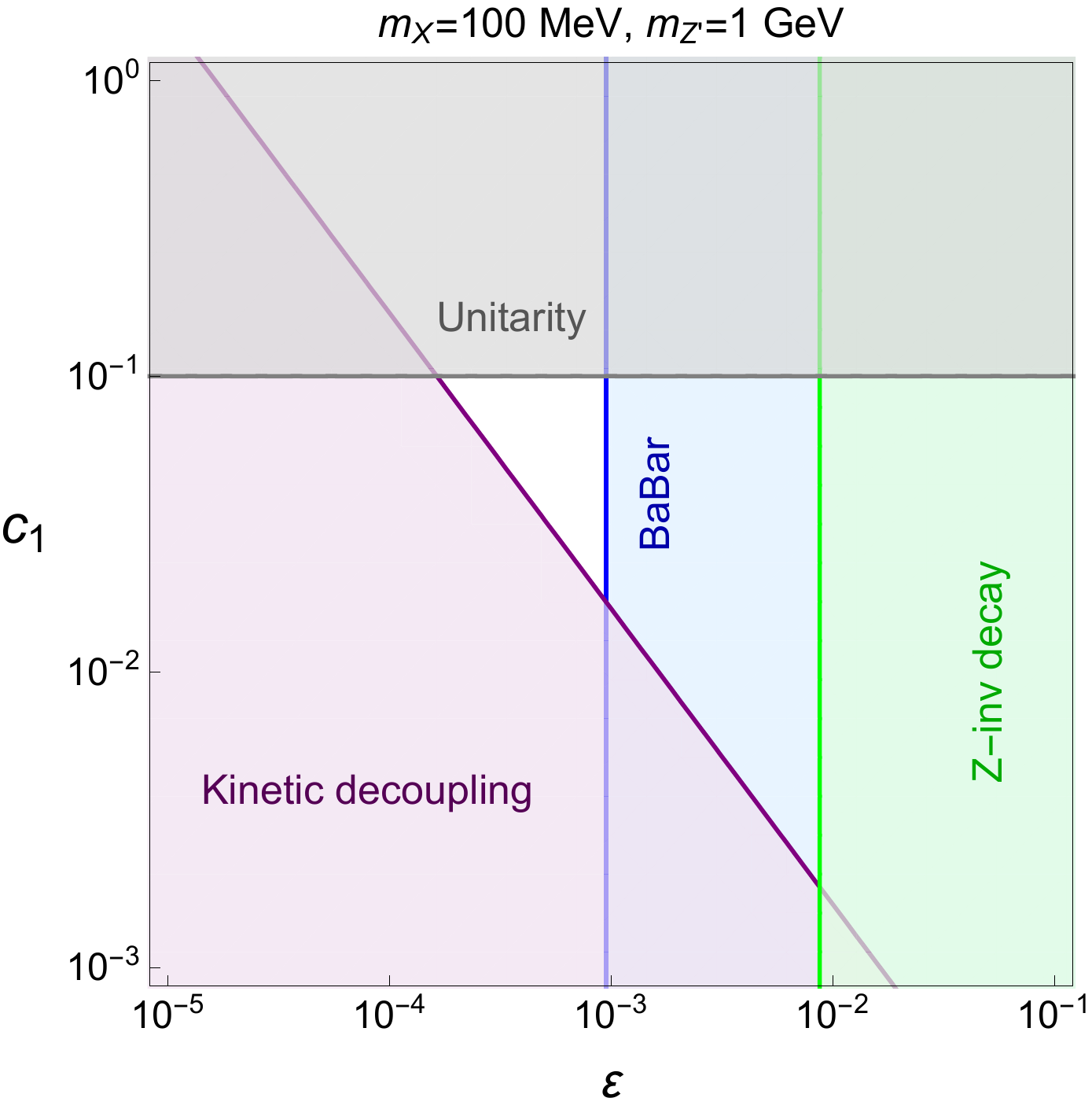}
       \includegraphics[height=0.42\textwidth]{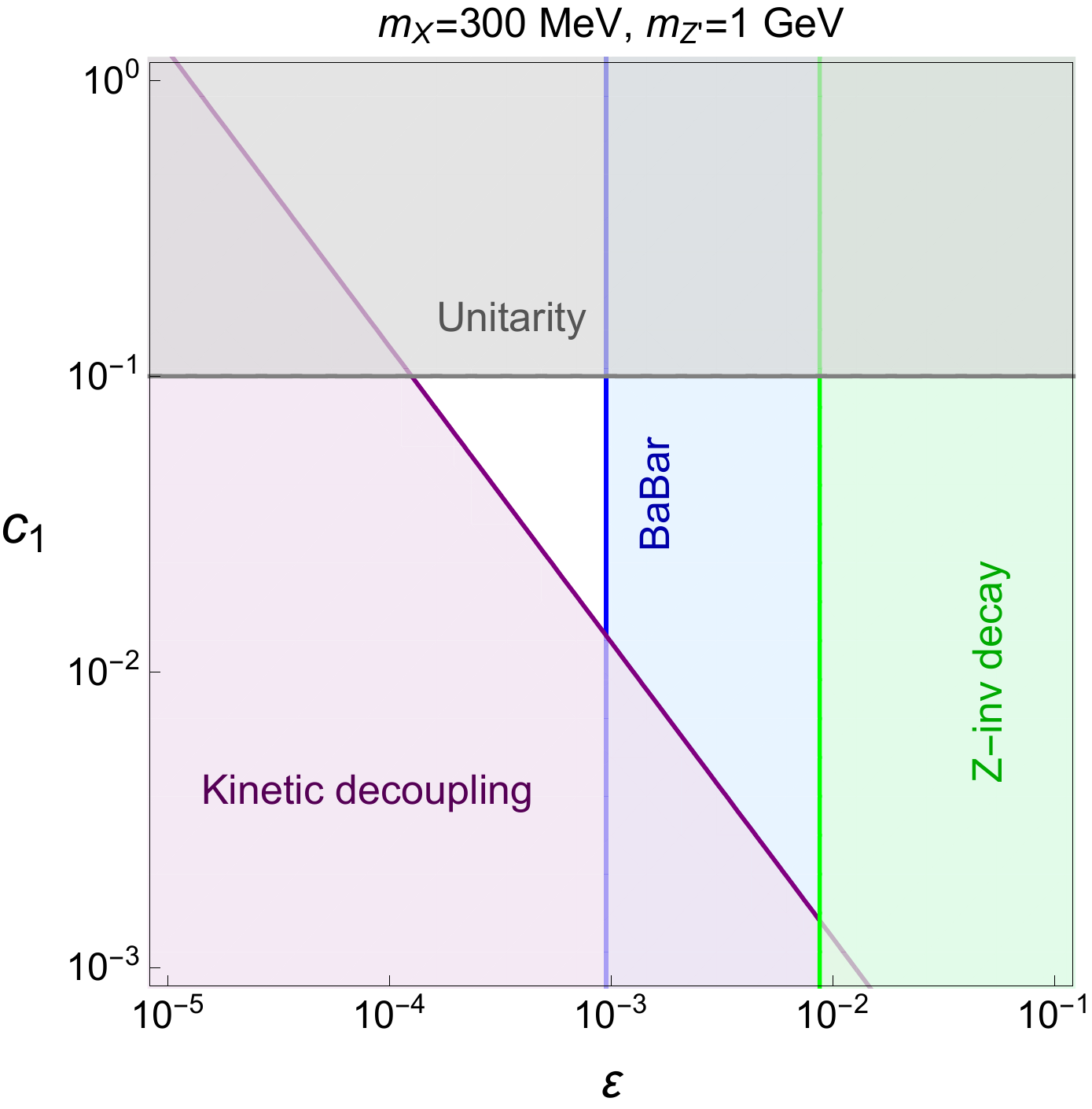}
   \end{center}
   \caption{Allowed parameter space for vector SIMPs with a $Z'$ portal in the $\varepsilon$ and $ c_1$ plane, for fixed values of DM and $Z'$ masses. We show the bounds from unitarity (brown), kinetic equilibrium (purple), the invisible width of the $Z$ boson (green)~\cite{ALEPH:2005ab} and BaBar monophoton+MET (blue)~\cite{Lees:2017lec}. Here, we took $N_f \alpha_{Z'}=1$ in Eq.~(\ref{Z-inv}) for $Z$-boson invisible decay bounds.
   }
  \label{c1-ep}
\end{figure}

 \begin{figure}
  \begin{center}
   \includegraphics[height=0.48\textwidth]{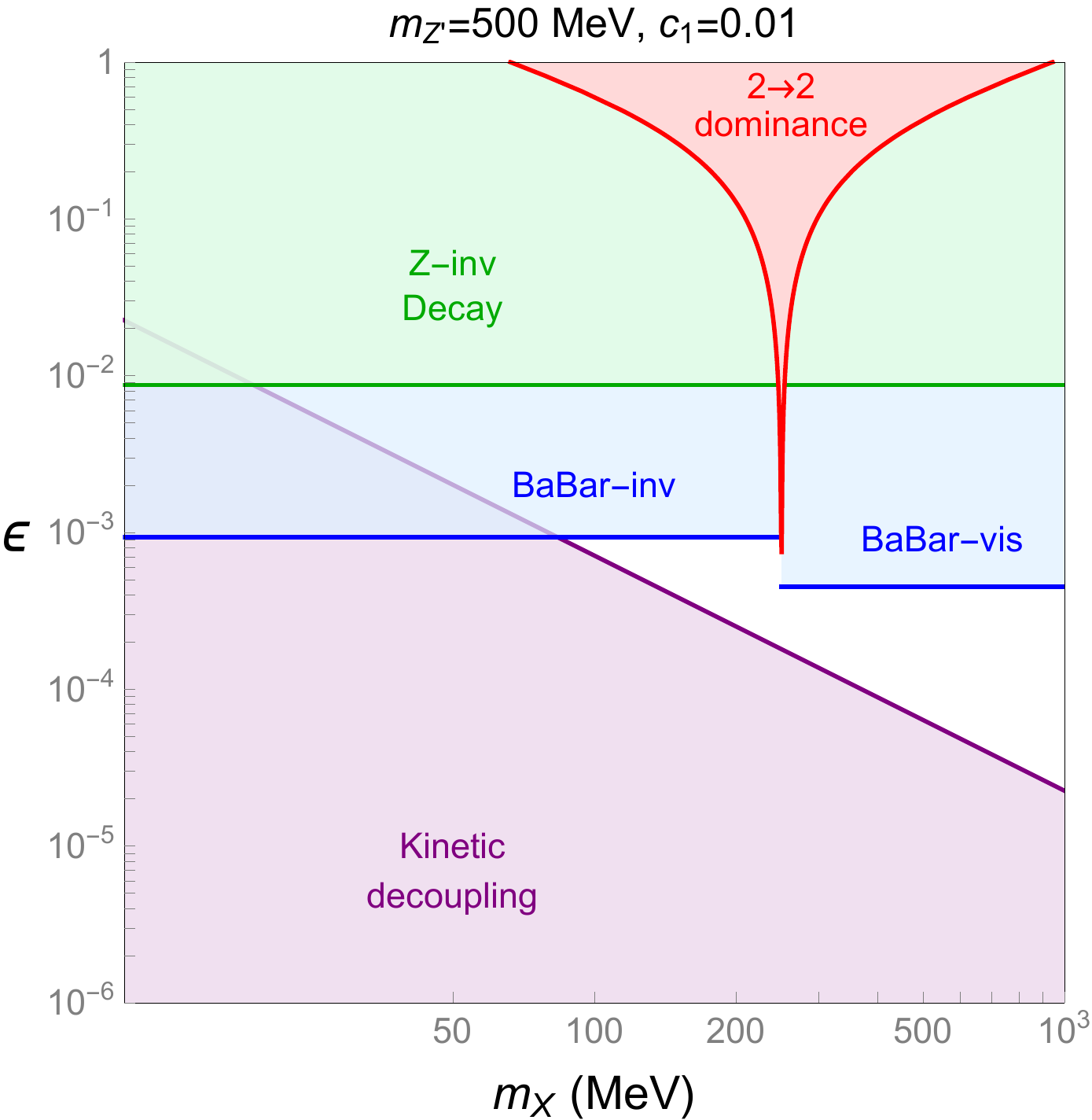}
      \includegraphics[height=0.48\textwidth]{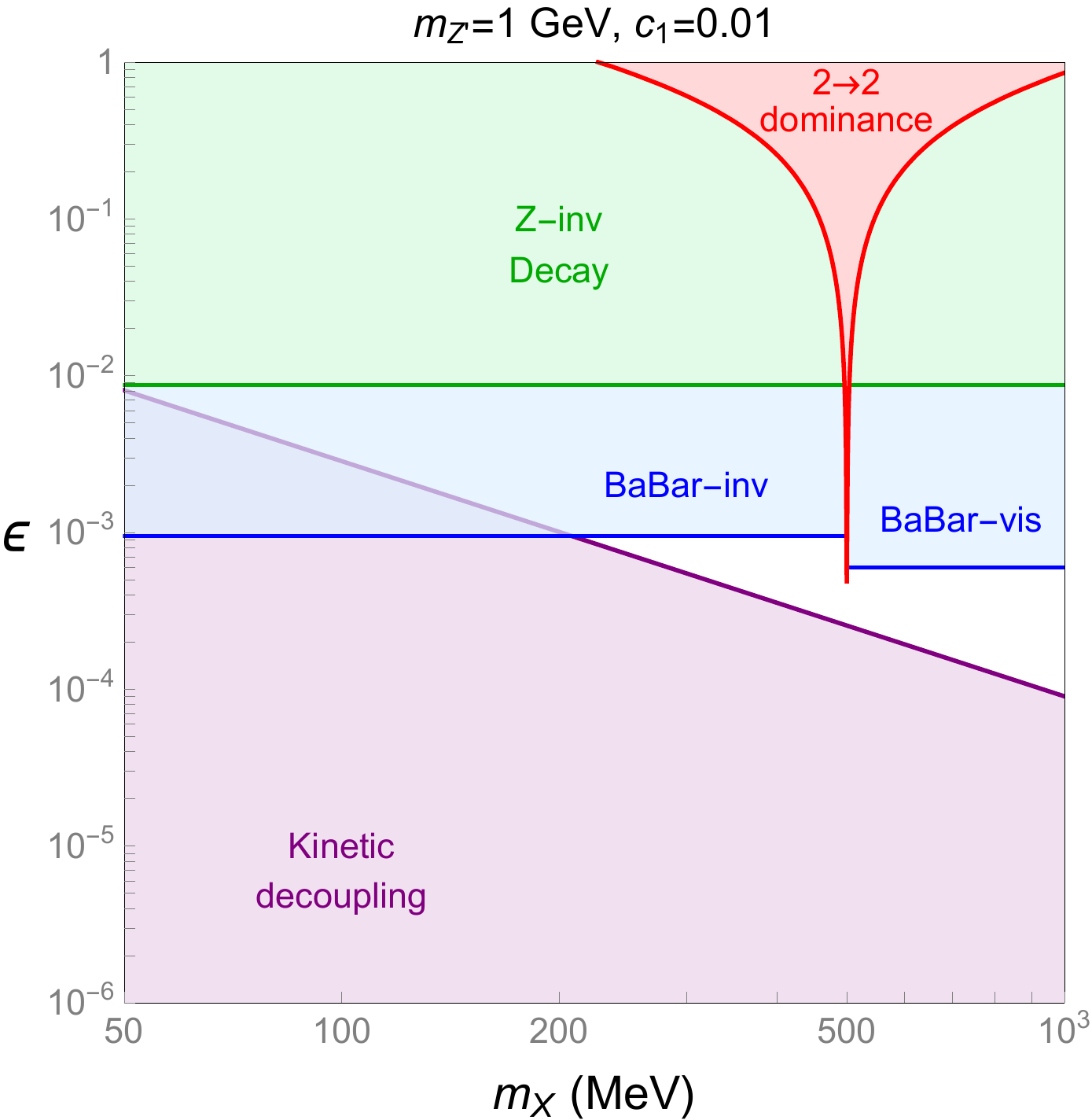} \vspace {0.5cm} \\
      \includegraphics[height=0.48\textwidth]{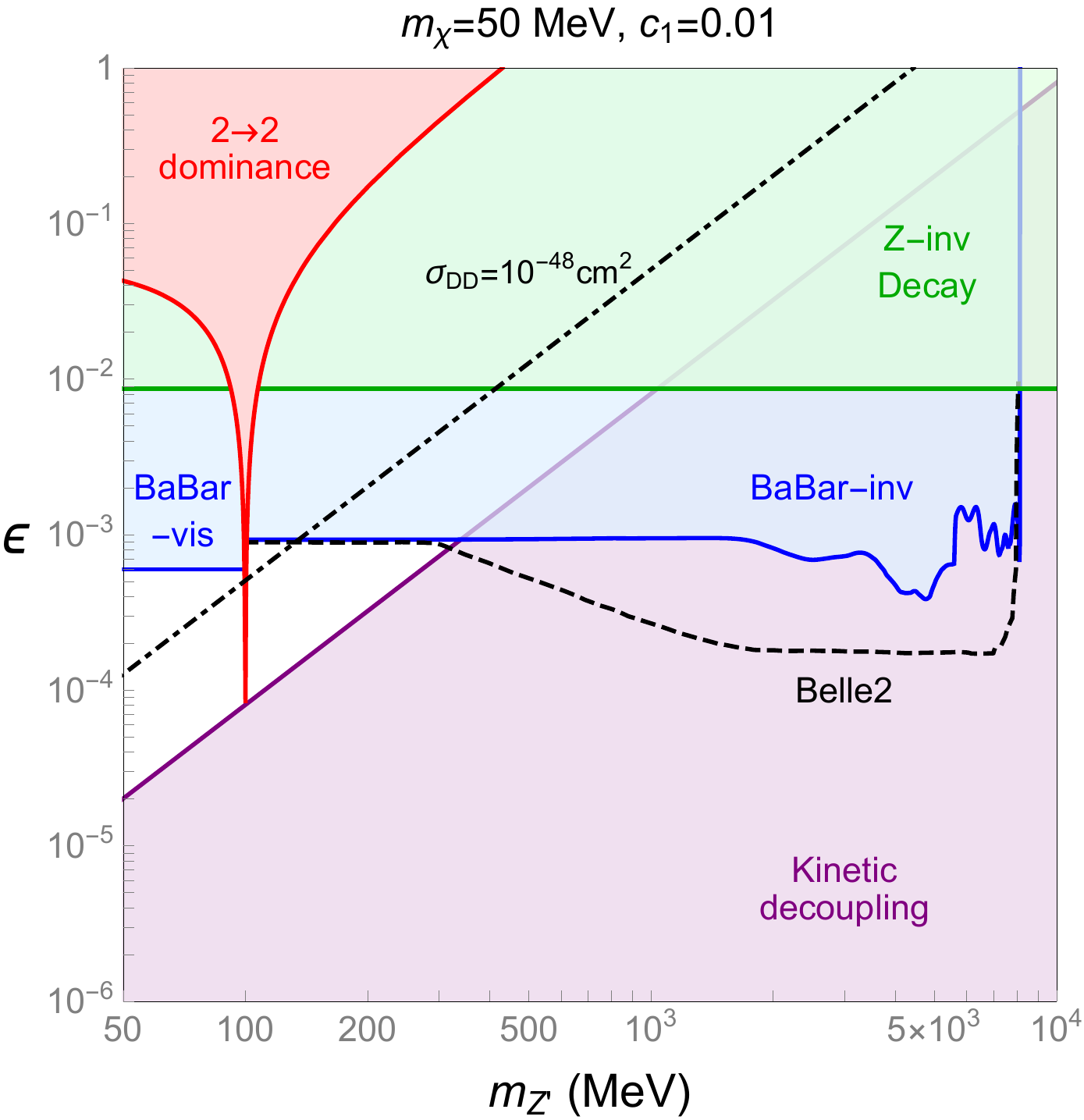}
         \includegraphics[height=0.48\textwidth]{figures/VSIMP/ep-mz2}
   \end{center}
  \caption{Allowed parameter space for vector SIMPs with a $Z'$ portal. {\bf Top:} Parameter space of $m_X$ vs. $\varepsilon$ for $c_1=0.01$, for $m_{Z'}=500\,{\rm MeV}$ ({\it left}) and $1\,{\rm GeV}$ ({\it right}).
  {\bf Bottom:} Parameter space of $m_{Z'}$ vs. $\varepsilon$ for $c_1=0.01$, for $m_X=50\,{\rm MeV}$ ({\it left}) and $100\,{\rm MeV}$ ({\it right}). Here, we took $N_f \alpha_{Z'}=1$ in Eq.~(\ref{Z-inv}) for the $Z$-boson invisible decay bound.
  {\bf In all panels:} the purple region indicates where the kinetic equilibrium condition fails; the green region is excluded by the $Z$-boson invisible decay~\cite{ALEPH:2005ab}; and the red region is where the $2\rightarrow 2$ annihilation becomes dominant. BaBar searches for monophotons with MET~\cite{Lees:2017lec} and with dileptons~\cite{Lees:2014xha} exclude the blue region. The projected Belle-II reach for monophoton+MET~\cite{Essig:2013vha} is depicted in dashed blue curve.  Contours for DM-electron scattering cross section with $\sigma_{\rm DD}=10^{-48\ (49)}{\rm cm}^2$ are also shown in dot-dashed lines on the left (right) panels.  }
  \label{Zportal}
\end{figure}

Next, we explore the kinetically mixed $Z'$ portal for mediation between the SIMP and visible sectors.
We use the CS terms of Eq.~\eqref{CS1}  to couple the vector DM to the $Z'$, together with kinetic mixing between the $Z'$ and the SM hypercharge. The momentum relaxation rate for vector DM scattering with electrons via the $Z'$ portal is given by
\bea
\gamma(T)_{Z^\prime} =\frac{1240\pi^3   c_1^2e^2\varepsilon^2}{567 m_X m_{Z'}^4}\,T^6\,, \label{gTeven}
\eea
and one imposes Eq.~\eqref{eq:eqcond} for kinetic equilibrium.

The resulting allowed parameter space is depicted in Fig.~\ref{c1-ep} as a function of kinetic mixing $\varepsilon$, for fixed DM and $Z'$ masses. The gray region is excluded by the unitarity bound on the CS term, and the kinetic equilibrium condition fails in the purple region. The LEP bound on the invisible decay width of the $Z$-boson, $\Gamma_{\rm inv}<3\,{\rm MeV}$~\cite{ALEPH:2005ab} is shown in green, where we have assumed the dominant mode is into dark fermions that generate the CS coupling [as would be the case in generic UV completions with $N_f \alpha_{Z'}=1$ in Eq.~(\ref{Z-inv})] in both plots. The BaBar constraint from invisible decays~\cite{Lees:2017lec} is shown in blue (with a similar-sized constraint from the beam dump experiment NA64 at CERN SPS~\cite{Banerjee2017}).

In Fig.~\ref{Zportal} we further show the allowed parameter space in $\varepsilon$ and $m_X$ (top panels) or $m_{Z'}$ (lower panels), for fixed values of the CS coefficient and of $m_{Z'}$ or $m_X$, respectively. Here, kinetic equilibrium is not maintained in the purple region; $2\to2$ processes are dominant over $3\to2$ processes in the red region; invisible $Z$-decay limits~\cite{ALEPH:2005ab} are imposed in green [where we have assumed the dominant mode is into dark fermions with $N_f \alpha_{Z'}=1$ in Eq.~(\ref{Z-inv})]; and constraints from BaBar invisible~\cite{Lees:2017lec} and visible~\cite{Lees:2014xha} searches are shown in blue. The projected reach of Belle-II into the parameter space is shown in the dashed blue curve~\cite{Essig:2013vha}.
As is evident, vector SIMPs through the $Z'$ portal can be achieved in an experimentally viable parameter space.

Concerning direct-detection, we note that the $Z'$ portal coupling of vector SIMPs via the CS term gives rise to a $p$-wave velocity-suppressed elastic cross section off electrons. As a result, the spin-independent cross section between vector SIMPs and electrons via the $Z'$ portal is highly suppressed, in contrast to the case of scalar SIMPs~\cite{Choi:2015bya}. For $m_e, m_X,m_{Z'}\gg p_{\rm DM}\simeq m_X v_{\rm DM}$, the DM-electron scattering cross section with $Z'$ portal is given by
\bea
\sigma_{\rm DD}&=&\frac{16 c_1^2 \varepsilon^2 \alpha_{\rm em} m^2_e }{3 m^4_{Z'}} \frac{(m^2_X+2m_e m_X-m^2_e) m_X^2}{(m_X+m_e)^4}\, v^2_{\rm DM} \nonumber \\
&\approx & 6\times 10^{-51}\,{\rm cm}^2 \left(\frac{c_1}{0.01} \right)^2\left(\frac{\varepsilon}{10^{-3}}\right)^2 \left(\frac{500\,{\rm MeV}}{m_{Z'}}\right)^4\left(\frac{v_{\rm DM}}{10^{-3}}\right)^2\,.
\eea
For illustration, contours of DM-electron scattering with $\sigma_{\rm DD}=10^{-48\ (49)}{\rm cm}^2$ are depicted in the lower left (right) panel of Fig.~\ref{Zportal}.

We learn that a kinetically mixed $Z'$ with CS couplings can successfully mediate interactions between the SIMP and SM sectors, consistent with all experimental constraints. We expect that future experiments such as Belle-II~\cite{Essig:2013vha} and potentially measurements at LHCb~\cite{Ilten:2016tkc}
can further probe the allowed parameter space for vector SIMPs with a vector portal.

\section{Conclusion}\label{sec:conc}

We have considered a spontaneously broken $SU(2)_X$ gauge theory in the hidden sector as an economical realization of vector SIMP Dark Matter. Kinetic equilibrium between the dark and visible sectors can be obtained via a Higgs portal in a minimal model or through a $Z'$-portal in an extended model with an additional $U(1)_{Z'}$ and its non-abelian Chern-Simons term. We have identified the parameter space for the $SU(2)_X$ gauge coupling and Dark Matter mass by taking into account the observed relic density as well as the self-scattering cross section. The kinetic equilibrium condition in combination with a variety of experimental constraints restrain the Higgs mixing or gauge kinetic mixing to a region that could be probed in current and planned experiments at the intensity frontiers.

%% file: parts/spin2.tex
\section{Introduction}
\begin{center}
\begin{figure}[h!]
  \begin{minipage}[c]{0.5\textwidth}
\includegraphics[width=0.95\linewidth]{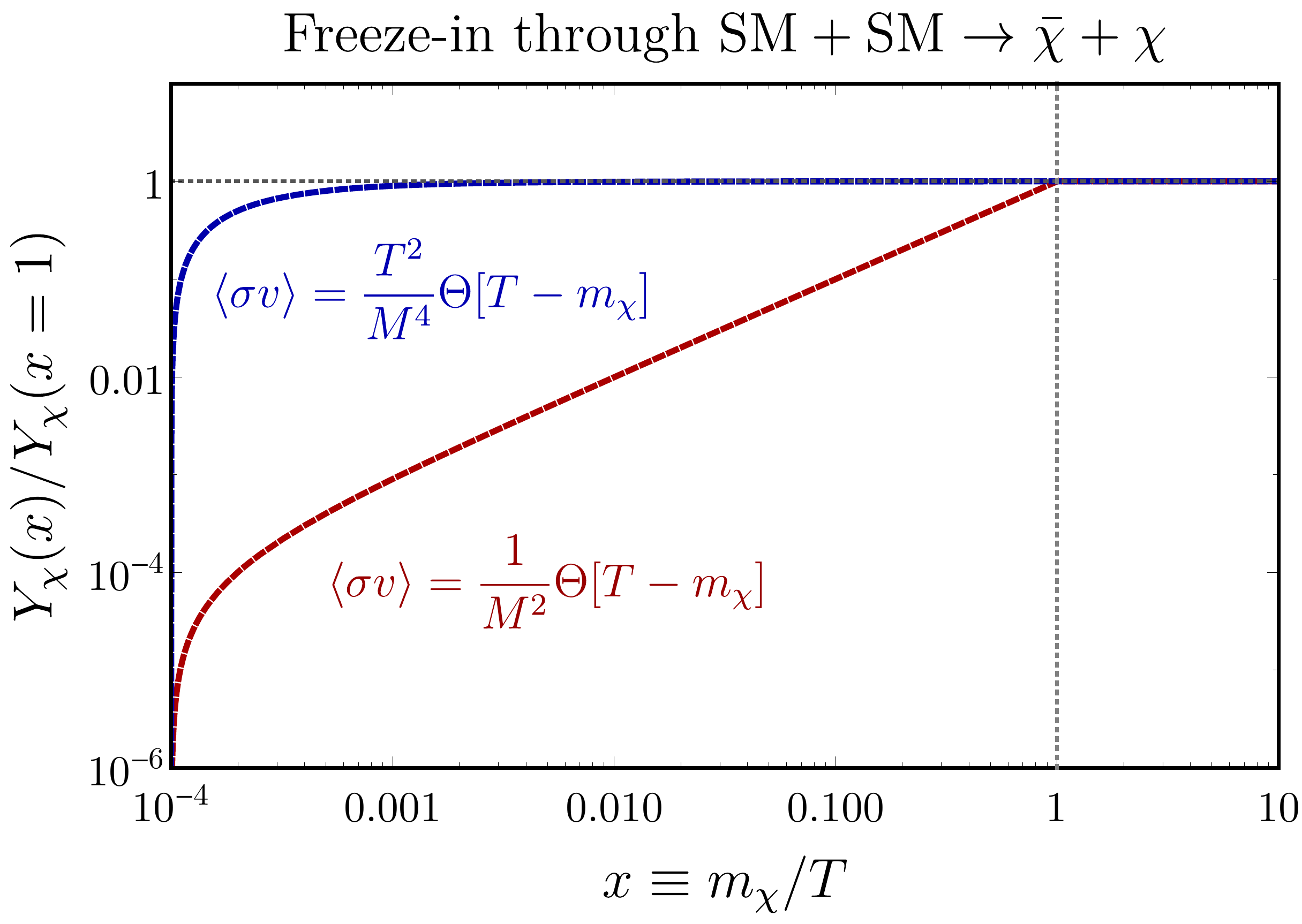}
   \end{minipage}\hfill
   \begin{minipage}[c]{0.5\textwidth}   
\includegraphics[width=0.95\linewidth]{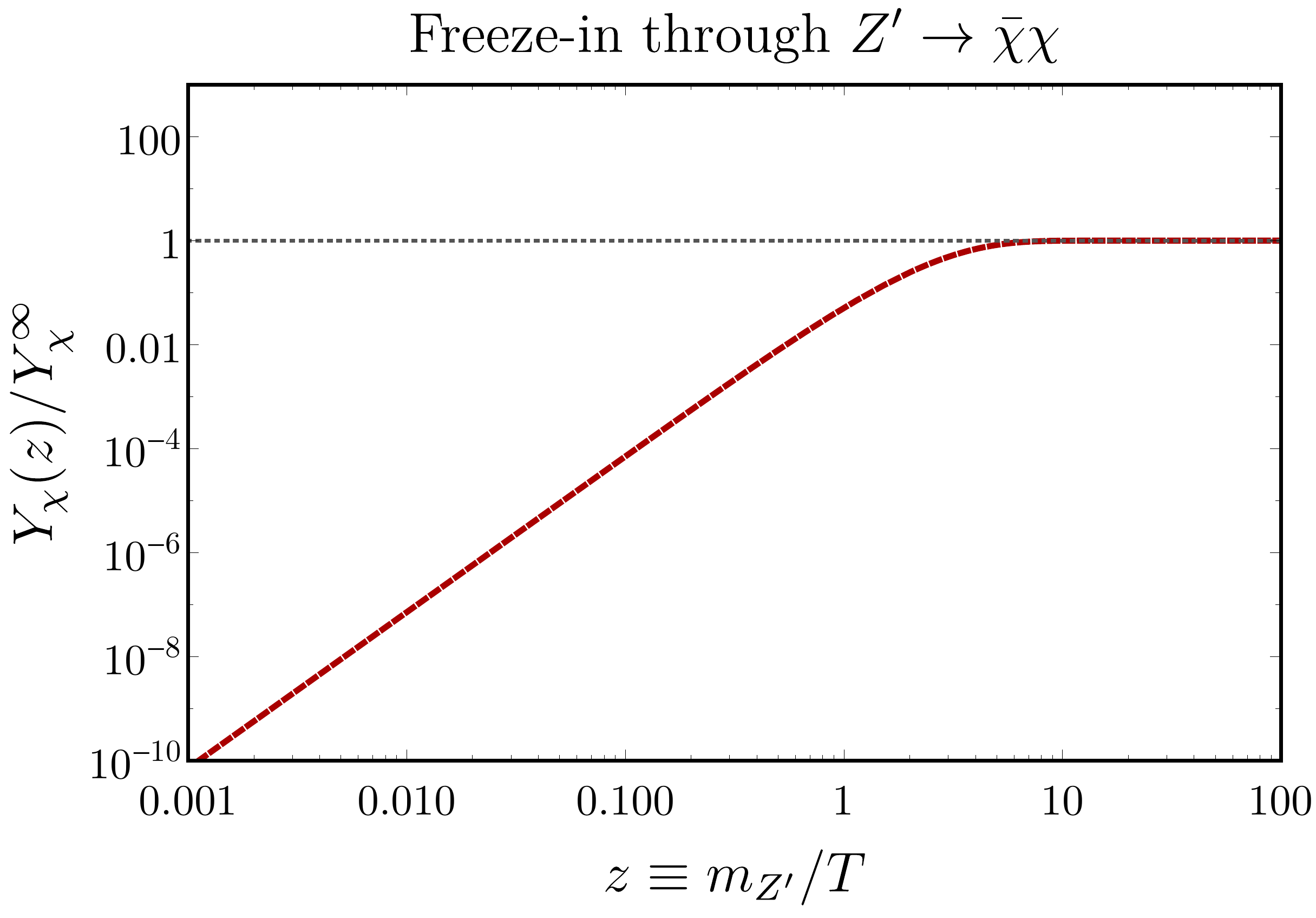}
   \end{minipage}
   \caption{Production of the Dark Matter yield $Y_\chi$ via freeze-in through decay of a heavy $Z^\prime$ mediator ({\bf right}) and through annihilation of a pair of SM particles ({\bf left}). The value $x_{\rm RH}=10^{-4}$ is used for illustration purposes.}
   \label{fig:freezein_solution}
\end{figure}
\end{center}

In this previous part we discussed the impact of recent results of direct detection experiments such as LUX \cite{Akerib:2016vxi}, PANDAX-II \cite{Cui:2017nnn} or XENON1T \cite{Aprile:2017iyp} that constrain a large part of the WIMP parameter space, and are close to
excluding the simplest extensions of the Standard Model (SM) such as the Higgs-portal model
\cite{Silveira:1985rk,McDonald:1993ex,Burgess:2000yq,Davoudiasl:2004be,Djouadi:2011aa,Han:2015hda,Mambrini:2016dca}, the $Z$-portal \cite{Arcadi:2014lta,Escudero:2016gzx} or even the $Z'$-portal \cite{Arcadi:2017jqd} 
(see \cite{Arcadi:2017kky} for a recent review on the subject). In the WIMP paradigm, the key element for this mechanism to work, is to introduce a sizable coupling between the dark sector and the standard model particle content. This is obviously a strong assumption motivated by the elegant simplicity of the WIMP miracle. However, facing the strong constraints from a plethora of experiments, the question of the effective viability of this mechanism has arose over the last years.
Consider the alternative case where the coupling between the SM and the dark sector is extremely small, in such a case the thermal equilibrium state cannot be reached in the early universe. There is the possibility to still generate non-thermally the correct Dark Matter density by producing it through decays or annihilations of SM particles, via the so-called \textit{freeze-in} mechanism~\cite{Hall:2009bx}~\footnote{See~\cite{fimp} for a review.}, while assuming a vanishing DM density at the reheating temperature. In order to illustrate the main idea of this mechanism, consider the following situations:
\begin{itemize}
\item The Dark Matter $\chi$ is weakly coupled to a heavy mediator $Z^\prime$ that thermalized with the SM particles in the early universe. If $Z^\prime$ is unstable, it will slowly decay, eventually to SM particles, but also to a Dark Matter pair $Z^\prime \rightarrow \bar{\chi} \chi$ if kinematically allowed. As shown in Sec.~\ref{sec:nonthermal_freezein}, in this case the Boltzmann equation for the DM number density reads:
\begin{equation}
\dfrac{\diff n_\chi}{\diff t}+3Hn_\chi=2 \Gamma_{Z^\prime \rightarrow \bar{\chi} \chi} \dfrac{K_1(z)}{K_2(z)}n_{Z^\prime}^{\rm eq}~,
\end{equation}
where $z\equiv m_{Z^\prime}/T$. The numerical solution of the Boltzmann equation for the DM yield $Y_\chi \equiv n_\chi /s$ is shown in Fig.~\ref{fig:freezein_solution} on the right pannel. In this case most of the Dark Matter would be produced when the $Z^\prime$ becomes non-relativistic around $z\sim1$ and the corresponding relic density is given by:
\begin{equation}
\Omega_\chi h^2\simeq \dfrac{m_\chi s_0 h^2}{\rho_c^0} \dfrac{3}{4 \pi}\dfrac{m_{Z^\prime}^3 \Gamma_{Z^\prime}}{H(m_{Z^\prime})s(m_{Z^\prime})}~.
\end{equation}
\item Another possibility is to produce a DM pair from relativistic SM particles annihilations $\text{SM}+\text{SM}\rightarrow \bar{\chi} \chi$. In this case the Boltzmann equation can be written as\footnote{See Sec.~\ref{sec:nonthermal_freezein} for a detailed derivation.}:
\begin{equation}
\dfrac{\diff n_\chi}{\diff t}+3Hn_\chi=\la \sigma v \ra (n_{\rm SM}^{\rm eq})^2~,
\end{equation}
where $\la \sigma v \ra$ is the production cross section and $n_{\rm SM}^{\rm eq}$ is the thermal density of the initial SM states which typically behaves as $n_{\rm SM}^{\rm eq} \propto T^3$ as they are assumed to be relativistic. We can define the production rate $R(T)$ as: 
\begin{equation}
R(T)\equiv \la \sigma v \ra (n_{\rm SM}^{\rm eq})^2~.
\end{equation}
Considering the following parametrization of the cross section:
\begin{equation}
\la \sigma v \ra = \dfrac{T^n}{M^{2+n}}\Theta[T-m_\chi]~,
\label{eq:sigmavTtothen}
\end{equation}
in this case the rate is simply given by a power law of the temperature and as shown in Sec.~\ref{sec:nonthermal_freezein}, the Boltzmann equation can be solved analytically in this case:
\begin{equation}
\Omega_\chi h^2 \propto \frac{ m_\chi^{n+2} M_{P} }{ M^{n+2}}    \dfrac{ x_{\text{RH}}^{-(n+1)}-1^{-(n+1)} }{(n+1)}~.
\end{equation}
where $M_{P}$ is the reduced Planck mass and $x$ is the usual variable $x\equiv m_\chi/T$. Assuming the reheating to occur at high temperatures $x_{\rm RH} \ll 1$, we can distinguish two interesting regimes:
\begin{itemize}
\item The light mediator regime ($n=-2$): where the only relevant scale in $\la \sigma v \ra$ is the temperature. The behavior of the yield in this case in shown in red on the left pannel of Fig.~\ref{fig:freezein_solution} and the relic density behaves as:
\begin{equation}
\Omega_\chi h^2 \propto M_{P}~.
\end{equation}
In this case the DM is mostly produced in the \textit{infrared} regime when $T\sim m_\chi$ and typically one would invoke very small couplings $\sim 10^{-10}$ to compensate the Planck mass dependence.
\item The heavy mediator regime ($n=0$): In this case most of the Dark Matter is produced around the reheating temperature as the cross section is not suppressed but enhanced by powers of the temperature, as shown on the left pannel of Fig.~\ref{fig:freezein_solution}. The relic density is given by:
\begin{equation}
\Omega_\chi h^2 \propto \frac{ m_\chi M_P }{ M^{2}} T_{\rm RH}~,
\end{equation}
which shows the importance of the \textit{ultra-violet} regime in this case via the reheating temperature dependence.
\end{itemize}
\end{itemize}
Therefore, in order to achieve the correct relic density in the simplest realization of the freeze-in mechanism, one has to invoke either tiny couplings or very large mass scales. It turns out there are many possible constructions where the Dark Matter is secluded from SM particle content by intermediate mass scale messengers.
Indeed, intermediate scales are naturally present in many well-motivated extensions of 
the SM, including Grand Unified Theories, models with a see-saw mechanism, string constructions, inflation and reheating or leptogenesis. 
In all of these frameworks, the presence of an intermediate mass scale generates a heavy particle
spectrum which can in principle mediate interactions between a possible dark sector and the SM.
Some specific examples of these frameworks are Grand Unified SO(10) models \cite{so10} or high-scale supersymmetry \cite{Benakli:2017whb,gravitino}, where Dark Matter candidates respecting the Planck/WMAP constraints \cite{Hinshaw:2012aka,Ade:2015xua} are present.
The effective superweak coupling generated by 
the exchange of an intermediate mass or even a superheavy mediator much heavier than the reheating temperature after inflation, $T_{\text{RH}}$,  allows for
the production of Dark Matter directly from the thermal bath in the same way that gravitinos are produced during reheating by Planck suppressed operators \cite{Nanopoulos:1983up,Ellis:1983ew,Khlopov:1984pf}. 
Often it is sufficient to approximate the details of 
particle production during reheating with the instantaneous reheating approximation. Namely, that all particle production occurs at the end of the reheating
process characterized by the reheating temperature $T_{\text{RH}}$.
However, depending on the specific production process, 
the instantaneous approximation may or may not be a good approximation.
For example, considering the parameterization of the production cross section of Eq.~(\ref{eq:sigmavTtothen}), this approximation has been shown to be reasonable for $n< 6$ \cite{Chung:1998rq,Giudice:2000ex,Kolb:2003ke,Garcia:2017tuj}.
However, for $n \ge 6$, the production rate is sensitive to the maximum temperature during the reheating process \cite{Ellis:2015jpg,Garcia:2017tuj} as we describe in 
more detail below. Thus the detailed mechanism for Dark Matter
production during reheating will in general be sensitive to 
the form of the coupling of the dark sector to the SM.

In the case of the gravitino, as noted above, from the strict observational point of view, Dark Matter only requires gravitational coupling. That is, communication between the two sectors (dark and SM) is mediated only through gravity, which couples to the energy--momentum tensor
of the Standard Model and Dark Matter. Because this coupling is fixed by the equivalence principle, the only free parameter in this model is the mass of the Dark Matter.
It has been shown in \cite{Garny:2015sjg,Tang:2017hvq,Garny:2017kha} that independent of the nature of Dark Matter, there exists a possibility to populate the relic abundance through a freeze-in mechanism via the exchange of a massless spin-2 graviton. 

In this work we generalize this to the case in which the exchanged spin-2 particle is massive. 
We borrow concepts from the theory of massive gravity. The Lorentz invariant linear theory of a massive graviton was formulated by Fierz and Pauli in \cite{Fierz:1939ix}, where it was shown that only one specific choice for the mass term is free from singularities. At the linear level, the massive graviton has 5 possible polarization states, as expected for a massive spin-2 particle. It was then shown that a generic nonlinear completion introduces a sixth state, that is a ghost \cite{Boulware:1973my}. Ref.~\cite{deRham:2010kj} provided a nonlinear construction that is ghost free.~\footnote{In fact, such a model was already formulated in \cite{GrootNibbelink:2004hg,Nibbelink:2006sz}, using the vielbein formalism. It was noted in these works that the ghost does not appear in the scalar sector of the theory. A conclusive proof that the ghost is absent in the full theory was later obtained in \cite{Hassan:2011hr}, also using the vielbein formulation.} We are not interested here in the nonlinear self-interactions of the massive spin-2 field, so we will simply employ the Fierz and Pauli  linear term, implicitly assuming a ghost-free nonlinear completion.

The coupling of the spin-2 mediator is expected to be universal,
but it might couple more strongly to the SM (and the dark sector) than a 
Planck suppressed gravitational coupling. Thus we consider here,
a massive spin-2 mediator coupling via the energy momentum tensor
but with an intermediate mass scale, thus enhancing its couplings 
relative to gravity. We generalize spin-2 couplings to Dark Matter and study the production mechanism of Dark Matter through a massive spin-2 portal. 

This chapter is organized as follows. In the next section, we lay out the model which includes a massive spin-2 mediator which is coupled to 
both the Dark Matter sector and the SM. In Sec.~\ref{sec:spin2relicabundance},
we discuss our computation of the relic Dark Matter abundance
and our results are given in Sec.~\ref{sec:spin2results} where we consider separately
the cases of heavy and light mediators and discuss the impact of dropping the instantaneous reheating approximation and our conclusions are given in Sec.~\ref{sec:spin2conclusion}.

\section{The model}

The model we consider is a relatively minimal 
extension of the SM which includes (in addition to the massless graviton $h_{\mu \nu}$) 
a Dark Matter candidate $X$, and a massive spin-2 mediator $\tilde h_{\mu \nu}$ and is described by
the Lagrangian
\beq
{\cal L} = {\cal L}_{\text{SM}} + {\cal L}_{\text{DM}} + {\cal L}_{\text{EH}} + {\cal L}_{\tilde h}+{\cal L}^1_{\text{int}}
+ {\cal L}^2_{\text{int}} \, ,
\eeq
with ${\cal L}_{\text{SM}}$ (${\cal L}_{\text{DM}}$) the Standard Model 
(Dark Matter) Lagrangian. ${\cal L}_{\text{EH}}$ is the Einstein-Hilbert sector\footnote{See Eq.~(\ref{eq:EHaction}).} which contains the kinetic terms of the massless graviton obtained after expanding the metric around flat space, $g_{\mu \nu} \simeq \eta_{\mu \nu} + h_{\mu \nu}/M_P$. ${\cal L}_{\tilde h}$ is the ghost-free Fierz-Pauli Lagrangian which contains the kinetic and mass terms for the massive spin-2 field. A general mass term for a spin-2 field can be written as:
\begin{equation}
{\cal L}_{\tilde h} \supset - \dfrac{1}{2}( a {\tilde h}_{\mu \nu} {\tilde h}^{\mu \nu} -b {\tilde h}^2)~,
\end{equation}
where ${\tilde h}\equiv {\tilde h}^\mu_\nu$. However, as briefly discussed in the introduction, one would expect at first sight that a massive spin-2 would carry ten degrees of freedom, which is reduced to six by considering a minimal coupling to matter with $\partial_\mu T^{\mu \nu}=0$. It was realized, when trying to quantize such a field, that one mode possesses kinetic terms with a wrong sign and therefore describing a ghost. The trick of Fierz and Pauli to avoid such a situation is to set $a=-b$. In this case the propagator describes five degrees of freedom corresponding to the five polarization states of the massive spin-2. Using this condition, the complete Fierz-Pauli Lagrangian is given by:
\begin{equation}
{\cal L}_{\tilde h} = -\dfrac{1}{2} \partial_\sigma {\tilde h}_{\mu \nu} \partial^\sigma {\tilde h}^{\mu \nu}+ \partial_\mu {\tilde h}_{\nu \sigma} \partial^{\nu} {\tilde h}^{\mu \sigma}-\partial_\mu {\tilde h}^{\mu \nu} \partial_\nu {\tilde h} + \dfrac{1}{2}\partial_\sigma {\tilde h} \partial^\sigma {\tilde h} -\dfrac{1}{2} m_{\tilde h}^2 ( {\tilde h}_{\mu \nu} {\tilde h}^{\mu \nu} -{\tilde h}^2)~.
\end{equation}
The massive spin-2 propagator $\Pi^{\alpha \beta,\rho \sigma}_2(\mathbf{p})$ and polarization sum rule can be built from the quantity $\Pi_{\alpha \beta} \equiv \eta_{\alpha \beta}+p_\alpha p_\beta/m^2$ as:
\begin{equation}
\Pi_2^{\alpha \beta,\rho \sigma}(\mathbf{p})=\dfrac{-i}{p^2-m^2}\Big[\dfrac{1}{2} (\Pi^{\alpha \rho} \Pi^{\beta \sigma}+\Pi^{\alpha \sigma} \Pi^{\beta \rho})-\dfrac{1}{3} \Pi^{\alpha \beta} \Pi^{\rho \sigma} \Big]~,
\label{eq:propmassivespin2}
\end{equation}
and
\begin{equation}
\sum_\lambda \epsilon^{\alpha \beta} (\mathbf{p},\lambda) \epsilon^{* \rho \sigma} (\mathbf{p},\lambda) =\dfrac{1}{2} (\Pi^{\alpha \rho} \Pi^{\beta \sigma}+\Pi^{\alpha \sigma} \Pi^{\beta \rho})-\dfrac{1}{3} \Pi^{\alpha \beta} \Pi^{\rho \sigma}~,
\end{equation}
where $\epsilon^{\mu \nu} (\mathbf{p},\lambda)$ is the polarization tensor of a spin-2 propagating with a momentum $\mathbf{p}$ and a polarization state $\lambda$. The polarization tensors satisfy the following relations:
\begin{equation}
p_\mu \epsilon^{\mu \nu} (\mathbf{p},\lambda) =0~, \quad \eta_{\mu \nu} \epsilon^{\mu \nu} (\mathbf{p},\lambda)=0~,\quad  \epsilon^{\mu \nu} (\mathbf{p},\lambda) \epsilon^*_{\mu \nu} (\mathbf{p},\lambda^\prime)= \delta_{\lambda \lambda^\prime}~. 
\end{equation}
The propagator of the massless graviton can be written in the Lorentz gauge as:
\begin{equation}
\Pi_1^{\alpha \beta,\rho \sigma}(\mathbf{p})=\dfrac{-i}{p^2}\Big[\dfrac{1}{2} (\eta^{\alpha \rho} \eta^{\beta \sigma}+\eta^{\alpha \sigma} \eta^{\beta \rho})-\dfrac{1}{2} \eta^{\alpha \beta} \eta^{\rho \sigma} \Big]~.
\label{eq:propgraviton}
\end{equation}
The final two terms present in the Lagrangian,
${\cal L}^i_{\text{int}}$ are the interaction terms with the massless graviton 
$h_{\mu \nu}$ ($i=1$) and the massive spin-2 mediator $\tilde h_{\mu \nu}$ ($i=2$)
that can be written, from the equivalence principle:
\begin{equation}
{\cal L}^1_{\text{int}} = \frac{1}{2 M_{P}}h_{\mu \nu}~(T^{\mu \nu}_{\text{SM}}+ T^{\mu \nu}_{\text{X}})
\end{equation}
\begin{equation}
{\cal L}^2_{\text{int}} =  \frac{1}{\Lambda} \tilde h_{\mu \nu}~ 
(\gsm T^{\mu \nu}_{\text{SM}}+ \gdm T^{\mu \nu}_{\text{X}})
\end{equation}
where $M_P$ is the reduced Planck mass $M_P \simeq 2.4\times 10^{18}$ GeV, and $\Lambda \lesssim M_P$ is an
intermediate scale and governs the strength of the new spin-2 interaction. The couplings, $\gsm$ ($\gdm$) of the messenger to the Standard Model (Dark Matter) allow us to distinguish interactions between the two sectors. Of course only 2 of the three parameters ($\Lambda, \gsm, \gdm$) are independent. The form of the stress-energy tensor of a field, $T^a_{\mu \nu}$ depends on its spin $a=0,1/2,1$.\footnote{We assume real scalars and Dirac 
fermions throughout our work.} In general, we can write
\bea
T^0_{\mu \nu} &=& \frac{1}{2} \left( \partial_\mu \phi~\partial_\nu \phi + \partial_\nu \phi ~\partial_\mu \phi -g_{\mu \nu} \partial^\alpha \phi ~\partial_\alpha \phi \right) \,, \nonumber\\
T^{1/2}_{\mu \nu} &=& \frac{i}{4}
\bar \psi \left( \gamma_\mu \partial_\nu + \gamma_\nu \partial_\mu \right) \psi
-\frac{i}{4} \left( \partial_\mu \bar \psi \gamma_\nu + \partial_\nu \bar \psi \gamma_\mu \right)\psi \,, 
\nonumber\\
T^{1}_{\mu \nu} & = & \frac{1}{2} \left[ F_\mu^\alpha F_{\nu \alpha} + F_\nu^\alpha F_{\mu \alpha} - \frac{1}{2} g_{\mu \nu} F^{\alpha \beta} F_{\alpha \beta} \right] \,.
\eea
The amplitudes relevant for the computation of the processes $\text{SM}^a(p_1)+\text{SM}^a(p_2) \rightarrow \text{DM}^b(p_3)+\text{DM}^b(p_4)$ can be parametrized by 
\begin{equation}
\mathcal{M}^{ab} \propto \sum_{i=1,2} \langle p_1^a p_2^a | \mathcal{L}_{\text{int}}^i | p_3^b p_4^b \rangle \propto \sum_{i=1,2} M_{\mu \nu}^a \Pi^{\mu \nu,\rho \sigma}_i M_{\rho \sigma}^b \;, 
\end{equation}
where $a$ and $b$ denote respectively the spin of the SM and DM particles involved in the process $a,b=0,1/2,1$. $\Pi^{\mu \nu,\rho \sigma}_i$ denotes the propagators of the graviton ($i=1$) and massive spin-2 ($i=2$) given respectively in Eq.~(\ref{eq:propgraviton}) and Eq.~(\ref{eq:propmassivespin2}).
The partial amplitudes, $M_{\mu \nu}^a$, can be expressed as
\bea 
M_{\mu \nu}^0 &=& \frac{1}{2}(p_{1\mu} p_{2\nu} + p_{1\nu} p_{2\mu} - g_{\mu \nu}p_1.p_2) \,, \nonumber\\ 
M_{\mu \nu}^{1/2} &=&  \frac{1}{4} {\bar v}(p_2) \left[ \gamma_\mu (p_1-p_2)_\nu + \gamma_\nu (p_1-p_2)_\mu \right] u(p_1) \;,  \nonumber\\ 
M_{\mu \nu}^{1} &=&  \frac{1}{2}\Bigg[ 
\epsilon_2^*.\epsilon_1(p_{1\mu} p_{2\nu}+ p_{1 \nu} p_{2 \mu})- \epsilon_2^*.p_1
(p_{1 \mu} \epsilon_{1 \nu} + \epsilon_{1 \mu} p_{2 \nu}) \nonumber
\\
&&
-\epsilon_1.p_2 (p_{1 \nu} \epsilon^*_{2 \mu} + p_{1 \mu} \epsilon^*_{2 \nu})
+p_1.p_2(\epsilon_{1 \mu} \epsilon^*_{2 \nu}+ \epsilon_{1 \nu} \epsilon^*_{2 \mu})
\nonumber
\\
&&
+\eta_{\mu \nu}(\epsilon_2^*.p_1 \epsilon_1.p_2 - p_1.p_2~ \epsilon^*_2.\epsilon_1)
\Bigg] \,, 
\eea
with similar expressions in terms of the Dark Matter momenta, $p_3, p_4$. The total amplitude squared implied in the processes SM SM $\rightarrow$ DM DM will be a sum of the three contributions, weighted by the Standard Model content in fields:
\beq
|{\cal M}|^2= 4 |{\cal M}^0|^2 + 45 |{\cal M}^{1/2}|^2 + 12 |{\cal M}^1|^2.
\eeq
Further details regarding these amplitudes are found in Sec.~\ref{sec:appendixspin2}.

%
%
%

\section{The relic abundance}
\label{sec:spin2relicabundance}

As noted in the introduction, reheating after inflation is often assumed to occur instantaneously, on a timescale given by the inflaton decay rate $\Gamma_\phi$. This results in a thermal bath of initial temperature (see \cite{Ellis:2015jpg,Garcia:2017tuj} for a detailed discussion) 
\begin{equation}
T_{\rm RH} = \left( \frac{40}{g_{\rm RH} \, \pi^2} \right)^{1/4} \left( \frac{\Gamma_\phi \, M_p}{c} \right)^{1/2} \,, 
\end{equation}
where $g_{\rm RH}$ is the number of effective degrees of freedom in the thermal bath of temperature $T_{\rm RH}$, and where $c$ is an order one parameter that depends on when precisely reheating is assumed to take place (setting $\Gamma_\phi^{-1}$ equal to the reheating time, $\Gamma_\phi^{-1} = t_{\rm RH}$, leads to $c=1$; 
setting it instead equal to the Hubble rate, $\Gamma_\phi = H \left( t_{\rm RH} \right)$, leads to $c=\frac{2}{3}$). Numerical solutions to particle yields during reheating agree with the
instantaneous approximation if $c \approx 1.2$ \cite{Pradler:2006hh,Rychkov:2007uq,Ellis:2015jpg}.

In reality reheating is a finite-duration process. The inflaton decay products thermalize on a much shorter timescale than $\Gamma_\phi^{-1}$, so it is appropriate to assume the co-existence of a decaying inflaton, and of a thermal bath arising from the decay. In this context, the reheating temperature is conventionally defined as the temperature of the thermal bath when it starts to dominate over the inflaton. However, the thermal bath reaches its maximum temperature while still subdominant to the inflaton. One finds (see for instance Ref.~\cite{Ellis:2015jpg}) 
\begin{equation}
T_{\rm max} \simeq 0.5 \left( \frac{m_\phi}{\Gamma_\phi} \right)^{1/4} \, T_{\rm RH} \,, 
\end{equation}
where $m_\phi$ is the inflaton mass. Perturbativity requires $\Gamma_\phi < m_\phi$, and it is typical to have $\Gamma_\phi \ll m_\phi$. Therefore the maximum temperature of the thermal bath can be many orders of magnitude greater than $T_{\rm RH}$. 

Particle production at temperatures $T > T_{\rm RH}$ can be significant.  There are two reasons, in fact,  why assuming an instantaneous reheating can lead to a significant underestimation of the Dark Matter abundance. The first case, which has extensively been pointed out in the literature, occurs when the (thermally averaged) Dark Matter production cross section times velocity given in Eq.~(\ref{eq:sigmavTtothen})  has a strong temperature dependence. 
Such a relation applies for instance when the Dark Matter is produced from quanta in the thermal bath by the exchange of a heavy mediator of mass $M \gg T$. If $n< 6$ in this relation, the Dark Matter abundance is mostly produced at the end of reheating, when $T \simeq T_{\rm RH}$. In contrast, the quanta produced at $T \simeq T_{\rm max}$ dominate the final abundance if $n > 6$. For $n=6$, particles produced at any temperature equally contribute to the final abundance. This results in a logarithmic $\ln \frac{T_{\rm max}}{T_{\rm RH}}$ enhancement of the abundance with respect to  the naive estimate based on instantaneous reheating. As shown in \cite{Garcia:2017tuj}, $n=6$ is obtained in the case of gravitino Dark Matter in high scale supersymmetry models. As we show in the present work, this also applies to the cases in which the mediator is a heavy spin-2 particle. While we do not consider it here, the same strong temperature dependence is found if the mediator is a pseudo-scalar particle, and both the Dark Matter and the quanta in the thermal bath are spin-1 particles. 

A second reason why the instantaneous reheating case can lead to a significant underestimation of the Dark Matter abundance, and which we explore in the present work, is if the mass $M$ of the mediator is between $T_{\rm RH}$ and $T_{\rm max}$. In this case, it is possible that the final Dark Matter abundance is due to the quanta produced on resonance, taking place at $T \simeq M$. 
The resonance is missed if one simply assumes that the thermal bath is instantaneously formed with $T = T_{\rm RH}$. 

The numerical analysis in this work takes into account the total set of Boltzmann equations for the time evolution of a system whose energy density is in the form of unstable massive particles $\phi$ (the inflaton for instance), stable massive DM particles $X$, and radiation $R$ (ie SM particles).
For the exact computation, 
we assumed that $\phi$ decays into radiation with a rate $\Gamma_\phi$, and that the DM particles are created and annihilate into radiation with a thermal-averaged cross section times velocity $\langle\sigma v\rangle$.
The corresponding energy and number densities satisfy the differential equations~\cite{Chung:1998rq, Giudice:2000ex}
\begin{align}
\frac{\text{d}n_X}{\diff t}&=-3H\,n_X-\langle\sigma v\rangle\left[n_X^2-(n_X^{\text{eq}})^2\right]\,, \nonumber \\
\frac{\text{d}\rho_R}{\diff t}&=-4H\,\rho_R+\Gamma_\phi\,\rho_\phi+2\langle\sigma v\rangle\langle E_X\rangle\left[n_X^2-(n_X^{\text{eq}})^2\right]\,, \nonumber \\
\frac{\text{d}\rho_\phi}{\diff t}&=-3H\,\rho_\phi-\Gamma_\phi\,\rho_\phi\,.
\label{Eq:setboltzmann}
\end{align}
We assumed that each $X$ has energy $\langle E_X\rangle\simeq\sqrt{m_X^2+ 9T^2}$ and the factor $\langle E_X\rangle$ is the average energy released per $X$ pair annihilation.
The Hubble expansion parameter $H$ is given by $H^2=\frac{1}{3M_P^2}(\rho_\phi+\rho_R+\rho_X)$. Thermalization of the SM radiation produced by inflaton decays is rapid \cite{Davidson:2000er,Harigaya:2013vwa,Mukaida:2015ria,Ellis:2015jpg}. However, as shown in~\cite{Garcia:2018wtq}, a sizable amount of DM could be produced before the SM particles thermalize, we do not take into account this effect here. As such, we can define a 
radiation temperature, 
\beq\label{inst_tempe}
T = \left(\frac{30\rho_{R}}{\pi^2 g(T)}\right)^{1/4}\,,
\eeq
and $T_{\text{\scriptsize{max}}}$ corresponds to the maximum temperature 
attained during the reheating process. When the Dark Matter number density is far below its equilibrium abundance (and when inflaton decays do not directly produce $X$) the Dark Matter density $n_X$ is given by the approximate Boltzmann equation
\beq
\frac{\text{d} n_X}{\text{d}t} = -3 H n_X + (n_X^{\text{eq}})^2 \langle \sigma v \rangle  \,, 
\eeq
where $H$ is the Hubble rate and $n_X^{\text{eq}}$ is the number density that the dark-matter would have in thermal equilibrium. This relation assumes that the Dark Matter is produced through $2 \rightarrow 2$ processes from quanta in the thermal bath, and that the Dark Matter abundance is well below its thermal equilibrium value, which is the case in the models considered here. This relation can be rewritten as 
\beq
\frac{\text{d} Y_X}{\text{d}T} = - \frac{R(T)}{H T s} \,, 
\label{Eq:y}
\eeq
where $Y_X=n_X/s$ is the Dark Matter yield, 
$s=\frac{2 \pi^2}{45} g_s(T) T^3$ is the 
entropy density in the thermal bath with $g_s(T)$ effective number of degrees of freedom. The production rate, $R(T)= (n_X^{\text{eq}})^2 \langle \sigma v \rangle$ for the 1 + 2 $\rightarrow$ 3 + 4 process is obtained from
\beq
R(T) = \int f_1 f_2 \frac{E_1 E_2 \text{d}E_1 \text{d}E_2 ~\text{d}\cos \theta_{12}}{1024 \pi^6} \int |{\cal M}|^2 \text{d} \Omega_{13} \,,  
\label{Eq:rt}
\eeq
where $f_1$ and $f_2$ are the distribution functions of the initial (SM) particles.

The two processes we consider contributing to the relic abundance are the exchange of the (massless) graviton $h_{\mu \nu}$, with Planck mass suppressed couplings, and the exchange of the massive spin-2 mediator $\tilde h_{\mu \nu}$. Different results are obtained in the heavy mediator ($m_{\tilde h} > T_{RH}$) and light mediator ($m_{\tilde h} < T_{RH}$) cases which are discussed separately below.

\section{Results}
\label{sec:spin2results}


As noted above, all of the results in this work are obtained via a numerical calculation using the complete set of the Boltzmann equations (\ref{Eq:setboltzmann}), and are not based on the instantaneous reheating approximation. However, it is useful to give approximate solutions in order to perform simple analytical estimates, as the difference between the instantaneous and the non-instantaneous reheating is often an overall multiplicative factor \cite{Garcia:2017tuj}. 
In Sec.~\ref{sec:appendixpreheating}, we derive in detail the computation of the relic abundance. We obtained the following expression for the relic density in the instantaneous approximation:
\begin{equation}
\begin{split}
\frac{\Omega h^2_{\text{RH}}}{0.1} \approx \left( \frac{m_X}{1 ~\text{GeV}} \right) \Big\{ & 8 \times 10^{-17} \Big(\frac{\alpha}{\alpha^0 }\Big) \Big(\frac{T_{\text{RH}}}{10^{12}~\text{GeV}}\Big)^3  \\ 
& + \Theta[T_{\text{RH}} - \mh] \Big[  4 \times 10^{-6} \Big(\frac{\beta_1}{\beta_1^0 }\Big) \Big(\frac{T_{\text{RH}}}{10^{12}~\text{GeV}}\Big)^3 \Big(\frac{10^{16}~\text{GeV}}{\Lambda}\Big)^4  \\
 & \hspace{2.8cm} + 2 \Big(\frac{\beta_2}{\beta_2^0 }\Big) \Big(\frac{\mh}{10^{10}~\text{GeV}}\Big) \Big(\frac{10^{16}~\text{GeV}}{\Lambda}\Big)^2 \Big]  \\
& +  \Theta[\mh - T_{\text{RH}}] \Big[ 70  \Big(\frac{\beta_3}{\beta_3^0}\Big) \Big(\frac{T_{\text{RH}}}{10^{12}~\text{GeV}}\Big)^7  \Big(\frac{10^{16}~\text{GeV}}{\Lambda}\Big)^4  \Big(\frac{10^{11}~\text{GeV}}{\mh}\Big)^4  \Big] \Big\} \,, 
\label{Eq:omega}
\end{split}
\end{equation}
where we fixed $\gsm=\gdm=1$, $g_s=100$ and $\Theta$ is the Heaviside step function.
The first term in this expression is due to ordinary graviton exchange in the processes producing Dark Matter. We see that this contribution is completely negligible unless the Dark Matter is very heavy. The following terms are due to the exchange of a massive spin-2 state in three different mass regimes, namely mass greater, smaller but comparable, and much smaller than the reheating temperature. The numerical factors $\alpha, \beta_1, \beta_2$ and $\beta_3$ are normalized by their values for the case of a scalar Dark Matter candidate and would change by a factor of one order of magnitude if the Dark Matter is a fermion or a vector. The superscript of the coupling coefficients denotes the 
spin of the Dark Matter candidate and values of these coefficients are tabulated in a
table given in Sec.~\ref{sec:appendixspin2}. Note that the parametric dependence in the above expressions for $\Omega h^2$ can be seen directly from the rates with 
$\Omega h^2 \sim R/nH \sim R M_P/T^5$.
Overall, we can distinguish 4 regimes corresponding to 4 different process leading to the production of a sufficient abundance of Dark Matter.
\begin{itemize}
\item{{\it Super-heavy mediator regime}, or decoupling regime ($\mh \gg T_{\text{RH}}$):
When $\mh \gtrsim 3000 (\beta_3 \alpha^0/\beta_3^0 \alpha)^{1/4} (10^{16} {\rm GeV}/\Lambda) T_{\text{RH}}$, the dominant production mode is through the exchange of a massless graviton. For very large masses, the rate mediated by the spin-2 propagator is suppressed even relative to the Planck suppressed rate mediated by gravity. In this regime, the first term of Eq.~(\ref{Eq:omega}) dominates. The production rate in such regime is very sensitive to the temperature of the thermal bath and is proportional to $R(T) \propto \frac{T^8}{M_P^4}$~\footnote{The exact expression of the rates can be found the Sec.~\ref{sec:appendixspin2} for the separate cases of scalar,
fermionic, and vector Dark Matter.}.}

\item{{\it Heavy mediator regime} ($\mh > T_{\text{RH}}$): When the condition for the super-heavy mediator regime above is not satisfied, yet the mediator mass still exceeds the reheating temperature, the dominant mode is massive spin-2 mediator with rate given by the final term (proportional to $\beta_3$) in Eq.~(\ref{Eq:omega}). Despite the massive mediator, the coupling and hence the rate are enhanced over the gravitational rate by the fact that $\Lambda < M_P$. The rate is in this case $highly$ dependent on the temperature and is proportional to $R(T) \propto \frac{T^{12}}{\Lambda^4 \mh^4}$.}

\item{{\it Narrow Width Approximation} (NWA) ($\mh \lesssim T_{\text{RH}}$): this regime dominates when the temperature of the thermal bath approaches the mediator mass $\mh$. $\tilde h_{\mu \nu}$ is then produced on-shell in resonance, and if the width $\Gamma_{\tilde h}$ is sufficiently small, this process will dominate. 
The expression of the width can be found in the Sec.~\ref{sec:appendixspin2} and scales as $\Gamma_{\tilde h} \sim \mh^3/\Lambda^2$. 
The rate is mildly dependent on the temperature and is proportional to $R(T)\propto \frac{\mh^9}{\Lambda^4\Gamma_{\tilde h}} \frac{T}{\mh} ~K_1\left(\frac{\mh}{T}\right) 
\propto \frac{\mh^6}{\Lambda^2} \frac{T}{\mh} K_1 \left(  \frac{\mh}{T}\right)$~\footnote{See the Sec.~\ref{sec:appendixspin2} for details.}. For $T \sim \mh$ we obtain the term proportional to $\beta_2$ in 
Eq.~(\ref{Eq:omega}).} 

\item{{\it Light mediator regime} ($\mh \ll T_{\text{RH}}$):
this regime is very similar to the well studied case of a light gravitino. Indeed, the behavior and couplings are exactly the same except the coupling, proportional to $1 / \Lambda$ and not $1/ M_P$, is much larger. This regime is the one for which the particle production is greatest and the rate is proportional to $R(T)\propto \frac{T^8}{\Lambda^4}$ }.
\end{itemize}

\begin{figure}[h!]
\centering
\includegraphics[width=0.7\textwidth]{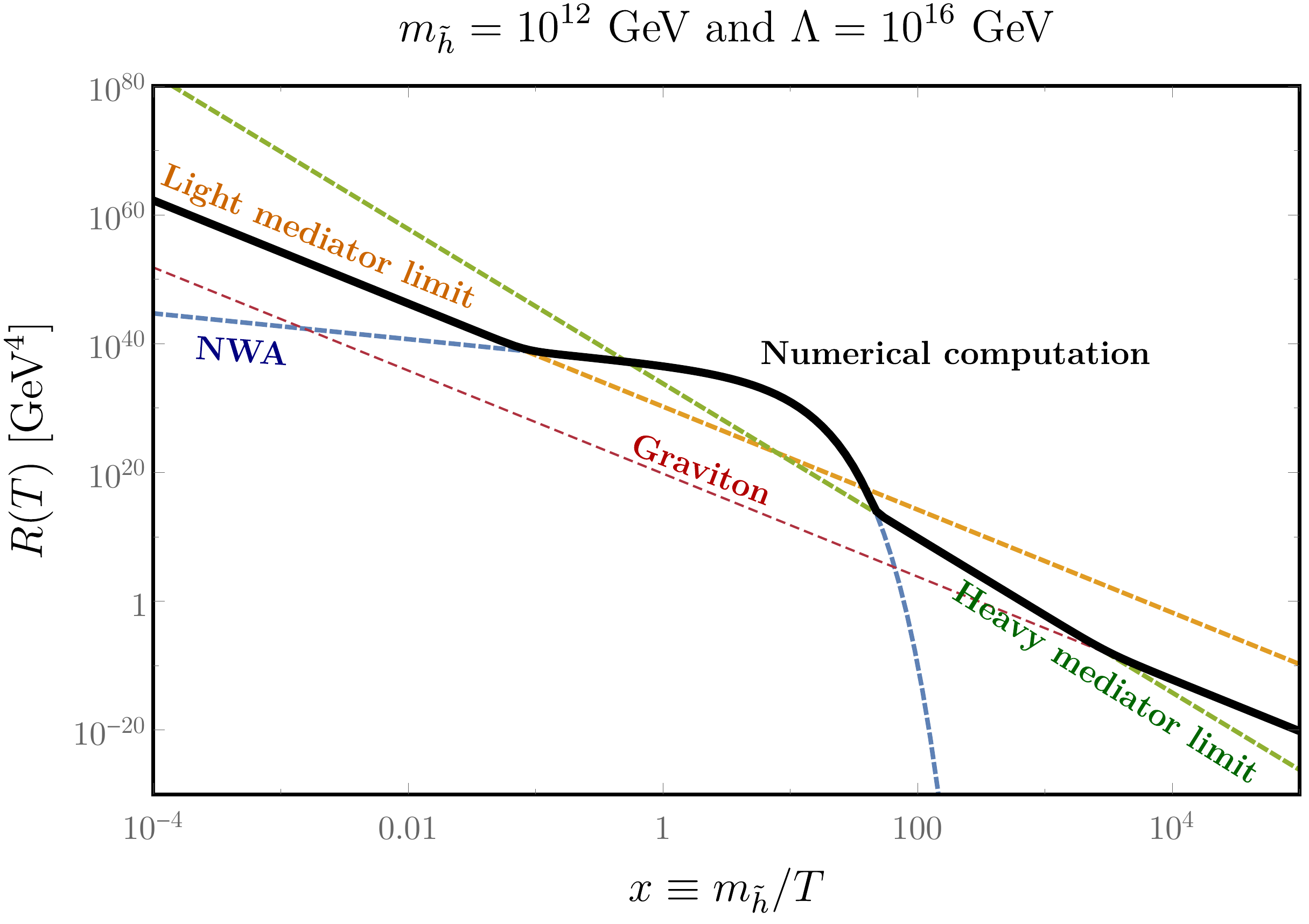}
\caption{Evolution of the production rate $R(T)$ from Eq.~(\ref{Eq:rt}) as function of $\mh /T$ for $\Lambda =10^{16} $ GeV and $\mh = 10^{12}$ GeV.}
\label{Fig:rateplot}
\end{figure}

The exact dependence on the temperature of each rate is detailed in Sec.~\ref{sec:appendixspin2}, and is fundamental in order to understand the behavior of the relic abundance as function of the reheating temperature. For an illustration, we show in Fig.~\ref{Fig:rateplot} the production rate $R(T)$ as function 
of the dimensionless parameter $x = \mh /T$ for $\mh=10^{12}$ GeV and $\Lambda=10^{16}$ GeV.

We can clearly distinguish 3 main regimes (in addition, the superheavy mediator regime is seen as
a slight bend in the curve at large $\mh/T$) in this figure. To guide the eye, we have plotted with dashed lines all four regimes as if they were valid at all values of $x$. The line labeled as graviton corresponds to the superheavy mediator with rate proportional to $\alpha$. 
The solid black line corresponds to the full calculation valid for all values of $x$. When $T \gg \mh $, the rate is dominated by the light mediator limit, $R(T)$ decreasing with the temperature at a rate proportional to $T^8$. Then, when the temperature approaches the mass of the mediator, the NWA regime dominates, giving a rate very mildly dependent on the temperature ($R(T) \propto \frac{T}{\mh} K_1(\frac{\mh}{T})$). 
It is only when $T$ drops below $\mh$ that the exponential behavior of the Bessel function dominates. At larger $x$, the rate then drops abruptly to enter in the heavy mediator regime, with a strong dependence on the temperature $R(T)\propto T^{12}$. At still lower temperatures, eventually graviton exchange dominates and the rate again falls as $R\sim T^8$.

In the following subsections, we compute the relic abundance of the Dark Matter, integrating the production rate in each of these regimes. The integration was made numerically, using the set of equations (\ref{Eq:setboltzmann}), taking into account the effect of non-instantaneous reheating on the relic abundance. However, as it was shown in \cite{Garcia:2017tuj}, the difference induced by the exact non-instantaneous reheating treatment is a multiplicative factor, independent of the model, except when the resonance is important. The analytical expressions in Eq.~(\ref{Eq:omega}) are based on the instantaneous reheating approximation and are used only as an aid to describe the results below.

\subsection{Heavy mediator regime}

\begin{figure}[h!]
\centering
\includegraphics[width=0.7\textwidth]{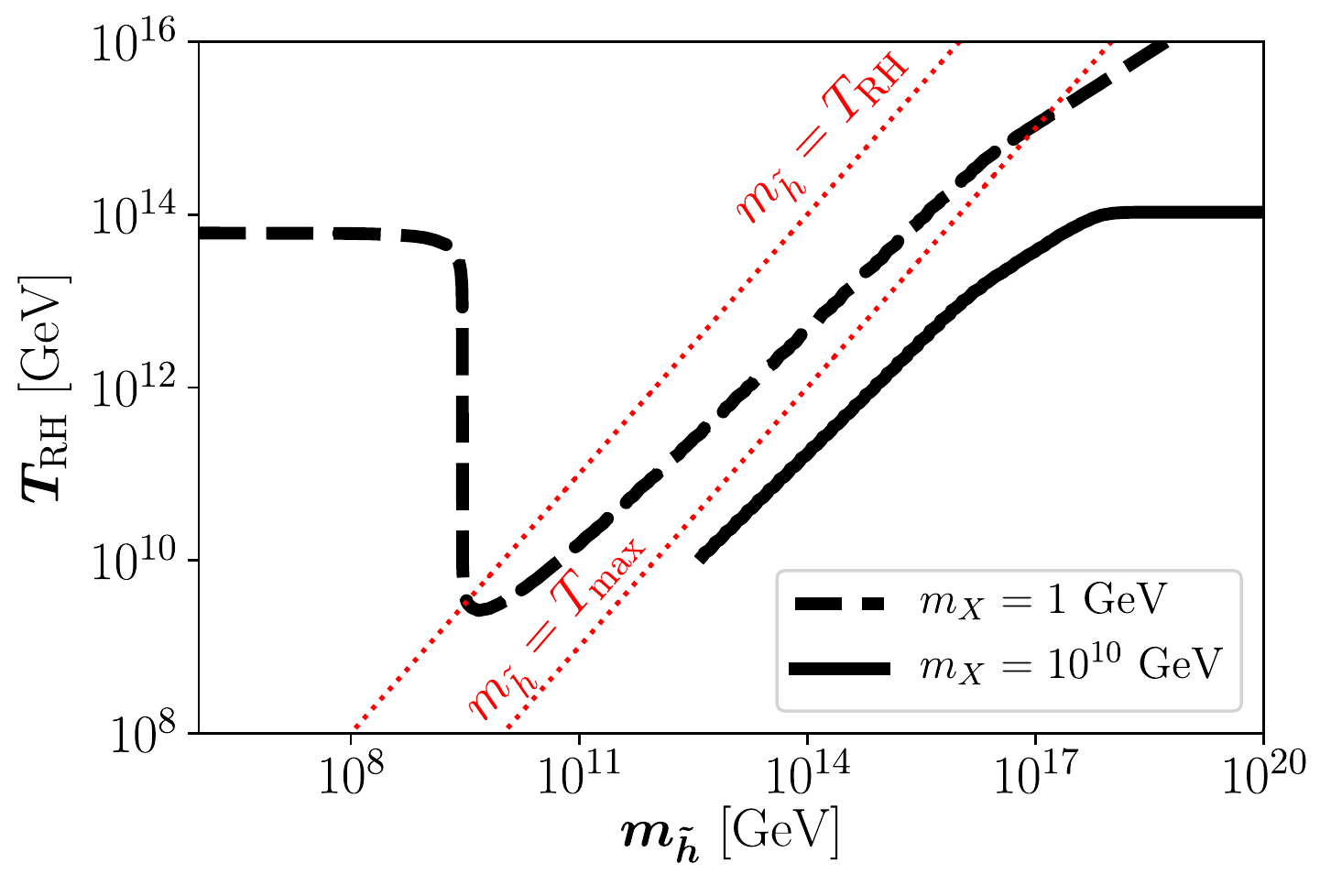}
\caption{Values of $T_{\text{RH}}$ and $\mh$ giving rise to the observed scalar DM relic abundance for $m_X=1$ GeV and $10^{10}$~GeV, $\gdm=\gsm=1$ and $\Lambda=10^{16}$~GeV.
The dotted diagonal lines with $\mh = T_{\text{RH}}$ and $\mh = T_{\text{\scriptsize{max}}} = 100 ~T_{\text{RH}}$  are shown for reference. 
}
\label{Fig:mtrh}
\end{figure}

In the heavy mediator scenario, for scalar Dark Matter\footnote{Sec.~\ref{sec:appendixspin2} also includes the exact formulae for fermionic and vectorial Dark Matter. However, these differ only by a factor of order one to ten.} one can extract from Eq.~(\ref{Eq:omega}) 
the expression for $\Omega h^2$ in the term
proportional to $\beta_3$. This result (based on instantaneous reheating) should be 
multiplied by a ``boost" factor, $B_F$, to account for non-instant reheating. 
It was calculated in \cite{Garcia:2017tuj}
to be $B_F = f(n) \frac{56}{5} \ln \left( \frac{T_{\text{max}}}{T_{\text{RH}}} \right)\simeq 20$ for 
$T_{\text{max}}/T_{\text{RH}} \sim 100$ and numerically $f(6) \approx 0.4$.
We plot in Fig.~\ref{Fig:mtrh} the values of $T_{\text{RH}}$ and $\mh$ required to obtain
a relic density of $\Omega h^2 \simeq 0.1$ 
for two choices of Dark Matter masses (1 GeV and $10^{10}$ GeV)  and\footnote{When not specified, we will fix $\gdm=\gsm=1$.} $\Lambda=10^{16}$ GeV. It is important to underline that, to produce this figure, we took into account the enhancement of the production rate due to non-instantaneous reheating. Indeed, as it was shown in \cite{Garcia:2017tuj}, such a high power-law dependence on the reheating temperature implied that the majority of Dark Matter is produced at the beginning of the reheating process and the approximation of instant reheating is not valid anymore. However, this enhancement {\it does not} depend on the production process of Dark Matter but only on the ratio $T_{\text{\scriptsize{max}}}/T_{\text{RH}}$.

In the heavy mediator regime, we can verify the fact that the relic abundance is compatible with WMAP/Planck data when $m_X \simeq 10^{10}$ GeV and $\Lambda=10^{16} $ GeV, from the analytical expression (\ref{Eq:omega}). Agreement between the curve
and our analytical expression requires the boost factor of about 20.
Thus for $m_{\tilde h}=10^{17}$ GeV, we find $T_{\text{RH}} \simeq 3 \times 10^{13}$ GeV. 
Note that for this value of $m_X$,
the solid curve cuts off at $T_{\text{RH}}=10^{10}$ GeV as we must require $T_{\text{RH}} > m_X$ so that the production of the Dark Matter is kinematically allowed.    
At lower value of $\mh$ the analytical expression (\ref{Eq:omega}) would require a
significantly larger boost factor as the effect of the pole can not be neglected
and this effect is not accurately taken into account in the analytical expression.
In fact, under close examination of the solid line in Fig.~\ref{Fig:mtrh}, 
we see a change in slope at $\mh \approx 10^{16}$ GeV. At higher masses,
the effect of the pole is safely neglected and the term propotional to $\beta_3$ in Eq.~(\ref{Eq:omega}) describes the numerical result reasonably well. 

At still higher $\mh$, the curve flattens out, when the term proportional to $\alpha$ in Eq.~(\ref{Eq:omega}) dominates, corresponding to graviton exchange. 
Indeed, Eq.~(\ref{Eq:omega}) shows that when $\mh > 3000 B_F^{1/4} T_{\text{RH}}$ (for the parameters shown in the figure), graviton exchange dominates and the 
necessary reheat temperature is independent of $\mh$ and is $T_{\text{RH}} \simeq 10^{14}$ GeV as seen in Fig.~\ref{Fig:mtrh}.  This is easily understood once one notices that, even if massless, graviton exchange is highly suppressed by Planck mass couplings to the standard-model and dark sector. Note that in the case of graviton exchange there is effectively no boost factor as the rate depends on $T^8$ rather than $T^{12}$.  A large reheating temperature is needed to compensate the weakness of the coupling. Then for all masses $\mh > 7 \times 10^{17}$ GeV, graviton exchange dominates.

For $m_X = 1$ GeV (as seen by the dashed line), the heavy mediator is only important 
at extremely high values of $\mh$ as seen by the slight bend in the curve at
the upper right of the figure. This bend corresponds to the point where
the effect of the pole ceases to dominate as we previously saw for $m_X = 10^{10}$ GeV
and discussed above.

\subsection{Light mediator regime}

As we discussed above, if $m_{\tilde h}$ is lighter than the reheating temperature $T_{RH}$, there is the possibility of resonant production of the mediator $\tilde h_{\mu \nu}$ \cite{blennow_freeze-through_2014}. 
One can easily understand that once the temperature of the thermal bath $T$ dropped to
the value $T\simeq m_{\tilde h}/2$, Dark Matter production will be enhanced by the rapid s-channel cross section on resonance. 
The important parameter in this case is the width of $\tilde h$. 
Within the narrow width approximation, one can compute the rate and relic density (see the Sec.~\ref{sec:appendixspin2} for details)
which is given in Eq.~(\ref{Eq:omega}) by the term proportional to $\beta_2$. 
This expression is obviously independent of $T_{RH}$ because it corresponds to rapid Dark Matter production around $T\sim \mh$. This pole-phenomena is clearly visible in Fig.~\ref{Fig:mtrh}, represented by the vertical line (for $m_X=1$ GeV) corresponding to the value $\mh \simeq 5 \times 10^9$ in  good agreement with our analytical computation Eq.~(\ref{Eq:omega}).

For lower values of $\mh$, the pole process occurs at lower temperatures and the Dark Matter production rate is not sufficient to obtain the correct relic density. The dominant production mode becomes the exchange of the spin-2 mediator offshell. Its contribution is given in Eq.~(\ref{Eq:omega}) by the term proportional to $\beta_1$. 
In this case, the rate does not scale with the mediator mass and we
expect a specific value of $T_{\rm RH}$ necessary to obtain the correct relic density for a given Dark Matter mass. We obtain the right amount Dark Matter with mass, $m_X = 1$ GeV for $T_{\text{RH}} \simeq 6\times 10^{13}$~GeV, which corresponds to the plateau observed on the left hand side of Fig.~\ref{Fig:mtrh}.

\subsection{Non-instantaneous reheating}

\begin{figure}[h!]
\centering
\includegraphics[width=0.7\textwidth]{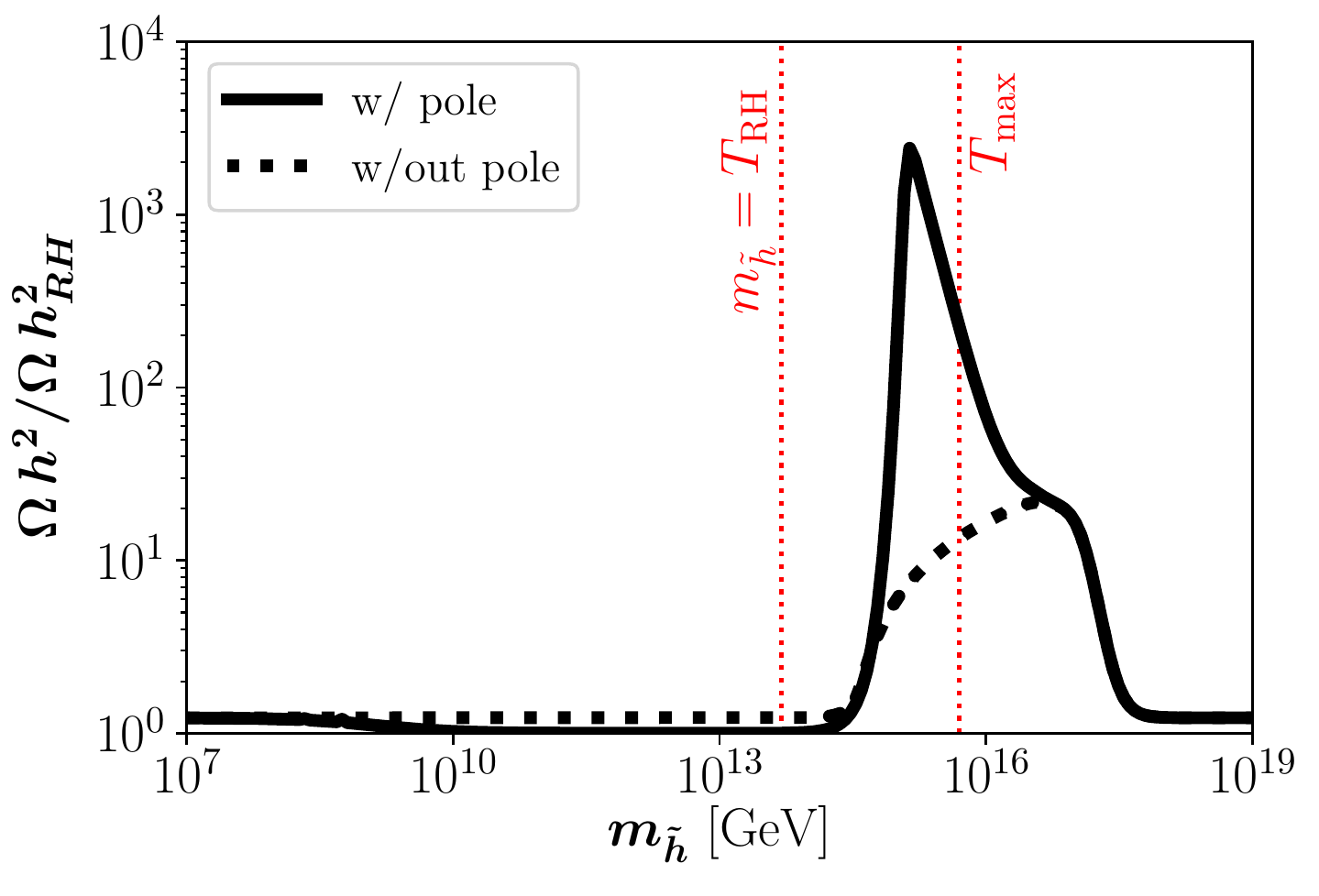}
\caption{Boost factor for scalar Dark Matter with $m_X = 10^{10}$ GeV due to non-instantaneous reheating as function of $\mh$ for $\Lambda= 10^{16}$ GeV. 
The red dotted-dashed lines correspond to $\mh = T_{\text{RH}}=5 \times 10^{13}$ GeV and $\mh = T_{\text{\scriptsize{max}}}= 100~T_{\text{RH}}$.}
\label{Fig:boost}
\end{figure}

The effects of non-instantaneous reheating on the relic abundance is shown in Fig.~\ref{Fig:boost}, where we plot the ratio of the relic abundance computed with the the exact numerical solution compared to instant-reheating approximation, $\Omega h^2 / \Omega h^2_{\text{RH}}$. 
There are several effects of non-instantaneous 
reheating, depending on the mass of the mediator $\mh$
relative to $T_{\text{RH}}$ and $T_{\text{\scriptsize{max}}}$:

\begin{itemize}
\item{If $\mh \ll T_{RH}$, the Dark Matter production process is dominated by the exchange of the (light) spin-2 mediator, $\tilde h$. Since the cross section is proportional to $T^2$ ($n=2$), the boost factor $B_F=\Omega h^2 / \Omega h^2|_{\text{RH}}$ is marginal (of the order of 1.5 in accordance with what was found in \cite{Garcia:2017tuj}). This is seen by the horizontal part of the solid line at the smallest values of $\mh$ shown. }

\item{At slightly higher $\mh$, we notice that the boost factor is unity. This is a consequence of the weak influence of the temperature on the production  rate. Indeed, at these values of $\mh$, the rate $R(T)$ is dominated by the pole-production
and  $R(T) \propto T K_1\left(  \frac{\mh}{T} \right)$. 
If we ignore the pole, then the abundance is characterized by a rate
which is proportional to $T^8$ (as it is for very large $\mh$)
and we obtain a boost factor of $\approx 1.5$ as seen by the dotted line.
 }

\item{If $T_{\text{RH}} < \mh < T_{\text{\scriptsize{max}}}$, there is a huge enhancement of the cross section depicted in Fig.~\ref{Fig:boost} due to the presence of the pole between $T_{\text{RH}}$ and $T_{\text{\scriptsize{max}}}$. This boost factor $B_F$ can even reach a few $\times 10^3$ for this value of $T_{\text{RH}}$. For lower $T_{\text{RH}}$, the boost factor can
be much higher as it scales as $\Lambda^2/T_{\text{RH}}^2$
for a fixed ratio of $\mh/T_{\text{RH}}$.}

\item{Although  $\mh \gtrsim T_{\text{max}}$, there is a large boost factor 
associated with the pole, an effect we already saw in the discussion of Fig.~\ref{Fig:mtrh}.
At higher $\mh$,
the boost factor drops from its peak due to the pole,
to the shoulder at around $\mh = 10^{17}$ GeV, where the rate is proportional to 
$T^{12}$ (corresponding to $n=6$ in the cross section due to the exchange of the 
off-shell spin-2 mediator). 
Here, the boost factor is approximately 20 as discussed above. 
The dotted line
ignores the effect of the pole and shows the smooth transition between
rates which vary as $T^8$ to $T^{12}$ to $T^8$. 
}

\item{Finally, at the largest values of $\mh$ shown in the figure,
the production rate is dominated by graviton exchange, and the rate
again varies as $T^8$. The pole can be safely ignored, and the boost 
factor is approximately 1.5 as was found for very low $\mh$.}

\end{itemize}

\section{Conclusion}
\label{sec:spin2conclusion}

We have shown that Dark Matter can naturally be produced through a spin-2 portal. We generalized our study to any massive spin-2 state with stress-energy tensor couplings to the Standard Model and the dark sector. In a large part of the parameter space, the massive spin-2 portal dominates over the (Planck-suppressed) graviton exchange. We have performed an exhaustive analysis, considering cases where the spin-2 field is both heavier and lighter than the reheating temperature. In both cases, the freeze-in process dominates the production, while enhanced during the reheating phase (heavy mediator case) or through its resonant production (light mediator case). We have also shown that our results are greatly influenced by taking into account the effects of non-instantaneous reheating. Not only do we recover boost factors in the production due to the large dependence on the temperature of the rate $R(T)$, but we have also shown that the presence of a mediator between $T_{\rm RH}$ and $T_{\rm max}$ strongly enhances the relic abundance due to rapid s-channel production when $T \simeq \mh$.

%% file: parts/conclusion.tex
In this PhD thesis we reviewed some theoretical elements required to understand how the combination of cosmological observations including (between others) studies of
the cosmic microwave background, distant supernovae, large samples of galaxy clusters,
baryon acoustic oscillation measurements and gravitational lensing has firmly established a standard cosmological
model where the Dark Matter, a new yet-to-be discovered form of matter, accounts for about 85$\%$ of the matter content of the Universe, and about $27\%$ of the global energy budget.

We reviewed the theoretical foundations of the WIMP paradigm as an attractive solution to this invisible mass issue since the Dark Matter abundance is set to the observed value by a new physics scale that is well motivated and only require one major assumption, i.e. the primordial thermal contact between the dark sector and the Standard Model bath. As a
result, concrete realizations of WIMP models had been developed in different Beyond-the-Standard-Model (BSM) frameworks, accessible to several search strategies such as direct, indirect and collider searches whose current status and prospects have been discussed.

In particular, we have studied the simplest WIMP constructions, i.e., the Higgs portal and the $Z$ portal, and we concluded that these models will be substantially ruled out, ad exception of the case of fermionic DM with only axial couplings with the $Z$ boson (e.g., Majorana DM), in absence of signals in next generation of Direct Detection experiments. The tension with direct detection constraints can be relaxed in somehow next-to minimal
scenarios for instance by introducting of a BSM scalar s-channel mediator. However, the introduction of such extra degree of freedom seems quite contrived and would require some theoretical arguments. 

The BSM framework of extended gauge symmetries is well suited to the WIMP paradigm as both the Dark Matter stability and the origin of the mediator can naturally arise in such constructions.
We studied specific models in which the connection between a vectorial Dark Matter candidate and the massive gauge field of a broken new BSM symmetry group $Z'$ is mediated by a Chern-Simons (CS) interaction, whose origin is motivated by anomaly cancellation mechanisms. In particular, we explored the possibility of connecting the $Z^\prime$ with the Standard Model via a kinetic mixing term with the hypercharge field $B$ as well as the possibility of having a second CS-like term involving the $Z^\prime$ and the hypercharge field strength tensor. We performed a complete phenomenological analysis of such models and proposed a framework for the generation of Chern-Simons coupling from a UV complete model as well as a radiative origin of the kinetic mixing.
The existing and near future experimental setups would favor a value of the CS coupling hard to explain 
with a radiative origin unless more families of the BSM fermions
are included or in the case where strong couplings are considered. The radiative origin of the kinetic mixing between the vectorial DM and the mediator could be avoided, to ensure the Dark Matter stability, by suitedly choosing the charge assignment of the heavy fermions, but would be quite contrived. The phenomenological 
analysis of these models showed the possibility to account for the Dark Matter abundance in our universe. However, a more detailed theoretical analysis based on UV completion of this framework showed that the viable parameter space compatible with the previous statement would require some tuning of the parameters of the theory.

In a different chapter, we scrutinized a flavourful model, where a fourth vector-like family is introduced and charged under an extra $U(1)^\prime$ symmetry, as a possible solution to the $R_{K^{(*)}}$ flavour anomalies. The massive $Z^\prime$ associated to this broken gauge symmetry plays the role of mediator between the Standard Model and the Dark Matter, i.e. the fourth-family singlet
Dirac neutrino. In the absence of mixing, the $Z^\prime$ is fermiophobic, having no couplings to the three chiral families, but does couple to a fourth vector-like family. The presence of Yukawa couplings between some new singlet scalars charged under the extra $U(1)^\prime$ and the four families of fermions induces mixing between generations. Such mixing effects induces $Z^\prime$ couplings to second family left-handed lepton doublets and third family left-handed quark doublets. This model can simultaneously account for the measured $B$-decay ratios $R_{K^{(*)}}$ for the observed relic abundance of Dark Matter. Facing a plethora of constraints from flavour physics, collider and Dark Matter searches, this analysis hints at a particular viable corner of the parameter space where the mediator has a mass of the order of 500 GeV. The allowed window can be further squeezed by better precision measurements of
the trident $\nu_\mu N \rightarrow \mu^+ \mu^- \nu_\mu N$ process, and by improving the theoretical precision of the
SM prediction for the $B_s$ meson mass difference. The preferred range of Dark Matter masses and couplings could be probed by the future generation of direct detection experiments.

Alternative thermal Dark Matter production mechanisms have been discussed in the third part of this thesis, such as the Strongly Interactive Massive Particles (SIMP) and ELastic Decoupling Relic (ELDER) mechanisms. In these frameworks, the role of interactions inside the dark sector as well as the effect of an early decoupling between the Dark Matter and the SM bath imply a different phenomenology. In particular, such frameworks provide the possibility of having a sizable Dark Matter self-interaction cross section that could alleviate tensions between observations and N-body simulations based on $\Lambda$CDM. In this thesis we considered an explicit realization of the SIMP mechanism in the form of vector SIMPs arising from an $SU(2)_X$ hidden non-abelian gauge theory, where the accidental custodial symmetry protects the stability of the Dark Matter. We propose several ways of equilibrating the dark and visible sectors in this setup. In particular, we showed that a light dark Higgs portal can maintain thermal equilibrium between the two sectors, as can a massive dark vector portal kinetically mixing the hyperchage, with its generalized Chern-Simons couplings to the vector SIMPs, all while remaining consistent with experimental constraints.

In the last chapter we discussed the possibility of producing the Dark Matter non-thermally from annihilation or decay of particles in thermal contact with the Standard Model bath. In particular, we studied the case where a heavy spin-2 messenger can efficiently play the role of a portal between the dark and Standard Model sectors. In a large part of the parameter space compatible by the requirement of generating the correct relic density, production through the exchange of a massive spin-2 mediator dominates over processes involving a graviton with Planck suppressed couplings. We studied the impact of the reheating stage of the universe on the Dark Matter production, and we showed that our results are greatly influenced by taking into account the effects
of non-instantaneous reheating. Not only do we recover
boost factors in the production due to the large dependence
on the temperature of the production rate, but we have
also shown that the presence of a mediator with a mass of the order of the reheating temperature might significantly enhance the production due to resonance effects of the massive spin-2 state. 

The work presented in this thesis aimed at tackling one of the open questions of modern physics regarding our understanding of the fundamental laws of nature. The precise nature of the Dark Matter still remains unknown to this day, but it is clear that most of the WIMP models will be scrutinized in the next decades. This thesis highlighted the paramount role of the next generation of experiments and its complementarity with theoretical considerations as a way to solve the disturbing puzzle of the presence of Dark Matter in our universe.

%% file: parts/appendices/boltzmann.tex
\vspace{0.3cm}

\noindent
In this chapter we provide some details regarding the derivation of the Boltzmann equation expressing the time evolution of the Dark Matter density and its resolution in the specific cases considered in this thesis.

\section{Derivation of the Boltzmann equation : the WIMP case}
\label{sec:Boltzmann}
In this section we derive the Boltzmann equation, relating the time evolution of the number density of some species in the universe to its interaction processes. The Liouville operator $\hat{\mathbf{L}}$ represents the time evolution of the phase space distribution function $f(x^\mu,p^\mu)$.
In the context of general relativity, the Liouville operator can be written as:
\begin{equation}
\hat{\mathbf{L}}[f] =p^{\mu}\frac{\partial f}{\partial x^{\mu}}-
\Gamma^{\mu}_{\alpha \beta}p^{\alpha}p^{\beta}\frac{\partial f}{\partial p^{\mu}}~.
\label{eq:Liouville}
\end{equation}
In the specific case of the FLRW metric, according to the homogeneity and isotropy hypotheses, the distribution function does not depend on space-time coordinates and depends only on the momentum modulus $f(\vec{p},\vec{r},t)=f(|\vec{p}|,t)$. Thus the Liouville operator then takes the form simplified form:
\begin{equation}
\hat{\mathbf{L}}[f] =E\frac{\partial f}{\partial t}-\frac{\dot{a}}{a}
p^{2}\frac{\partial f}{\partial E}~.
\end{equation}
The number $n(t)$ density can be defined as:
\begin{equation}
n(t)\equiv \dfrac{g}{(2\pi)^3}\int f(p,t) \diff ^3p~,
\end{equation}
where $g$ is the number of degrees of freedom of the species considered. Integrating of the momentum gives:
\begin{equation}
\frac{g}{(2\pi)^3}\int \diff^3 p \frac{\mathbf{L}[f]}{E}=\dot{n}-H\frac{g}{(2\pi)^3}\int  \diff^3 p \frac{p^2}{E}\frac{\partial f}{\partial E}=\dot{n}+3Hn=\frac{1}{a^3}\frac{\diff (na^3)}{\diff t}~.
\end{equation}
In a non-interacting universe, the momentum integral of the Liouville operator would vanish 
\begin{equation}
\frac{1}{a^3}\frac{\diff (na^3)}{\diff t}=0~,
\end{equation}
expressing the conservation of the number of particles per comoving volume. In the case of number changing microscopic processes such as scattering or annihilations, the evolution of the phase space distribution is ruled by the Boltzmann equation
\begin{equation}
\hat{\mathbf{L}}[f] \ = \ \hat{\mathbf{C}}[f]~,
\label{eq:LiouvilleequalsCollision}
\end{equation}
where $\hat{\mathbf{C}}[f]$ is the collision operator. Consider the microscopic process:
\begin{equation}
1+2+\ldots \leftrightarrow a + b + \ldots~,
\end{equation}
For this specific process, the collision operator can be defined as:
\begin{equation}
\begin{split}
\hat{\mathbf{C}}[f_1]\equiv & -\dfrac{1}{2}\int \d\Pi_2 \ldots \d \Pi_a \d\Pi_b(2\pi)^4\delta^{(4)}(p_1 + p_2 + \ldots - p_a - p_b -\ldots)\\
& \times \Big[ \overline{|\mathcal{M}|^2}_{1+2+\ldots \leftrightarrow a + b + \ldots} f_1 f_2 \ldots (1\pm f_a) (1\pm f_b) ... \\ & \hspace{0.5cm}- \overline{|\mathcal{M}|^2}_{a+b+\ldots \leftrightarrow 1 + 2 + \ldots} f_a f_b \ldots (1\pm f_1) (1\pm f_2)\ldots \Big]
\end{split}
\end{equation}
where $+,-$ respectively apply to bosons and fermions. The amplitude squarred $\overline{|\mathcal{M}|^2}$ is averaged over initial and final spin states and divided by symmetry factors accounting for identical initial or final states. We introduced the Lorentz invariant quantity \d \Pi:
\begin{equation}
\diff \Pi_i \equiv g_i \dfrac{1}{(2\pi)^3}\dfrac{ \diff ^3p_i}{2E_i}~.
\end{equation}
Integrating over the momentum of $1$ gives:
\begin{equation}
\begin{split}
\dot{n}_1+3Hn_1= & -\int \d\Pi_1 \d\Pi_2 \ldots \d \Pi_a \d\Pi_b(2\pi)^4\delta^{(4)}(p_1 + p_2 + \ldots - p_a - p_b -\ldots)\\
& \times \Big[ \overline{|\mathcal{M}|^2}_{1+2+\ldots \leftrightarrow a + b + \ldots} f_1 f_2 \ldots (1\pm f_a) (1\pm f_b) ... \\  & \hspace{0.5cm} - \overline{|\mathcal{M}|^2}_{a+b+\ldots \leftrightarrow 1 + 2 + \ldots} f_a f_b \ldots (1\pm f_1) (1\pm f_2)\ldots \Big]
\end{split}
\end{equation}
In order to consider all the possible collision terms that are affecting the phase space distribution $f_1$, one has to sum on the right-hand side, over all the possible processes involving the particle $1$, including processed involving a larger number of particles. In the following, we will focus on how to rewrite the collision operator in a more convenient way, for few specific cases.

\subsection{The Liouville operator in the radiation era}

In the radiation era entropy is conserved and the expansion rate is dominated by radiation components, therefore entropy density and $H$ can be expressed as a function of the temperature

\begin{equation}
s(T)=g_{\star,s} \dfrac{2\pi^2}{45}T^3 \qquad \text{and} \qquad H(T)= \Big( \dfrac{g_\star \pi^2}{90} \Big)^{1/2} \dfrac{T^2}{M_{\text{Pl}}},
\end{equation}
Where $M_{\text{Pl}}$ is the reduced Planck mass.

We define the yield $Y_i=n_i/s$ as a quantity proportional to the number of particle when entropy is conserved, allowing to write the left-hand side of the Boltzmann equation as:

\begin{equation}
\dot{n}_1+3Hn_1= - H(T)s(T)T\dfrac{\d Y_1}{\d T} \left(\dfrac{1}{3}\dfrac{\diff \log g_{\star,s}}{\diff \log T}+1\right)^{-1}~.
\end{equation}
However, the term on the RHS corresponding to the temperature variation of the effective numbers of relativistic species can sometimes be neglected. In this case the Boltzmann equation has the simple form:

\begin{equation}
\dot{n}_1+3Hn_1= - H(T)s(T)T\dfrac{\d Y_1}{\d T} ~.
\end{equation}

\subsection{Kinetic equilibrium and Maxwell-Boltzmann statistics}
Considering of the process $ i + j \leftrightarrow a + b$, one can write the right-hand side of the Boltzmann equation as a function of a physical quantity that can experimentally be measured and theoretically be computed assuming a specific microscopic model. In order to remain general we write every introduced quantity in a Lorentz invariant way, hence define the Lorentz invariant cross section $\sigma_{ij}$\footnote{As $\overline{|\mathcal{M}|^2}$ includes symmetry and spin averaging factors for both initial and final states, this definition would have to be rescaled accordingly to match the usual definition of a cross section in particle physics. This defintion differs to the particle physics standard definition where Lorentz invariance is satisfied but only when a Lorentz boost is considered along the beam axis.} as:

\begin{equation}
\sigma_{ij}\equiv \frac{1}{2E_i2E_j(1-\vec{\beta}_i\cdot \vec{\beta_j})v_{ij}}\int \d\Pi_a \d\Pi_b (2\pi)^4\delta^{(4)}(p_i+p_j-p_a-p_b) \overline{|\mathcal{M}_{i+j\rightarrow a + b}|^2} ,
\end{equation}
where $\vec{\beta}_i \equiv v_i/c$ and we introduced the Lorentz invariant relative velocity :
\begin{equation}
v_{ij}\equiv \dfrac{ \sqrt{(p_i\cdot p_j)^2-m_i^2m_j^2}}{p_i\cdot p_j}~,
\end{equation}
Considering the relation
\begin{equation}
4(p_i\cdot p_j)v_{ij}=4E_i E_j(1-\vec{\beta}_i\cdot \vec{\beta}_j)v_{ij}~,
\end{equation}
assuming thermal distributions and neglecting the Pauli blocking and stimulation factors\footnote{equivalent to consider Maxwell-Boltzmann distributions} $1\pm f_i \simeq 1$, we can write the right-hand side of the Boltzmann equation in a more convenient way:
\begin{align*}
\dfrac{\d n_i}{\d t}+3Hn_i=& - \int \d \Pi_i f_i(p_i) \int \d \Pi_j f_j(p_j) 4(p_i \cdot p_j) \sigma_{ij \to ab} v_{ij} \\ & \hspace{0.5cm}+ \int \d\Pi_a f_a(p_a) \int \d\Pi_b f_b(p_b) 4(p_a \cdot p_b) \sigma_{ab \to ij} v_{ab} \\ = &   - n_a n_b \la \sigma_{ab \to ij} v_{ab} \ra + n_i n_j \la \sigma_{ij \to ab} v_{ij} \ra~,
\end{align*}
we used the \textit{velocity averaged cross section} defined as follow:
\begin{equation}
\la \sigma_{ij \to ab} v_{ij} \ra \equiv \frac{1}{n_in_j} \int \d \Pi_i f_i(p_i) \int \d \Pi_j f_j(p_j) 4(p_i \cdot p_j) \sigma_{ij \to ab} v_{ij} ~.
\end{equation}
Alternatively, it is common to find definition of $\la \sigma_{ij \to ab} v_{ij} \ra$ involving the \textit{M\o ller velocity} $v_{\text{M\o l}}$, more adapted to the non-relativistic regime that can be defined as
\begin{equation}
v_{\text{M\o l}} \equiv (1-\vec{\beta}_i\cdot \vec{\beta}_j)v_{ij}  ~.
\end{equation}
In the following we will assume, unless stated explicitely, CP invariance such that $\overline{|\mathcal{M}_{i+j\rightarrow a + b}|^2} =\overline{|\mathcal{M}_{a+b\rightarrow i + j}|^2} $.
Assuming that species $a,b$ are in kinetic equilibrium allows to write $f_{a,b} \propto f_{a,b}^{\text{eq}}=e^{-E_{a,b}/T}$. In this case, using energy conservation during the $ i + j \leftrightarrow a + b$ process we have 
\begin{equation}
f_a f_b=f_a^{\text{eq}} f_b^{\text{eq}}=e^{-E_a/T }e^{-E_b/T }=e^{-E_i/T }e^{-E_j/T }=f_i^{\text{eq}} f_j^{\text{eq}} ~.
\end{equation}
Rewriting the product $f_i f_j=\Big( \dfrac{f_i^{\text{eq}} f_j^{\text{eq}}}{n_i^{\text{eq}} n_j^{\text{eq}}} \Big) n_i n_j$ we derive a compact formulation of the Boltzmann equation:

\begin{align}
\dfrac{\d n_i}{\d t}+3Hn_i= -\la \sigma_{ij \to ab} v_{ij} \ra(n_i n_j - n^{\text{eq}}_i n^{\text{eq}}_j) ~.
\end{align}
It is important to remind the few assumptions implied in the derivation of this expression :
\begin{itemize}
\item We assumed CP invariance: using the previous derivation in CP violating processes would not be valid and one would have to start from the definition of the collision operator when considering such cases, relevant in the context of Baryogenesis for instance.
\item $a$ and $b$ are always in kinetic equilibrium throughout the freeze-out process: This condition can be estimated using simple estimate or more exact treatment as described further on.
\item Maxwell-Boltzmann distributions: we assumed that Pauli blocking or enhancement factor can be neglected, this is generally a reasonable assumption in the non-relativistic regime but these factors can play a role when relativistic decoupling is considered for instance.
\item Chemical potentials can be neglected: this condition is not always satisfied, one has to verify that the rate of number changing processes is sufficiently high to ensure chemical equilibrium. 
\item Reference frame: we neglected the fact that temperature is defined in a cosmic comoving frame while usually computation of cross sections are performed in the centre-of-mass frame, one would need to define a boost factor from one frame to another to express the apparent "redshift" of the temperature in the centre-of-mass frame. In this case factorization of the several terms in the collision operator is not trivial and would require a specific treatment. However, in the non-relativistic treatment, as a first approximation, it is a reasonable approximation to neglect this effect.
\end{itemize}

\subsection{Computation of $\la \sigma v \ra$}

\subsubsection{Non-relativistic expansion}
In the WIMP case, an important point is that the DM is still in kinetic equilibrium while the decoupling occurs, allowing to trade the distribution $f_i/n_i$ by $f_i^{\text{eq}}/n_i^{\text{eq}}$ in $\la \sigma v \ra$:

\begin{equation}
\la \sigma v \ra_{ij} = \iint \frac{  \d \Pi_i f^{\text{eq}}_i(p_i)  \d \Pi_j f^{\text{eq}}_j(p_j) 4(p_i \cdot p_j) \sigma_{ij} v_{ij}}{n^{\text{eq}}_i n^{\text{eq}}_j} ~.
\end{equation}
In order to have an analytical approximation of this quantity, one can perform an expansion, in the non-relativistic limit of $ \sigma v $, as a power series of the DM velocity squared:

\begin{equation}
\sigma v \simeq (\sigma v)_0 + (\sigma v)_2 v^2 +(\sigma v)_4 v^4 + \ldots ~.
\end{equation}
Hence $\la \sigma v \ra$ can be written in the non relativistic limit by considering a Maxwellian velocity distribution as:

\begin{equation}
\la \sigma v \ra =  \frac{ \iint  \sigma v e^{-(p_1^2/2m_\chi) T}  e^{-(p_2^2/2m_\chi) T} \diff^3 p_1  \diff^3 p_2 }{\iint e^{-(p_1^2/2m_\chi) T}  e^{-(p_2^2/2m_\chi) T} \diff^3 p_1  \diff^3 p_2}=\frac{x^{3/2}}{2\sqrt{\pi}}\int_0^\infty (\sigma v)v^2e^{-xv^2/2}\diff v ~,
\end{equation}
where $1$ and $2$ denote the DM particles. Injecting the expansion of $\sigma v$ in the previous expression gives :

\begin{equation}
\la \sigma v \ra \simeq (\sigma v)_0 + \dfrac{3}{2x}(\sigma v)_2 + \dfrac{15}{8x^2}(\sigma v)_4 + \ldots ~.
\end{equation}
Even though this result relies on several approximations, as long as the DM is in the non-relativistic regime (i.e. $x \gg 1$) the first term should give a rather reliable analytical approximation. Notice that this expansion is not valid around pole regions or near thresholds, in this cases one has to compute numerically the quantity $\la \sigma v \ra$.

\subsubsection{Near a resonance}

In the case where the DM can annihilate into SM particles through s-channel exchange of some heavy field, and if this field is unstable, the cross section expressed as a function of the center-of-mass energy follows a Breit-Wigner distribution and presents a resonance when the heavy mediator is produced on shell. In this case the usual velocity expansion of $\la \sigma v \ra$ is not longer valid and must be treated differently~\cite{Griest1991c,Gondolo:1990dk}. In particular, it can be computed analytically in the Narrow Width Approximation (NWA), assuming that $(\sigma v)$ can be expressed as:

\begin{equation}
(\sigma v)=\sum_{l=0}^\infty \dfrac{b_l}{l!}\eta^l \dfrac{\gamma_R^2}{(\epsilon_R-\eta)^2+\gamma_R^2}~,
\end{equation}
where $\eta \equiv \dfrac{v^2}{4}$, $\epsilon_R\equiv\dfrac{m_R^2-4m_\chi^2}{4m_\chi^2}$ and $\gamma_R$ is the width of the resonance normalized by its mass. $\la \sigma v \ra $ can then be expressed in the following way:
\begin{align}
\la \sigma v \ra =&\frac{x^{3/2}}{2\sqrt{\pi}}\int_0^\infty \sum_{l=0}^\infty \dfrac{b_l}{l!}\eta^l \dfrac{\gamma_R^2}{(\epsilon_R-\eta)^2+\gamma_R^2}v^2e^{-xv^2/2}\d v \nonumber \\
=& \frac{2x^{3/2}}{\sqrt{\pi}}\int_0^\infty \sum_{l=0}^\infty \dfrac{b_l}{l!}\eta^l \dfrac{\gamma_R^2}{(\epsilon_R-\eta)^2+\gamma_R^2}\eta^{1/2}e^{-x\eta}\d \eta ~.
\end{align}
In the NWA limit $\gamma_R \ll 1$ the relation:

\begin{equation}
\lim_{\gamma_R \rightarrow 0}\dfrac{\gamma_R^2}{(\epsilon_R-\eta)^2+\gamma_R^2}=\pi \gamma_R \delta(\epsilon_R-\eta)~,
\end{equation}
can be injected in the previous integral, giving:

\begin{align}
\la \sigma v \ra =2x^{3/2}\pi^{1/2} \epsilon_R^{1/2} e^{-x\epsilon_R} \sum_{l=0}^\infty \dfrac{b_l}{l!}\epsilon_R^l~.
\end{align}
This expression depends mostly on the parameter $\epsilon_R$ why quantify the mass difference between the DM and the mediator, where thermal effects can have sizable effect on $\la \sigma v \ra$.

\subsubsection{General case}
In the relativitic case, near threshold and around poles, the velocity expansion of $\la \sigma v \ra$ is not valid and one has to evaluate the full numerical integration which can be expressed, by the change of variable $x_i=m_i/T$ as a function of a single integral:

\begin{align}
\la \sigma v \ra_{ij} = \dfrac{1}{8T\Pi_i m_i^2 K_2(x_i)} \int_{(m_i+m_j)^2}^\infty \sigma _{ij}\dfrac{\lambda(s,m_i,m_j)}{\sqrt{s}}K_1(\sqrt{s}/T)\d s~,
\end{align}
where $\lambda(s,m_i,m_j)=[s-(m_i+m_j)^2][s-(m_i-m_j)^2]^2$. This expression can be simplified in the case of the identical initial states~\cite{Gondolo:1990dk} by defining a dimensionless variable $z=\sqrt{s}/T$ :
\begin{align}
\la \sigma v \ra = \frac{1}{4x^4K^2_2(x)} \int_{2x}^\infty \sigma (z^2-4x^2)z^2K_1(z) \diff z ~.
\label{eq:sigmavfullintegral}
\end{align}

\section{Forbidden channels}

\subsection{General treatment}
When computing the Boltzmann equation for the DM density, one has to take into account all the DM number changing physical processes. In the case where a DM candidate $\chi$ is slightly lighter than some other particle, at zero temperature the DM annihilation in this channel would be kinematically forbidden but if the difference between their masses is small enough, the DM particles with velocities corresponding to the most energetic part of the distribution could have sufficient kinetic energy to produce this other particle on-shell~\cite{Griest1991c}. This can be the case for instance in the context of SSB where the higgs $h$ generating the DM mass can be just slightly more massive. If the higgs is thermalized with the SM for instance,  the forbidden channel sector of the Boltzmann equation for the DM density would read :

\begin{equation}
\dfrac{\d n_\chi}{\d t}+3Hn_\chi\simeq -2n_\chi^2\la \sigma v \ra_{\chi \chi \rightarrow h h}+2(n^{\text{eq}}_h)^2\la \sigma v \ra_{h h \rightarrow \chi \chi}~.
\end{equation}
When the right-hand side of this equation vanishes, the DM evolves only according to the Hubble expansion rate and the DM would be in thermal equilibrium with the SM bath, giving the following balance relation:

\begin{align}
\la \sigma v\ra_{\chi \chi \rightarrow h h}=&\dfrac{(n^{\text{eq}}_h)^2}{(n^{\text{eq}}_\chi)^2}\la \sigma v\ra_{hh \rightarrow \chi \chi} \nonumber\\
=&(1+\Delta)^{3}e^{-2\Delta x}\la \sigma v\ra_{hh \rightarrow \chi \chi}~,
\end{align}
in the case of non-relativistic collisions and assuming the same number of degrees of freedom for the Higgs and the DM, where we introduced the variable $\Delta=(m_h-m_\chi)/m_\chi$. This relation can be injected in the Boltzmann equation:

\begin{equation}
\dfrac{\d Y_\chi}{\d x}  = \Big(2(Y_h^{\text{eq}})^2 - 2Y_\chi^2(1+\Delta)^{3}e^{-2\Delta x} \Big)  \la \sigma v\ra_{hh \rightarrow \chi \chi} \dfrac{s}{Hx}~.
\end{equation}
The Higgs density should be suppressed by a exponential Boltzmann factor and then can be neglected. Assuming a s-wave contribution of $\la \sigma v \ra$ for simplicity and integrating over $x$ gives:

\begin{align}
Y_{\chi,\infty}^{-1} \simeq 2 \lambda (1+\Delta)^3\la \sigma v \ra_{hh \rightarrow \chi \chi} \int_{x_{\text{F}}}^\infty \d x x^{-2} e^{-2\Delta x}~.
\end{align}
Where we defined $\lambda \equiv xs /H$.
The relic density at the present time becomes:

\begin{align}
\Omega_\chi h^2=\frac{M_\chi s_0 Y_{\chi,\infty} h^2}{\rho_c^0}\approx \frac{M_\chi s_0  h^2}{\rho_c^0} 2 \lambda (1+\Delta)^3 x_{\rm F} e^{2\Delta x_{\rm F}}\la \sigma v \ra_{hh \rightarrow \chi \chi} \Big( 1-2\Delta x_{\text{F}} \Gamma (0,2\Delta x_{\text{F}}) \Big)~.
\end{align}
Where $\Gamma$ is the Euler function. Notice that this expression is extremely sensitive to the mass splitting parameter $\Delta$ because of the exponential Boltzmann factor and thus can significantly enhance the DM annihilation and, as a result, the relic density.

\subsection{Contributions in the VSIMP case}
\label{sec:VSIMPforbidden}
Based on the discussion in Sec.~\ref{ssec:forb}, the $2\rightarrow 2$ forbidden (semi-)annihilation cross sections (with notations, $h_1 h_1\rightarrow  ii$ meaning that $ h_1 h_1 \rightarrow X_i X_i$ and $i h_1\rightarrow jk$ meaning that $X_i h_1\rightarrow X_j X_k$) are also given by
\bea
\langle \sigma v\rangle_{h_1 h_1 \rightarrow ii} &=&\frac{m_{h_1}^2}{512\pi m_X^4}\sqrt{1-\frac{m_X^2}{m_{h_1}^2}}\Big[64\lambda_\phi^2\frac{m_X^4}{m_{h_1}^4}\Big(4-4\frac{m_X^2}{m_{h_1}^2}+3\frac{m_X^4}{m_{h_1}^4}\Big) \nonumber \\
&&-16g_X^2\lambda_\phi\frac{m_X^2}{m_{h_1}^2}\Big(4+8\frac{m_X^2}{m_{h_1}^2}-15\frac{m_X^4}{m_{h_1}^4}+12\frac{m_X^6}{m_{h_1}^6}\Big)  \nonumber \\
&&+g_X^4\Big(4+20\frac{m_X^2}{m_{h_1}^2}+11\frac{m_X^4}{m_{h_1}^4}-56\frac{m_X^6}{m_{h_1}^6}+48\frac{m_X^8}{m_{h_1}^8}\Big)\Big]~,
\eea
\begin{align}
\langle\sigma v\rangle_{i h_1\rightarrow jk} =&\frac{g_X^4m_{h_1}^3}{384\pi m_X^5}\Big(1+3\frac{m_X}{m_{h_1}}\Big)^{3/2}\Big(1-\frac{m_X}{m_{h_1}}\Big)^{3/2}\Big(1+\frac{m_X}{m_{h_1}}\Big)^{-1}\Big(1+2\frac{m_X}{m_{h_1}}\Big)^{-2} \nonumber \\
&\times \Big(1+4\frac{m_X}{m_{h_1}}-4\frac{m_X^2}{m_{h_1}^2}-10\frac{m_X^3}{m_{h_1}^3}+144\frac{m_X^4}{m_{h_1}^4}+396\frac{m_X^5}{m_{h_1}^5}+297\frac{m_X^6}{m_{h_1}^6}\Big)~,
\end{align}
with $i\neq j\neq k$ in the latter case.

\section{Strongly Interacting Massive Particle}
\label{sec:SIMPappendix}
\subsection{General treatment}
Assuming that the DM density evolution is triggered by a $ 1 + 2 + 3 \rightarrow 4 + 5$ process, we can write the Boltzmann equation as follow:
\begin{equation}
\dfrac{\diff n_1}{\diff t}+3Hn_1=-\int \d \Pi_1 \d \Pi_2 \d \Pi_3 \d \Pi_4 \d \Pi_5 (2\pi)^4 \delta^4(p) \overline{|\mathcal{M}|^2}(f_1 f_2 f_3 - f_4 f_5)~.
\end{equation}
where $p=p_1+p_2+p_3-p_4-p_5$. By defining $\la \sigma v^2 \ra $ in the following way:
\begin{equation}
\la \sigma v^2 \ra \equiv \dfrac{1}{n^{\text{eq}}_1 n^{\text{eq}}_2 n^{\text{eq}}_3} \int \d \Pi_1 \d \Pi_2 \d \Pi_3 \d \Pi_4 \d \Pi_5  (2\pi)^4 \delta^4(p) \overline{|\mathcal{M}|^2} f^{\text{eq}}_1 f^{\text{eq}}_2 f^{\text{eq}}_3~,
\end{equation}
then we can write the Boltzmann equation as:
\begin{equation}
\dfrac{\diff n_1}{\diff t}+3Hn_1= - \la \sigma v^2 \ra(n_1 n_2 n_3 - \dfrac{n_4 n_5}{n_4^{\text{eq}} n_5^{\text{eq}}}n_1^{\text{eq}} n_2^{\text{eq}}n_3^{\text{eq}})~.
\end{equation}
The quantity $\langle \sigma v^2 \rangle $ can be understood as the generalization of the notion of cross section for a collision process 
involving three particles in the initial state. In the case where all the particles have the same mass $m_\chi$, in the non-relativistic regime , $\langle \sigma v^2 \rangle $ can be related to the amplitude squared as:
\begin{equation}
\la \sigma v^2 \ra = \dfrac{1}{n^{\text{eq}}_1 n^{\text{eq}}_2 n^{\text{eq}}_3} \int \dfrac{\diff^3 p_1}{(2 \pi)^3} \dfrac{\diff^3 p_2}{(2 \pi)^3} \dfrac{\diff^3 p_3}{(2 \pi)^3}  f^{\text{eq}}_1 f^{\text{eq}}_2 f^{\text{eq}}_3 (\sigma v^2)_{3 \rightarrow 2}~,
\end{equation}
with
\begin{equation}
(\sigma v^2)_{3 \rightarrow 2} \equiv \dfrac{1}{2E_12E_22E_3}\int \dfrac{\diff^3 p_4}{(2\pi)^3 2 E_4}\dfrac{\diff^3 p_5}{(2\pi)^3 2E_5}(2\pi)^4\delta^{4}(p_1+p_2+p_3-p_4-p_5) \overline{|\mathcal{M}|^2}~.
\end{equation}
In the non-relativistic limit, energy conservation gives $3m_\chi=2\sqrt{m_\chi^2+p^2}$ with $p$ the external momentum of an outgoing particle. The quantity $(\sigma v^2)_{3 \rightarrow 2}$ can be simplified if the matrix element $\mathcal{M}$ does not depend on the direction of the outgoing particles as:
\begin{align}
(\sigma v^2)_{3 \rightarrow 2}= \dfrac{\overline{|\mathcal{M}|^2}}{64 \pi m_\chi^3} \int \dfrac{p \diff p}{\sqrt{m_\chi^2+p^2}} \delta(p-\dfrac{\sqrt{5}m_\chi}{2}) =\dfrac{\sqrt{5}}{192 \pi m_\chi^3}\overline{|\mathcal{M}|^2}~.
\end{align}
Therefore in the specific regime mentioned above, the following relation holds:
\begin{equation}
\la \sigma v^2 \ra=\dfrac{\sqrt{5}}{192 \pi m_\chi^3}\overline{|\mathcal{M}|^2}~.
\end{equation}
\subsection{Solution of the Boltzmann equation}
Assuming a $3\rightarrow2$ process involving identical particles, one can write the relevant Boltzmann equation as:
\begin{equation}
\dfrac{\diff n_\chi}{\diff t}+3Hn_\chi=-\la \sigma v^2 \ra (n_\chi^3-n_\chi^{\text{eq}}n_\chi^2)~.
\end{equation}
A non-relatisitic freeze-out, similar to the WIMP case, occurs as the second term on the RHS becomes exponentially suppressed after the freeze-out time $x_{\rm F}$, where $x\equiv m_\chi/T$. One can derive an analytical approximation of the relic density by integrating the Boltzmann equation, which can be written as:
\begin{equation}
Hsx\dfrac{\diff Y_\chi}{\diff x}\simeq-\la \sigma v^2 \ra Y_\chi^3 s^3~,
\end{equation}
with $Y_\chi \equiv n_\chi/s$. This equation can be integrated straightforwardly, giving a relic density:
\begin{equation}
\Omega_\chi h^2=\dfrac{m_\chi s_0 h^2}{\rho_c^0}\dfrac{\sqrt{2 H(m_\chi)}}{s(m_\chi)}x_{\rm F}^2 \la \sigma v^2 \ra^{-1/2}~,
\label{eq:omegahsqSIMP}
\end{equation}
where the freeze-out time $x_{\rm F}$ can be estimated by comparing the interaction rate and the Hubble expansion rate, which gives:
\begin{equation}
H(x_{\rm F})n(x_{\rm F})\sim \la \sigma v^2 \ra n^3(x_{\rm F})~,
\end{equation}
for which $x_{\rm F}\sim20$ is a solution, as in the WIMP case. Plugging this value is Eq.~\ref{eq:omegahsqSIMP} provides an estimation of the relic abundance generated by the SIMP mechanism.

\subsection{Kinetic equilibrium condition}
In order to check if kinetic equilibrium is maintained between the dark sector and SM thermal baths, we need to understand how energy is transferred between the two baths. In order to do so, consider the Boltzmann equation~\ref{eq:LiouvilleequalsCollision} associated to the evolution of a DM candidate $\chi$ with a phase space distribution $f$. Integrating over all possible momenta after plugging~\ref{eq:Liouville} gives:
\begin{align}
\frac{g}{(2\pi)^3}\int \diff^3 p ~\mathbf{L}[f]=&\frac{g}{(2\pi)^3}\int  \Big(  E \dfrac{\partial f}{\partial t}- Hp^2 \dfrac{\partial f}{\partial E} \Big)  \diff^3 p ~, \nonumber \\
=&\frac{g}{2\pi^2}  \int  \Big( Ep^2 \dfrac{\partial f}{\partial t}+ H \dfrac{\partial (p^3 E)}{\partial p} f(p) \Big)  \diff p ~,
\end{align}
where the second term has been obtained after integrating by parts. Using the expressions of the energy density and pressure as defined in Sec.~\ref{sec:entropyandenergydensity}, the LHS of the Boltzmann equation reads:
\begin{equation}
\frac{g}{(2\pi)^3}\int \diff^3 p ~\mathbf{L}[f]=\dfrac{\diff \rho_\chi}{\diff t}+3H \Big( \rho_\chi + P_\chi \Big)~.
\end{equation}
The RHS of the Boltzmann equation corresponds to the integral of the collision operator. In order to evaluate this term, one would have to sum over all possible processes that can affect the DM phase space distribution. Assuming a scenario where the DM could scatter with SM particles $\psi$ through processes such as $\chi_1 + \psi_1 \rightarrow \chi_2 + \psi_2$, the RHS of the Boltzmann equation would read:
\begin{align}
\frac{g}{(2\pi)^3}\int \diff^3 p ~\mathbf{C}[f]=\int & \diff \Pi_{\chi_1} \diff \Pi_{\psi_1} \diff \Pi_{\chi_2} \diff \Pi_{\psi_2} (2 \pi)^4 \delta^{(4)} (p_{\chi_1}+p_{\psi_1}-p_{\chi_2}-p_{\psi_2}) ~,\nonumber \\
& \times (E_{\chi_2}-E_{\chi_1})f_{\chi_1} f_{\chi_2} \overline{|\mathcal{M}_{\chi_1+\psi_1 \rightarrow \chi_2+\psi_2}|^2} ~,
\end{align}
The integral of the collision operator can be understood as the typical energy transfer between $\chi$ and $\psi$ per unit of time. Therefore we define the quantity $\la \sigma_{\rm el} v \cdot \delta E \ra$ as:
\begin{align}
n_\chi n_\psi \la \sigma_{\rm el} v \cdot \delta E \ra \equiv \int & \diff \Pi_{\chi_1} \diff \Pi_{\psi_1} \diff \Pi_{\chi_2} \diff \Pi_{\psi_2} (2 \pi)^4 \delta^{(4)} (p_{\chi_1}+p_{\psi_1}-p_{\chi_2}-p_{\psi_2}) ~,\nonumber \\
& \times  (E_{\chi_1}-E_{\chi_2})f_{\chi_1} f_{\chi_2} \overline{|\mathcal{M}_{\chi_1+\psi_1 \rightarrow \chi_2+\psi_2}|^2} ~,
\end{align}
and the Boltzmann equation takes the simple from:
\begin{equation}
\dfrac{\diff \rho_\chi}{\diff t}+3H \Big( \rho_\chi + P_\chi \Big)=-n_\chi n_\psi \la \sigma_{\rm el} v \cdot \delta E \ra ~.
\label{eq:Botlzmannenergydensity}
\end{equation}
From this equation we see that in absence of interaction (i.e. $\la \sigma_{\rm el} v \cdot \delta E \ra =0$), if the DM particles are non-relativistic, the pressure can be neglected and this implies:
\begin{equation}
\dfrac{\diff \rho_\chi}{\diff t}+3H \rho_\chi =0~,
\end{equation}
which allows to recover the DM energy density evolution with the scale factor: $\Omega_\chi \propto \Omega_\chi^0 a^{-3}$. The quantity $\la \sigma_{\rm el} v \cdot \delta E \ra$ can be related to the momentum relaxation rate $\gamma(T)$ as:
\begin{equation}
\la \sigma_{\rm el} v \cdot \delta E \ra = \dfrac{4}{3} n_\psi^{-1} \gamma(T) \Big( T_\chi - T \Big)~,
\end{equation}
where $T$ and $T_\chi$ are the temperatures of the SM and dark sector respectively. In the limit where a non-relativistic DM particle scatters with a relativistic SM particle of momentum $k$, the momentum relaxation rate can be expressed as a sum over all possible SM particles $i$ as:
\begin{equation}
\gamma(T) \equiv \dfrac{1}{T}\sum_i \frac{g_i}{6m_{\chi}} \int^\infty_0 \frac{\diff^3 {\vec k}}{(2\pi)^3} \, f_i (1\pm f_i) \frac{|{\vec k}|}{\sqrt{{\vec k}^2+m^2_i}}\,\int^0_{-4k^2} \diff t(-t) \frac{\diff\sigma_{\chi + \psi \rightarrow \chi + \psi}}{\diff t}~, \label{elasticloss}
\end{equation}
where $t$ is the squared momentum transfer between DM and the relativistic species. The differential elastic scattering cross section is given by
\begin{align}
 \frac{\diff \sigma_{\chi + \psi \rightarrow \chi + \psi}}{\diff t} =\frac{1}{64\pi m^2_{\chi}k^2}\, \overline{|{\cal M}_{\chi + \psi \rightarrow \chi + \psi }|^2}~.
\end{align}
Going back to Eq.~\ref{eq:Botlzmannenergydensity} and considering that the DM density evolves according to its thermal equilibrium function gives:
\begin{equation}
\dfrac{\partial T_\chi}{\partial T}=\dfrac{3 T_\chi^2}{m_\chi T}+\dfrac{4 T_\chi^2}{3m_\chi^2}\dfrac{\gamma(T)}{HT}\Big( T_\chi - T \Big)~.
\end{equation}
The first term on the RHS of this equation drives the DM temperature to evolve grossly as $T_\chi \appropto 1/(-\log T)$ while the second term tends to moderate the temperature difference between the two sectors with an efficiency proportional to the momentum relaxation rate. The decoupling temperature $T_{\rm d}$ can be estimated as the temperature for which the second term becomes of the order of one, yielding:
\begin{equation}
\dfrac{4 T_{\rm d}^2}{3 m_\chi^2} \dfrac{\gamma( T_{\rm d})}{H} \sim 1~.
\end{equation}

\subsection{The VSIMP case}
\label{sec:VSIMP3to2}
Based on the discussion in Sec.~\ref{ssec:SIMP}, the $3\rightarrow 2$ annihilation cross sections including only $SU(2)_X$ gauge interactions (with notations, $123\rightarrow 11$ meaning $X_1 X_2 X_3\rightarrow X_1 X_1$, etc) are, in the non-relativistic limit,
\bea
\langle\sigma v^2\rangle_{ijk}&\equiv&\langle\sigma v^2\rangle_{123\rightarrow 11}=\langle\sigma v^2\rangle_{123\rightarrow 22}=\langle\sigma v^2\rangle_{123\rightarrow 33} \nonumber \\
&=&\frac{5\sqrt{5}g_X^6}{331776\pi m_X^5(m_{h_1}^2+m_X^2)^2}\,(347 m^4_{h_1}+586 m_{h_1}^2 m_X^2+707 m^4_X) \nonumber \\
&&+\frac{19\sqrt{5} g^6_X}{1152\pi m_X(9m^2_X-m^2_{h_1})^2}\, \langle( v^2_1+v^2_2+v_1 v_2 \cos\theta_{12})\rangle~,
\eea
\bea
\langle\sigma v^2\rangle_{iij}&\equiv &\langle\sigma v^2\rangle_{112\rightarrow 13}=\langle\sigma v^2\rangle_{113\rightarrow 12}=\langle\sigma v^2\rangle_{221\rightarrow 23}=\langle\sigma v^2\rangle_{223\rightarrow 12}  \nonumber \\
&=&\langle\sigma v^2\rangle_{331\rightarrow 23}=\langle\sigma v^2\rangle_{332\rightarrow 13} \nonumber \\
&=&\frac{5\sqrt{5}g_X^6}{2654208\pi m_X^5}\Big(14377+\frac{6m_X^2(157m_{h_1}^2-763m_X^2)}{(m_{h_1}^2-4m_X^2)(m_{h_1}^2+m_X^2)} \nonumber \\
&&+\frac{3m_X^4(5281m_{h_1}^4-18558m_{h_1}^2m_X^2+32561m_X^4)}{(m_{h_1}^2-4m_X^2)^2(m_{h_1}^2+m_X^2)^2}\Big)~,
\eea
\bea
\langle\sigma v^2\rangle_{iii}&=&\langle\sigma v^2\rangle_{111\rightarrow 23}=\langle\sigma v^2\rangle_{222\rightarrow 13}=\langle\sigma v^2\rangle_{333\rightarrow 12} \nonumber \\
&=&\frac{25\sqrt{5}g_X^6}{2654208\pi m_X^5}\Big(8375+\frac{362m_X^2}{m_{h_1}^2-4m_X^2}+\frac{1713m_X^4}{(m_{h_1}^2-4m_X^2)^2}\Big)~.
\eea
Here, we have included the $p$-wave terms in $\langle\sigma v^2\rangle_{iii}$ as they have a resonance at $m_{h_1}=3m_X$. We note that $v_1, v_2$ are the speeds of two Dark Matter particles in the initial states and $\theta_{12}$ is the angle between the two in the center of mass frame.

On the other hand, the $3\rightarrow 2$ annihilation cross sections including the dark Higgs (with notations, $122\rightarrow 1 h_1$ meaning $X_1 X_2 X_2\rightarrow X_1 h_1$, etc) are, in the non-relativistic limit,
\bea
\langle\sigma v^2\rangle^h_{iii}&\equiv& \langle\sigma v^2\rangle_{111\rightarrow 1 h_1}=\langle\sigma v^2\rangle_{222\rightarrow 2 h_1}=\langle\sigma v^2\rangle_{333\rightarrow 3 h_1}  \nonumber \\
&=& \frac{\sqrt{5}g_X^6 m^{16}_{h_1} C_1 (1-m^2_{h_1}/(16m^2_X))^{1/2}}{17915904\pi m_X^{10}(4m_X^2-m_{h_1}^2)^{7/2}(2m_X^2+m_{h_1}^2)^2}~,
\eea
with 
\begin{align}
\begin{split}
C_1\equiv &\frac{1}{m^{16}_{h_1}}\Big( 3m_{h_1}^{16}-270m_{h_1}^{14}m_X^2+9917m_{h_1}^{12}m_X^4-187056m_{h_1}^{10}m_X^6+1952400m_{h_1}^{8}m_X^8\\
&-11318848m_{h_1}^{6}m_X^{10}+35045232m_{h_1}^{4}m_X^{12}-52110336m_{h_1}^{2}m_X^{14}+30261248m_X^{16} \Big) ~, \label{C1}
\end{split}
\end{align}
and 
\begin{align}
\langle\sigma v^2\rangle^h_{ijj}&\equiv&& \langle\sigma v^2\rangle_{122 \rightarrow 1 h_1}=\langle\sigma v^2\rangle_{133\rightarrow 1 h_1}=\langle\sigma v^2\rangle_{211\rightarrow 2 h_1} =\langle\sigma v^2\rangle_{233\rightarrow 2 h_1} \nonumber \\
&=&& \langle\sigma v^2\rangle_{311\rightarrow 3 h_1}=\langle\sigma v^2\rangle_{322\rightarrow 3 h_1}  \nonumber \\
&=&& \frac{\sqrt{5}g_X^6m^{20}_{h_1} C_2(1-m^2_{h_1}/(16m^2_X))^{1/2}}{17915904\pi m_X^{10}(4m_X^2-m_{h_1}^2)^{7/2}(2m_X^2+m_{h_1}^2)^2(7m^2_X-m^2_{h_1})^2}~,
\end{align}
with
\begin{equation}
\begin{split}
C_2\equiv&\frac{1}{m^{20}_{h_1}}\Big( 13m_{h_1}^{20}-568m_{h_1}^{18}m_X^2+33204m_{h_1}^{16}m_X^4-724140m_{h_1}^{14}m_X^6+6743931m_{h_1}^{12}m_X^8\\
&-26087280m_{h_1}^{10}m_X^{10}+48284736m_{h_1}^{8}m_X^{12}-166749984m_{h_1}^{6}m_X^{14}+806289168m_{h_1}^4m_X^{16}\\
&-2275720192m_{h_1}^2m_X^{18}+3442229248m_X^{20} \Big)~.
\label{C2}
\end{split}
\end{equation}
We note that the factor $1/(4m^2_X-m^2_{h_1})^4$ in the above results is the squared product of the dark Higgs propagator in $s$-channel  and the Dark Matter propagator in $t$-channel, which are regularized at $m_{h_1}=2m_X$ by the finite width of the dark Higgs and a nonzero Dark Matter velocity, respectively. 
The $3\rightarrow 2$ annihilation cross sections including two dark Higgs bosons such as $XXX\rightarrow h_1 h_1$ are $p$-wave suppressed and sub-dominant. 

\section{Non-thermal production : the freeze-in case}
\label{sec:nonthermal_freezein}
\subsection{Dark Matter production in the radiation domination era}
In the case where the coupling strength between the SM particles and the dark sector is not sufficient to reach a primordial thermal equilibrium state, the DM could be produced non-thermally assuming that at the end of reheating, Dark Matter is not already produced in our universe.
\subsubsection{Production through decay of a heavy field}
One possibility could be that the Dark Matter is produced from annihilations of SM particles through the exchange of some heavy $Z^\prime$ mediator for instance or produced directly by the decay of this mediator. The  full Boltzmann equation relevant in the former case involves the amplitude $\mathcal{M}$ of the $\chi(p_1) + \chi(p_2) \longleftrightarrow Z^\prime (p_3)$ process where $\chi$ denotes a DM candidate:
\begin{align}
-HsT \frac{\diff Y_{\chi}}{\diff T} =  2 \int  \frac{\diff^3 \vec p_1}{(2\pi)^3 2 E_1} \frac{\diff^3 \vec p_2}{(2\pi)^3 2 E_2} \frac{\diff^3 \vec p_3}{(2\pi)^3 2 E_3}  \Big[   |{\cal M}_{\chi\bar \chi \to Z^\prime}|^2 f_1(\vec p_1) f_2(\vec p_2) &  \nonumber \\   -     
|{\cal M}_{Z^\prime \to \chi\bar \chi}|^2 f^{\rm eq}_3(\vec p_3)  \Big] (2\pi)^4 \delta^4(p_1+p_2 - p_3) & ~.
\end{align}
From the freeze-in approximation, the initial Dark Matter number density vanishes, we can neglect the term proportional to $f_1(\vec p_1) f_2(\vec p_2)$. We can use the definition of the decay width
\be
\Gamma_{Z^\prime} = \dfrac{1}{2 m_{Z^\prime}} \int  \frac{\diff^3 \vec p_1}{(2\pi)^3 2 E_1} \frac{\diff^3 \vec p_2}{(2\pi)^3 2 E_2} |{\cal M}_{Z^\prime \to \chi \bar \chi}|^2 
        (2\pi)^4 \delta^4(p_1+p_2 - p_3)~,
\ee
to write the Boltzmann equation as:
\begin{equation}
-HsT \frac{\diff Y_{\chi}}{\diff T} = \int \dfrac{\diff^3 p_3}{(2 \pi)^3 E_3} \Gamma_{Z^\prime} e^{-E_3/T}~,
\end{equation}
where we considered Maxwell-Boltzmann statistics for the $Z^\prime$. Computing the integral over all possible momenta and using the variable $z \equiv m_{Z^\prime}/T$ gives:
\begin{equation}
Hsz\frac{\diff Y_{\chi}}{\diff z} = \dfrac{m_{Z^\prime}^3}{2 \pi^2}\Gamma_{Z^\prime} \dfrac{K_1(z)}{z}~.
\end{equation}
The yield $ Y_{\chi}(z)$ can be expressed as a function of the following integral:
\begin{equation}
Y_\chi(z)=\dfrac{m_{Z^\prime}^3}{2 \pi^2}\dfrac{\Gamma_{Z^\prime}}{H(m_{Z^\prime})s(m_{Z^\prime})} \int^{z}_0 K_1(z^\prime)z^{\prime 3} \diff z^\prime~.
\end{equation}
The yield at the present time $Y_\chi^{\rm now} \simeq Y_\chi^\infty $ can be estimated analytically as:
\begin{equation}
Y_\chi^\infty \simeq \dfrac{1}{2\pi^2}\dfrac{m_{Z^\prime}^3 \Gamma_{Z^\prime}}{H(m_{Z^\prime})s(m_{Z^\prime})} \int_0^\infty K_1(z)z^3 \diff z \simeq \dfrac{3}{4 \pi}\dfrac{m_{Z^\prime}^3 \Gamma_{Z^\prime}}{H(m_{Z^\prime})s(m_{Z^\prime})}~.
\end{equation} 
In this case most of the DM is produced at a late time $T\sim m_{Z^\prime}$. As a result the relic density is given by the following analytical formula:
\begin{equation}
\Omega_\chi h^2\simeq \dfrac{m_\chi s_0 h^2}{\rho_c^0} \dfrac{3}{4 \pi}\dfrac{m_{Z^\prime}^3 \Gamma_{Z^\prime}}{H(m_{Z^\prime})s(m_{Z^\prime})}~.
\end{equation}~.
\subsubsection{Production through annihilation of SM particles}
Another possibility is to produce Dark Matter particles via processes such as $\text{SM(1) + SM(2)} \rightarrow\text{DM(3) + DM(4)}$. In that case the Boltzmann equation can be written as:
\begin{align}
\dfrac{\diff n_\chi}{\diff t}+ 3Hn_\chi=2 \int \diff \Pi_1 \diff \Pi_2 \diff \Pi_3 \diff \Pi_4 (2\pi)^4 \delta^{(4)}(p_1+p_2-p_3-p_4)\overline{|\mathcal{M}|^2}_{1+2\rightarrow 3+4}f_1^{\rm eq} f_2^{\rm eq} ~,
\end{align}
where we neglected the backreaction term by using the freeze-in approximation. Assuming two initial identical SM particles with internal degrees of freedom equals to the ones of the DM for simplicity and Maxwell-Boltzmann distributions, one can rewrite the Boltzmann equation as a function of the velocity averaged annihilation cross section $\la \sigma v \ra $\footnote{as defined in Eq.~(\ref{eq:sigmavfullintegral}).} as:
\begin{equation}
\dfrac{\diff n_\chi}{\diff t}+ 3Hn_\chi=2 \la \sigma v \ra (n_{\rm SM}^{\rm eq})^2~,
\end{equation}
where $n_{\rm SM}^{\rm eq}=n_{\rm 1}^{\rm eq}=n_{\rm 2}^{\rm eq}$. We can define the production rate $R(T)$ as being the RHS of the Boltzmann equation:
\begin{equation}
R(T)\equiv 2 \la \sigma v \ra (n_{\rm SM}^{\rm eq})^2~.
\end{equation}
The Boltzmann equation can be written as:
\begin{equation}
Hsx\dfrac{\diff Y_\chi}{\diff x}=R(x)~,
\end{equation}
with the usual variables $x\equiv m_\chi/T$ and $Y_\chi \equiv n_\chi /s$. Parametrizing the cross section as a power law:
\begin{equation}
\la \sigma v \ra =\dfrac{T^n}{M^{2+n}}\Theta[T-m_\chi]~,
\end{equation}
where $M$ is some mass scale and $\Theta$ the Heaviside step function, gives the following:
\begin{align}
Hsx\dfrac{\diff Y_\chi}{\diff x}= m_\chi^{n+6} \dfrac{x^{-n}}{M^{2+n}} \Big(g_i \dfrac{\zeta(3)}{\pi^2} x^{-3} \Big)^2~,
\end{align}
where $g_i$ are the number of internal degrees of freedom of the initial states, leading to:
\begin{equation}
\dfrac{\diff Y}{\diff x}= \frac{135 \sqrt{\frac{5}{2}}\zeta (3)^2 g_{i}^2 M_{\text{Pl}}  m_{\chi }^{n+1}}{\pi^7 \sqrt{g_{\star }} g_{\star,s} M^{n+2} x^{n+2}}~,
\end{equation}
Integrating this equation~\footnote{Assuming $n\neq 1$, otherwise a logarithmic behavior is expected.} from the reheating temperature $x_{\rm RH}$ to some temperature $x_F$ gives:
\begin{equation}
Y_\chi(x_F)=\frac{135 \sqrt{\frac{5}{2}} \zeta (3)^2 g_{i}^2 m_\chi^{n+1} M_{\text{Pl}} }{\pi ^7  \sqrt{g_{\star }} g_{\star,s}  M^{n+2}}    \dfrac{ x_{\text{RH}}^{-(n+1)}-x_{F}^{-(n+1)} }{(n+1)}~,
\end{equation}
assuming $g_{\star }$ and $g_{\star,s}$ constant. Since we considered that the production rate vanishes for temperatures larger than the DM mass, the yield at the present time will be given by taking $x_F=1$. 
Therefore, the relic density can be derived in a straightforward way:
\begin{equation}
\Omega_\chi h^2=\frac{135 \sqrt{\frac{5}{2}} h^2 s_0 \zeta (3)^2 g_{i}^2 m_\chi^{n+2} M_{\text{Pl}} }{  \rho_c^0 \pi^7  \sqrt{g_{\star }} g_{\star,s}  M^{n+2}}    \dfrac{ x_{\text{RH}}^{-(n+1)}-1^{-(n+1)} }{(n+1)}~.
\end{equation}

\subsection{The preheating contribution}
\label{sec:appendixpreheating}
As it was discussed in Sec.~\ref{sec:endofinflation}, the reheating is not expected to be a instantaneous process and the physics occuring at temperatures above the reheating temperature $T_{\rm RH}$ plays a role in the production of Dark Matter. In the following we call the period of time occuring before the reheating temperature as \textit{preheating}. Assuming that the inflaton $\phi$ mostly decays into SM particles, the evolution of energy densities of radiation $R$, Dark Matter $X$ and $\phi$ are given by the following set of Boltzmann equations:
\begin{align}
\frac{\text{d}n_X}{\diff t}&=-3H\,n_X-\langle\sigma v\rangle\left[n_X^2-(n_X^{\text{eq}})^2\right]\,, \nonumber \\
\frac{\text{d}\rho_R}{\diff t}&=-4H\,\rho_R+\Gamma_\phi\,\rho_\phi+2\langle\sigma v\rangle\langle E_X\rangle\left[n_X^2-(n_X^{\text{eq}})^2\right]\,, \nonumber \\
\frac{\text{d}\rho_\phi}{\diff t}&=-3H\,\rho_\phi-\Gamma_\phi\,\rho_\phi\,.
\end{align}
where $\Gamma_\phi$ denotes the inflaton decay width and $\la \sigma v \ra$ the DM velocity averaged annihilation cross section. The Hubble parameter can be written:
\begin{equation}
H^2=\dfrac{8 \pi}{3 M_{\text{Pl}}^2}(\rho_\phi+\rho_R + \rho_X)\simeq \dfrac{8 \pi}{3 M_{\text{Pl}}^2}(\rho_\phi+\rho_R)~.
\end{equation}
We can write the set of Boltzmann equations using dimensionless variables for convenience:
\begin{equation}
\Phi \equiv \rho_\phi T_{\text{RH}}^{-1} a^3;\quad R \equiv \rho_R a^4; \quad X\equiv n_X a^3; \quad A \equiv a T_{\text{RH}}~,
\end{equation}
which gives:
\begin{eqnarray}
\dfrac{\text{d} \Phi}{\diff A}&=&-\left( \dfrac{\pi^2 g_\star}{30} \right)^{1/2} \dfrac{A^{1/2}\Phi}{\sqrt{\Phi + R/A}} ~,\nonumber \\
\dfrac{\text{d} R}{\diff A}&=&\left( \dfrac{\pi^2 g_\star}{30} \right)^{1/2} \dfrac{A^{3/2}\Phi}{\sqrt{\Phi + R/A}}+ \left( \dfrac{3}{8\pi} \right)^{1/2} \dfrac{ A^{-3/2} 2\langle\sigma v\rangle\langle E_X\rangle M_{\text{Pl}}}{\sqrt{\Phi+R/A}}(X^2-X_{\text{eq}}^2) ~,\nonumber \\
\dfrac{\text{d} X}{\diff A}&=& - \left( \dfrac{3}{8\pi} \right)^{1/2} \dfrac{ A^{-5/2} 2\langle\sigma v\rangle\langle E_X\rangle M_{\text{Pl}}}{\sqrt{\Phi+R/A}}(X^2-X_{\text{eq}}^2)~,
\end{eqnarray}
with $T_{\text{RH}}$ defined as follow
\begin{equation}
\Gamma_\Phi = \dfrac{\sqrt{4\pi^3g_\star}}{45} \dfrac{T_{\text{RH}}^2}{M_{\text{Pl}}}~,
\end{equation}
where $g_\star$ denotes the effective number of degrees of freedom of the SM thermal bath at the reheating temperature. This set of equation can be solved assuming that the inflaton is the only species present at the initial time $t_I$:
\begin{equation}
\Phi_I = \dfrac{3}{8 \pi}\dfrac{M_{\text{Pl}}^2 H_I}{T_{\text{RH}}^4}, \quad R_I=0, \quad X_I=0, \quad A_I=1~,
\end{equation}
where the condition $A_I=1$ is arbitrary and will not impact the results. Following~\cite{Giudice:2000ex} we can infer the following analytical formula assuming a constant $\Phi \simeq \Phi_I$ during the reheating process:
\begin{equation}
T \equiv \left( \dfrac{30}{\pi^2 g_\star(T)}\right)^{1/4} \dfrac{R^{1/4}}{A}T_{\text{RH}}~.
\end{equation}
The maximum temperature achieved in the universe can be expressed as a function of the Hubble parameter at the initial time and the reheating temperature:
\begin{equation}
T_{\text{max}}=\left( \dfrac{3}{8} \right)^{2/5} \left( \dfrac{5}{\pi^3} \right)^{1/8} \dfrac{1}{g_\star^{1/4}}M_{Pl}^{1/4} H_I^{1/4} T_{\text{RH}}^{1/2}~,
\end{equation}
and the relation between the normalized scale factor and the temperature is given by:
\begin{equation}
T\simeq \left( \dfrac{9}{5 \pi^3 g_\star} \right)^{1/8} M_{\text{Pl}}^{1/4} H_I^{1/4}T_{\text{RH}}^{1/2}A^{-3/8}~,
\end{equation}
where we considered the effective number of relativistic species as constant $g_\star(T)\simeq g_\star$. An important result is the temperature evolution of the Hubble parameter which is very specific to this preheating stage of the universe:
\begin{equation}
H(T)=\left( \dfrac{5 \pi^3 g_\star}{9}\right)^{1/2} \dfrac{T^4}{T^2_{\text{RH}} M_{\text{Pl}}} \equiv \hat{H}T^4~.
\end{equation}
The Boltzmann equation relevant for DM production in the context of freeze-in can be simplified by neglecting the backreaction term:
\begin{equation}
\dfrac{\text{d}n_X}{\text{d}t}+3Hn_X=R(T)~,
\end{equation}
where $R(T)=\langle \sigma v \rangle (n_X^{\text{eq}})^2$ is the production rate. Considering the relation $a^{-3/8}\propto T$ and $H(T)=\hat{H}T^4$ we can express the time derivative as:
\begin{equation}
\dfrac{1}{\text{d}t}=\dfrac{a}{\text{d}a}H=-\dfrac{3 \hat{H}}{8}\dfrac{T^5}{\text{d}T}~,
\end{equation}
Therefore, the Boltzmann equation for the DM number density is give by:
\begin{equation}
\dfrac{\text{d}n_X}{\text{d}T}-8\dfrac{n_X}{T}=-\dfrac{8}{3 \hat{H}}\dfrac{R(T)}{T^5}~,
\end{equation}
whose LHS can be written as a total derivative:
\begin{equation}
\dfrac{\text{d}(n_XT^{-8})}{\text{d}T}=-\dfrac{8}{3 \hat{H}}\dfrac{R(T)}{T^{13}}~.
\end{equation}
This formula shows that the quantity $n_X T^{-8}$ is conserved in absence of interactions and is not limited to the DM case, it remains valid for any species that does not interact sufficiently enough with the SM to thermalize. This equation can be integrated in order to deduce the DM density at a temperature $T_\alpha$ with $T_{\text{RH}}<T_\alpha<T_{\text{max}}$:
\begin{equation}
(n_XT^{-8})|_{\text{max}}-(n_XT^{-8})|_{\alpha}=-\int_{T_\alpha}^{T_{\text{max}}} \dfrac{8}{3 \hat{H}}\dfrac{R(T)}{T^{13}}\text{d}T~.
\end{equation}
The first term on the LHS of the initial condition for the DM density vanishes from the initial condition used to derive the results in the previous section, coming from the freeze-in hypothesis. Assuming that the rate can be written as a power law and vanishes when the temperature is lower than the DM mass\footnote{In a more realistic description, the rate would not vanish for $T<m_X$ but would be exponentially suppressed due to the high-energy tail of the thermal distribution of the initial states.}
\begin{equation}
R(T)=\dfrac{T^n}{\Lambda^{n-4}}\Theta[T-m_X]~,
\end{equation}
allows for an analytical estimation of the density in the preheating phase. First, if we assume that the DM is lighter than the reheating temperature:
\begin{equation}
(n_XT^{-8})|_{\text{RH}}=\int_{T_{\text{RH}}}^{T_{\text{max}}} \dfrac{8}{3 \hat{H}}\dfrac{R(T)}{T^{13}}\text{d}T=\int_{T_{\text{RH}}}^{T_{\text{max}}} \dfrac{8}{3 \hat{H}}\dfrac{T^{n-13}}{\Lambda^{n-4}}\text{d}T~,
\end{equation}
giving:
\begin{equation}
n_X|_{\text{RH}}\simeq \dfrac{8}{3 \hat{H}} \dfrac{T_{\text{RH}}^8}{\Lambda^{n-4}}  \left \{ \begin{array}{ccr}
\dfrac{T_{\text{max}}^{n-12}}{(n-12)} & \hspace{2cm} &[n>12 ] ~\\ 
\ln \left( \dfrac{T_{\text{max}}}{T_{\text{RH}}} \right) & \hspace{2cm} &[n=12 ] ~\\
\dfrac{1}{(12-n)}\dfrac{1}{T_{\text{RH}}^{12-n}}  & \hspace{2cm}   &[n<12 ] ~ 
\end{array} \right.
\end{equation}
assuming $T_{\text{max}} \gg T_{\text{RH}}$. Now if we assume that $m_X>T_{\text{RH}}$ we have:
\begin{equation}
n_X|_{m_X}\simeq \dfrac{8}{3 \hat{H}} \dfrac{m_X^8}{\Lambda^{n-4}}  \left \{ \begin{array}{ccr}
\dfrac{T_{\text{max}}^{n-12}}{(n-12)} & \hspace{2cm} &[n>12 ] ~\\ 
\ln \left( \dfrac{T_{\text{max}}}{m_X} \right) & \hspace{2cm} &[n=12 ] ~\\
\dfrac{1}{(12-n)}\dfrac{1}{m_X^{12-n}}  & \hspace{2cm}   &[n<12 ] ~ 
\end{array} \right.
\end{equation}
since below $m_X$ the rate vanishes the quantity $(n_XT^{-8})$ is conserved during the preheating phase and we can express the density at the reheating temperature as:
\begin{equation}
n_X|_{\text{RH}}=n_X|_{m_X} \dfrac{T_{\text{RH}}^8}{m_X^8}~.
\end{equation}
This formula shows how strongly the DM density is redshifted during the preheating phase, much stronger than in a radiation-dominated universe. After the reheating, the universe is dominated by radiation and we can use the usual conservation of the yield $Y_X \equiv n_X/s$ which lead to:
\begin{equation}
\Omega_X h^2 = m_X Y_X \dfrac{s_0 h^2}{\rho_0^c}= m_X \dfrac{n_X|_{\text{RH}}}{s_{\text{RH}}} \dfrac{s_0 h^2}{\rho_0^c}=m_X \dfrac{n_X|_{\text{RH}}}{g_\star T_{\text{RH}^3}} \dfrac{g_{\star,0}T_0^3 h^2}{\rho_0^c}~.
\end{equation}
As a result, analytical formulas can be derived in the case where $m_X<T_{\text{RH}}$ as:
\begin{equation}
\Omega_X h^2= \dfrac{8}{3 \hat{H}} \dfrac{m_X}{g_\star} \dfrac{g_{\star,0}T_0^3 h^2}{\rho_0^c}  \dfrac{T_{\text{RH}}^5}{\Lambda^{n-4}}  \left \{ \begin{array}{ccr}
\dfrac{T_{\text{max}}^{n-12}}{(n-12)} & \hspace{2cm} &[n>12 ] ~\\ 
\ln \left( \dfrac{T_{\text{max}}}{T_{\text{RH}}} \right) & \hspace{2cm} &[n=12 ] ~\\
\dfrac{1}{(12-n)}\dfrac{1}{T_{\text{RH}}^{12-n}}  & \hspace{2cm}   &[n<12 ] ~ 
\end{array} \right.
\end{equation}
If $m_X>T_{\text{RH}}$ the relic density can be written:

\begin{equation}
\Omega_X h^2= \dfrac{8}{3 \hat{H}}   \dfrac{m_X}{g_\star } \dfrac{g_{\star,0}T_0^3 h^2}{\rho_0^c} \dfrac{T_{\text{RH}}^5}{\Lambda^{n-4}}  \left \{ \begin{array}{ccr}
\dfrac{T_{\text{max}}^{n-12}}{(n-12)} & \hspace{2cm} &[n>12 ] ~\\ 
\ln \left( \dfrac{T_{\text{max}}}{m_X} \right) & \hspace{2cm} &[n=12 ] ~\\
\dfrac{1}{(12-n)}\dfrac{1}{m_X^{12-n}}  & \hspace{2cm}   &[n<12 ] ~ 
\end{array} \right. 
\end{equation}

%% file: parts/appendices/computation_chernsimons.tex
\label{appendix:CS}
\vspace{0.3cm}

\noindent
In this section of the appendices we provide some detailed elements related to the computation of the Chern-Simons couplings considered in this thesis.
\section{Computation of the abelian Chern-Simons coupling}
\label{sec:appendixC}

In this section of the appendix we show how to derive the effective Lagrangian considered in Chapter~\ref{ch:CS} from a UV complete model framework. To achieve this goal, we review computations performed in Ref.~\cite{Anastasopoulos:2006cz} of potential anomalous diagrams of the theory which involve three external gauge 
bosons or axions, interacting through a triangular loop of massive fermions. We can classify the possible diagrams in three categories: 
(1) diagrams without any mass insertions. Such diagrams are linearly divergent and proportional to the usual anomaly trace~\cite{Adler:1969gk,Bell:1969ts} (see Ref.~\cite{Bilal:2008qx} also for an overview). (2) Diagrams involving three gauge bosons with two mass insertions which give a finite result.
They are connected to the so-called  "CS" interaction (see, for example Ref.~\cite{Anastasopoulos:2006cz})
and, (3) diagrams involving axions and 
two gauge bosons with one mass insertion. Just like the former
class of diagrams, these diagrams are also finite.
We will consider examples where the incoming state for the triangle loop
diagrams with heavy fermions is either a gauge boson $A^\mu_i(\vec{k}_3)$
or an axion $\theta_i$ while two outgoing states are two gauge fields,
denoted as $A^\rho_k(\vec{k}_2)$ and $A^\nu_j(\vec{k}_1)$, respectively. 
After evaluating these loop diagrams, we also compute the gauge transformations of the effective 
Lagrangian to ensure an anomaly free setup.

\subsection{Diagrams with two mass insertions : "Chern-Simons" contribution}

\begin{figure}[h!]
\includegraphics[width=7.5cm]{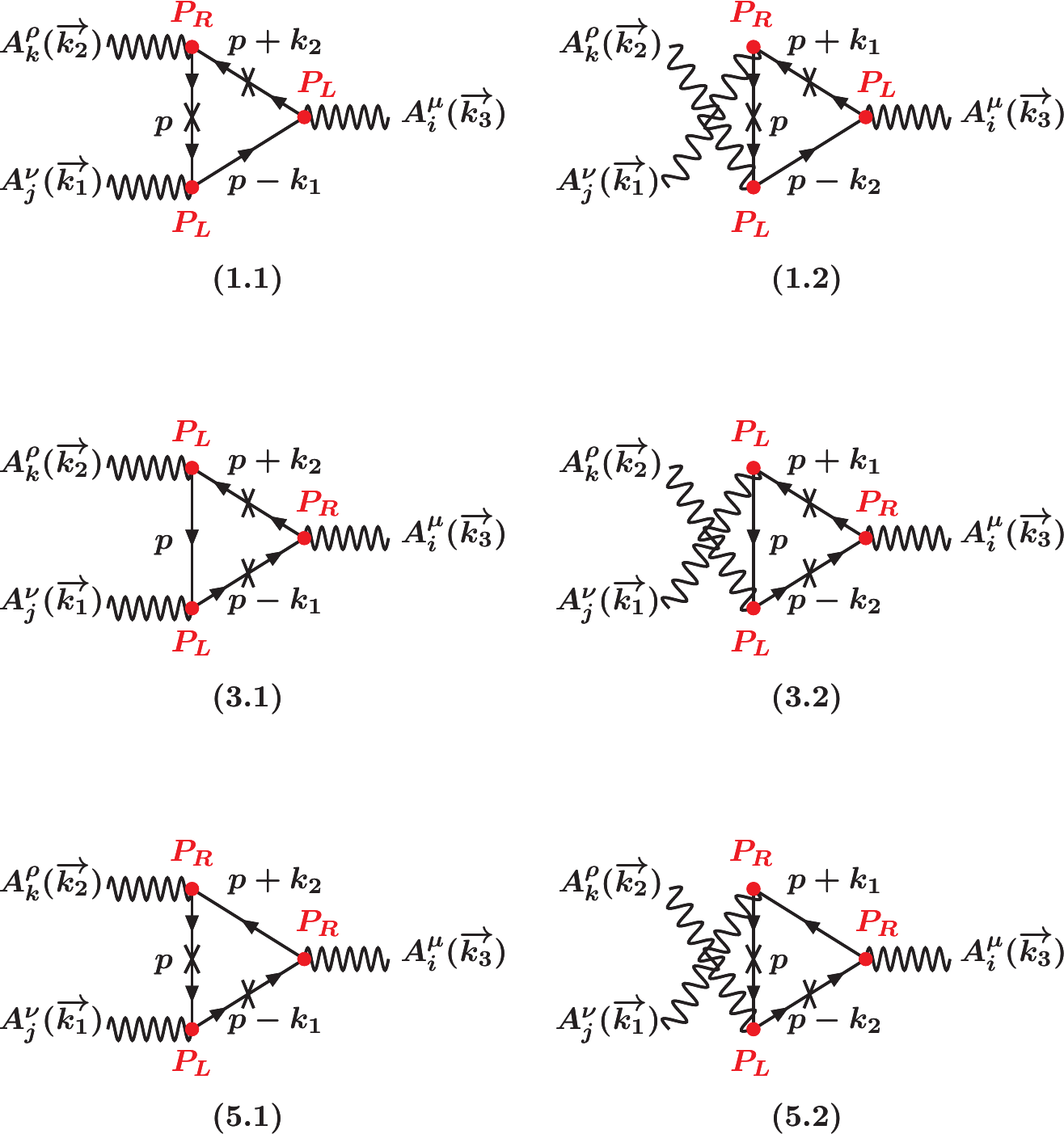}
\includegraphics[width=7.5cm]{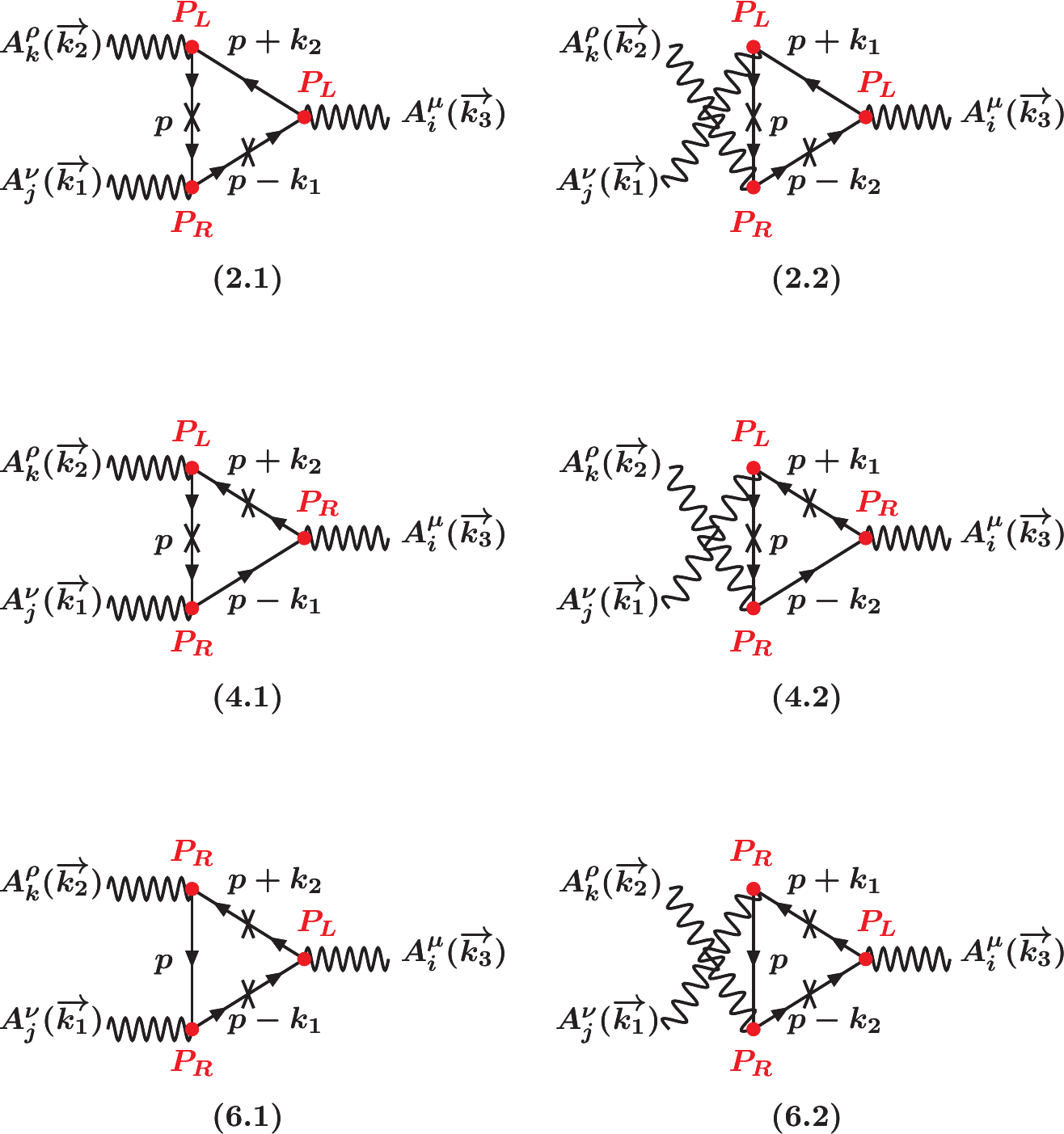}
\caption{Possible set of diagrams for triangular loops of heavy BSM fermions giving
rise to an effective CS interaction. Here {\boldmath `$\times$'} denotes
a mass insertion and the arrows represent the direction of fermion flow. Directions 
of the momentum, i.e., $p,\,p\pm k_1,\,p\mp k_2$, etc. are opposite to the fermion
flow.}
\label{fig:CSgenerate}
\end{figure}

We initiate our analysis for diagrams having gauge fields in the three external legs
and two mass insertions, which is equivalent to have two chirality flips for each diagram. Thus, we can chose one dominant chirality over the remaining two others in the fermionic loop and have three  possibilities to place the 
mass insertions for each dominant chirality. Further, considering the fact
that we can contract the external legs of the 
outgoing gauge fields in two different ways, we end up with twelve
diagrams as shown in Fig.~\ref{fig:CSgenerate}. 

For a systematic analysis we start with the following Lagrangian:
%
\begin{equation}
\mathcal{L}\supset \sum_{a=i,j,k} i\bar{F}\partial_\mu \gamma^\mu F-m\bar{F}F-Q^a_L 
\bar{F}_L\gamma_\mu F_L A^\mu_a-Q^a_R \bar{F}_R\gamma_\mu F_R A^\mu_a,
\end{equation} 
where $F=F_L+F_R$ are the BSM heavy fermions running in the triangle
loops with $Q^a_L,\,Q^a_R$ as the relevant gauge charges
associated with $A^a_\mu$ field for the left- and right-chiral BSM fermions, respectively. We can write contributions of the twelve diagrams without contractions on 
any external legs as an integral over the momentum $p$ as:
\begin{equation}
\label{eq:gammafracCS}
\int \dfrac{\diff ^4p}{(2\pi)^4}\Gamma^{\mu \nu \rho}(p,k_1,k_2),
\end{equation}
where $\Gamma^{\mu \nu \rho}$, after integrating out heavy fermionic
degrees of freedom, can be expanded in the powers of external
momentums to achieve the low-energy effective Lagrangian. The 
expansion goes as:
\begin{align}
\label{eq:Fullgammapart}
\Gamma^{\mu \nu \rho}(p,k_1,k_2)&\simeq \Gamma^{\mu \nu \rho}(p,0,0)
+\sum^{2}_{i=1} k_i^\alpha \left.\left(\dfrac{\partial \Gamma^{\mu \nu \rho}(p,k_1,k_2)}{\partial k_i^\alpha}\right)\right\vert_{k_i=0}
\nonumber\\
&+\frac{1}{2!} \sum^{2}_{i,j=1} k_i^\alpha k_j^\beta 
\left.\left(\dfrac{\partial^2 \Gamma^{\mu \nu \rho}(p,k_1,k_2)}{\partial k_i^\alpha \partial k_j^\beta}\right)\right\vert_{k_{i,j}=0}
+ \mathcal{O} (k^3_{i,\,j,\,k}).  
\end{align}
Now from Fig.~\ref{fig:CSgenerate} it is apparent that $\Gamma^{\mu \nu \rho}(p,k_1,k_2)$
can be decomposed as a product of the two terms,
i.e., $\Gamma^{\mu \nu \rho}(p,k_1,k_2)=\bold{\Pi} \cdot \bold{TR}^{\mu \nu \rho}$ where $\bold{TR}^{\mu \nu \rho}$ includes possible couplings and traces over gamma matrices 
while $\bold{\Pi}$ is defined in the following way:
\begin{align}
\label{eq:picalc}
\bold{\Pi}=&\dfrac{1}{p^2-m^2}\dfrac{1}{(p+k_2)^2-m^2}\dfrac{1}{(p-k_1)^2-m^2} \quad \text{for diagrams
like x.1 with x=1,2,...,6}, \nonumber\\ 
\bold{\Pi}=&\dfrac{1}{p^2-m^2}\dfrac{1}{(p-k_2)^2-m^2}\dfrac{1}{(p+k_1)^2-m^2} \quad \text{for diagrams 
like x.2 with x=1,2,...,6}. 
\end{align}
The trace in the leading term $\Gamma(p,0,0)$ 
(see Eq.~(\ref{eq:Fullgammapart}))
appears to be proportional to odd powers of $p$ for  the 
numerator, which vanishes after $\int d^4p$ integration. 
The linear and quadratic terms in $k_i^\alpha$ can be computed 
straightforwardly from Eq.~(\ref{eq:Fullgammapart}) as:
\begin{align}
\label{eq:kalphaexp}
k_i^\alpha \left. \left( \dfrac{\partial \Gamma^{\mu \nu \rho}(p,k_1,k_2)}{\partial k_i^\alpha}  
\right) \right\vert_{0} &=k_i^\alpha \left. \left( \dfrac{\partial \bold{TR}^{\mu \nu \rho}}{\partial k_i^\alpha} 
\cdot \bold{\Pi} \right) \right\vert_{0} +k_i^\alpha \left. \left( \dfrac{\partial \bold{\Pi}}{\partial k_i^\alpha} 
\cdot \bold{TR}^{\mu \nu \rho} \right) \right\vert_{0},\nonumber\\
k_i^\alpha k_j^\beta \left. \left( \dfrac{\partial^2 \Gamma^{\mu \nu \rho}(p,k_1,k_2)}
{\partial k_i^\alpha \partial k_j^\beta} \right) \right\vert_{0} &=
k_i^\alpha k_j^\beta \left. \left( \dfrac{\partial^2 \bold{TR}^{\mu \nu \rho}}{\partial k_i^\alpha \partial k_j^\beta} 
\cdot \bold{\Pi} \right) \right\vert_{=0} +k_i^\alpha k_J^\beta 
\left. \left( \dfrac{\partial^2 \bold{\Pi}}{\partial k_i^\alpha \partial k_j^\beta} 
\cdot \bold{TR}^{\mu \nu \rho} \right) \right\vert_{0}\nonumber\\
&+
k_i^\alpha k_j^\beta \left. \left( \dfrac{\partial \bold{TR}^{\mu \nu \rho}}{\partial k_i^\alpha} 
\cdot \dfrac{\partial \bold{\Pi}}{\partial k_j^\beta}  \right) \right\vert_{0} + 
k_i^\alpha k_j^\beta \left. \left( \dfrac{\partial \bold{TR}^{\mu \nu \rho}}{\partial k_j^\beta} 
\cdot \dfrac{\partial \bold{\Pi}}{\partial k_i^\alpha}  \right) \right\vert_{0}.
\end{align}
The contributions from denominators $\bold{\Pi}$ are shown
in Eq.~(\ref{eq:picalc}) while trace contributions 
$\bold{TR}^{\mu\nu\rho}$ from the twelve diagrams are:
\begin{align}
\label{eq:trcalccs}
\bold{TR}^{\mu \nu \rho}_{1.1}&=Q_L^iQ_L^jQ_R^k\text{Tr}[(\cancel{p}+m)\gamma^\rho P_R (\cancel{p}+\cancel{k}_2+m)\gamma^\mu P_L (\cancel{p}-\cancel{k}_1)\gamma^\nu P_L],\nonumber\\
\bold{TR}^{\mu \nu \rho}_{1.2}&=Q_L^iQ_R^jQ_L^k\text{Tr}[(\cancel{p}+m)\gamma^\nu P_R (\cancel{p}+\cancel{k}_1+m)\gamma^\mu P_L (\cancel{p}-\cancel{k}_2)\gamma^\rho P_L],\nonumber\\
\bold{TR}^{\mu \nu \rho}_{2.1}&=Q_L^iQ_R^jQ_L^k\text{Tr}[(\cancel{p}+m)\gamma^\rho P_L (\cancel{p}+\cancel{k}_2)\gamma^\mu P_L (\cancel{p}-\cancel{k}_1+m)\gamma^\nu P_R],\nonumber\\
\bold{TR}^{\mu \nu \rho}_{2.2}&=Q_L^iQ_L^jQ_R^k\text{Tr}[(\cancel{p}+m)\gamma^\nu P_L (\cancel{p}+\cancel{k}_1)\gamma^\mu P_L (\cancel{p}-\cancel{k}_2+m)\gamma^\rho P_R],\nonumber\\
\bold{TR}^{\mu \nu \rho}_{3.1}&=Q_R^iQ_L^jQ_L^k\text{Tr}[\cancel{p}\,\gamma^\rho P_L (\cancel{p}+\cancel{k}_2+m)\gamma^\mu P_R (\cancel{p}-\cancel{k}_1+m)\gamma^\nu P_L],\nonumber\\
\bold{TR}^{\mu \nu \rho}_{3.2}&=Q_R^iQ_L^jQ_L^k\text{Tr}[\cancel{p}\,\gamma^\nu P_L (\cancel{p}+\cancel{k}_1+m)\gamma^\mu P_R (\cancel{p}-\cancel{k}_2+m)\gamma^\rho P_L],\nonumber\\
\bold{TR}^{\mu \nu \rho}_{4.1}&=Q_R^iQ_R^jQ_L^k\text{Tr}[(\cancel{p}+m)\gamma^\rho P_L (\cancel{p}+\cancel{k}_2+m)\gamma^\mu P_R (\cancel{p}-\cancel{k}_1)\gamma^\nu P_R],\nonumber\\
\bold{TR}^{\mu \nu \rho}_{4.2}&=Q_R^iQ_L^jQ_R^k\text{Tr}[(\cancel{p}+m)\gamma^\nu P_L (\cancel{p}+\cancel{k}_1+m)\gamma^\mu P_R (\cancel{p}-\cancel{k}_2)\gamma^\rho P_R],\nonumber\\
\bold{TR}^{\mu \nu \rho}_{5.1}&=Q_R^iQ_L^jQ_R^k\text{Tr}[(\cancel{p}+m)\gamma^\rho P_R (\cancel{p}+\cancel{k}_2)\gamma^\mu P_R (\cancel{p}-\cancel{k}_1+m)\gamma^\nu P_L],\nonumber\\
\bold{TR}^{\mu \nu \rho}_{5.2}&=Q_R^iQ_R^jQ_L^k\text{Tr}[(\cancel{p}+m)\gamma^\nu P_R (\cancel{p}+\cancel{k}_1)\gamma^\mu P_R (\cancel{p}-\cancel{k}_2+m)\gamma^\rho P_L],\nonumber\\
\bold{TR}^{\mu \nu \rho}_{6.1}&=Q_L^iQ_R^jQ_R^k\text{Tr}[\cancel{p}\,\gamma^\rho P_R (\cancel{p}+\cancel{k}_2+m)\gamma^\mu P_L (\cancel{p}-\cancel{k}_1+m)\gamma^\nu P_R],\nonumber\\
\bold{TR}^{\mu \nu \rho}_{6.2}&=Q_L^iQ_R^jQ_R^k\text{Tr}[\cancel{p}\,\gamma^\nu P_R (\cancel{p}+\cancel{k}_1+m)\gamma^\mu P_L (\cancel{p}-\cancel{k}_2+m)\gamma^\rho P_R],
\end{align}
where mass insertions are properly taken into account. One can use Eq.~(\ref{eq:kalphaexp}) to extract contributions
from the twelve diagrams shown in Fig.~\ref{fig:CSgenerate}.
For example, the contribution proportional to $Q_L^iQ_L^jQ_R^k$ involving diagrams $(1.1)$ and $(2.2)$, in the linear terms of Eq.~(\ref{eq:kalphaexp}), gives:
\begin{align}
\int \dfrac{\diff ^4p}{(2\pi)^4}\Big( \Gamma^{\mu \nu \rho}_{(1.1)}+\Gamma^{\mu \nu \rho}_{(2.2)} 
\Big)= Q_L^iQ_L^jQ_R^k \epsilon^{\mu \nu \rho \sigma} 
\dfrac{1}{24\pi^2}(k_3+k_1)_\sigma,
\end{align}
where we used the known expressions for different momentum
integrals over $p$ (see Ref.~\cite{Peskin:1995ev} for example) 
and $k_1+k_2=k_3$.
In a similar way the contributions from all the twelve
diagrams of Fig.~\ref{fig:CSgenerate} can be grouped
as shown in Table~\ref{tab:CScont}.
\begin{table}[h]
\begin{center}
\begin{tabular}{|c||c|}
\hline 
Diagrams & Contribution to $\Gamma^{\mu \nu \rho}$ \\ 
\hline 
(1.1)+(2.2) & $Q_L^iQ_L^jQ_R^k \epsilon^{\mu \nu \rho \sigma} (k_3+k_1)_\sigma /(24\pi^2)$ \\ 
\hline
(2.1)+(1.2) & $-Q_L^iQ_R^jQ_L^k \epsilon^{\mu \nu \rho \sigma} (k_3+k_2)_\sigma/(24\pi^2)$ \\ 
\hline
(3.1)+(3.2) &  $Q_R^iQ_L^jQ_L^k \epsilon^{\mu \nu \rho \sigma} (k_2-k_1)_\sigma/(24\pi^2)$ \\ 
\hline
(4.1)+(5.2) &  $-Q_R^iQ_R^jQ_L^k \epsilon^{\mu \nu \rho \sigma} (k_3+k_1)_\sigma/(24\pi^2)$ \\
\hline
(5.1)+(4.2) &  $Q_R^iQ_L^jQ_R^k \epsilon^{\mu \nu \rho \sigma} (k_3+k_2)_\sigma/(24\pi^2)$ \\
\hline
(6.1)+(6.2) &  $Q_L^iQ_R^jQ_R^k \epsilon^{\mu \nu \rho \sigma} (k_1-k_2)_\sigma/(24\pi^2)$ \\  
\hline
\end{tabular} 
\end{center}
\caption{Resultant contributions of the twelve diagrams of figure
\ref{fig:CSgenerate}, clubbed according to the same pre-factor.}
\label{tab:CScont}
\end{table}

From Table~\ref{tab:CScont}, one can factorize the sum of all contributions proportional to the external momentum $k_3$ as:

\begin{align}
\int \dfrac{\diff ^4p}{(2\pi)^4}\Gamma^{\mu \nu \rho} &\supset \dfrac{1}{24\pi^2} \epsilon^{\mu \nu \rho \sigma} 
k_{3 \sigma} (Q_L^iQ_L^jQ_R^k-Q_L^iQ_R^jQ_L^k-Q_R^iQ_R^jQ_L^k+Q_R^iQ_L^jQ_R^k),\nonumber\\
&\supset\dfrac{1}{24\pi^2} \epsilon^{\mu \nu \rho \sigma} k_{3 \sigma} (Q_L^i+Q_R^i)(Q_L^jQ_R^k-Q_L^kQ_R^j).
\end{align}
The same factorization can be done for $k_2$ and $k_1$
to produce the following effective Lagrangian :
\begin{equation}
\label{eq:LeffCS}
\mathcal{L}^{\text{eff}}_{\rm CS} = \dfrac{1}{96\pi^2} (Q_L^k+Q_R^k)(Q_L^iQ_R^j-Q_L^jQ_R^i) 
\epsilon_{\mu \nu \rho \sigma} A_i^\mu A_j^\nu F_k^{\rho \sigma},
\end{equation}
where summation over all the possible combinations of the 
gauge fields is implied. 

\subsection{Diagrams with axions}

\begin{figure}[h!]
\begin{center}
\includegraphics[width=8.65cm]{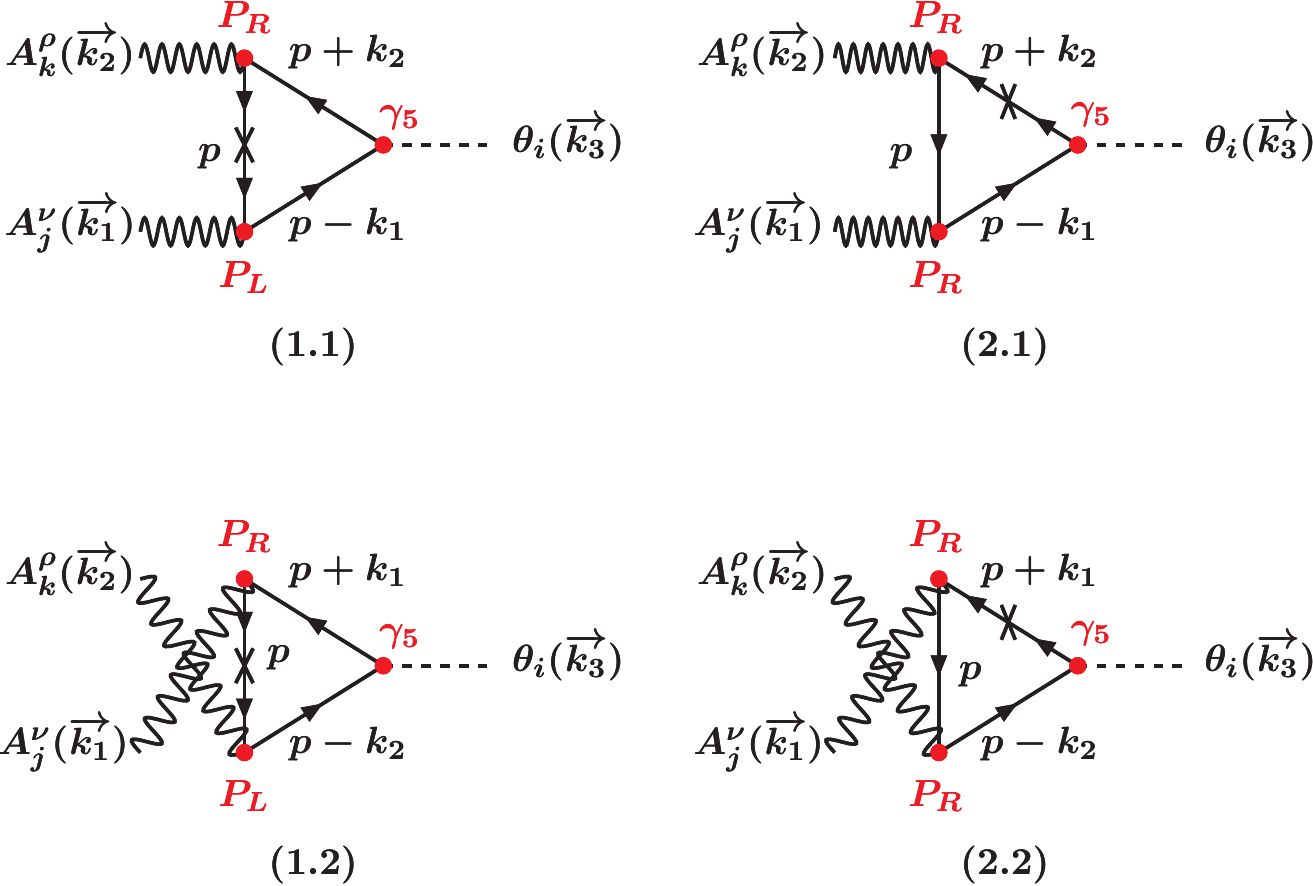}
\includegraphics[width=4cm]{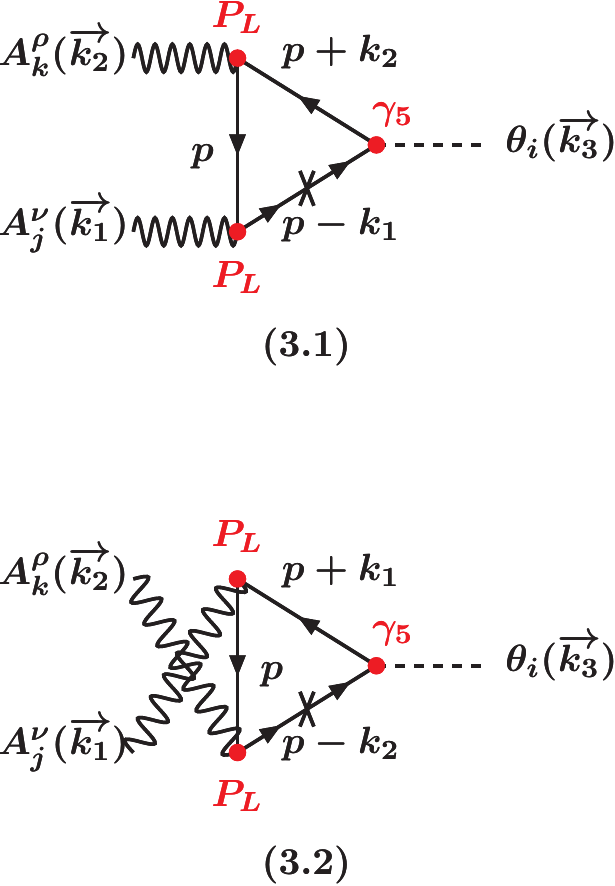}
\caption{Possible set of diagrams for triangular loops of heavy BSM fermions with
an external axion field. Here {\boldmath `$\times$'} denotes
a mass insertion and the arrows represent the direction of fermion flow. Directions 
of the momentum, i.e., $p,\,p\pm k_1,\,p\mp k_2$, etc. are opposite to the fermion
flow.}
\label{fig:CSaxion}
\end{center}
\end{figure}

The diagrams involving an axion field $\theta_i$ include only one mass insertion since vertices with two fermionic legs and an axion field
flips the chirality, as evidenced from the following Lagrangian:
\begin{equation}
\mathcal{L}^{\text{axion}}=-iy_F \theta_i \bar{F}_L  F_R +\text{h.c.}-m\bar{F}F
=-i y_F \theta_i \bar{F}\gamma_5 F -m\bar{F}F,
\end{equation}
where  $y_F$ is the associated Yukawa coupling. For the chosen Lagrangian we have three possible ways to place a mass insertion on the
fermionic propagators and two different ways to connect the external lines with the vertices, giving a total of six diagrams as shown in Fig.~\ref{fig:CSaxion}.
Once again, like the CS case (see Eq.~(\ref{eq:gammafracCS})),
we can write the sum over the
six diagrams without contractions on the external legs as an 
integral over the momentum $p$ as:
\begin{equation}
\int \dfrac{\diff ^4p}{(2\pi)^4}\Gamma^{\nu \rho}(p,k_1,k_2),
\end{equation}
where $\Gamma^{\nu \rho}(p,k_1,k_2)$ is decomposed as
$\bold{\Pi}\cdot\bold{TR}^{\nu \rho}$ with $\bold{\Pi}$
as already defined in Eq.~(\ref{eq:picalc}) and 
the trace factors $\bold{TR}^{\nu \rho}$ written as:
%
\begin{align}
\label{eq:traceano}
\bold{TR}^{\nu \rho}_{1.1}&=y_FQ_L^jQ_R^k  \text{Tr}[\gamma_5(\cancel{p}-\cancel{k}_1)\gamma^\nu 
P_L (\cancel{p} +m) \gamma^\rho P_R (\cancel{p}+\cancel{k}_2) ],\nonumber\\
\bold{TR}^{\nu \rho}_{1.2}&=y_FQ_R^jQ_L^k  \text{Tr}[\gamma_5(\cancel{p}-\cancel{k}_2)\gamma^\rho P_L 
(\cancel{p} +m) \gamma^\nu P_R (\cancel{p}+\cancel{k}_1)],\nonumber\\
\bold{TR}^{\nu \rho}_{2.1}&=y_F Q_R^jQ_R^k  \text{Tr}[\gamma_5(\cancel{p}-\cancel{k}_1)\gamma^\nu P_R 
\,\cancel{p}\,\gamma^\rho P_R (\cancel{p}+\cancel{k}_2+m)],\nonumber\\
\bold{TR}^{\nu \rho}_{2.2}&=y_F Q_R^jQ_R^k  \text{Tr}[\gamma_5(\cancel{p}-\cancel{k}_2)\gamma^\rho P_R 
\,\cancel{p}\,\gamma^\nu P_R (\cancel{p}+\cancel{k}_1+m)],\nonumber\\
\bold{TR}^{\nu \rho}_{3.1}&=y_F Q_L^jQ_L^k  \text{Tr}[\gamma_5
(\cancel{p}-\cancel{k}_1+m) \gamma^\nu P_L \,\cancel{p}\,\gamma^\rho P_L (\cancel{p}+\cancel{k}_2) ],\nonumber\\
\bold{TR}^{\nu \rho}_{3.2}&=y_F Q_L^jQ_L^k  \text{Tr}[\gamma_5 (\cancel{p}-\cancel{k}_2+m)
\gamma^\rho P_L \,\cancel{p}\,\gamma^\nu P_L (\cancel{p}+\cancel{k}_1)],
\end{align}
with proper mass insertion. Now if we consider expansion of $\Gamma^{\nu \rho}(p,k_1,k_2)$ 
in powers of the external momentums $k_1,\,k_2$, just like the already studied CS
scenario, the zeroth and linear order terms of the expansion vanish.
This happens as the former is proportional to $k_i$ while the latter
yields contributions $\propto p$ and thus, disappears after performing $\int d^4 p$
over an odd function. The leading contribution thus, comes from the second order. Considering a 
CP invariant UV complete theory, we keep only the CP-odd contribution in $\Gamma^{\nu \rho}(p,k_1,k_2)$ since axions are CP-odd fields. 
With this approach one can compute the detail expressions for all the six
diagrams of Fig.~\ref{fig:CSaxion}, for example, for the diagram $(3.1)$ one gets:
\begin{align}
\int \dfrac{\diff ^4p}{(2\pi)^4}\Gamma^{\nu \rho}_{(3.1)}(p,k_1,k_2)&= im y_F Q_L^j Q_L^k 
\epsilon^{\mu \nu \rho \sigma}k_{1 \rho} k_{2 \sigma} \int \dfrac{\diff ^4p}{(2\pi)^4} \dfrac{p^2}{(p^2-m^2)^4},\nonumber\\
&=\frac{1}{v_i}\frac{Q_L^jQ_L^k}{48\pi^2} \epsilon^{\nu \rho \alpha \beta} k_{1 \alpha} k_{2 \beta},
\end{align}
where in the last step we have used the known momentum integral as of Ref.~\cite{Peskin:1995ev}
as well as $m=y_F v_i$ with $v_i$ being the VEV of the 
scalar field giving masses to the BSM fermions $F$ and the gauge field $A^\mu_i$. In a similar way the contributions from all the six diagrams of
Fig.~\ref{fig:CSaxion} can be evaluated as tabulated
in Table~\ref{tab:Axioncont}.
\begin{table}[h]
\begin{center}
\begin{tabular}{|c||c|}
\hline 
Diagrams & Contribution to $\Gamma^{\nu \rho}$ \\ 
\hline 
(1.1) & $Q_L^jQ_R^k \epsilon^{\nu \rho \alpha \beta} k_{1 \alpha} k_{2 \beta}/(48\pi^2v_i)$ \\ 
\hline
(2.1) & $Q_R^jQ_R^k \epsilon^{\nu \rho \alpha \beta} k_{1 \alpha} k_{2 \beta}/(48\pi^2v_i)$ \\ 
\hline
(3.1) &  $Q_L^jQ_L^k \epsilon^{\nu \rho \alpha \beta} k_{1 \alpha} k_{2 \beta}/(48\pi^2v_i)$ \\ 
\hline
(1.2) &  $Q_R^jQ_L^k \epsilon^{\nu \rho \alpha \beta} k_{1 \alpha} k_{2 \beta}/(48\pi^2v_i)$ \\
\hline
(2.2) &  $Q_R^jQ_R^k \epsilon^{\nu \rho \alpha \beta} k_{1 \alpha} k_{2 \beta}/(48\pi^2v_i)$ \\
\hline
(3.2) &  $Q_L^jQ_L^k \epsilon^{\nu \rho \alpha \beta} k_{1 \alpha} k_{2 \beta}/(48\pi^2v_i)$ \\  
\hline
\end{tabular} 
\end{center}
\caption{Final contributions of the six diagrams of figure
\ref{fig:CSaxion}.}
\label{tab:Axioncont} 
\end{table}
Summing all the six contributions from Table~\ref{tab:Axioncont} yields :
\begin{equation*}
\int \dfrac{\diff^4p}{(2\pi)^4}\Gamma^{ \nu \rho} = \dfrac{1}{v_i} \dfrac{1}{48\pi^2} 
\epsilon^{\nu \rho \alpha \beta} k_{1 \alpha} k_{2 \beta} [2(Q_L^jQ_L^k+Q_R^jQ_R^k )+Q_R^jQ_L^k +Q_L^jQ_R^k ],
\end{equation*}
which finally produce the following effective Lagrangian:
\begin{equation}
\label{eq:Leffaxion}
\mathcal{L}^{\text{eff}}_{\rm axion} = \dfrac{1}{192\pi^2}[2(Q_L^jQ_L^k+Q_R^jQ_R^k )+Q_R^jQ_L^k +Q_L^jQ_R^k ] 
\epsilon_{\mu \nu \rho \sigma} \frac{\theta_i}{v_i} F_j^{\mu \nu} F_k^{\rho \sigma},
\end{equation}
where, once again summation over all the possible combinations of the 
gauge fields is implied.

\subsection{Anomaly cancellation}
In the last two subsections we discussed about the three different anomalous
contributions, namely, (1) diagrams without any chirality flip which
are linearly divergent and giving contributions proportional to the anomaly
traces. (2) Diagrams with one chirality flip involving an axion field
that are finite and, (3) the so-called "CS" contributions
which are finite and invoke two chirality flips. In this subsection
we show that gauge transformation of the effective CS
and axion Lagrangians (see Eq.~(\ref{eq:LeffCS}) and Eq.~(\ref{eq:Leffaxion}))
is proportional to the ``usual'' anomaly trace. Hence, a vanishing
anomaly trace, with appropriate distribution of the charges of BSM fermions,
assures an anomaly free theory construction. Given the following
gauge transformations of an axion field $\theta_i$ and a gauge field
$A^\mu_i$: 
\beq
\label{eq:CSaxionGT}
\theta_i  \rightarrow \theta_i + v_i (Q_L^i-Q_R^i) \alpha_i, \,\,\,\,
A^\mu_i  \rightarrow A^\mu_i + \partial^\mu \alpha_i,
\eeq
where $\alpha_i$ is the parameter of gauge transformation,
the variation of the effective CS Lagrangian
(see Eq.~(\ref{eq:LeffCS})) becomes:
\begin{equation}
\label{eq:LCSvarn}
\delta\mathcal{L}^{\text{CS}} =  -\dfrac{1}{192\pi^2} \Big[(Q_L^k+Q_R^k)(Q_L^iQ_R^j-Q_L^jQ_R^i)
+(Q_L^j+Q_R^j)(Q_L^iQ_R^k-Q_L^kQ_R^i)\Big] \epsilon_{\mu \nu \rho \sigma} \alpha_i F_j^{\mu \nu}  F_k^{\rho \sigma},
\end{equation}
where we have used the advantages of integrating by parts as well
as Bianchi identity and included all possible combinations of the $i,\,j,\,k$ indices. The change in effective axion Lagrangian (see Eq.~(\ref{eq:Leffaxion})) is given by:
\begin{equation}
\label{eq:Laxionvarn}
\delta \mathcal{L}^{\text{axion}} = \dfrac{1}{192\pi^2}[2(Q_L^jQ_L^k+Q_R^jQ_R^k )+Q_R^jQ_L^k 
+Q_L^jQ_R^k ]  (Q_L^i-Q_R^i) \alpha_i \epsilon_{\mu \nu \rho \sigma} F_j^{\mu \nu} F_k^{\rho \sigma}.
\end{equation}

Combining Eq.~(\ref{eq:LCSvarn}) and Eq.~(\ref{eq:Laxionvarn})
the resultant variation, given the transformations of 
Eq.~(\ref{eq:CSaxionGT}), is written as:
\begin{equation}
\label{eq:LCSaxiontotvarn}
\delta \mathcal{L}  = \delta \mathcal{L}^{\text{axion}} + \delta \mathcal{L}^{\text{CS}} = 
\dfrac{1}{96\pi^2} \alpha_i \Big[Q_L^iQ_L^jQ_L^k-Q_R^iQ_R^jQ_R^k \Big] 
\epsilon_{\mu \nu \rho \sigma} F_j^{\mu \nu} F_k^{\rho \sigma}.
\end{equation}
It is now apparent from Eq. (\ref{eq:LCSaxiontotvarn}) that
gauge transformation of the total Lagrangian, i.e., axion and
CS Lagrangians is
proportional to the ``usual'' anomaly trace $Q_L^iQ_L^jQ_L^k-Q_R^iQ_R^jQ_R^k $. 
Hence, with proper choice of the charges for the heavy fermions
one can ensure an anomaly free theory setup where the anomaly trace
$Q_L^iQ_L^jQ_L^k-Q_R^iQ_R^jQ_R^k $ vanishes for all $i,j,k$.
Further, when this anomaly trace disappears with proper choice of $Q^i_L,\,Q^i_R,\,Q^j_L,\,Q^j_R,\,Q^k_L$ and $Q^k_R$,
the combination of the axion and CS effective Lagrangians
(i.e., Eq.~(\ref{eq:LeffCS}) + Eq.~(\ref{eq:Leffaxion})) can be 
embedded into a dimension-six operator as:
\begin{equation}
\epsilon_{\mu \nu \rho \sigma} D^\mu \theta_i D^\nu \theta_j F_k^{\rho \sigma},
\end{equation}
where $D^\mu,\,D^\nu$ are co-variant derivatives for 
axion fields $\theta_i,\,\theta_j$, respectively. One can always
consider the case of unitary gauge when the axion Lagrangian
(see Eq.~(\ref{eq:Leffaxion})) vanishes and the total Lagrangian
is simply the CS one, as given by Eq.~(\ref{eq:LeffCS}).
This is the scenario which we studied in this work. Recasting
Eq.~(\ref{eq:LeffCS}) for the specific case of $U(1)_X\times U(1)_V$,
as considered in this work, one can generate
%
\begin{equation}
\mathcal{L} = \mathcal{L}^{\rm eff}_{\rm CS} = \dfrac{1}{48\pi^2} (Q_L^X+Q_R^X)(Q_L^XQ_R^V-Q_L^VQ_R^X) 
\epsilon_{\mu \nu \rho \sigma} X^\mu \wt V^{\nu} X^{\rho \sigma} \equiv 
\alpha_{\text{CS}} \epsilon_{\mu \nu \rho \sigma} X^\mu \wt V^{\nu}  X^{\rho \sigma},
\end{equation}
where terms $\propto \epsilon_{\mu \nu \rho \sigma} \wt V^\mu X^{\nu}  \wt V^{\rho \sigma}$
is effaced with suitable choice of associated charges and the parameter
$\alpha_{\rm CS}$ is given by
\begin{equation}
\label{eq:effctivealphaCS}
\alpha_{\text{CS}} \equiv \dfrac{1}{48\pi^2} (Q_L^X+Q_R^X)(Q_L^XQ_R^V-Q_L^VQ_R^X),
\end{equation}
which we have already used to derive Eq.~(\ref{eq:alphaCSwork})
for the charges given in Table~\ref{tab:charges}.

\section{Generation of non-abelian Chern-Simons couplings}
\label{sec:CSnonabelianappendix}

In this section, we discuss the origin of the generalized CS terms in a concrete VSIMP model, considered in Sec.~\ref{ssec:Zprimeportal}, with dark fermions for a UV completion.
Furthermore, we show that the effective CS terms and the general $Z'-X-X$ interactions can be derived from manifestly gauge invariant operators at low energy. Suppose that there is a set of light fermions charged under $SU(2)_X\times U(1)_{Z'}$ such as
\beq
l=(2,+1), \quad {\tilde l}=(2,+1), \quad e^c=(1,-1),  \quad {\tilde e}^c=(1,-1).
\eeq
along with a heavy dark fermions with opposite $U(1)_{Z'}$ charges ($L,\tilde{L},E^c, \tilde{E}^c$) that cancel the anomalies. With dark Higgs fields of charges $\Phi=(2,0)$ and $S=(1,-2)$ , then $SU(2)$ vector-like and chiral masses from terms
\beq
S \,l \,{\tilde l}+ S^* e^c\,{\tilde e}^c +
\Phi\, l \, e^c+ {\tilde \Phi} \,{\tilde l}\, {\tilde e}^c
\eeq
where ${\tilde \Phi}=i\tau^2\Phi^*$,  are generated after $SU(2)_X\times U(1)_{Z'}$  spontaneous symmetry breaking\footnote{The $SU(2)_X$ gauge bosons masses are degenerate at tree level and receive small loop corrections due to the mass splitting between the members of each doublet fermion, that is proportional to chiral fermion mass.  If the mass splitting between $SU(2)_X$ gauge bosons is smaller than $10\%$ of DM mass, all the SIMP processes are still active and dominant and the vector dark matter remains stable for heavy fermions. One can check explicitly in the example with vector-like dark fermions that there is no $X_3-Z'$ mixing generated at loop level, so there is no issue of Dark Matter instability. }.

When integrating out the light fermions, the non-decoupling portion of the one-loop triangle diagrams gives an effective CS term
\beq
{\cal L}_{\rm CS,EFT} = \frac{N_f g_{Z'}\alpha_X}{4\pi} \frac{m^2_X}{m^2_f}  \epsilon^{\mu\nu\rho\sigma} Z'_\mu {\vec X}_\nu\cdot (\partial_\rho {\vec X}_\sigma - \partial_\sigma {\vec X}_\rho) ,
\eeq
where  $N_f$ being the number and mass of light fermion generations of mass, $m_f$.

For instance, for $\alpha_X=1(4)$, $N
_f=4(1)$, $g_{Z'}\sim 0.3-3$, and $m_f\sim 4m_X-10 m_X$, we find the coefficient of the operator of Eq.~\eqref{CS1}, $c_1\simeq 0.01$.
Therefore, for $m_f \gtrsim m_X$, we can avoid additional $2\rightarrow 2$ annihilations of vector Dark Matter into light dark fermions, such as $XX\rightarrow f {\bar f}$, and a sizable CS term required for kinetic equilibrium can be consistently realized.

Notice here that the values of $c_1 \gtrsim 10^{-2}$ required for achieving the correct relic density in this setup imply a large multiplicity of the dark fermions or sizable gauge couplings which might drive the theory toward its non perturbative regime or the unstability of the dark higgs potential vacuum for energies of the order of the GeV scale. One could invoke more elaborate mechanisms in order to solve this potential issues but those are beyond the phenomenological considerations of this work.

If one considers only the light fermions
$l=(2,+1), e^c=(1,-1)$ and their heavy partners for anomaly cancellation, then are only chiral fermion masses due to the $SU(2)_X$ breaking. In this case, the needed CS terms are not generated. Instead, a nonzero dimension-6 interaction \bea
{\cal L}_{\rm D6}=\frac{c_3}{M^2}\, \epsilon^{\mu\mu\rho\sigma} \partial^\lambda Z'_{\mu\nu} (X_{1,\rho\sigma}X_{2,\lambda}-X_{2,\rho\sigma}X_{1,\lambda}). \label{d6}
\eea
appears, which can also be sufficient for equilibrating the two sectors

Similarly, the effective dimension-6 operator in Eq.~(\ref{d6}) can be derived from another gauge invariant dimension-8 operator,
\bea
{\cal L}_{\rm D8}= \frac{1}{M^4} \, \epsilon^{\mu\nu\rho\sigma} (\Phi^\dagger X_{\mu\nu} D_\lambda \Phi) \partial^\lambda Z'_{\rho\sigma}.
\eea
Then, in both cases with dimension-6 and dimension-8 operators, after the $SU(2)_X$ is broken by the VEV of the scalar doublet $\Phi$, the needed $Z'XX$ interactions are generated.

The effective approach considered in this work would be valid only for processes involving energies below the lightest dark fermion mass. Our approach is then justified for the DM freeze-out process which occurs when the DM becomes non-relativistic (i.e. for processes occuring at energies $\sim m_X \ll m_f$) and the dark fermions have already decoupled for the thermal bath. However, considering the invisible decay of the $Z$ boson leads the effective approach to fail and one has to consider the complete dark fermions degrees of freedom in the computation.

%% file: parts/appendices/spin2appendix.tex
\label{sec:appendixspin2}
\vspace{0.3cm}

\noindent
In this part of the appendices, we provide analytical expressions of the amplitude squared, partial decay widths as well as rates relevant for the spin-2 portal model.

\section{Decay rate of the massive spin-2}
The decay modes of the spin-2 state are
\begin{equation}
\Gamma_{\tilde h \rightarrow \varphi \varphi} = N_\varphi \frac{g^2_\varphi}{960 \pi}\frac{\mh^3}{\Lambda^2} (1-4r_\varphi)^{5/2}~,
\end{equation}
\begin{equation}
\Gamma_{\tilde h \rightarrow \psi \psi} = N_\psi \frac{g^2_\psi}{160 \pi}\frac{\mh^3}{\Lambda^2} \left(1+\frac{8}{3}r_\psi\right) (1-4r_\psi)^{3/2}~,
\end{equation}
and
\begin{equation}
\Gamma_{\tilde h \rightarrow V V} = N_V \frac{g^2_V}{960 \pi}\frac{\mh^3}{\Lambda^2}(13+56r_V + 48 r_V^2) (1-4r_V)^{1/2}~,
\end{equation}
where $r_i \equiv m_i^2/\mh^2$. The total decay width of the massive spin-2 to SM state is given by:

\begin{equation}
\Gamma_{\tilde h \rightarrow \text{SM}}=  4\Gamma_{\tilde h \rightarrow \varphi \varphi}+45\Gamma_{\tilde h \rightarrow \psi \psi}+12\Gamma_{\tilde h \rightarrow V V} = \dfrac{43g_{\rm SM}^2 m_{\tilde h}^3}{96 \pi \Lambda^2}~.
\label{Eq:width}
\end{equation}

\section{Amplitudes and production rates}

\subsection{Scalar Dark Matter}

The amplitudes involving a scalar Dark Matter and massless spin-2 mediator (graviton) are:
\bea
&&
|{\mathcal M}^0|^2_{h_{\mu \nu}} = \frac{1}{16 M_P^4} \frac{t^2(s+t)^2}{s^2}~,
\\
&&
|{\mathcal M}^{1/2}|^2_{h_{\mu \nu}} = \frac{1}{32 M_P^4} \frac{(-t(s+t))(s+2t)^2}{s^2}~,
\\
&&
|{\mathcal M}^{1}|^2_{h_{\mu \nu}} = \frac{1}{8 M_P^4} \frac{t^2(s+t)^2}{s^2} ~.
\eea
The corresponding amplitudes assuming a massive spin-2 propagator are:
\bea
&&
|{\mathcal M}^0|^2_{\tilde h_{\mu \nu}} = \frac{g_{\rm DM}^2 g_{\rm SM}^2}{36 \Lambda^4}\frac{\left[6t(s+t)+s^2\right]^2}{(s-m_{\tilde h}^2)^2 +  \Gamma_{\tilde h}^2m_{\tilde h}^2}~,
\\
&&
|{\mathcal M}^{1/2}|^2_{\tilde{h}_{\mu\nu}} = \frac{g_{\rm DM}^2 g_{\rm SM}^2}{2 \Lambda^4} \frac{(-t(s+t))(s+2t)^2}{(s-\mh^2)^2+\mh^2\Gh^2}~,
\\
&&
 |{\mathcal M}^1|^2_{\tilde{h}_{\mu\nu}} = \frac{2 g_{\rm DM}^2 g_{\rm SM}^2}{\Lambda^4} \frac{t^2(s+t)^2}{(s-\mh^2)^2+\mh^2\Gh^2} ~.
\eea
From these amplitudes, the rate in the graviton domination regime is given by:
\bea
&&
R^{0}_{h_{\mu \nu}}(T) = \frac{3997\pi^3}{663552000} \frac{T^8}{M^4_P} \equiv \alpha \frac{T^8}{M^4_P} ~.
\label{R0a}
\eea
In the case where the massive spin-2 state is exchanged, we can distinguish three regimes, depending on the relative value of $\mh$ with respect to typical exchanged momentum $\sim T$
\begin{equation}\begin{split}
& R^0_{\tilde h_{\mu \nu}}\Big|_{\mh\ll T}= \frac{g_{\rm DM}^2 g_{\rm SM}^2 11351 \pi^3}{124416000}\frac{T^8}{\Lambda^4} \equiv \beta_1  \frac{T^8}{\Lambda^4}~,
\label{R0b1}
\end{split}\end{equation}
\begin{equation}\begin{split}
& R^0_{\tilde h_{\mu \nu}}\Big|_{\mh \sim T} = \frac{g_{\rm DM}^2 g_{\rm SM}^2 209}{184 320 \pi^4}\frac{\mh^8}{\Lambda^4} \frac{T}{\Gamma_{\tilde h}} K_1\Big(\frac{\mh}{T}\Big) \equiv g_{\rm DM}^2 g_{\rm SM}^2 \beta_2 \frac{\mh^8}{\Lambda^4} \frac{T}{\Gamma_{\tilde h}} K_1\Big(\frac{\mh}{T}\Big) \\
& \hspace{1.8cm} = \frac{160\pi \beta_2}{(g_{\rm DM}^2/6 + 215 g_{\rm SM}^2/3)} \frac{\mh^5 T}{\Lambda^2}K_1\Big(\frac{\mh}{T}\Big)~,
\label{R0b2}
\end{split}\end{equation}
\begin{equation}
R^0_{\tilde h_{\mu \nu}}\Big|_{\mh\gg T}= \frac{g_{\rm DM}^2 g_{\rm SM}^2 205 511 \pi^7}{57153600}\frac{T^{12}}{\Lambda^4 \mh^4} \equiv g_{\rm DM}^2 g_{\rm SM}^2 \beta_3 \frac{T^{12}}{\Lambda^4 \mh^4}~.
\label{R0b3}
\end{equation}
In the last line of Eq.~(\ref{R0b2}), we expressed the width $\Gamma_{\tilde h}$ using Eq.~(\ref{Eq:width}). The values of 
$\alpha$, $\beta_1, \beta_2$ and $\beta_3$ are collected in Table~\ref{tab:re} which also includes the corresponding coefficients for fermionic and vectorial Dark Matter.

\subsection{Fermionic Dark Matter}

\begin{equation}
|{\mathcal M}^0|^2_{h_{\mu\nu}} = \frac{(-t(s+t))(s+2t)^2}{32 M_P^4 s^2} ~,
\end{equation}

\begin{equation}
|{\mathcal M}^{1/2}|^2_{h_{\mu\nu}} = \frac{s^4+10s^3t+42s^2t^2+64st^3+32t^4}{128 M_P^4 s^2} ~,
\end{equation}

\begin{equation}
|{\mathcal M}^1|^2_{h_{\mu\nu}} = \frac{(-t(s+t))(s^2+2t(s+t))}{8 M_P^4 s^2}~,
\end{equation}

\begin{equation}
|{\mathcal M}^0|^2_{\tilde{h}_{\mu\nu}} = \frac{g_{\rm DM}^2 g_{\rm SM}^2}{2 \Lambda^4} \frac{(-t(s+t))(s+2t)^2}{(s-\mh^2)^2+\mh^2\Gh^2}~,
\end{equation}

\begin{equation}
|{\mathcal M}^{1/2}|^2_{\tilde{h}_{\mu\nu}} = \frac{g_{\rm DM}^2 g_{\rm SM}^2}{8 \Lambda^4} \frac{s^4+10s^3t+42s^2t^2+64st^3+32t^4}{(s-\mh^2)^2+\mh^2\Gh^2}~,
\end{equation}

\begin{equation}
|{\mathcal M}^1|^2_{\tilde{h}_{\mu\nu}} = \frac{2g_{\rm DM}^2 g_{\rm SM}^2}{\Lambda^4} \frac{(-t(s+t))(s^2+2t(s+t))}{(s-\mh^2)^2+\mh^2\Gh^2}~.
\end{equation}
The corresponding rates are:
\begin{equation}\begin{split}
& R^{1/2}_{ h_{\mu \nu}} = \frac{11351 \pi ^3}{331776000 } \frac{T^8}{M_P^4}~,
\end{split}
\end{equation}

\begin{equation}
R^{1/2}_{\tilde h_{\mu \nu}}\Big|_{\mh\ll T}= \frac{g_{\rm DM}^2 g_{\rm SM}^2 11351 \pi^3}{20736000}\frac{T^8}{\Lambda^4}~,
\end{equation}

\begin{equation}
\begin{split}
& R^{1/2}_{\tilde h_{\mu \nu}}\Big|_{\mh \sim T} = \frac{g_{\rm DM}^2 g_{\rm SM}^2 209}{30720 \pi^4}\frac{\mh^8}{\Lambda^4} \frac{T}{\Gamma_{\tilde h}} K_1\Big(\frac{\mh}{T}\Big)~,
\end{split}
\end{equation}

\begin{equation}
R^{1/2}_{\tilde h_{\mu \nu}}\Big|_{\mh\gg T}= \frac{g_{\rm DM}^2 g_{\rm SM}^2 205 511 \pi^7}{9525600}\frac{T^{12}}{\Lambda^4 \mh^4}~.
\end{equation}

\subsection{Vectorial Dark Matter}

\begin{equation}
|{\mathcal M}^0|^2_{h_{\mu\nu}} = \frac{3 t^2(s+t)^2}{16 M_P^4 s^2}~,
\end{equation}

\begin{equation}
|{\mathcal M}^{1/2}|^2_{h_{\mu\nu}} = \frac{(-t(s+t))(5s^2+12t(s+t))}{32 M_P^4 s^2} ~,
\end{equation}

\begin{equation}
|{\mathcal M}^1|^2_{h_{\mu\nu}} = \frac{(s^2+t(s+t))(s^2+3t(s+t))}{8 M_P^4 s^2}~,
\end{equation}

\begin{equation}
|{\mathcal M}^0|^2_{\tilde{h}_{\mu\nu}} = \frac{g_{\rm DM}^2 g_{\rm SM}^2}{36 \Lambda^4} \frac{s^4+12s^3t+120s^2t^2+216st^3+108t^4}{(s-\mh^2)^2+\mh^2\Gh^2}~,
\end{equation}

\begin{equation}
|{\mathcal M}^{1/2}|^2_{\tilde{h}_{\mu\nu}} = \frac{g_{\rm DM}^2 g_{\rm SM}^2}{2 \Lambda^4} \frac{(-t(s+t))(5s^2+12t(s+t))}{(s-\mh^2)^2+\mh^2\Gh^2}~,
\end{equation}

\begin{equation}
|{\mathcal M}^1|^2_{\tilde{h}_{\mu\nu}} = \frac{2g_{\rm DM}^2 g_{\rm SM}^2}{\Lambda^4} \frac{(s^2+t(s+t))(s^2+3t(s+t))}{(s-\mh^2)^2+\mh^2\Gh^2}~.
\end{equation}
The corresponding rates are:
\begin{equation}
\begin{split}
& R^{1}_{ h_{\mu \nu}} = \frac{5489 \pi ^3}{73728000} \frac{T^8}{M_P^4} 
\end{split}~,
\end{equation}

\begin{equation}
R^{1}_{\tilde h_{\mu \nu}}\Big|_{\mh\ll T}= \frac{g_{\rm DM}^2 g_{\rm SM}^2 147563 \pi^3}{124416000}\frac{T^8}{\Lambda^4}~,
\end{equation}

\begin{equation}
\begin{split}
& R^{1}_{\tilde h_{\mu \nu}}\Big|_{\mh \sim T} = \frac{g_{\rm DM}^2 g_{\rm SM}^2 2717}{184320 \pi^4}\frac{\mh^8}{\Lambda^4} \frac{T}{\Gamma_{\tilde h}} K_1\Big(\frac{\mh}{T}\Big)~,
\end{split}
\end{equation}

\begin{equation}
R^{1}_{\tilde h_{\mu \nu}}\Big|_{\mh\gg T}= \frac{g_{\rm DM}^2 g_{\rm SM}^2 2671643 \pi^7}{57153600}\frac{T^{12}}{\Lambda^4 \mh^4}~.
\end{equation}

\begin{table}[ht]
\begin{center}
\begin{tabular}{|l|c|c|c|c|}
\hline
 Spin & $\alpha$ & $\beta_1$ & $\beta_2$ & $\beta_3$   \\
\hline
0 & $1.9 \times 10^{-4}$ & $2.8 \times 10^{-3}$ & $1.2 \times 10^{-5}$ & 10.9 \\
\hline
1/2 & $1.1 \times 10^{-3}$ & $1.7 \times 10^{-2}$ & $7.0 \times 10^{-5}$ & 65.2 \\
\hline
1 & $2.3 \times 10^{-3}$ & $3.7 \times 10^{-2}$ & $1.5 \times 10^{-4}$ & 141 \\
\hline
\end{tabular}
\caption{Numerical values of the coupling coefficients depending on the Dark Matter spin.
\label{tab:re}
}
\end{center}
\end{table}